# Determination of Mass Loss and Mass Transfer Rates of Algol (Beta Persei) from the Analysis of Absorption Lines in the UV Spectra Obtained by the IUE Satellite

by

Kristen Wecht

Presented to the Graduate and Research Committee
of Lehigh University
in Candidacy for the Degree of
Doctor of Philosophy

in

Physics

Lehigh University

April 2006

# CERTIFICATE OF APPROVAL

Approved and recommended for acceptance as a dissertation to partial fulfillment of the requirements for the degree of Doctor of Physics.

______________________
Date

______________________                                    ______________________________
Accepted Date                                                            George E. McCluskey, Jr. (Advisor)

                              Special Committee Directing the Doctoral Work of Kristen Wecht

                                                           ______________________________
                                                           Gary G. DeLeo

                                                           ______________________________
                                                           A. Peet Hickman

                                                           ______________________________
                                                           John P. Huennekens

                                                            ______________________________
                                                           Daniel Zeroka

# CONTENTS

# LIST OF TABLES

# LIST OF FIGURES

# ABSTRACT

The International Ultraviolet Explorer (IUE) archive of high-resolution ultraviolet spectra of the eclipsing semi-detached binary star, Algol (β Persei, HD 19356), taken from September 1978 to September 1989, is analyzed in order to characterize the movement of gas within and from this system. Light curves are constructed, using a total of 1647 continuum level measurements. These results support the semidetached status of this interacting binary star. Radial velocities, residual intensities, full width half maxima (FWHM), line asymmetries, and equivalent widths of UV absorption lines for aluminum, magnesium, iron, and silicon in a range of ionization states are determined and analyzed. For selected epochs, we were able to isolate gas stream and photospheric contributions by an examination of the differences between spectral line shapes.

We observed variations in line shape and strength, with orbital phase and epoch, indicating the presence of stable gas streams and circumstellar gas, and periods of increased mass-transfer activity associated with transient gas streams. The 1989 data indicates moderate activity. This epoch was examined most closely since it provides the greatest phase coverage. Spectral line profiles in 1978 and 1984 have the strongest gas-flow absorption components, indicating that these are the epochs of the greatest activity. The dense phase coverage in September 1989 allows us to measure the mass loss rate from Algol B into Algol A which is of order ~$10^{-14}$ $M_\odot$/yr. Since the highest gas-flow velocities are in the 100 kilometer per second range, well below escape velocity, we conclude that systemic mass loss due to gas flow is small for the Algol system.

# 1. INTRODUCTION

A grain of sand held at arm's length covers a patch of sky that holds 1,000 galaxies (StarGaze DVD 2000). When multiplied across the entire sky, this means the universe contains at least 120 billion galaxies, and each galaxy hosts an average of 100 billion stars.

Our current understanding of stars derives almost entirely from the detection and examination of electromagnetic radiation. The physical and evolutionary properties of stars are revealed when these spectroscopic studies are combined with theoretical models based on applications of fundamental physical laws. In particular, it was in large part the combination of the classification scheme of Annie Jump Cannon, the quantification of temperature-luminosity relationships by Henry Norris Russell, and the development of quantum mechanics that led to our current understanding. We now recognize that nearly all stars are composed mostly of hydrogen and helium, and that differences in color and brightness are related to both mass and evolutionary stage, but ultimately to mass.

It is estimated that 40-60% of the 100 billion stars in each of the 120 billion galaxies are actually gravitationally bound collections of two or more stars – *binary stars* or *multiple stars*. This fact of nature is essential to our understanding of all stars since stellar masses can be directly determined only by observing the motions of binary star components under their mutual gravitational interactions. Of course, once the relationship between mass and other observed properties is established, this scheme can then be used to estimate the masses of stars that are not components of binary systems.

In some cases, the separations between the binary-star components are so small that the stars experience mutual structural disturbances. These interactions lead to some very interesting phenomena not found in systems with greater separations, including orbital perturbations, stellar distortions and mass flow. The characteristics of such *interacting binaries* can be revealed by spectroscopic studies in a variety of wavelength ranges.

In this work we study the ultraviolet light from the variable star Algol, the second brightest star in the constellation of Perseus. This star, β Persei, is actually a multiple star system. Two of its component stars are so close together that there is movement of gas from one star to the other. The orientation of the orbit is such that one star partially eclipses the other every 2.9 days, resulting in the observable variability of total light from the system with the naked eye.

Ultraviolet observations of Algol enable us to study in more detail the hot circumstellar gas moving within and from the system. Instrumentation aboard the International Ultraviolet Explorer (IUE) spacecraft recorded 103 exposures in UV light from Algol between September 1978 and September 1989. Each exposure corresponds to a specific point in time and related configuration of the relative positions of the component stars with respect to our line of sight from Earth. We see the manifestation of mass flow in the UV light variations across these exposures.

A detailed analysis of the ultraviolet spectral lines reveals noticeable variations in spectral features with orbital phase and time frame (epoch) that correspond to the properties of mass flow within and from the Algol system. From the differences in spectral line widths at different phases with and without the gas stream, we estimate the number of atoms in the gas stream. From the shifts in wavelength from the rest wavelength, we determine the radial velocities as functions of time for each spectral line. From these quantities we estimate the mass loss rate and infer other physical characteristics of this interacting binary system.

# 2. BACKGROUND

## 2.1 *Stellar Properties, Evolution, and Classification*

In order to fully appreciate the strange phenomenon known as the Algol Paradox, we first examine our current understanding of the properties and evolution of non-interacting stars. These include isolated (single) stars and components of multiple-star systems that are well separated.

The classification of stars according to their surface temperature and luminosity is represented by the Hertzsprung-Russell (HR) diagram, FIG. 2.1.1. The physical properties of a star are correlated with position on the diagram. The horizontal axis is labeled in order of decreasing surface temperature (Kelvin) with the letters O, B, A, F, G, K, and M (subtypes 0 and 5 are labeled as space permits). The vertical axis indicates increasing luminosity (in fractions of solar luminosity $L_\odot$*). Fainter stars appear near the bottom, brighter ones near the top. *Diagonal lines* indicate approximate radii in units of solar diameters ($2R_\odot$). Broad grey curves indicate the locations of the luminosity classes Ia (Very luminous supergiants), Ib (Less luminous supergiants), II (Luminous giants), III (Giants), IV (Subgiants), V (Main sequence stars or dwarf stars). The position of the Sun indicates its identification as a G2 V star.

---

* $L_\odot = 3.826 \times 10^{33}$ ergs/s

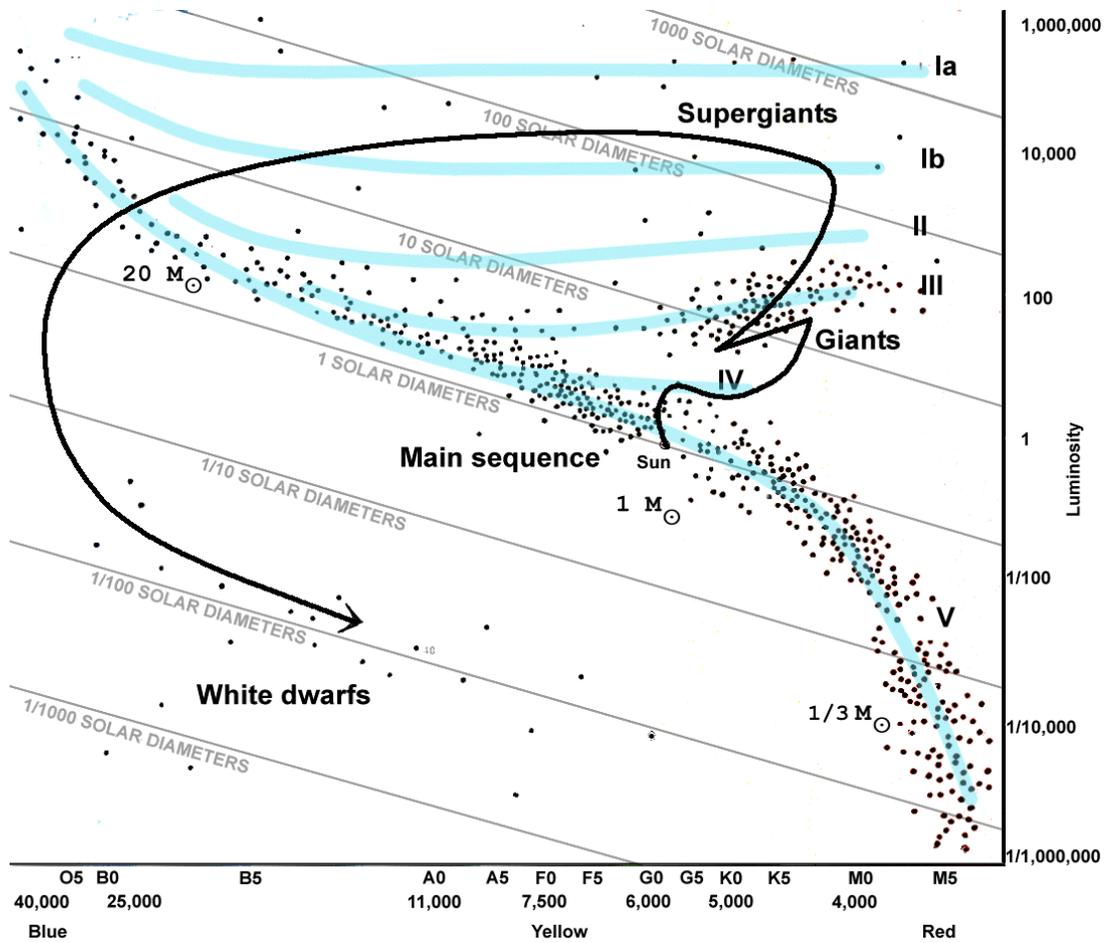

FIG. 2.1.1–*Hertzsprung-Russell Diagram (HRD)*. Adapted from a figure in "The Universe and Beyond, 3rd Ed., T. Dickinson, Firefly, 1999. Also shown is the evolutionary path of the sun.

The broad diagonal band, the main sequence, includes more than 90% of all stars (including our Sun) and associates increasing luminosity with increasing surface temperature. For stars on the main sequence, both temperature and luminosity increase monotonically with increasing mass. The most massive stars, such as O and B stars, are very hot and very bright. The low mass M stars are cooler and dimmer.

Off the main sequence, however, surface area, or size, rather than temperature, is the stronger influence on luminosity. Some red giants and supergiants are only half as hot as the Sun, but they can be a hundred to a million times more luminous. Planet-size white dwarfs, although generally hotter than the Sun, are much dimmer. Luminous but cool giants fall in the upper right, hot but dim white dwarfs in the lower left.

A star's evolution can be described by its change of position on the diagram, as shown for our sun by the solid line in FIG. 2.1.1. A star of solar mass will spend roughly 10 billion years on the main sequence and an additional 2 billion years along the various stages of becoming a white dwarf. Stellar masses also diminish as the stars evolve beyond the main sequence. For example, by the time the sun completes its evolution, its mass will have diminished to about 2/3 of its current value.

An important element of our understanding of stellar evolution is that the lifetime of a star varies inversely with its main-sequence mass. For example, even with its great mass, an O 5 star burns its fuel so aggressively that it has a main sequence lifetime of only about one million years. Yet a typical M star has a theoretical lifetime of 500 billion years. (see TABLE 2.1.1)

TABLE 2.1.1
Role of Mass of a Star

| Mass | Temperature | Spectral Type | Lifetime (yrs) |
|------|-------------|---------------|----------------|
| 40   | 40,000      | O             | 1 million      |
| 7    | 15,000      | B             | 80 million     |
| 2    | 8,200       | A             | 2 billion      |
| 1.3  | 6,600       | F             | 5 billion      |
| 1.0  | 5,800       | G             | 10 billion     |
| 0.8  | 4,300       | K             | 20 billion     |
| 0.2  | 3,300       | M             | 500 billion    |

Although FIG. 2.1.1 only shows the evolutionary pathway for a solar-mass star, these pathways and the corresponding elapsed times are known for all main sequence masses. However, these are appropriate to isolated or non-interacting stars. As we shall see, these evolutionary trajectories can be significantly altered by a close binary partner.

2.2 *Binary Star Characteristics and Classification*

The stars that compose binary star systems have a variety of separations, from one to two times the radius of our Sun,* to many times the distance between Sun and Pluto.† The components of an interacting binary are so close together that the evolutionary history of each of the two components at some stage begins to depart appreciably from the evolution of single (isolated) stars (Sahade and Wood 1978).

Systems that contain stars with small separations have shorter orbital periods and are difficult or impossible to resolve using current methods. The more widely separated the component, the longer the period. Orbital periods can range from minutes to thousands of years.

FIG. 2.2.1 is a schematic binary system showing the elliptical orbits of components $S_1$ (the primary star) and $S_2$ (the secondary star) about their center of mass (X) with semimajor axes $a_1$ and $a_2$, and masses $m_1$ and $m_2$. The eccentricity and period of both orbits are the same. Some systems have eccentric orbits, others are circular. Binary systems where the component stars are close together tend to have circularized orbits due to tidal interactions between the two stars.

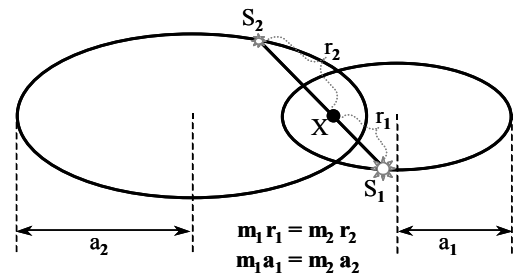

FIG. 2.2.1–*Schematic binary system*

---

* Radius of the Sun: $R_\odot = 6.9599 \times 10^{10}$ cm
† Distance between Sun and Pluto: 39.4821 AU, AU = $1.4960 \times 10^{13}$ cm, the average Earth-Sun distance.

Seven orbital elements "...define the form and size of the true orbit, the position of the orbit plane, the position of the orbit within that plane, and the position of the companion star in the orbit at any specified time" (Aitken 1935).

FIG. 2.2.2 is a schematic of a binary system inclined with respect to the plane of our sky showing the orbit of $S_2$ (the secondary star) relative to $S_1$ (the primary star).

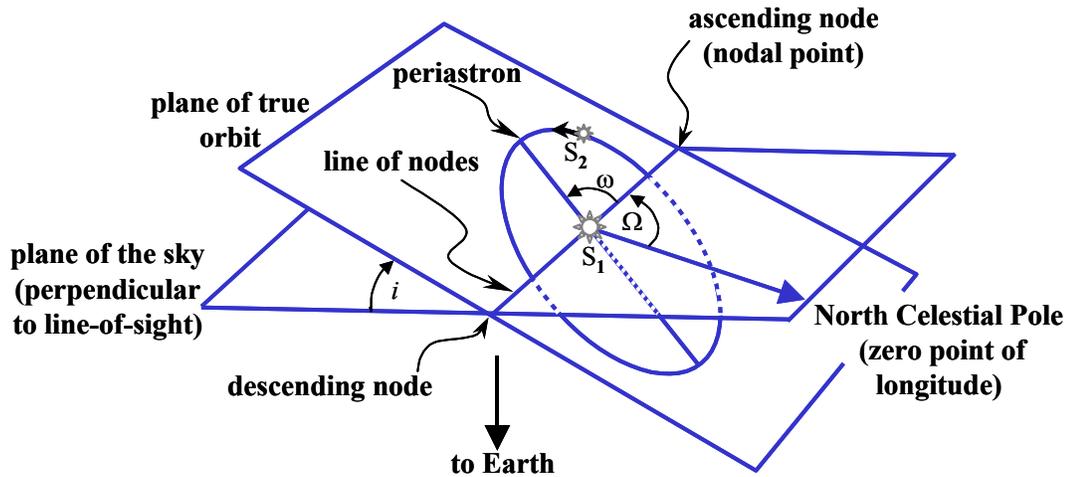

FIG. 2.2.2–*Orbital Plane of a Binary Star, Inclined (adapted from Fig. 25 p. 60 Binnendijk 1960, and http://scienceworld.wolfram.com/physics/OrbitalElements.html, 2006).*

The orbital elements are defined as: (1) the *period* of revolution P, (2) the *time of periastron passage* T, (3) the *eccentricity* e of the true ellipse, (4) the *semimajor axis* a of the true ellipse, (5) the *position angle of nodal point* Ω, (6) the *longitude of periastron* ω, and (7) the *orbital inclination* i. (See TABLE 2.2.1).

TABLE 2.2.1
Orbital Elements of Binary Systems

| | | |
|---|---|---|
| P | Period of revolution | Time it takes for one orbit |
| T | Time of periastron passage | A time at which the body passed through pericenter, the time of maximum orbital velocity |
| e | Eccentricity of the true ellipse | Eccentricity of the ellipse in the plane of the orbit (not of the projection of the orbit on the sky) |
| $\Omega$ | Position angle of nodal point which lies between $0°$ and $180°$ | It is also the position angle of the line of nodes. (The line of nodes is the line of intersection of the orbit plane with the plane perpendicular to the line of sight). |
| $\omega$ | Longitude of periastron, or the angle in the plane of the true orbit between the line of nodes and the major axis | It is measured from the nodal point to the point of periastron passage in the direction of the companion's motion and may have any value from 0 to 360. It should be stated whether the position angles increase or decrease with time. |
| i | Inclination of the orbit plane | The angle between the orbit plane and the plane at right angles to the line of sight. Its value lies between $0°$ and $\pm 90°$ depending on whether the orbital motion at the nodal point is carrying the companion star away from (+), or toward (−) the observer. |

*(Source: Aitken 1935 p.77-78; Binnendijk 1960 p.59-60)*

Binary stars are often classified according to mode of observation and special characteristics. The categories are not mutually exclusive and include visual, astrometric, spectroscopic, eclipsing, and interacting.

A *visual binary* is a bound system that can be resolved into two stars telescopically. If the orbital period is not too long, it is possible to track the motion of each member of the system, though usually only relative motion is observable. The mutual orbits of these stars are observed to have periods ranging from about one year to thousands of years (decades are ideal for observation).

Fig. 2.2.3 is a representation of the motions of Sirius A and B, a visual binary. The left figure (A) shows the apparent motions relative to background stars of Sirius A, its companion B and the center of mass of the system C. The right figure (B) shows the orbital motions of Sirius A and B relative to the system's center of mass.

In an *astrometric binary* only one star is discernible, but its oscillatory motion in the sky reveals the presence of a hidden companion orbiting a common center of mass.

A *spectroscopic binary* is often an unresolved system whose binary nature is revealed by the periodic oscillations of lines in its spectrum. In the case of a *double-lined spectroscopic binary*, two sets of spectral features are observed (one set for each star), oscillating with opposite phases. In the case of a *single-lined spectroscopic binary* one of the stars is comparatively too faint to be detected, and only one set of oscillating spectral lines dominates. Typical orbital periods range from hours to months.

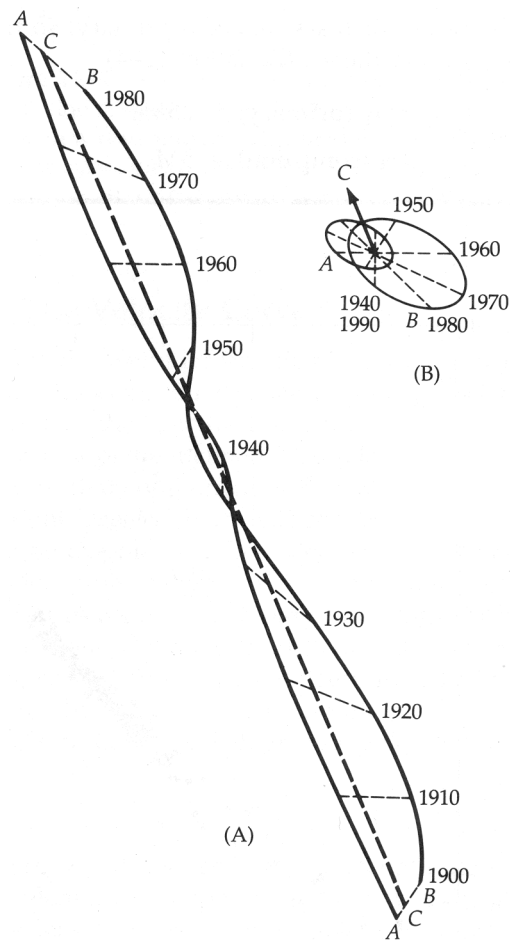

Fig. 2.2.3–*Visual Binary*
(*Source: Fig. 12-2 p. 237 Zeilik 1998*)

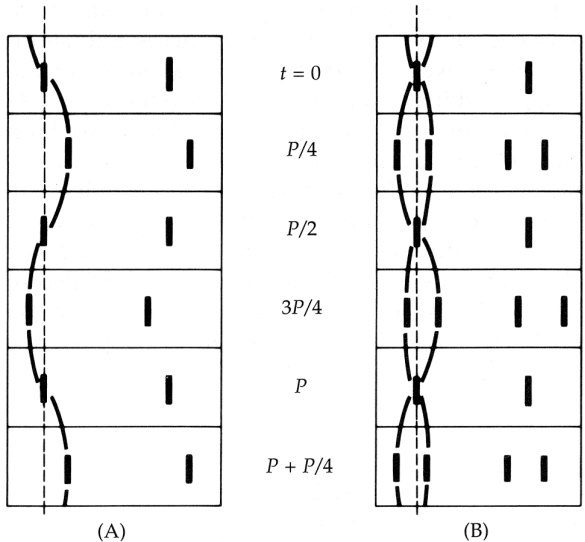

FIG. 2.2.4–*Spectroscopic Binaries*
(*Source: Fig. 12-4 p. 239 Zeilik 1998*)

FIG. 2.2.4 are schematic spectra of spectroscopic binaries. In a single-line system (A), only one set of lines shows an oscillation in wavelength from the Doppler shift. In a double-line system (B), two sets of lines oscillate out of phase as the two stars revolve around their common center of mass.

The degree to which a spectral feature shifts from its rest wavelength (the Doppler shift) is a measure of the velocity of its source along our line of site, the *radial velocity*. A plot of these line-of-sight velocities as a function of time is called a *radial velocity curve*.

FIG. 2.2.5 is a schematic representation of a radial velocity curve for a double-lined spectroscopic binary system with a circular orbit. The left figure (A) shows the circular path of primary star P and companion star C about the center of mass X, with an orbital inclination of 90° ("edge on" with respect to our line of sight from Earth). The figure on the right (B) shows the component radial velocities, where the center of mass of the system is receding at a constant speed CM relative to the Sun. V and v represent the speeds relative to the CM of the primary and companion (secondary), respectively, as they move through one orbital period, P. According to convention, the primary is generally taken as the brightest star in the system.

FIG. 2.2.5–*Radial Velocity Curve*
(*Source: Fig. 12-5 p. 240 Zeilik 1998*)

A significant fraction of binary systems have inclinations in the range of about 70-90°; that is, their orbits are almost "edge on" with respect to our line of sight from Earth. If the component separations are sufficiently small, the stars will undergo eclipses, one star blocking all or part of the light from the other as viewed from Earth. FIG. 2.2.6 illustrates the photometric orbit (or light curve) of a hypothetical partially eclipsing binary. In this illustration, the smaller star is assumed to be the hotter and brighter star. Hence, it is identified as the primary. During the "primary eclipse," the hotter primary star

FIG. 2.2.6–*Light Curve of Partially Eclipsing Binary*
(*Source: Fig. 12-10 p. 244 Zeilik 1998*)

is (partially) eclipsed by the secondary star, leading to a large reduction in binary-system flux.

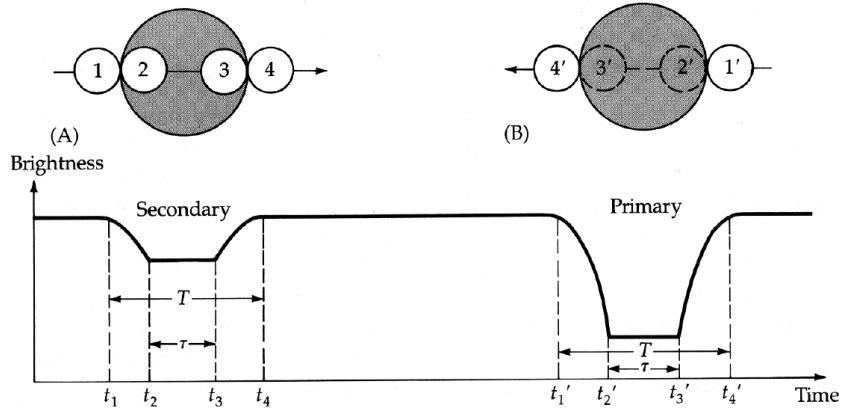

FIG 2.2.7–*Light curve of eclipsing binary (primary is total)*

FIG. 2.2.7 represents the light curve of an eclipsing system where the small star, the hotter component, is completely eclipsed. The four numbered contact points define the duration of the eclipse. These central eclipses would have flat bottoms if the limb darkening effect were excluded. During secondary eclipse (A), the primary passes in front of the secondary (i.e., the secondary is eclipsed). During primary eclipse (B), one-half an orbital period later, the primary star passes behind the larger secondary.

In close binary systems the gravitational field of one component can distort the shape of its companion. In some cases, the interaction produces a flow of mass out of one star, which either impacts the other star directly, forms an accretion disk or envelope around one or both components (circumstellar or circumbinary, respectively), or leaves the system altogether, resulting in a variety of interesting phenomena. For instance, the components may be so close together, and highly distorted, that they share a common atmosphere.

The shape of the component stars in a binary star system depends on the degree to which limiting equipotential surfaces called Roche lobes (described later in this section) are filled by the stars (for more information, see Hilditch 2001, p. 688). This is shown in FIG. 2.2.8. The components of *detached* systems (A) have radii much less than their separation and are nearly spherical in these cases; the two stars evolve nearly independently. In a s*emidetached* system (B) one of the stars is so distorted and distended by its proximity to the companion that mass may flow away from the larger star. If both stars fill or expand beyond their Roche lobes, the two stars will share a common atmosphere bounded by a dumbbell-shaped equipotential surface. These are called *contact* systems (C).

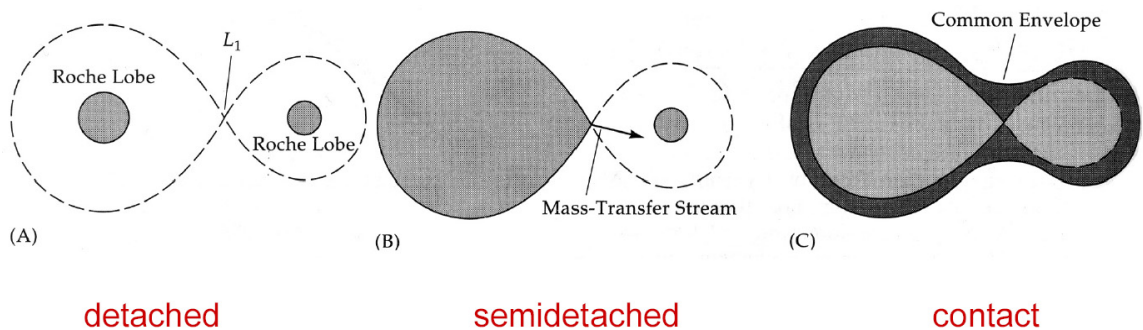

FIG. 2.2.8–S*hape of Binary Components (Source: Fig. 12-12 p. 246 Zeilik 1998)*

Semidetached and contact systems are *interacting binaries*. *Algol-type* binaries (discussed Section 2.4) belong to the semidetached subclass of interacting binaries. Algol itself, the focus of this work, is the prototype of this class.

Roche lobe geometry is derived from the *Restricted three-body problem*: Two bodies $m_1$ and $m_2$ move in circular orbits about their common center of mass, while a third body of infinitesimal mass moves in the gravitational field of the other two bodies (for further information, see Hilditch 2001, p. 157). Masses $m_1$ and $m_2$ are considered centrally condensed or point masses, and are in circular and synchronous orbits so that the orbital period equals the rotational period.

FIG. 2.2.9 illustrates the geometry of the restricted three-body problem. Masses $m_1$ and $m_2$ are in orbit about their mutual center of mass, X (with $m_1 > m_2$). The distances from $m_1$ and $m_2$ to the infinitesimal point mass at (x, y, z) are $r_1$ and $r_2$ respectively. The semimajor axis, a, is the separation of $m_1$ and $m_2$. The Roche equipotentials (indicated by the contours) are equipotential surfaces in this co-rotating frame of the restricted three-body problem. These effective potential-energy surfaces include both true gravitational and fictitious centrifugal contributions. If we let the semimajor axis a = 1 then the normalized potential $\phi$ (in the co-rotating frame) experienced at any position (x, y, z) is the sum of the two point-mass potentials and the rotational potential (*Equation 4.46 p.157 Hilditch 2001*):

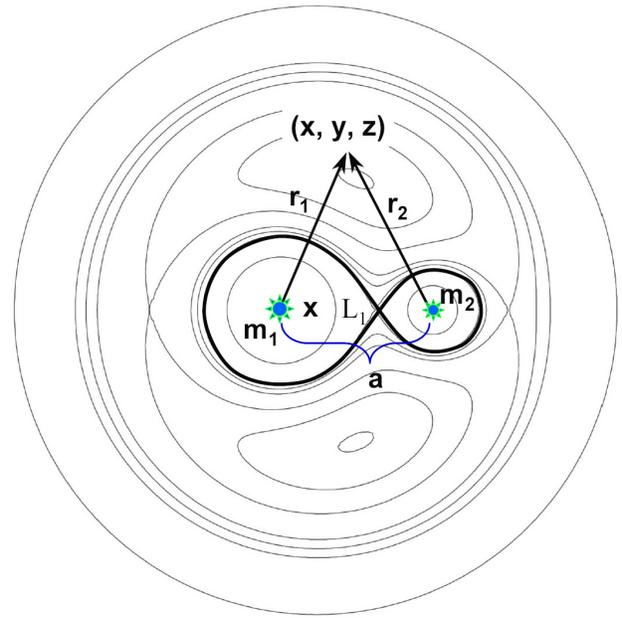

FIG. 2.2.9–*Geometry of Restricted 3-body problem*

$$\phi = -\frac{Gm_1}{r_1} - \frac{Gm_2}{r_2} - \frac{\omega^2}{2}\left[\left(x - \frac{m_2}{m_1 + m_2}\right)^2 + y^2\right] \quad (2.2.1)$$

where G is the Gravitational constant[*] and ω is the angular velocity of the co-rotating coordinate system. The equipotentials that intersect (dark, thick teardrop-shaped contours) at a point along the line of centers that joins the two masses are called *Roche lobes*. The point of intersection is called the $L_1$ point, or the *inner Lagrangian point*.

Initial conditions restrict the third body to a certain volume. This means the particle cannot cross the equipotential surface defined by its initial conditions; that is, a small test mass, viewed within the rotating coordinate system, will move within (not cross) the equipotential surface corresponding to the test particle's total energy (i.e., positions where the test mass would have no kinetic energy – no velocity – with respect to the rotating frame).

The Roche equipotentials provide constraints on the motion of matter moving under the gravitational influence of $m_1$ and $m_2$. In particular, we are concerned with the flow of stellar gases in the case of a semidetached interacting binary. Observational evidence for such mass flow is provided in the following section.

## 2.3 *Observational Indications of Mass Flow*

The activity levels of Algol-type binaries vary from one system to another and within the same system across epochs. Gas stream effects accompanying periods of higher activity are observed in the radial velocity curves of Algol-type systems. FIG. 2.3.1a is a schematic radial velocity curve of an eclipsing binary *without* the gas stream.

---

[*] $G = 6.67259 \times 10^{-8}$ dyne cm$^2$/g$^2$

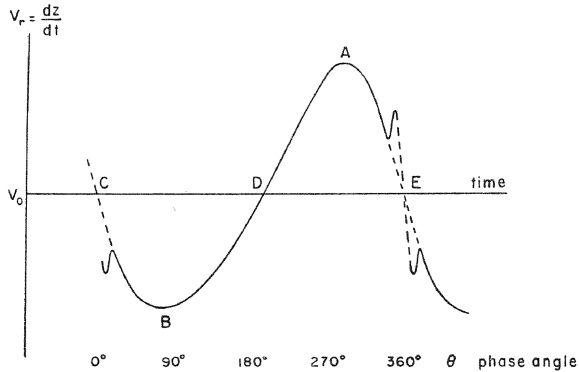 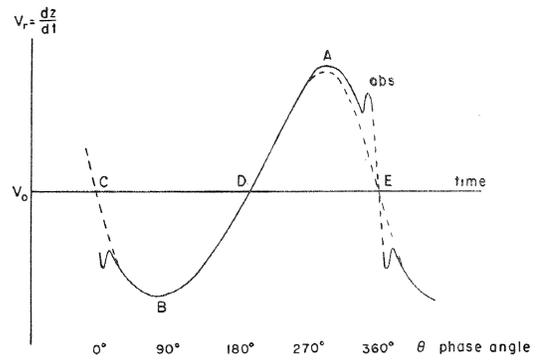

FIG. 2.3.1a–*Radial velocity curve with rotation effect only*
(Source: Fig. 88 p.173 Binnendijk 1960)

FIG. 2.3.1b–*Radial velocity curve with gas stream and rotation effects*
(Source: Fig. 89.p.174 Binnendijk 1960)

Primary eclipse occurs at 0° and 360° (points C and E). The extra oscillation at phase = 360° (primary minimum), is due to the rotation effect.[*] FIG. 2.3.1b, on the other hand, is a typical radial velocity curve of an Algol-type system undergoing an epoch of mass transfer. The location (in phase) of the excess radial velocity is indicative of mass flowing from the secondary, through the inner Lagrangian point, toward the facing hemisphere of the primary. This type of asymmetry in radial velocity curves was noted as early as 1908 when Barr described these curves as having an ascending branch of greater length than the descending branch. The explanation of a gas stream, however, was not suggested until Struve (1941) postulated[†] such a gas stream in β Lyrae[*§].

---

[*] "When the bright star is entering eclipse, one limb is gradually covered by the eclipsing star and consequently the lines from the bright star are fully broadened on one side only because of the velocity of the one wholly visible limb. When these lines are measured for determination of radial velocity, the center of density of the line will be shifted toward the broadened edge and away from the center of the symmetrical line that would be observed if both lines were visible." (Rossiter 1924)

[†] "...[V]iolet satellites are interpreted as a stream flowing from the B9 star, along the side of the invisible second component which becomes uncovered after minimum light." (Struve 1941)

Epochs of active mass transfer are also evident in the light curves of Algol-type systems. Dugan (1924) was the first to conclude that the "anomalous hump"[†] in the light curve of R Canis Majoris[‡] at the end of primary minimum[§] "has some real significance." He indicates its transient nature when he says, "The constancy of the hump, however, is placed in doubt by the observations of other nights." That is, the hump was not present on other nights. FIG. 2.3.2 is the R CMa light curve containing the "hump," as indicated by the solid circles (between 4-6h on the figure).

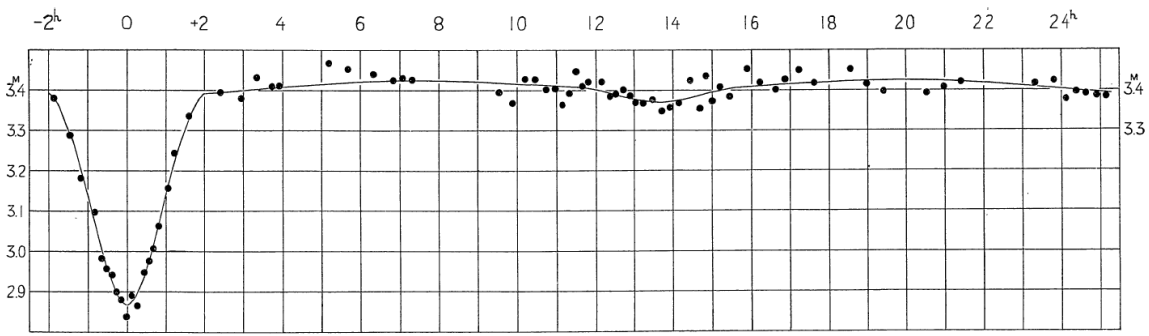

FIG. 2.3.2–*Light Curve of R Canis Majoris*
*(Source: Fig. 8 p.68 Dugan 1924)*

The solid circles are averaged data points. The solid curve on the other hand is the expectation for the light curve in the absence of mass flow effects. Data are from observations in epoch 1898-1899.

Another example of gas stream effects is provided by observations of U Cep.[**] Olson (1980) compares the light curves of U Cep, from different epochs during ingress and egress, which become "... grossly distorted during mass-transfer events." FIG. 2.3.3a is the U Cep U-band light curve of primary eclipse ingress. The solid line is the mean undisturbed egress curve reflected around mid eclipse; *filled circles*, 1978 September 27 (low activity); *squares*, 1976 (moderate activity); *triangles*, October 26 (high activity). FIG. 2.3.3b is the U Cep U-band light curve of primary eclipse egress. The *filled circles* are data from five nights of low photometric contamination; *triangles*, active night 1975 October 31.

---

[*§] "There is weak evidence that β Lyrae is semidetached but often this is simply assumed to be the case" (Sahade, McCluskey, and Kondo 1993 p.40).
[†] This was discovered independently in 1898 by Pickering (1904) and Wendel (1909) according to Sahade and Wood (1978).
[‡] R Canis Majoris (R CMa) is a bright, short period, semidetached eclipsing binary of the Algol type (Verricatt and Ashok, 1999) with the lowest known mass of Algol systems hosting the least massive secondary star (Ribas et al., 2002).
[§] Hump observed 1898-1899.
[**] U Cephei, an eclipsing semidetached binary, is the most active of short period interacting Algol-type binaries. (McCluskey et. al 1988) with a period of 2.49 days (Olson et al.1981).

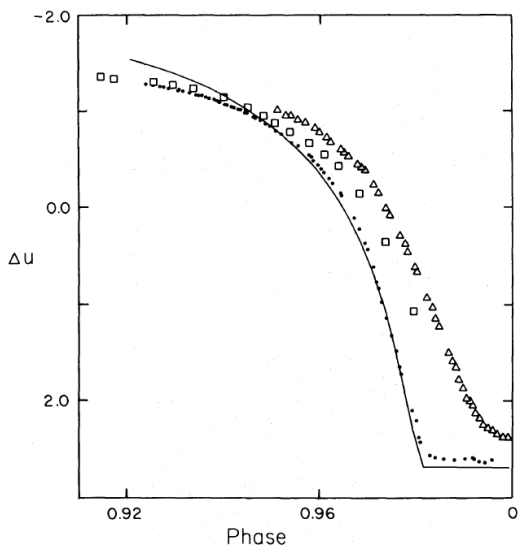 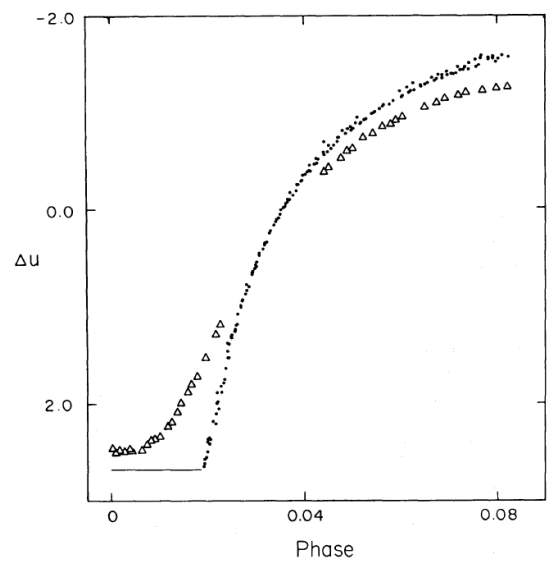

FIG. 2.3.3a—*U Cep U-band light curve of primary eclipse ingress*
*(Source: Fig. 2 p.258 Olson 1980)*

FIG. 2.3.3b—*U Cep U-band light curve of primary eclipse egress*
*(Source: Fig. 1 p. 258 Olson 1980)*

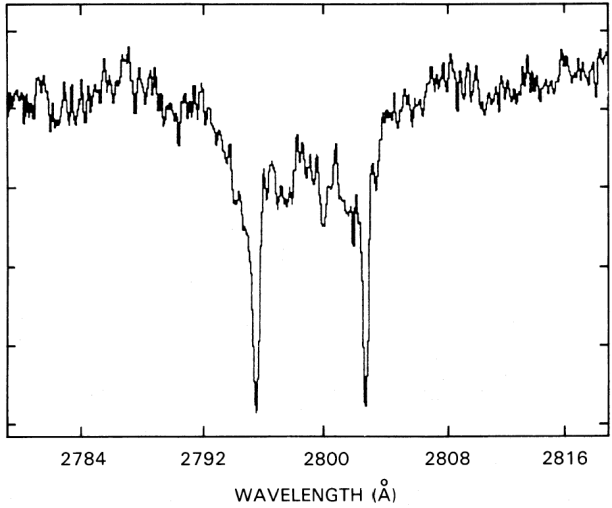

FIG 2.3.4a–*U Cep UV Mg II resonance doublet 1986 Oct 22 (Source: Fig. 2d p.1025 McCluskey et al. 1988)*

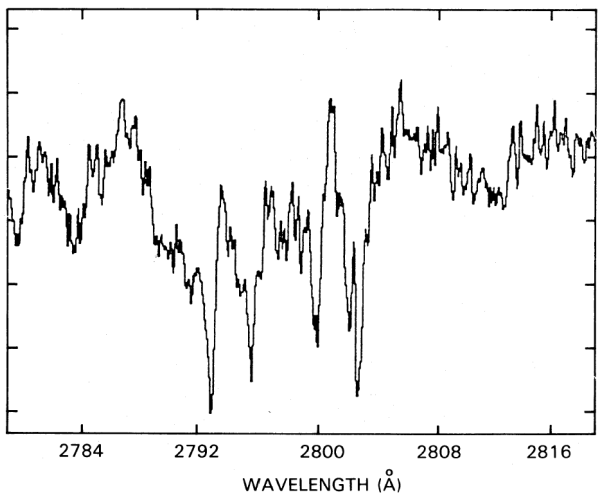

FIG 2.3.4b–*U Cep UV Mg II resonance doublet 1986 Jun 4 (Source: Fig. 2a p.1024 McCluskey et al. 1988)*

McCluskey et al. (1988) detected gas stream effects in U Cephei by examining ultraviolet Mg II line profiles from different epochs. FIG. 2.3.4a shows the Mg II resonance doublet, presumably during a quiescent period on 22 October 1986 at phase 0.180. The spectral lines at 2796 Å and 2803 Å are well resolved. FIG. 2.3.4b shows the same spectral range on 4 June 1986 at a similar phase (0.132). The pair of extra, narrow lines, blueshifted by 2.7 Å at ~2793 Å and ~2800 Å are interpreted as a transient high-mass flow event with a radial mass velocity of -286 km/s.

Walter (1973) presents a general picture of the gas stream effects. FIG. 2.3.5 illustrates hypothetical distortions of light curves outside primary eclipse caused by absorption (cross-hatched) and emission (dotted) for systems with impact regions of the gas streams near the equator (A and B) and near the poles (D) of the small bright component. These are consistent with the features already discussed of FIGs. 2.3.1 and 2.3.2.

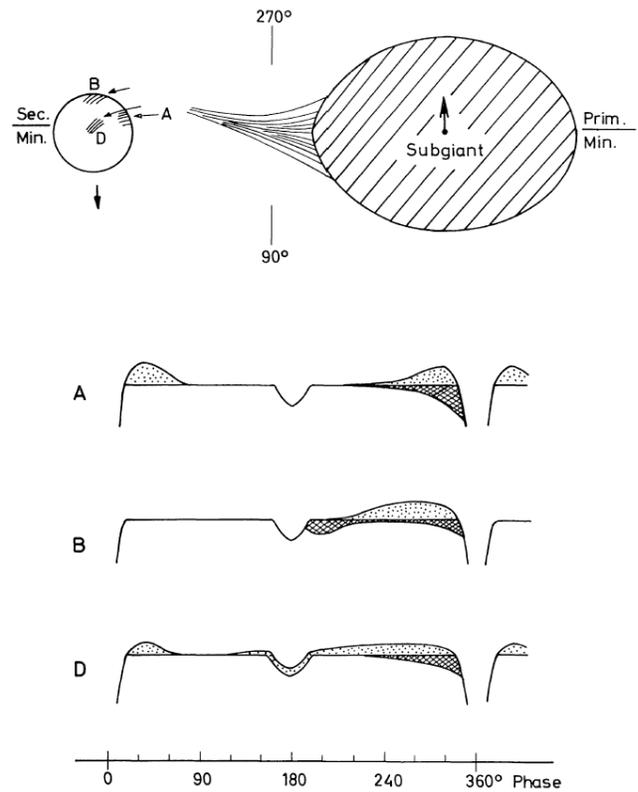

FIG. 2.3.5–*Hypothetical photometric distortions from gas stream effects (Source: Fig. 3 p. 292 Walter 1973)*

2.4 *Algol (Beta Persei), Algol-Type Systems and the Algol Paradox*

Algol-type binary star systems have been studied actively since 1783, when Englishman John Goodricke conjectured that the cause of the light variations in Beta

Persei[*] "... could hardly be accounted for otherwise than ... by the interposition of a large body revolving round Algol ..." (Goodricke 1783).

Ancient people had already noticed the dramatic decrease in Algol's light (1.2 magnitudes within five hours) that gave rise to the name *Al Ghūl*, which means "changing spirit" to the inhabitants of the Arabian Peninsula. The ancient Hebrews named this bright star *Rōsh-ha-Satan*, which means "Satan's Head," and the Chinese named it *Tseih She*, "the Piled-up Corpses" (Kopal 1959). Other descriptive names include "the eye of the Medusa,"[†] and "the demon star."

Algol (HD 19356)[‡] is the brightest eclipsing binary system in the sky and the prototype of its class (Drake 2003). It is the first variable system to be identified as an eclipsing binary. Algol is actually a triple system where the earliest reference to a third component is credited to Stebbins in 1924 (Sahade and Wood 1978). This ternary stellar system consists of an eclipsing close-binary with a period of 2.87 days, orbiting a main-sequence A7m star in 1.86 years (Lestrade et al. 1993). The primary star of the eclipsing close-binary is a nearly spherical B8 V star and the less massive secondary is a highly distorted K0-3 IV star in contact with its Roche equipotential surface (Guinan et al. 1989). The masses of Algol A, B and C are 3.7 $M_\odot$, 0.81 $M_\odot$, and 1.6 $M_\odot$, respectively. The Algol system is located a distance of 28 pc from Earth (1pc = 3.25 light years).

A comprehensive set of system parameters, as reported by numerous investigators, is collected in Appendix A, Tables 1-59. A selection of these parameters is provided in TABLE 2.4.1.

---

[*] Algol itself is also known as Beta Persei.
[†] In Greek mythology Medusa is a Gorgon, a vicious female monster with sharp fangs and hair of living venomous snakes (en.wikipedia.org 2006).
[‡] Also designated β Per; 26 Persei; HR 936; SAO 38592; ADS 2362

TABLE 2.4.1
Selected Algol Parameters

| Algol A | Algol B | Algol C |
|---|---|---|
| B8 V | K2 IV | F2 V |
| $T_A = 12500 \pm 500$ K | $T_B = 4500 \pm 300$ K | $T_C = 7000 \pm 200$ K |
| $M_A = 3.7 \pm 0.3\ M_\odot$ | $M_B = 0.81 \pm 0.05\ M_\odot$ | $M_C = 1.7 \pm 0.2\ M_\odot$ |
| $R_A = 2.90 \pm 0.04\ R_\odot$ | $R_B = 3.5 \pm R_\odot$ | $R_C = 1.6 \pm 0.2\ R_\odot$ |
| $P_{A\text{-}B} = 2.8673^d$ | | $P_{AB\text{-}C} = 679.9^d$ |
| $\text{sep}_{A\text{-}B} = 14.0\ R_\odot$ | | $\text{sep}_{AB\text{-}C} = 2.67 \pm 0.08$ AU |

FIG. 2.4.1 is a diagram of the Algol binary as seen from above the orbital plane. The Roche lobes and components of the system are drawn to scale. The + sign indicates the position of the center of mass of the binary. The zero-velocity surface passes through the inner Lagrangian point $L_1$.

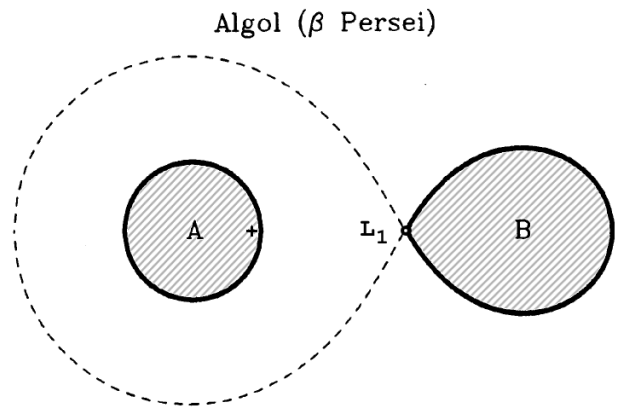

*FIG. 2.4.1–Scale diagram of Algol (adapted from Fig. 7 p. 335 Richards et al. 1988)*

Like Algol itself, Algol-type binaries are composed of an early-type main-sequence (O V - B V) primary that lies well within its Roche lobe, and a less massive late-type subgiant (F IV - K IV) secondary that fills its Roche lobe and loses mass to the primary.

The eclipses of Algol-type binaries are dramatic because the hotter (and brighter) primary star of the eclipsing pair contributes about 90 percent to the total visible light of the system, and most of this light is blocked when the cooler (and fainter) secondary star eclipses the primary. The eclipse of the secondary star produces a small diminution in the visible light from the pair. The light curve of Algol demonstrates this signature difference in the depths of primary and secondary minimum.

FIG. 2.4.2 is a light curve of Algol at 3428 Å, using Copernicus satellite data (Chen et al. 1977). Secondary minimum occurs at phase = 0.5, when Algol B is partially eclipsed by Algol A. Primary minimum occurs at phase = 0.0 (equivalent to 1.0) and is noticeably deeper since most of the light from the system, which comes from Algol A, is being attenuated by the dimmer Algol B as it passes between us and the primary.

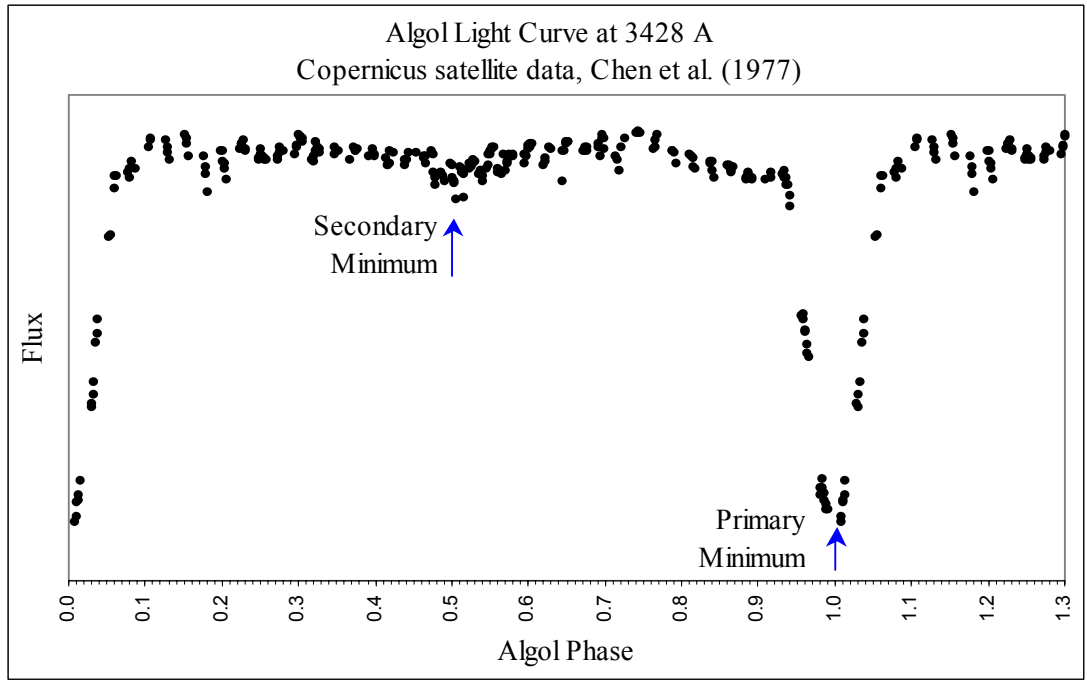

FIG. 2.4.2—*Algol light curve at 3428 Å (Source: Copernicus satellite data, Chen et al. 1977)*

Outside of the eclipses, the spectrum of an Algol-type binary shows absorption lines from the primary, while the spectrum of the fainter secondary star can be detected only during primary eclipse (Richards 2002 Web site ).

The secondaries of Algol-type systems are peculiar because they do not follow the expected mass-luminosity relation; that is, "... the secondary components possess masses which are as a rule too small for their observed luminosities" (Kopal 1955). Crawford (1955) describes these peculiar secondaries as "... combining subgiant character with small mass (<1$\odot$)". This circumstance, the well known *Algol Paradox*, is described by Crawford: "It is not possible to understand how stars with such small mass could have had enough time in five billion years to reach the advanced stage of evolution suggested by their subgiant characteristics..." In Kopal's words, "... as both components are no doubt of equal age (and, presumably, of initially the same chemical composition), the subgiant secondaries cannot manifestly be 'old' on the same time scale."

Crawford explains this seemingly contradictory set of circumstances by relating "the 'advanced age' of these stars to the fact that they fill the inner zero-velocity surface"[*] Kopal, independently, makes this same ingenious connection:

> "If ... these stars are secularly expanding, there is a compelling reason why their size cannot exceed that of their Roche limit: for no larger *closed* equipotential exists which would contain the whole mass of the respective configuration. ...[O]nce this maximum distension permissible on dynamical grounds has been attained, a continuing tendency to expand is bound to bring about a *secular loss of mass* ... [T]here remains but little room for doubt that the relative smallness of their present masses is but the consequence of a secular loss caused by continued expansion."

Crawford's evolutionary scenario, which solves the Algol paradox, is illustrated in FIG. 2.4.3.

---

[*] This is equivalent to filling the Roche lobe.

| Algol "A" | Algol "B" | | Crawford (1955) p.75 |
|---|---|---|---|

**Initially**  P = 1.61 d

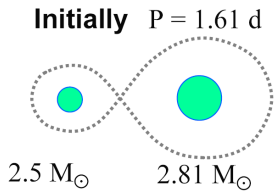

2.5 M$_\odot$    2.81 M$_\odot$

"Let us consider in a general way the evolution of a close binary system subsequent to its formation at a given date by some unknown mechanism. One of the stars will, in general, be more massive than the other."

**Begin mass transfer**

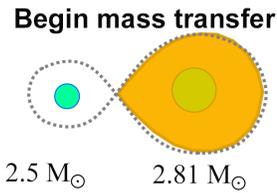

2.5 M$_\odot$    2.81 M$_\odot$

"The more massive star will, from the mass-luminosity relation, be much more luminous and will evolve more rapidly. At a certain point, it will acquire an unstable depleted core (Schöberg and Chandrasekhar 1942) and begin to move very rapidly to the right in the H-R diagram (Sandage and Schwarzschild 1952), expanding its radius as it does so. It is then very probable that this expansion will continue until the star fills the inner zero-velocity surface."

**Begin reversal of mass ratio**

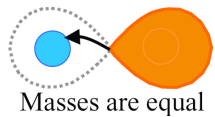

Masses are equal

"The star will then continue to expand, but will not be able to hold on to the mass that passes out of the surface. This mass will either be collected by the companion or be ejected from the system by centrifugal force, probably depending on whether or not fairly violent ejection mechanisms, such as prominences, dominate. The material lost, lying at the surface of the star, will be rich in hydrogen; and, as mass is lost, the ratio of the mass contained in the depleted core to the total mass of the star will increase. This ratio is a measure of the "biological age" of the star. As this age increases, the expansion will continue."

**Present**    P = 2.87 d

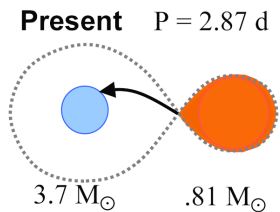

3.7 M$_\odot$    .81 M$_\odot$

"By this mechanism, the star can gradually become less massive than its companion and eventually less luminous, becoming the secondary component."

FIG. 2.4.3 - *Crawford's Evolutionary Scenario (1955, p.75) illustrated using Algol (Beta Persei)*

## 2.5 *Previous Observations and Models for Algol*

Algol was first studied in the visual range, then in other wavelength ranges as technologies became available. Budding (1986) compiled a table of "advances in observational coverage" of Algol (see TABLE 2.5.1).

TABLE 2.5.1

Some advances in observational coverage of Algol

| Year | Authors | Innovation | Result |
| --- | --- | --- | --- |
| 1966 | Chen and Reuning | IR (1.6 $\mu$) light curve | Pronounced secondary minimum; photometric peculiarities |
| 1966 | Cristaldi *et al.* | H$\alpha$ photometry | Evidence of spectal instabilities |
| 1967 | Glushneva and Esipov | IR (0.7–1.1 $\mu$) spectroscopy | Transient emission features |
| 1972 | Hjellming *et al.* | Radio observations (2695 and 8085 MHz) | Emission with short time variations discovered |
| 1973 | Ryle and Elsmore | Radio-interferometry | Accurate location of barycentre |
| 1974 | Labeyrie *et al.* | Speckle interferometry | Resolution of Algol C. |
| 1975 | Chen and Wood | UV spectroscopy | Detailed scans of L$\alpha$ at different orbital phases |
| 1975 | Eaton | UV light curves | Physical parameter values in UV |
| 1975 | Smyth *et al.* | IR (2.2 $\mu$) light curves | Apparent discrepancies with models |
| 1975 | Longmore and Jameson | IR (2.2, 3.6, 4.8, 8.6 $\mu$) light curves | IR excess – attributed to gas stream |
| 1975 | Bachmann and Hershey | Inclusion of accurate astrometry | More confident element specification |
| 1975 | Gibson *et al.* | Radio observations (1400, 2695, 6000, and 8085 MHz) | Large flare ($>$ 1 Jy) detected. |
| 1976 | Guinan *et al.* | H$\alpha$ and H$\beta$ index monitoring | Presence of circumstellar emission region revealed |
| 1976 | Schnopper *et al.* | X-ray observation | Emission in 2–6 keV band detected |
| 1978 | Rudy and Kemp | Polarimetry | Independent determination of orbital inclination |
| 1978 | Tomkin and Lambert | Reticon spectroscopy | Detection of Na II *D* lines of secondary |
| 1978 | Nadeau *et al.* | IR (10 $\mu$) data | Anomalous shape of secondary minimum |
| 1979 | Bonneau | Speckle interferometry | Revised absolute elements of triple system |
| 1979 | White *et al.* | X-ray spectrometry | No eclipses. Active corona of subgiant inferred |
| 1980 | Zeilik *et al.* | IR (JHKLM) light curves | IR luminosity ratios |
| 1979 / 1981 | Cugier / Chen *et al.* | UV Spectrometry (Copernicus) | High resolution Mg II line profiles allow detailed modelling |
| 1981 | Kemp *et al.* | Upgraded polarimetry | Perpendicularity of AB and AC orbit planes inferred |
| 1982 | Zirin and Liggett | IR coude spectrometry | Deep absorption at $\lambda$10830 due to Algol B |
| 1983 | Kemp *et al.* | Polarimetry | Discovery of eclipse polarization effect |

*(Source: Budding 1986, Table 1 page 244)*

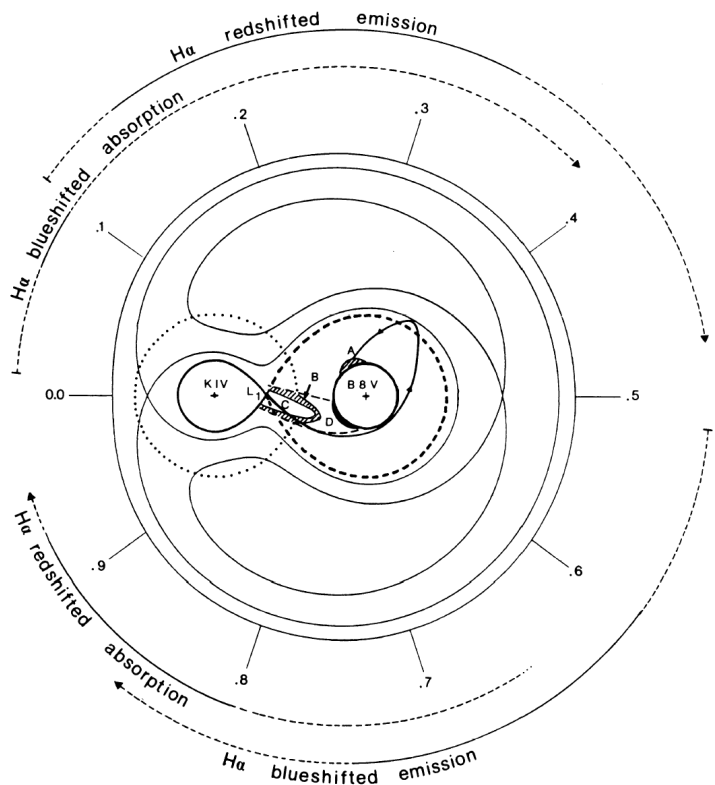

FIG. 2.5.1—*Model of Algol deduced from $H_\alpha$ observations (Source: Fig. 3 p. 225 Gillet et al. 1989)*

FIG. 2.5.1 is a model of Algol (Gillet et al. 1989) deduced from $H_\alpha$ (6562.8 Å) observations in 1985 and 1986. They observed the greatest blueshifted $H_\alpha$ emission between phases 0.6 and 0.75. The greatest redshifted $H_\alpha$ emission is between phases 0.13 and 0.34. These phase regions are identified by the outer circular segments in FIG. 2.5.1. Gillet et al. (1989) explained the two emission regions as (1) the stream itself (region B in the figure) and (2) a localized region near the photosphere of Algol A (region A). The thin continuous lines are the Roche equipotentials. The two thin dashed lines mark the extension of the stream of gas from Algol B near the $L_1$ point to Algol A (deduced from the hydrodynamic model of Prendergast and Taam 1974). These trajectories create the first shock, a hot spot on the photosphere of Algol A. The surrounding region, identified by "D" on the figure, is completely ionized by the hot spot. The second shock is created by the impact of the stream that passes around the primary. This hot spot is identified as "A." The dotted circle is the approximate position of the corona associated with the cool secondary star (as assumed by White et al. 1986; and Lestrade et al. 1988).

FIG. 2.5.2 is a model for the circumstellar material in Algol. Like the 1989 model of FIG. 2.5.1, this 1993 model is also developed from an analysis of $H_\alpha$ difference profiles. The 1993 analysis, however, used observations made in 1972, 1976, 1977, and 1985. According to this study, Richards (1993) observed strong single-peaked emission features outside of primary eclipse and absorption features during primary eclipse which she interpreted as a transient, high density ($N_e \geq 10^{11}$ /cm$^3$), low mass (M ~$10^{-13}$ M$_\odot$), localized region situated close to the photosphere of the primary but high above the orbital plane. She interpreted the weak double-peaked emission profiles seen outside of primary eclipse and the emission seen during primary eclipse as a transient rotating accretion disk which surrounded the primary. She observed broad, shallow absorption features, in the late 1976 to early 1977 data, at phases throughout the orbit of the binary and attributed this to a high-velocity component of the disk which surrounded the primary at that time. The region *lr* is the "localized region," *hrvr* is the "high-rotational-velocity region." Both regions are part of a "weak transient rotating accretion disk." The arced line drawn between the secondary (K2 IV star) and the primary (B8 V star) is the Lubow-Shu (1975, 1976) gas stream.

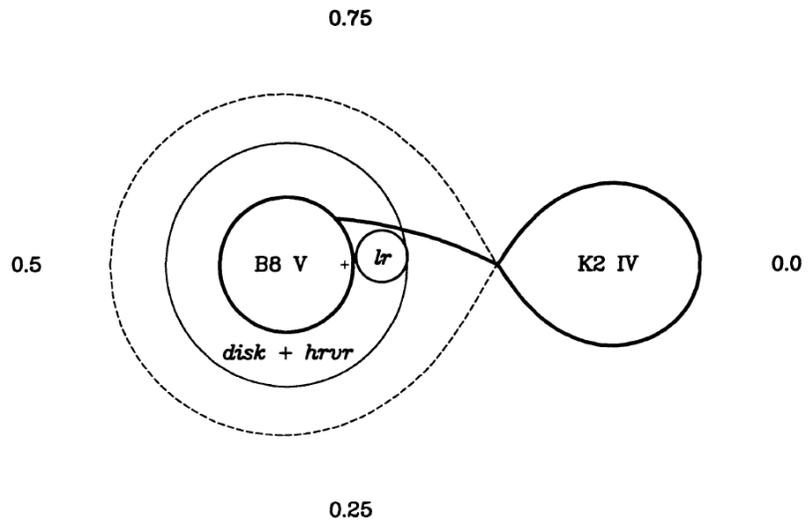

FIG. 2.5.2–*Model for the circumstellar material in Algol deduced from $H_\alpha$ observations*
*(Source: Fig. 20 p. 290, Richards 1993)*

The VLBA (Radio) observations presented by Mutel and Molner (1998) provided "...the first evidence for double-lobed structure in the radio corona of an active late-type star in the quiescent (basal) state." FIG. 2.5.3a is a schematic drawing of their "simple polar cap model." The basal components fill the extended region above the polar caps, while the flare emission regions are concentrated in the smaller regions of the higher magnetic field closer to the stellar surface. The lobes are separated by 1.6 mas[*], or 2.8 $R_B$. FIG. 2.5.3b contains a scale drawing of the binary system with respect to the radio source. It is assumed that the centroid of the radio emission is coincident with the center of the K star.

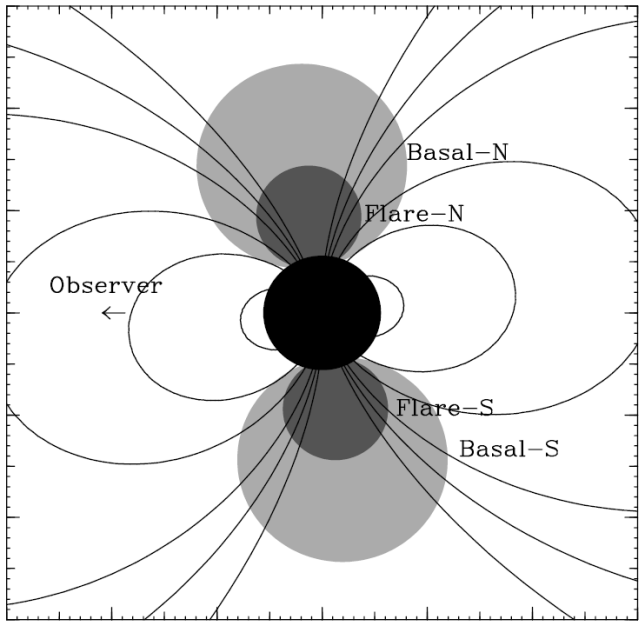

FIG. 2.5.3a–*Radio Model (Source: Fig.3 p. 373 Mutel and Molner 1998)*

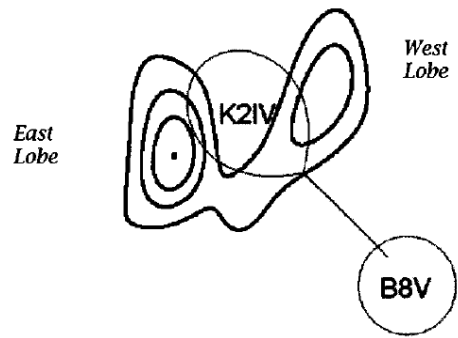

FIG. 2.5.3b–*Schematic of Algol with radio source (Source: Fig. 13 p. 381 Mutel and Molner 1998)*

---

[*] An angular measurement. There are 60 arc-seconds in one arc-minute, 60 arc-minutes in one degree, and 360 degrees in a full circle. One thousandth of an arc-second = 1 milliarc-second (mas)

Harnden et al. (1977) detected Algol in the soft X-ray region and interpreted the X-ray flux as "...thermal emission produced by the direct accretion of mass from the K star to the B8 member of the triple star system." The 0.15 to 2 KeV X-ray luminosity is $\sim 10^{30}$ ergs/s. FIG. 2.5.4 shows a simulated restricted three-body trajectory corresponding to Roche lobe overflow in Algol for gas leaving the surface of the K star at a thermal velocity $\sim 7$ km/s from the inner Lagrangian point $L_1$. The stream intercepts the B8 V star at an angle of $43°$ with respect to the line of centers and with a velocity of 470 km/s.

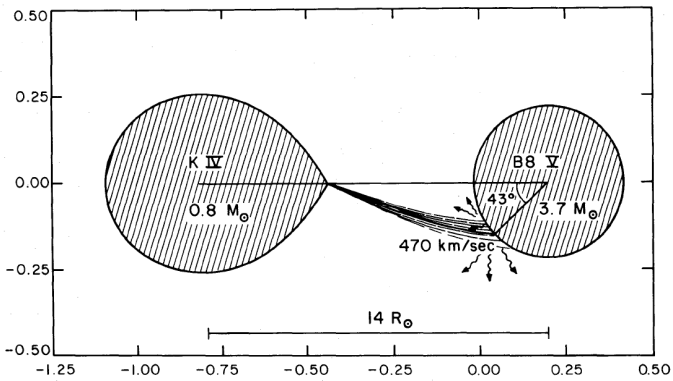

FIG. 2.5.4–*Soft X-ray Model*
*(Source: Fig. 2 p. 420 Harnden et al. 1977)*

Cugier (1982) analyzed UV observations of Algol from the Copernicus satellite. From the additional absorption components of Mg II near 2800 Å, he concludes that "...the absorbing region in which these components originate is optically thick and covers at least 20 per cent of the disc of the primary star, Algol A."

FIG. 2.5.5 is the geometry of the region of effective formation of the additional absorption components. The regions indicated as (1) and (2) correspond to Mg II 2795.5 Å and Mg II 2802.7 Å lines, respectively.

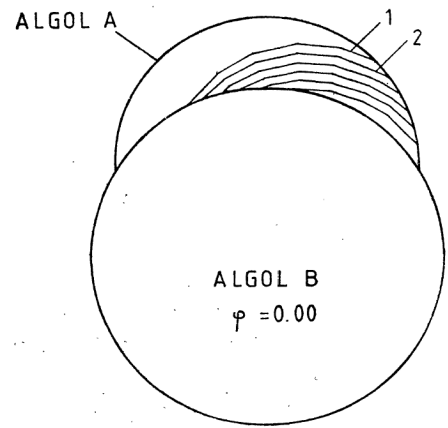

FIG. 2.5.5–Mg II Absorption Model
*(Source: Fig. 10 p.397 Cugier 1982)*

# 3. IUE OBSERVATIONS

3.1 *IUE Spacecraft and Database*

The IUE satellite,[*] launched on January 26, 1978 and shown in FIG. 3.1.1, was the first observatory type satellite totally dedicated to ultraviolet astronomy (Boggess et al. 1978).

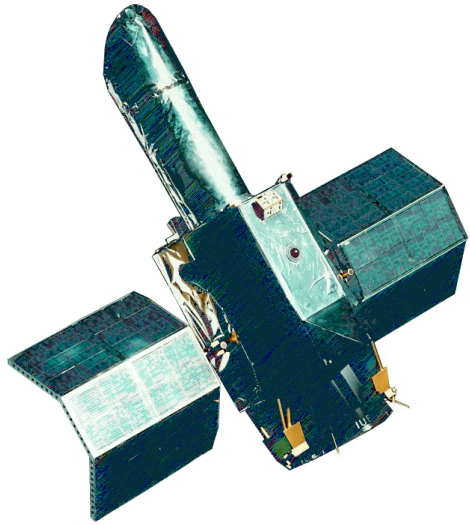

FIG. 3.1.1–*IUE Satellite*. (Source: http://sp dext. estec.esa.nl/content/doc/af/2479_.htm)

Guest astronomers were able to operate the satellite in real time (with a joystick) similar to the operation of a ground-based observatory (IUE, a). Initially, these remote operations were performed 16 hours a day from NASA/GSFC[†] and eight hours a day from VILSPA[‡] (IUE, b). Later, scientists could make their observations remotely from their home institutions (IUE, c). Although scheduling was facilitated by placing the spacecraft in a geosynchronous orbit, the demand for observation time consistently exceeded availability (by two to three times) during its 18.5 years of operation (IUE, b). (It's interesting to note that IUE's estimated lifetime was 3 to 5 years) (IUE, d). Currently IUE is considered to be the most productive and successful satellite in the history of space astronomy (IUE, e). IUE made, on average, one one-hour observation every 90

---

[*] International Ultraviolet Explorer (IUE) satellite
[†] Goddard Space Flight Center (GSFC), Greenbelt, Maryland
[‡] Villafranca Satellite Tracking Station (VILSPA), Villanueva de la Canada, Madrid, Spain.

minutes, around the clock (IUE, a). The durability of IUE enabled astronomers to monitor longer term changes by revisiting many targets. IUE remained operational until 30 September 1996 (IUE, b). The 104,000 images obtained by IUE were transformed into 111,000 spectral files, available to scientists throughout the world (IUE, e). "The IUE Data Archive remains the most heavily used astronomical archive in existence." (IUE, f)

The five mission goals are (IUE, g): (1) to obtain high-resolution spectra of stars of all spectral types to determine their physical characteristics, (2) to study gas streams in and around binary star systems, (3) to observe faint stars, galaxies and quasars, (4) to observe the spectra of planets and comets, and (5) to make repeated observations of objects with variable spectra. (6) This work addresses goals (1), (2), and (5) above.

Here are just a few "firsts" (selected from the 27 listed) (IUE, h): (a) the first evidence for strong magnetic fields in chemically peculiar stars, (a) the first observational evidence for semi-periodic mass loss in high mass stars, (b) the first discovery of high velocity winds in stars other than the Sun, (c) the discovery of starspots on late type stars through the Doppler mapping techniques, (d) the discovery of large scale motions in the transition regions of low gravity stars, (e) the discovery of the effect of chemical abundance on the mass loss rate of stars, (f) the first determination of a temperature and density gradient in a stellar corona outside the Sun, and (g) the creation of the first worldwide astronomical reduced-data archive delivering 44,000 spectra per year (5 spectra per hour) to astronomers in 31 countries.

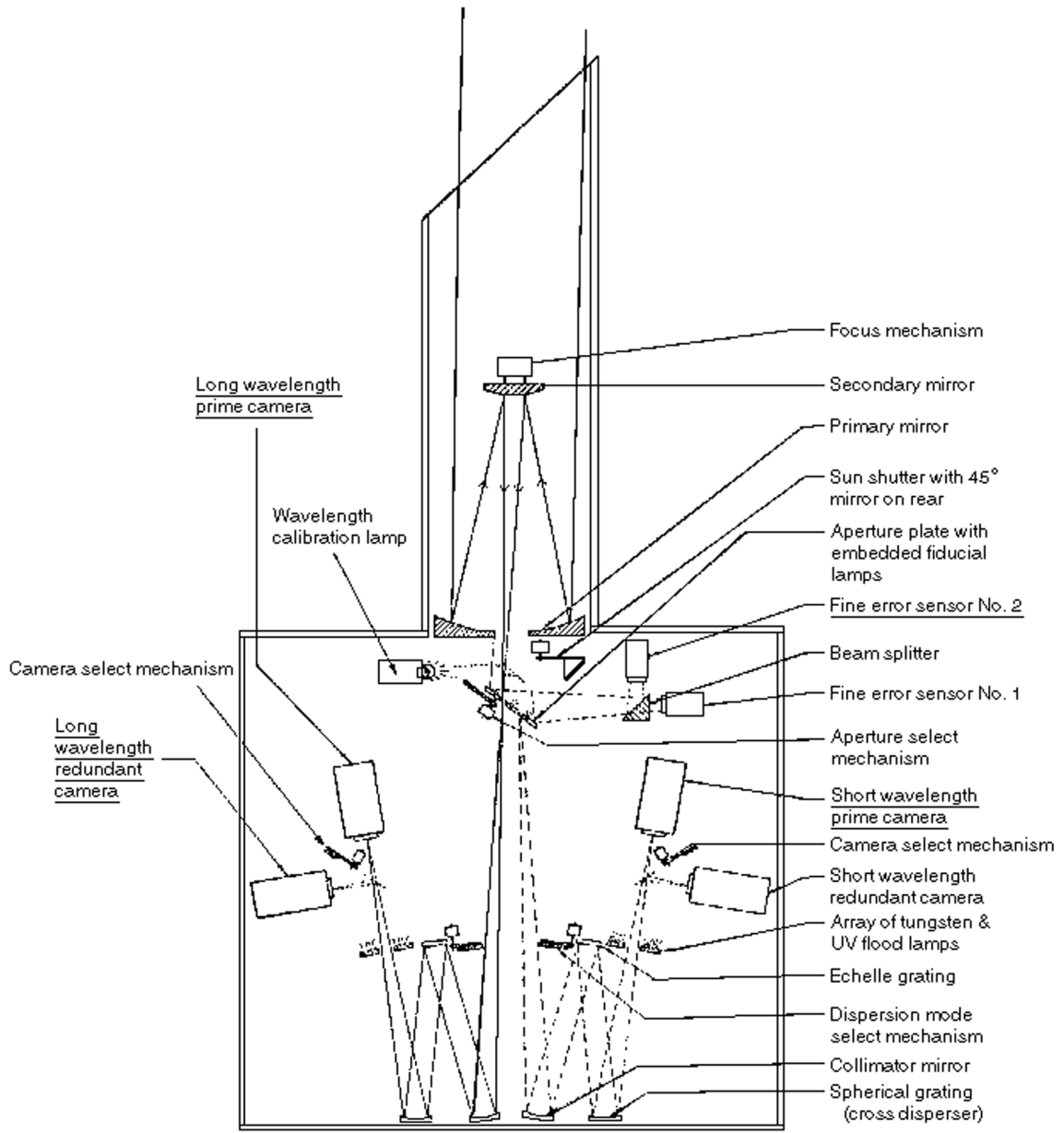

FIG. 3.1.2—*Optical Schematic of the IUE Scientific Instrument.* (Source: online document cited 4/22/06 http://archive.stsci.edu/ iue/ instrument/ obs_guide/ fig2.1.gif)

The IUE scientific instrument consists of a telescope, acquisition camera, two ultraviolet spectrographs, and four cameras. (See FIG. 3.1.2 and TABLE 3.1.1.) The telescope is a 45-cm aperture, f/15 telescope of Ritchey-Chretien design. The Fine Error Sensor (FES), is used in camera mode to obtain a visual star field image for target identification and acquisition. It is then used in tracking mode for guiding, providing a pointing stability of about 1/2 arcsec.

TABLE 3.1.1
SCIENCE PAYLOAD AND CAMERAS

| Science Payload | Description[m,n,o] |
|---|---|
| Telescope | 45 cm, f/15 Ritchey-Chretien Cassegrain |
| Spectrographs | Echelle (1150 - 1980 Å and 1800 - 3200 Å) |
| Apertures | SMALL (3 arcsec diameter) circle<br>LARGE (10 x 20 arcsec) slot |
| Resolution | HIGH 0.08 Å @ 1400 Å (17 km/s)<br>HIGH 0.17 Å @ 2600 Å (20 km/s)<br>LOW 2.7 Å @1500 Å (600 km/s)<br>LOW 14.0 Å @2700 Å (1500 km/s) |

| Cameras | Description[m,n,o] |
|---|---|
| SWP (1097-2097 Å) | Short Wavelength Prime<br>(sensitivity: $2 \times 10^{-15}$ ergs/s/cm$^2$/Å) |
| LWP (1808-3359 Å) | Long Wavelength Prime<br>(sensitivity: $1 \times 10^{-15}$ ergs/s/cm$^2$/Å) |
| LWR (1810-3456 Å) | Long Wavelength Redundant<br>(sensitivity: $2 \times 10^{-15}$ ergs/s/cm$^2$/Å) |
| SWR | Short Wavelength Redundant<br>(Never operational) |
| FES #1 | Fine Error Sensor #1<br>(Has not been used in operations.) |
| FES #2 | Fine Error Sensor #2<br>(16 arcmin field of view and effective image resolution of about 8 arcsec) |

Each spectrograph may be used with either of two apertures. The large aperture is a slot (approximately 10 by 20 arcsec) and the small aperture is a circle about 3 arcsec in diameter. The image quality of the IUE telescope's optics yields a roughly 3 arcsec image, so observations using the small aperture result in some light loss (IUE, h). The large aperture is used most frequently since it gives photometric reliability with little or no loss of resolution (IUE, d). Approximately 27% of the IUE exposures of Algol are taken in the small aperture mode.

The spectrograph design permits operation of each spectrograph in either of two modes: high dispersion/high resolution and low dispersion /low resolution (IUE, d). These produce an approximate velocity resolution between 10 and 25 km/sec for high resolution and between 600 and 1500 km/sec for low resolution (IUE, d). The long-wavelength spectrograph (LWP/LWR) operates in a wavelength range of 1850 to 3300 Å. The short-wavelength spectrograph (SWP) operates in a wavelength range of 1150 to 2000 Å (IUE, d). See TABLE 3.1.2 for the IUE spectrograph parameters. The high dispersion mode was used for all IUE exposures of Algol.

There are four cameras, two for each spectrograph. One is designated prime and the other backup (or redundant). Both the prime (LWP) and redundant

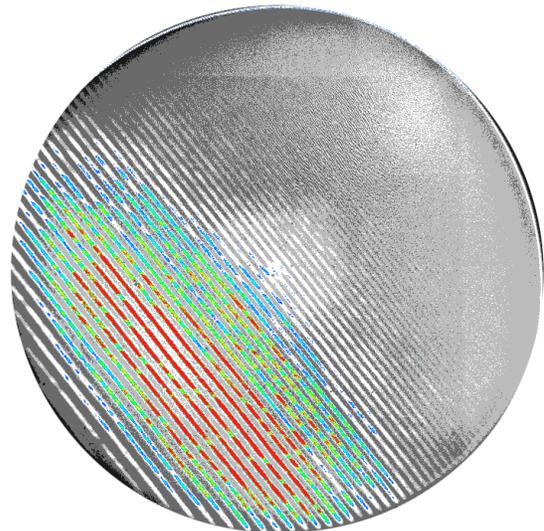

FIG. 3.1.3–*The First IUE Image.* High dispersion LWP image of Capella taken on February 9, 1978. (Source: http://www.vilspa.esa.es/ iue/ last / first.html)

(LWR) cameras were used during the mission. In the short wavelength range, only the prime camera (SWP) was fully functional. Each camera consists of an SEC Vidicon with an ultraviolet converter. The signal is digitized and transmitted to the ground via the spacecraft telemetry stream. FIG. 3.1.3 is a sample raw image. (Notice the parallel lines which are the echelle orders that have been spacially separated via the cross disperser).

TABLE 3.1.2

IUE SPECTROGRAPH PARAMETERS

| Parameter | LWP | LWR | SWP |
|---|---|---|---|
| Wavelength Range (Å) | 1808-3359 | 1810-3456 | 1097-2097 |
| Order Range | 69-127 | 67-127 | 66-125 |
| Abs. Calib. Wavelength Range (Å) | 1850-3350 | 1850-3350 | 1150-1980 |
| Abs. Calib. Order Range | 69-125 | 69-125 | 70-120 |
| Coll. Focal Length (mm) | 950 | 950 | 950 |
| Cam. Focal Length (mm) | 684 | 684 | 684 |
| Coll.-to-Cam. Angle (deg.) | 20.42 | 20.42 | 20.37 |
| Image Scale ($\mu m\ pix^{-1}$) | 36.4 | 36.4 | 35.7 |
| X-Disp. Ruling Freq. ($gr\ mm^{-1}$) | 241.50 | 241.50 | 369.233 |
| X-Disp. Order | 1 | 1 | 1 |
| Ech. Ruling Freq. ($gr\ mm^{-1}$) | 63.207 | 63.207 | 101.947 |
| Ech. Blaze Wavel. ($\mu m$) | 23.19 | 23.19 | 13.76 |
| Ech. Blaze Angle (deg.) | 48.126 | 48.126 | 45.449 |
| Ech. Grating Angle (deg.) | 0 | 0 | 0 |

*(Source of Table 3.1.2: cited 4/22/06*
*http://archive.stsci.edu/iue/manual/newsips/node94.html)*

The main data product that now resides in the final archive has been processed with the New Spectral Image Processing System (NEWSIPS). The resulting downloadable files have the extension MXHI (for high dispersion spectra) or MXLO (for low dispersion spectra). Since all IUE spectra of Algol were taken in the high dispersion mode the archived spectra of Algol are in the MXHI format only.

In order to extract the wavelength and flux vectors from these files we processed the MXHI files with an IDL (Interactive Data Language) subroutine written by Jess Johnson (February 2004). This custom program exports each MXHI file as a text file that is easily imported into commonly used software packages and manipulated.

### 3.2 *IUE Observations of Algol*

We present here the complete list of Algol IUE observations from the online MAST[*] IUE archive, 103 in all, 59 SWP images in the far-ultraviolet (1150-1900 Å) and 44 LWP/LWR images in the mid-ultraviolet (1900-3200 Å). Algol IUE observations were made during intermittent observing runs, the first starting on September 13, 1978, and the last ending on September 4, 1989 (or HJD[†] 2,443,765.24896 – 2,447,783.64315). The 1989 September data are especially useful in this work since the density of orbital coverage is the highest. TABLE 3.2.1 is a summary of the observing runs of Algol with their corresponding densities of orbital coverage.

---

[*] Multimission Archive at Space Telescope (MAST), http://archive.stsci.edu/
[†] Julian date, or JD is the number of days since 12:00 on 1 January, 4713 BC; the basis of time for astrodynamics. HJD, or the heliocentric Julian date, is the JD at the Sun. In this work JD is assumed to be heliocentric and is used interchangeably with HJD.

TABLE 3.2.1

IUE OBSERVING RUNS OF ALGOL

| Year 1900+ | # of LWP spectra per observing run | Approx. span of days | # of SWP spectra per observing run | Approx. span of days | Approx. density of data points |
|---|---|---|---|---|---|
| 78 Sept. | 1 | - | 1 | - | 1/(orbit) |
| 78 Nov. | 3 | 3 | 6 | 3 | 4.3/orbit |
| 78 Dec-79 Jan | 6 | 7 | 5 | 7 | 2.3/orbit |
| 79 Nov. | 3 | < 1 (~2 hr) | 3 | < 1 (~2 hr) | 3/(<⅓ orbit) |
| 83 Nov. | 6 | < 1 (~6 hr) | 7 | < 1 (~6 hr) | 6.5/(<⅓ orbit) |
| 84 Mar. | 1 | - | 4 | 2 | 5/(⅔orbit) |
| 85 Jul. | 7 | < 1 (~6.5 hr) | 7 | < 1 (~6.5 hr) | 7/(<⅓ orbit) |
| 86 Feb.-Mar. | 2 | 6 | 9 | 7 | 4.5/orbit |
| 89 Sept. | 15 | 3.3 | 17 | 3.3 | 13.9/orbit |

The orbital phase, $\phi$, of each image was calculated using the equation

$$\phi = NP - INT\,(NP) \qquad (3.2.1)$$

where NP is the number of orbital periods since the time of primary minimum,

$$NP = \frac{HJD_{MIDEXP} - t_{Pr.Min\,A-B}}{P_{A-B}}, \qquad (3.2.2)$$

and INT (NP) is the integer part of NP. $HJD_{MIDEXP}$ is the heliocentric Julian date of mid exposure, $t_{Pr.Min\,A-B} = 2{,}445{,}641.5135$ days (Al-Naimiy et al. 1984) is the time of a reference primary minimum of the eclipsing pair, and $P_{A-B} = 2^d.86731077$ is the orbital period (Gillet et al. 1989).

The entire set of Algol IUE exposures were taken in the high-resolution mode ($\lambda/\Delta\lambda \approx 10^4$). Of the 59 SWP images and 44 LWP/LWR images, 15 and 13 were taken in the small aperture modes, respectively. Tables 3.2.2 and 3.2.3 list the pertinent information for each SWP and LWP/LWR exposure in order of increasing phase.

TABLE 3.2.2

IUE SWP Observations of Algol

| SWP ID | Data ID | Aper | Month | Day | Year | Hour | min | JD at mid exposure | Phase |
|---|---|---|---|---|---|---|---|---|---|
| 1 | SWP03772 | SMALL | 1 | 1 | 1979 | 18 | 5 | 2,443,875.2543 | 0.0015 |
| 2 | SWP03262 | SMALL | 11 | 8 | 1978 | 6 | 42 | 2,443,820.7800 | 0.0031 |
| 3 | SWP36983 | LARGE | 9 | 11 | 1989 | 1 | 0 | 2,447,780.5423 | 0.0052 |
| 4 | SWP07111 | SMALL | 11 | 7 | 1979 | 10 | 23 | 2,444,184.9348 | 0.0053 |
| 5 | SWP37019 | LARGE | 9 | 13 | 1989 | 22 | 26 | 2,447,783.4350 | 0.0141 |
| 6 | SWP07112 | SMALL | 11 | 7 | 1979 | 11 | 17 | 2,444,184.9720 | 0.0183 |
| 7 | SWP36984 | LARGE | 9 | 11 | 1989 | 2 | 3 | 2,447,780.5858 | 0.0204 |
| 8 | SWP36985 | LARGE | 9 | 11 | 1989 | 3 | 15 | 2,447,780.6356 | 0.0378 |
| 9 | SWP03303 | SMALL | 11 | 11 | 1978 | 8 | 56 | 2,443,823.8726 | 0.0816 |
| 10 | SWP37021 | LARGE | 9 | 14 | 1989 | 3 | 21 | 2,447,783.6397 | 0.0855 |
| 11 | SWP36989 | LARGE | 9 | 11 | 1989 | 9 | 43 | 2,447,780.9050 | 0.1317 |
| 12 | SWP22417 | LARGE | 3 | 5 | 1984 | 17 | 58 | 2,445,765.2489 | 0.1538 |
| 13 | SWP22418 | LARGE | 3 | 5 | 1984 | 18 | 34 | 2,445,765.2739 | 0.1625 |
| 14 | SWP36992 | LARGE | 9 | 11 | 1989 | 13 | 39 | 2,447,781.0689 | 0.1889 |
| 15 | SWP36996 | LARGE | 9 | 11 | 1989 | 22 | 44 | 2,447,781.4474 | 0.3209 |
| 16 | SWP03752 | SMALL | 12 | 30 | 1978 | 19 | 52 | 2,443,873.3282 | 0.3297 |
| 17 | SWP37001 | LARGE | 9 | 12 | 1989 | 9 | 37 | 2,447,781.9008 | 0.4790 |
| 18 | SWP03818 | SMALL | 1 | 6 | 1979 | 0 | 43 | 2,443,879.5302 | 0.4927 |
| 19 | SWP37003 | LARGE | 9 | 12 | 1989 | 14 | 11 | 2,447,782.0911 | 0.5454 |
| 20 | SWP27781 | LARGE | 2 | 24 | 1986 | 19 | 36 | 2,446,486.3168 | 0.6326 |
| 21 | SWP02643 | SMALL | 9 | 13 | 1978 | 17 | 58 | 2,443,765.2490 | 0.6361 |
| 22 | SWP27782 | LARGE | 2 | 24 | 1986 | 20 | 28 | 2,446,486.3529 | 0.6452 |
| 23 | SWP27783 | LARGE | 2 | 24 | 1986 | 20 | 57 | 2,446,486.3731 | 0.6522 |
| 24 | SWP27784 | LARGE | 2 | 24 | 1986 | 21 | 27 | 2,446,486.3939 | 0.6595 |
| 25 | SWP27785 | LARGE | 2 | 24 | 1986 | 22 | 12 | 2,446,486.4251 | 0.6704 |
| 26 | SWP27786 | LARGE | 2 | 24 | 1986 | 22 | 40 | 2,446,486.4446 | 0.6772 |
| 27 | SWP03794 | SMALL | 1 | 3 | 1979 | 22 | 48 | 2,443,877.4504 | 0.7674 |
| 28 | SWP37011 | LARGE | 9 | 13 | 1989 | 6 | 13 | 2,447,782.7592 | 0.7784 |
| 29 | SWP27830 | LARGE | 3 | 2 | 1986 | 23 | 51 | 2,446,492.4939 | 0.7869 |
| 30 | SWP27831 | LARGE | 3 | 3 | 1986 | 0 | 47 | 2,446,492.5328 | 0.8005 |
| 31 | SWP27832 | LARGE | 3 | 3 | 1986 | 1 | 16 | 2,446,492.5529 | 0.8075 |
| 32 | SWP37014 | LARGE | 9 | 13 | 1989 | 13 | 18 | 2,447,783.0543 | 0.8813 |
| 33 | SWP26472 | LARGE | 7 | 25 | 1985 | 11 | 55 | 2,446,271.9967 | 0.8866 |
| 34 | SWP21452 | LARGE | 11 | 5 | 1983 | 13 | 35 | 2,445,644.0665 | 0.8904 |
| 35 | SWP26473 | LARGE | 7 | 25 | 1985 | 12 | 54 | 2,446,272.0376 | 0.9009 |
| 36 | SWP21453 | LARGE | 11 | 5 | 1983 | 14 | 44 | 2,445,644.1142 | 0.9070 |
| 37 | SWP26474 | LARGE | 7 | 25 | 1985 | 13 | 52 | 2,446,272.0779 | 0.9149 |
| 38 | SWP22411 | LARGE | 3 | 5 | 1984 | 2 | 0 | 2,445,764.5835 | 0.9217 |
| 39 | SWP21454 | LARGE | 11 | 5 | 1983 | 15 | 45 | 2,445,644.1565 | 0.9218 |
| 40 | SWP37016 | LARGE | 9 | 13 | 1989 | 16 | 17 | 2,447,783.1786 | 0.9247 |
| 41 | SWP26475 | LARGE | 7 | 25 | 1985 | 14 | 54 | 2,446,272.1210 | 0.9299 |
| 42 | SWP21455 | LARGE | 11 | 5 | 1983 | 16 | 43 | 2,445,644.1968 | 0.9358 |
| 43 | SWP22439 | LARGE | 3 | 7 | 1984 | 23 | 59 | 2,445,767.4996 | 0.9388 |
| 44 | SWP36979 | LARGE | 9 | 10 | 1989 | 20 | 30 | 2,447,780.3543 | 0.9397 |
| 45 | SWP26476 | LARGE | 7 | 25 | 1985 | 16 | 5 | 2,446,272.1703 | 0.9471 |
| 46 | SWP21456 | LARGE | 11 | 5 | 1983 | 18 | 11 | 2,445,644.2578 | 0.9571 |
| 47 | SWP36980 | LARGE | 9 | 10 | 1989 | 21 | 43 | 2,447,780.4050 | 0.9573 |
| 48 | SWP26477 | LARGE | 7 | 25 | 1985 | 17 | 13 | 2,446,272.2175 | 0.9636 |

| SWP ID | Data ID | Aper | Month | Day | Year | Hour | min | JD at mid exposure | Phase |
|---|---|---|---|---|---|---|---|---|---|
| 49 | SWP03298 | SMALL | 11 | 11 | 1978 | 1 | 2 | 2,443,823.5436 | 0.9669 |
| 50 | SWP21457 | LARGE | 11 | 5 | 1983 | 19 | 9 | 2,445,644.2983 | 0.9712 |
| 51 | SWP36981 | LARGE | 9 | 10 | 1989 | 23 | 7 | 2,447,780.4634 | 0.9777 |
| 52 | SWP03259 | SMALL | 11 | 8 | 1978 | 5 | 3 | 2,443,820.7111 | 0.9791 |
| 53 | SWP26478 | LARGE | 7 | 25 | 1985 | 18 | 17 | 2,446,272.2621 | 0.9791 |
| 54 | SWP21458 | LARGE | 11 | 5 | 1983 | 19 | 44 | 2,445,644.3226 | 0.9797 |
| 55 | SWP03260 | SMALL | 11 | 8 | 1978 | 5 | 37 | 2,443,820.7349 | 0.9874 |
| 56 | SWP03771 | SMALL | 1 | 1 | 1979 | 17 | 9 | 2,443,875.2158 | 0.9880 |
| 57 | SWP07110 | SMALL | 11 | 7 | 1979 | 9 | 28 | 2,444,184.8953 | 0.9915 |
| 58 | SWP36982 | LARGE | 9 | 11 | 1989 | 0 | 17 | 2,447,780.5122 | 0.9947 |
| 59 | SWP03261 | SMALL | 11 | 8 | 1978 | 6 | 9 | 2,443,820.7572 | 0.9951 |

In each of these tables, the first column identifies for use in this work an abbreviated designation used for cross-referencing and indexing. The second column is the unique identifying label assigned by the IUE mission. The third column indicates the aperture (large or small). Columns 4-8 are the month, day, year, hour, and minute of the exposure. Column 9 is the heliocentric JD at mid exposure, which is a simple average of the JD's representing the start and end times of each exposure. The last column is the phase, calculated using equations (3.2.1) and (3.2.2) above.

TABLE 3.2.4 is a comparison of our calculated phases with those calculated in other work. The differences in phases stem from the use of different ephemeredes; i.e., slightly different values of orbital parameters, such as time of primary minimum and period.

TABLE 3.2.3
IUE LWP/LWR Observations of Algol

| LWP/LWR ID | Data ID | Aper | Month | Day | Year | Hour | min | JD at mid exposure | Phase |
|---|---|---|---|---|---|---|---|---|---|
| 1 | LWR03350 | SMALL | 1 | 1 | 1979 | 18 | 12 | 2,443,875.2588 | 0.0030 |
| 2 | LWP16323 | LARGE | 9 | 11 | 1989 | 0 | 53 | 2,447,780.5371 | 0.0034 |
| 3 | LWR06047 | SMALL | 11 | 7 | 1979 | 10 | 17 | 2,444,184.9295 | 0.0034 |
| 4 | LWR02856 | SMALL | 11 | 8 | 1978 | 6 | 50 | 2,443,820.7851 | 0.0049 |
| 5 | LWR06048 | SMALL | 11 | 7 | 1979 | 11 | 11 | 2,444,184.9668 | 0.0165 |
| 6 | LWP16324 | LARGE | 9 | 11 | 1989 | 2 | 10 | 2,447,780.5905 | 0.0220 |
| 7 | LWP16325 | LARGE | 9 | 11 | 1989 | 3 | 20 | 2,447,780.6390 | 0.0390 |
| 8 | LWR02911 | SMALL | 11 | 11 | 1978 | 8 | 50 | 2,443,823.8683 | 0.0801 |
| 9 | LWP16356 | LARGE | 9 | 14 | 1989 | 3 | 26 | 2,447,783.6432 | 0.0867 |
| 10 | LWP16327 | LARGE | 9 | 11 | 1989 | 9 | 37 | 2,447,780.9008 | 0.1303 |
| 11 | LWP02895 | LARGE | 3 | 5 | 1984 | 18 | 2 | 2,445,765.2514 | 0.1547 |
| 12 | LWP16329 | LARGE | 9 | 11 | 1989 | 13 | 45 | 2,447,781.0730 | 0.1903 |
| 13 | LWP16333 | LARGE | 9 | 11 | 1989 | 22 | 39 | 2,447,781.4438 | 0.3196 |
| 14 | LWR03328 | SMALL | 12 | 30 | 1978 | 19 | 12 | 2,443,873.3003 | 0.3200 |
| 15 | LWR03329 | SMALL | 12 | 30 | 1978 | 19 | 45 | 2,443,873.3231 | 0.3280 |
| 16 | LWP16338 | LARGE | 9 | 12 | 1989 | 9 | 32 | 2,447,781.8973 | 0.4778 |
| 17 | LWR03398 | SMALL | 1 | 6 | 1979 | 0 | 48 | 2,443,879.5335 | 0.4939 |
| 18 | LWP16340 | LARGE | 9 | 12 | 1989 | 14 | 15 | 2,447,782.0939 | 0.5463 |
| 19 | LWP07717 | LARGE | 2 | 24 | 1986 | 19 | 43 | 2,446,486.3216 | 0.6343 |
| 20 | LWR02345 | SMALL | 9 | 13 | 1978 | 19 | 4 | 2,443,765.2946 | 0.6520 |
| 21 | LWR03374 | SMALL | 1 | 3 | 1979 | 22 | 53 | 2,443,877.4536 | 0.7685 |
| 22 | LWP16347 | LARGE | 9 | 13 | 1989 | 6 | 19 | 2,447,782.7633 | 0.7798 |
| 23 | LWP07738 | LARGE | 3 | 2 | 1986 | 23 | 57 | 2,446,492.4980 | 0.7884 |
| 24 | LWP16350 | LARGE | 9 | 13 | 1989 | 13 | 23 | 2,447,783.0577 | 0.8825 |
| 25 | LWP06484 | LARGE | 7 | 25 | 1985 | 12 | 0 | 2,446,272.0001 | 0.8878 |
| 26 | LWP02226 | LARGE | 11 | 5 | 1983 | 13 | 40 | 2,445,644.0698 | 0.8915 |
| 27 | LWP06485 | LARGE | 7 | 25 | 1985 | 12 | 58 | 2,446,272.0404 | 0.9018 |
| 28 | LWP02227 | LARGE | 11 | 5 | 1983 | 14 | 49 | 2,445,644.1176 | 0.9082 |
| 29 | LWP06486 | LARGE | 7 | 25 | 1985 | 13 | 59 | 2,446,272.0827 | 0.9166 |
| 30 | LWP02228 | LARGE | 11 | 5 | 1983 | 15 | 49 | 2,445,644.1591 | 0.9227 |
| 31 | LWP16352 | LARGE | 9 | 13 | 1989 | 16 | 22 | 2,447,783.1820 | 0.9259 |
| 32 | LWP06487 | LARGE | 7 | 25 | 1985 | 15 | 4 | 2,446,272.1279 | 0.9323 |
| 33 | LWP02229 | LARGE | 11 | 5 | 1983 | 16 | 47 | 2,445,644.1994 | 0.9367 |
| 34 | LWP16320 | LARGE | 9 | 10 | 1989 | 20 | 35 | 2,447,780.3577 | 0.9409 |
| 35 | LWP06488 | LARGE | 7 | 25 | 1985 | 16 | 12 | 2,446,272.1751 | 0.9488 |
| 36 | LWP02230 | LARGE | 11 | 5 | 1983 | 18 | 15 | 2,445,644.2605 | 0.9581 |
| 37 | LWP16321 | LARGE | 9 | 10 | 1989 | 21 | 50 | 2,447,780.4098 | 0.9590 |
| 38 | LWP06489 | LARGE | 7 | 25 | 1985 | 17 | 19 | 2,446,272.2216 | 0.9650 |
| 39 | LWR02906 | SMALL | 11 | 11 | 1978 | 1 | 8 | 2,443,823.5476 | 0.9683 |
| 40 | LWP02231 | LARGE | 11 | 5 | 1983 | 19 | 13 | 2,445,644.3009 | 0.9721 |
| 41 | LWP16322 | LARGE | 9 | 10 | 1989 | 23 | 14 | 2,447,780.4682 | 0.9794 |
| 42 | LWP06490 | LARGE | 7 | 25 | 1985 | 18 | 21 | 2,446,272.2647 | 0.9801 |
| 43 | LWR03349 | SMALL | 1 | 1 | 1979 | 17 | 16 | 2,443,875.2199 | 0.9895 |
| 44 | LWR06046 | SMALL | 11 | 7 | 1979 | 9 | 23 | 2,444,184.8913 | 0.9901 |

TABLE 3.2.4
Comparison of IUE Phases to Other Work

| | SWP Exposure Information | | | | SWP Phase | | | |
|---|---|---|---|---|---|---|---|---|
| SWP ID | Data ID | Aper | Month | Year | Cugier & Molaro (1984) | Sahade & Hernandez (1985) | Brandi et al. (1989)` | This work (2006) |
| 1 | SWP03772 | SMALL | 1 | 1979 | 0.00216 | 0.018 | ... | 0.0015 |
| 2 | SWP03262 | SMALL | 11 | 1978 | 0.00424 | ... | ... | 0.0031 |
| 4 | SWP07111 | SMALL | 11 | 1979 | 0.00636 | ... | ... | 0.0053 |
| 6 | SWP07112 | SMALL | 11 | 1979 | 0.01920 | ... | ... | 0.0183 |
| 9 | SWP03303 | SMALL | 11 | 1978 | 0.08298 | ... | ... | 0.0816 |
| 16 | SWP03752 | SMALL | 12 | 1978 | 0.33060 | 0.346 | ... | 0.3297 |
| 18 | SWP03818 | SMALL | 1 | 1979 | 0.49347 | 0.509 | ... | 0.4927 |
| 20 | SWP27781 | LARGE | 2 | 1986 | ... | ... | 0.63 | 0.6326 |
| 21 | SWP02643 | SMALL | 9 | 1978 | 0.63637 | ... | ... | 0.6361 |
| 22 | SWP27782 | LARGE | 2 | 1986 | ... | ... | 0.64 | 0.6452 |
| 23 | SWP27783 | LARGE | 2 | 1986 | ... | ... | 0.65 | 0.6522 |
| 24 | SWP27784 | LARGE | 2 | 1986 | ... | ... | 0.65 | 0.6595 |
| 25 | SWP27785 | LARGE | 2 | 1986 | ... | ... | 0.66 | 0.6704 |
| 26 | SWP27786 | LARGE | 2 | 1986 | ... | ... | 0.67 | 0.6772 |
| 27 | SWP03794 | SMALL | 1 | 1979 | 0.76818 | 0.784 | ... | 0.7674 |
| 29 | SWP27830 | LARGE | 3 | 1986 | ... | ... | 0.78 | 0.7869 |
| 30 | SWP27831 | LARGE | 3 | 1986 | ... | ... | 0.79 | 0.8005 |
| 31 | SWP27832 | LARGE | 3 | 1986 | ... | ... | 0.80 | 0.8075 |
| 49 | SWP03298 | SMALL | 11 | 1978 | 0.96818 | ... | ... | 0.9669 |
| 52 | SWP03259 | SMALL | 11 | 1978 | 0.98026 | ... | ... | 0.9791 |
| 55 | SWP03260 | SMALL | 11 | 1978 | 0.98849 | ... | ... | 0.9874 |
| 56 | SWP03771 | SMALL | 1 | 1979 | 0.98860 | 0.004 | ... | 0.9880 |
| 57 | SWP07110 | SMALL | 11 | 1979 | 0.99280 | ... | ... | 0.9915 |
| 59 | SWP03261 | SMALL | 11 | 1978 | 0.99624 | ... | ... | 0.9951 |

| | LWP/LWR Exposure Information | | | | LWP/LWR Phase | | | |
|---|---|---|---|---|---|---|---|---|
| LWP/ LWR ID | Data ID | Aper | Month | Year | Cugier & Molaro (1984) | Sahade & Hernandez (1985) | Brandi et al. (1989)` | This work (2006) |
| 1 | LWR03350 | SMALL | 1 | 1979 | 0.00386 | 0.019 | ... | 0.0030 |
| 3 | LWR06047 | SMALL | 11 | 1979 | 0.99280 | ... | ... | 0.0034 |
| 4 | LWR02856 | SMALL | 11 | 1978 | 0.00617 | ... | ... | 0.0049 |
| 5 | LWR06048 | SMALL | 11 | 1979 | 0.01775 | ... | ... | 0.0165 |
| 8 | LWR02911 | SMALL | 11 | 1978 | 0.08153 | ... | ... | 0.0801 |
| 14 | LWR03328 | SMALL | 12 | 1978 | 0.32092 | 0.336 | ... | 0.3200 |
| 15 | LWR03329 | SMALL | 12 | 1978 | 0.32891 | 0.344 | ... | 0.3280 |
| 17 | LWR03398 | SMALL | 1 | 1979 | 0.49463 | 0.510 | ... | 0.4939 |
| 20 | LWR02345 | SMALL | 9 | 1978 | 0.65235 | ... | ... | 0.6520 |
| 21 | LWR03374 | SMALL | 1 | 1979 | 0.76939 | 0.785 | ... | 0.7685 |
| 39 | LWR02906 | SMALL | 11 | 1978 | 0.96964 | ... | ... | 0.9683 |
| 43 | LWR03349 | SMALL | 1 | 1979 | 0.99030 | 0.006 | ... | 0.9895 |
| 44 | LWR06046 | SMALL | 11 | 1979 | 0.99159 | ... | ... | 0.9901 |

## 3.3 *Line-of-sight Geometry for Each Algol IUE Observation*

The IUE spectra contain two types of changing features: those associated with periodic changes in our line of sight and identified by phase, and others related to actual changes in binary system activity, identified by time period, or "epoch." Each IUE observation of Algol is correlated with a specific date and time which is in turn related to a specific orbital phase (see Section 3.2). It is therefore important to visualize the line-of-sight geometry of each phase as well as the phases observed during each epoch.

FIGS. (3.3.1a-h) and FIGS. (3.3.2a-h) illustrate the line-of-sight geometry for each Algol IUE SWP and LWP/LWR observation, respectively, grouped by epoch. The epochs are grouped as follows: 1978 Sept. - 1978 Nov., 1978 Dec. -1979 Jan., 1979 Nov., 1983 Nov., 1984 March, 1985 July, 1986 Feb. - March, and 1989 Sept. where we have combined the lone 1978 September observation (in each of the wavelength ranges) with the 1978 November observations, keeping in mind that marked changes can (and did) occur in the system across that time span.

The vertical dashed line in each figure is the axis of rotation, the sense of which is up, and passes through the center of mass of the system. The large dark spot on Algol A marks the impact region of the stream of gas according to the model of Harnden et al. 1977 (See FIG. 2.5.4). It is located at an angle of $43^o$ from a line connecting the centers of the two components, originating from Algol A. The four smaller spots on Algol A are simply markers to help the eye keep track of which parts of the hemispheres are visible at each phase. These four markers are located on the equator at $0^o$, $90^o$, $180^o$, and $270^o$ with respect to a line connecting the center of Algol A to the center of Algol B. The + marks indicate the positions of the center of mass of Algol A, Algol B, and Algol A-B.

**SWP**                                                         **1978 Sep - Nov**

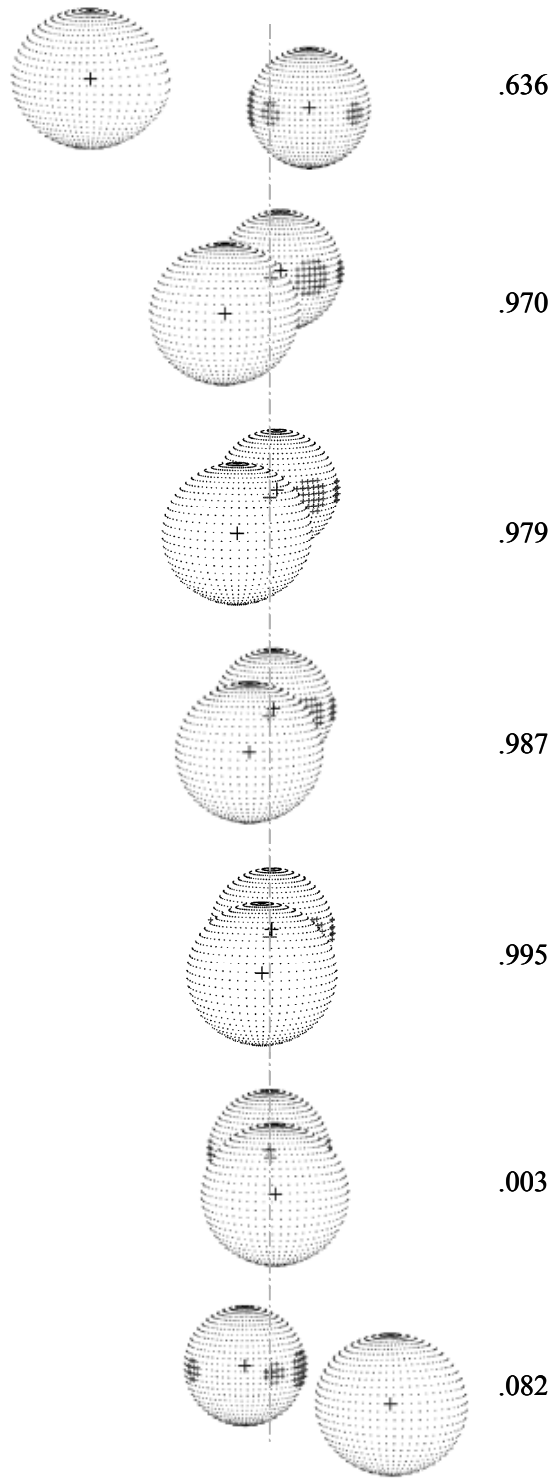

.636

.970

.979

.987

.995

.003

.082

FIG. 3.3.1a–*Line-of-sight Geometry for Epoch 1978 Sept. – Nov. SWP phases*

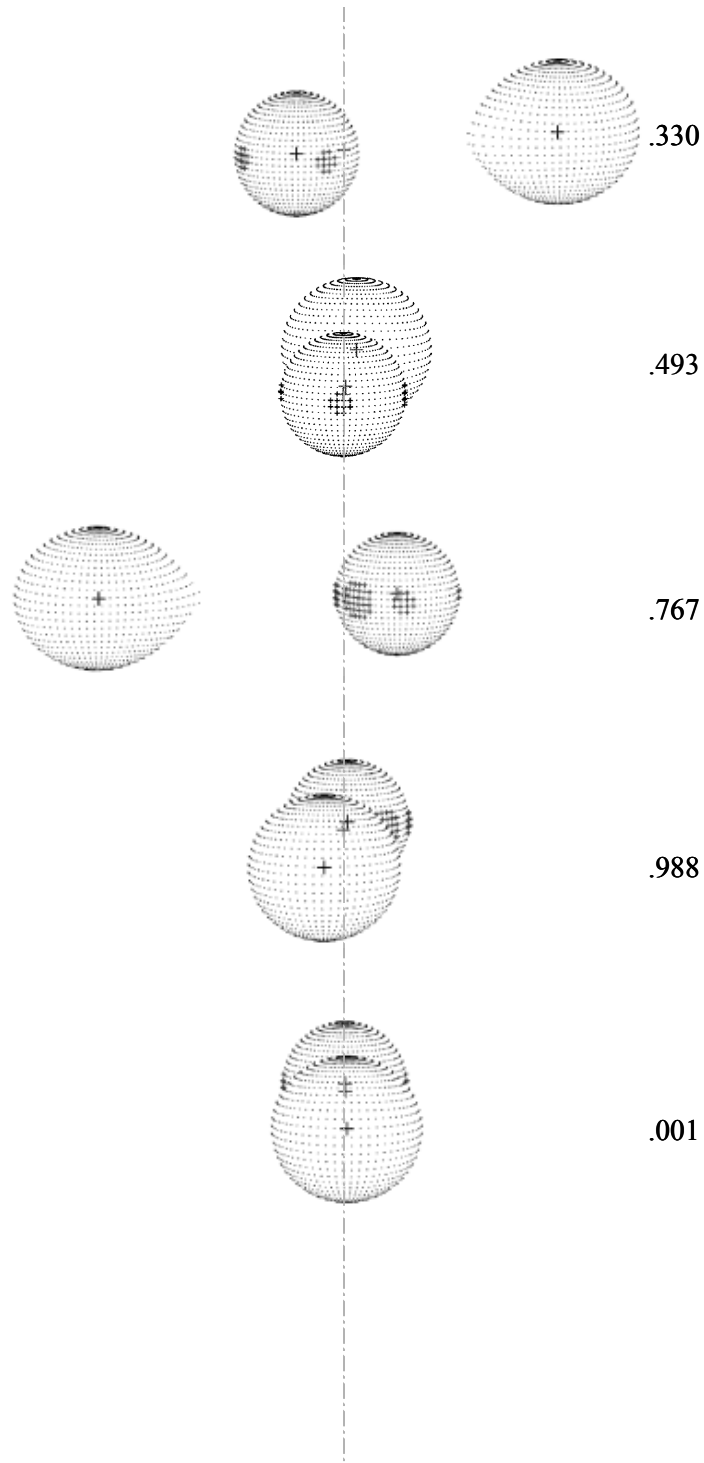

FIG. 3.3.1b–*Line-of-sight Geometry for Epoch 1978 Dec. – 1979 Jan. SWP phases*

**SWP** **1979 Nov**

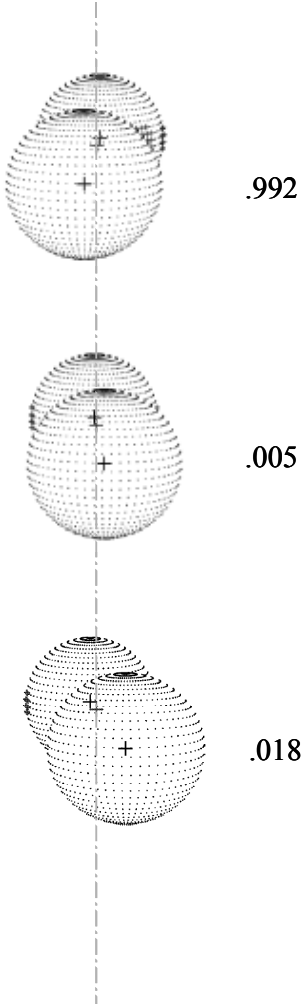

.992

.005

.018

FIG. 3.3.1c—*Line-of-sight Geometry for Epoch 1979 November SWP phases*

**SWP** **1983 Nov**

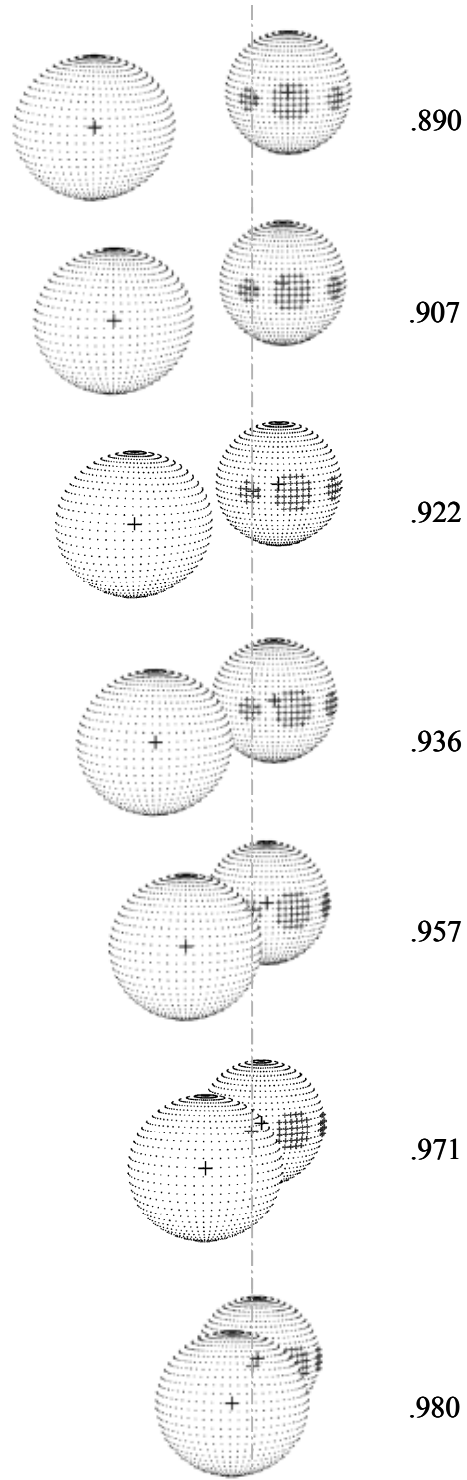

.890

.907

.922

.936

.957

.971

.980

FIG. 3.3.1d—*Line-of-sight Geometry for Epoch 1983 November SWP phases*

**SWP**                                                        **1984 Mar**

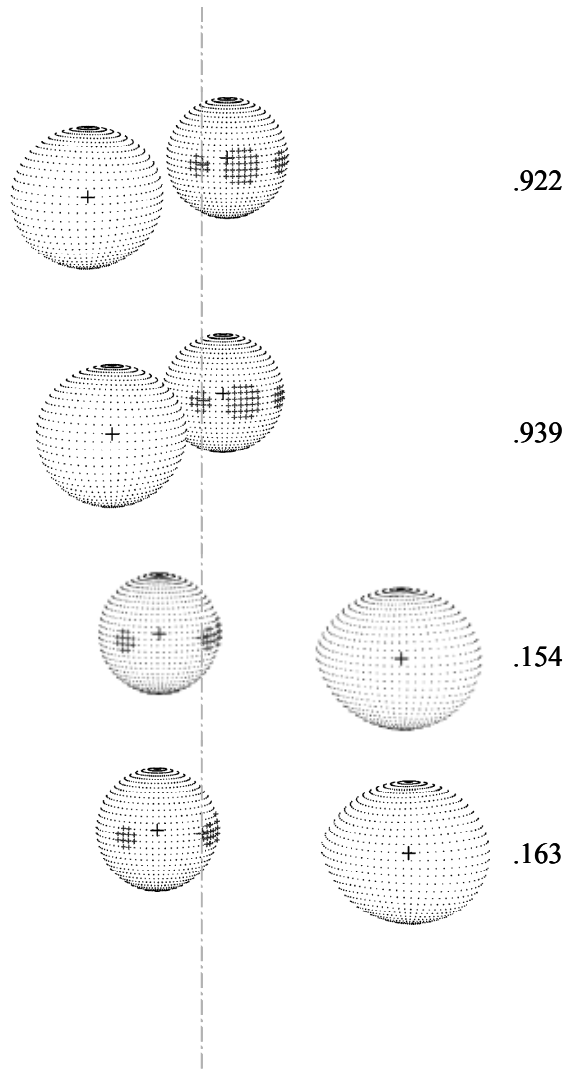

FIG. 3.3.1e–*Line-of-sight Geometry for Epoch 1984 March SWP phases*

**SWP**                                                      **1985 Jul**

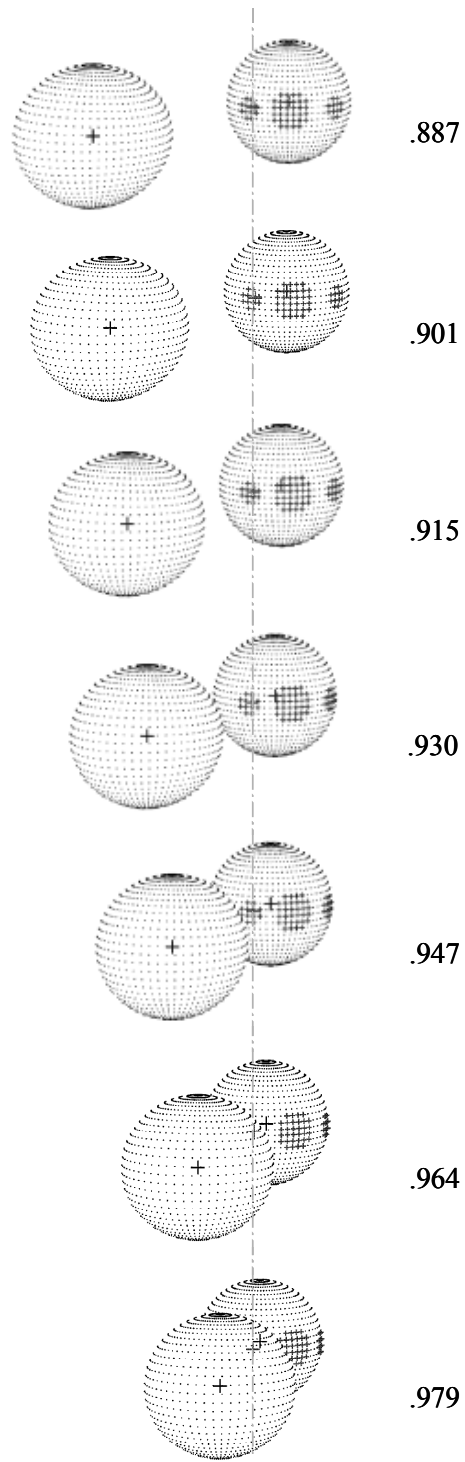

.887

.901

.915

.930

.947

.964

.979

FIG. 3.3.1f–*Line-of-sight Geometry for Epoch 1985 July SWP phases*

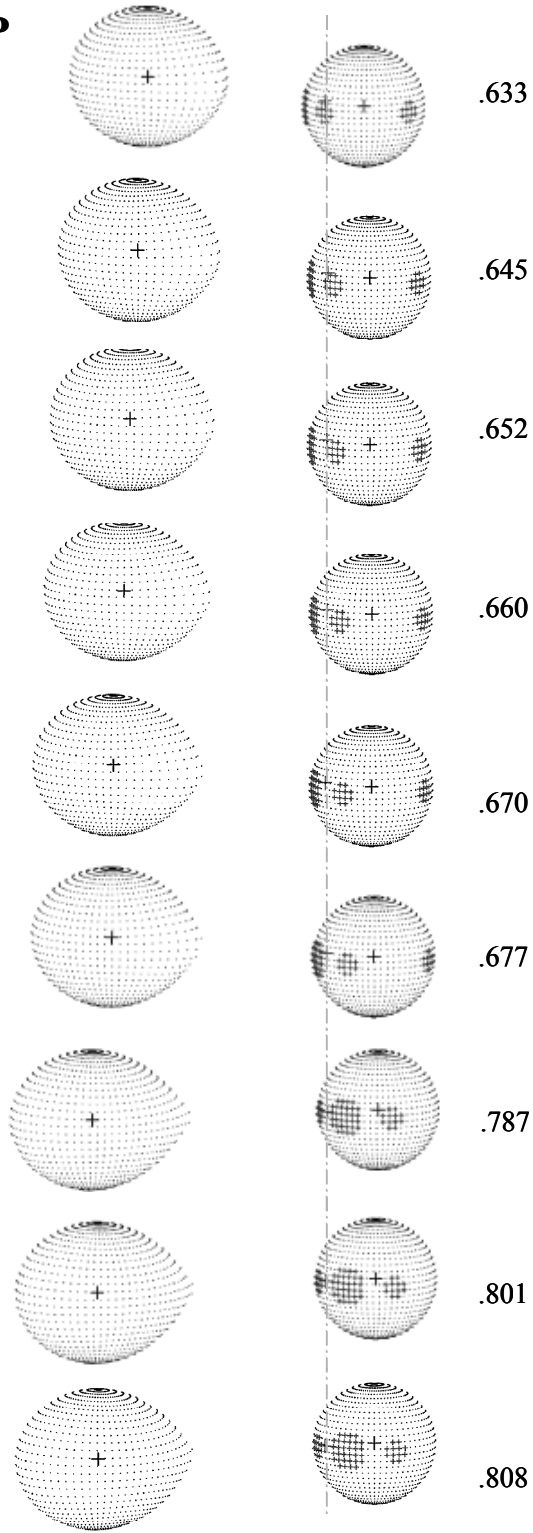

FIG. 3.3.1g–*Line-of-sight Geometry for Epoch 1986 Feb. - March SWP phases*

**SWP**            **1989 Sep**

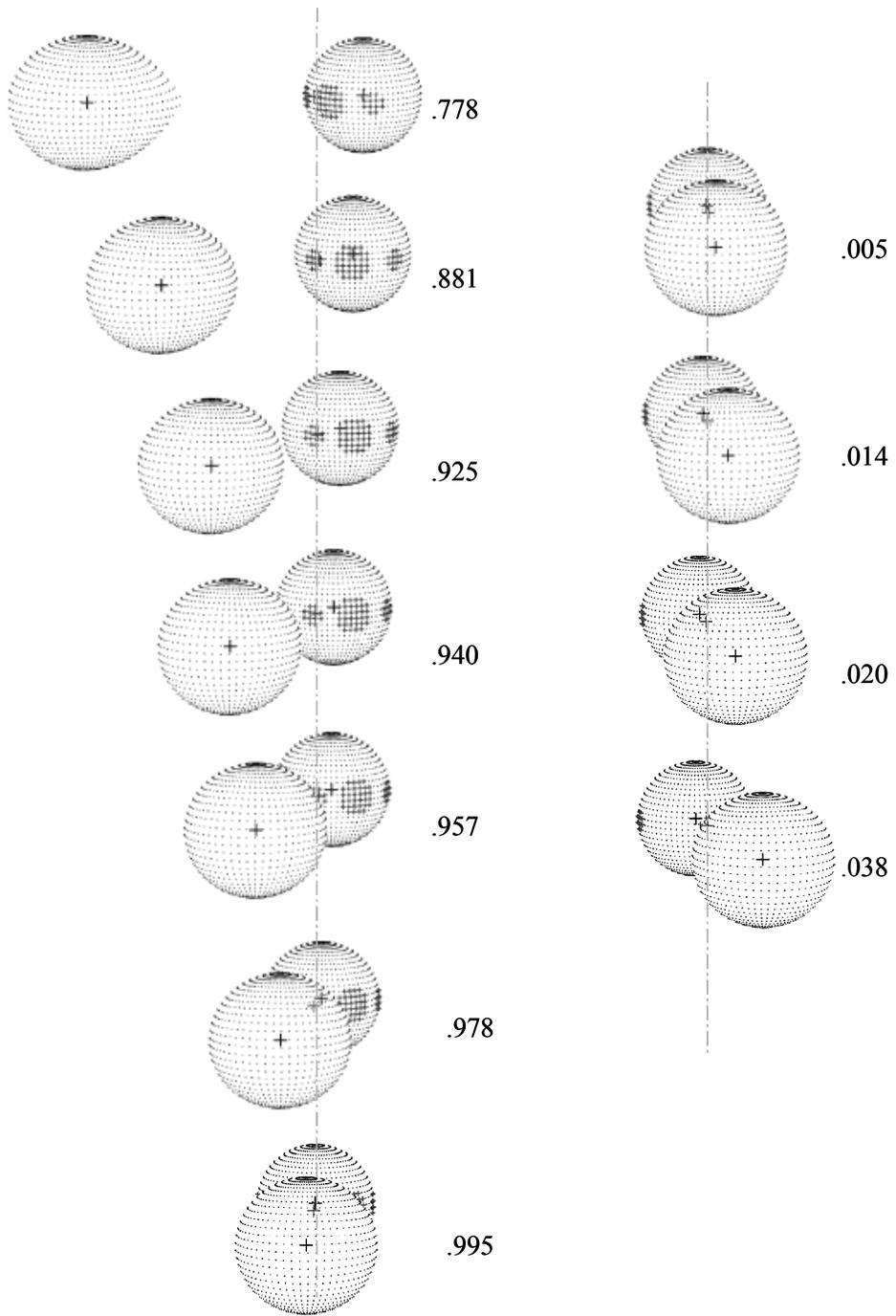

FIG. 3.3.1h–*Line-of-sight Geometry for Epoch 1989 September SWP phases*

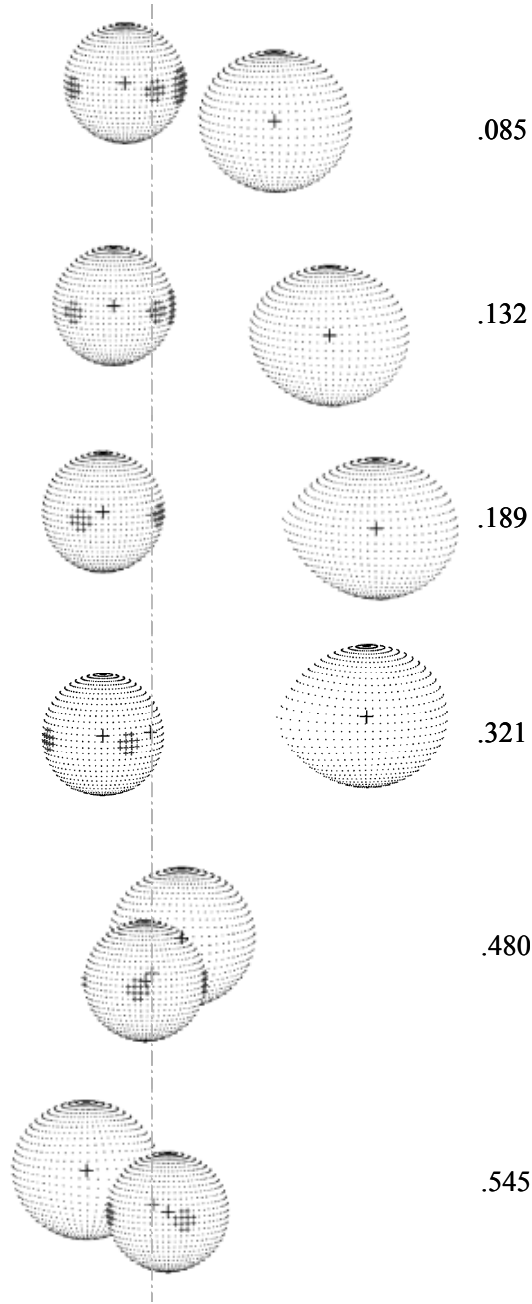

.085

.132

.189

.321

.480

.545

FIG 3.3.1h–*continued*

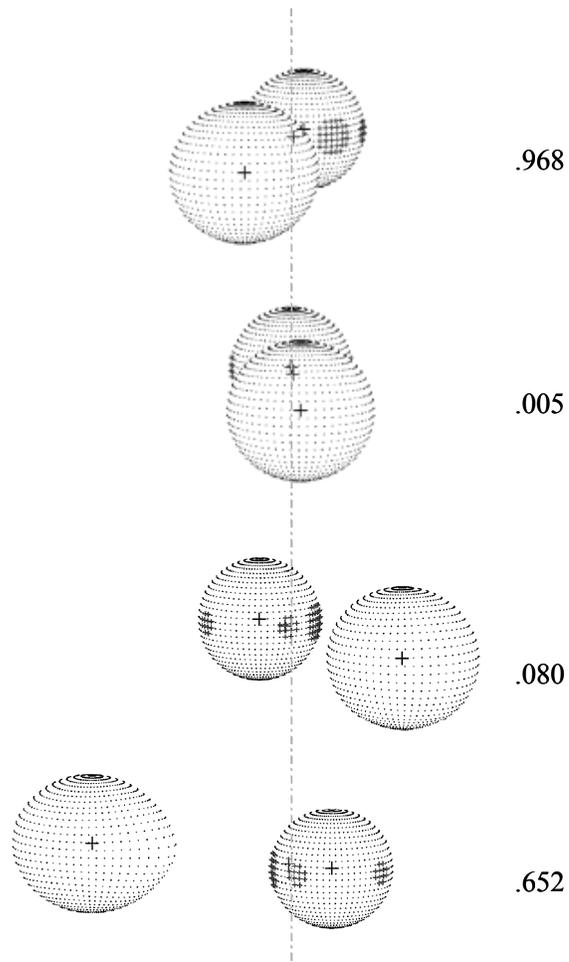

.968

.005

.080

.652

FIG. 3.3.2a–*Line-of-sight Geometry for Epoch 1978 Sept. – Nov. LWP/LWR phases*

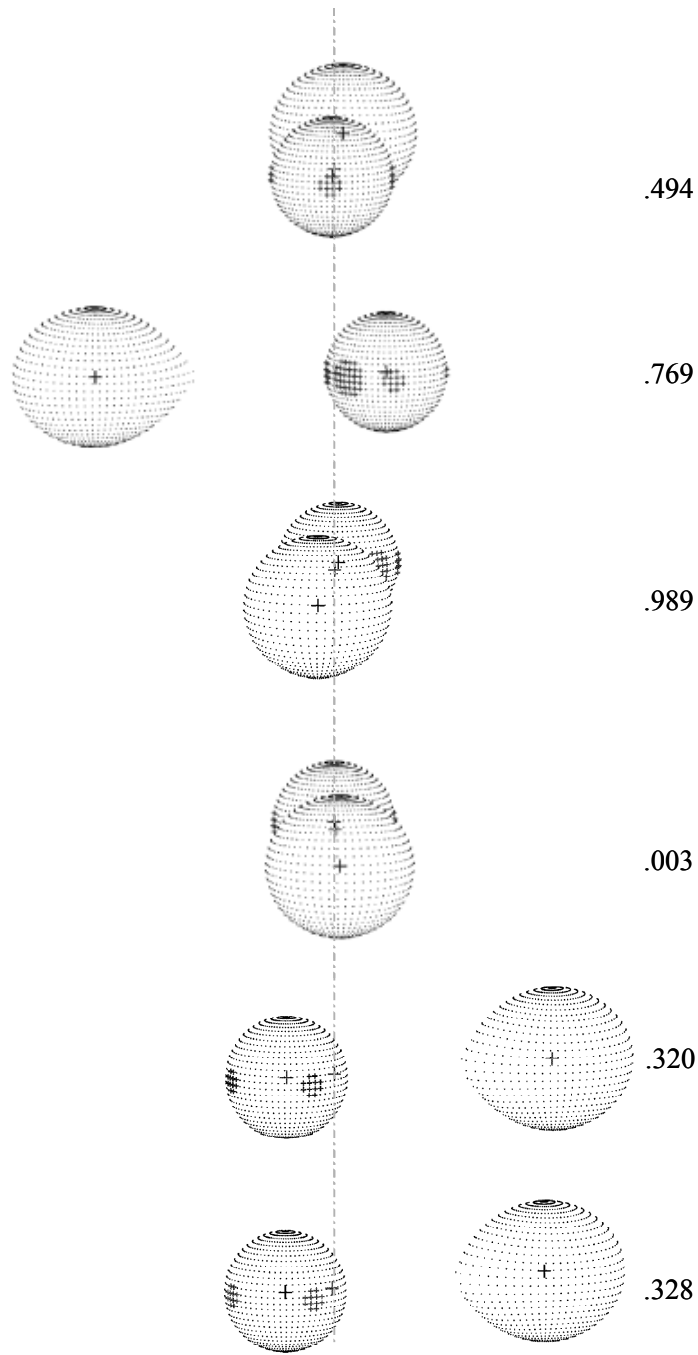

.494

.769

.989

.003

.320

.328

FIG. 3.3.2b—*Line-of-sight Geometry for Epoch 1978 Dec. – 1979 LWP/LWR phases*

**LWP/LWR**                                    **1979 Nov**

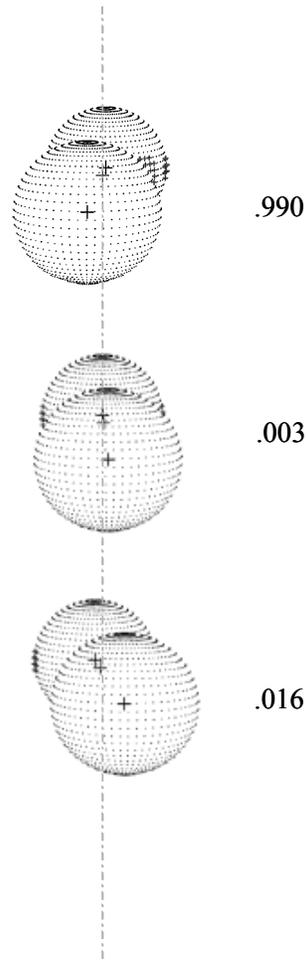

FIG. 3.3.2c—*Line-of-sight Geometry for Epoch 1979 November LWP/LWR phases*

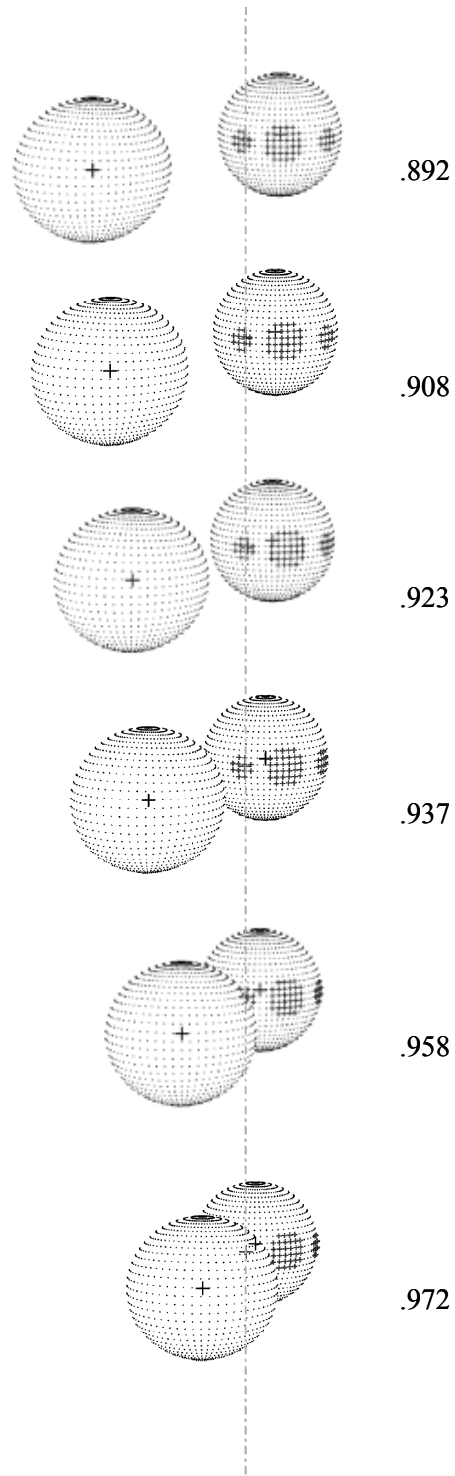

.892

.908

.923

.937

.958

.972

FIG. 3.3.2d–*Line-of-sight Geometry for Epoch 1983 November LWP/LWR phases*

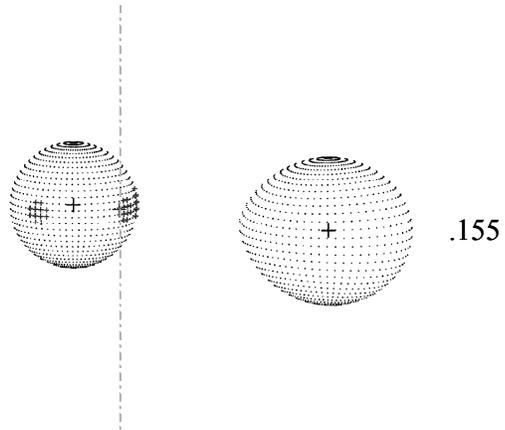

FIG. 3.3.2e–*Line-of-sight Geometry for Epoch 1984 March LWP/LWR phases*

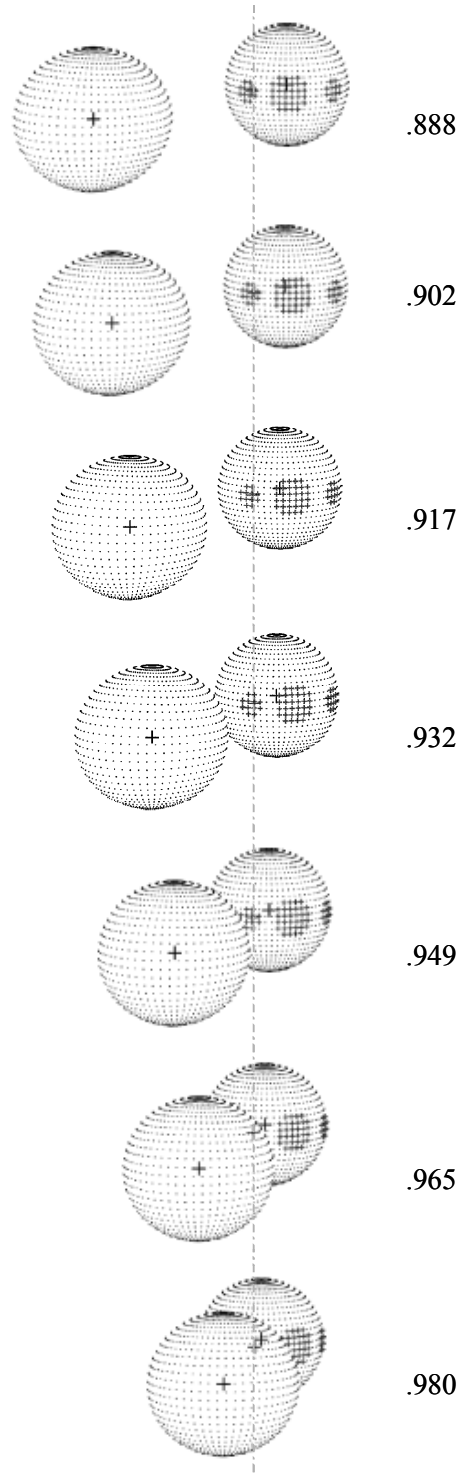

FIG. 3.3.2f–*Line-of-sight Geometry for Epoch 1985 July LWP/LWR phases*

**LWP/LWR** **1986 Feb - Mar**

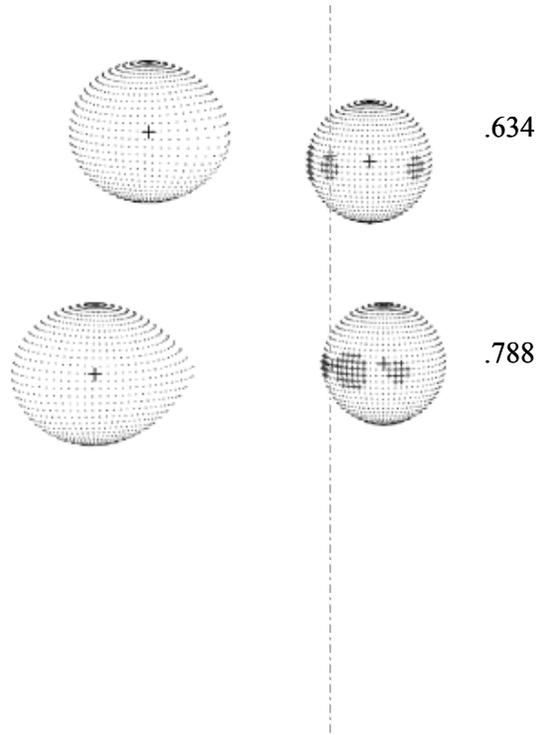

.634

.788

FIG. 3.3.2g–*Line-of-sight Geometry for Epoch 1986 Feb. - March LWP/LWR phases*

**LWP/LWR**  **1989 Sep**

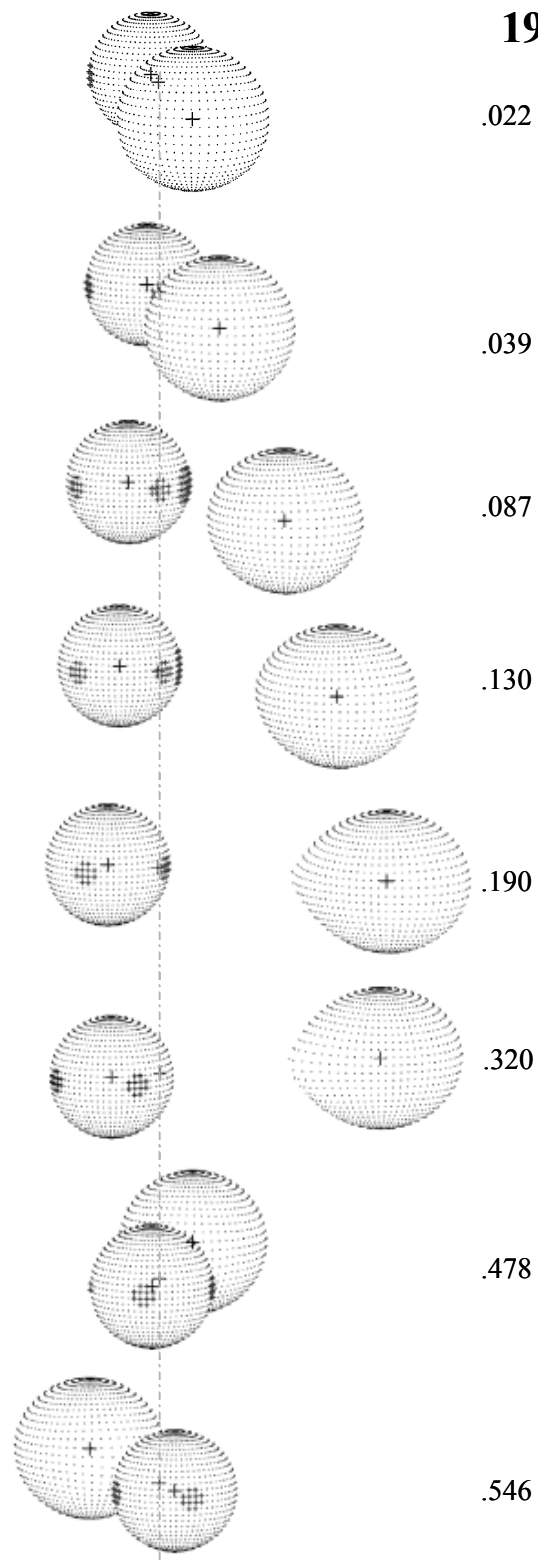

.022

.039

.087

.130

.190

.320

.478

.546

FIG. 3.3.2h–*Line-of-sight Geometry for Epoch 1989 September LWP/LWR phases*

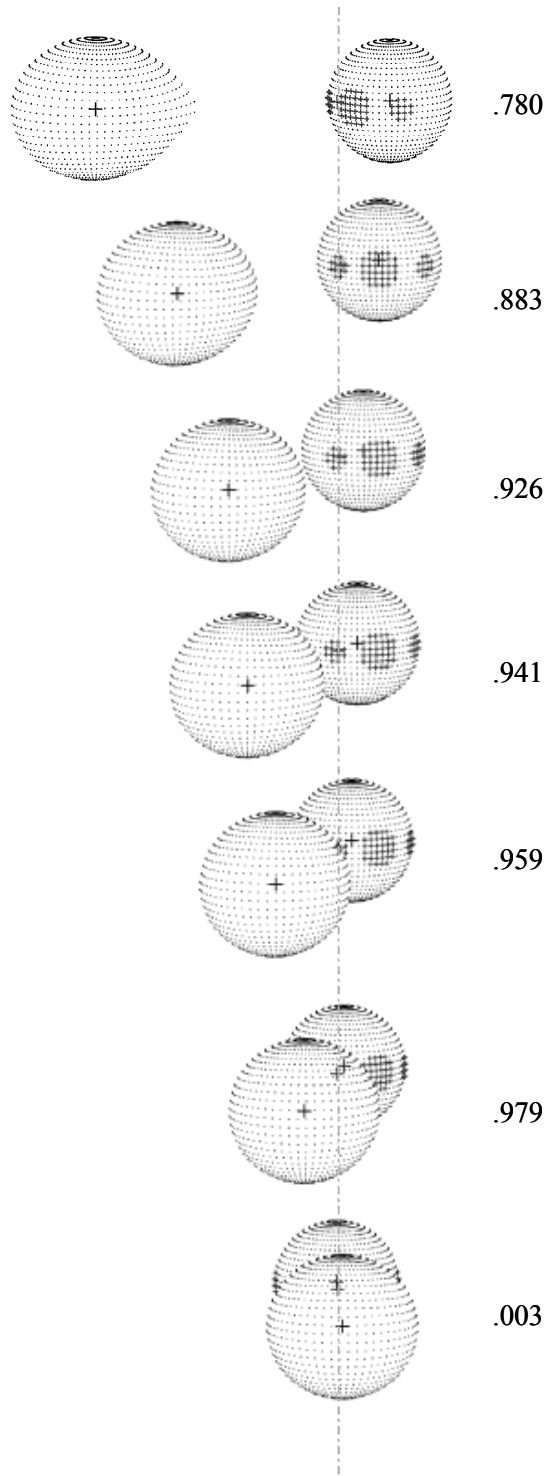

.780

.883

.926

.941

.959

.979

.003

FIG. 3.3.2h—*continued*

These figures (FIGS. 3.3.1a-h and 3.3.2a-h) were generated using a software package called Binary Maker 3.0.[*] Its engine is based on the Wilson-Divinney code[†]. This program outputs light curves, radial velocity curves, and orbit geometry. We varied the input parameters, selecting combinations of the values listed in the Tables in Appendix A, until the simulated radial velocity curve from the program matched our best measured radial velocity curve, that of Al II λ1670 (discussed in Section 4). The final input parameters are listed in TABLE 3.3.1. (These values are provided in this table for completeness, but play no role in this work other than the generation of the figures.)

TABLE 3.3.1
INPUT PARAMETERS FOR BINARY MAKER 3.0

| | |
|---|---|
| mass ratio input = 0.218900 | mass ratio < 1 = 0.218900 |
| Omega 1 = 5.088680 | Omega 2 = 2.279096 |
| Omega inner = 2.279097 | Omega outer = 2.140058 |
| C 1 = 8.381879 | C 2 = 3.771847 |
| C inner = 3.771848 | C outer = 3.543710 |
| Fillout 1 = -0.550000 | Fillout 2 = 0.000002 |
| Lagrangian L1 = 0.650349 | Lagrangian L2 = 1.451217 |
| AG = r1(back) = 0.206784 | AS = r2(back) = 0.281108 |
| BG = r1(side) = 0.206255 | BS = r2(side) = 0.248632 |
| CG = r1(pole) = 0.205160 | CS = r2(pole) = 0.238990 |
| DG = r1(point) = 0.206960 | DS = r2(point) = 0.349510 |
| Surface area 1 = 0.533939 | Surface area 2 = 0.839262 |
| Volume 1 = 0.036403 | Volume 2 = 0.070632 |
| Mean radius 1 = 0.206066 | Mean radius 2 = 0.256243 |
| Mean radius 1 (vol) = 0.205597 | Mean radius 2 (vol) = 0.256431 |
| Eccentricity = 0.00000 | Longitude of Periastron = 133.0000 |
| Phase of periastron = 0.00000 | Phase of conjunction = 0.00000 |
| Angular Rotation F1 = 1.0000 | Angular Rotation F2 = 1.0000 |
| Normalization Phase = 0.25000 | Normalization Factor = 1.00000 |
| inclination = 81.400 | wavelength = 2000.00 |
| temperature 1 = 12500.00 | temperature 2 = 5000.00 |
| luminosity 1 = 0.9997 | luminosity 2 = 0.0003 |
| gravity coefficient 1 = 0.250 | gravity coefficient 2 = 0.080 |
| limb darkening 1 = 0.850 | limb darkening 2 = 0.560 |
| reflection 1 = 1.000 | reflection 2 = 0.500 |
| Third light = 0.0000 | Period = 2.86731000 |
| K1 = 44.000000 | K2 = 201.000000 |
| V0 = 0.000000 | |
| Absolute Parameters | |
| Mass 1 = 3.721372 solar masses | Mass 2 = 0.814629 solar masses |
| Semi-major axis = 9.769090 million km | Semi-major axis = 14.036050 solar radii |
| Mean radius 1 = 2.013081 million km | Mean radius 1 = 2.892357 solar radii |
| Mean radius 2 = 2.503266 million km | Mean radius 2 = 3.596646 solar radii |
| Mean density 1 = 0.218199 grams/cm^3 | Mean density 2 = 0.024618 grams/cm^3 |

---

[*] © Contact Software, Binary Maker 3, D.H. Bradstreet, D.P. Steelman, dbradstr@eastern.edu
[†] R. E. Wilson and E. J. Divnney (1971)

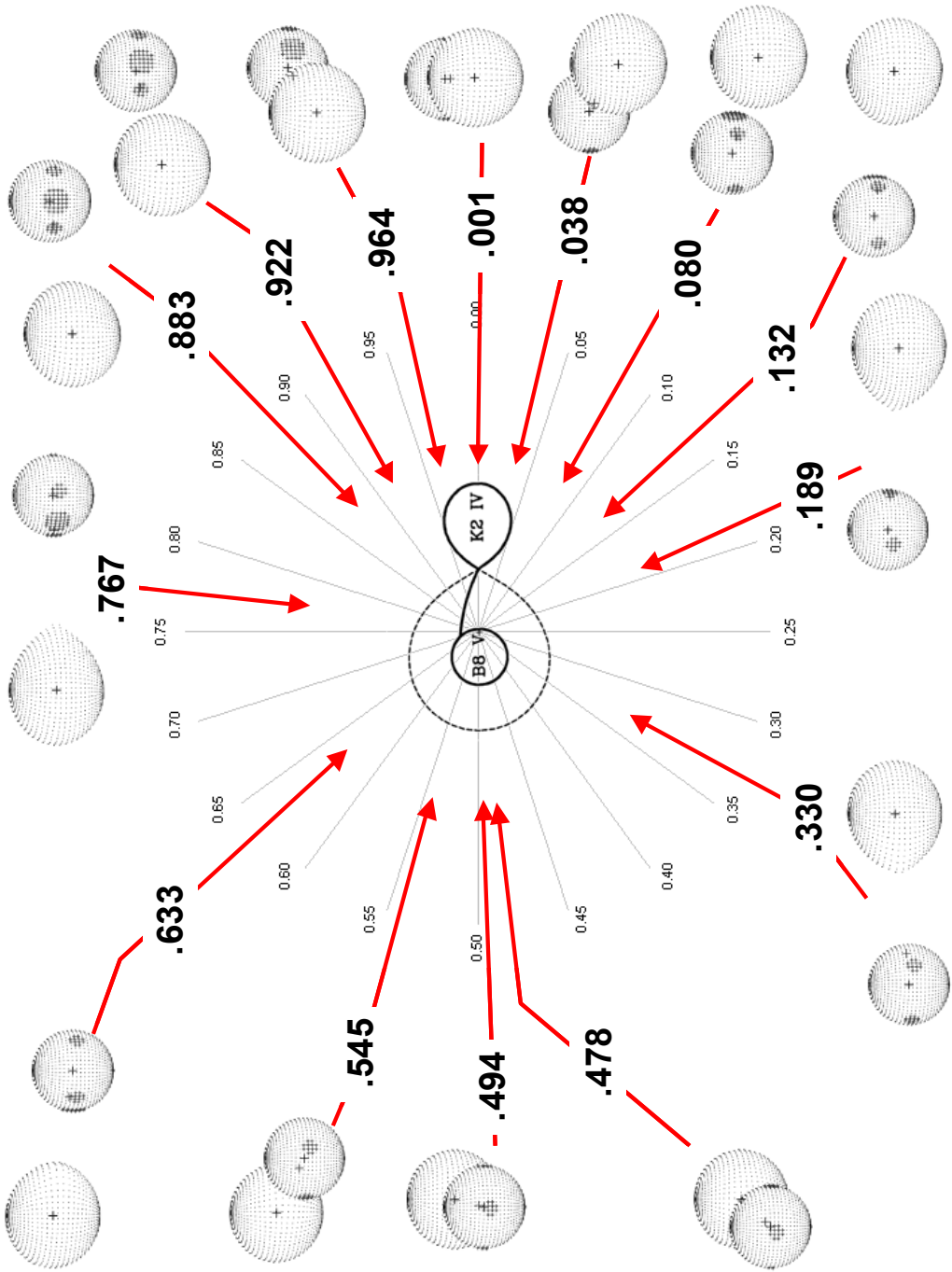

FIG. 3.3.3—Selected Line of Sight Geometries of Algol IUE Observations

3.4  *Sample IUE Data for Algol*

Before proceeding into a detailed analysis of the IUE data in the following chapter, we present a sample of the data in order to clarify the meaning of essential features, especially in the context of the phases described above. The samples in FIG. 3.4.1a and FIG. 3.4.1b represent spectra in the SWP UV range 1390 Å to 1405 Å taken 1.1 orbits apart, on 3/5/1984 ($\phi$ = 0.922) and 3/7/1984 ($\phi$ = 0.939) respectively. In each of the figures the binary-system orientations at the corresponding phases

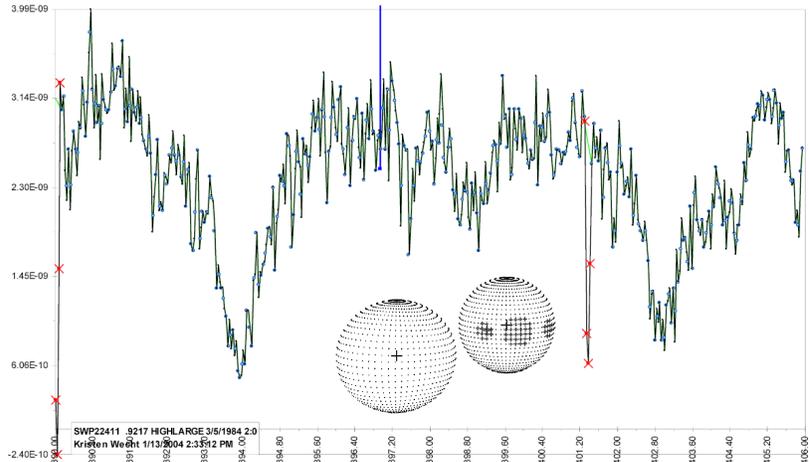
FIG. 3.4.1a—Sample IUE data SWP 22411 Si IV 1393 Å and Si IV 1402 Å ($\phi$=0.9217 β Per 3/5/1984)

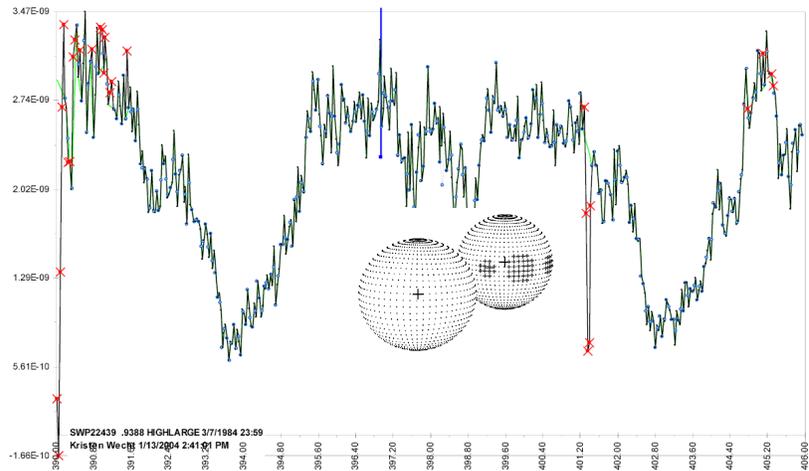
FIG. 3.4.1b–Sample IUE data SWP 22439  Si IV 1393 Å and Si IV 1402 Å ($\phi$=0.9388 β Per 3/7/1984)

are shown as viewed along our line of sight from the Earth.  (These orientations are just those presented in the previous section.)  The small x's mark the flagged points.  The small dots mark the "preview" data points. The black thin jagged line is the best output product from the NEWSIPS processing.

These spectra primarily show absorption features, the strongest of which is the resonance absorption doublet of Si IV at 1393 and 1402 Å. These 1984 line profiles of Si IV become increasingly asymmetric, (additional red absorption) in going from phase 0.922 to 0.939. (This phase interval is the beginning of primary eclipse.) The increasing red absorption is due to a change in our viewing angle of the gas stream, from looking diagonally across the stream to looking more along the stream.

It is important to understand that only the B8 V (primary) star is emitting significantly in the UV spectral range. Hence, the spectral features represent the absorption of ultraviolet light emitted from the primary by constituents located between the B8 V star and the Earth. Therefore, the change in phase (orientation) alters (i) the quantity of intervening gas located in a cylinder between the primary and Earth, and (ii) the radial components of the corresponding gas velocities. Both of these effects contribute to the spectral line change in going from FIG. 3.4.1a to FIG. 3.4.1b which is due to gas flowing from the secondary to the primary. FIG. 3.4.2 shows the direction of the gas stream as well as the lines of site, as viewed from above.

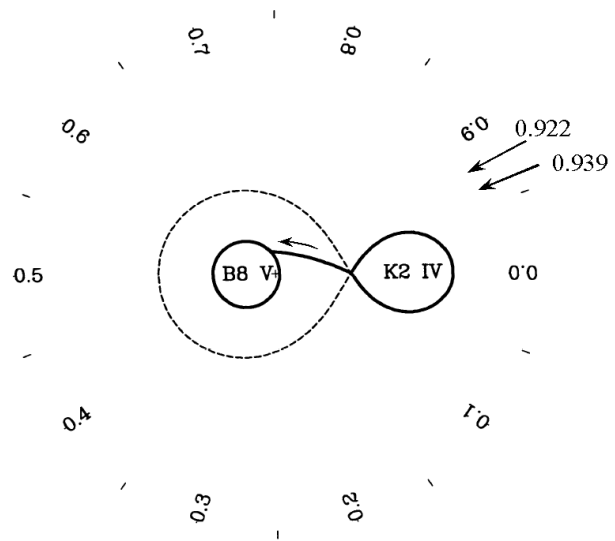

FIG. 3.4.2–Lines of sight as viewed from above for the 1984 exposures at phases 0.922 and 0.939 (Adapted from Fig. 16 p. 285 Richards 1993)

# 4. ANALYSIS OF THE DATA

## 4.1 *Data Analysis Overview*

Spectra from the MAST IUE archive (see Section 3.1) are processed and analyzed in a series of stages, and described in detail in the following sections. An overview of the process is described here and supported by a flow-chart representation in FIG. 4.1.1. Numerical results generated by the various levels of data analysis are identified by rectangular boxes in the figure. Operations are identified by ovals and circles. The titles appearing in the rectangular and oval shapes are italicized in the description below.

Data from the MAST IUE archive ("IUE Spectra" for our purposes) suffer from several deficiencies, including a jaggedness that renders them poorly suited for certain types of mathematical processing and flagged data points indicative of suspected errors. We subject the IUE Spectra to a *Preliminary Processing* step to deal with these issues, thereby producing our *Smoothed Spectra*. This is described in Section 4.3.1.

By *Identifying the Continuum* levels in small UV ranges, as described in Section 4.4.1, we generate the *Continuum Flux and Light Curves* reported in Section 4.4.2. Continuum flux curves represent UV intensities as functions of wavelength, whereas light curves display these intensities as functions of phase.

The remaining analyses involve the examination of spectral line features. In order to facilitate those treatments, two preliminary steps are taken. Our methods for *Spectral-Line Identification and Systematic Velocity Determination* and a *Correction for the Influence of Algol C* are described in Sections 4.3.2 and 4.3.3, respectively.

*Line Feature Analyses* are then performed, as described in Section 4.5.1, resulting in *Line Positions, Line Depths, Line Widths, and Line Asymmetries*. The latter two results are described in Section 4.5.3. A straightforward calculation results in our reported *Radial Velocity Curves*, described in Section 4.5.2. Further calculations produce the *Residual Intensities* described in Section 4.5.4 and the *Equivalent Widths* and *Gas Densities* described in Section 4.5.5.

The use of *Difference Spectra* to remove photosphere effects is explored in Section 4.5.6. The *Mass Loss Rate* from Algol B is computed in Section 4.5.7.

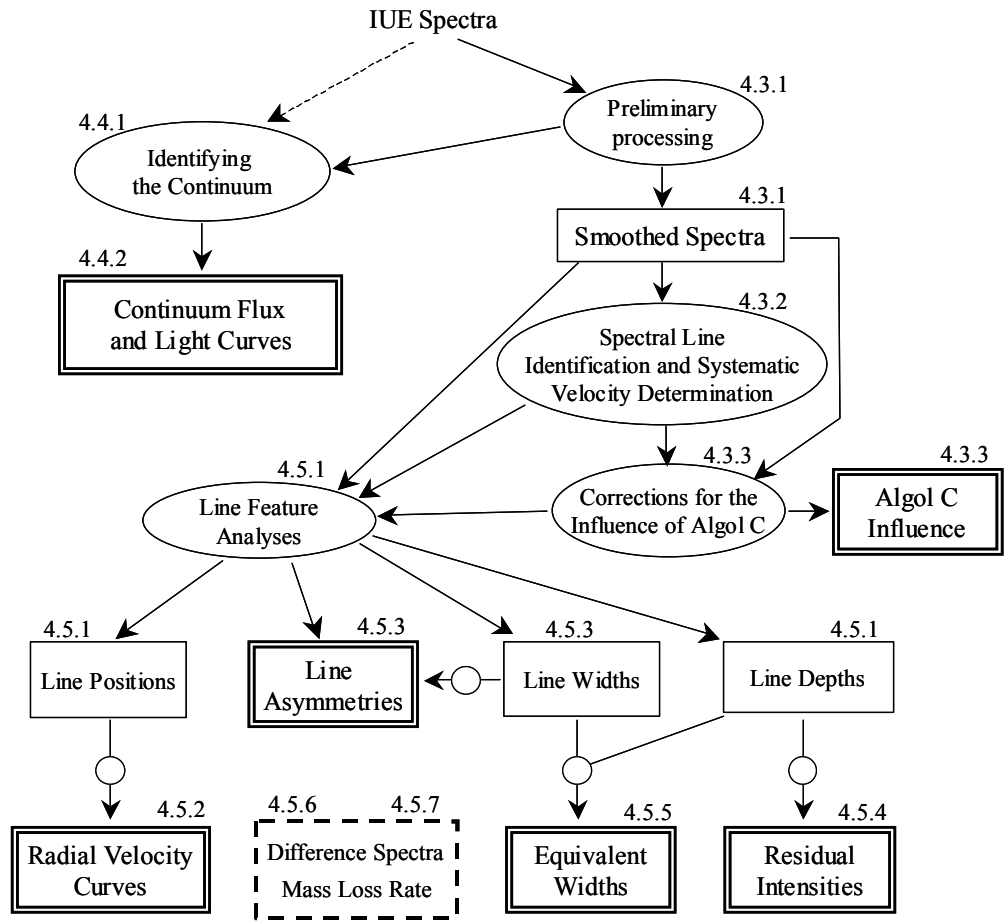

FIG. 4.1.1–*IUE Spectra Analysis Flow-Chart*

## 4.2 *Gas Composition, Excitation, and Ionization*

### 4.2.1 *Ion Types and Electronic-Transitions*

The IUE spectral features are produced mostly by the selective absorption of UV light, emitted by the primary, due to intervening gases. The atoms and ions constituting the intervening gas could be located at the photosphere of the primary, associated with Algol-system gas flow, or present in the interstellar region between Algol and Earth. Gases occupying the various regions are generally characterized by different constituent atoms, temperatures, electron pressures, and radial velocities. In some cases, the transitions are sufficiently strong and isolated so that definitive identifications are straightforward. In other cases, however, the transitions may be weak and/or "blended" with nearby and perhaps unrelated spectra features.

We developed an algorithm to facilitate the identification of electronic transitions corresponding to various absorption features. This procedure, described in Section 4.3.2, provides an additional benefit by producing an estimate of the center of mass velocity of the Algol system, called the "systematic velocity."

Using this algorithm along with experiences based upon previous examinations of similar stellar systems, we identified many spectral lines with ion type and electronic transition. Characteristics of these ions and transitions are provided in TABLE 4.2.1.1.

TABLE 4.2.1.1 – Electronic Transitions (Source: NIST and Kurucz online databases)

| SWP Line ID | $\lambda_{obs}$ Vac (Å) | Configurations | Terms | $J_i$ | $J_k$ | $g_i$ | $g_k$ | $f_{ik}$ | $\log(g_i f_{ik})$ | $A_{ki}$ (s$^{-1}$) |
|---|---|---|---|---|---|---|---|---|---|---|
| **Al II 1670** | 1670.7867 | $2p^63s^2 \rightarrow 3s3p$ | $^1S \rightarrow ^1P°$ | $0 \rightarrow 1$ | | $1 \rightarrow 3$ | 1.83e+00 | 0.263 | 1.46e+09 |
| **Al III 1854** | 1854.7164 | $2p^63s \rightarrow 2p^63p$ | $^2S \rightarrow ^2P°$ | $1/2 \rightarrow 3/2$ | | $2 \rightarrow 4$ | 5.57e-01 | 0.047 | 5.40e+08 |
| **Al III 1862** | 1862.7895 | $2p^63s \rightarrow 2p^63p$ | $^2S \rightarrow ^2P°$ | $1/2 \rightarrow 1/2$ | | $2 \rightarrow 2$ | 2.77e-01 | -0.256 | 5.33e+08 |
| **C II 1334** | 1334.532 | $2s^22p \rightarrow 2s2p^2$ | $^2P° \rightarrow ^2D$ | $1/2 \rightarrow 3/2$ | | $2 \rightarrow 4$ | 1.27e-01 | -0.597 | 2.37e+08 |
| **C II 1335** | 1335.708 | $2s^22p \rightarrow 2s2p^2$ | $^2P° \rightarrow ^2D$ | $3/2 \rightarrow 5/2$ | | $4 \rightarrow 6$ | 1.14e-01 | -0.341 | 2.84e+08 |
| **C IV 1548** | 1548.185 | $1s^22s \rightarrow 1s^22p$ | $^2S \rightarrow ^2P°$ | $1/2 \rightarrow 3/2$ | | $2 \rightarrow 4$ | 1.90e-01 | -0.419 | 2.65e+08 |
| **C IV 1550** | 1550.774 | $1s^22s \rightarrow 1s^22p$ | $^2S \rightarrow ^2P°$ | $1/2 \rightarrow 1/2$ | | $2 \rightarrow 2$ | 9.52e-02 | -0.720 | 2.64e+08 |
| **Fe II 1608** | 1608.45106 | $(5D)4s$ a6D $\rightarrow$ (6S)sp y6P | | $4.5 \rightarrow 3.5$ | | | | | -0.145 | 2.31E+08 |
| **Fe II 1635** | 1635.4010 | $3d^6(^5D)4s \rightarrow 3d^5(^6S)4s4p(^3P°)$ | a $^4D \rightarrow$ x $^4P°$ | $7/2 \rightarrow 5/2$ | | $8 \rightarrow 6$ | 7.2e-02 | -0.24 | 2.4e+08 |
| **Fe II 1639** | 1639.40124 | $(5D)4s$ a6D $\rightarrow$ (6S)sp y6P | | $0.5 \rightarrow 1.5$ | | | | | -0.874 | 8.29E+07 |
| **Fe II 1640** | 1640.15205 | D7 a4F $\rightarrow$ (3F)4p y4G | | $1.5 \rightarrow 2.5$ | | | | | -0.717 | 7.93E+07 |
| **Fe III 1895** | 1895.456 | (6S)4s 7S $\rightarrow$ (6S)4p 7P | | $3 \rightarrow 4$ | | | | | 0.461 | 5.96E+08 |
| **Fe III 1914** | 1914.056 | (6S)4s 7S $\rightarrow$ (6S)4p 7P | | $3 \rightarrow 3$ | | | | | 0.344 | 5.74E+08 |
| **Si II 1260** | 1260.42 | $3s^23p \rightarrow 3s^23d$ | $^2P° \rightarrow ^2D$ | $1/2 \rightarrow 3/2$ | | $2 \rightarrow 4$ | 9.5e-01 | 0.28 | 2.0e+09 |
| **Si II 1264** | 1264.737 | $3s^23p \rightarrow 3s^23d$ | $^2P° \rightarrow ^2D$ | $3/2 \rightarrow 5/2$ | | $4 \rightarrow 6$ | 8.3e-01 | 0.52 | 2.3e+09 |
| **Si II 1304** | 1304.37 | $3s^23p \rightarrow 3s3p^2$ | $^2P° \rightarrow ^2S$ | $1/2 \rightarrow 1/2$ | | $2 \rightarrow 2$ | 9.2e-02 | -0.74 | 3.6e+08 |
| **Si II 1305** | 1305.592 | s3p2 2D $\rightarrow$ 3d' 2F | | $2.5 \rightarrow 3.5$ | | | | | 0.71 | 2.51E+09 |
| **Si II 1309** | 1309.277 | $3s^23p \rightarrow 3s3p^2$ | $^2P° \rightarrow ^2S$ | $3/2 \rightarrow 1/2$ | | $4 \rightarrow 2$ | 9.0e-02 | -0.44 | 7.0e+08 |
| **Si II 1526** | 1526.708 | $3s^23p \rightarrow 3s^24s$ | $^2P° \rightarrow ^2S$ | $1/2 \rightarrow 1/2$ | | $2 \rightarrow 2$ | 1.30e-01 | -0.584 | 3.73e+08 |
| **Si II 1533** | 1533.432 | $3s^23p \rightarrow 3s^24s$ | $^2P° \rightarrow ^2S$ | $3/2 \rightarrow 1/2$ | | $4 \rightarrow 2$ | 1.3e-01 | -0.28 | 7.4e+08 |
| **Si II 1808** | 1808.012 | $3s^23p \rightarrow 3s3p^2$ | $^2P° \rightarrow ^2D$ | $1/2 \rightarrow 3/2$ | | $2 \rightarrow 4$ | 3.6e-03 | -2.14 | 3.7e+06 |
| **Si II 1816** | 1816.98 | $3s^23p \rightarrow 3s3p^2$ | $^2P° \rightarrow ^2D$ | $3/2 \rightarrow 5/2$ | | $4 \rightarrow 6$ | 5.9e-03 | -1.63 | 7.9e+06 |

TABLE 4.2.1.1 SWP - continued

| SWP Line ID | $\lambda_{obs}$ Vac (Å) | Configurations | Terms | $J_i$ | $J_k$ | $g_i$ | $g_k$ | $f_{ik}$ | $\log(g_i f_{ik})$ | $A_{ki}$ (s$^{-1}$) |
|---|---|---|---|---|---|---|---|---|---|---|
| **Si III 1294** | 1294.545 | $3s3p \rightarrow 3p^2$ | $^3P^o \rightarrow {}^3P$ | $1 \rightarrow 2$ | | $3 \rightarrow 5$ | | 2.27e-01 | -0.167 | 5.42e+08 |
| **Si III 1296** | 1296.726 | $3s3p \rightarrow 3p^2$ | $^3P^o \rightarrow {}^3P$ | $0 \rightarrow 1$ | | $1 \rightarrow 3$ | | 5.44e-01 | -0.265 | 7.19e+08 |
| **Si III 1298** | 1298.89 | $3s3p \rightarrow 3p^2$ | $^3P^o \rightarrow {}^3P$ | $1 \rightarrow 1$ | | $3 \rightarrow 3$ | | 1.36e-01 | -0.391 | 5.36e+08 |
| **Si III 1301** | 1301.149 | $3s3p \rightarrow 3p^2$ | $^3P^o \rightarrow {}^3P$ | $1 \rightarrow 0$ | | $3 \rightarrow 1$ | | 1.80e-01 | -0.267 | 2.13e+09 |
| **Si III 1302** | 1302.2 | | | | | | | | | |
| **Si III 1303** | 1303.323 | $3s3p \rightarrow 3p^2$ | $^3P^o \rightarrow {}^3P$ | $2 \rightarrow 1$ | | $5 \rightarrow 3$ | | 1.35e-01 | -0.170 | 8.85e+08 |
| **Si IV 1393** | 1393.755 | $2p^63s \rightarrow 2p^63p$ | $^2S \rightarrow {}^2P^o$ | $1/2 \rightarrow 3/2$ | | $2 \rightarrow 4$ | | 4.50e-01 | -0.046 | 7.73e+08 |
| **Si IV 1402** | 1402.77 | $2p^63s \rightarrow 2p^63p$ | $^2S \rightarrow {}^2P^o$ | $1/2 \rightarrow 1/2$ | | $2 \rightarrow 2$ | | 2.24e-01 | -0.349 | 7.58e+08 |

TABLE 4.2.1.2 – LWP/LWR

| LWP/LWR Line ID | bl. | $\lambda_{obs}$ Vac (Å) | Configurations | Terms | $J_i$ | $J_k$ | $g_i$ | $g_k$ | $f_{ik}$ | $\log(g_i f_{ik})$ | $A_{ki}$ (s$^{-1}$) |
|---|---|---|---|---|---|---|---|---|---|---|---|
| **LWP/LWR** | | | | | | | | | | | |
| **Fe II 2586** | | 2586.6500 | $3d^6(^5D)4s \rightarrow 3d^6(^5D)4p$ | $a\,^6D \rightarrow z\,^6D^o$ | $9/2 \rightarrow 7/2$ | | $10 \rightarrow 8$ | | 6.5e-02 | -0.19 | 8.1e+07 |
| **Fe II 2599** | | 2599.147 | $3d^6(^5D)4s \rightarrow 3d^6(^5D)4p$ | $a\,^6D \rightarrow z\,^6D^o$ | $7/2 \rightarrow 5/2$ | | $8 \rightarrow 6$ | | 9.9e-02 | -0.10 | 1.3e+08 |
| **Fe II 2600** | | 2600.173 | $3d^6(^5D)4s \rightarrow 3d^6(^5D)4p$ | $a\,^6D \rightarrow z\,^6D^o$ | $9/2 \rightarrow 9/2$ | | $10 \rightarrow 10$ | | 2.2e-01 | 0.35 | 2.2e+08 |
| **Fe II 2750** | (a) | 2750.13 | $3d^6(^5D)4s \rightarrow 3d^6(^5D)4p$ | $a\,^4D \rightarrow z\,^4F^o$ | $5/2 \rightarrow 7/2$ | | $6 \rightarrow 8$ | | 3.2e-01 | 0.28 | 2.1e+08 |
| **Fe II 2750** | (b) | 2750.29 | $3d^6(^5D)4s \rightarrow 3d^6(^5D)4p$ | $a\,^4D \rightarrow z\,^4D^o$ | $1/2 \rightarrow 1/2$ | | $2 \rightarrow 2$ | | 1.2e-01 | -0.60 | 1.1e+08 |
| **Mg II 2796** | | 2796.352 | $2p^63s \rightarrow 2p^63p$ | $^2S \rightarrow {}^2P^o$ | $1/2 \rightarrow 3/2$ | | $2 \rightarrow 4$ | | 6.08e-01 | 0.085 | 2.60e+08 |
| **Mg II 2803** | | 2803.531 | $2p^63s \rightarrow 2p^63p$ | $^2S \rightarrow {}^2P^o$ | $1/2 \rightarrow 1/2$ | | $2 \rightarrow 2$ | | 3.03e-01 | -0.218 | 2.57e+08 |
| **Mg II 2791** | | 2791.599 | $2p^63p \rightarrow 2p^63d$ | $^2P^o \rightarrow {}^2D$ | $1/2 \rightarrow 3/2$ | | $2 \rightarrow 4$ | | 9.37e-01 | 0.273 | 4.01e+08 |
| **Mg II 2798** | (a) | 2798.823 | $2p^63p \rightarrow 2p^63d$ | $^2P^o \rightarrow {}^2D$ | $3/2 \rightarrow 5/2$ | | $4 \rightarrow 6$ | | 8.44e-01 | 0.528 | 4.79e+08 |
| **Mg II 2798** | (b) | 2798.754 | $3p\,^2P \rightarrow 3d\,^2D$ | | 1.5 | 1.5 | | | | -0.42 | 8.09E+07 |

Since we are concerned with gas flow associated with the Algol system, it is important to distinguish the Algol gas-flow contributions from the photospheric absorption. Since the photospheric conditions of the primary are relatively well understood, we can, to some extent, distinguish photospheric contributions by examining the predicted ionization fractions as functions of temperature and electron pressure. This analysis is described in the following section.

### 4.2.2 *Ionization Fractions*

In our preliminary examination of the spectra, we noticed several lines that are known to be indicators of hot plasma and circumstellar material, (Kempner and Richards, 1999); namely, Si IV λλ1393.755, 1402.770 and C IV λ1548.85. "The presence of Si IV and C IV indicates the existence of regions considerably hotter than a normal B8V photosphere," (McCluskey and Kondo, 1984).

The Saha equation can be used to estimate the ionization fraction of each atomic species as functions of temperature and electron pressure, assuming conditions of thermodynamic equilibrium. The Saha equation is given below in ordinary and logarithmic forms

$$\frac{N^{(r+1)}}{N^{(r)}} = \frac{2kT}{P_e} \left( \frac{2\pi \, m_e \, kT}{h^2} \right)^{\frac{3}{2}} \frac{Z^{(r+1)}}{Z^{(r)}} e^{-\chi^{(r)}/kT} \qquad (4.2.2.1)$$

$$\log \frac{N^{(r+1)}}{N^{(r)}} = \log \frac{Z^{(r+1)}}{Z^{(r)}} + \frac{5}{2} \log T - \frac{5040}{T} \chi^{(r)} - \log P_e - 0.179 \qquad (4.2.2.2)$$

The terms in these equations are defined as follows:

| | |
|---|---|
| $r$ | ionization state, where $r = 0$ indicates un-ionized |
| $N^{(r)}$ | number of atoms in ionization state $r$ |
| $Z^{(r)}$ | atomic partition function for ionization state $r$ |
| $\chi^{(r)}$ | ionization energy in eV for ionization state $r$ |
| $P_e$ | electron pressure, in dynes/cm$^2$ |
| $T$ | temperature in Kelvin |

The partition functions are given by

$$Z^{(r)} = \sum_{i=1}^{i_{max}} g_i^{(r)} e^{-\varepsilon_i^{(r)}/kT} \qquad (4.2.2.3)$$

where

| | |
|---|---|
| $i$ | atomic energy level |
| $g_i^{(r)}$ | statistical weight (degeneracy) of the $i^{th}$ level for ionization state $r$ |
| $i_{max}$ | last bound level for ionization state $r$ (depends on $T$ and $P_e$) |
| $\varepsilon_i^{(r)}$ | energy of $i^{th}$ level for ionization state $r$, relative to the ground state (i.e., $\varepsilon_i = 0$) |

The ionization fraction is the ratio of $N^{(r)}$ to the total number of atoms of that type; hence,

$$X^{(r)} = \frac{N^{(r)}}{N^{(0)} + N^{(1)} + \cdots N^{(r)} + \cdots N^{(r_{max})}} \qquad (4.2.2.4)$$

or

$$X^{(r)} = \frac{1}{\frac{N^{(0)}}{N^{(r)}} + \frac{N^{(1)}}{N^{(r)}} + \cdots + 1 + \cdots \frac{N^{(r_{max})}}{N^{(r)}}} \qquad (4.2.2.5)$$

Since the Saha equation gives ratios of consecutive ionization stages, terms such as $\frac{N^{(2)}}{N^{(5)}}$ were constructed as follows:

$$\frac{N^{(2)}}{N^{(5)}} = \left(\frac{N^{(2)}}{N^{(3)}}\right)\left(\frac{N^{(3)}}{N^{(4)}}\right)\left(\frac{N^{(4)}}{N^{(5)}}\right) \tag{4.2.2.6}$$

We developed a Visual Basic for Application to automate this process. The input parameter values are from the NIST[*] and Kurucz[†] online databases.

The effective temperature $T_e$ of Algol is ~ 12,300 K with an electron pressure $P_e$ ~ 316.2 dynes/cm$^2$, or log($P_e$) ~2.5 (Aller 1953; Underhill 1972, Table 8, gives a value of $P_e$ = 3.9709x10$^2$ dynes/cm$^2$ for $T_e$ = 12370 K). The solutions of the Saha equation for the B8 V component of Algol are plotted in FIGS. 4.2.2.1 and 4.2.2.2. We see in the primary's photosphere the presence of C II, N II, and O II; and silicon is almost evenly split between Si II and Si III.

---

[*] http://physics.NIST.gov/cgi-bin/AtData/main_asd
[†] http://cfa-www.Harvard.edu/amdata/Kurucz23/sekur.html

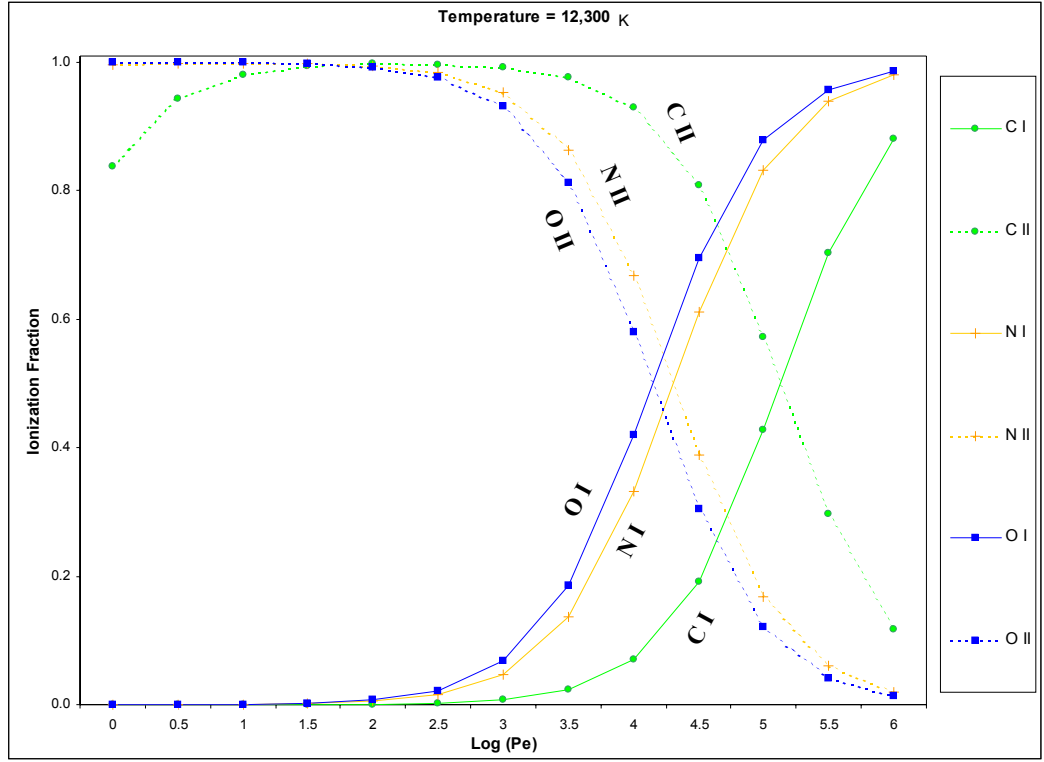

FIG. 4.2.2.1 - *Ionization Fraction of Carbon, Nitrogen, and Oxygen at 12,300 K*

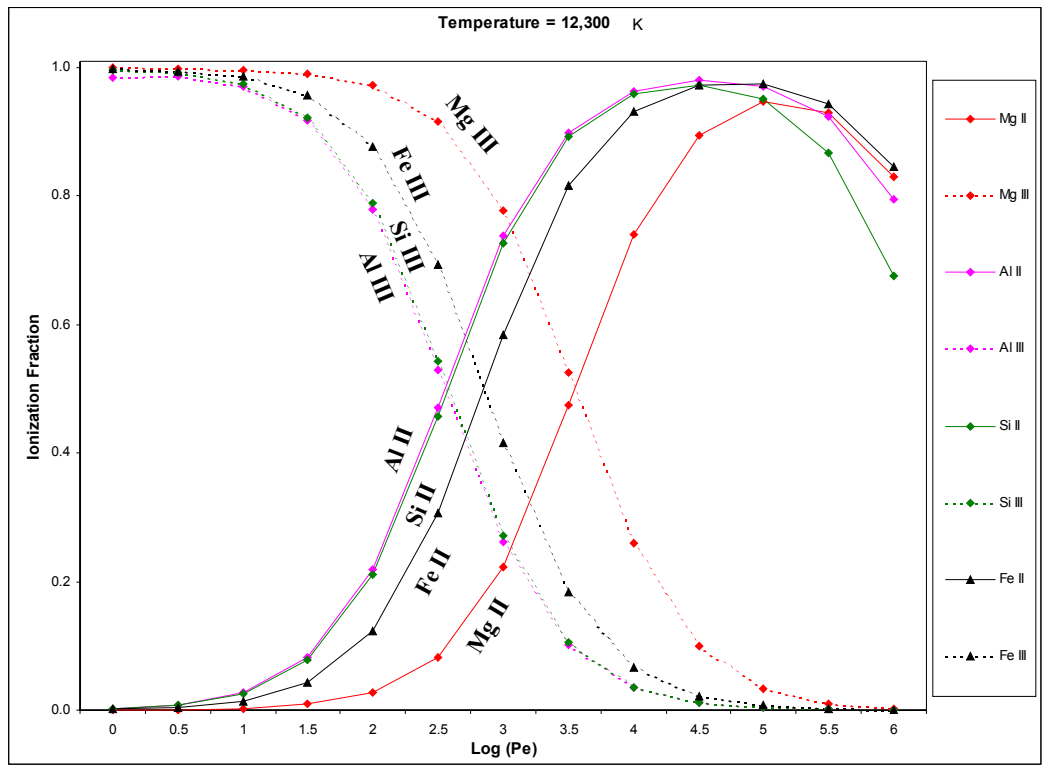

FIG. 4.2.2.2 - *Ionization Fraction of Mg, Al, Si, and Fe at 12,300 K*

Anticipating the presence of gases at non-photospheric temperatures, we examine additional solutions of the Saha equation, revealing the formation of C IV at ~ 25,000 K (for electron pressures ≤ 1000 dynes/cm$^2$; see F$_{IG}$. 4.2.2.3), Si IV at ~ 20,000 K (F$_{IG}$. 4.2.2.4), O IV at ~ 30,000 K (F$_{IG}$. 4.2.2.5), and N V at ~ 45,000 K (F$_{IG}$. 4.2.2.6).

For completeness, a set of these figures corresponding to temperatures of 8,000 K, 10,000 K, 12,300 K, 15,000 K, 20,000 K, 25,000 K, 28,500 K, 30,000 K, 35,000 K, 45,000 K, and 50,000 K are provided in Appendix D.1.

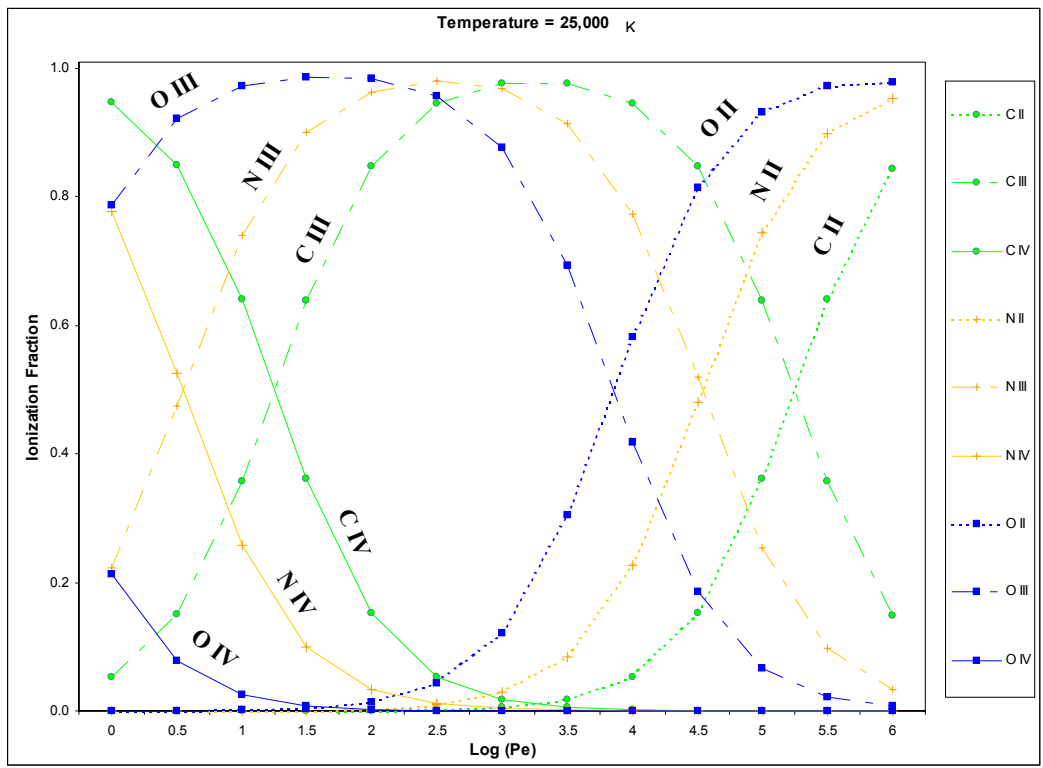

F$_{IG}$. 4.2.2.3 - *Ionization Fraction of Carbon, Nitrogen, and Oxygen at 25,000 K*

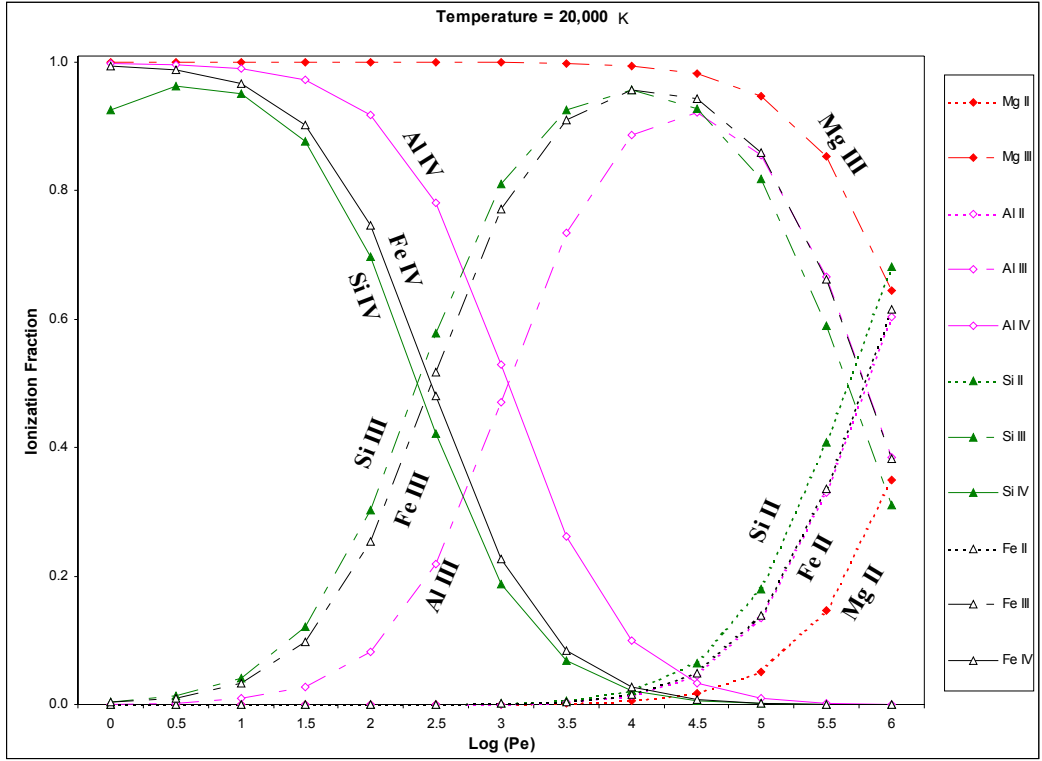

FIG. 4.2.2.4 - *Ionization Fraction of Mg, Al, Si, and Fe at 20,000 K*

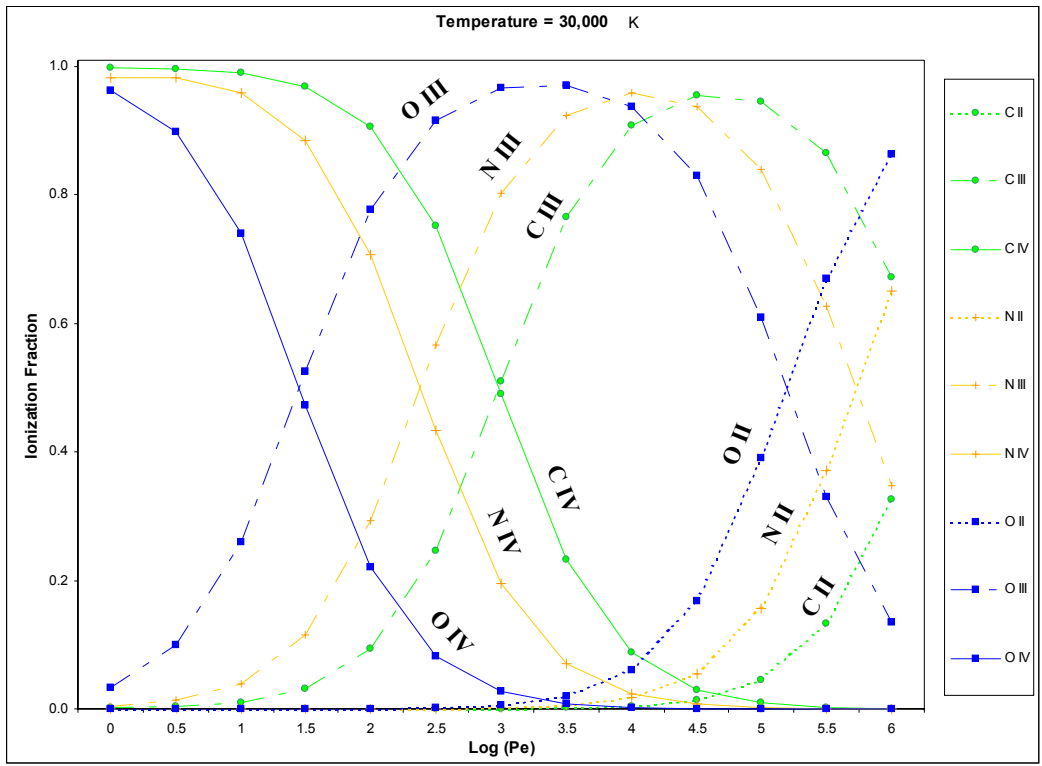

FIG. 4.2.2.5 - *Ionization Fraction of Carbon, Nitrogen, and Oxygen at 30,000 K*

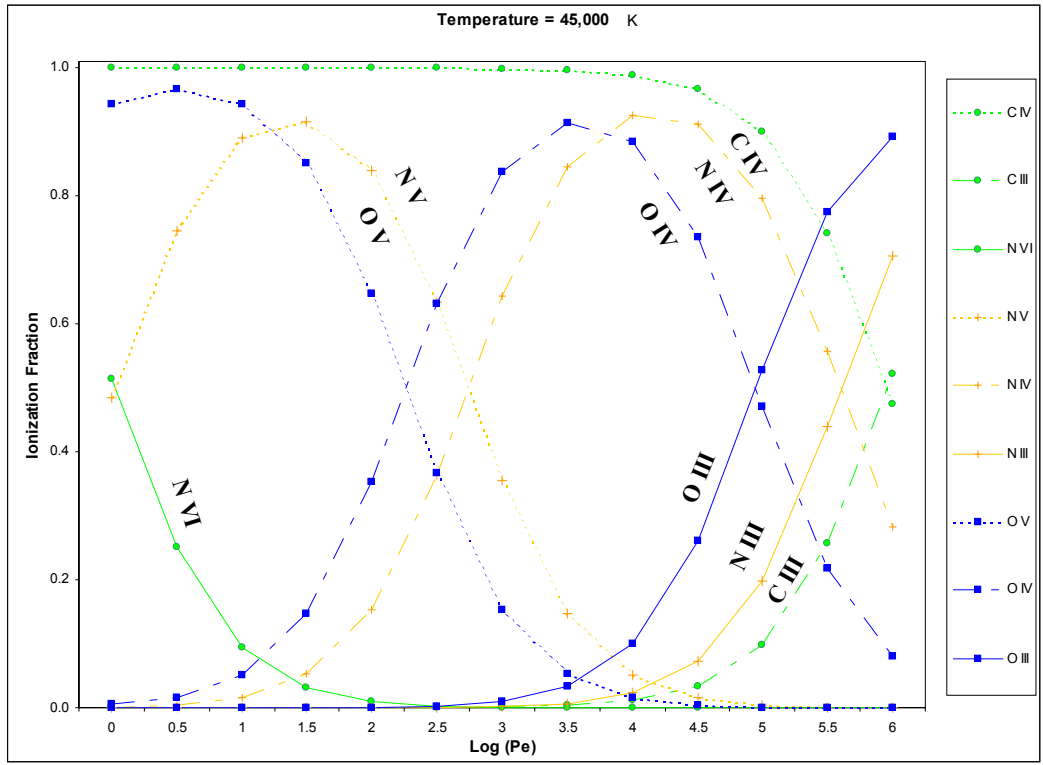

FIG. 4.2.2.6 - *Ionization Fraction of Carbon, Nitrogen, and Oxygen at 45,000 K*

We did not see difinitive evidence for the presence of C IV λ1550.7742, N V λλ1238.821, 1242.804, or O IV λ1397.20 in any of the Algol spectra, which indicate plasma temperatures ranging from $2 \times 10^4$ to $5 \times 10^4$ K. These were too blended or too weak to detect. Chemical depletion was previously suggested. The Si IV doublet is relatively strong, compared to the C IV λ1548 line, which is quite weak and difficult to measure accurately. In the SWP analysis, we focus on the Si IV lines since they show a nice variation across the orbital phases, as well as across the epochs.

## 4.3. *Preliminary Considerations*

### 4.3.1 *Managing the IUE Data*

The IUE dataset contains many points that have been flagged to indicate suspected measurement or recording errors. We replaced these values with numbers computed by a simple linear interpolation using the first point on either side of the flagged point(s). Symbols ("x") were retained in our plots to denote the locations of flagged data points. We did not use regions of spectra characterized by large sequences of contiguous flagged data points.

These adjusted datasets were further processed to remove the jaggedness that renders them poorly suited for certain types of mathematical processing. In order to accomplish this, we used a moving average procedure with Pascal Triangle weighting, which is commonly used for this application.

A 12-step Pascal Triangle scheme (N = 12) was applied to each vector of 21,000 data points for the SWP measurements and 23,000 data points for LWP/LWR. Hence, a vector of smoothed data points F was constructed from the corresponding IUE vector F′ as follows (for even N):

$$F_i = \frac{1}{2^N} \sum_{n=0}^{N} \left[ \frac{N!}{(N-n)!n!} \right] F'_{-\frac{N}{2}+n+i} \quad (4.3.1.1)$$

This approach is particularly appropriate when one wishes to reduce noise while preserving Gaussian features.

4.3.2 *Spectral-Line Identification and Systematic Velocity*

The smoothed spectra show many absorption features, even over wavelength ranges of only several Angstroms. In order to facilitate the process of identifying the spectral features with ion types and transitions, we developed a custom interactive visualization program using VBA (Visual Basic for Applications). The program connects to our smoothed-spectra database and automatically locates the wavelengths of local minima. Upon specifying a wavelength range, all of the local minima in that range are plotted as a function of orbital phase. An illustration of the output is shown by the data circles of FIG. 4.3.2.1, which was computed for a wavelength range of 1636.76 – 1642.76 Å. The three strips on the right are duplicates of the left-hand plot, but compressed and periodically reproduced so as to assist the eye in detecting patterns. A variety of sinusoidal patterns are easily seen.

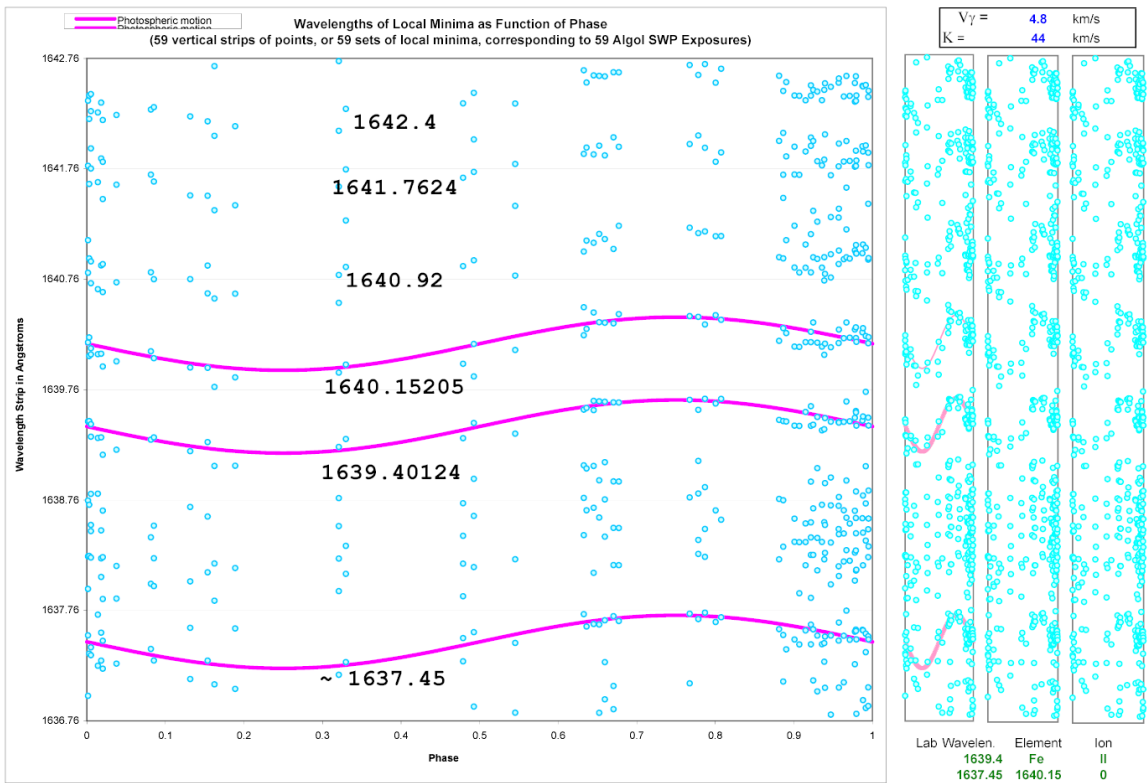

FIG. 4.3.2.1. —*Wavelength of local minima as a function of phase.* The 59 vertical strips of points, or 59 sets of local minima, correspond to the 59 Algol IUE SWP exposures.

These patterns represent Doppler shifts associated with the most fundamental motions of the gases producing the absorptions, for instance (1) the photospheric motion of Algol A relative to the center of mass of the Algol A-B system, and (2) the motion of the center of mass of the Algol system relative to the Sun. The factors leading to the scatter in the data (and broadening of the spectral lines) include the rotation of Algol A, the revolution of the Algol A-B system about Algol C, and gas flow relative to the photosphere of Algol A. The effect of Algol C is addressed in the next section, and non-photospheric gas flow in all subsequent sections.

This first order analysis allows us to identify prominent spectral lines, and also refine previous measurements of the systematic velocity, $V_0$, and the semiamplitude, $K_A$, which represents the maximum radial velocity of Algol A relative to the center of mass. In the simple model described above, Algol A moves at radial velocity $V(\varphi)$ with respect to the Sun, where $\varphi$ is the orbital phase and:

$$V(\varphi) = V_0 - K_A \sin(2\pi\phi) \tag{4.3.2.1}$$

Accordingly, the predicted absorption wavelengths can be expressed in terms of $V_0$, $K_A$, and the vacuum wavelength, $\lambda_0$:

$$\lambda(\varphi) = \lambda_o + \frac{1}{c}\lambda_o V(\varphi) \tag{4.3.2.2}$$

Three of these lines are shown in FIG. 4.3.2.1 for the vacuum rest wavelengths indicated and values of $K_A = 44$ km/s and $V_0 = 4.8$ km/s. These values represent best fits to our data, and we include them in Table 9 and Table 1, respectively, of Appendix A as a contribution from this work.

The process of identifying spectral lines is illustrated for the spectral range of 1636.76 – 1642.76 Å. The corresponding smooth spectra are shown in FIG. 4.3.2.2 for phase 0.0015 corresponding to one of the 59 SWP exposures. Compressed plots, such as those shown on the right side of FIG. 4.3.2.1 are used to identify strong photospheric features. The sinusoidal lines of FIG. 4.3.2.1 (left figure) are then interactively adjusted by varying the rest wavelength until a best fit to the data points is achieved. This method has in some cases allowed us to predict the wavelength of a line to within 0.001 Å even before we have identified it.

In the spectrum of FIG. 4.3.2.2 the local minima (hollow triangles) correspond to the first vertical strip of points, at phase 0.0015, in FIG. 4.3.2.1. Our program automatically outputs 59 such plots, each containing a set of local minima. This collection of 59 plots enables us to select the strongest lines, at a glance, in the wavelength range of interest. The strongest lines are the ones we want to measure since there will be less effects from blending and less scatter in the measurements.

Fe II λλ1639.40124,

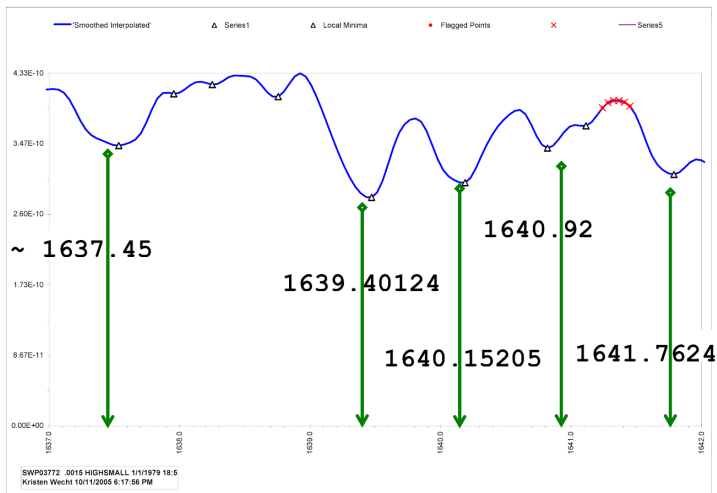

FIG. 4.3.2.2. —Rest wavelengths identified with the interactive program we developed. Fe II 1639.40124 and Fe II 1640.15205 are the strongest. Hollow triangles indicate the local minima. The arrows locate the positions of the lab vacuum wavelengths of the identified lines.

1640.15205 are consistently the strongest across the 59 exposures in this wavelength range. The 59 sets of minima correspond to the 59 vertical strips of points that are automatically plotted against phase in FIG 4.3.2.1.

### 4.3.3. *Correction for Influence of Algol C*

The binary motions of Algol A and B are complicated by the presence of a third component, Algol C. Since the Algol A, B, and C masses are 3.7$M_\odot$, 0.81$M_\odot$, and 1.6$M_\odot$, respectively, none of the components are sufficiently small to justify ignoring their effect on the others. However, an approximate solution is suggested by the small separation of A and B compared to the distance from either of those to Algol C. The A-B system is characterized by separation 14 $R_\odot$ (Richards et al. 2003) and orbital period 2.86731077 days (Gillet et al. 1989), and the AB-C system has separation 2.67 AU (Pan et al. 1993) and orbital period 680.08 days. (Hill et al. 1971)

Our goal here is to correct the radial velocities of Algol A to account for the presence of Algol C. We accomplish this by treating the three stars as point masses. The motion of Algol A and B about their common center of mass is taken to be independent of Algol C. Algol C is assumed to influence Algol A and B equally; hence, motion involving Algol C is treated as a two-body problem, where Algol A and B are replaced by a fictitious mass equal to the sum of the A and B masses located at the A-B center of mass. Hence, the three-body problem is divided into two, two-body problems and the radial velocity of Algol A is given by:

$$V_{A:\odot} = V_o + V_{A\text{-}B:AB\text{-}C} + V_{A:A\text{-}B} \qquad (4.3.3.1)$$

where we use the following notation:

The center of mass (c.m.) of A with respect to the sun ≡ A: ☉.

The c.m. of AB-C with respect to the sun ≡ AB-C: ☉ ≡ $V_0$

The c.m. of A-B with respect to the c.m. AB-C ≡ A-B:AB-C

The c.m. of A with respect to the c.m. of A-B ≡ A:A-B

Although the trajectories of binary star components can be determined analytically, no one has succeeded in developing an analytical expression for the time dependences of the coordinates. Hence, this must be solved numerically.

As described in Section 2.2, the true orbit is characterized by semimajor axis *a* and eccentricity, *e*. The position of one binary component relative to the other is identified by the two time-dependent quantities, separation *r* and true anomaly (angle) θ, as shown in FIG. 4.3.3.1. $r(\theta)$ traces the well known elliptical shape, where

$$r(\theta) = \frac{a(1-e^2)}{1+e\cos\theta} \tag{4.3.3.2}$$

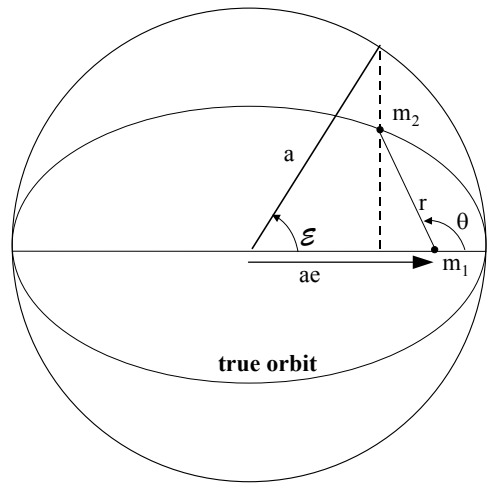

FIG. 4.3.3.1 – *Elliptical Geometry (Source: original work)*

The time dependence of the true anomaly can be calculated in principle from the following equation:

$$t(\theta) - T = \frac{P(1-e^2)^{3/2}}{2\pi} \int_0^\theta \frac{d\theta'}{(1+e\cos\theta')^2} \qquad (4.3.3.3)$$

where T is the time of periastron passage (time when θ = 0, when the components are closest) and P is the period.

This equation was reformulated by Kepler so that it could be written

$$\mathcal{E} - e\sin\mathcal{E} = \mathcal{M} \qquad (4.3.3.4)$$

(See, for example, Escobal, 1976)

The mean anomaly, $\mathcal{M}$, is proportional to time (Eq. 4.3.3.5), and the eccentric anomaly, $\mathcal{E}$, is related to the true anomaly (Eq. 4.3.3.6).

$$\mathcal{M} = \left(\frac{2\pi}{P}\right)(t-T) \qquad (4.3.3.5)$$

$$\tan\frac{\mathcal{E}}{2} = \left[\frac{1-e}{1+e}\right]^{1/2} \tan\frac{\theta}{2} \qquad (4.3.3.6)$$

A geometric interpretation of $\mathcal{E}$ is provided in FIG. 4.3.3.1

Since $\mathcal{E}$ cannot be determined analytically, we apply the Newton-Raphson iterative procedure (Hilditch 2001, p. 38), according to which a solution of f(x) = 0 is given by

$$x_n = x_{n-1} - \frac{f(x_{n-1})}{f'(x_{n-1})} \qquad (4.3.3.7)$$

When applied to Kepler's equation, we construct

$$f(\mathcal{E}) = \mathcal{E} - e\sin\mathcal{E} - \frac{2\pi}{P}(t-T) \tag{4.3.3.8}$$

Taking the derivative with respect to $\mathcal{E}$ we have

$$f'(\mathcal{E}) = 1 - e\cos\mathcal{E} \tag{4.3.3.9}$$

Substituting into the Newton-Raphson iterative solution (Molnar and Mutel 1998), we have

$$\mathcal{E}_n = \mathcal{E}_{n-1} - \frac{f(\mathcal{E}_{n-1})}{f'(\mathcal{E}_{n-1})} \tag{4.3.3.10}$$

$$\mathcal{E}_n = \mathcal{E}_{n-1} - \frac{\mathcal{E}_{n-1} - e\sin\mathcal{E}_{n-1} - \frac{2\pi}{P}(t-T)}{1 - e\cos\mathcal{E}_{n-1}} \tag{4.3.3.11}$$

The iterative solutions of Eq. 4.3.3.11 are calculated separately for the A-B and AB-C motions, yielding solutions $\mathcal{E}_{A\text{-}B}$ and $\mathcal{E}_{AB\text{-}C}$ which satisfy

$$\mathcal{E}_{A\text{-}B} - e_{A\text{-}B}\sin\mathcal{E}_{A\text{-}B} = \mathcal{M}_{A\text{-}B} \tag{4.3.3.12}$$

and

$$\mathcal{E}_{AB\text{-}C} - e_{AB\text{-}C}\sin\mathcal{E}_{AB\text{-}C} = \mathcal{M}_{AB\text{-}C} \tag{4.3.3.13}$$

where

$$\mathcal{M}_{A\text{-}B} = 2\pi\frac{(t-T_{A-B})}{P_{A-B}} \tag{4.3.3.14}$$

$$\mathcal{M}_{AB-C} = 2\pi\frac{(t-T_{AB-C})}{P_{AB-C}} \tag{4.3.3.15}$$

and $e_{x\text{-}y}$ is the eccentricity and $T_{x\text{-}y}$ is the period of the x-y pair where x-y is either A-B or AB-C.

These results are used to calculate the true anomalies (in radians), $\theta_{A\text{-}B}$ and $\theta_{AB\text{-}C}$, for the A-B and AB-C systems, respectively

$$\theta_{A\text{-}B} = 2 \arctan\left[\sqrt{\frac{1+e_{A-B}}{1-e_{A-B}}} \tan\frac{\mathcal{E}_{A-B}}{2}\right], \qquad (4.3.3.16)$$

$$\theta_{AB\text{-}C} = 2 \arctan\left[\sqrt{\frac{1+e_{AB-C}}{1-e_{AB-C}}} \tan\frac{\mathcal{E}_{AB-C}}{2}\right]. \qquad (4.3.3.17)$$

The computation of radial velocities requires the inclusion of geometrical factors, as shown in FIG. 2.2.2. Hence, the radial velocity of the A-B system's center of mass with respect to AB-C's center of mass, $V_{AB:AB\text{-}C}$, is then given by

$$V_{AB:AB\text{-}C} = K_{AB\text{-}C}\{e_{AB\text{-}C}\cos\omega_{AB\text{-}C} + \cos(\theta_{AB\text{-}C} + \omega_{A\text{-}B})\} \qquad (4.3.3.18)$$

The radial velocities of Algol A, B and C with respect to the sun are given respectively by

$$V_{A:\odot} = V_0 + V_{AB:AB\text{-}C} + K_A\{e_{A\text{-}B}\cos\omega_{A\text{-}B} + \cos(\theta_{A\text{-}B} + \omega_{A\text{-}B})\} \qquad (4.3.3.19)$$

$$V_{B:\odot} = V_0 + V_{AB:AB\text{-}C} + K_b\{e_{A\text{-}B}\cos(\omega_{A\text{-}B} + \pi) + \cos(\theta_{A\text{-}B} + \omega_{A\text{-}B} + \pi)\} \qquad (4.3.3.20)$$

and

$$V_{C:\odot} = V_0 + K_C\{e_{AB\text{-}C}\cos(\omega_{AB\text{-}C} + \pi) + \cos(\theta_{AB\text{-}C} + \omega_{AB\text{-}C} + \pi)\} \qquad (4.3.3.21)$$

The phase of the inner binary, Algol A-B, is calculated using

$$\phi_{A\text{-}B} = \left| INT\left[\frac{(t - t_{pr.\min A-B})}{P_{A-B}}\right] - \frac{(t - t_{pr.\min A-B})}{P_{A-B}} \right| \qquad (4.3.3.22)$$

where $t_{pr.\min A\text{-}B}$ is the time of primary minimum for the eclipsing pair.

In a similar manner the phase of the outer binary, the AB-C system, is calculated using

$$\phi_{AB\text{-}C} = ABS \left| INT\left[\frac{(t - t_{AB-C})}{P_{AB-C}}\right] - \frac{(t - T_{AB-C})}{P_{AB-C}} \right| \qquad (4.3.3.23)$$

We wrote a program in VBA that iteratively computes the predicted radial velocity of the Algol AB center of mass (Eq. 4.3.3.18) and the predicted radial velocity of Algol A (Eq. 4.3.3.19) as functions of HJD and phase. In order to test the program and refine input parameters, we selected a "best" set of starting parameters, provided in Table 4.3.3.1, and compared our predictions for the radial velocities of Algol A with the results of a study by Hill, et al. (1971). The Hill study used 95 spectroscopic observations of the Algol system, performed over a period of about four years from August 1937 to February 1941, to estimate the radial velocities.

Our calculated Algol A radial velocities as a function of Algol A-B phase are shown in FIG. 4.3.3.2 along with the measured values of Hill. The same results are plotted with respect to the Algol AB-C phase in FIG. 4.3.3.3. The sinusoidal pattern so apparent in both curves on FIG. 4.3.3.2 is due to the nearly circular motion of A about B. The jagged appearance is a consequence of the fact that many points with similar phases actually correspond to times associated with different epochs. Hence, neighboring points may be separated by significant time intervals, where Algol C is imparting a very different radial velocity. The phase shift between the computed and observed velocity curves seen in both figures is a result of a variation in the period of the A-B system, possibly caused by system activity.

TABLE 4.3.3.1 – INITIAL ALGOL PROPERTIES ADOPTED FOR THIS COMPARISON

| Description of Property | Property Symbol | Initial Input Value | Units |
|---|---|---|---|
| Systematic Velocity (km/s) – Should always refer to the entire known system, ie., velocity of Algol AB-C's CM wrt our Sun (Gillet et al. 1989 p.228) | $V_o$ | 4 | km/s |
| Period of A-B system (days) (Gillet et al. 1989 p.221) | $P_{A-B}$ | 2.86731077 | Days |
| Eccentricity of A-B system (literature ranges between 0.00 and 0.015) | $e_{A-B}$ | 0.01 | |
| Angle between line of nodes & periastron of A-B system (degrees) (Gillet et al. 1989 p.228) | $\omega_{A-B}$ | 163 | degrees |
| Semi-amplitude of A wrt A-B system's CM (km/s) (Tompkin & Lambert 1978) | $K_A$ | 44 | km/s |
| Semi-amplitude of B wrt A-B system's CM (km/s) (Tompkin & Lambert 1978) | $K_B$ | 201 | km/s |
| Semi-amplitude of C wrt AB-C's CM (km/s) (Hill et al. 1971 p.450) | $K_C$ | 31.6 | km/s |
| Time of Periastron passage of A-B system (days) - Time of maximum orbital velocity (Gillet et al. 1989 p.228) | $T_{A-B}$ | 2,445,639.2146 | Days |
| Period of AB-C system (days) (Pan et al. 1993 p.L131) | $P_{AB-C}$ | 680.05 | Days |
| Eccentricity of AB-C system (Pan et al. 1993 p.L131) | $e_{AB-C}$ | 0.225 | |
| Angle between line of nodes & periastron of AB-C system (degrees) (Molnar & Mutel 1998 p.17 note that Pan et al. 1993 lists 310.29) | $\omega_{AB-C}$ | 130.29 | degrees |
| Semi-amplitude of AB wrt AB-C system's CM (km/s) - Should follow the equality: $M_C/(M_A+M_B+M_C) = K_{ab-c}/K_c \sim .27$ (Hill et al. 1971 p.451) | $K_{AB-C}$ | 12 | km/s |
| Time of Periastron passage of AB-C system (days) - Time of maximum orbital velocity (Pan et al. 1993 p.L131) | $T_{AB-C}$ | 2,446,931.4000 | Days |
| Time of primary minimum of Algol A-B (days) (Gillet et al. 1989 p.221) | $t_{PRMIN\_A-B}$ | 2,445,641.5135 | Days |

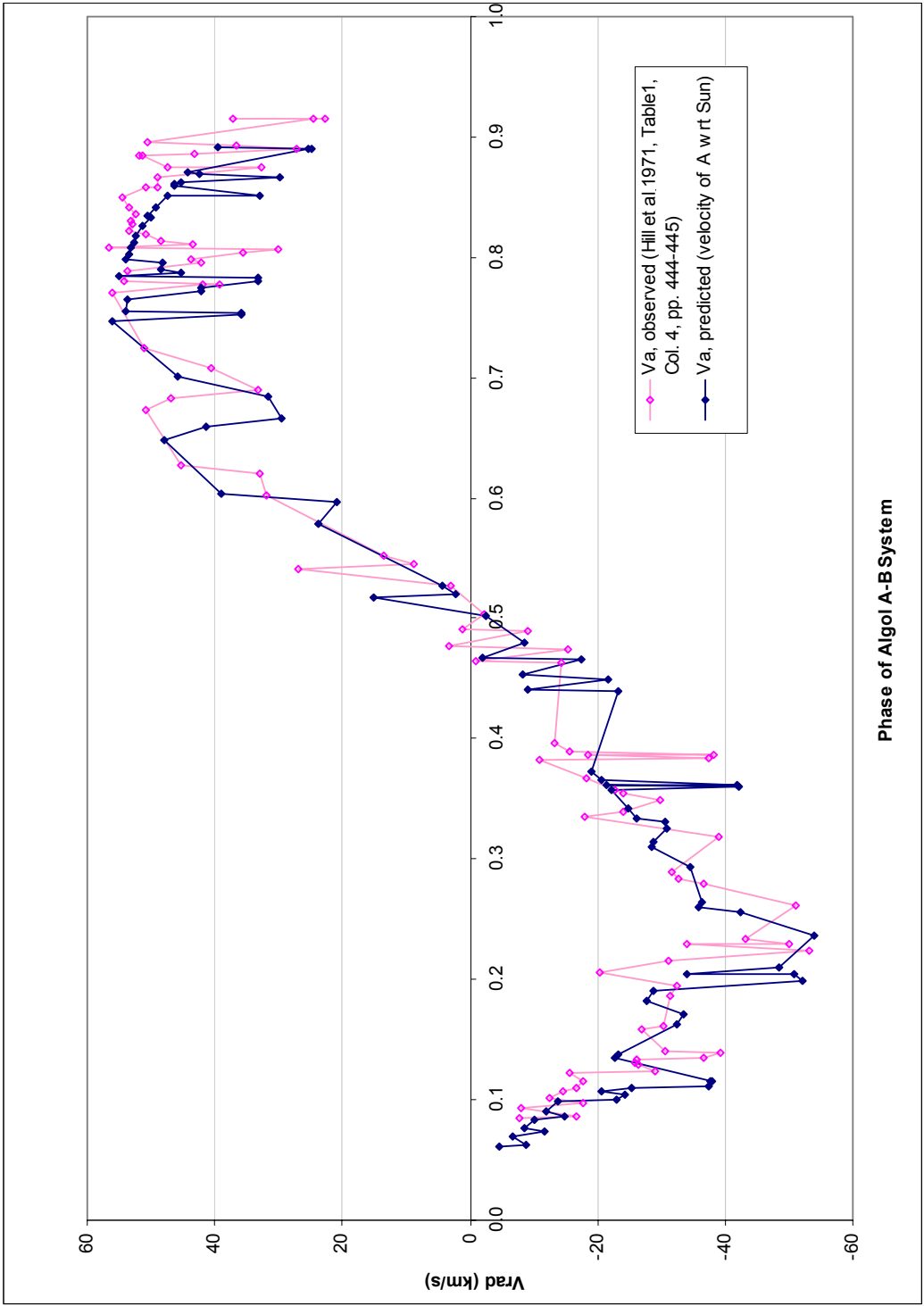

FIG. 4.3.3.2 – *Radial Velocity Comparison of Algol A wrt the Sun vs. Phase of Algol A-B*

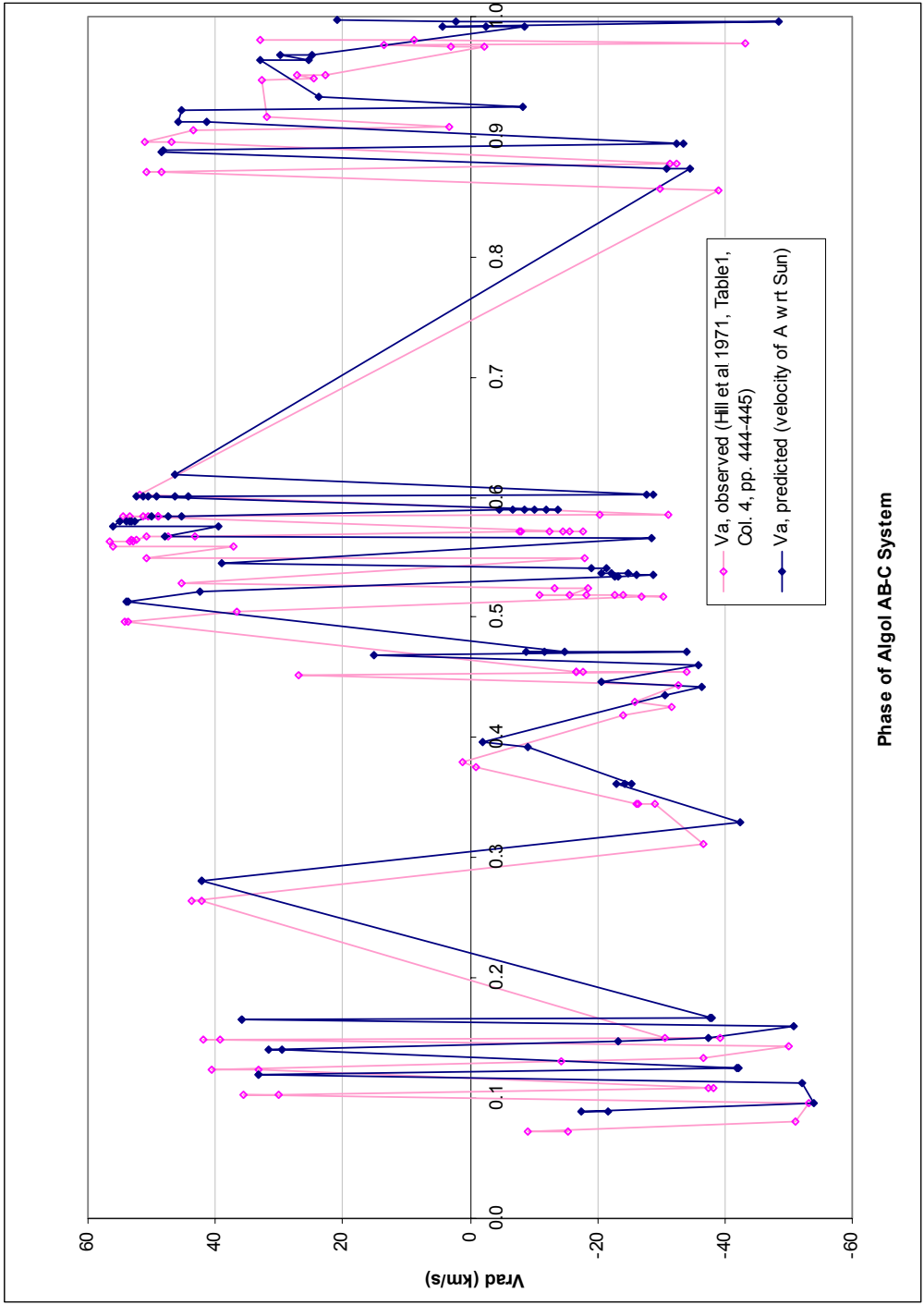

FIG. 4.3.3.3 - RADIAL VELOCITY COMPARISON of ALGOL A wrt the SUN vs. PHASE of ALGOL AB-C

Ignoring the slight phase differences, the departures from a sinusoidal form for both the computed and observed results shown in FIG. 4.3.3.2 are mostly consequences of the motion of the AB center of mass due to the influence of Algol C. Since the Hill velocities were based upon measurements taken over many periods of the AB system (and we simulate their measurements using the same JDs), a given A-B phase can correspond to different positions of Algol C (different AB-C phases); hence, different AB center of mass contributions to the radial velocity of A. The similarities in the computed and measured velocity curves show clearly that our calculations are capturing the effects of Algol C.

Before applying this approach to the IUE UV data, we make two improvements to our method: a correction for the variation in the A-B line of apsides, and an adjustment of parameters. The line of apsides connects the periastron and apastron points (at opposite sides of the major axis of the ellipse) of the A-B system. The orientation of this line, defined by the longitude of periastron, is found to vary in time, rotating through $360°$ in about 32 years. (Hill, et al., 1971, Fig. 4, p. 454; also see Hilditch, (2001), Fig. 4.2, p. 134 for an illustration)

In order to incorporate the effect of the time dependence of the longitude of periastron, we follow the approach of Hill, et al. (1971) in assuming a linear dependence on time. Note that this also implies a time dependence for the time of periastron passage. With this adjustment to the computational program, we adjusted parameters to improve the fit to the Hill data. These new parameters are presented in Table 4.3.3.2 and the corresponding radial velocity results in FIG. 4.3.3.4. We estimated best fit parameters for the linear time dependence of the longitude of periastron of a = 0.0308°/d and b = 521.392°, compared to Hill's parameters of a = 0.03079°/d and b = 494.327°, where

$$\omega(t)_{AB} = at + b$$

and t represents the heliocentric Julian date.

As a final test, we correct the Hill radial velocity measurements by removing the affect of Algol C. That is, we subtract the radial velocity of the AB center of mass (Eq. 4.3.3.18) from the measured radial velocities. These results, shown in FIG. 4.3.3.5, indicate that many of the non-sinusoidal features have been removed.

This is the process that we use to "correct" our IUE results for the presence of Algol C. However, we need to use a parameter set which is more appropriate to the period of the IUE observations. These parameters are provided in TABLE 4.3.3.3.

TABLE 4.3.3.2 – FINAL ALGOL PROPERTIES ADOPTED FOR THIS COMPARISON

| Description of Property | Property Symbol | Final Input Value | Units |
|---|---|---|---|
| Systematic Velocity (km/s) – Should always refer to the entire known system, ie., velocity of Algol AB-C's CM wrt our Sun | $V_o$ | 4.9 | km/s |
| Period of A-B system (days) (Hill et al. 1971 p.444) | $P_{A-B}$ | 2.86730807 | Days |
| Eccentricity of A-B system (Hill et al. 1971 p.451) | $e_{A-B}$ | 0.015 | |
| Angle between line of nodes & periastron of A-B system (degrees) However, the best constant value was 359° from trial#3 Research Vol. 10 | $\omega_{A-B}$ | (function of time) | degrees |
| Semi-amplitude of A wrt A-B system's CM (km/s) (Hill et al. 1971 p.451) | $K_A$ | 44 | km/s |
| Semi-amplitude of B wrt A-B system's CM (km/s) (Tompkin & Lambert 1978) | $K_B$ | 201 | km/s |
| Semi-amplitude of C wrt AB-C's CM (km/s) (Hill et al. 1971 p.450) | $K_C$ | 31.6 | km/s |
| Time of Periastron passage of A-B system (days) - Time of maximum orbital velocity (Hill et al. 1971 p.451) | $T_{A-B}$ | 2,428,482.7390 | Days |
| Period of AB-C system (days) (Hill et al. 1971 p.444) | $P_{AB-C}$ | 680.08 | Days |
| Eccentricity of AB-C system (Hill et al. 1971 p.450) | $e_{AB-C}$ | 0.23 | |
| Angle between line of nodes & periastron of AB-C system (degrees) | $\omega_{AB-C}$ | 133 | degrees |
| Semi-amplitude of AB wrt AB-C system's CM (km/s) - Should follow the equality: $M_C/(M_A+M_B+M_C) = K_{ab-c}/K_c \sim .27$ (Hill et al. 1971 p.451) | $K_{AB-C}$ | 12 | km/s |
| Time of Periastron passage of AB-C system (days) - Time of maximum orbital velocity [calculated from 1952.05 (= 2,434,029.76) listed in Hill et al. 1971 p.451 then adjusted by eye] | $T_{AB-C}$ | 2,434,022.8500 | Days |
| Time of primary minimum of Algol A-B (days) (Hill et al. 1971 p.444 & 450 which refers to Kopal et al. 1960) | $t_{PRMIN\_A-B}$ | 2,428,483.4560 | Days |
| Slope of linear equation for $\omega_{A-B}(t)$ (degrees/day) (Hill's slope = .03079 degrees/day calculated from his estimate of 360° per 32 years. See Fig. 4 of Hill et al. 1971 p.454) | $m_{PRMIN\ A-B}$ | 0.03080 | degrees/day |
| y-intercept of linear equation for $\omega_{A-B}(t)$ (degrees) (Hill's y-intercept ~ -494.327° estimated from Fig. 4 of Hill et al. 1971 p.454) | $b_{\omega A-B}$ | -521.392 | degrees |

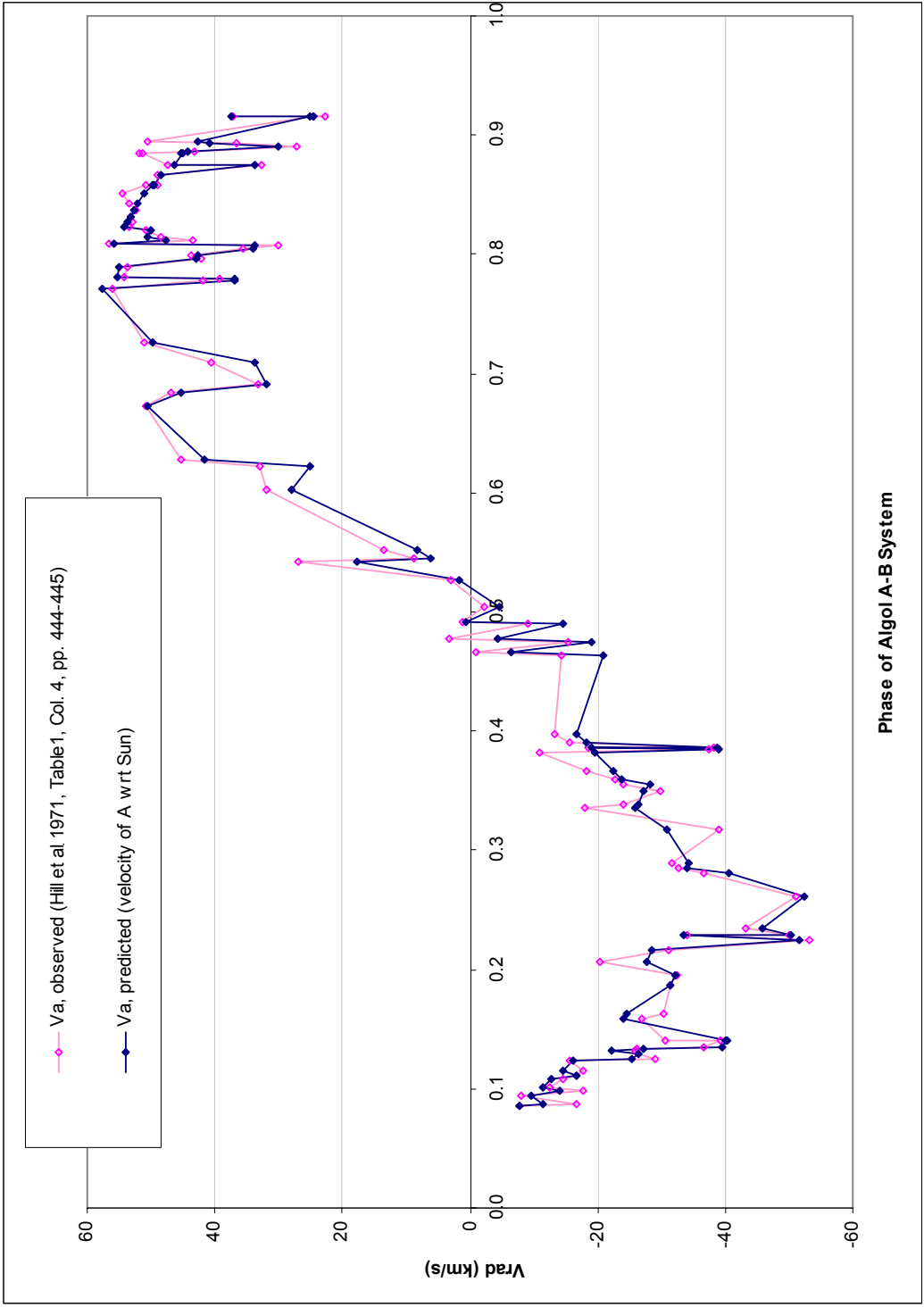

FIG. 4.3.3.4 - *Radial Velocity Comparison of ALGOL A wrt the SUN vs. PHASE of ALGOL A-B for* $\omega = \omega(t)$

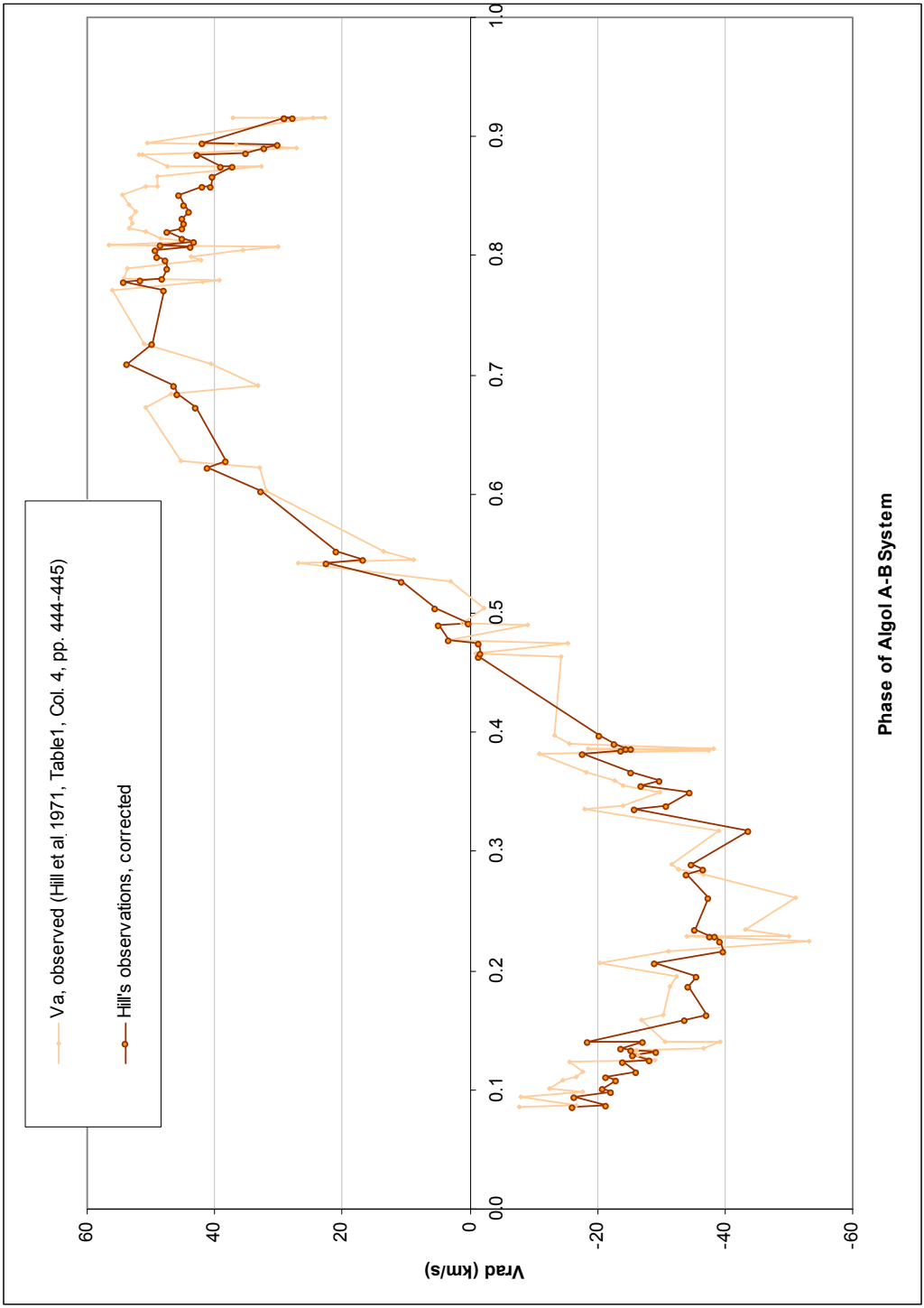

FIG. 4.3.3.5 – *Hill's Observed RADIAL VELOCITY COMPARISON of ALGOL A wrt the SUN With Corrections Applied for* ALGOL A-B *for* $\omega = \omega(t)$

TABLE 4.3.3.3 – FINAL ALGOL PROPERTIES ADOPTED FOR IUE CORRECTIONS

| Description of Property | Property Symbol | Final Input Value | Units |
|---|---|---:|---|
| Systematic Velocity (km/s) – Should always refer to the entire known system, ie., velocity of Algol AB-C's CM wrt our Sun (This work) | $V_o$ | 4.8 | km/s |
| Period of A-B system (days) (Gillet et al. 1989 p.221) | $P_{A-B}$ | 2.86731077 | Days |
| Eccentricity of A-B system (This work) | $e_{A-B}$ | 0.001 | |
| Semi-amplitude of A wrt A-B system's CM (km/s) (Hill et al. 1971 p.451) | $K_A$ | 44 | km/s |
| Semi-amplitude of B wrt A-B system's CM (km/s) (Tompkin & Lambert 1978) | $K_B$ | 201 | km/s |
| Semi-amplitude of C wrt AB-C's CM (km/s) (Hill et al. 1971 p.450) | $K_C$ | 31.6 | km/s |
| Time of Periastron passage of A-B system (days) - Time of maximum orbital velocity (Gillet et al.1989 p.228) | $T_{A-B}$ | 2,445,639.2146 | Days |
| Period of AB-C system (days) (Hill et al. 1971 p.444) | $P_{AB-C}$ | 680.08 | Days |
| Eccentricity of AB-C system (Pan et al. 1993 p.L131) | $e_{AB-C}$ | 0.225 | |
| Angle between line of nodes & periastron of AB-C system (degrees) (Molnar & Mutel 1998 p.17 note that Pan et al. 1993 lists 310.29) | $\omega_{AB-C}$ | 130.29 | degrees |
| Semi-amplitude of AB wrt AB-C system's CM (km/s) - Should follow the equality: $M_C/(M_A+M_B+M_C) = K_{ab-c}/K_c \sim .27$ (Hill et al. 1971 p.451) | $K_{AB-C}$ | 12 | km/s |
| Time of Periastron passage of AB-C system (days) - Time of maximum orbital velocity (Pan et al. 1993 p.L131) | $T_{AB-C}$ | 2,446,931.4000 | Days |
| Time of primary minimum of Algol A-B (days) (Gillet et al. 1989 p.221) | $t_{PRMIN\_A-B}$ | 2,445,641.5135 | Days |
| Slope of linear equation for $\omega_{A-B}(t)$ (degrees/day) (Hill's slope = .03079 degrees/day calculated from his estimate of 360° per 32 years. See Fig. 4 of Hill et al. 1971 p.454) | $m_{PRMIN\ A-B}$ | 0.03080 | degrees/day |
| y-intercept of linear equation for $\omega_{A-B}(t)$ (degrees) (This work) | $b_{\omega A-B}$ | -1,242.734 | degrees |

## 4.4 Determination of UV Light Curves

### 4.4.1 Method for Determining Continuum Flux

In order to estimate the continuum flux levels we first coplotted the raw and smoothed data in 50 Å intervals as illustrated by FIG. 4.4.1.1. Since the continuum flux is expected to vary only slightly over these intervals, the flux levels were estimated by drawing a "best" horizontal line by eye for each interval. Some degree of subjective decisions have to be made due to noise and other instrumental effects. Consistency in the determination of the continuum over all spectra is very important and was one determining factor in the final values obtained in each 50 Å interval. The flux level was then recorded in units of $1 \times 10^{-10}$ ergs/cm$^2$/s/Å. In FIG. 4.4.1.1 the level is $5.58 \times 10^{-10}$ ergs/cm$^2$/s/Å.  The SWP and LWP/LWR continuum flux level estimates are reported as described in the following section, and used to determine the UV light curves, showing intensity as function of phase.

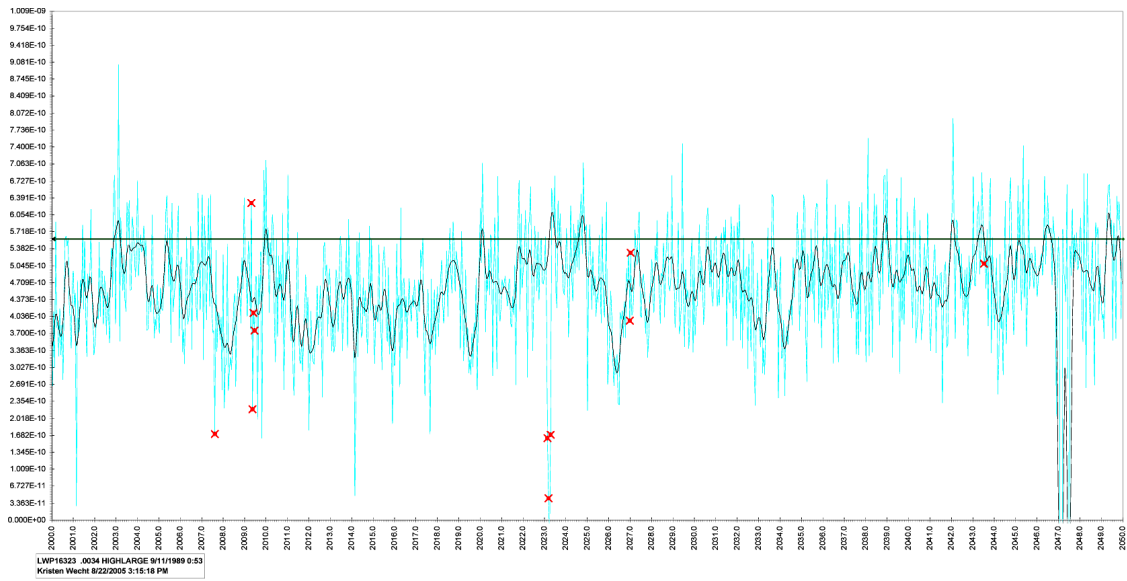

FIG. 4.4.1.1– *Flux vs. wavelength in the interval 2000-2050 Å.*
LWP16323 ϕ = 0.003 9/11/89 large aperture

### 4.4.2  *Continuum Flux and Light Curve Results*

FIGS. 4.4.2.1 and 4.4.2.2 show the wavelength dependence of the continuum flux for SWP and LWP/ LWR UV ranges, respectively.  In both figures, a range of phases was selected corresponding to the 1989 epoch.  Our complete set of continuum flux results are tabulated in Appendix D.2.

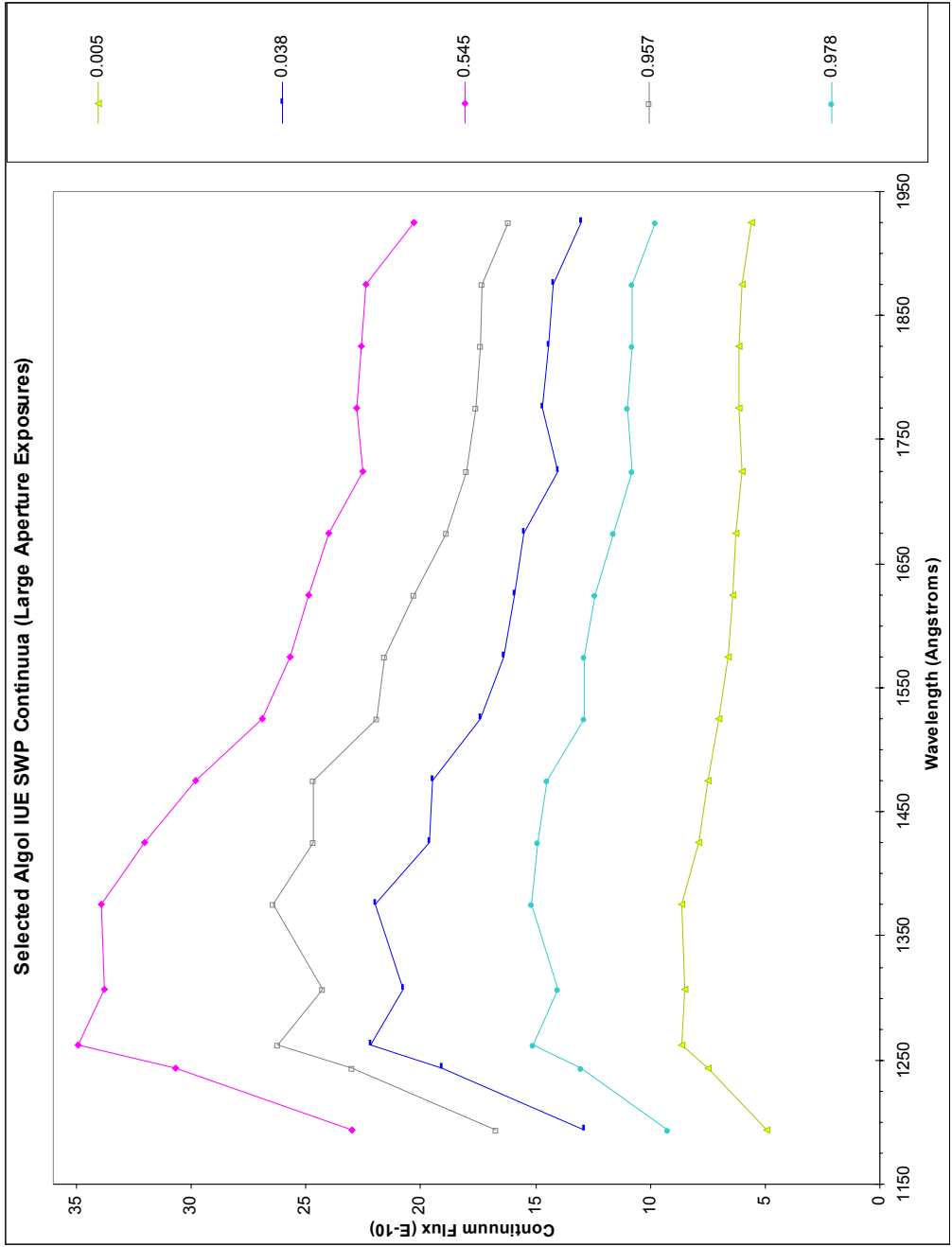

FIG. 4.4.2.1 – *Selected Algol IUE SWP Continuua (Large Aperture Exposures)*

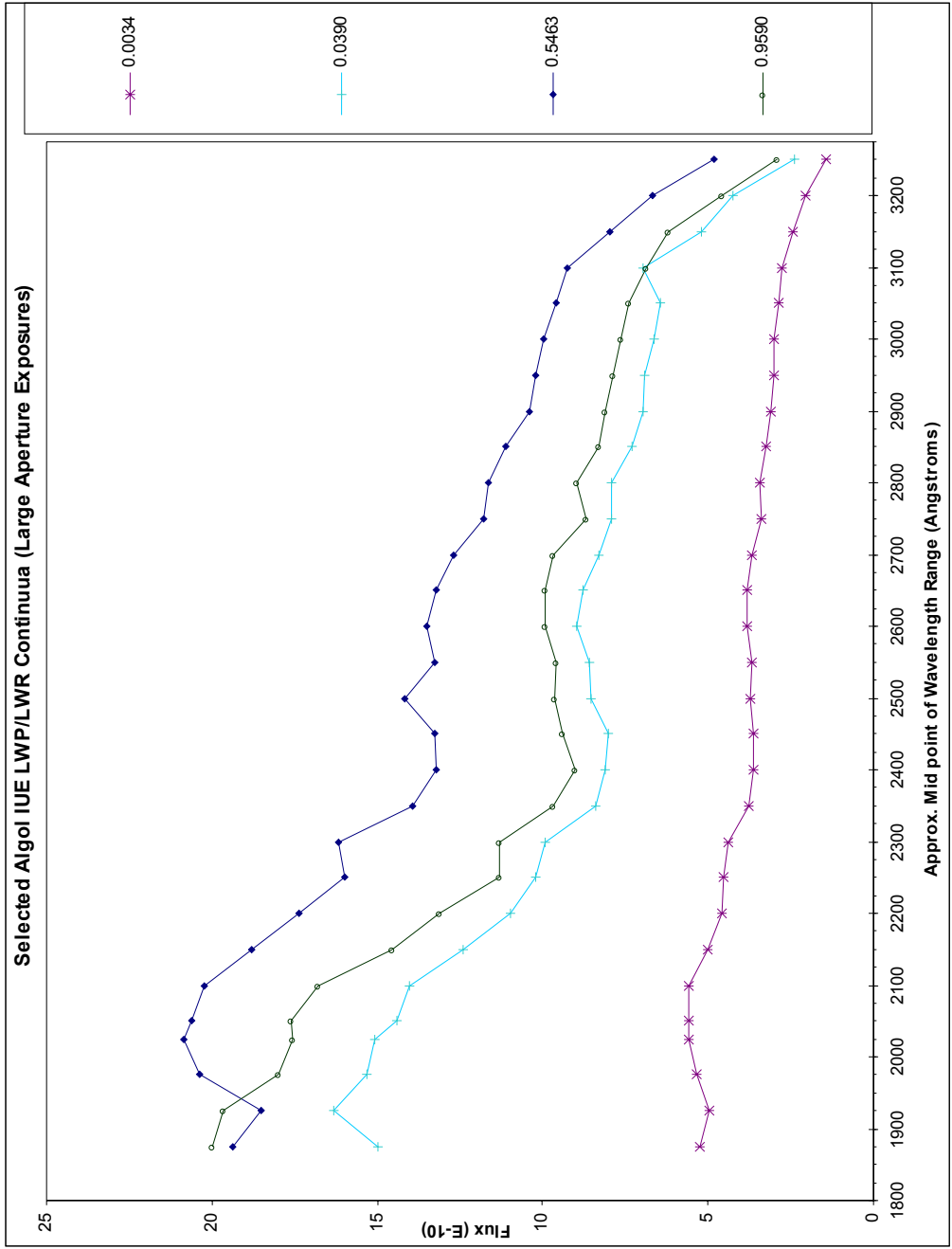

FIG. 4.4.2.2 – *Selected Algol IUE LWP/LWR Continuua (Large Aperture Exposures)*

FIG. 4.4.2.3 compares two SWP continuum flux curves corresponding to about the same phase, but one during an epoch of high activity and the other of moderate activity.

The SWP and LWP/LWR continuum curves of FIGs. 4.4.2.1 and 4.4.2.2 indicate an overall decrease in flux with increasing wavelength beyond approximately 1350 Å. The peak emission for a 12,500 K B8 star should be located at about 2300 Å. Hence, although black-body dependencies contribute to the curves, other factors are more important.

The abrupt flux increase in the 1200 Å region is combination of instrumental effects since this is near the SWP wavelength limit and Lyman alpha absorption. We do not use data near the SWP or LWP/LWR wavelength limits due to such effects. The generally decreasing flux with increasing wavelength is due primarily to the instrumental response.

Most of the LWP/LWR flux curves show a broad absorption feature in the 2100 – 2500 Å spectral image. These broad features might be composed of two overlapping absorptions, in the ranges of 2200 – 2300 Å and 2350 – 2500 Å. The 2100 – 2500 Å range corresponds to 0.21 – 0.25 μm or 4.8 – 4.0 μm$^{-1}$.

This wavelength range is interesting in the context of interstellar extinction and scattering due to small particles. The average interstellar extinction curve has a significant "bump" centered at ~ 4.6 μm$^{-1}$, or 2175 Å, and we speculate that this might be related to our observed features. The origin of the feature is a source of continued study and controversy; however, various models include contributions from particles composed of silicon-carbon composites and graphite along with polycyclic aromatic hydrocarbons (PAHs). These observations and models are discussed in Whittet (2003; Chapter 3). We will not pursue this interesting topic of interstellar grains further since it is outside the scope of this work.

FIGS. 4.4.2.1 and 4.4.2.2 both show expected phase dependencies. The flux is lowest near primary minima (near phases of 0.0 or 1.0). Among these curves the largest flux occurs at phase 0.55, and they should be slightly higher at a phase of around 0.6, for example, when the system is fully out of the secondary minimum. (See FIGS. 3.3.2.)

FIG. 4.4.2.3 shows a comparison of two epochs, where one shows evidence of greater activity. Notice the overall lower flux during the more active period.

FIGS. 4.4.2.4 and 4.4.2.5 represent SWP and LWP/LWR light curves, showing the phase dependencies of the flux, but presented in a 3-D perspective to cover many wavelength ranges. The curve closest to the "Algol Phase" axis represents the 3225 – 3275 Å continuum flux averages. Primary minimum is clearly seen, and secondary is weak but distinguishable in some curves. After normalization, the light curves in FIGS. 4.4.2.7 and 4.4.2.8 are replotted and stacked in FIGS. 4.4.2.6 and 4.4.2.7.

As expected, the depth of primary minimum decreases as the wavelength increases. The light curve of Algol at 3428 Å (Chen et al. 1977) is coplotted in FIG. 4.4.2.7 for comparison. From this comparison, we see that the light curves we extracted from the IUE data are consistent with the Copernicus satellite data.

A sample normalized SWP light curve, averaged over a 200 Å interval is plotted in FIG. 4.4.2.8, where different epochs are distinguished. Likewise, a sample normalized LWP/LWR light curve averaged over a 400 Å interval is plotted in FIG. 4.4.2.9. FIGS. 4.4.2.10 and 4.4.2.11 are the same, but focused on a narrow range of phases near primary minimum. The final set of normalized light curves show several interesting features. In particular, the 1984 data is unusual in almost every spectral range, often distinctly different from all other epochs. For example, notice especially in FIG. 4.4.2.10 that the March 1984 feature during ingress at phase 0.9388 is significantly below the light curve appropriate to other epochs (1989 at phase 0.9397 for example). This depression in the continuum can be seen more clearly in the continuum flux comparison of FIG. 4.4.2.3.

Other features are recurring and appear in the same range of phases at a number of epochs; these include dips or bumps in the approach to eclipse. In particular, note the absorption feature at phase 0.89-0.90 in the 1750 – 1950 Å spectral range, shown enlarged in FIG. 4.4.2.10.

Other features will be examined as appropriate in the Discussion and Conclusions chapter.

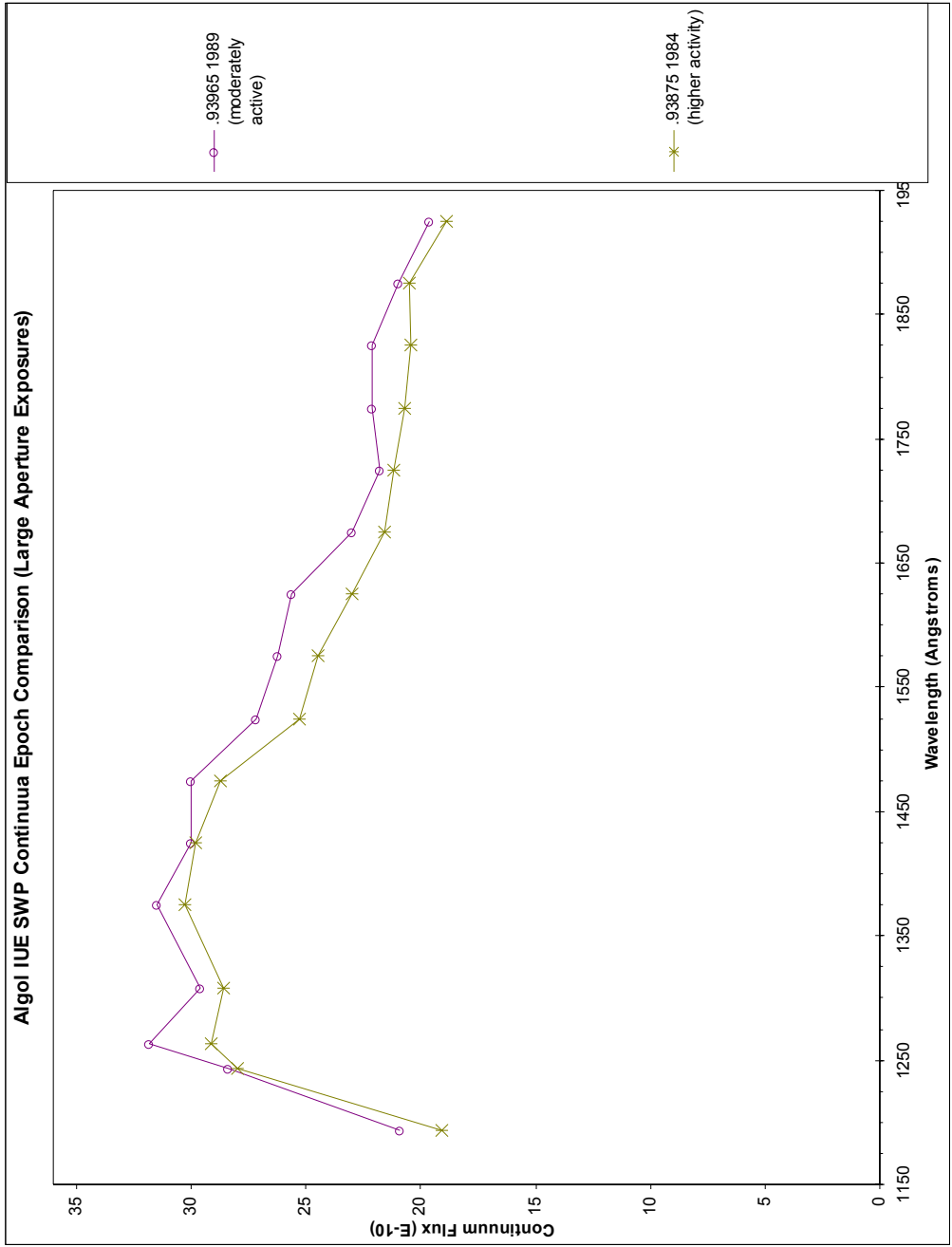

FIG. 4.4.2.3 – *Algol IUE SWP Continuua Epoch Comparison (Large Aperture Exposures)*

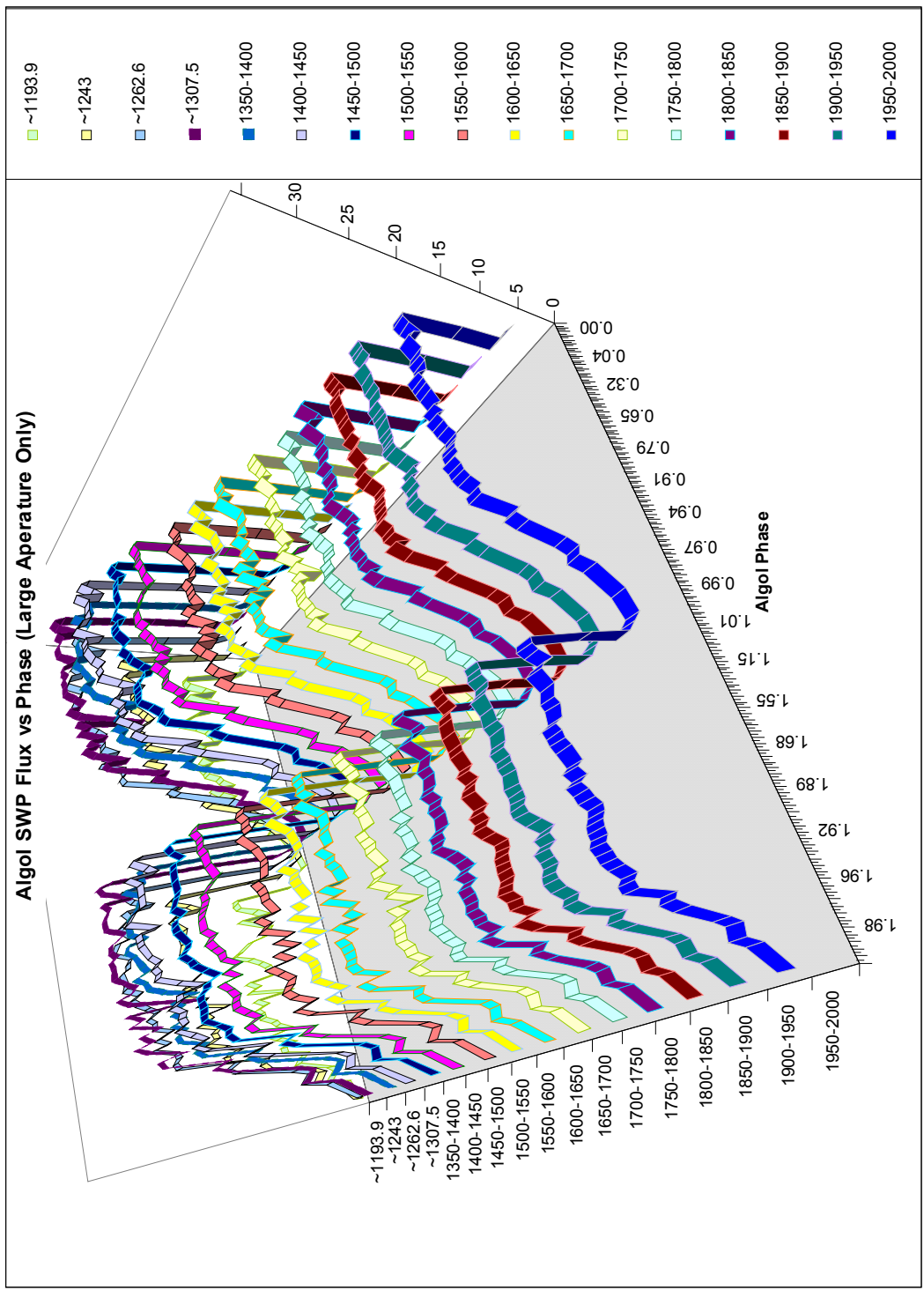

FIG. 4.4.2.4 – *Algol SWP vs. Phase (Large Aperture Only)*

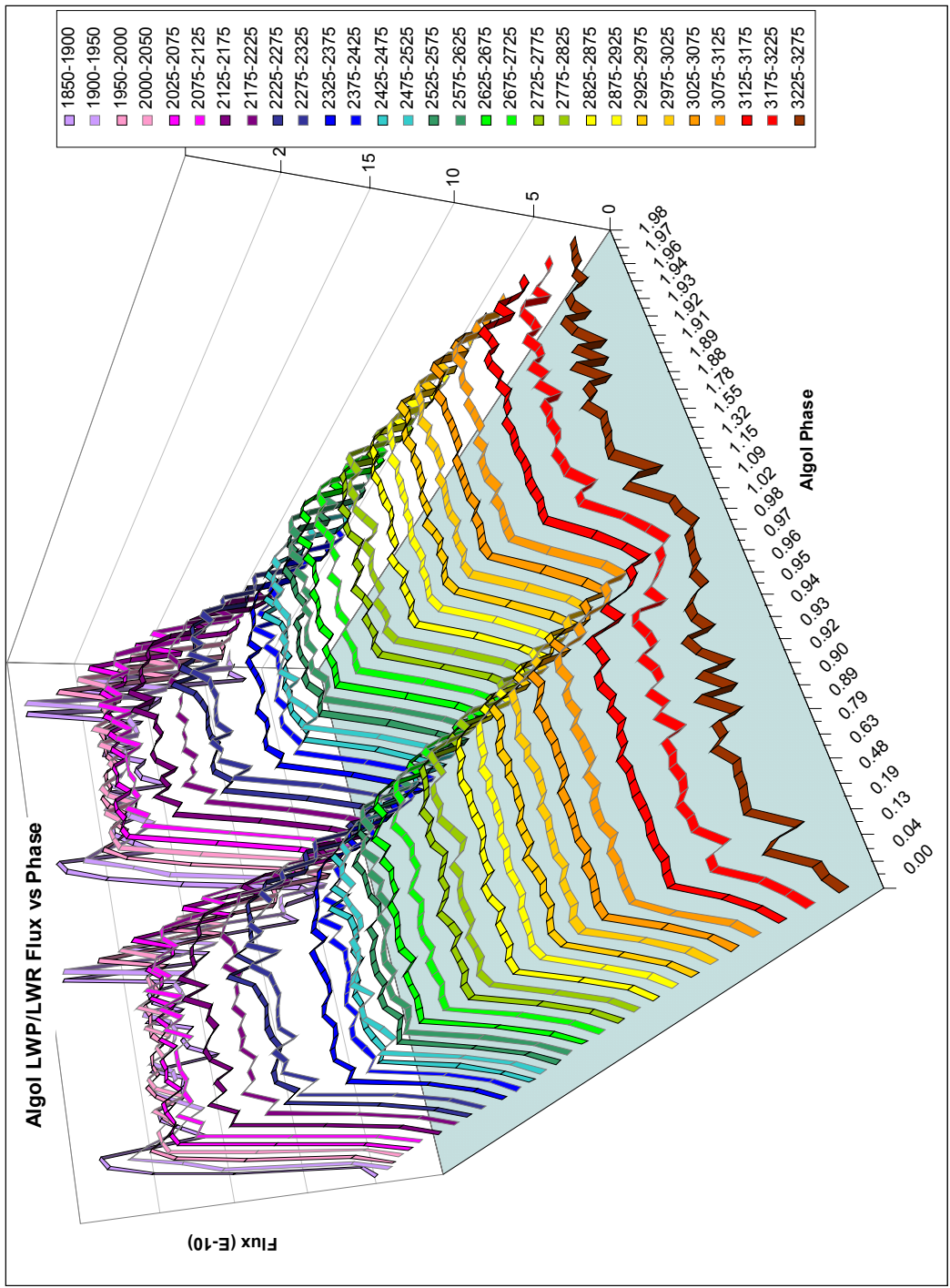

FIG. 4.4.2.5 – Algol LWP/LWR Flux vs. Phase

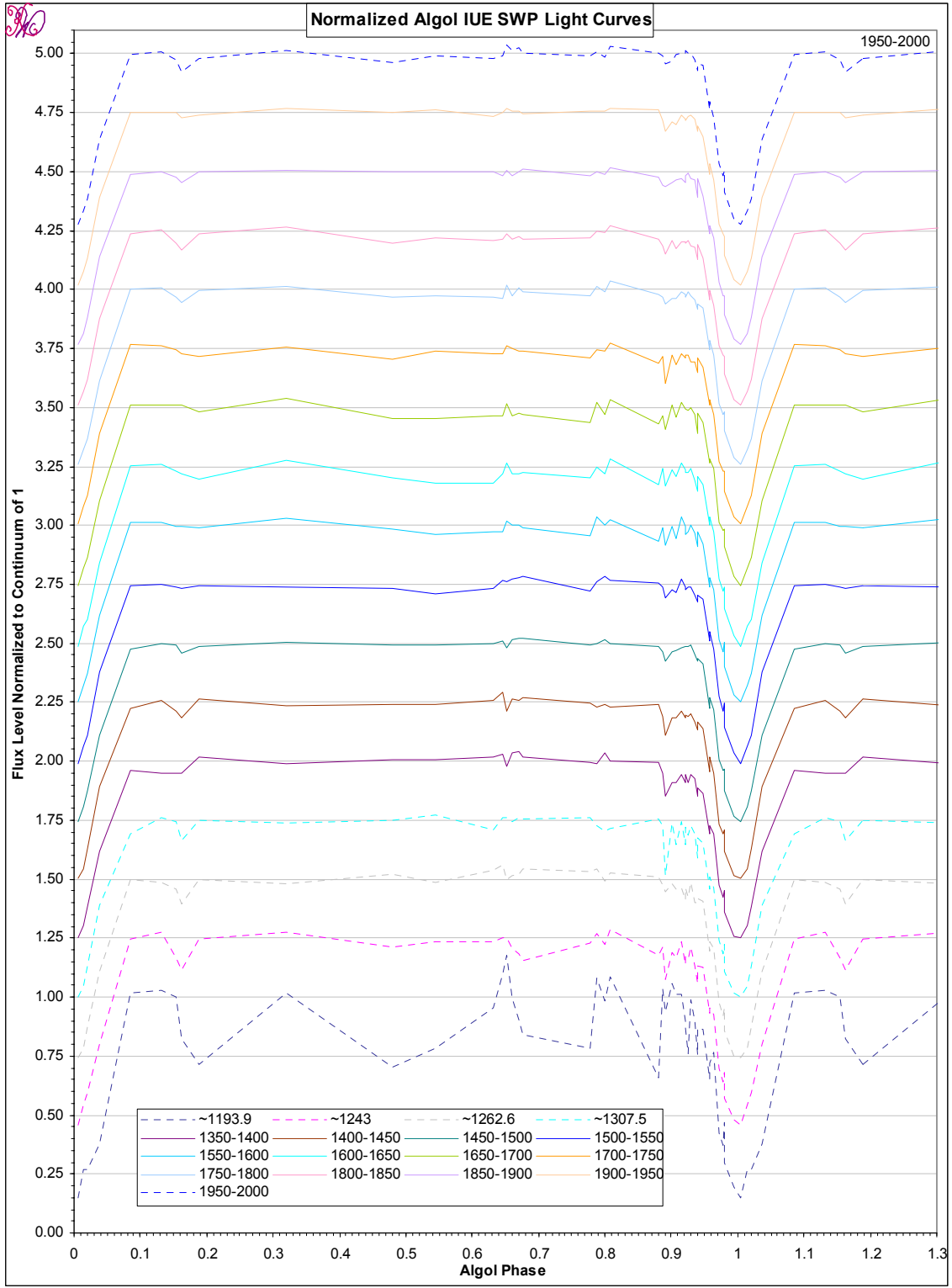

FIG. 4.4.2.6 – *NORMALIZED ALGOL IUE SWP LIGHT CURVES*. The curves are stacked by adding 0.25 to each successively.

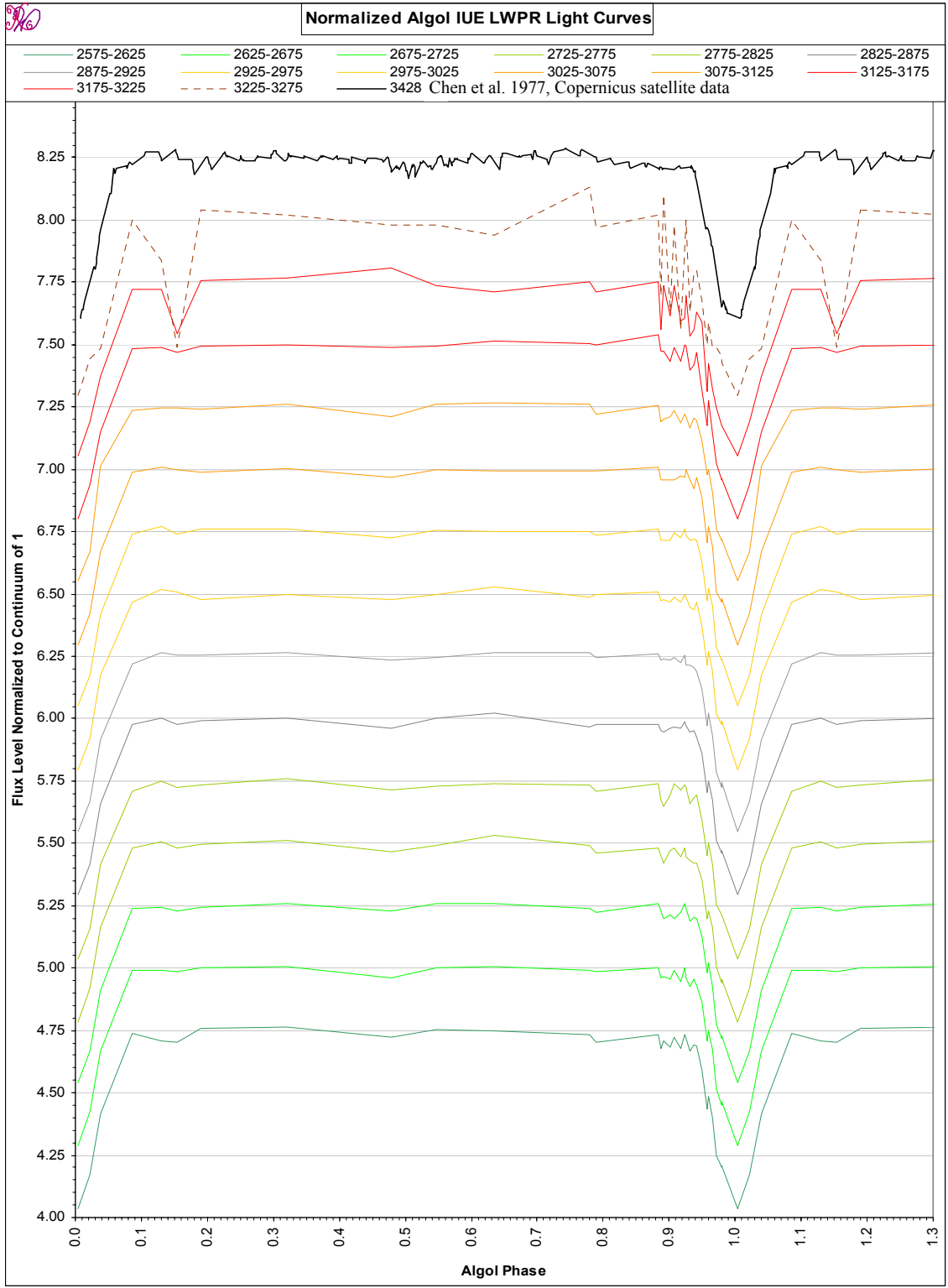

FIG. 4.4.2.7a – *NORMALIZED ALGOL IUE LWP/LWR LIGHT CURVES.* The curves are stacked by adding 0.25 to each successively.

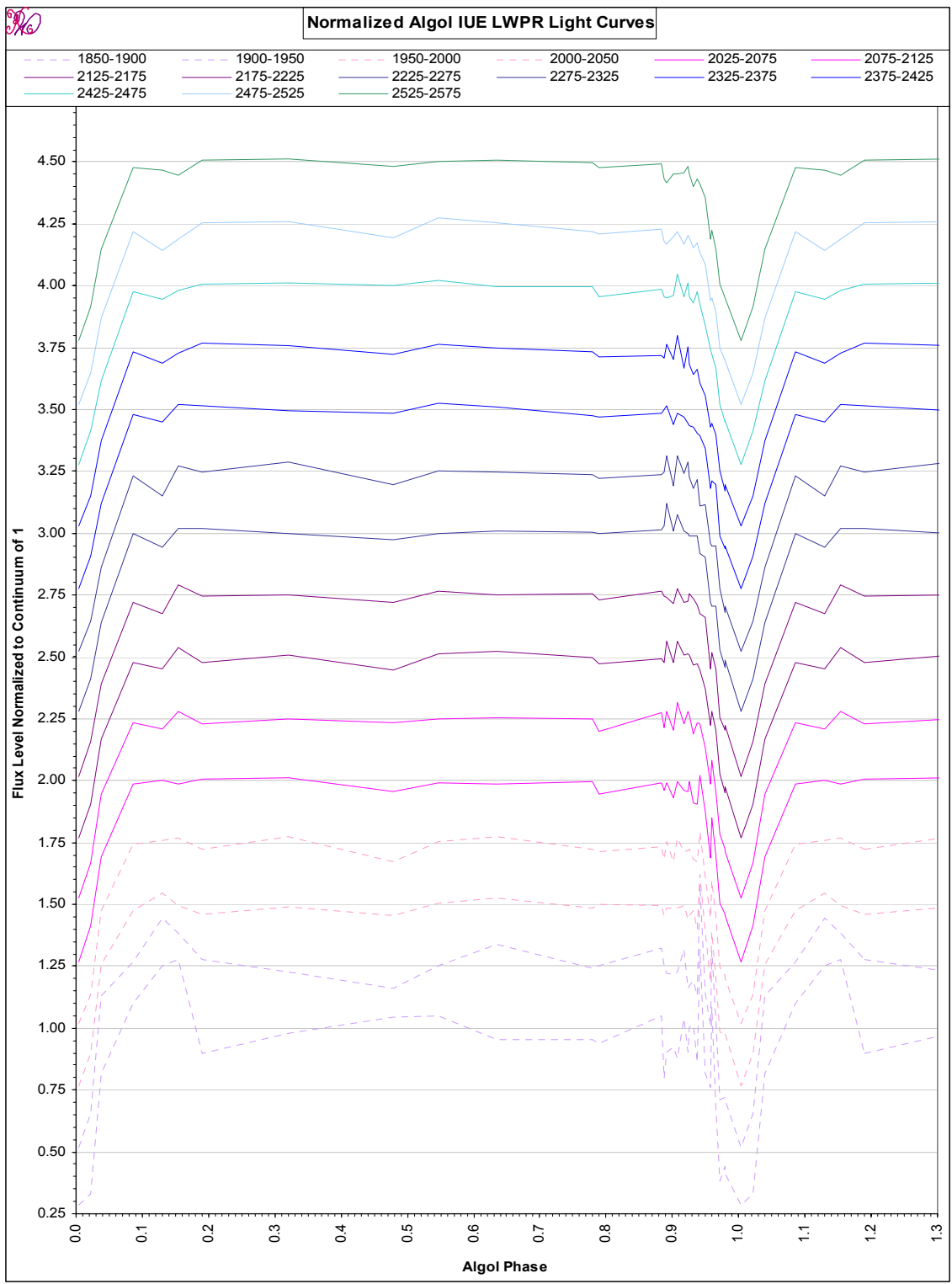

FIG. 4.4.2.7b – *NORMALIZED ALGOL IUE LWP/LWR LIGHT CURVES*

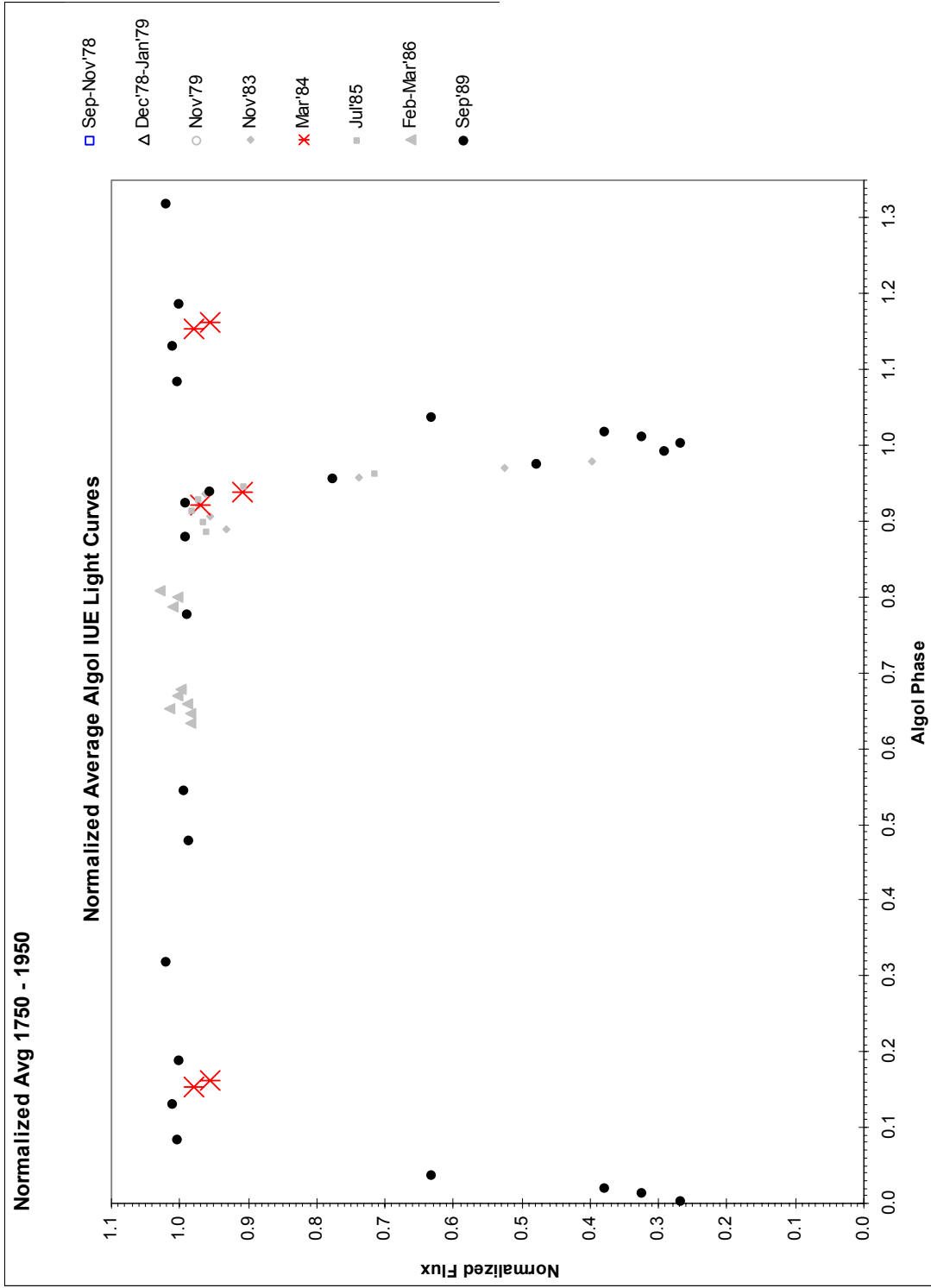

FIG. 4.4.2.8 – SWP *Normalized Average Algol IUE Light Curves. 1550 - 1750 Å*

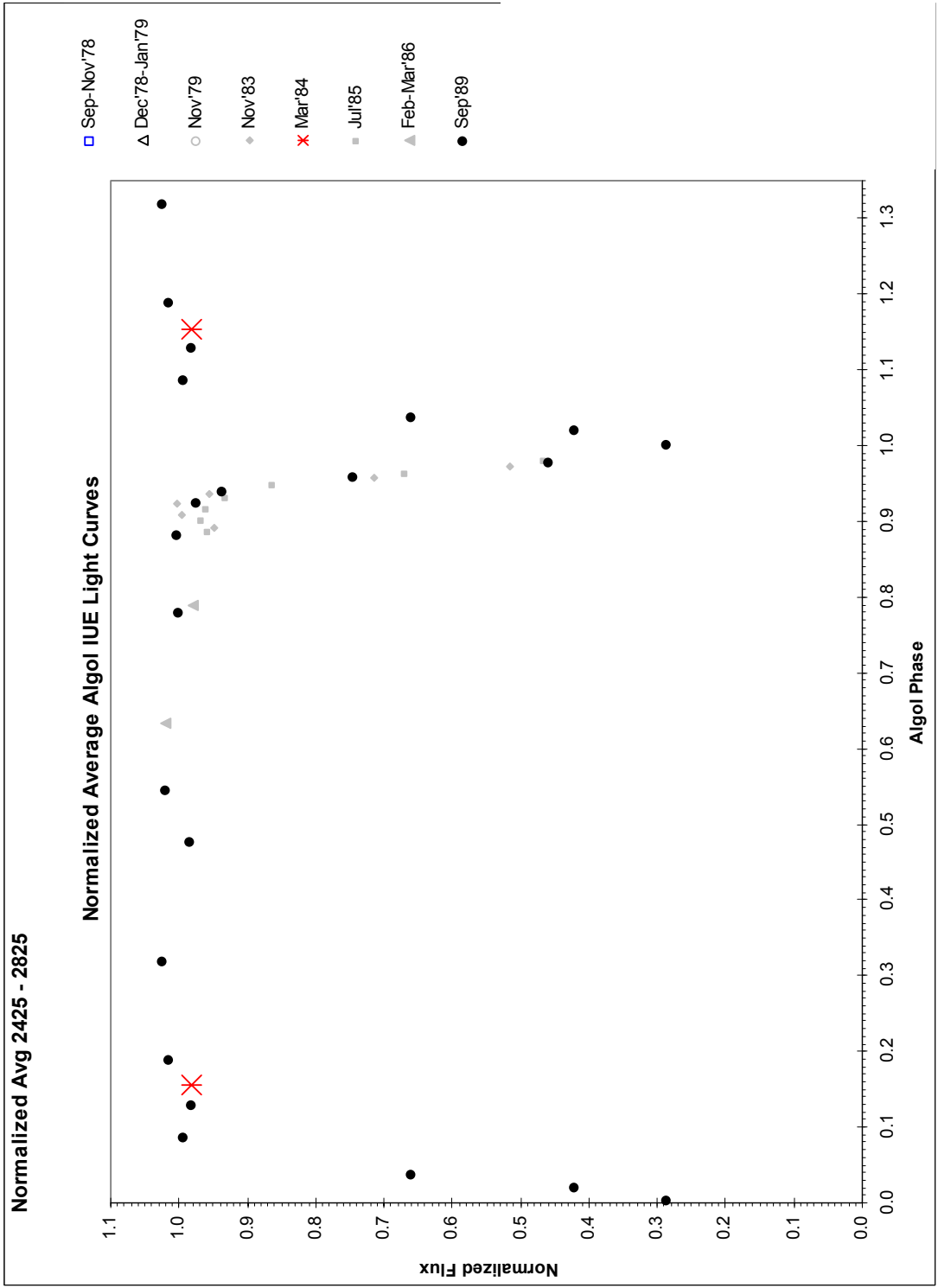

FIG. 4.4.2.9 – *LWP/LWR Normalized Average Algol IUE Light Curves. 2425 - 2825 Å*

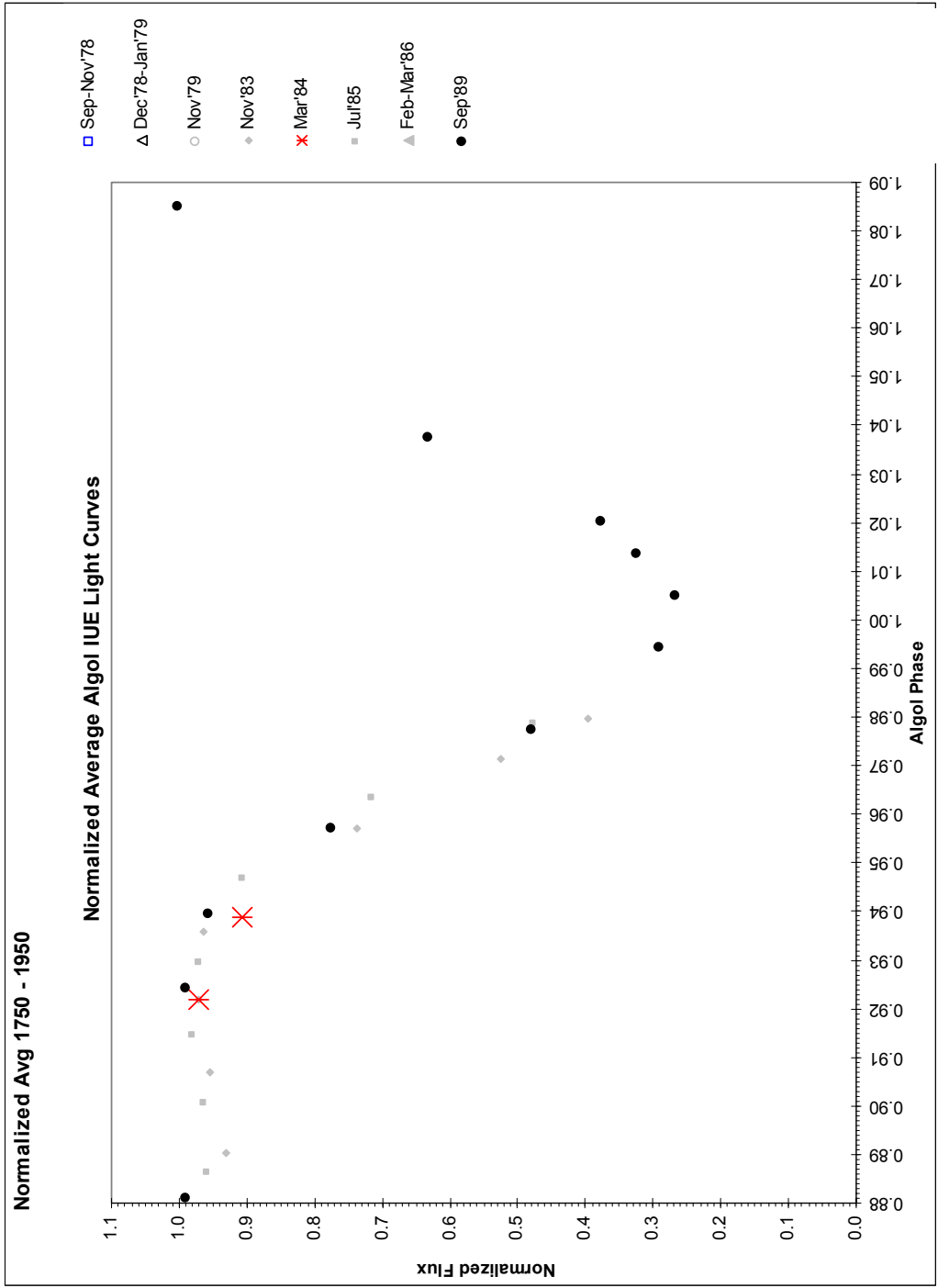

FIG. 4.4.2.10 – SWP *Normalized Average Algol IUE Light Curves Near Primary Eclipse. 1550 - 1750 Å*

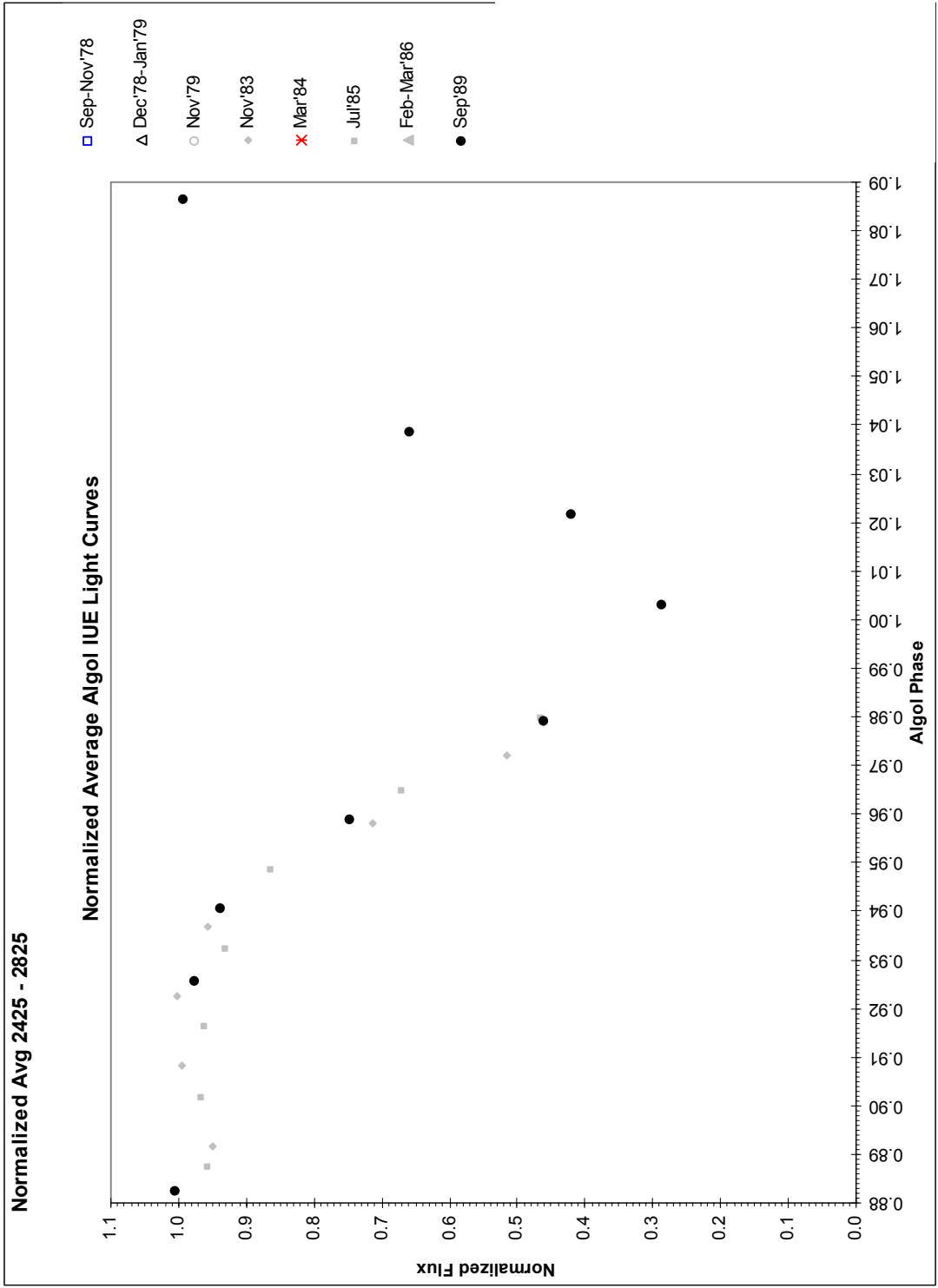

FIG. 4.4.2.11 – *LWP/LWR Normalized Average Algol IUE Light Curves Near Primary Eclipse. 2425 - 2825 Å*

## 4.5 *Determination of Radial Velocities and Absorbing-Gas Characteristics from Spectral Line Analyses*

### 4.5.1 *Methods of Spectral Line Analyses*

After identifying a spectral line, we determined its position, depth, and width for each exposure. Si IV, for instance, appears in the SWP range, where there are 59 SWP exposures; hence, we determined 59 positions, depths, and widths for each of Si IV λλ1393, 1402 lines. Since Mg II appears in the LWP/LWR range, we determine 44 positions, depths, and widths for Mg II λλ2796, 2803 lines and subordinate lines Mg II λλ2791, 2798.

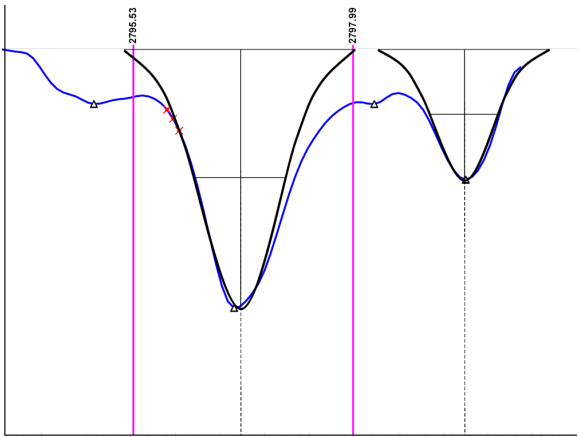
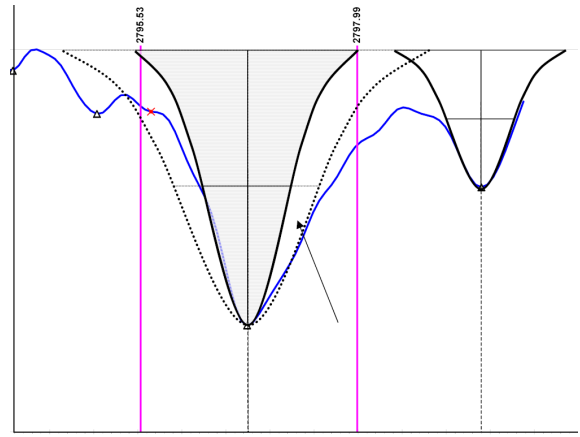

FIG. 4.5.1.1.—Mg II 2796 and Mg II 2798  
ϕ=0.7798  9/13/1989 -- *symmetric* -- LWP16347

FIG. 4.5.1.2.—Mg II 2796 and Mg II 2798  
ϕ=0.7884  3/2/1986 -- *asymmetric* -- LWP07738

The position, depth, and width for each line were determined simultaneously by estimating the best inverted Gaussian shape that fits the line, as shown in FIG. 4.5.1.1. A perpendicular was dropped from the center of the Gaussian shape (narrow dashed line) to the wavelength axis and recorded. The rest wavelength was then subtracted from the recorded value to determine the Doppler shift, and ultimately, the radial velocity. The depth of the Gaussian shape was measured at its deepest point as a fraction of the continuum level. For the purpose of these calculations, new continuum levels were estimated for each spectral line examined. We used the same method as that described in Section 4.4; however, these were identified using narrower wavelength ranges.

The difference between the continuum flux values and the line depths provided us with measures of residual intensities. These measure the amount of UV flux that is not extinguished along the line of sight.

The width of the Gaussian shape at half of its depth was measured in angstroms and recorded. If the line profile was slightly asymmetric, we captured the essence of the line shape using a symmetrical Gaussian shape as described above, fitting what appeared to be the most unblended half.

If the line profile was noticeably asymmetric, we approximated the line shape with two Gaussian shapes having the same centers and depths, differing only in width. The width of one Gaussian was adjusted to fit the left side of the absorption well, and the width of the other was adjusted to the right side. The position of the line is coincident with the centers of the two overlaid Gaussian shapes. The depth of the line is represented by either Gaussian shape since their depths are also equal. The FWHM was determined by adding the HWHM of the wider Gaussian to the HWHM of the narrower Gaussian.

FIG. 4.5.1.2 is an example of an asymmetric Mg II λ2796 line at phase $\phi=0.7884$ from March 1986. Its shape was approximated with two Gaussian shapes. The narrower, shaded Gaussian is overlaid and centered on the wider Gaussian shape (dotted curve) that captures the extra red absorption indicated by the arrow.

A measure of the degree of line asymmetry was captured by recording the additional red-shifted or blue-shifted Gaussian half-widths required to fit the line.

### 4.5.2 *Radial Velocity Calculations and Results*

We obtained radial velocity curves as functions of phase, $\varphi$, for each spectral line by using our measured line centers, $\lambda(\varphi)$, and calculated Doppler shifts from vacuum rest wavelengths, $\lambda_o$. The radial velocities, not yet "corrected" to remove the effects of systematic Algol-system motion and the presence of Algol C, are given by Eq. 4.5.2.1, where "uc" indicates "uncorrected."

$$V_{uc}(\varphi) = \frac{c(\lambda(\varphi) - \lambda_o)}{\lambda_o} \qquad (4.5.2.1)$$

These results are tabulated in Appendix D.3.

The results presented in this section remove the effects due to systematic velocity and the presence of Algol C, as developed in Section 4.3.3. Hence, the "corrected" radial velocities are given by

$$V(\varphi) = V_{uc} - V_o - V_{AB:AB-C} \qquad (4.5.2.2)$$

FIGs. 4.5.2.1- 4.5.2.11 show radial velocity curves for the following spectral lines: Al II $\lambda1670$, Al III $\lambda1854$, C II $\lambda1334$, Fe II $\lambda1640$, Si III $\lambda1294$, Si IV $\lambda 1393$, Si IV $\lambda1402$, Fe II $\lambda2586$, Fe II $\lambda2600$, Mg II $\lambda2796$, Mg II $\lambda2803$, respectively.

The solid lines represent our simulation of the radial velocity of the center of mass of Algol A relative to the center of mass of the Algol A-B system. In other words, the systematic velocity and the velocity of the A-B center of mass (i.e., the effect of Algol C) have been removed in producing our solid line, as they have been removed from the data points.

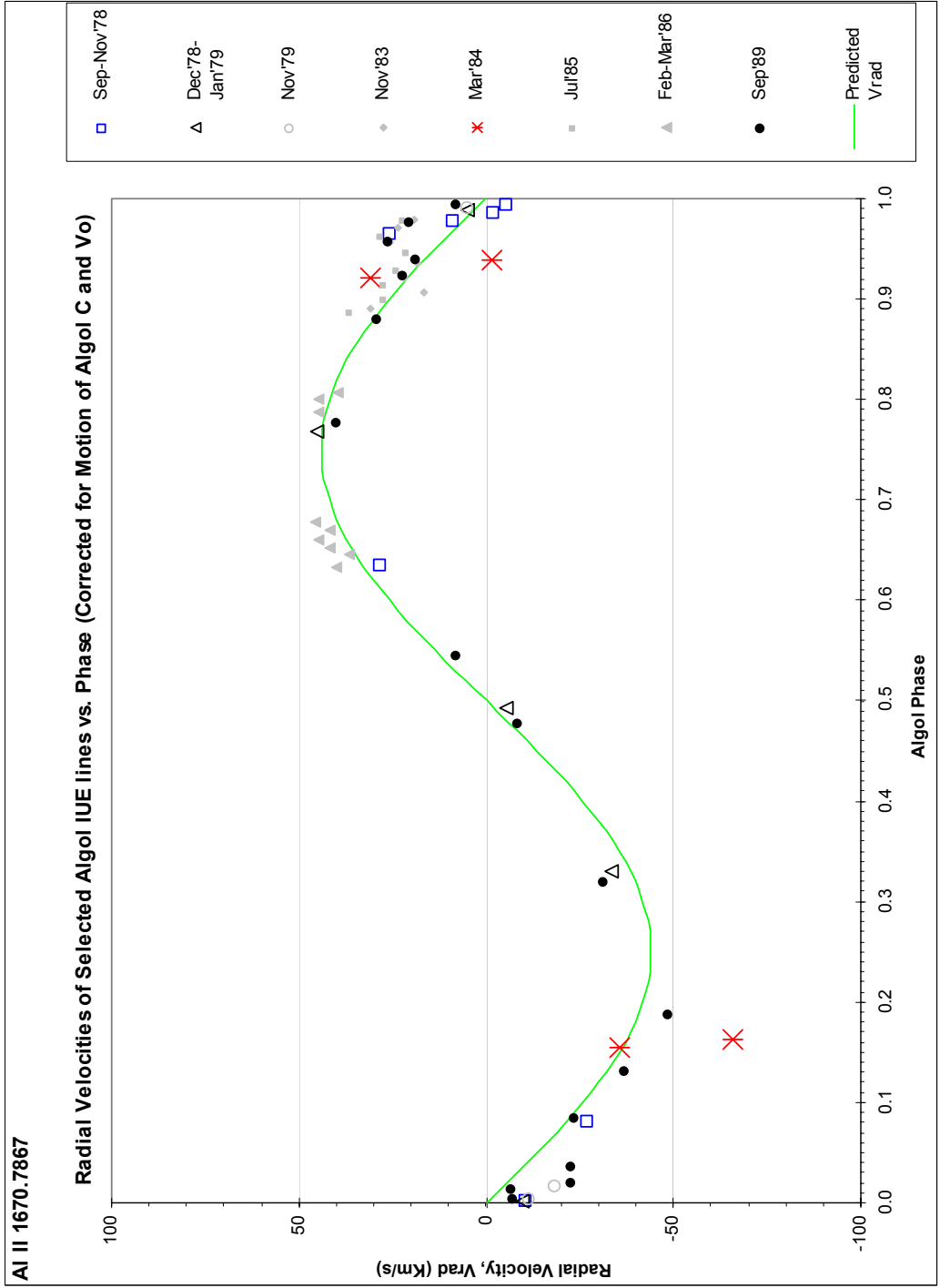

FIG. 4.5.2.1 – *Radial Velocity vs Phase Al II 1670*

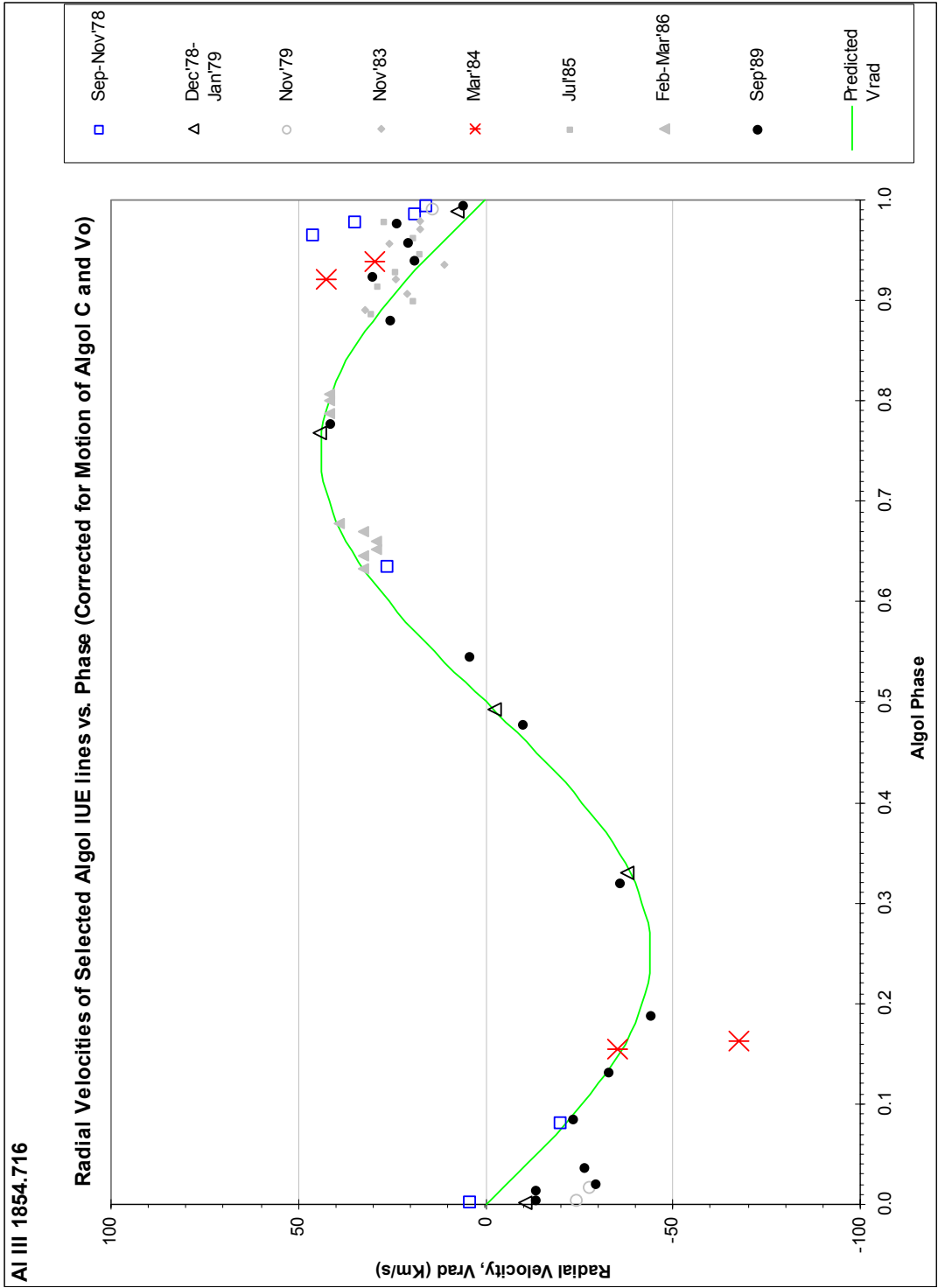

FIG. 4.5.2.2 – *Radial Velocity vs Phase Al III 1854*

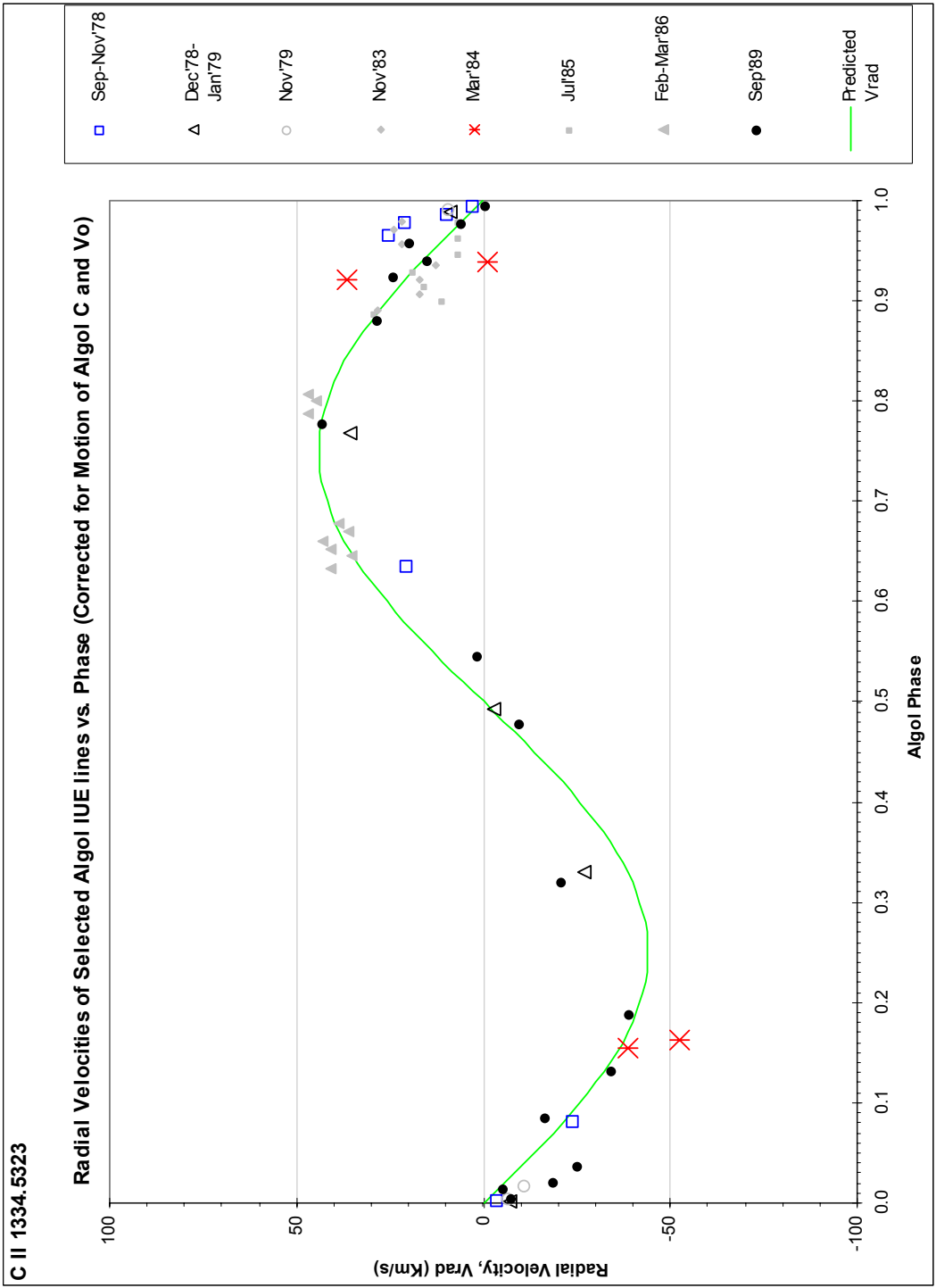

FIG. 4.5.2.3 – *Radial Velocity vs Phase C II 1334*

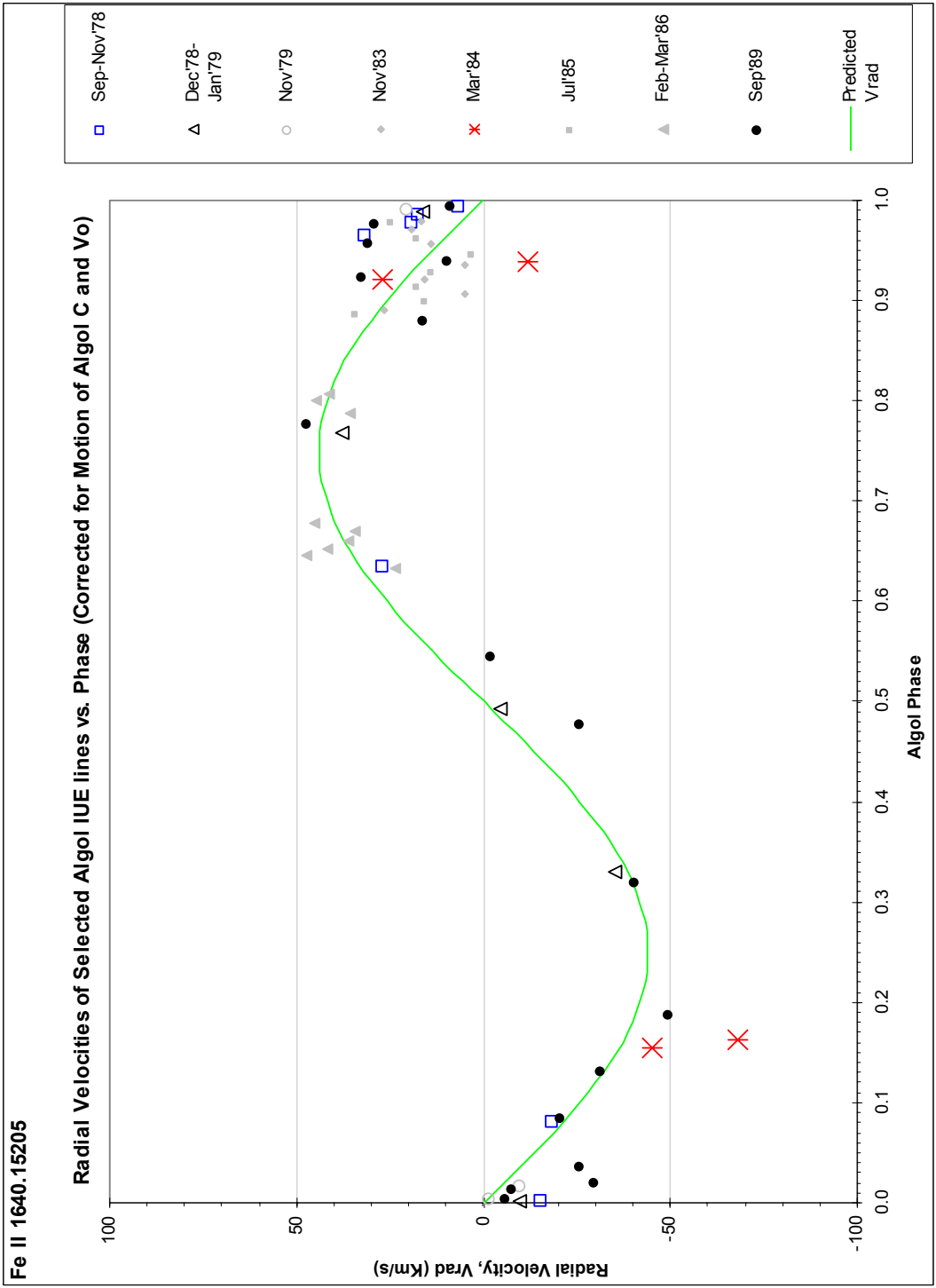

FIG. 4.5.2.4 – *Radial Velocity vs Phase Fe II 1640*

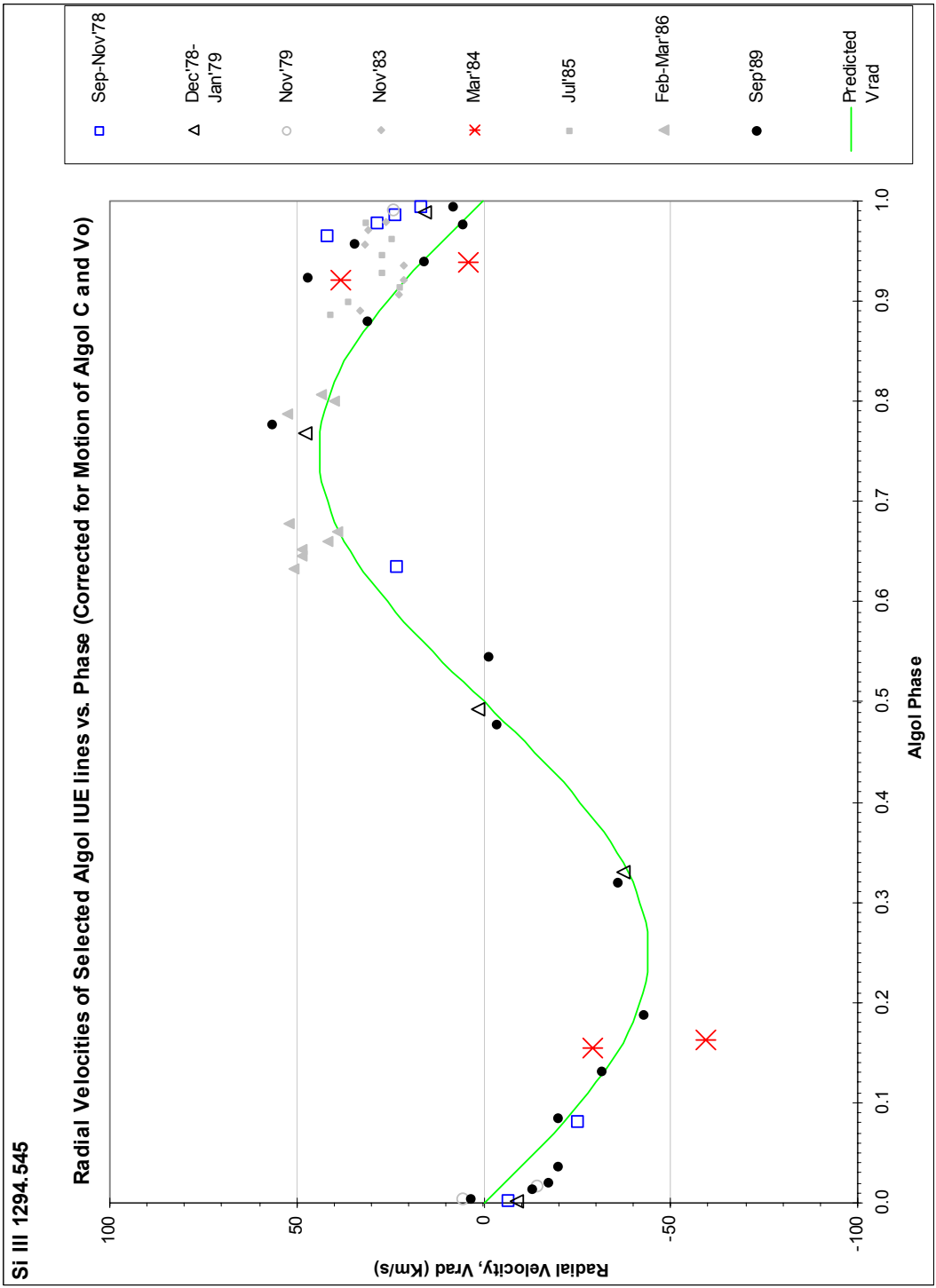

FIG. 4.5.2.5 – *Radial Velocity vs Phase Si III 1294*

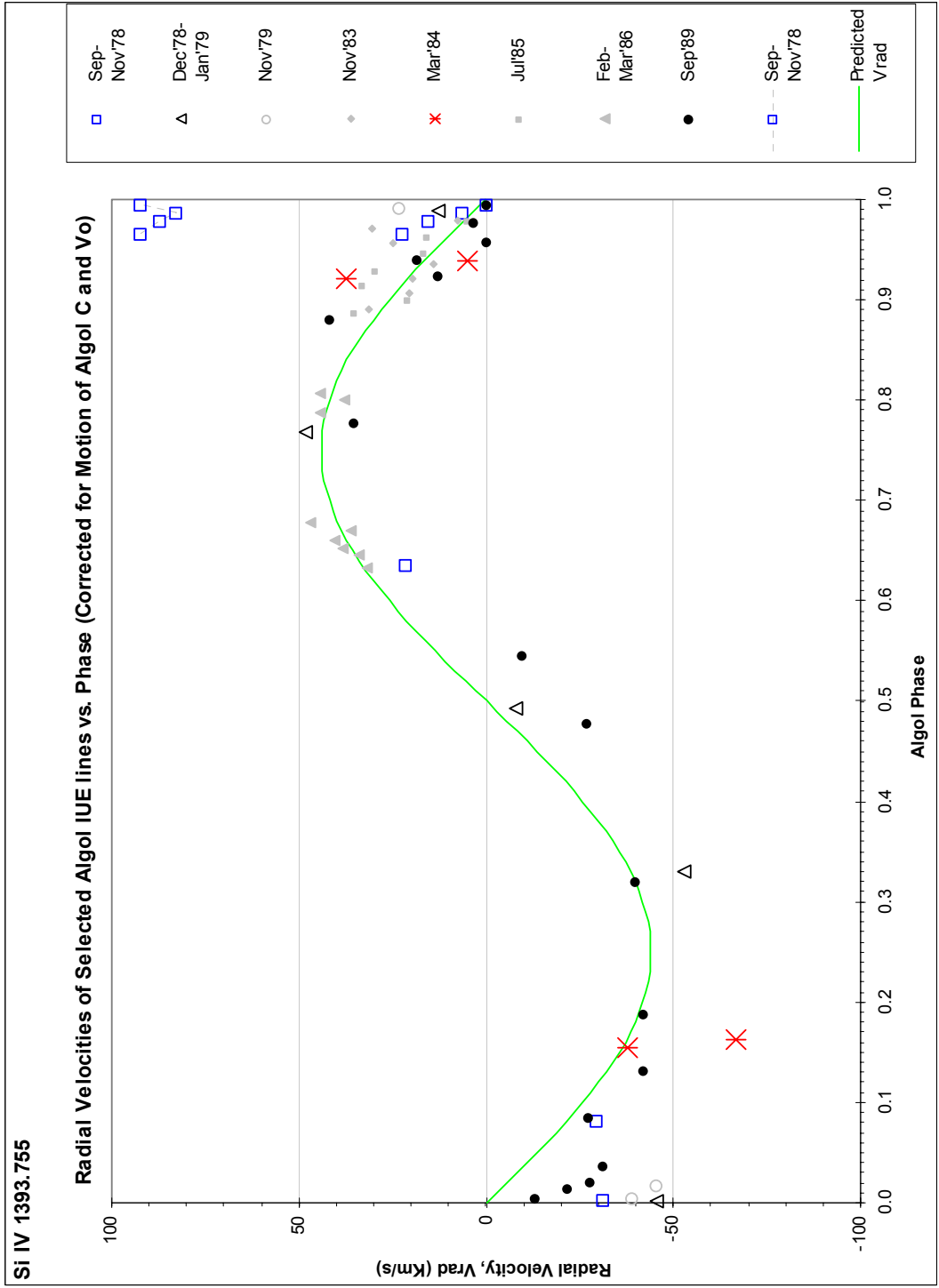

FIG. 4.5.2.6 – *Radial Velocity vs Phase Si IV 1393*

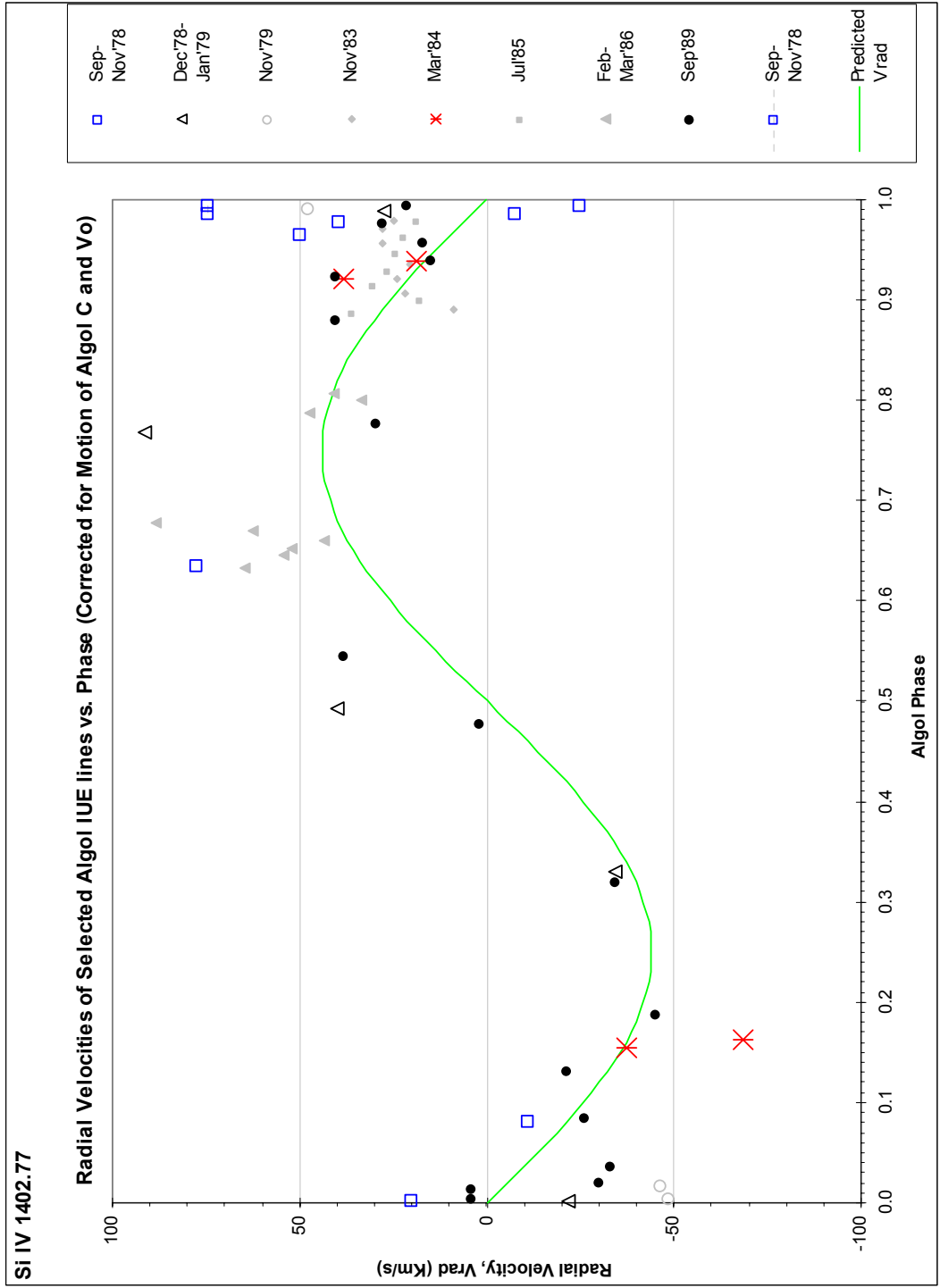

FIG. 4.5.2.7 – *Radial Velocity vs Phase Si IV 1402*

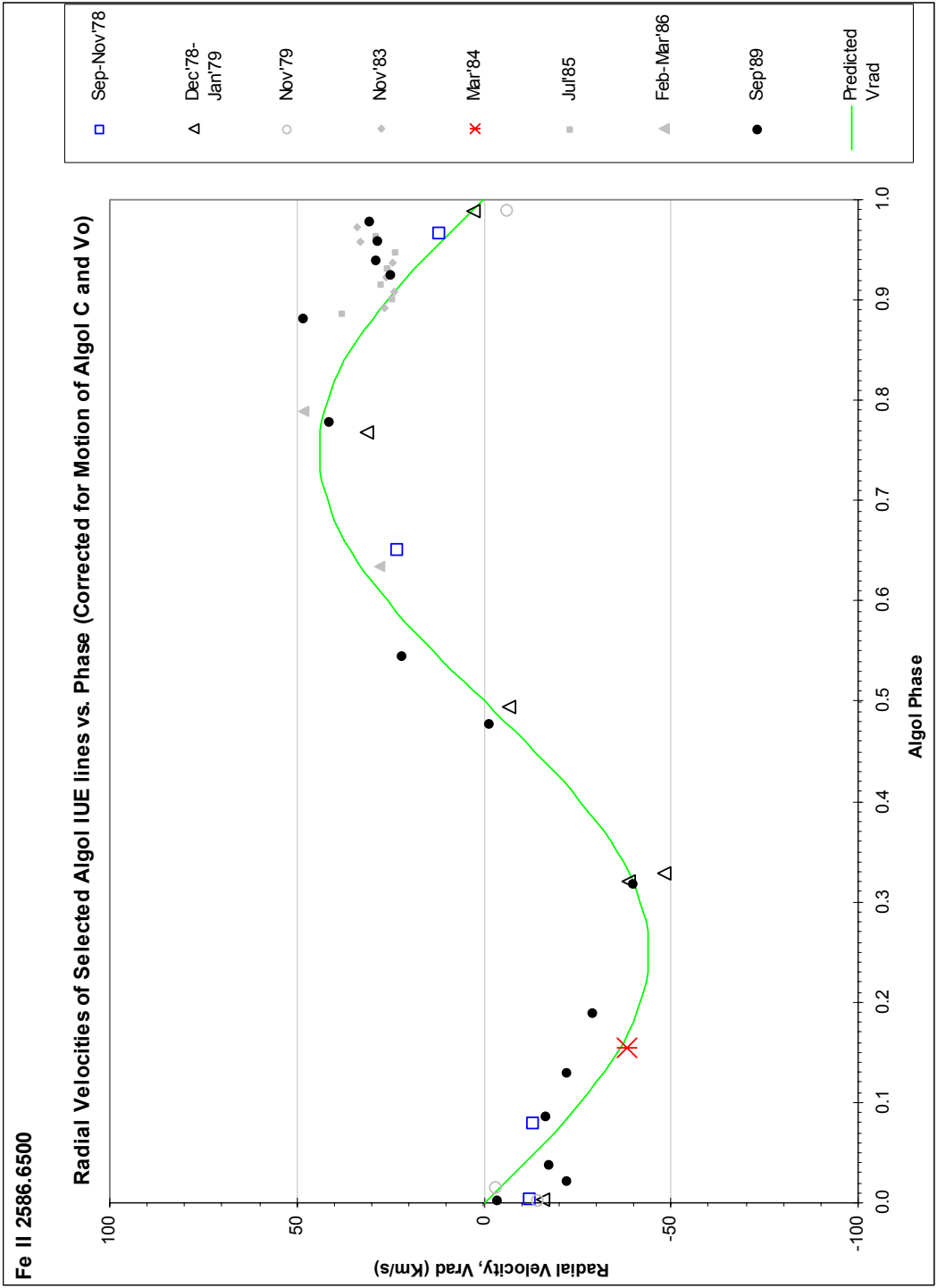

FIG. 4.5.2.8 – *Radial Velocity vs Phase Fe 2586*

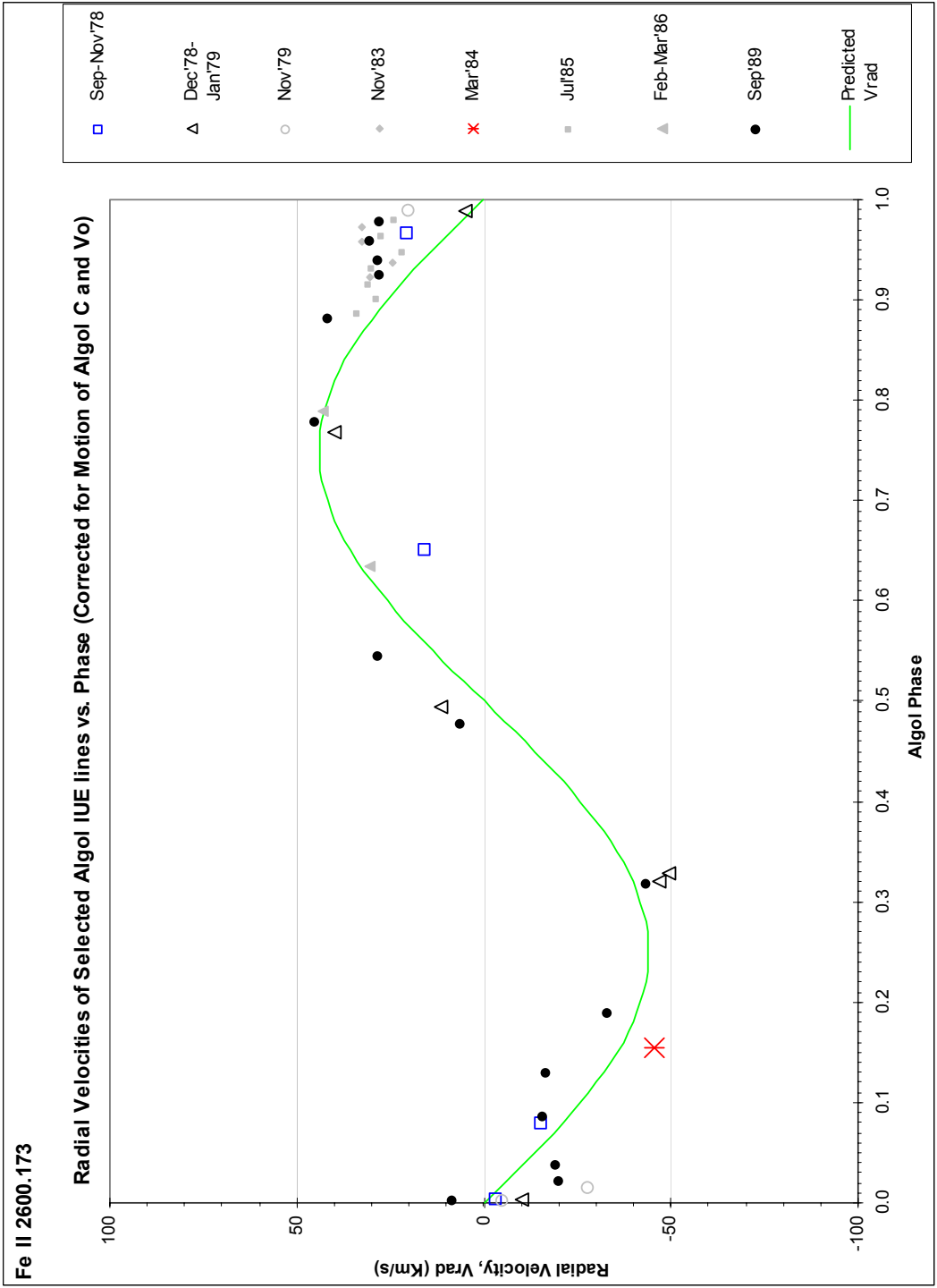

FIG. 4.5.2.9 – *Radial Velocity vs Phase Fe II 2600*

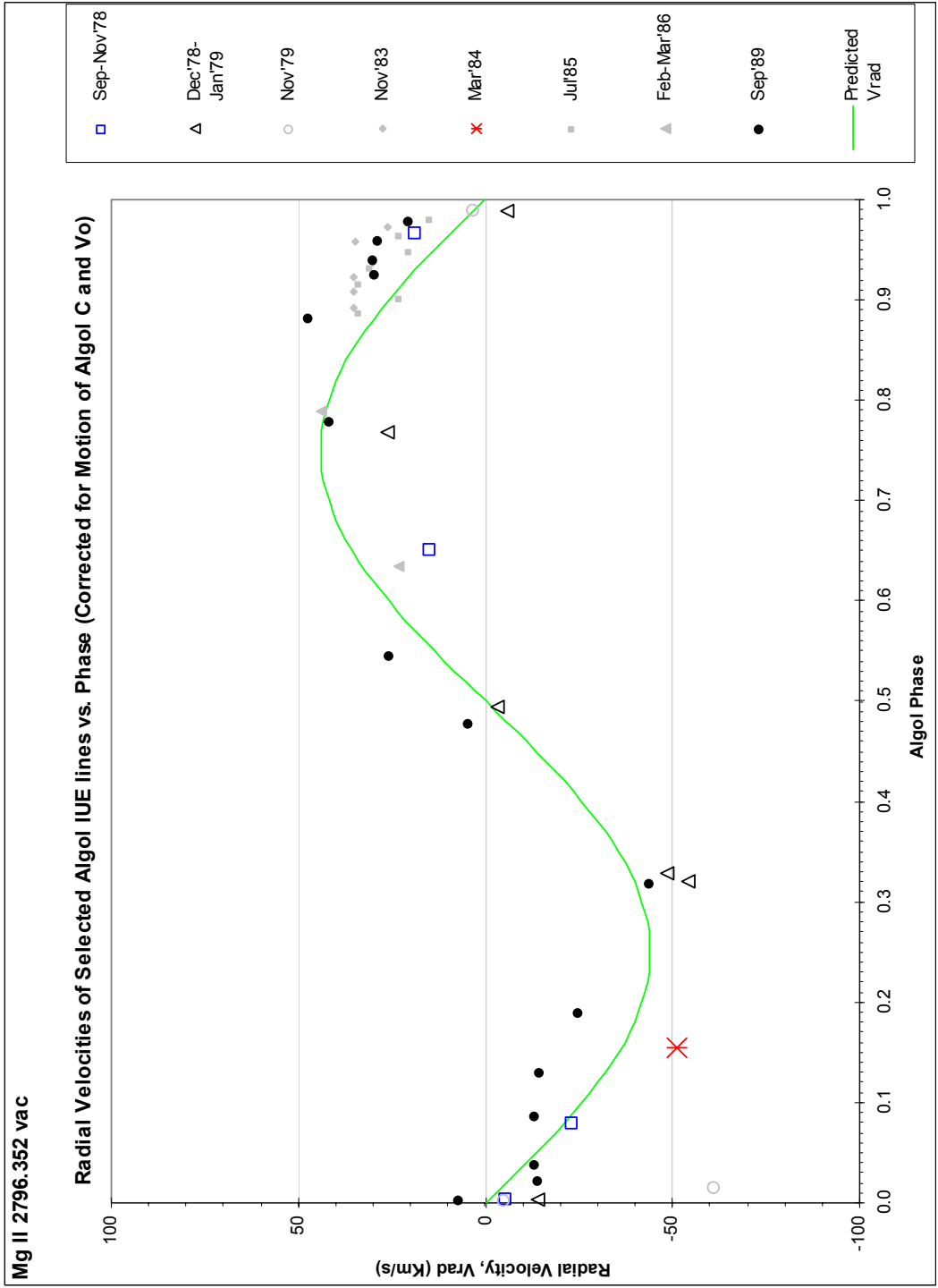

FIG. 4.5.2.10 – *Radial Velocity vs Phase Mg II 2796*

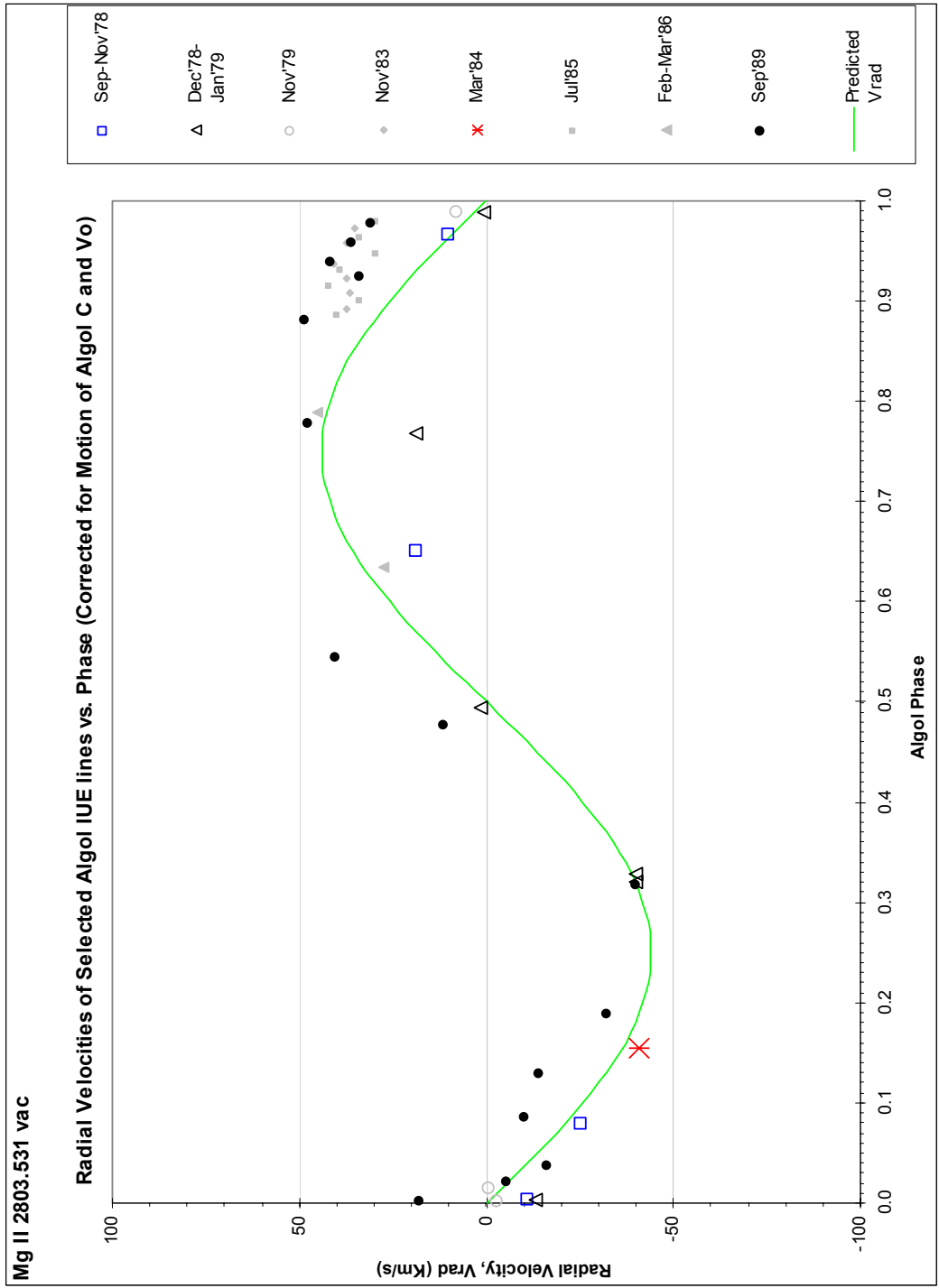

FIG. 4.5.2.11 – *Radial Velocity vs Phase Mg II 2803*

With the exception of the March '84 epoch, the Al II λ1670 radial velocity data matches closely the phase dependence expected for photospheric gases on Algol A. The data points lie along the predicted curve, except in the phase ranges just before and just after primary minima, $\varphi \simeq 0.92 - 0.08$. This effect is associated with the partial eclipsing of Algol A, revealing the hemisphere with rotational motion directed towards us (egress; $\varphi \simeq 0.0 - 0.08$) or away from us (ingress; $\varphi \simeq 0.92 - 1.0$). This is the Rossiter effect, described earlier.

Another feature common to several plots is particularly conspicuous for Al III λ1862. The observations appear to be uniformly displaced downward to the predicted (photospheric) curve. This occurs when a pair of spectral lines are blended and undistinguishable. In this case, Al III λ1862 is blended with Al II λ1862. Therefore, these near-uniform downward displacements, or upward displacements in other cases, are not indicative of gas flowing toward us.

We now consider spectral features where red-shifted or blue-shifted data might be indicative of true gas-flow phenomena. Some of these features are associated with certain epochs, indicative of secular changes in the system: outbursts or periods of unusual activity.

One of the most interesting features in the LWP/LWR RV (radial velocity) curves in the 1989 epoch in the phase range of ~0.1 – 0.3. There is a persistent arc-like trajectory with opposite curvature to the photospheric RV trajectory. It is most pronounced in Mg II λ2796 where the peak of the deviation is redshifted almost 20 km/s from the photospheric RV at phase 0.130. This feature also appears in the Fe II λ 2599, Fe II λ2600, and Fe II λ 2750 spectra.

Some spectral lines seem to show blue shifts in the phase range 0.65 - 0.8, such as those shown in FIG. 4.5.2.8 for Fe II λ2586.

There appears to be a considerable amount of epoch dependence in the 0.9 – 1.0 phase regions, particularly for the '78 and '84 epochs.

Measurements during the '84 epoch were only taken at phases near 0.15 and 0.93, and most of these values depart significantly for the photospheric values. Other features will be considered as appropriate in the Discussion and Conclusions chapter.

### 4.5.3 *Line Widths and Asymmetries*

A plot of the line width as a function of phase gives us an idea of how the velocity gradient of the circumstellar gas varies along different lines of sight and from epoch to epoch. This is illustrated schematically in FIG. 4.5.3.1. The spectral line shown in FIG. 4.5.3.1a is broadened in part by photospheric rotation. This schematic could also represent the motion of circumstellar gases near the photosphere, sometimes referred to as a "pseudo-photosphere." Since a pseudo-photosphere may have a different temperature than the true photosphere, the two regions may be represented by different ion abundances, hence, different line strengths. In either case, one expects that contributions are mostly symmetrical with line widths relatively constant over phase and epoch changes. On the other hand, UV absorptions due to gas streaming effects could show increased line widths and asymmetries, as suggested by one possible scenario, shown in FIG. 4.5.3.1b. In the case shown, the gas is moving away from the observer and increasing in velocity. Also, in this representation, the line strength is decreasing, as though the line strengths are diminishing as the gas stream approaches Algol A, indicating a temperature gradient along the stream. As seen in the figure, the gas streaming result would clearly depend on viewing angle (phase) and degree of system activity (epoch), as well as velocity and temperature gradients.

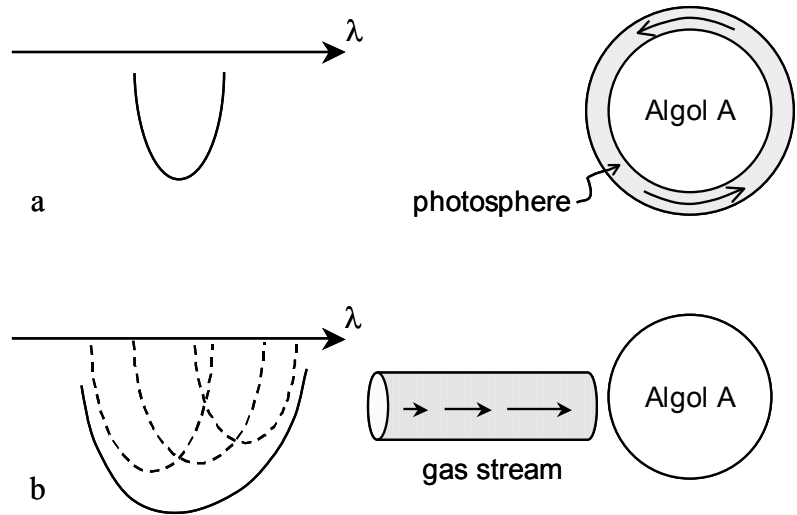

FIG. 4.5.3.1—*Variation in line widths.* Symmetrical *(a)* and asymmetrical *(b)* velocity distributions.

In examining FIG. 4.5.3.2 we see that the width of Al II λ1670 is relatively constant across phase and epoch, supporting its photospheric status.

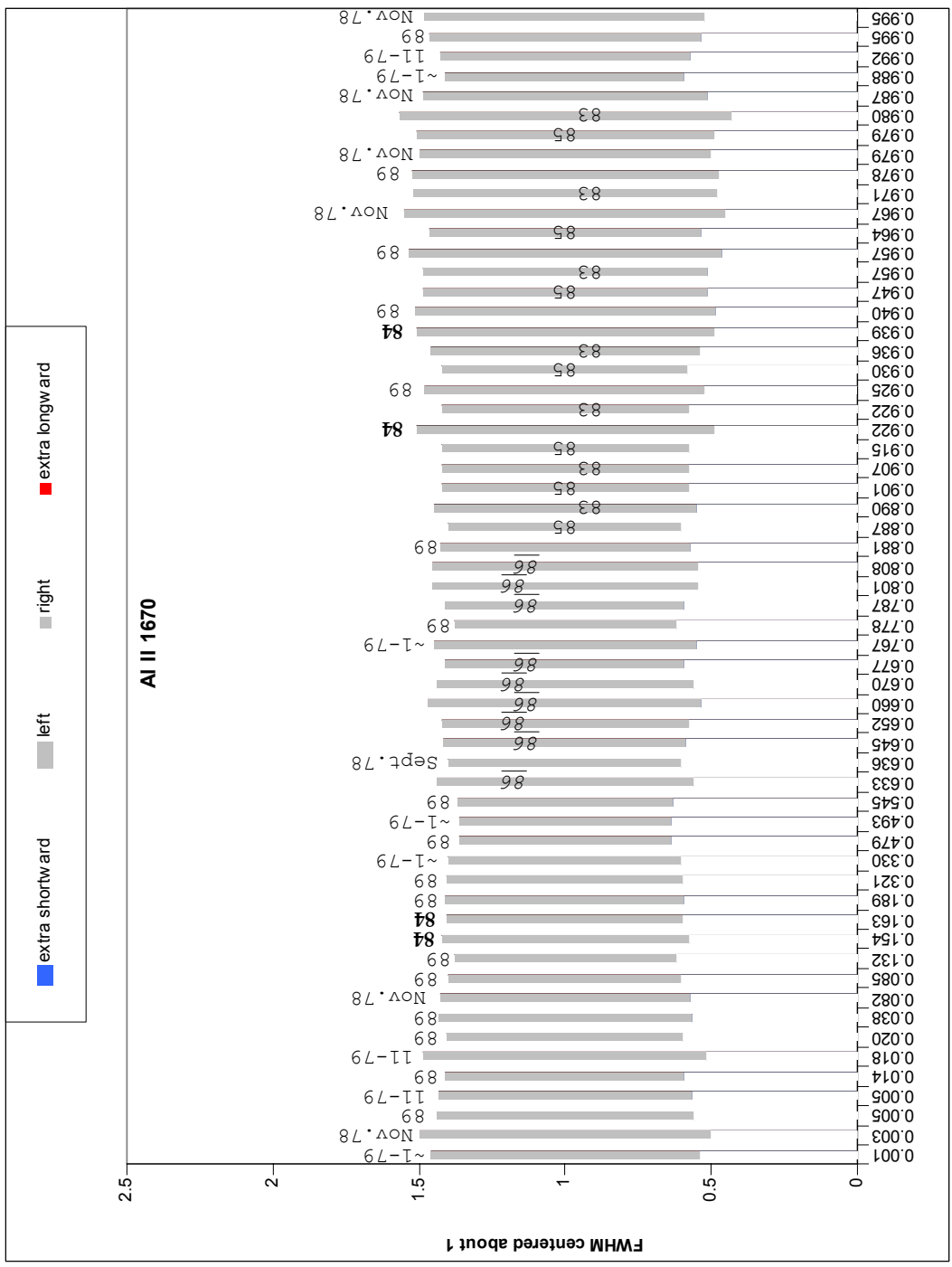

FIG. 4.5.3.2 — *Line width as a function of phase.* Al II 1670

FIG. 4.5.3.3 – *Line width as a function of phase.* Si IV 1393

The width of Si IV on the other hand shows dramatic changes over both phase and epoch. The line width increases dramatically on average during primary eclipse ingress near phase 0.922, decreases before primary eclipse, and continues to decrease on average through primary eclipse egress, reaching a minimum near phase 0.5 (secondary minimum). The 1989 results are mostly symmetrical, and show increasing widths as ingress progresses. In the phase range 0.922 – 1.0, there are significant epoch dependences, as evidenced by the larger widths and asymmetries in the 1978 and 1984 data. During these epochs, there is on average more longward (redshifted) width during the first half of primary eclipse, which is indicated by the (darker) top portions of the bars in FIG. 4.5.3.3. This asymmetry is an indicator of gas streaming from the secondary toward the primary, suggesting that these are epochs of increased activity.

A tabulation of all measured line widths are included in Appendix D.3.

### 4.5.4 *Residual Intensities*

The normalized residual intensities (transmission at line center) of the resonance absorption lines have a clear pattern of variation across the orbital phases especially in the 1989 data shown in FIGs. 4.5.4.1 through 4.5.4.7, where the 1989 data are connected by a smooth line for ease of viewing. The most prominent feature is the distorted asymmetric "W" shape situated about primary eclipse. The first "V" of the "W" has a bump at the base near phase 0.925, especially noticeable in the Al III $\lambda\lambda$1854, 1862 plots (FIGs. 4.5.4.2 and 4.5.4.3). As discussed in section 4.5.1 the residual intensity (RI) is a measure of the attenuation of the incident intensity after passing through an absorbing region, or the amount of flux that is not extinguished along the line of sight. This bump might be interpreted as an emission region competing with a broader absorbing region.

An outstanding attribute of the "W" shape is its asymmetry, the first "V" being broader, i.e., extending through a larger phase range before primary eclipse than the second "V" does after primary eclipse. This indicates that we are looking through more material before ingress than after egress. In comparison with the 1986 data, which suggests a more extended absorbing region between phases 0.6 and 0.8, 1989 is the less active of the two epochs. It is interesting to note the development of the RIs in the 1986 epoch, which increasingly depart from the 1989 levels across Al II, Al III, and Si IV (FIGs. 4.5.4.1 through 4.5.5). A tabulation of all measured RIs is located in D.4.

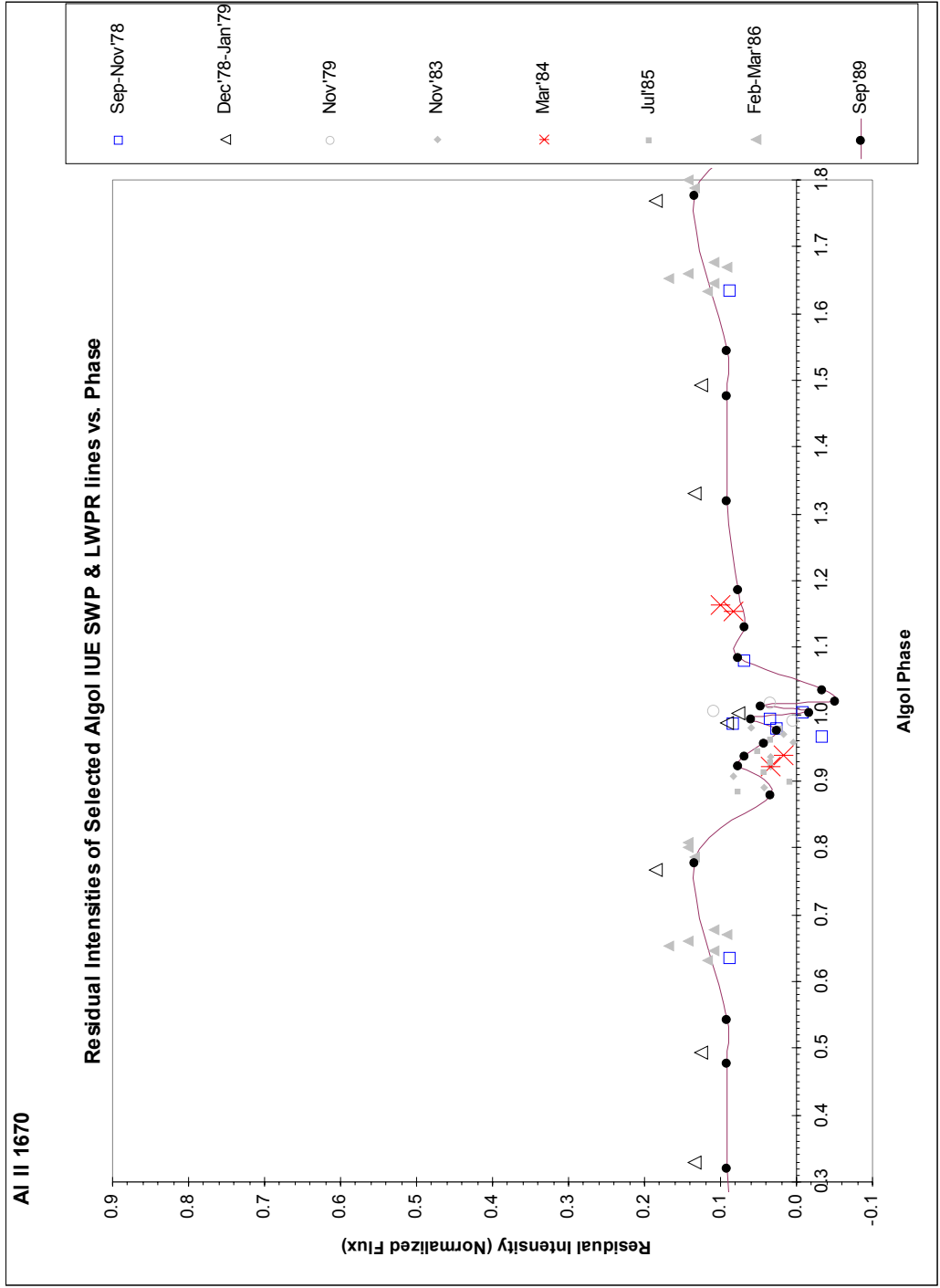

FIG. 4.5.4.1 — *Residual Intensities of Selected IUE SWP & LWP/LWR lines vs. Phase Al II 1670*

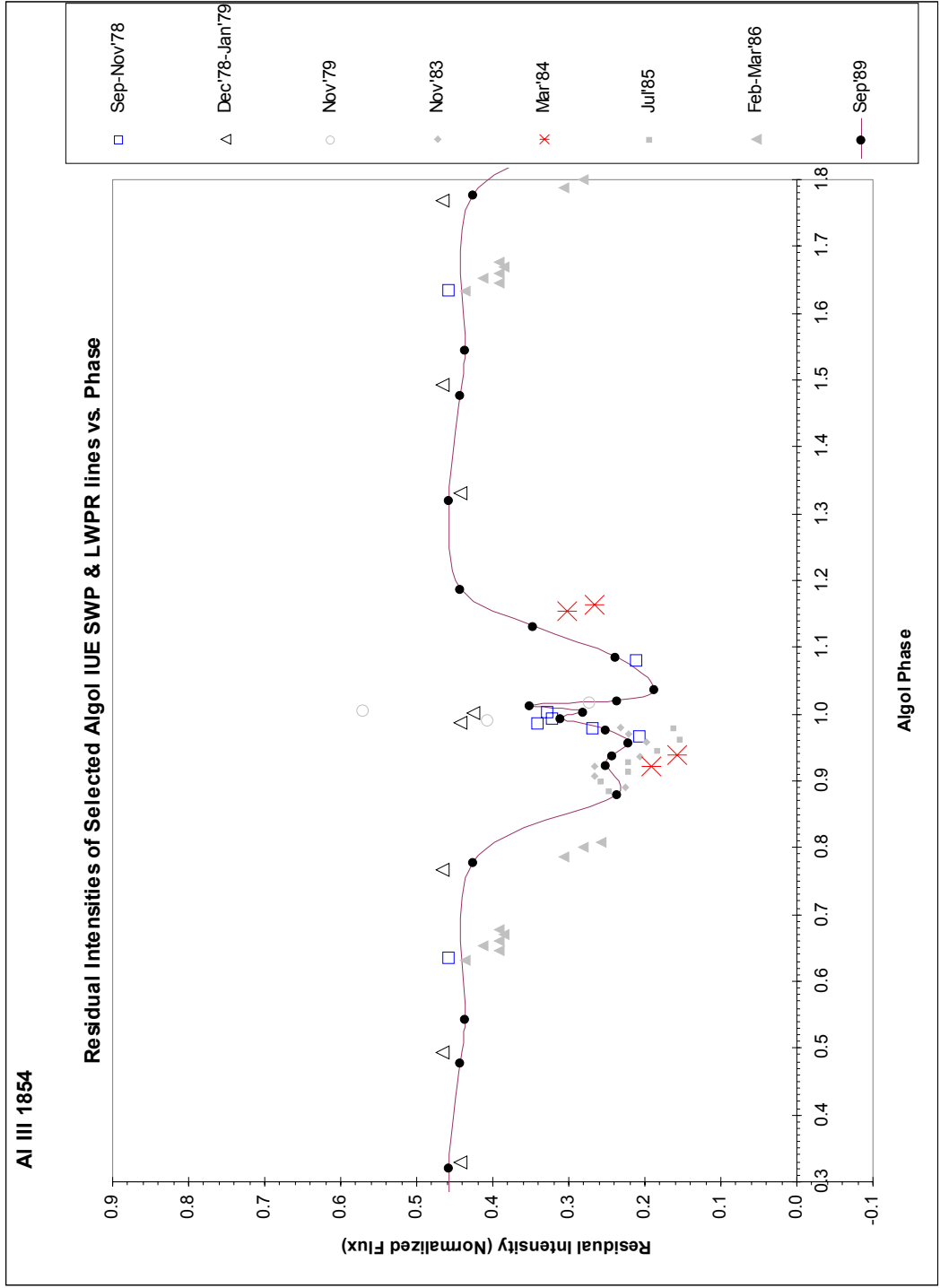

FIG. 4.5.4.2 – *Residual Intensities of Selected IUE SWP& LWP/LWR lines vs. Phase Al III 1854*

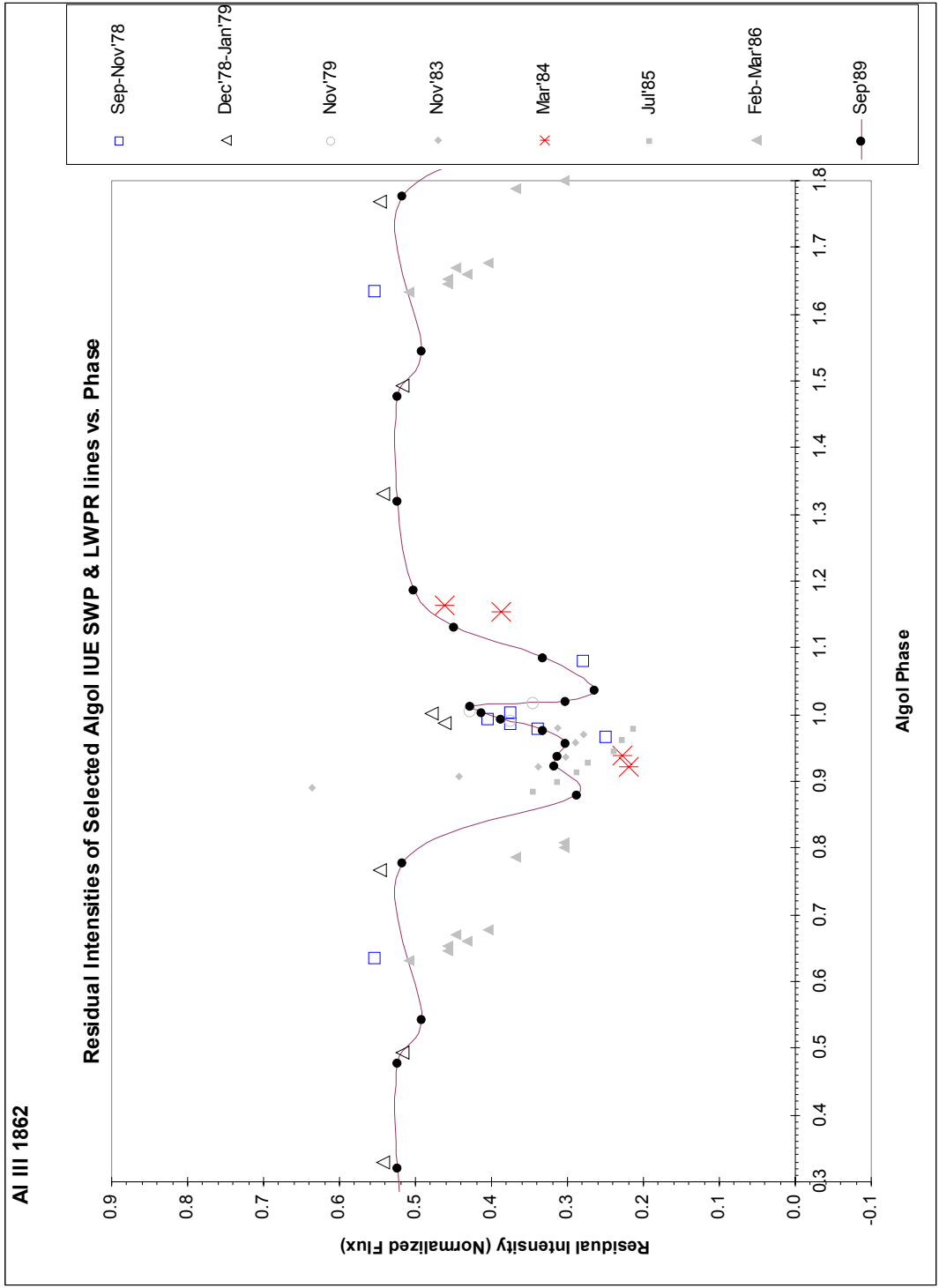

FIG. 4.5.4.3 — *Residual Intensities of Selected IUE SWP& LWP/LWR lines vs. Phase Al III 1862*

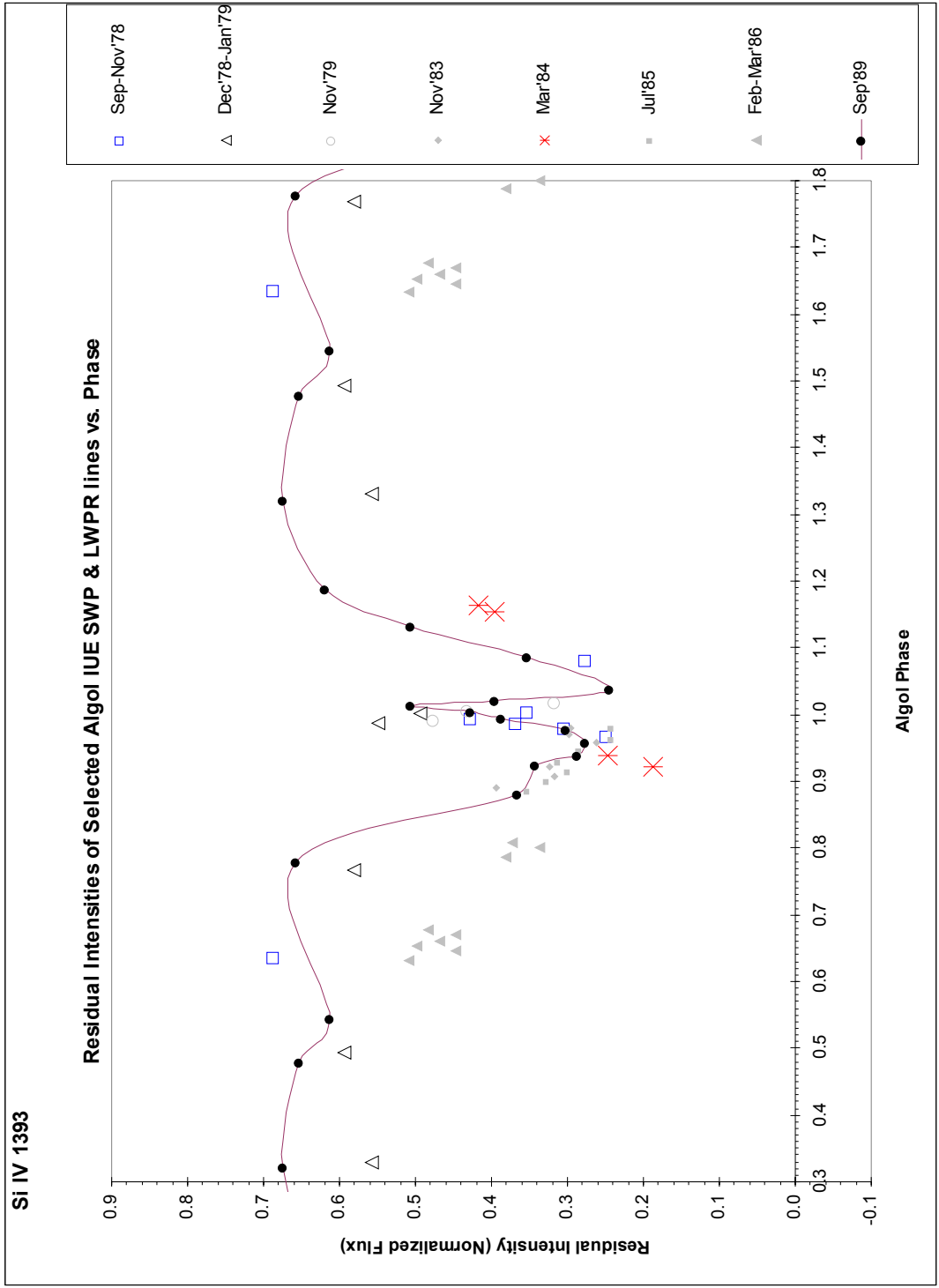

FIG. 4.5.4.4 – *Residual Intensities of Selected IUE SWP& LWP/LWR lines vs. Phase Si IV 1393*

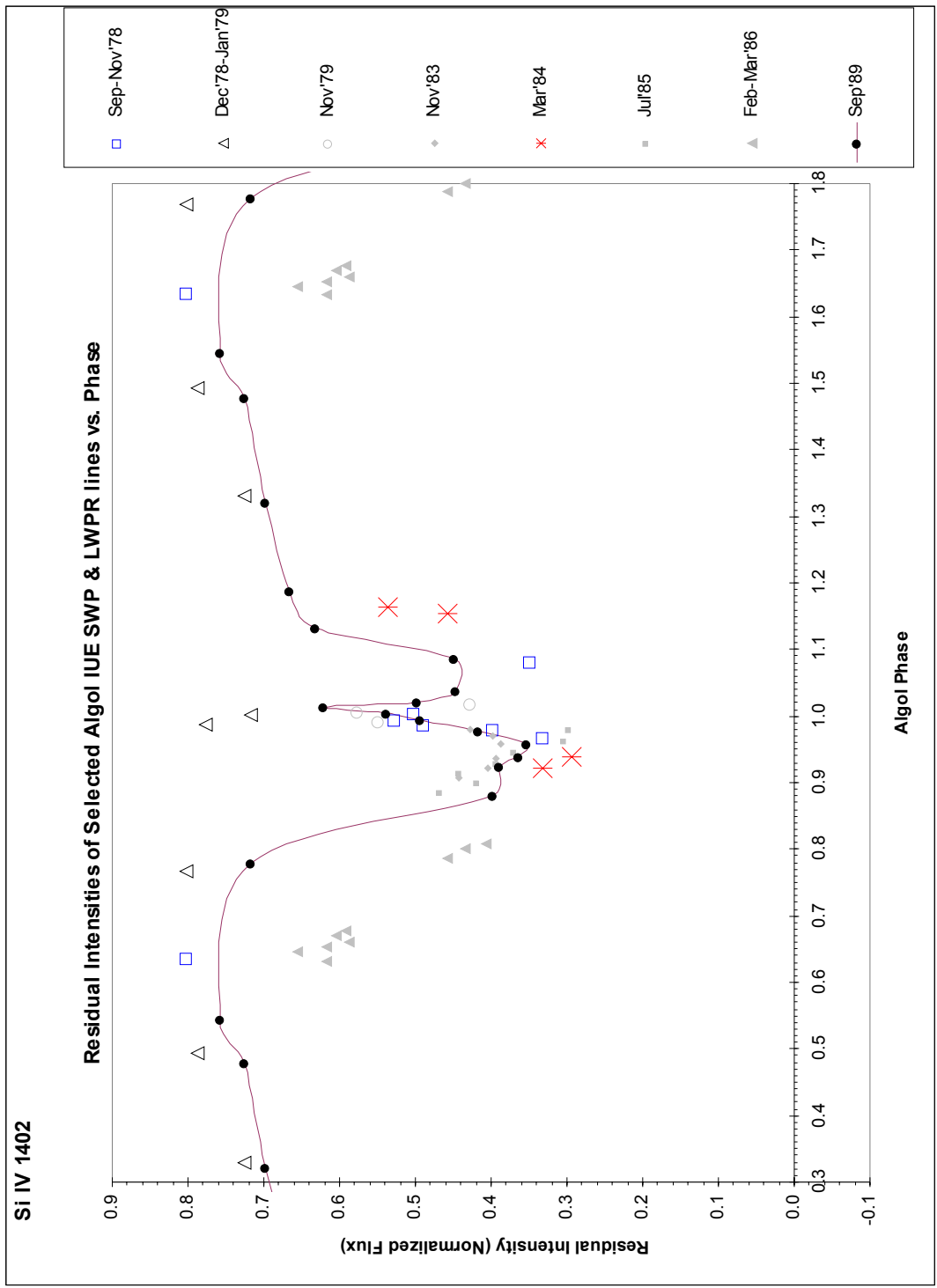

FIG. 4.5.4.5 – *Residual Intensities of Selected IUE SWP & LWP/LWR lines vs. Phase Si IV 1402*

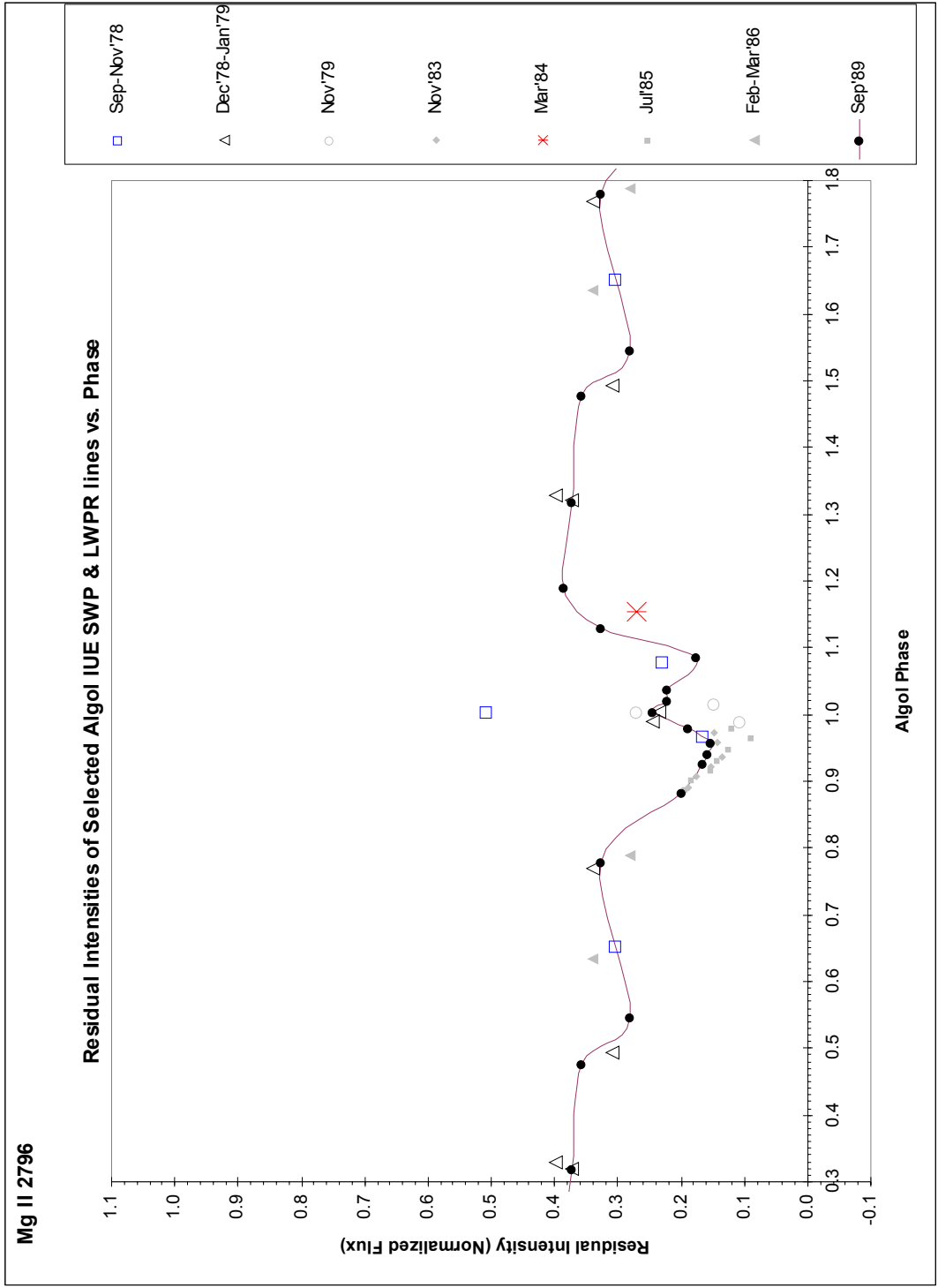

FIG. 4.5.4.6 – *Residual Intensities of Selected IUE SWP& LWP/LWR lines vs. Phase Mg II 2796*

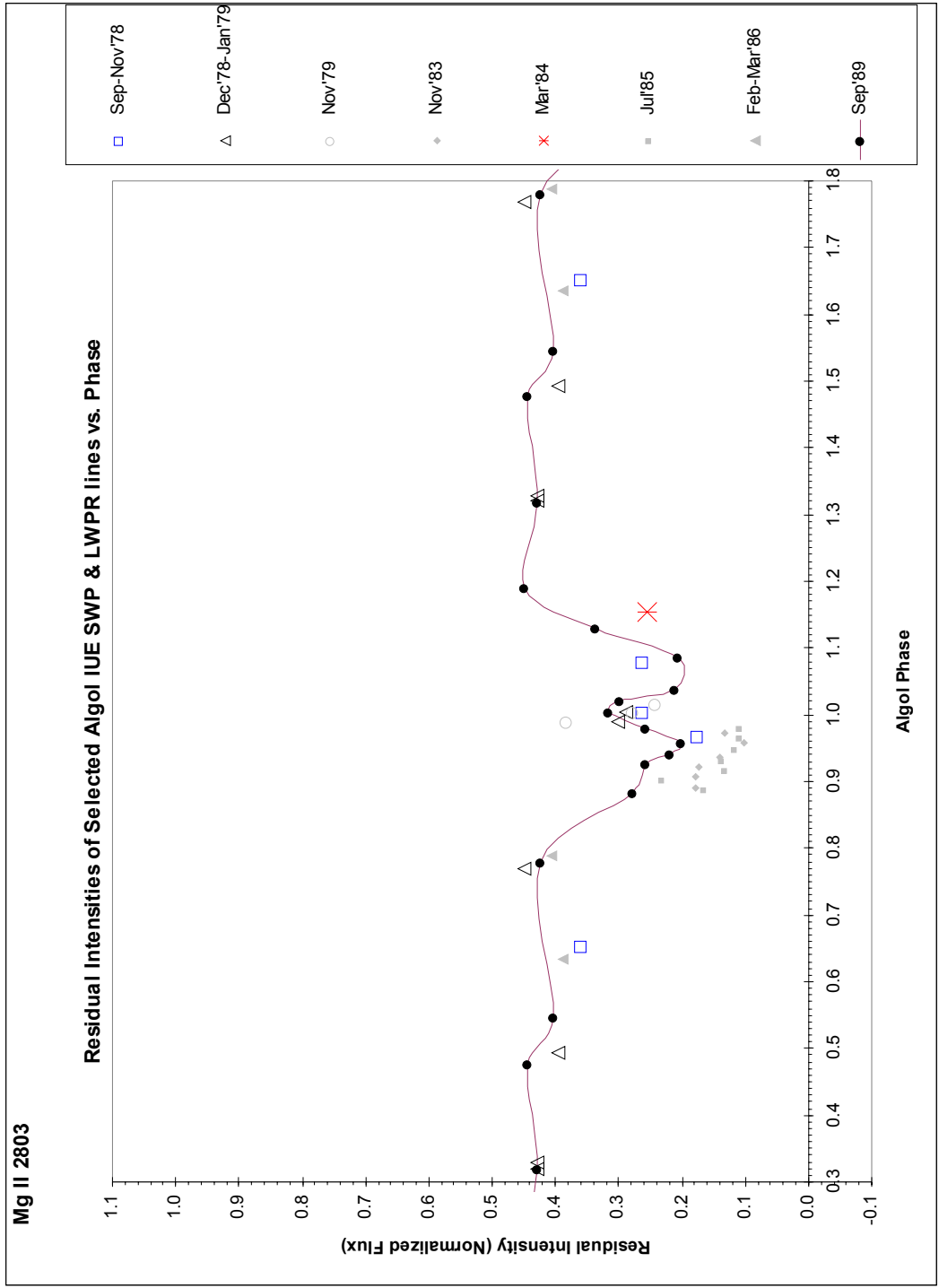

FIG. 4.5.4.7– *Residual Intensities of Selected IUE SWP& LWP/LWR lines vs. Phase Mg II 2803*

### 4.5.5 *Equivalent Widths and Column Densities*

The equivalent width $W_\lambda$ of a line is an indicator of its strength and is proportional to the number of atoms ("column density") along our line of sight $N_A$ if the absorptions are unsaturated. A well behaved photospheric line should vary little with phase since the photosphere is assumed to be uniform in thickness and density. Al II 1670 is the strongest photospheric line in our data, though it indicates an increase in $N_A$ between phases 0.881 and 0.038. (See FIG. 4.5.5.1.) Si IV ($\lambda\lambda$1393, 1402), however, shows a dramatic increase in $W_\lambda$ between phases 0.633 and 0.979 (FIGs. 4.5.5.2 and 4.5.5.3). The equivalent width is a minimum in the phase interval 0.321 to 0.545.

In FIGs. 4.5.5.2 and 4.5.5.3, we see that the region of formation of Si IV is visible to varying degrees in the orbit and across epochs. On average, the number of atoms along the line of sight is the greatest between phases 0.881 and 0.987 ($\phi \approx .9$) and the least between phases 0.321 and 0.545 ($\phi \approx 0.4$). This suggests that the hot structure that forms the Si IV is located in a region between Algol A and Algol B offset from the line connecting the centers of the components by ~ 0.1 x 360° = 36 °. The complete set of equivalent width measurements is found in Appendix D.2.

If we make an estimate of the gas column volume of the absorbing region we can make an estimate of the gas density, as described in Section 4.5.7.

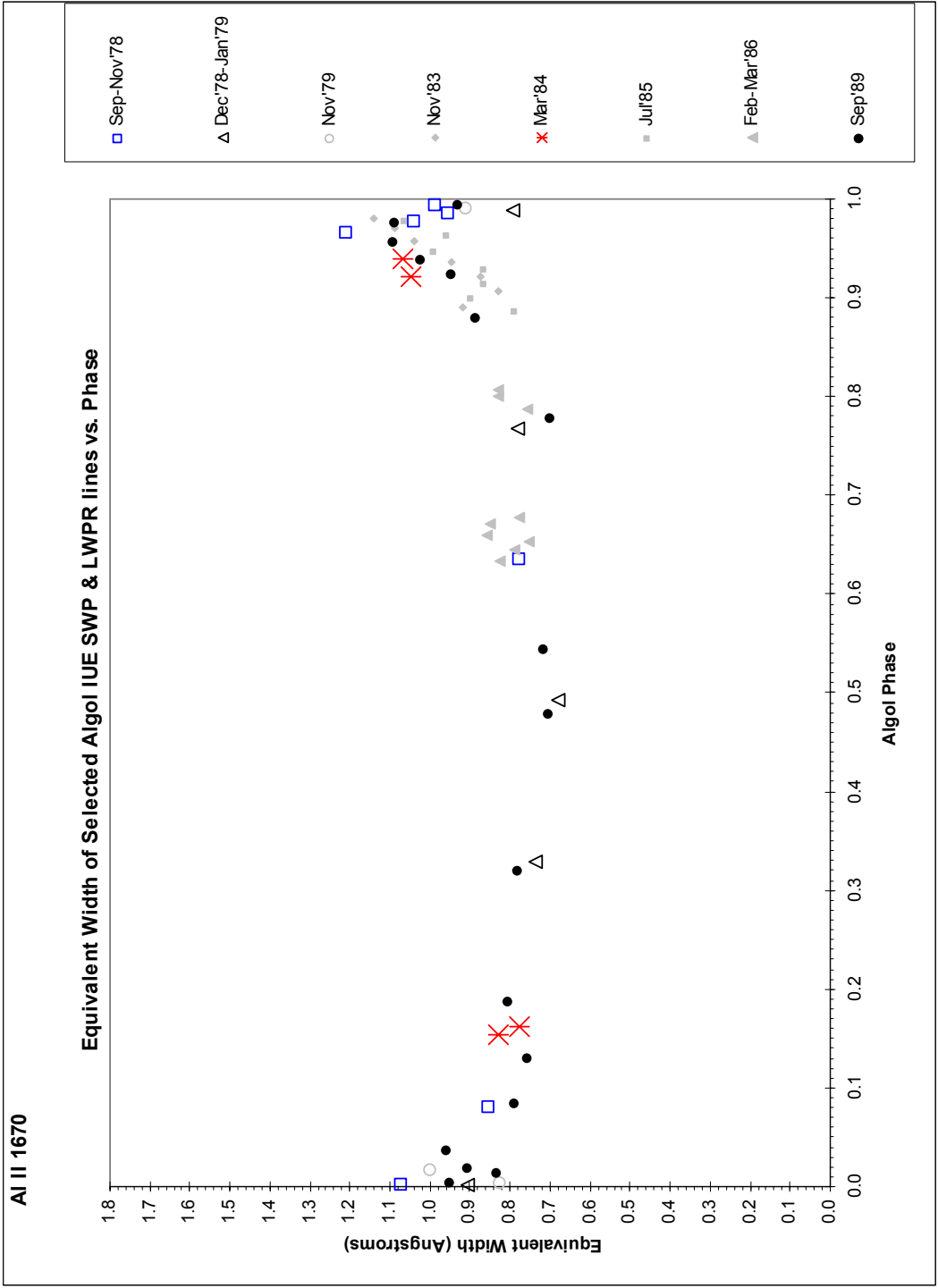

FIG 4.5.5.1 – *Equivalent Width Al II 1670 line*

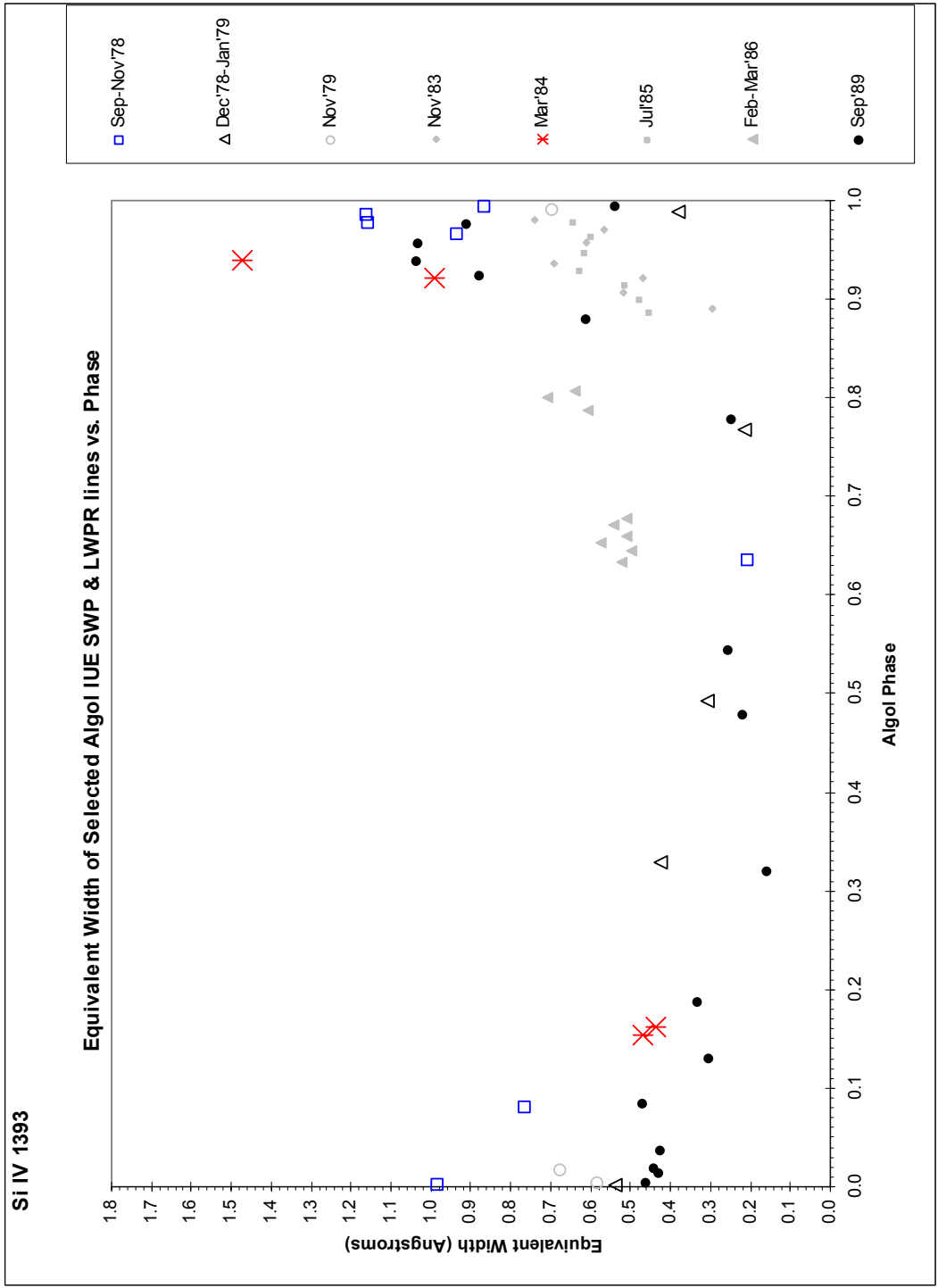

FIG. 4.5.5.2 - *Equivalent Width Si IV 1393*

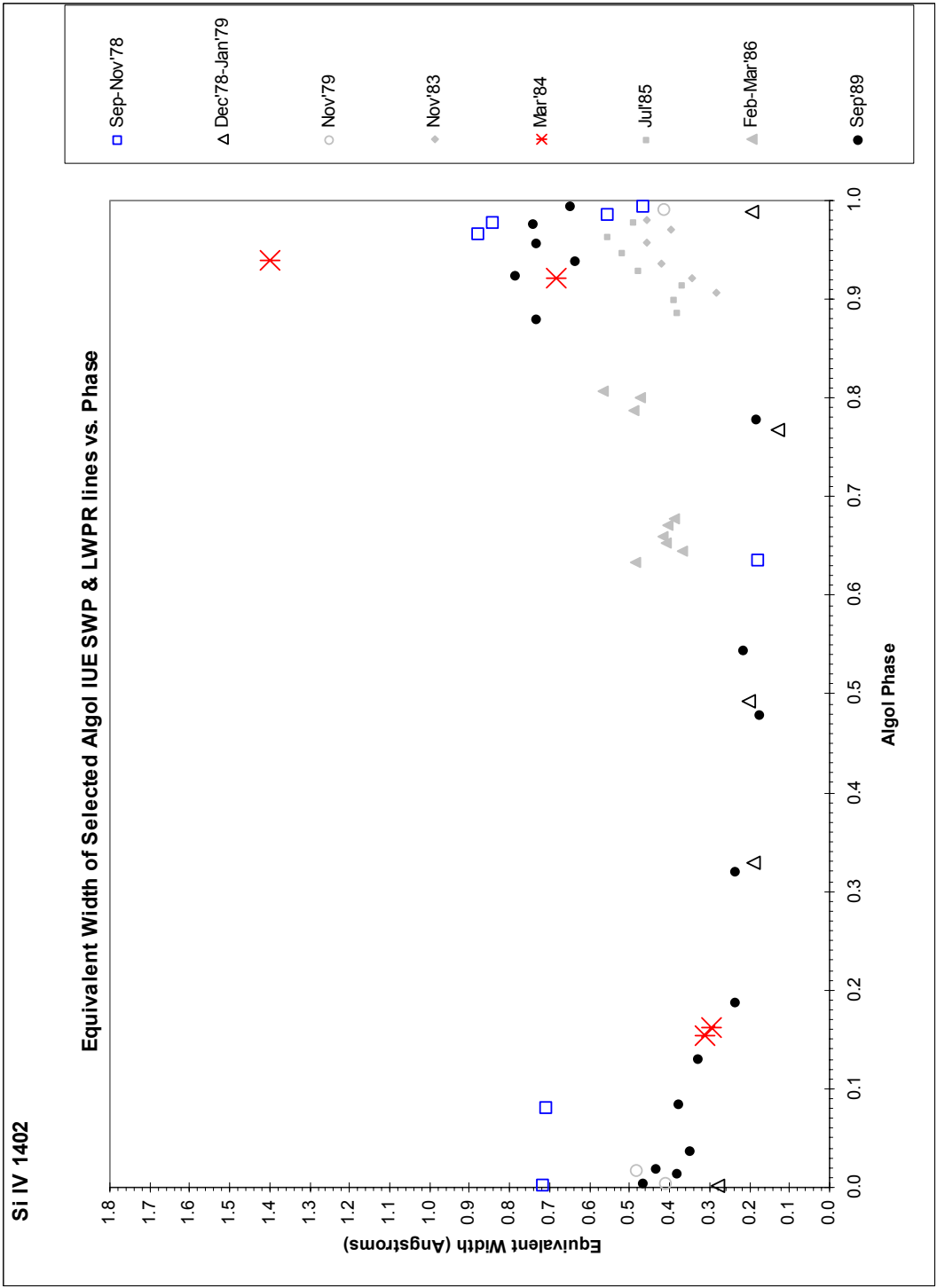

FIG. 4.5.5.3 - *Equivalent Width Si IV 1402*

### 4.5.6 *Difference Spectra*

In order to isolate gas flow effects from photospheric and perhaps pseudo-photospheric contributions, we developed an approach that removes the stellar contributions from the spectra. This represents the first time the entire IUE archive for Algol is self-differenced in this manner.

The stellar contribution is assumed to be that of September 1989 exposure at phase 0.3209 since it consistently shows the most well-behaved absorption across the elements. Moreover, as mentioned earlier, the epoch of 1989 has the most complete phase coverage, and based on other work on Algol-type systems, it is reasonable to assume that phases near, first quadrature will show the effects of secular variations the least (Walter 1973, 1980).

To eliminate the assumed photospheric contribution, we subtracted that spectrum from the spectra of all other phases. Kempner and Richards (1999) used the star HD 23432 as a standard B8 V star. Their UV analysis of U Sge was done by subtracting the spectrum of the B8 V standard star HD 23432 from the observed U Sge spectrum. For Algol, this may not be a suitable comparison star because its rotational velocity is 283 km/s (Battrick 1984) where Algol's rotational velocity is 55 km/s. That of U Sge, on the other hand, is estimated to be 100 km/s (Struve, 1949), which is less of a differential. Although extra line broadening is present in HD 23432 as compared with U Sge, they felt that "this extra absorption should be relatively small compared to any emission from the circumstellar gas" (Kempner & Richards 1999, p. 347.) However, since Algol is in a low-activity stage as compared to U Sge (McCluskey et al. 1991, p. 281), it would be preferable to find a more slowly rotating single B8 V star from the IUE archives taken in high dispersion mode. Unfortunately, no adequate comparison stars appropriate to Algol are included in the IUE data.

Before subtracting the spectra, we normalized them by eye, and then shifted the spectra to the rest frame of the center of mass (CM) of the primary star, since we expect the source of the UV observations to be largely relevant to structures at the surface of, near, and around the primary component. (See Section 4.3.3. Correction for Influence of Algol C).

By using the spectra of Algol itself for differencing, we eliminated the complication of the significant rotational broadening in the spectrum of the standard star, which in Kempner and Richards's words, "prevents us from truly eliminating the primary star's contribution to the spectrum since the standard star has shallower absorption features than desired." However, the choice of using a well-behaved spectrum from Algol itself does not guarantee success in "truly eliminating" the photospheric contribution of Algol A. But we proceeded with this method since it gives us an excellent comparison with Algol itself. The Si IV difference spectra from September 1978 date to September 1989 are presented in FIGs. 4.5.6.1a-c along with the corresponding original spectra.

The observed profiles show variations with orbital phase indicative of mass flow within and from the system.

Starting at phase ~0.54 and scanning through to phase ~0.88, we see a double peaked absorption feature developing which we interpret as an asymmetrically distributed disk-like structure. Then from phase ~0.88 to phase ~0.98, the redshifted absorption peak develops as our line of sight from Earth changes from looking across the gas stream to looking along the gas stream. As primary eclipse progresses the gas stream is, for the most part, eclipsed at phase 0.003 where we again see evidence of a disk-like structure with one absorption peak blueshifted and the other redshifted. As the eclipse approaches egress, the gas stream starts to become visible again (phase ~0.09) as it recedes from us along our line of sight, redshifted. Notice that the absorption after primary is not as strong as the absorption before primary indicative of gas streaming. At phase ~0.48, the gas stream is no longer evident, that is, it is hidden from view as it is occulted during secondary eclipse.

Radial velocities have been recalculated and column ion numbers have been calculated from the spectral features isolated using the difference spectrum method. These results are shown for Mg II $\lambda 2796$ in FIGs. 4.5.6.6 and 4.5.6.7, respectively. Note that there are two components to the gas stream, moving toward and away from us at about $\leq$ km/sec, as seen in FIG. 4.5.6.6. FIG. 4.5.6.7 shows the ion numbers appropriate to the two components. These results can be interpreted as representing the two sides of a circumstellar disc located about the primary.

A complete table of these results is provided in Appendix D.6.

FIG. 4.5.6.1a—*ALGOL IUE SWP SPECTRA Si IV 1393. Left*, smoothed spectra. *Right*, difference spectra.

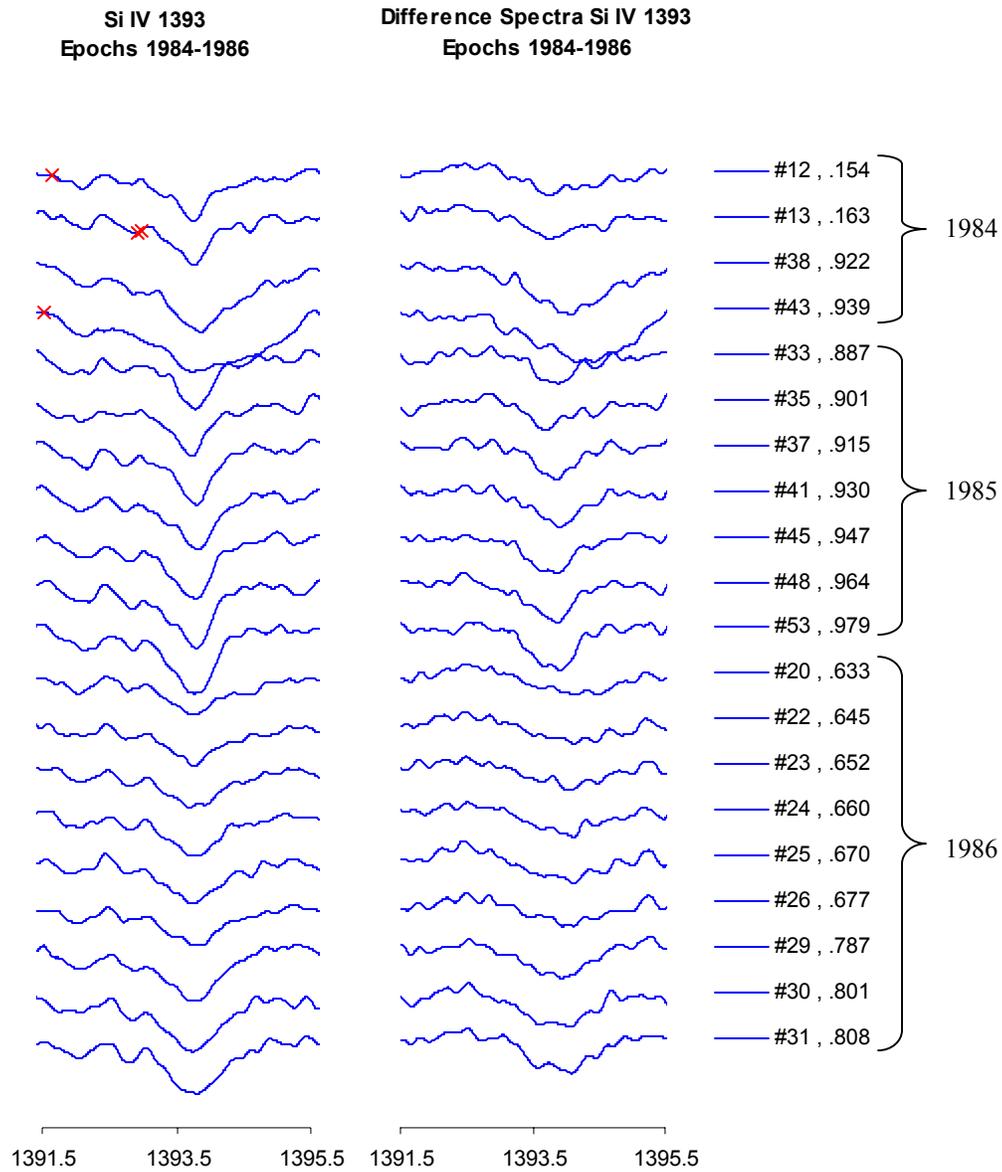

FIG. 4.5.6.1b—*ALGOL IUE SWP SPECTRA Si IV 1393.* *Left,* smoothed spectra. *Right,* difference spectra.

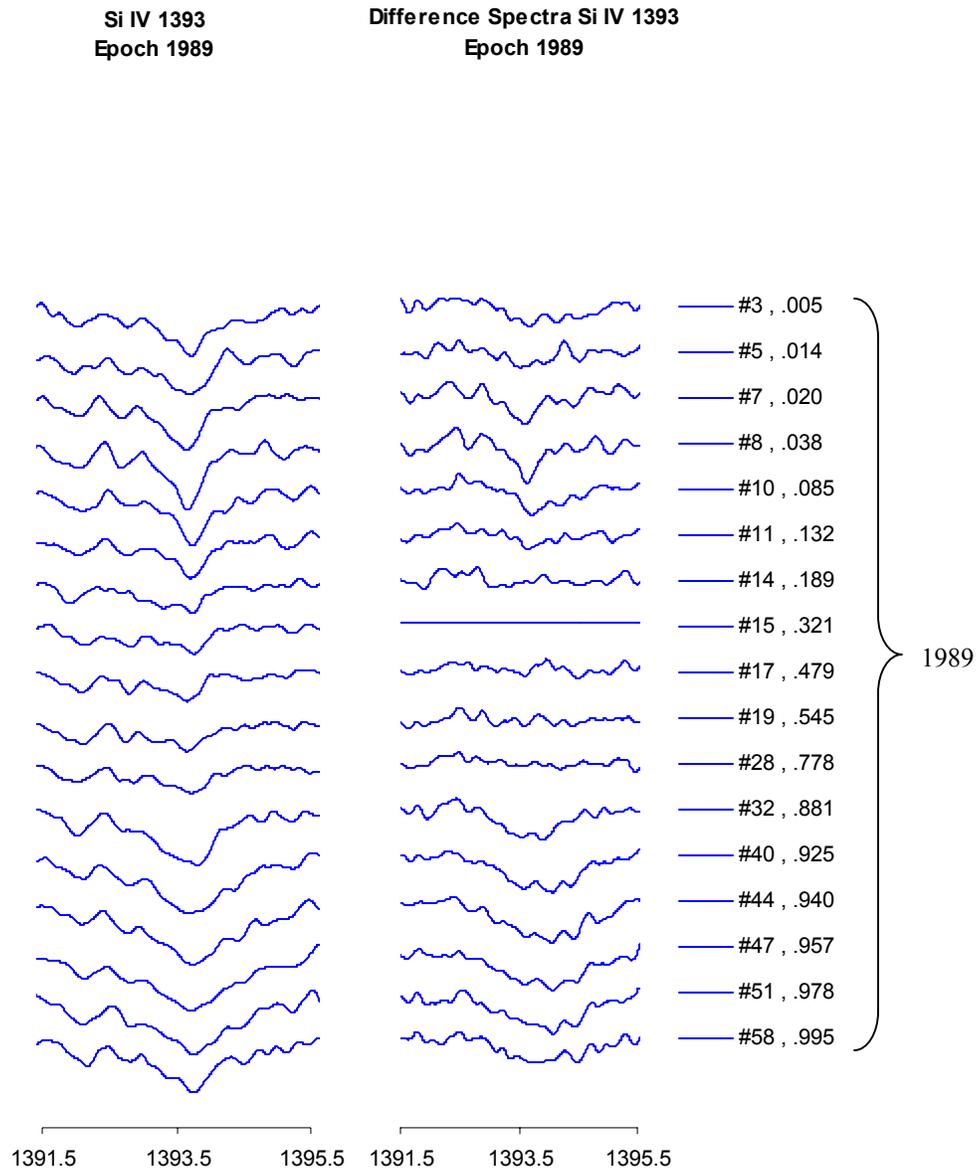

FIG. 4.5.6.1c—*ALGOL IUE SWP SPECTRA Si IV 1393. Left*, smoothed spectra. *Right*, difference spectra.

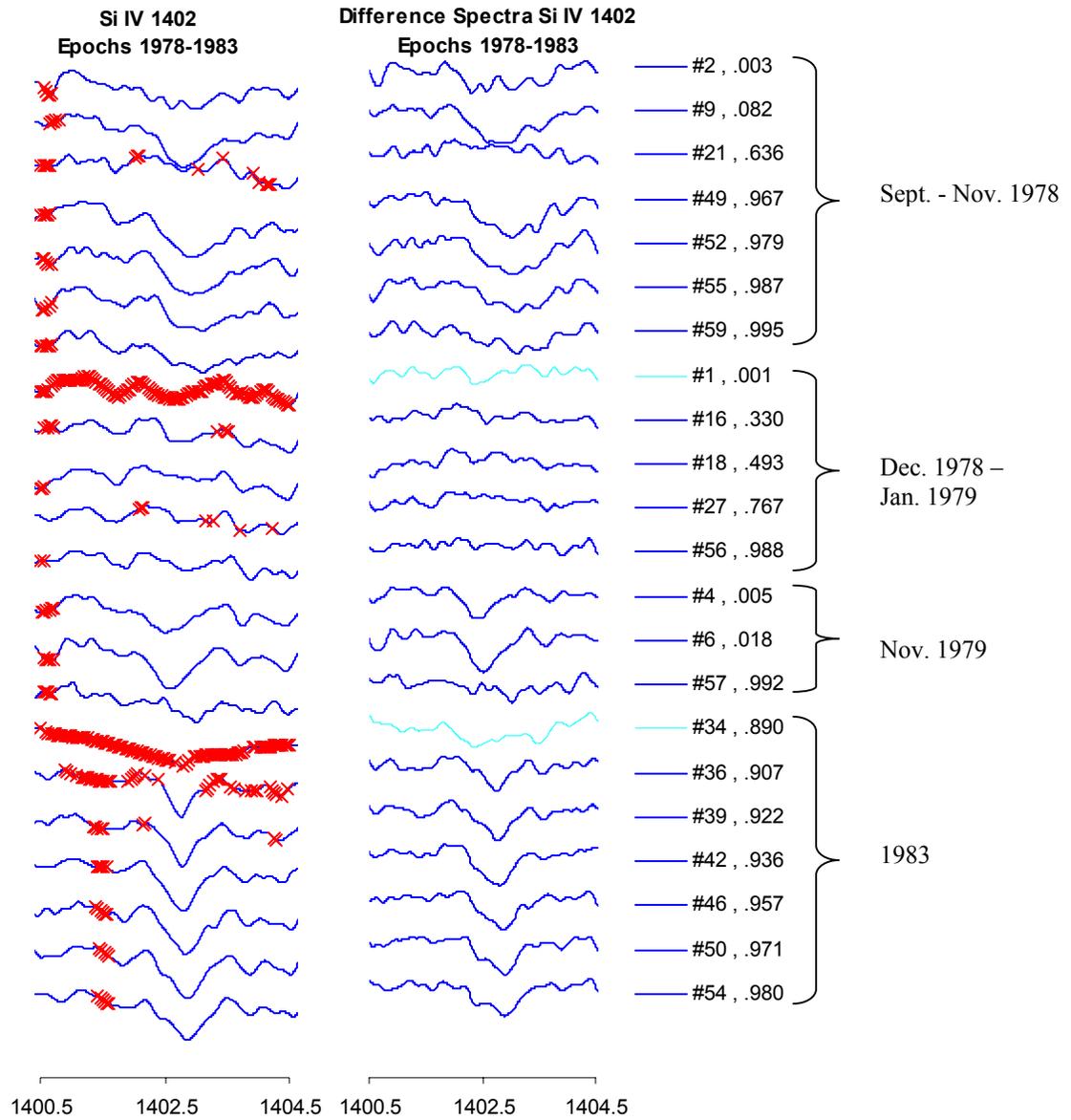

FIG. 4.5.6.2a—*ALGOL IUE SWP SPECTRA Si IV 1402. Left*, smoothed spectra. *Right*, difference spectra.

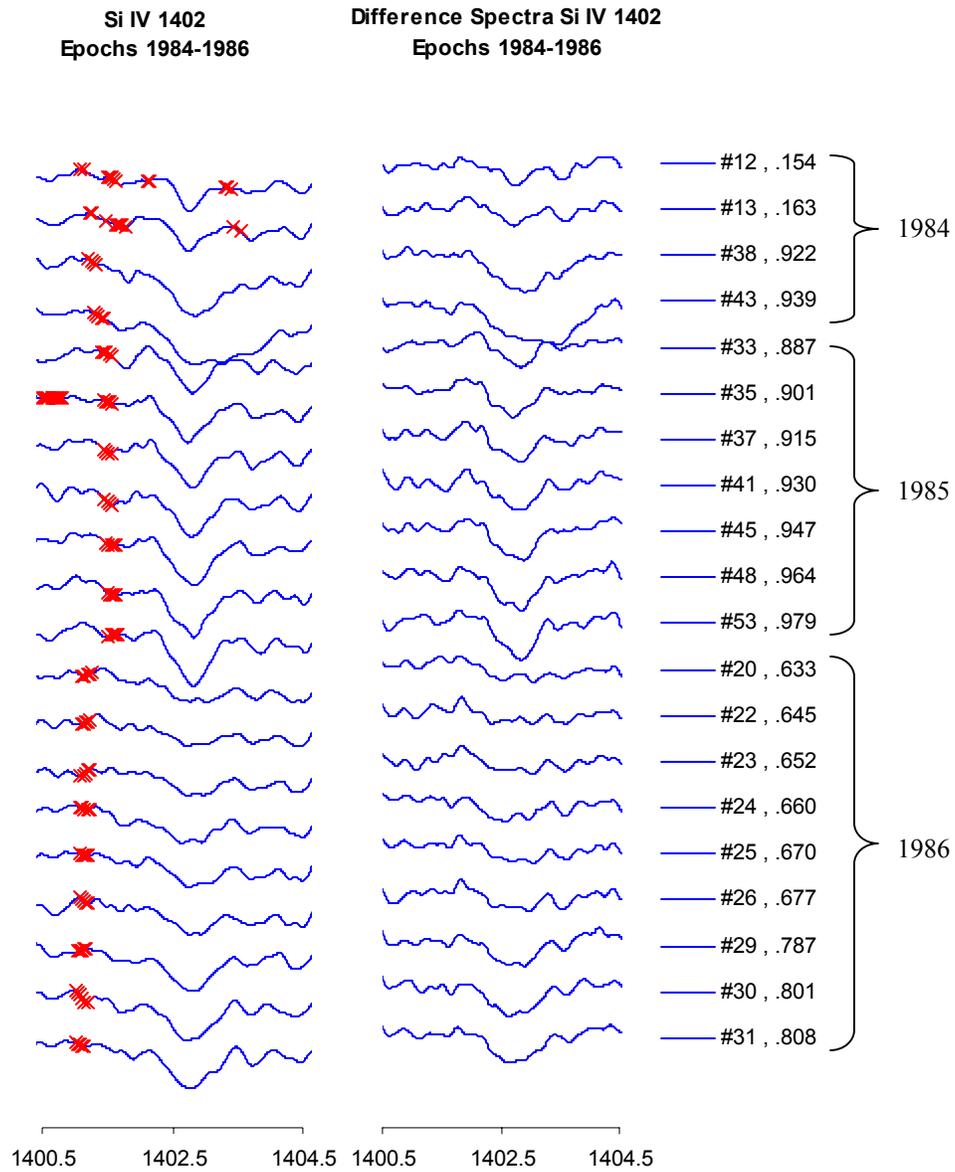

FIG. 4.5.6.2b—*ALGOL IUE SWP SPECTRA Si IV 1402*. *Left*, smoothed spectra. *Right*, difference spectra.

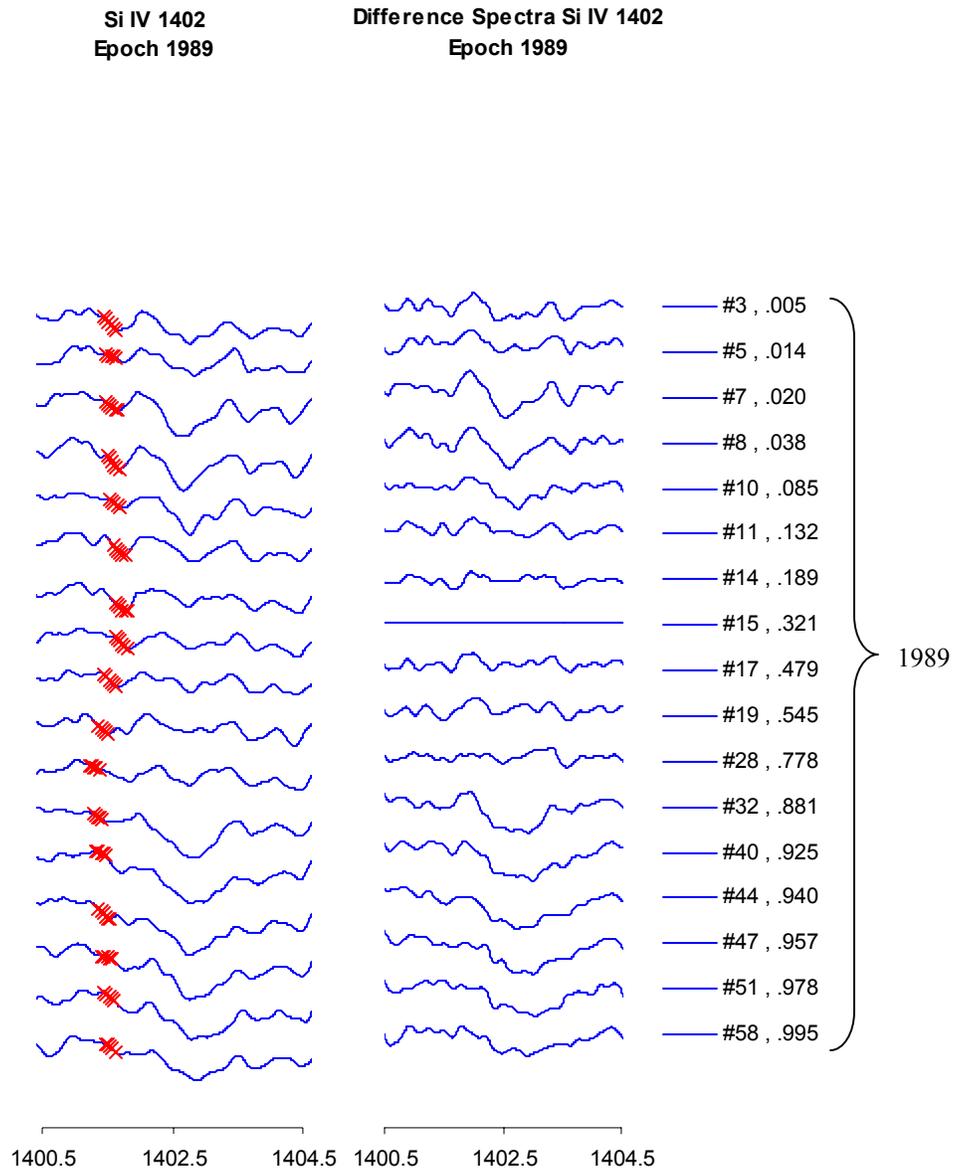

FIG 4.5.6.2c—ALGOL IUE SWP SPECTRA Si IV 1402. *Left*, smoothed spectra. *Right*, difference spectra.

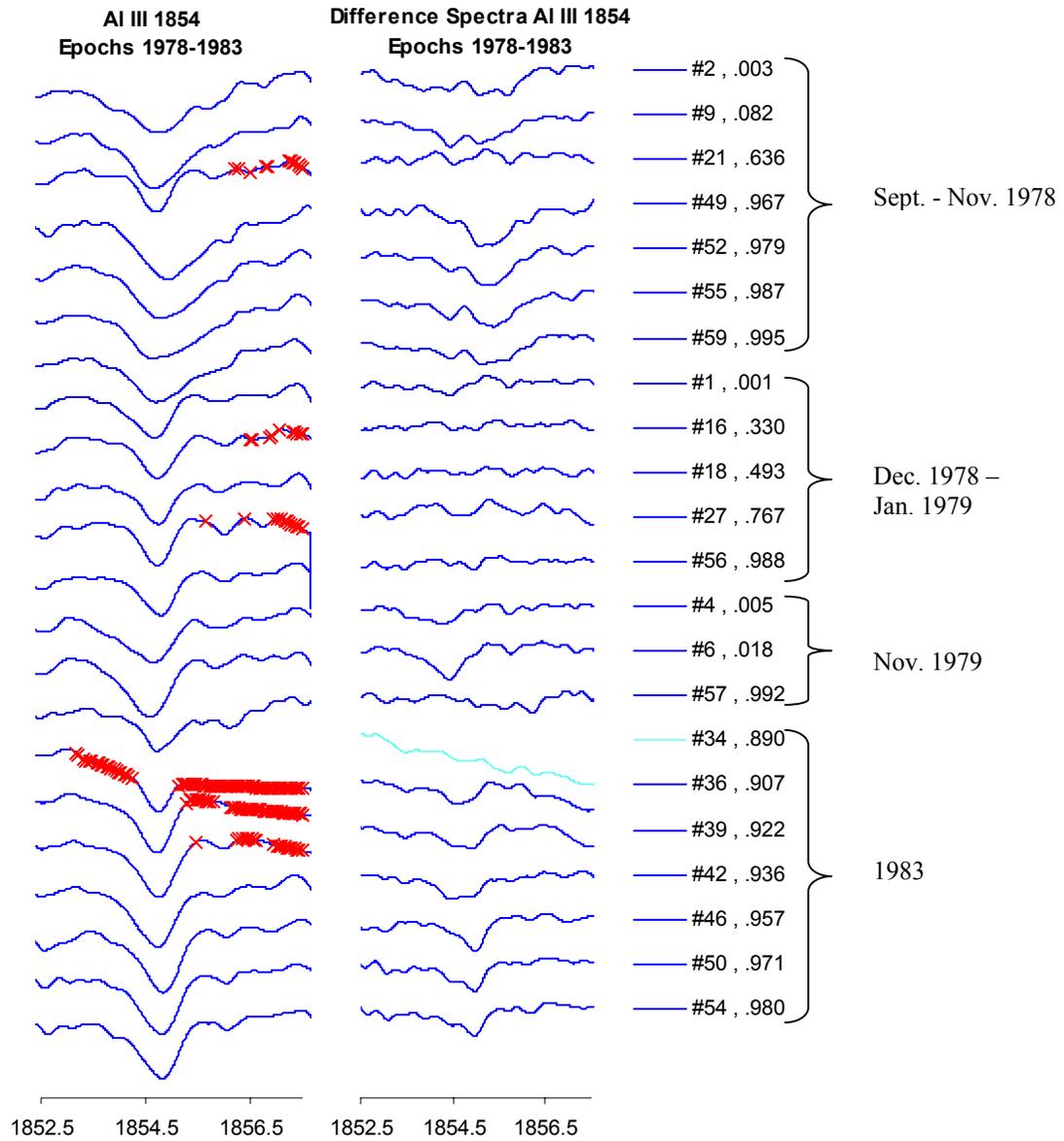

FIG. 4.5.6.3a—ALGOL IUE SWP SPECTRA Al III 1854. *Left*, smoothed spectra. *Right*, difference spectra.

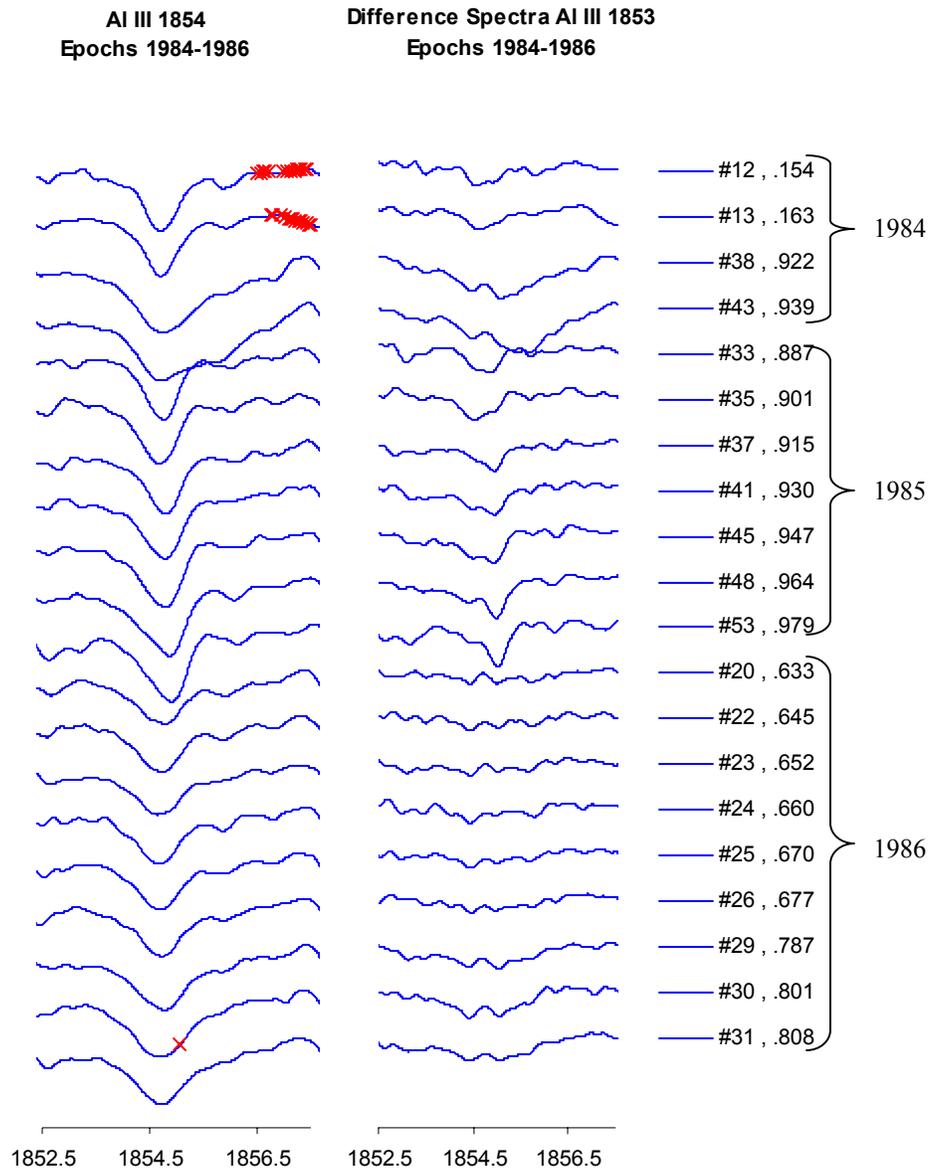

FIG. 4.5.6.3b—ALGOL IUE SWP SPECTRA Al III 1854. *Left*, smoothed spectra. *Right*, difference spectra.

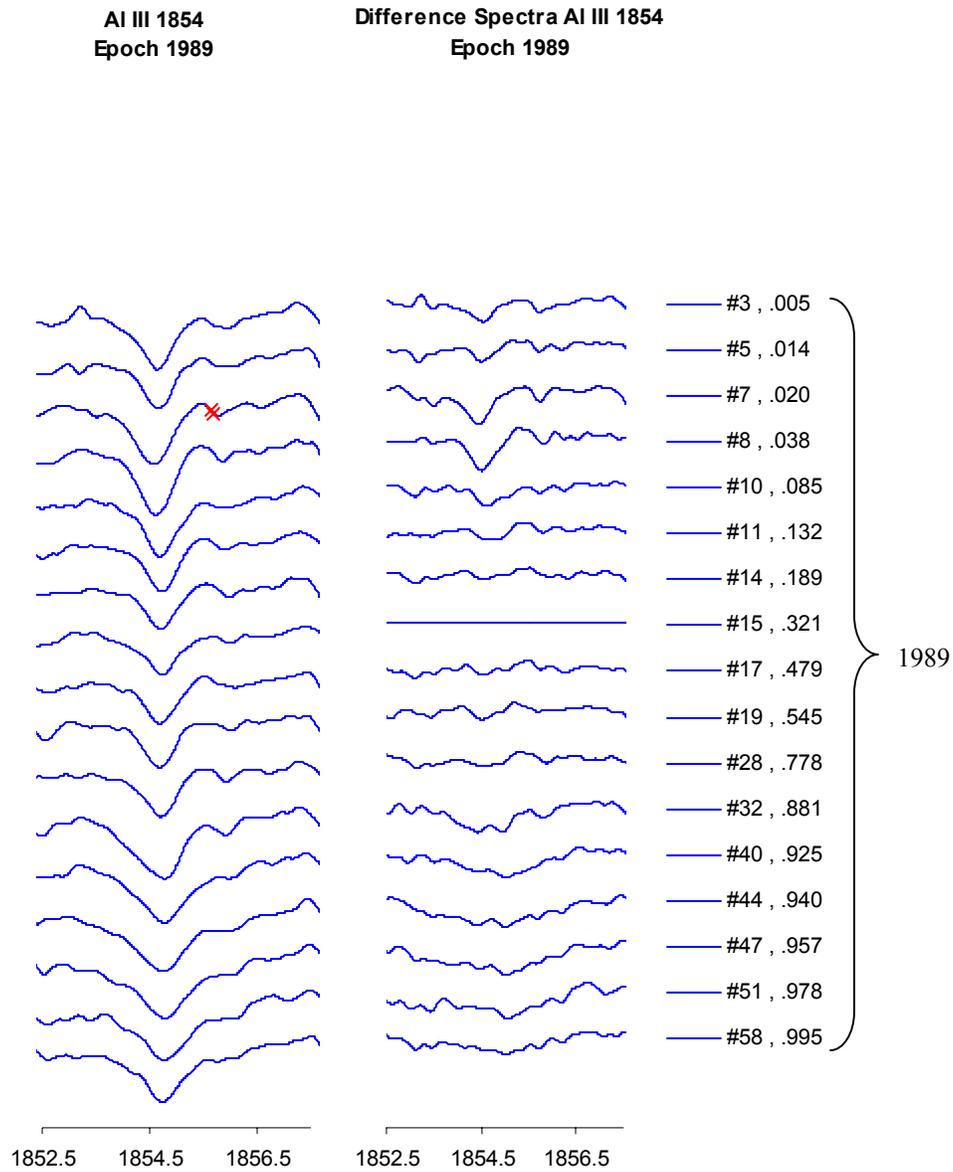

FIG. 4.5.6.3c—ALGOL IUE SWP SPECTRA Al III 1854. *Left*, smoothed spectra. *Right*, difference spectra.

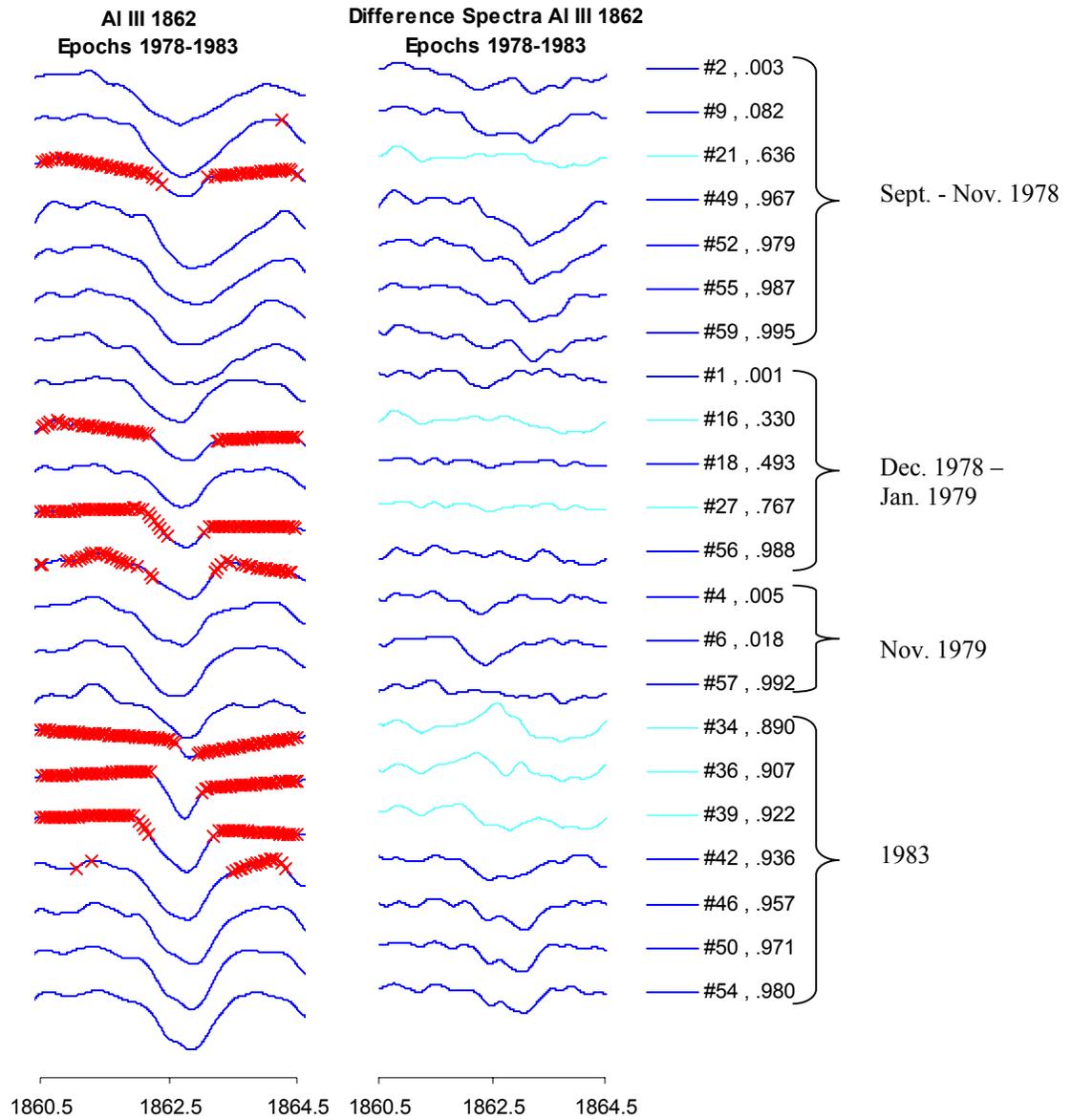

FIG 4.5.6.1.4a—ALGOL IUE SWP SPECTRA Al III 1862. *Left*, smoothed spectra. *Right*, difference spectra.

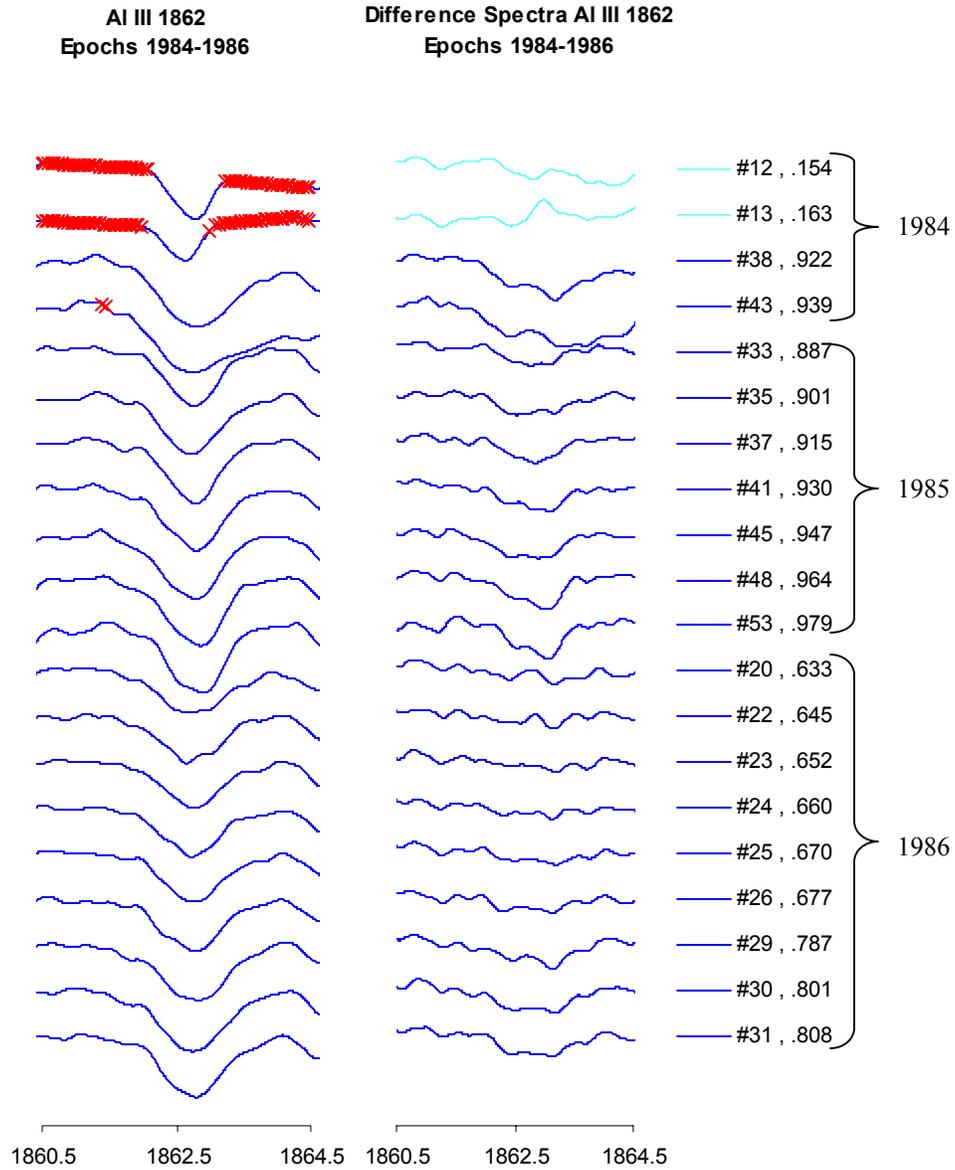

FIG, 4.5.6.4b—ALGOL IUE SWP SPECTRA Al III 1862. *Left*, smoothed spectra. *Right*, difference spectra.

**Al III 1862
Epoch 1989**  **Difference Spectra Al III 1862
Epoch 1989**

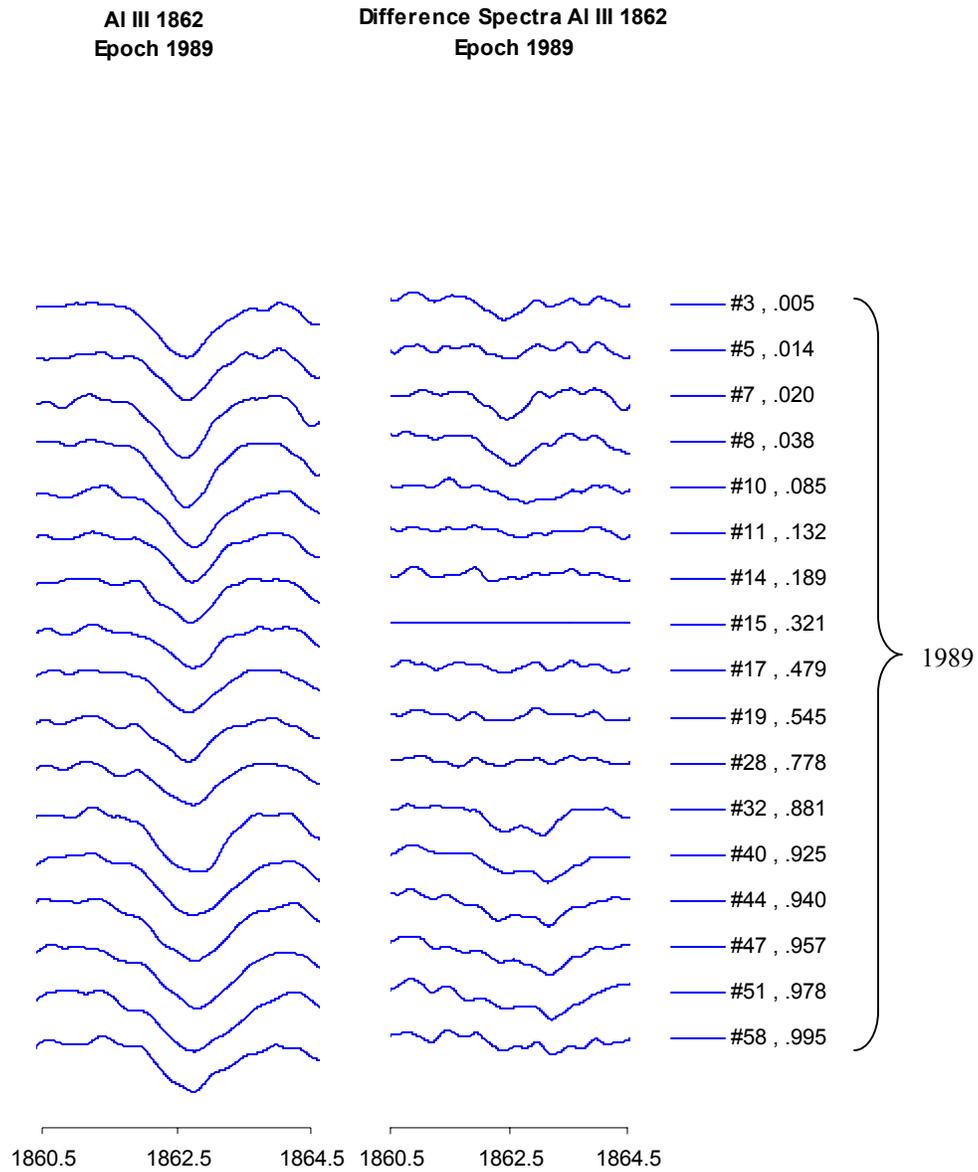

FIG. 4.5.6.4c—*ALGOL IUE SWP SPECTRA Al III 1862. Left*, smoothed spectra. *Right*, difference spectra.

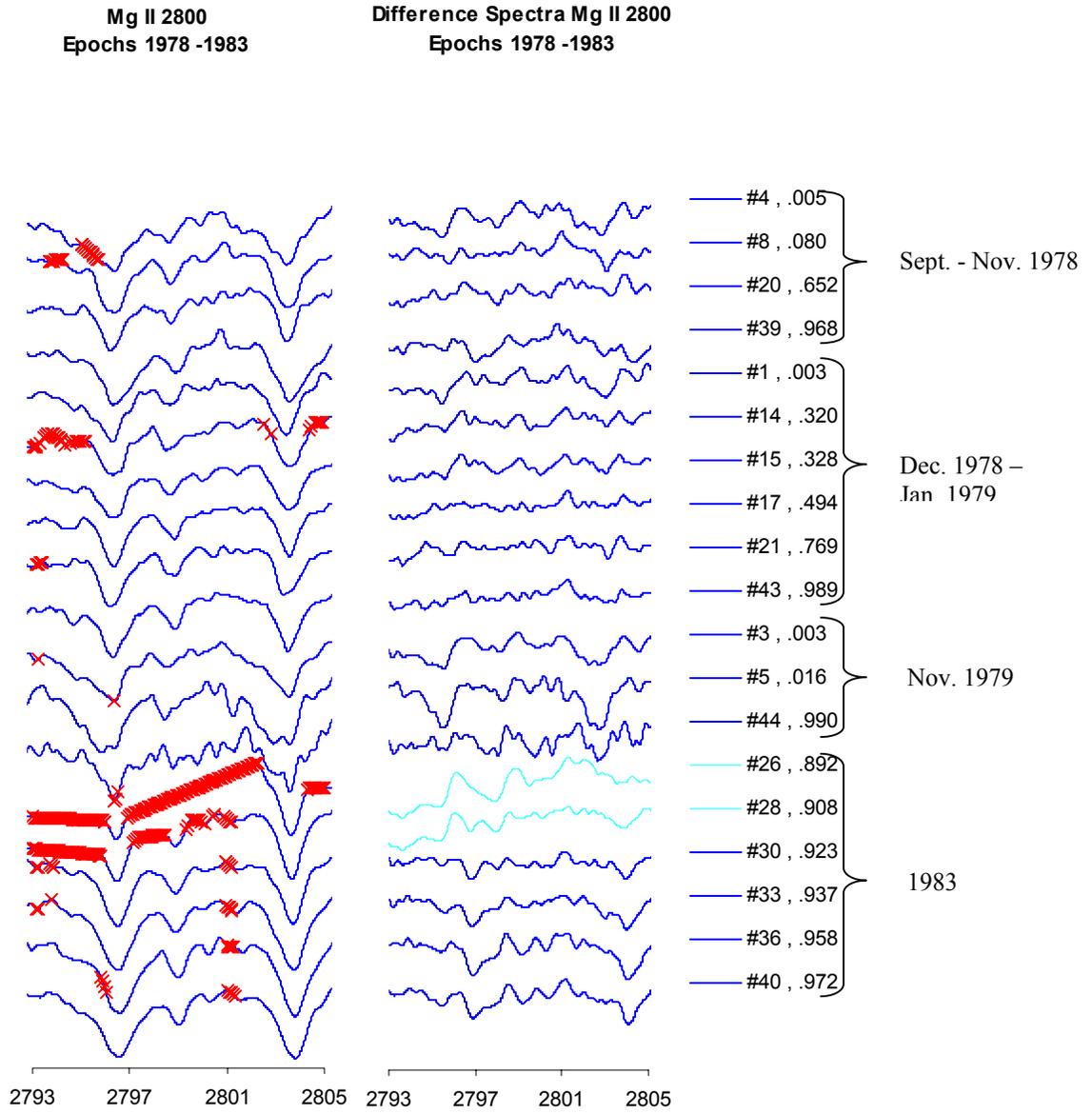

FIG. 4.5.6.5a—*ALGOL IUE LWP/LWR SPECTRA Mg II near 2800. Left*, smoothed spectra. *Right*, difference spectra.

**Mg II 2800**
**Epochs 1984 - 1986**

**Difference Spectra Mg II 2800**
**Epochs 1984 - 1986**

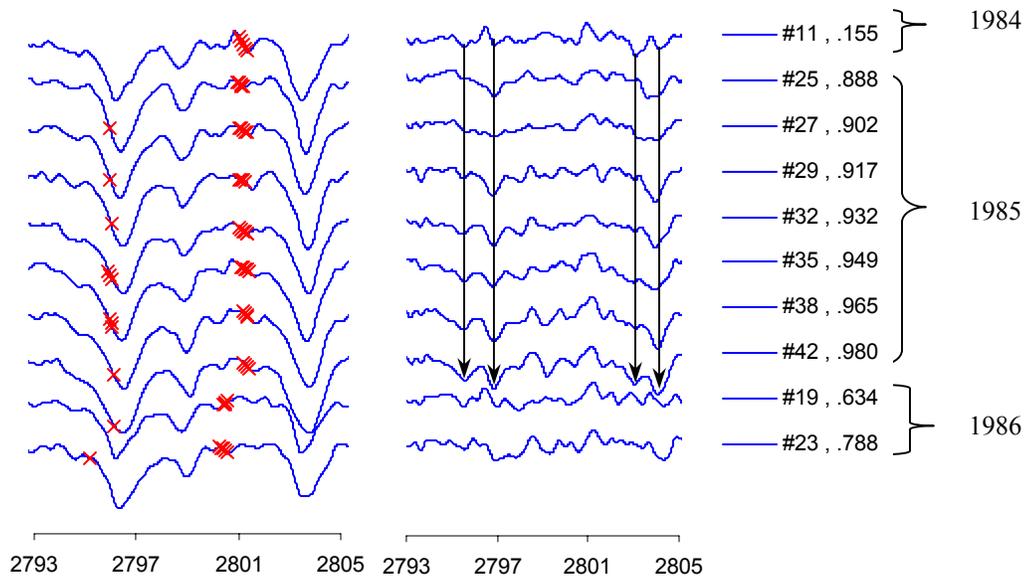

FIG. 4.5.6.5b—*ALGOL IUE LWP/LWR SPECTRA Mg II near 2800. Left*, smoothed spectra. *Right*, difference spectra. Arrows indicate sets of double peaked absorption features.

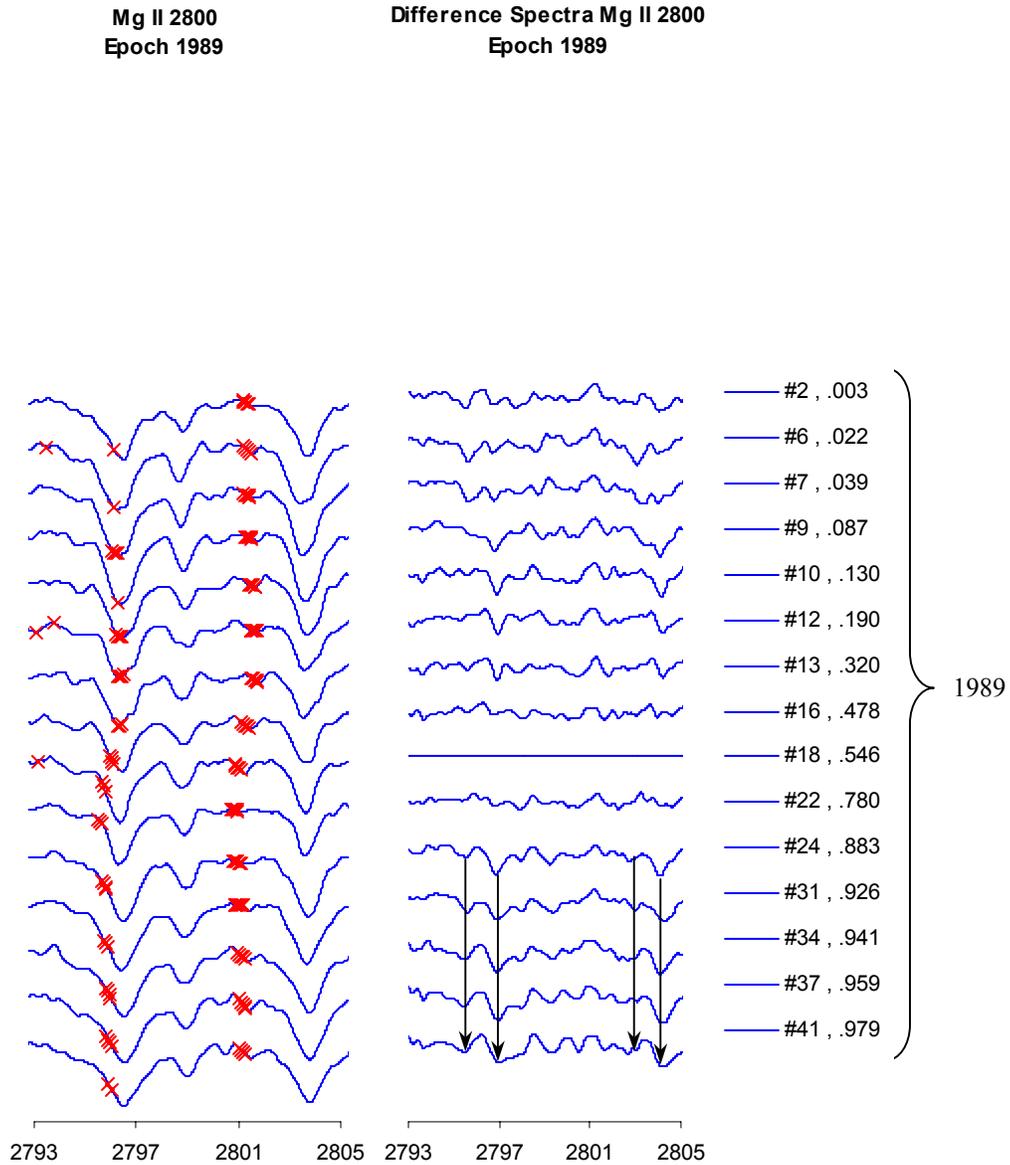

FIG. 4.5.6.5c—*ALGOL IUE LWP/LWR SPECTRA Mg II near 2800. Left, smoothed spectra. Right, difference spectra.* Arrows indicate sets of double peaked absorption features.

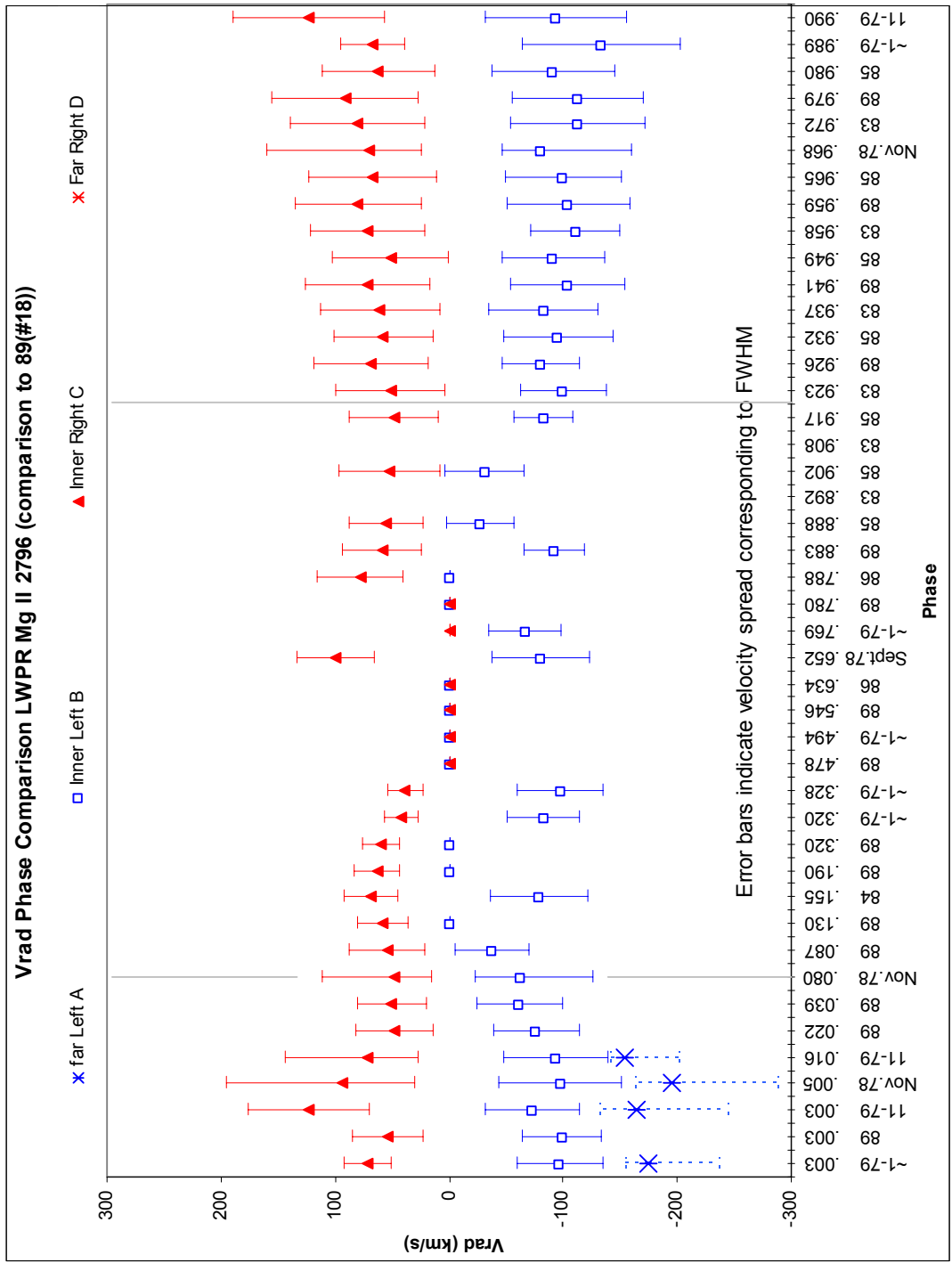

FIG. 4.5.6.6—Vrad Phase Comparison LWP/LWR Mg II 2796

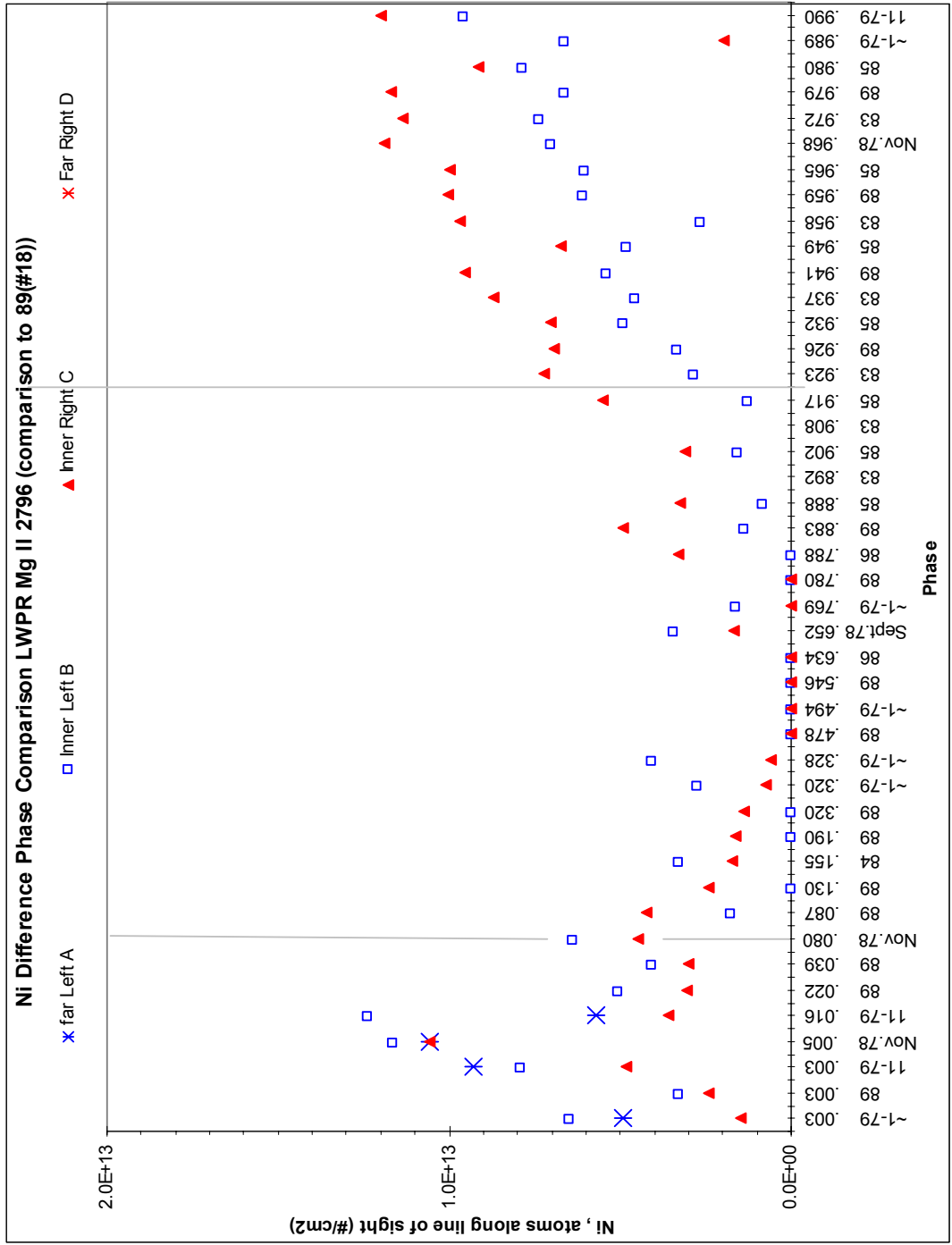

FIG. 4.5.6.7.–Ni Difference Phase Comparison LWP/LWR Mg II 2796

In anticipation of our mass flow calculation in the following sections, we must consider complications due to the Rossiter effect as they contribute to our difference spectra. Therefore, we follow the method of Cugier and Chen (1977) where they differenced in-eclipse exposures of Algol (taken with the Copernicus satellite 1974 January 4) situated approximately symmetrically about the time of mid-primary eclipse. FIG. 4.5.6.8 is a coplot of such a pair from their data at phases ~0.928 (solid line) and ~0.074 (broken line).

If primary eclipse is symmetric the difference between these two symmetrically placed exposures about primary eclipse should be zero. However, Cugier and Chen demonstrated that there is an additional absorption component evident in the difference spectrum, shown in FIG. 4.5.6.9, which indicates gas flowing from the secondary star. Our difference spectra (next section) are also consistent with a gas flow model.

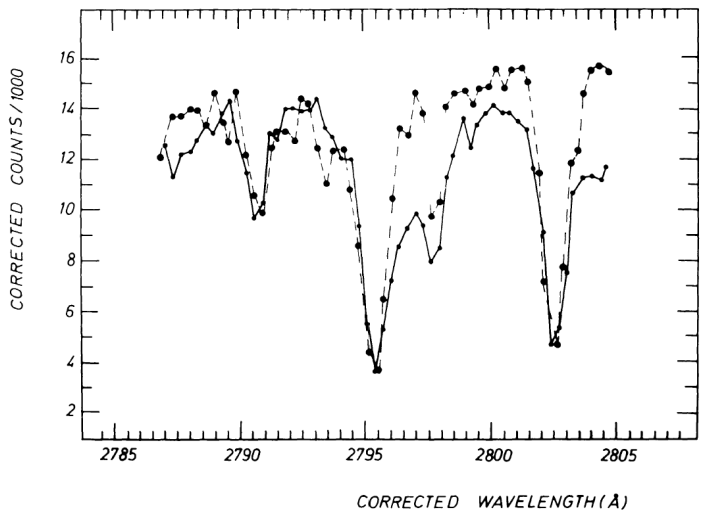

FIG 4.5.6.8—*Mg II λλ2796, 2803.* (*Source Fig. 1, p. 170, Cugier and Chen 1977*)

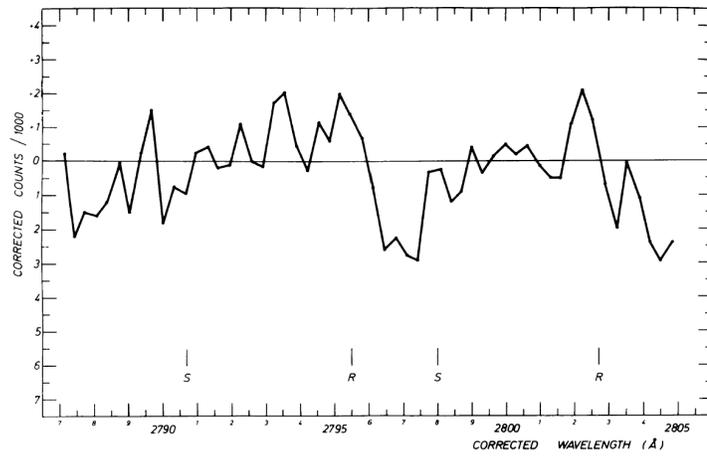

FIG 4.5.6.9—*Mg II λλ2796, 2803 Difference Spectrum.* (*Source Fig. 2, p. 171, Cugier and Chen 1977*)

### 4.5.7 *Mass Loss Rates*

Using the method of Cugier and Chen (1977), we estimate the mass loss rate from the secondary star. The first step is to identify pairs of spectra situated symmetrically with respect to the time of primary minimum ($\phi = 0.0$), where one of the pair contains extra absorption presumed to be the gas stream. Selected phase pairs in epoch 1989 are: $\phi = 0.022$ and 0.9794, 0.959 and 0.039, and 0.9259 and 0.0867. See Table 4.5.7.1.

TABLE 4.5.7.1
PHASES USED FOR MIRROR COMPARISON OF Mg II

| LWP/LWR Group I.D. | Phase | Epoch | Lines Used |
|---|---|---|---|
| 41 | 0.9794 | 1989 | Mg II 2796 and 2803 |
| 6 | 0.022 | | |
| 37 | 0.959 | 1989 | Mg II 2796 and 2803 |
| 7 | 0.039 | | |
| 31 | 0.9259 | 1989 | Mg II 2796 and 2803 |
| 9 | 0.0867 | | |

The difference spectrum is then calculated between each mirror pair, i.e., 41 minus 6, 37 minus 7, and 31 minus 9. FIG. 4.5.7.1. shows the 41-6 pair which corresponds to exposure LWP16322 ($\phi = 0.9794$, LWP/LWR ID: 41) minus exposure LWP16324 ($\phi = 0.022$, LWP/LWR ID: 6). See TABLE 3.2.3 for the index of LWP/LWR IDs. The two upper curves are the spectra before differencing. The lower curve is the difference spectrum.

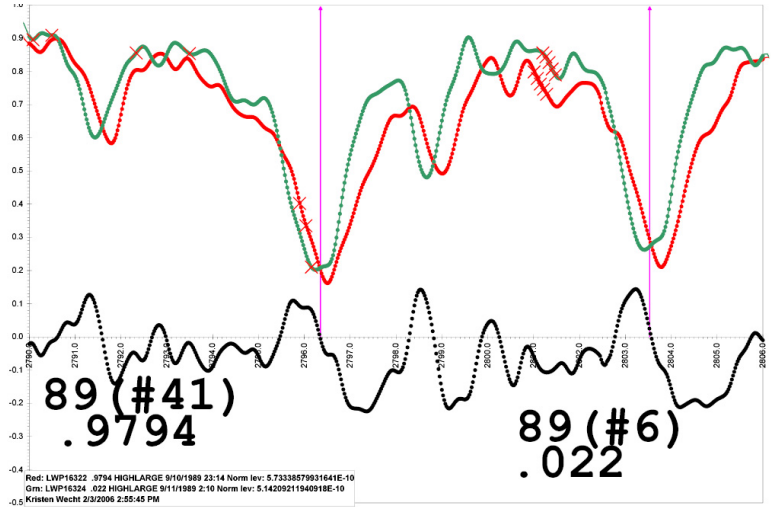

FIG. 4.5.7.1– *Difference Spectrum Corresponding Phases Symmetrically Situated About Primary Eclipse*

Equivalent widths and radial velocities of the extra absorption components are measured and the number of atoms along the line of sight is calculated using (in SI units)

$$N = \frac{4\varepsilon_\circ c^2 m_e}{e^2 \lambda^2 f_{ik}} W_\lambda \qquad (4.5.7.1)$$

where $f_{ik}$ is the oscillator strength (NIST online archive)

The mass loss rate may now be calculated using Equations (1), (2), and (3), Cugier and Chen (1977), page 173.

Their Equation (1) states that the equivalent width $W_\lambda$ is proportional to the product of the mean density of the gas stream $\rho$, the geometrical depth along the line of sight $\Delta Z$, and the ratio of the number of Mg II atoms to the total number of Mg atoms N (Mg II)/N(Mg).

$$W_\lambda \propto \rho \times \Delta Z \times \frac{N(Mg\,II)}{N(Mg)} \qquad (4.5.7.2)$$

For a rough estimate, we assume that most of the Mg is singly ionized, i.e.,

$$\frac{N(Mg\ II)}{N(Mg)} \approx 1 \quad (4.5.7.3)$$

And since

$$N_i \propto W_\lambda, \quad (4.5.7.4)$$

we have

$$N_i \propto \rho \times \Delta Z \quad (4.5.7.5)$$

But $N_i$ is in units of atoms/area and $\rho \times \Delta Z$ is in units of mass/ area. Therefore, we multiply $N_i$ by mass/atom, which for Mg is (24.305 amu) x (1.6606 x $10^{-24}$ g/amu)

$$= 4.036 \times 10^{-23} \text{ g/ Mg atom}$$

Therefore,

$$\rho = \frac{N_i (4.036 \times 10^{-23} g/Mg\ atom)}{\Delta Z} \quad (4.5.7.6)$$

Cugier and Chen (1977) estimate $\Delta Z \approx 1\ R_A$ and the cross section of the stream $\sigma \sim (R_A)^2$. We then let the velocity of the stream $v \approx V_{rad}$. The mass is then calculated using the relation

$$\frac{dM}{dt} = \rho \times \sigma \times v$$

$$\approx \left(4.036 \times 10^{-23} \frac{g}{Mg\ atom}\right) N_i R_A V_{rad} \left(\frac{M_\odot}{M_{mg}}\right) \quad (4.5.7.7)$$

where $R_A = 2.01837 \times 10^{11}$ cm (See Appendix A, Table 27.)

$$M_{Mg} = .00076\ M_\odot \text{ (assuming solar abundances)} \quad (4.5.7.8)$$

The final convenient equation we used to calculate the mass loss rate is

$$\frac{dM}{dt} = (1.295 \times 10^{-32}) N_i V_{rad} \qquad (4.5.7.9)$$

which gives the results in units of $M_\odot$/yr when $N_i$ is expressed as the number of Mg atoms per cm$^2$ and $V_{rad}$ is expressed in km/s. The results are listed in TABLE 4.5.7.2. The mirror difference pair 41 ($\phi = 0.9794$) minus 6 ($\phi = 0.022$) gives the highest mass loss rate $\frac{dM}{dt} = (1.9 \times 10^{-14})$ $M_\odot$/yr in the 1989 epoch. The mass loss rate estimated by Cugier and Chen (1977) from the Copernicus satellite data is $10^{-13}$ $M_\odot$/yr. This would indicate that January 1974 is an epoch of higher activity than September 1989.

There are several sources of error to consider. The spectra are noisy and although the smoothing process is helpful in distinguishing noise from real features it slightly changes the line width and depth. We are estimating the line profile assuming a Gaussian shape which is not necessarily the shape of the line if the instrumental broadening is significant. This does not affect the equivalent width measurements, however it may affect the estimate of the velocity gradients and thus the difference spectra. The in-eclipse spectra have a lower continuum and therefore the signal to noise ratio is less as well.

To obtain an estimate of the error the mass loss rate can be calculated using the difference spectra of other spectral lines.

TABLE 4.5.7.2
PROPERTIES OF LONGWARD MIRROR – DIFFERENCE COMPONENT

| MIRROR DIFFERENCE PAIR | $W_\lambda$ λ2796 λ2803 | $W_\lambda$ AVG (Å) | $N_i$ | $N_{i\,Avg}$ (#/cm²) | $V_{rad}$ | $V_{rad}$ Avg (km/s) | HWHM | HWHM$_{Avg}$ (km/s) | $\dfrac{dM}{dt}$ (M$_\odot$/yr) |
|---|---|---|---|---|---|---|---|---|---|
| 41 - 6 | .23 .33 | .28 | 5.58 x 10¹² 1.55 x 10¹³ | 1.054 x 10¹³ | 100.6 115.4 | 108 | 52.5 77.5 | 65 | 1.94 x 10⁻¹⁴ ± 1.17 x 10⁻¹⁴ |
| 37 - 7 | .15 .15 | .15 | 3.48 x 10¹² 6.99 x 10¹² | 5.235 x 10¹² | 73.8 80.1 | 77.45 | 32.1 30.5 | 31.3 | 6.9 x 10⁻¹⁵ ± 2.79 x 10⁻¹⁵ |
| 31 - 9 | .08 .09 | .085 | 1.99 x 10¹² 4.08 x 10¹² | 3.035 x 10¹² | 99.5 123.9 | 111.7 | 34.8 41.2 | 38 | 5.8 x 10⁻¹⁵ ± 1.965 x 10⁻¹⁵ |

# 5. DISCUSSION AND CONCLUSIONS

In the previous chapter, we presented several analyses of the IUE data resulting in various measures of Algol system properties. These measures pertain to the ionized gases flowing into and about Algol A and those constituting the photospheric region. And, they tell a story that is difficult to unravel. Although a degree of interpretation was provided with each analysis, it is the intention of this final chapter to expand on these interpretations, with the particular goal of integrating these measures into a coherent picture. Furthermore, we suggest steps that can be taken in subsequent studies to further expand our understanding.

Most of the results presented in the previous chapter (Sections 4.5.1 - 4.5.5 ) do not directly distinguish photospheric or photospheric-like regions from other ionized-gas contributions. It was our goal in the last two sections (4.5.6 and 4.5.7) to isolate and distinguish photosphere-related effects from other gas-flow contributions. This "Difference" approach has the advantage over the "Direct" approach of potentially unraveling several distinct contributions to the IUE data, particularly isolating those associated with the gas flow and circumstellar characteristics that define Algol-type systems. However, it also has several disadvantages: it can only be used where spectra are available that do not contain the gas-flow effects being isolated; and such difference spectra always contain the possibility of spurious features.

We begin this chapter with a discussion of the "Direct" (non-difference) approach, with particular attention to phase dependences of the system during the 1989 epoch.

## 5.1. *Direct Approach*

An examination of the 1989 phase dependences of the equivalent widths reveals common features among the different ion species. We focus on the absorption lines, Al II λ1670, Al III λ1854, Mg II λλ2796, 2803, and Si IV λλ1393, 1402. The Al II, Al III, and Mg II observations reveal equivalent width curves that are relatively flat and featureless at phases in the approximate range of 0.2 – 0.8, though Al II and Mg II are slightly elevated within the 0.2 – 0.4 range. The equivalent widths for all of these ions are elevated in the 0.0 – 0.1 (egress) and 0.8 – 1.0 (ingress) phase ranges, but higher during ingress.

The Si IV equivalent widths show a gradual but uneven decrease in the 0.0 – 0.8 range, then increasing sharply in the 0.8 – 1.0 range. Incidentally, equivalent widths for Si II and C II are uneven, but relatively level.

Al II and Al III show very photospheric radial velocities as functions of phase, indicating photospheric motions, or motions with symmetry or apparent symmetry about the photospheric values.

Mg II and Si IV show radial velocity features distinctly different from Al II and Al III, though both contain the rough, overall features of photospheric motion. Mg II is slightly red shifted during egress, and then shows a characteristic rise and reverse curvature in the 0.1 – 0.3 phase range. It is mostly red shifted in the remaining phase range as well, except for reductions to about photospheric velocities at phases of ~0.33 and ~0.8. Interestingly, Si IV λ1402 exhibits features qualitatively similar to those of Mg II. However, Si IV λ1393 shows blue shifts in the 0.0 – 0.2 and 0.5 – 0.8 ranges.

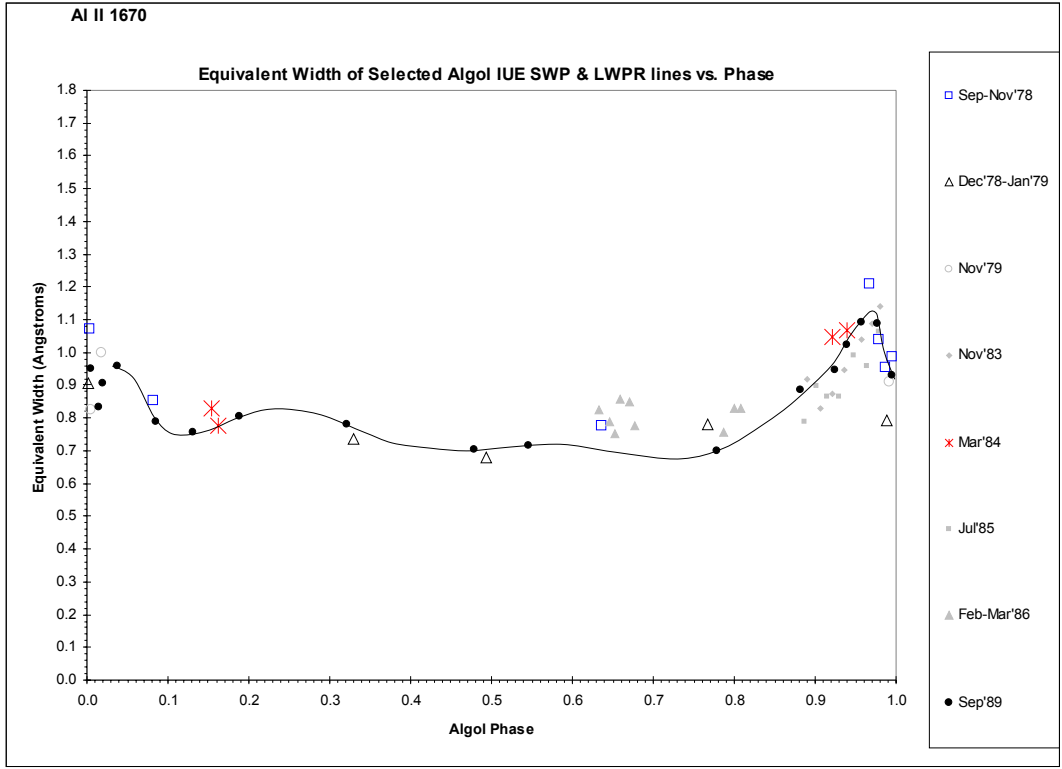

FIG. 5.1.1a–Al II 1670 Equivalent Width vs. Phase

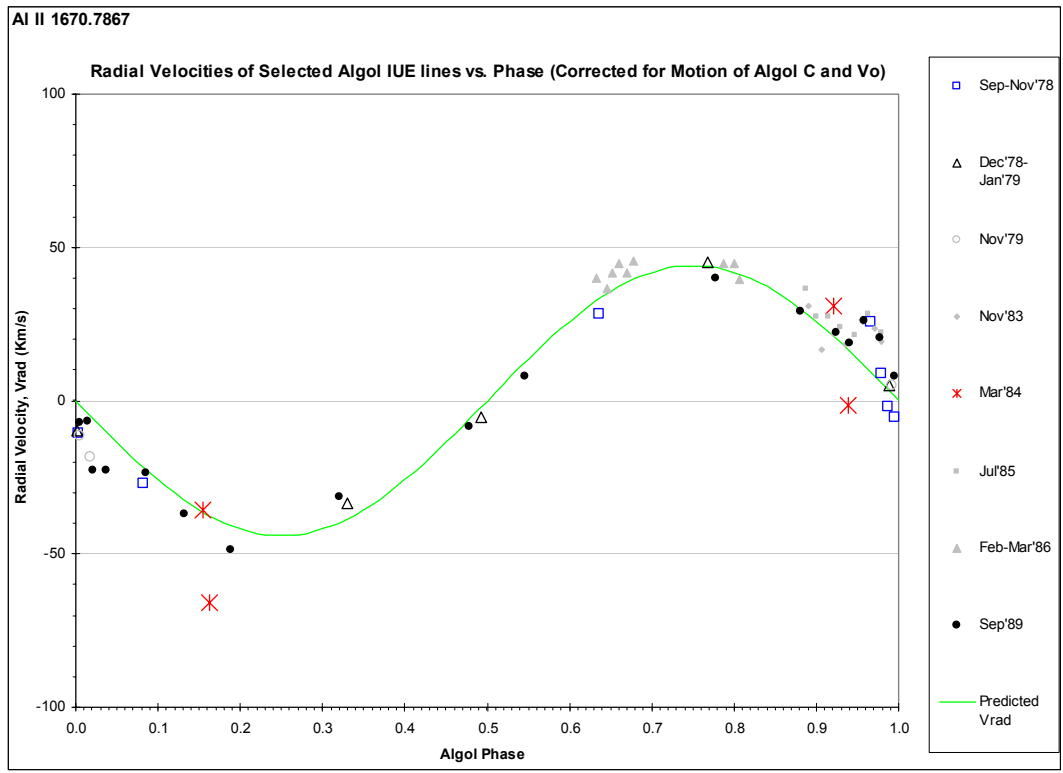

FIG. 5.1.1b–Al II 1670 Radial Velocity Curve

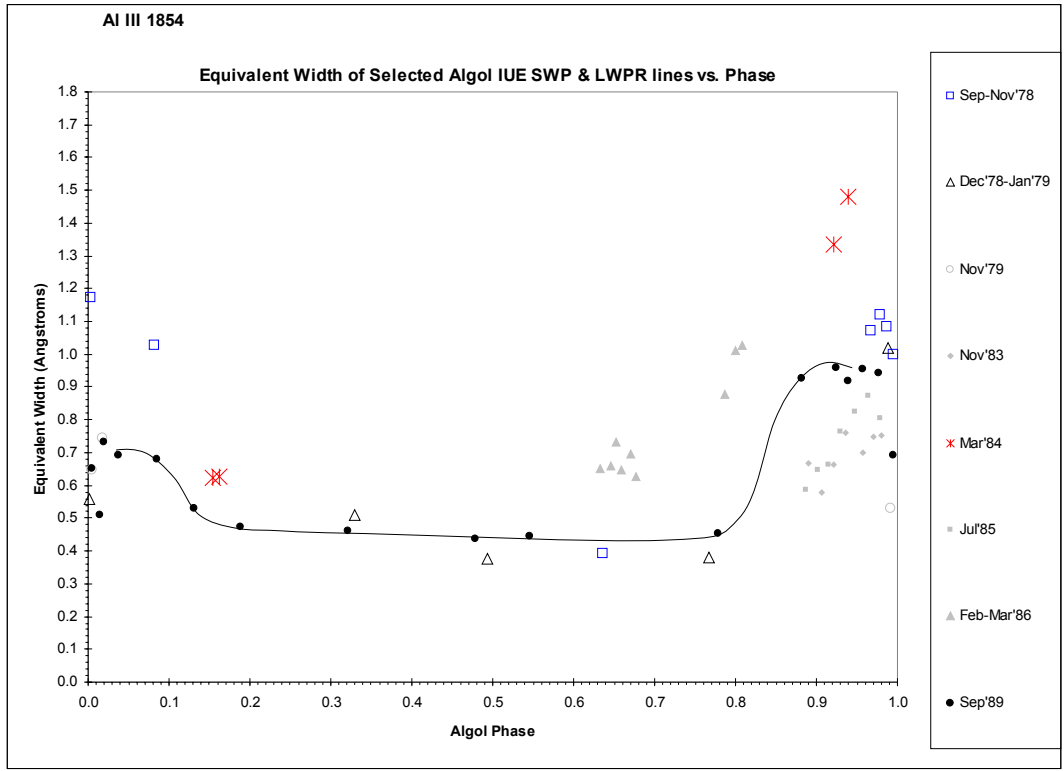

FIG. 5.1.2a–Al III 1854 Equivalent Width vs. Phase

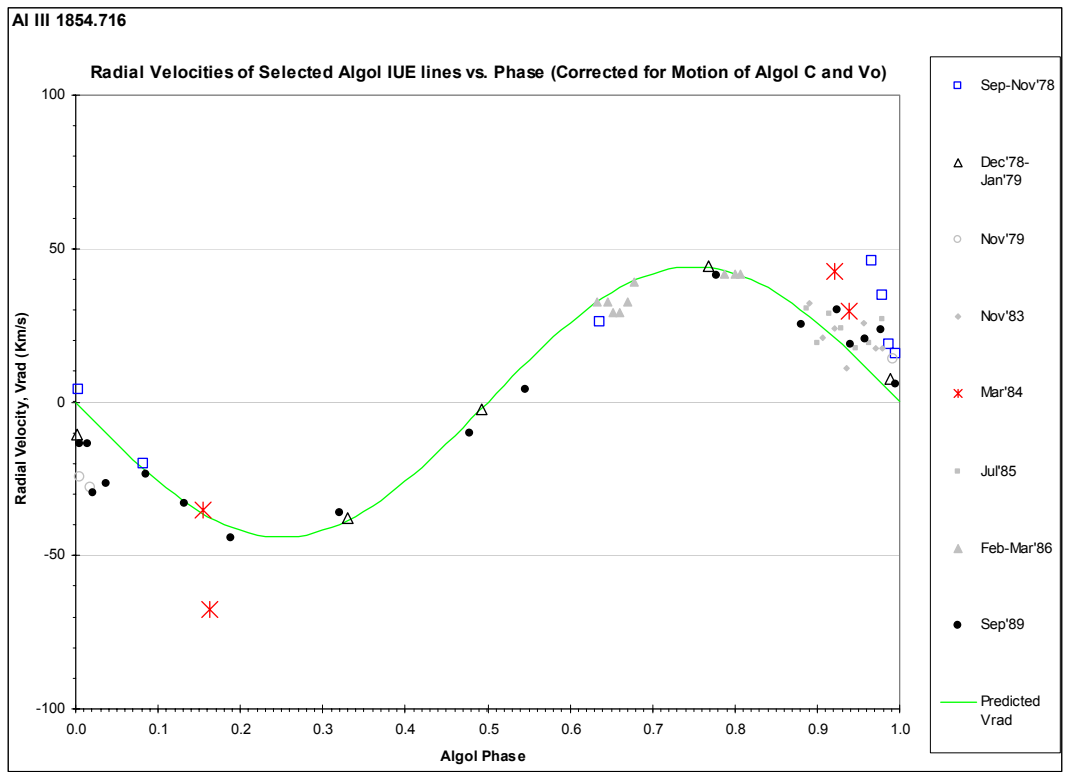

FIG. 5.1.2b–Al III 1854 Radial Velocity Curve

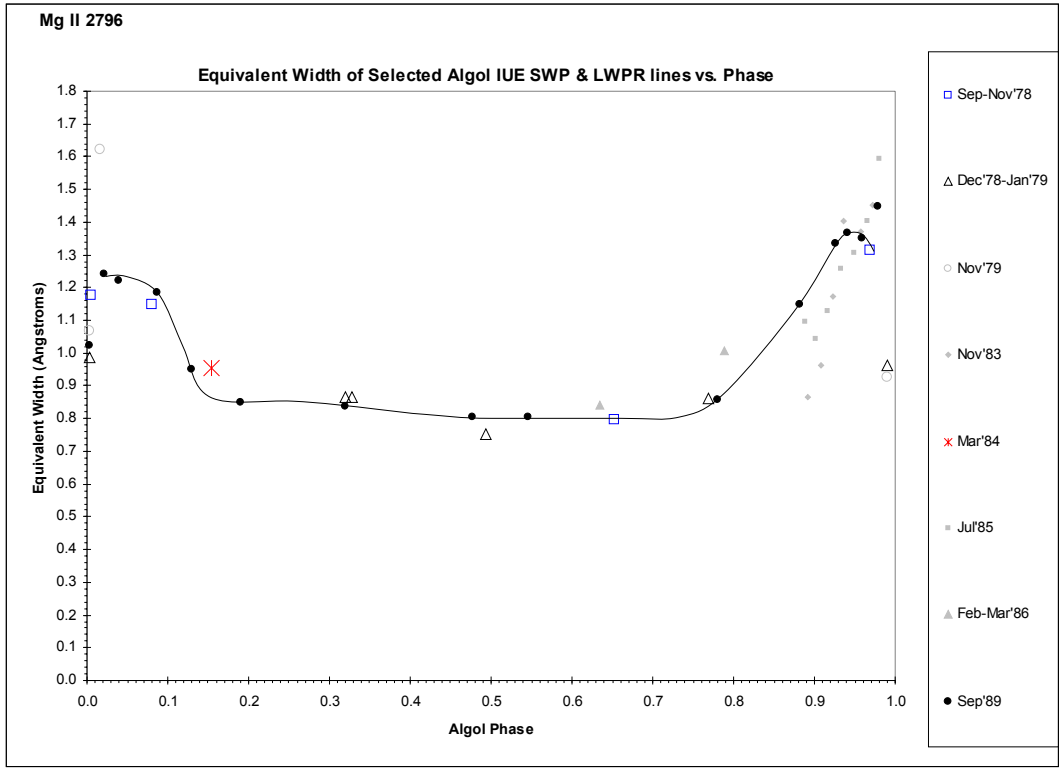

FIG. 5.1.3a–Mg II 2796 Equivalent Width vs. Phase

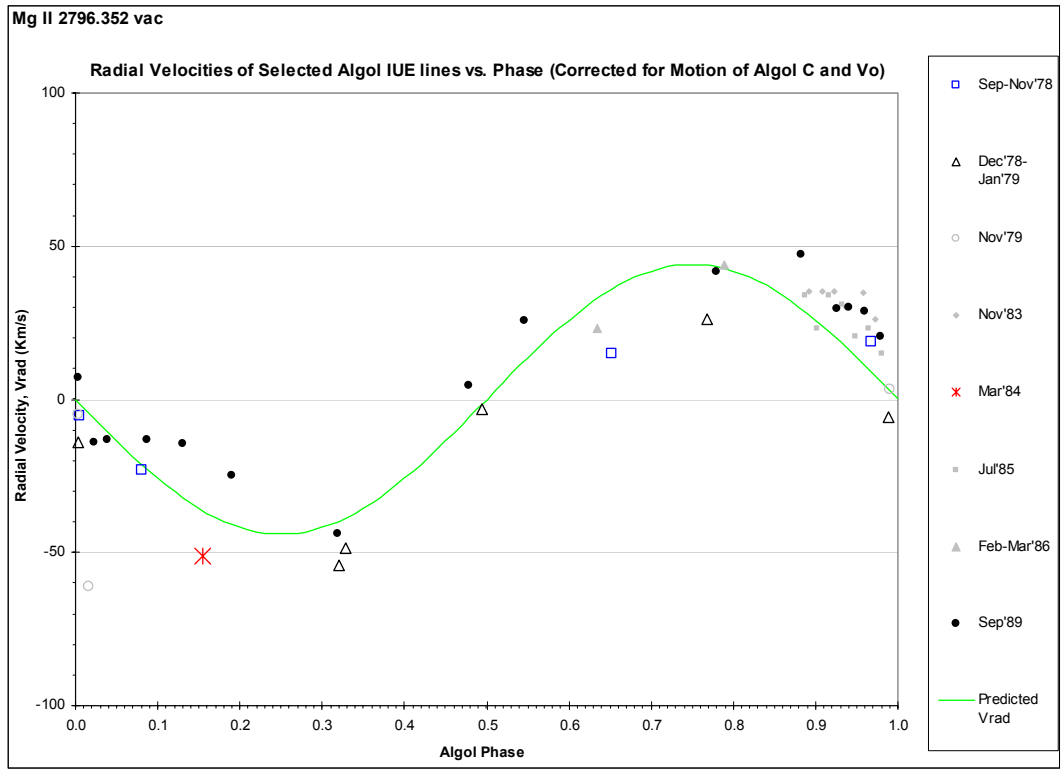

FIG. 5.1.3b–Mg II 2796 Radial Velocity Curve

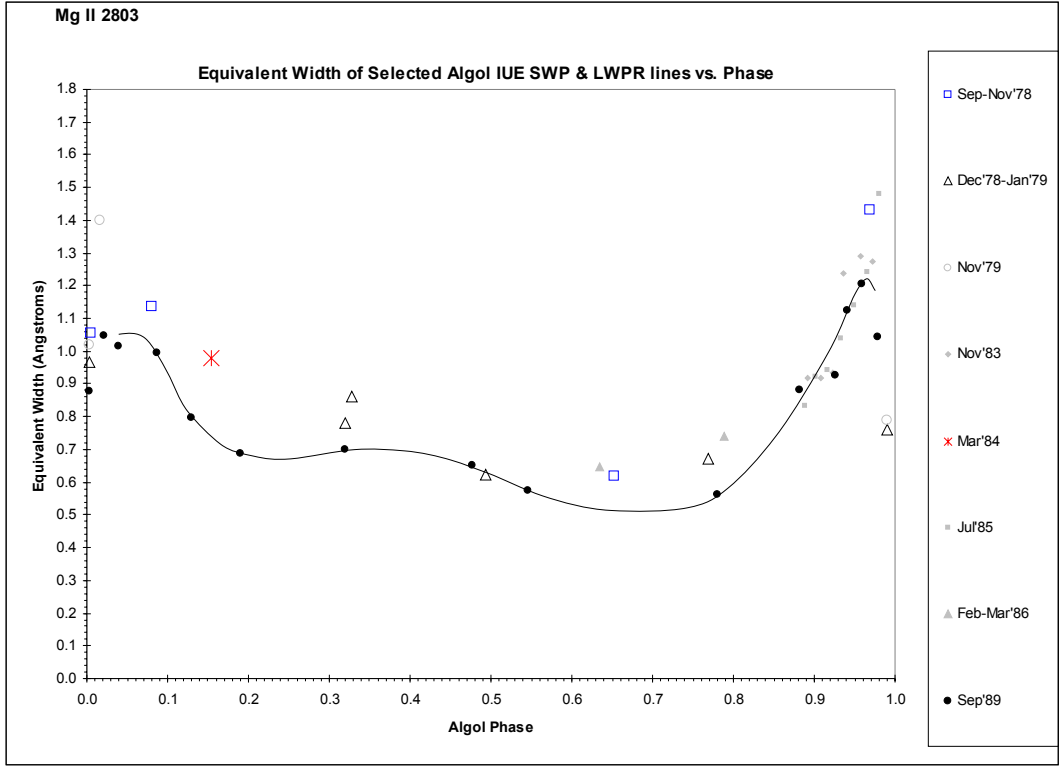

FIG. 5.1.4a–Mg II 2803 Equivalent Width vs. Phase

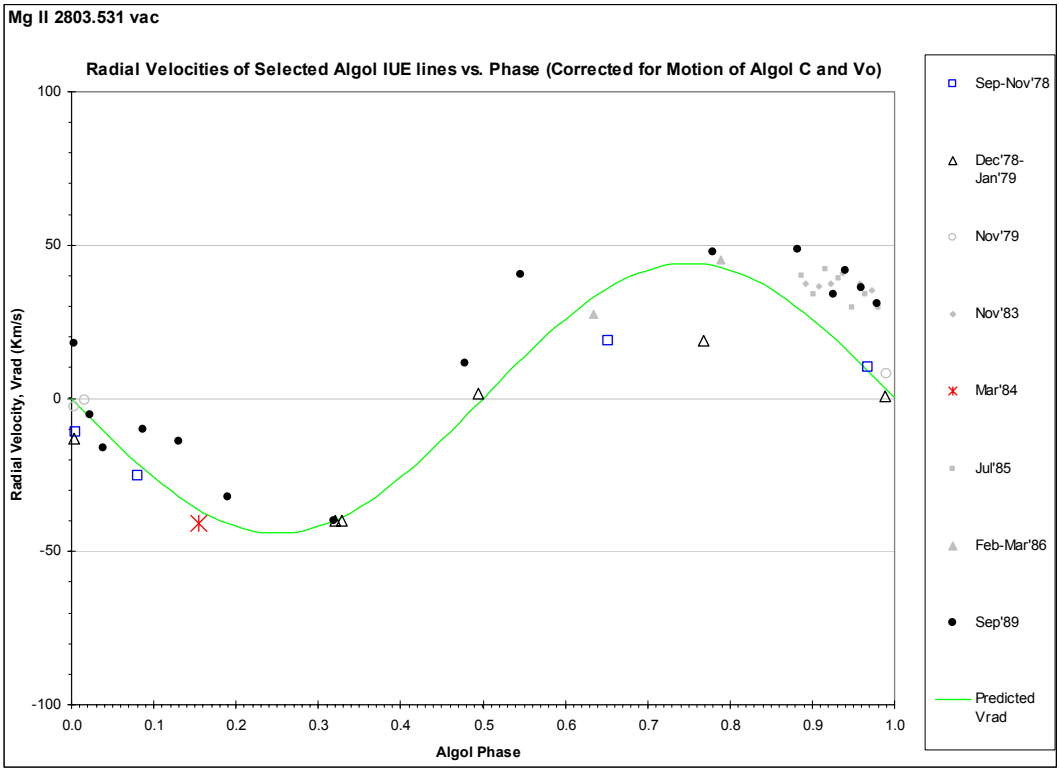

FIG. 5.1.4b–Mg II 2803 Radial Velocity Curve

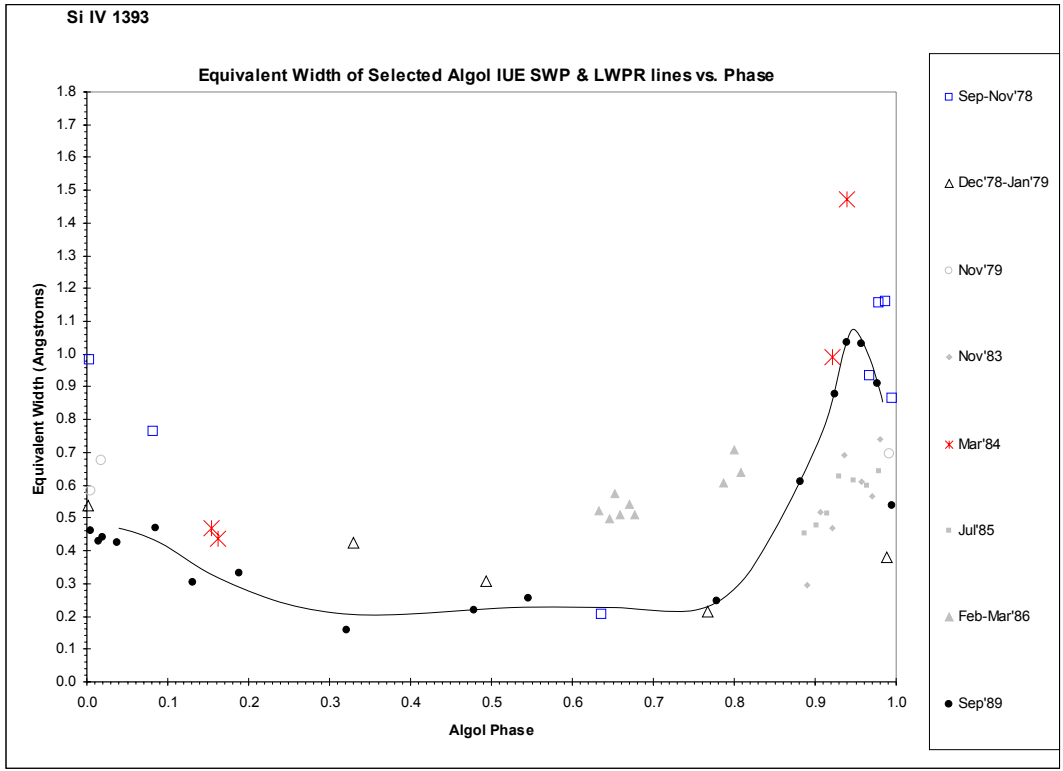

FIG. 5.1.5a–Si IV 1393 Equivalent Width vs. Phase

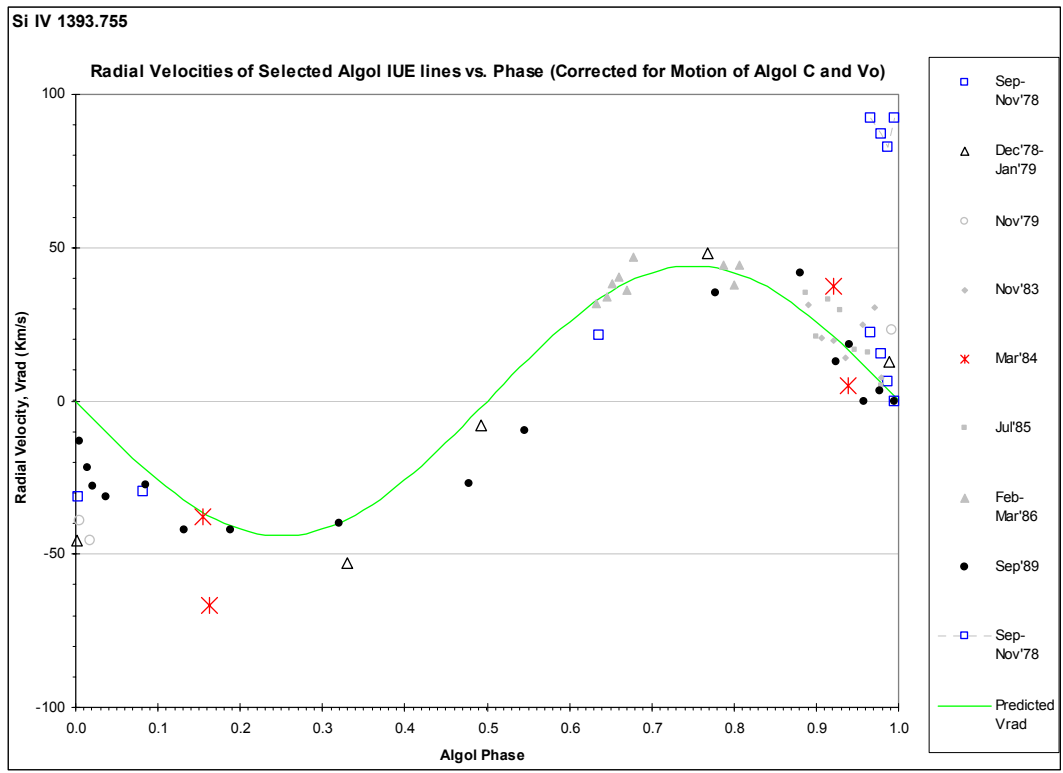

FIG. 5.1.5b–Si IV 1393 Radial Velocity Curve

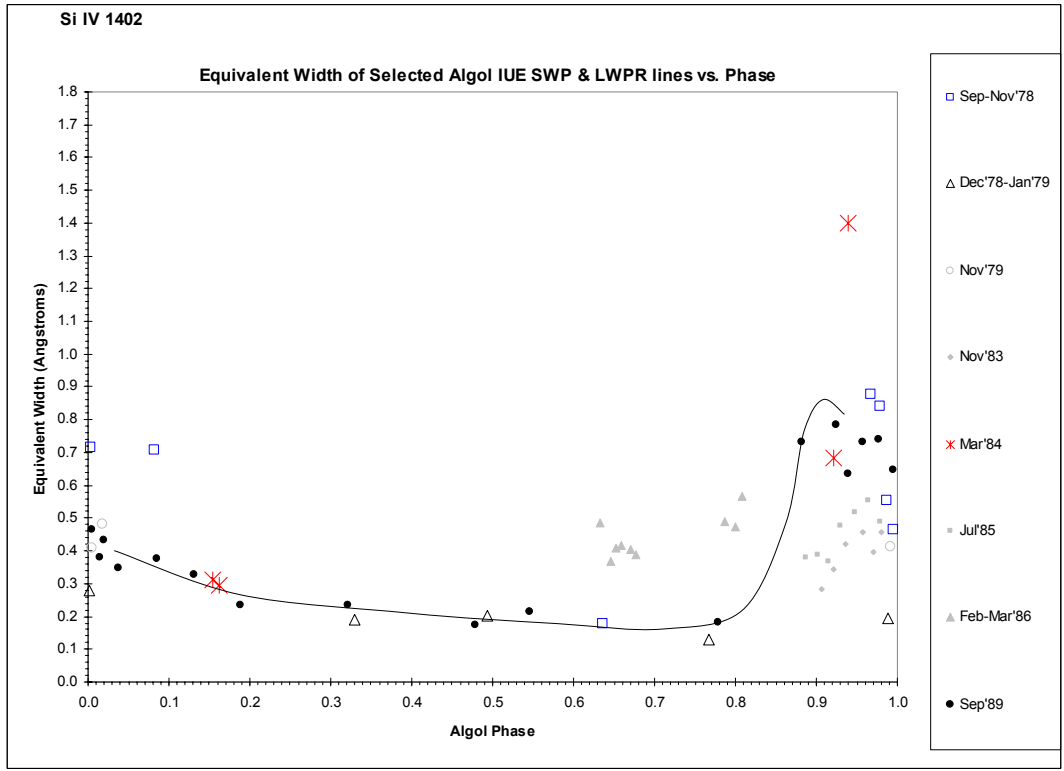

FIG. 5.1.6a—Si IV 1402 Equivalent Width vs. Phase

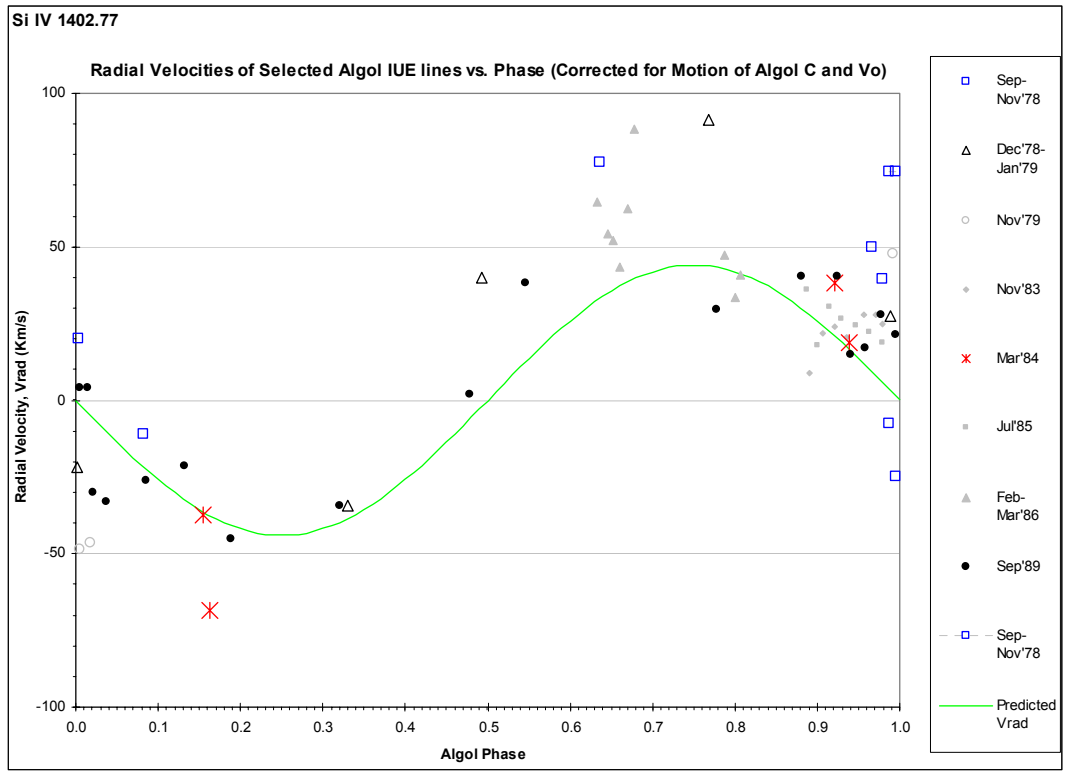

FIG. 5.1.6b—Si IV 1402 Radial Velocity Curve

These differences between the Si IV λλ1393, 1402 lines then must be associated with blending and significant velocity or density differences or gradients; this blending is probably most significant for Si IV λ1402.

The predominance of Al II and Al III photospheric characteristics can be understood by an examination of the ionization fractions. Ionization fractions appropriate to selected ions, temperatures, and electron pressures have been identified from the ionization fraction plots. The only candidates for normal photospheric features for Algol A (temperature 12,300 K; log of electron pressure 2.5) from among those examined in this work are singly ionized Al, Si, and Fe (Al II, Si II, and Fe II), doubly ionized Al, Mg, Si, and Fe (Al III, Mg III, Si III, and Fe III), and possibly Mg II. As seen in FIG. 4.2.2.2, Al II and Al III appear in equal concentrations under the photospheric conditions of Algol A.

The substantial presence of Mg II and Si IV, whose appearances respectively require lower and higher temperatures than the Algol A photosphere, indicate that competition with photospheric features is likely.

In order to explain these observations and those involving the Difference Approach to be described shortly, we propose the model shown in FIG. 5.1.7. Al II, Al III, Mg II, and Si IV all show high equivalent widths in the vicinity of ingress (phase range 0.8 – 0.92). This can be ascribed to the gas flow from Algol B to Algol A which, in this phase range, intervenes between Earth and Algol A and is moving away from us. The aluminum radial velocities are dominated by the photospheric contributions. However, Mg II is moving away in this region. This is consistent with the cool flow from the Lagrange point vicinity toward Algol A. The Si IV results indicate movement away from us in the 0.8 phase range; however, the results are very uneven in the 0.5 – 1.0 range, suggestive of competing effects. This is not surprising since the Si IV would only exist in a higher temperature region, where gas flow would be interacting strongly with the photospheric region of Algol A.

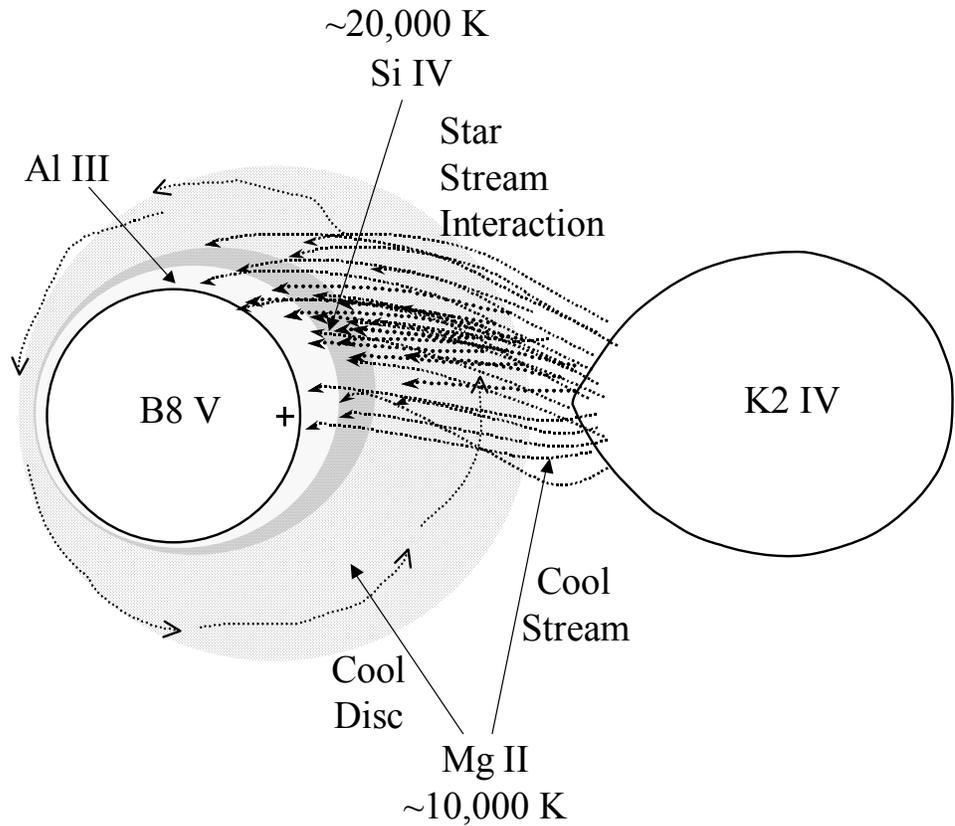

FIG 5.1.7–*Algol Model*

The Mg II results show high equivalent widths and receding gas motion also around egress. The motion of receding cool gas flow is consistent with the model. This effect is less pronounced for Al II and Al III, again due to the strong competition from photospheric contributions, and the fact that the abundance of aluminum is about an order of magnitude less that that of magnesium. The asymmetry between egress and ingress can be understood as a consequence of coriolis force deflection, which produces a greater density on the ingress side.

Si IV, however, shows a clear photospheric radial velocity curve at egress, including a very pronounced Rossiter effect. In fact, the Rossiter effect during egress is also exhibited by the Mg II. This suggests that there is gas flow that moves around Algol A in some form of circumstellar disk. However, its velocity is too low to constitute Keplerian orbital motion since that would require a velocity in the range of hundreds of kilometers per second.

This implies the presence of a disk about Algol A that rotates roughly with the star, and relatively close to its surface. The presence of both Mg II and Si IV is indicative of regions of different temperature, and presumably different densities (lower than the photospheric values).

The variations in the residual intensities, in particular the distorted "W" shape (FIGs. 4.5.4.1 - 4.5.4.7) situated about primary eclipse, are consistent with this model. The changes in residual intensities are due to the varying column densities projected onto the visible part of the disk as it is being eclipsed. The photosphere is a continuous source of absorption which is approximately uniform over the disk. The gas stream is projected against the disk during the later half of the orbit and does not cover the entire disk, however it is more extensive than the hot spot which is localized. The distorted "W" shape is the effect of the progressive occultation of these combined features that have different extents, densities, and temperatures.

## 5.2 *Difference Approach*

In order to examine these effects more thoroughly, we attempt to isolate the multiple contributions to the gas flow that make the previous plots difficult to interpret. We accomplish this by the Difference Approach. These calculations support the model presented in FIG. 5.1.7. However, we choose instead to focus here on a comparison across epochs. Since 1989 has the best phase coverage, we will use this epoch as the baseline to compare the other epochs.

In 1989 there is evidence of irregularly distributed circumstellar material around the B8 V star (primary) and gas streaming from the K2 IV star (secondary) toward the primary. We infer circumstellar gas (disk-like) absorption from the difference spectrum 32 – 15 ($\phi$ = 0.881) where there is a double peaked absorption feature in Al III $\lambda\lambda$1854, 1862 and Si IV $\lambda$1393. In Si IV $\lambda$1402, the absorption feature has a rectangular shape. The difference spectra of Mg II $\lambda\lambda$2796, 2803 show double peaked absorption from phase 0.883 through phase 0.0867.

There is evidence for a gas stream in the difference spectra in that the longward component of the double peaked absorption increases in strength and radial velocity as the line of sight changes from looking across the stream (near phase 0.75) to looking along the stream (near phase 0.95).

The 1983 difference spectra show a similar development of the gas stream from phase 0.9082 through phase 0.972, though the disk-like feature is not apparent (due to limited phase coverage) and the stream feature is not as strong, suggesting a period of lesser activity. Epoch 1985 also shows the development of a gas stream, from phase 0.8866 to 0.979. The stream is marginally stronger in 1985 than 1983 and there is evidence of a disk-like structure.

Epoch 1986 has evidence of a disk-like structure that is more extensive than 1989, with distinct double peaked absorption features across Al III and Si IV. Though the 1986 phase coverage does not extend beyond phase 0.808, the strength of the development of the gas stream component indicates that 1986 is more active than 1989. The phase coverage of 1984 is limited, however the spectra show direct evidence that 1984 is the epoch of greatest mass flow from the secondary toward the primary. In particular, exposure SWP ID: 43 (SWP22439, $\phi$ = 0.939) has the strongest and most asymmetric longward absorption features. The 1984 anomaly is substantial enough to depress continuum, evident in the light curve at primary eclipse ingress.

The November 1978 exposures (0.970 – 0.082) also show evidence of increase of mass flow compared to 1989. However direct comparison between 1984 and November 1978 is not available due to limited phase coverage. We will leave open the possiblility that November 1978 is more active than 1984.

The lone September 1978 spectrum at phase 0.636 appears to be normal. This indicates that Algol can undergo abrupt changes within a month.

The December 1978 – January 1979 epoch is the least active epoch. Evidence of gas streaming at phase 0.0015 is weak compared to 1989.

November 1979 appears to be more active than December 1978 – January 1979, less active than November 1978, and comparable to 1989.

TABLE 5.2.1- EPOCH ACTIVITY LEVEL COMPARISON

|  | Epoch |
|---|---|
| Least Active | Dec. 1978 – Jan. 1979 |
|  | 1983 |
|  | 1985 |
|  | 1989, Nov. 1979 |
|  | 1986 |
|  | 1978 |
| Most Active | 1984 |

Overall the radial velocity deviations from the orbital motion of Algol A extracted from the difference spectra range from ~ ±100 km/s for Mg II, Al III, and Si IV. These low velocities along with the low mass loss rate of ~ $10^{-14}$ $M_\odot$/yr indicate that the Algol system is currently undergoing an epoch of conservative mass transfer.

## 5.3. *Concluding Remarks*

Beta Persei (Algol), from our analysis, behaves as expected like other Algol-type binary systems. We have detected gas streaming effects with short term increases in activity. The light curves follow the expected pattern of increasing depth of primary minimum with decreasing wavelength. These results confirm previous results that at present, Algol is in a variable but relatively low activity state with respect to Algol-type binaries in general.

In the future the secondary star will become a white dwarf and a period of quiescence will follow until the B8 star begins to evolve. It will overflow its critical Roche lobe and accretion of H-He rich material by the white dwarf together with a decrease in orbital period due to systemic mass loss and viscous forces may lead to super-soft x-ray source activity on the white dwarf. Eventually, a pair of inactive, cooling white dwarfs will result. It is possible that the older white dwarf will accrete and retain sufficient mass to reach the Chandrasekhar limit and undergo a Type Ia supernova explosion. Detailed evolutionary calculations would have to be carried out to determine the likelihood of such and ending to the prototype of the Algol-binary systems.

IUE, f [online document: cited 05/08/06], http://sci.esa.int/science-e/www/area/index.cfm?fobjectid=31297

IUE, g [online document: cited 05/08/06], http://sci.esa.int/science-e/www/area/index.cfm?fobjectid=31285

IUE, h [online document: cited 05/08/06], http://esapub.esrin.esa.it/br/br200/Iue.pdf

IUE, i [online document: cited 05/08/06], http://sci.esa.int/science-e/www/area/index.cfm?fobjectid=31311

Johnson, J. 2004, Software, Unpublished,

Kallrath, J., & Milone, E. F. 1999, *Eclipsing Binary Stars: Modeling and Analysis* (New York: Springer)

Kamp, L. W. 1982, Astrophysical Journal Supplement Series, 48, 415

Kempner, J. C., & Richards, M. T. 1999, Astrophysical Journal, 512, 345

Keskin, V., & Pohl, E. 1989, Informational Bulletin on Variable Stars, 3355, 1

Kim, H.-I. 1989, Astrophysical Journal, 342, 1061

Kitchin, C. R. 1995, *Optical astronomical spectroscopy* (Bristol; Philadelphia: Institute of Physics Pub.)

Kitchin, C. R. 1998, *Astrophysical techniques* (3rd ed.; Philadelphia, Pa.: Institute of Physics Publishing)

Kizilirmak, A., & Pohl, E. 1971, Informational Bulletin on Variable Stars, 530, 1

Kizilirmak, A., & Pohl, E. 1974, Informational Bulletin on Variable Stars, 937, 1

Kondo, Y., McCluskey, G. E., Jr. , & Parsons, S. B. 1985, Astrophysical Journal, 295, 580

Kondo, Y., Modisette, J. L., & Morgan, T. H. 1977, Informational Bulletin on Variable

Zeilik, M., & Gregory, S. A. 1998, *Introductory Astronomy & Astrophysics* (4th ed.; Fort Worth: Saunders College Pub.)

# APPENDIX A

# Published Estimates of Algol Parameters

## TABLE OF CONTENTS

# APPENDIX A

## TABLE OF CONTENTS - continued

# APPENDIX A

## TABLE OF CONTENTS – continued

# APPENDIX A

Previously published ALGOL Identifying Terms

| β Persei | Sahade, J. & Hernandez, C. A. (1985), p. 257 (Introduction) | |
|---|---|---|
| β Per | | |
| 26 Persei | | |
| Algol | | |
| HR 936 | | |
| HD 19356 | | |
| SAO 38592 | | |
| BD + 40°0673 | | |
| ADS 2362 | | |
| $\alpha = 3^h\ 04^m\ 54^s$ | | |
| $\delta = +40°46'0$ (1950.0) | | |
| B = 2.0 – 3.3 mag. | | |
| The first eclipsing variable to be discovered (actually a Triple System). | | |

TABLE 1

Previously published estimates of Systemic Velocity
a.k.a. Gamma Velocity = Vγ =Vgamma (km/s should always refer to whole system, ie., velocity of Algol AB-C's) CM with respect to our Sun

| $V_o$ | Source | Notes/Comments |
|---|---|---|
| Vγ (km/sec) = -9.0 (3.9) | Mukherjee et al (1996) p. 620 | |
| $V_o$ (kms $^{-1}$) = 3.8 ± 1.7 | Hill et al. (1971)<br>Eaton (1975)<br>Ebbighausen & Gange (1963)<br>Stein & Beardsley (1977) | Richards, M. T. (1993) p. 257, Table 1A |
| $V_o$ = 4.9 ± 1.6 km/sec | Hill et al. (1971) | Richards et al. (1993) p. 262 |
| 4 km/s | Gillet et al. (1989), p. 221 | |
| 4.8 km/s | This work | |

TABLE 2

Previously published estimates of Period of A-B System in Days

| P$_{A-B}$ | Source | Notes/Comments |
|---|---|---|
| Pa-b 2.867315 | S&T (1993), 86, 79 | Richards et al. (2003), p. 338 |
| Period in Days 2.87 | Soderhjelm (1980) | Drake (2003), p. 497, Table I |
| 2.8673 | Soderhjelm (1980) Lestrade (1993) | Borkovits et al. (2003) p.1096, Table 1 |
| 2.8673 | Zavala et al. (2002) | p. 456 (β per), Column 8 |
| Days P = 2.87$^d$ | Richards (2002) | p. 7 |
| β per – Period in Days 2.8673 | Mukherjee et al. (1996), p. 620, Table 2 | & Richards, M. T. (1993) p. 257 Hill et al. (1971) |
| Period Days 2.87 | Gimenez, A. (1996) p. 23 #2 The Case of Algol | |
| 2.8673d | Soderhjelm (1980) | Sarna MJ (1993) p. 1 (Intro.) |
| β per P = 2.87 days | Stern et al. (1998) p. 1166 (Intro.) | |
| P = 2.86 days | Mutel et al. (1998), p. 371 | |
| β per 2.7 < P < 4.5 P=days  P = 2.87 days | Richards, M. T. (2001), pp. 284, 295 | See chart: p. 284, (Table 1) Single Peaked Emissions |
| 2$^d$.8673 | Pan, X. et al. (1993) p. L129 | |
| 2.9 days | Lestrade et al. (1993), p. 808 | |
| 2.8673 | Lestrade et al. (1993), p. 809 Table 1B | |
| 2.87 days | Guinan, E. F. (1989) p. 37 | |
| 2.87 days | Sahade, J. & Hernandez, C.A. (1985), p. 257 | |
| 2.867 days | Cugier, H. & Molaro, P. (1984) p. 105 | |
| Binary Computed | (observed trend of period Variation) | |
| β Per  Po/P  Trend  0.87  Increase | Year of Observation (Kriner) 1971 1783-1967 | Chaubey, U.S. (1984) p. 56, Table 1 |
| 2$^d$.87 & 2.8673 p. 110 | Soderhjelm, S. (1980) | Original |
| 2$^d$.8674 days | Bachmann & Hershey (1975) | Cugier, H. (1979), p. 449 |
| 2.8673 days | Tomkin & Lambert (1978), PL 119 | Concluded the 2$^{nd}$ filled its Roche Lobe |
| 2$^d$.867 | Cugier & Chen (1977) p. 169 | |
| 2.86730807 | Hill et al. (1971) p.444 | |
| 2$^d$.8673 | Frieboes-Conde, H et al. (1970), p.78 | |
| 2.87 days | Ida Barry (1923) | Ebbighausen, E.G. (1958) |
| 2.87 days | Struve, O. & Sahade, J. (1957) p. 41 | |
| 2.867 day | Eggen, O.J. (1948) p. 1 | |
| 2.86731077 | Gillet et al. 1989, p. 221 | |

TABLE 3

Previously published estimates of Period of AB-C System in Days

| $P_{AB-C}$ | Source | Notes/Comments |
|---|---|---|
| P 679.9 days | Borkovits et al. (2003) p. 1096, Table 2 | |
| (P =1.87 yr.) AB & C distant A star in a long period | Mutel et al. (1998), p. 371 | |
| Orbital period 1.87 yr. | Gimenez, A. (1996) p. 23, #2 The case of Algol | |
| 1.862 yr. orbit | Sarna, M. J. (1993) p. 1 (Intro.) | |
| 679.9 ± 02 days | Soderhjelm, S. (1980) | Sarna, M. J. (1993) p. 535, Table 1 |
| P = days 680.08 | Hill et al. (1971) Ebbighausen (1958) Frieboes-Conde et al. (1970) | Richards, M. T. (1993) p. 257, Table 1-A |
| $P = 680^d.05 \pm 0^d.06$ | Pan, X. et al. (1993), p. L129 | |
| (AB-C) 1.86 yr. | Pan, X. et al. (1993), p. L129 | |
| 1.86 yr. | Lestrade et al. (1993) p. 808 | |
| 679.9 days | Lestrade et al. (1993) p. 809, Table 1-B | |
| 1.87 yrs. | Guinan, E. F. (1989), p. 37 | |
| 680 days | Gillet, D. et al. (1989), p. 219 | P. 110  6.79.9 ±0.2 |
| 1.862 yrs. | Sahade, J. & Hernandez, C.A. (1985), p. 257 | |
| 1.862 yrs. | Cugier, H. & Molaro, P. (1984) p. 105 | |
| 680 days | Soderhjelm, S. (1980) original | |
| 1.862 | Cugier, H. (1979) p. 550 | |
| 1.862 yrs. | Tomkins & Lambert (1978) p L 119 | |
| 1.862 yrs. | Bachmann & Hershey (1975) original | With an astrometric half amplitude of 0".01 |
| 1.862 yrs. | Hill et al. (1971) p. 443 | |
| 1.862 yrs. 1.873 (an earlier quote) | Ebbinghausen (1958) | Frieboes-Conde, H. et al. (1970) |
| 1.733 | Blopolsky (1906) (1908) (1909) (1911) | Ebbighausen, E. G. (1958), p. 598, Paragraph 1 |
| 1.899 | Curis's (1908) | |
| 1.874 | Schlesinger (1912) | |
| 1.873 | McLaughlin (1937) | |
| 1.873 | Eggen (1948) | |
| 1.873 | Pavel (1950) | |
| 1.87 yrs. | Struve, O. & Sahade, J. (1957) p. 41 | |
| 1.873 yrs, | Eggen, O. J. (1948), p. 1 | A fourth component star with a period of 188.4 years is suggested by Eggen |

TABLE 4

Previously published estimates of Eccentricity of A-B System

| e A-B | Source | Notes/Comments |
|---|---|---|
| e a-b 0.0 | Borkovits et al. (2003), p. 1096, Table 1 | |
| e a-b = 0 | Soderhjelm (1980) | Sarna, M.J. (1993) p. 535, Table 1 |
| e ab = 0.0 | Richards, M. T. (1993) p. 257, Table 1-A | |
| e ab = 0.000 | Lestrade et al. (1993) | p. 809, Table 1-B |
| e ab = 0.000 | Soderhjelm, S. (1980) | p. 110 |
| e ab = 0.26 | McLaughlin (1937) | Ebbighausen, E.G. (1958) p. 600, Table 2 |
| e ab = 0.47 | Eggen (1949) | Ebbighausen, E.G. (1958) p. 600, Table 2 |
| 0.015 ± .008 | Hill et al. (1971), p. 451 | |

TABLE 5

Previously published estimates of Eccentricity of AB-C System in (Days)

| e AB-C | Source | Notes/Comments |
|---|---|---|
| e 0.23 | Borkovits et al. (2003), p. 1096, Table 1 | |
| e AB-C = 0.22 ± 0.02 | Soderhjelm (1980) | Sarna, M. J. (1993) p. 535, Table 1 |
| e AB-C = 0.23 ± 0.04 | Richards et al. (1988) Ebbighausen (1958) Bachmann & Hershey (1975) | Richards, M. T. (1993), p. 257, Table 1-A |
| e ab-c = 0.225 ± 0.005 | Pan, X. et al. (1993), p. L129 | |
| e ab-c = 0.22 days | Lestrade et al. (1993) p. 809, Table 1-B | |
| e ab-c = 0.22 ± 0.02 | Soderhjelm, S. (1980), p. 110 | |
| e c = 0.211 ± 0.005 | Ebbighausen, E. G. (1958) p. 600, Table 2 | |
| e = 0.2557 ± 66.10$^{-4}$ | Pavel, F. (1949), p. 59 | |
| e = 0.25 | Eggen, O. J. (1948), p. 1 | |
| e = 0.23 | Hill et al. (1971), p. 450 | |

TABLE 6

Previously published estimates of longitude of periastron, ie. Angle between the line of nodes and periastron of A-B System (degrees)

| $\omega_{A-B}$ | Source | Notes/Comments |
|---|---|---|
| $\omega_{Algol\ AB} = 133°$ | Hill et al. (1971) | Richards, M. T. (1993) p. 257, Table 1-A |
| 163° | Gillet et al. (1989), p. 228 | |
| $\omega_{A-B} = 359°$, if $\dot{\omega}_{A-B} = 0$, for epoch of Hill et al. (1971) data | This work | |
| $\omega(t) = (.0308°/day)t - 521.392$ if $\dot{\omega}_{A-B} \approx \frac{360°}{32\ year}$ | This work | |

TABLE 7

Previously published estimates of Angle between line of nodes & periastron of AB-C system (degrees)

| $\omega_{AB-C}$ | Source | Notes/Comments |
|---|---|---|
| $\omega_{Algol\ C} = 313°$ | Hill et al. (1971) | Richards, M. T. (1993) p. 257, Table 1-A |
| $\omega^{(o)}$ AB-C $125 \pm 4$ | Soderhjelm, S. (1980) p. 110 | |
| 130.29(8) ° (outer binary) | Molnar & Mutel (1998), p. 17 | Possible Misquote of Pan et al. (1993) |
| $\omega_{Algol\ C} = 313°$ | Hill et al. (1971), Table 5, p. 450 | |
| $\omega_{AB-C} = 310°.29 + 0.08$ | Pan, X. et al. (1993), p. L129 | |
| Periastron$_{AB-C} = 125°$ | Lestrade et al. (1993) p. 809, Table 1-B | |
| $\omega_{AB-C}$ $\omega c = 133°$ for the epoch of Hill et al. (1971) data | This work | |
| $\omega_{AB} = 133°$ | Hill et al. (1971) Table 6, p. 451 | Appears to be in conflict with his $\omega_{AB}$ 313° entry |
| $\omega_{AB} = 313°$ | Hill et al. (1971) p. 450, middle of the page | Appears to be in conflict with his Table 6 entry |

TABLE 8

Previously published estimates of Rate in Change of $\omega_{AB}$ with time (degrees/day) (aka $d\omega_{AB}/dt$, or $\dot{\omega}$

| $\Delta\omega_{A-B}/\Delta t = \dot{\omega}$ | Source | Notes/Comments |
|---|---|---|
| $\Delta\omega_{A-B}/\Delta t = 0.03079$ Degrees/day or equivalently, 32 years for one revolution | Hill et al. (1971), p.454 | Estimated from Figure 4 (360 degrees per 32 years) |
| $\dot{\omega}_{A-B} = .03080$ for Hill date | This work | |

TABLE 9

Previously published estimates of semi-amplitude of A with respect to A-B system CM, $K_A$ (km/s)

| $K_A$ | Source | Notes/Comments |
|---|---|---|
| $K_A$ = 44 km/s | This work | |
| K (Kms$^{-1}$) 44 + 0.4 (Algol A) | Hill et al. (1971) | Richards, M. T. (1993), p. 257 |

TABLE 10

Previously published estimates of semi-amplitude of B with respect to A-B system CM, $K_B$ (km/s)

| $K_B$ | Source | Notes/Comments |
|---|---|---|
| K (Kms$^{-1}$) 201±6 (Algol B) | Tomkin, J. & Lambert, D. (1978) | Richards, M. T. (1993), p. 257 |
| 201 ± 6 km$^{-1}$ | Tomkin, J. & Lambert, D. (1978), p. L119 | |

TABLE 11

Previously published estimates of semi-amplitude of C with respect to AB-C system CM, $K_C$ (km/s)

| $K_C$ | Source | Notes/Comments |
|---|---|---|
| K (Kms$^{-1}$) 31.6 ± 2 (Algol C) | Hill et al. (1971), p. 450 | Richards, M.T. (1993), p. 257 |
| Kc = 33.6 ± 0 2 km/sec | Ebbighausen (1958), p. 600 | |

TABLE 12

Previously published estimates of semi-amplitude of AB with respect to AB-C ($K_{AB-C}$) system CM (km/s) – Should follow the equality: Mc/(Ma+Mb+Mc) = $K_{AB-C}$/Kc ~.27

| $K_{AB-C}$ | Source | Notes/Comments |
|---|---|---|
| K (Kms$^{-1}$)12.0 ± 0.4 (Algol AB) | Hill et al. (1971), p. 451 | Richards, M. T. (1993), p. 257 |
| Algol AB 9.7 km/s  10.6 km/s | McLaughlin (1937) Eggen (1948) | Ebbighausen, E. G. (1958), p. 600 |

TABLE 13

Previously published estimates of time of Periastron passage of A-B System – Time of maximum positive radial velocity

| $T_{A-B}$ | Source | Notes/Comments |
|---|---|---|
| 2,445,639.2146 days | Gillet et al., p.228 | p. 419, 1st paragraph |
| 2,428,482.7390 | Hill et al. (1971), p. 451 | |

TABLE 14

Previously published estimates of Periastron passage of AB-C System (day) – Time of positive radial velocity

| $T_{AB-C}$ | Source | Notes/Comments |
| --- | --- | --- |
| T =J D 2.446.931.4 ± 1.5 | Pan, X. et al. (1993), p. L129 | |
| Periastron Passage = 2,434,009 days | Lestrade et al (1993), p. 809 Table 1-B | |
| 2,434,022.8500 days | This work | Calculated from 1952.05, then adjusted by eye for Hill data |
| 1952.007 ± 0.015 | Ebbighausen (1958) | Hill et al. (1971), p. 450 Appears that error is .005 greater than Ebbighausen tabulated value. |
| 2421780.3 | Hill et al. (1971), p. 444 (footnote) | |
| 1952.05 | Hill et al. (197), p. 451 | His final solution |
| 1952.007 ± 0.010 | Ebbighausen (1958) p. 600, Table 2 | |

TABLE 16

Previously published estimates of light elements (a.k.a. ephemeris, or epoch of primary minimum) (usually in days or equivalently JD)

| Pr.Min | Source | Notes/Comments |
| --- | --- | --- |
| HJD 2,441,773.894 + $2^d.8673285$E | Kim et al. (1989), p. 1061 | |
| JD = 2440953.4657 + $2^d.8673075$E | Ashbrook (1976) | Al-Naimiy, H.M.K. et al, (1984), p. 1 |
| β per epoch = 2434705.5493 Period = 2.86732973 | Frieboes-Conde et al. (1970) | Nha, I.S. & Jeong, J. H. (1979), p. 82, Table II |
| JD 2378497.7588 + $2^d$ 86731077E | J. Hellerich (1919) (He derived the extension JD). | From Eggen, O. J. (1948), p. 2 |
| JD 2378497.756 + $2^d.87631085$E | K. Ferrare (1934) | |
| JD (Hel) 2441934.0796 Pd. of $2^d.86733$ | Chen, K-Y et al. (1977), p. 68 | |
| HJD 2445641.5135 | Al-Naimiy et al., 1984 | |

TABLE 16A

Previously published estimates of rate of mass loss dm/dT solar mass per year.

| $M_\odot$/yr | Source | Notes/Comments |
|---|---|---|
| $(10^{-11}-10^{-7})$ $M_\odot$/year$^{-1}$ | Richardson, M. T. (2001), p. 276 | (Classical Algols are typically in a slow stage of mass transfer) |
| $|\dot{M}|$ ($M_\odot$/year) $\geq 10^{-7}$ | Soderhjelm (1980) | Sarna, M.J. (1993) p. 535, Table 1 |
| $\dot{M} = 10^{-8}$ $M_\odot$/year | Cugier, H. (1982) p. 393 | |
| $\dot{M} \approx 10^{10}$ $M_\odot$/year$^{-1}$ | Harnden et al. (1977) | |
| $\dot{M} \approx 10^{14}$ $M_\odot$/year (for September 1989 data only) | This work | |
| $10^{-13}$ $M_\odot$/year$^{-1}$ | Cugier & Chen (1977) | |

TABLE 18

Previously published estimates of Full Width at Half Maximum (FWHM) Å. (Specify the ion and the lab wavelength in Å.

| FWHM | | Source | Notes/Comments | |
|---|---|---|---|---|
| Light Curve FWHR (Å) | Wavelength (Å) | | # of Datapts. | Comparison Star |
| 260 | 1920.2 | Eaton (1975) | 53 | ----- |
| 100 | 4350.0 | Wilson et al. (1972) | 700 | π Per |
| 180 | 4861.3 | Guinan et al. (1976) | 203 | γ Per |
| 900 | 5500.0 | Wilson et al. (1972) | 700 | π Per |
| 295 | 6562.8 | Guinan et al. (1976) | 214 | γ Per |
| 410 | 12000.0 | Chen & Reuning (1966) | 848 | κ Per |
| | | | Richards, M. T. et al. (1988), p. 328, Table II | |
| FWHM 1.60 Å at 1.24 Å for 1.10 Å for .86 Å for | Mg II Lines λ2795.5 Å λ2802.7 Å λ2795.5 Å λ2802.7 Å | Cugier, H. (1979), p.569 | | |

TABLE 19

Previously published estimates of radial velocity, Vrad (or Vr), km/s.  (Specify the ion and the Lab wavelength in Å and phase.)

| $V_{rad}$ | Source | Notes/Comments |
|---|---|---|
| Journal of radial velocities (huge) | Hill, G. et al. (1993) p. 581, Table 1 | |
| Table of Radial Velocity Si IV range from + 29 to + 4 (kms$^{-1}$) | Brandi, E. et al (1989) p. 332, Table I | |
| *See Table 2 – Radial Velocities for (β Persei) with ions | Sahade, J. & Hernandez, C.A. 1985) pp. 260-261 | |
| Radial Velocity of (C) was –13 km/sec (on Sept. 1956) | Ebbighausen, E. G. Predicted | Struve, O. & Sahade, J. (1957) p. 41 |

TABLE 20

Previously published estimates of the orbital inclination $i_{A-B}$.

| Inclination $i_{A-B}$ | Source | Notes/Comments |
|---|---|---|
| 80.2 | Chen and Reuning (1966) | Richards et al. (1988) |
| 81.6 ± 0.2 | Hill & Hutchings (1970) | p. 335, Table IV |
| 82.42 ± 0.04 | Wilson et al. (1972) | |
| 81.4 ± 0.1 | Grygar & Horak (1974) | |
| 82.4 | Eaton (1972) | |
| 81.5, 81.1 | Guinan et al. (1976) | |
| 81.5 ± 0.5 | Al-Naimiy & Budding (1977) | |
| 81.2 | Chen et al (1977) | |
| 81 ± 4 | Rudy & Kemp (1978) | |
| 82.3 ± 0.3 | Soderhjelm (1980) | |
| 81.4 ± 0.2 | Richards (1986) | |
| $i$ 82.3 degrees | Borkovits et al. (2003) p. 1096, Table I | Hill et al. (1971) |
| $i$ 82.31 degrees | Mukherjee et al. (1996), p. 620 | |
| $i$ 82.2 degrees | Palubek, G. (1998), p.22 See Moden description, p. 19 | |
| $i$[deg]81.4 ± 0.2 degrees | Richards et al. (1998) and Soderhjelm (1980) | Sarna, M.J. (1993) p. 535, Table I |
| 81.4 ± 0.2 degrees | Tomkin & Tan (1985) | Richards, M.T. (1993) p. 257, Table 1-A |
| 82.°.3 | Lestrade et al. (1993) p. 809, Table 1-B | |
| 81.2 | Koch, Plavec & Wood (1970) | Sahade J. & Hernandez, C.A. (1985), p. 257 |
| 82.3 ± 0.3 | Soderhjelm, S. (1980) p. 110 (original) | |
| $i$ = 57 | Frieboes-Conde et al. (1970) | Bonneau (1979) |

TABLE 21

Previously published estimates of the orbital inclination of the AB-C ($i_{AB-C}$).

| $i_{AB-C}$ | Source | Notes/Comments |
|---|---|---|
| $i$ [deg] 83 ± 2 degrees | Richards et al. (1988) and Soderhjelm (1980) | Sarna, M. J. (1993) p. 535, Table 1 |
| $i_{abc}$ 83 ± 3 degrees | Soderhjelm (1980) | Richards, M. T. (1993), p. 257 |
| $i_{abc}$ 83°.98 ± 0°.09 | Pan, X et al. (1993) p. L129 | |
| $i$ 83° | Lestrade et al. (1993) p. 809, Table 1-B | |
| $i$ 83 ± 2 | Soderhjelm, S. (1980), p. 110 | |
| $i = 79$ | Soderhjelm, S. (1975) | Bonneau, D. (1979), p. L11 |
| ab-c $i = 78.5° ± 3.0°$ | Bonneau, D. (1979), p. L12 | |

TABLE 22

Previously published estimates of $V_{rot}$ Sin $i$ (Rotational Velocity).

| V sin $i$ | Source | Notes/Comments |
|---|---|---|
| β per Veq sin $i$ = 51.9 ± 1.6<br>V syn sin i = 49.9 ± 0.5 | Mukherjee et al. (1996) p. 623, Table 4 | See Table 4 for Rotation Values for various methods |
| V sin $i$ (kms$^{-1}$)<br>  Algol A = 53 ± 3<br>  Nothing for B<br>  Algol C = 46 ± 10 | Richards, M. T. (1993) p. 257 | Also cites:<br>  Rucinski (1979)<br>  Tomkin & Tan (1985) |
| Algol A = 50 ±5 kms$^{-1}$<br>        55 ± 4 kms$^{-1}$<br>synchronous values is = 51 kms$^{-1}$ | Heuvel (1970)<br>Hill et al. (1971)<br>Soderhjelm (1980), p. 103 | Soderhjelm, S. (1980), p. 103 |

TABLE 23

Previously published estimates of a sin *i* (km)

| a sin *i* | Source | Notes/Comments |
|---|---|---|
| a sin *i* (kms)<br>A = 1.73 x $10^6$<br>B = 7.93 x $10^6$<br>C = 2.88 x $10^8$<br>AB = 1.09 x $10^8$ | Hill et al. (1971)<br>Tomkin & Lambert (1978)<br>Hill et al. (1971)<br>Hill et al. (1971) | Richards, M. T. (1993),<br>p. 257 |
| a sin *i*<br>AB = 88 x $10^6$ km<br>AB = 87 x $10^6$ km<br>C = 307 x $10^6$ km | McLaughlin (1937)<br>Eggen (1948)<br>Ebbighausen (1958) | Ebbighausen, E. G. (1958)<br>p. 600, Table 2 |
| a sin *i* = 0.13867 ±<br>71.$10^{-5}$ = 24 AE | Pavel, F. (1949), p. 59 | |
| a sin *i* = 37 x $10^8$ km | Eggen, O. J. (1948), p. 1 | |

TABLE 24

Previously published estimates of mass of Algol A ($M_A$)

| $M_A$ | Source | Notes/Comments |
|---|---|---|
| M   3.7<br>      3.6<br>      3.8 | Hill et al. (1971)<br>Soderhjelm (1980)<br>Kim et al. (1989)<br>Present Study | Kim et al. (1989)<br>p. 1067, Table 4 |
| ($M_1$)  3.7 | Soderhjelm (1980)<br>Lestrade et al. (1993) | Borkovits et al. (2003)<br>p. 1096, Table 1 |
| $M_\odot$ 3.7 ± 0.30 | Kim (1989)<br>Richard et al. (1988)<br>Tomkin & Lambert (1978) | Sarna, M. J. (1993)<br>p. 535, Table 1 |
| M ($M_\odot$) 3.7 ± 0.3 | Tomkin & Lambert (1978) | Richards, M. T. (1993)<br>p. 257, 1-B |
| $M_{AB}$ = 3.98 ± 0.38 $M_\odot$ | Pan, X. et al. (1993)<br>p. L131, Table 2 | Mass of A+B Together |
| $M_A$ = 3.6 $M_\odot$ | Soderhjelm (1980) &<br>Kim (1989) | Lestrade et al. (1993) |
| $M_A$ = 3.7 $M_\odot$ | Guinan, E. F. (1989), p. 37 | |
| $M_A$ = 3.7 ± 0.2 $M_\odot$ | Soderhjelm (1980), p. 100 | |
| $M_A$ = 3.6 ± 0.1 $M_\odot$ | Soderhjelm (1980), p. 101 | After adjustments |
| $M_A$ = 3.7 ± 0.3 | Tomkin & Lambert (1978)<br>p. L119 | |
| Mass of AB 5.3 $M_\odot$ | Bachmann & Hershey (1975)<br>p. 836 | Combining elements independent<br>of the parallax |
| 3.7 $M_\odot$ | Hill et al. (1971) 459 | |
| Mass 1 = 5.8 | Eggen, O. J. (1948), p.1 | |

TABLE 25

Previously published estimates of mass of Algol B ($M_B$)

| $M_B$ | Source | Notes/Comments |
|---|---|---|
| 0.8 | Hill et al. (1971) | Kim et al. (1989) p. 1067, Table 4 |
| 0.79 0.82 | Soderhjelm (1980) (Present Study) Kim et al. (1989) | Kim et al. (1989) p. 1067, Table 4 |
| ($M_2$) 0.8 | Soderhjelm (1980) Lestrade et al. (1993) | Borkovits et al. (2003) p. 1096, Table 1 |
| $M_\odot$ 0.81 ± 0.05 | Kim (1989) Richards et al. (1988) Tomkin & Lambert (1978) | Sarna, M. J. (1993) p. 535, Table 1 |
| M ($M_\odot$) 0.81 ± 0.05 | Tomkin & Lambert (1978) | Richards, M. T. (1993) p. 257, 1-B |
| $M_B$ = 0.79 $M_\odot$ | Soderhjelm (1980); Kim (1989) | Lestrade et al. (1993) p. 809, Table 1 |
| $M_B$ = 0.8 $M_\odot$ | Guinan, E. F. (1989), p. 37 | |
| $M_B$ = 0.80 ± 0.03 $M_\odot$ | Soderhjelm, S. (1980), p. 100 | To be corrected by him |
| $M_B$ = 0.7980 ± 0.01 ($M_\odot$) | Soderhjelm, S. (1980), p.101 | |
| $M_B$ = 0.81 ± 0.05 | Tomkin & Lambert (1978) p. L119 | |
| 0.8 $M_\odot$ | Hill et al. (1971), p. 459 | |
| Mass 2 = 1.0$_\odot$ | Eggen, O. J. (1948), p.1 | |

TABLE 26

Previously published estimates of mass of Algol C ($M_C$)

| $M_C$ | Source | Notes/Comments |
|---|---|---|
| 1.7<br>1.6<br>1.8 | Hill et al. (1971)<br>Soderhjelm (1980)<br>Kim et al. (1989)<br>Present Study | Kim et al. (1989)<br>p. 1067, Table 4 |
| $M_\odot$ .8 | Richards et al. (1988)<br>Kim (1989) | Drake (2003)<br>p. 497, Table 1 |
| $M_\odot$ 1.70 ± 0.2 | Kim (1989)<br>Richards et al. (1988);<br>Tomkin & Lambert (1978) | Sarna, J. J. (1993)<br>p. 535, Table 1 |
| M ($M_\odot$) 1.6 ± 0.1 | Hill et al. (1971)<br>Tomkin & Lambert (1978)<br>Soderhjelm (1980) | Richards, M. T. (1993)<br>p. 257, 1-B |
| M($M_\odot$) 1.5 ± 0.11 $M_\odot$ | Pan, X. et al (1993) p. L129 | |
| $M_c$ = 1.6 $M_\odot$ | Soderhjelm (1980) &<br>Kim (1989) | Lestrade et al. (1993)<br>p. 809, Table 1 |
| $M_c$ = 1.6 ± 0.1 $M_\odot$ | Soderhjelm (1980), p. 101 | |
| $M_c$ ($M_\odot$) = 1.6 ± 0.2 | Bonneau, D. (1979), p. L12 | |
| $M_c$ = 1.7 ± 0.2 $M_\odot$ | Tomkin & Lambert (1978)<br>p. L119 | |
| Mass C 1.8 $M_\odot$<br><br>1.7 $M_\odot$ | Bachman & Hershey (1975)<br>p. 836<br>Hill et al. (1971), p. 459 | |
| Mass 3 = 1.2$_\odot$<br>Mass 4 = 3.8$_\odot$ | Eggen, O. J. (1948), p.1 | |

TABLE 27

Previously published estimates of Radius ($R_\odot$) A.

| $R_A$ | Source | Notes/Comments |
|---|---|---|
| 3.0<br>3.22<br>2.89<br>2.88 | Hill et al. (1971)<br>Eaton (1975)<br>Soderhjelm (1980)<br>Kim et al. (1989) Present study | Kim et al. (1989)<br>P. 1067, Table 4 |
| Stellar $R_\odot$ 2.9, 3.5 | Richards (1993) | Richards (2003)<br>Chart 338 |
| $R_1/A = .204$ | Mukherjee et al. (1996)<br>p. 620, Table 2 | |
| $R_\odot = 2.90 \pm 0.04$ | Kim (1989) and<br>Richards et al. (1988) | Sarna, M. J. (1993)<br>p. 535, Table 1 |
| $R(R_\odot)$ $2.9 \pm 0.04$ | Tomkin & Tan (1985) | Richards, M. T. (1993)<br>p. 257, 1-B |
| $R_A$ 2.89 $R_\odot$ | Soderhjelm (1980) &<br>Kim (1989) | Lestrade et al. (1993), p. 809 |
| Unperturbed fractional<br>Radius $R_A = .188$ | Gillet, D. et al. (1989)<br>p. 222, Table 3 | |
| RA ($R_\odot$) $2.90 \pm 0.04$ Roche<br>3.08 Roche<br>$3.2 \pm 0.2$ Russell-Merrile<br>3.0 Roche<br>2.89 Roche | Richards (1986)<br>Wilson et al (1972)<br>Eaton (1975)<br>Hill (1971)<br>Soderhjelm (1980) | Richards, M. T. et al. (1988)<br>p. 335, Table VIII |
| $2.89 \pm 0.04$ | Soderhjelm (1980)<br>p. 110, Table 9 | |
| Radii of Algol A = 3.2 | Eaton (1975) | Cugier & Chen (1977), p. 169 |
| 3.0 $R_\odot$ | Hill et al. (1971), p. 459 | |
| Semiaxes of Algol A<br>$a_A$ .188<br>$b_A$ .188<br>$c_A$ .187 | Gillet, D. (1989)<br>p. 222, Table 3 | |
| $R_A = r_s$ 0.201 - 0.209 at<br>3428 Å | Chen, K. –Y. et al. (1977)<br>Table III | In terms of fraction of separation |
| $R_A = r_s$ 0.205 at<br>3700 Å | Herczeg (1959) | Chen, K.-Y. et al. (1977)<br>Table III ( in terms of fraction of separation.) |
| $R_A = r_s$ 0.224 at<br>2980 Å | Eaton (1975) | Chen, K.-Y. et al (1977)<br>Table III (in terms of fraction of separation) |

TABLE 28

Previously published estimates of Radius ($R_\odot$) B.

| $R_B$ | Source | Notes/Comments |
|---|---|---|
| 3.4 | Hill et al. (1971) | Kim et al. (1989) |
| 3.57 | Eaton (1975) | P. 1067, Table 4 |
| 3.53 | Soderhjelm (1980) | |
| 3.54 | | |
| | Kim et al. (1989), Present Study | |
| 3.5 | Richards (1993) | Ness et al. (2002), p. 913 |
| $R_\odot = 3.50 \pm 0.10$ | Kim (1989) and Richards et al. (1988) | Sarna, M. J. (1993) p. 535, Table 1 |
| $R(R_\odot)$ 3.5 ± 0.1 | Tomkin & Tan (1985) | Richards, M. T. (1993) p. 257, 1-B |
| $R_B (R_\odot)$ 3.4 | Credited to Soderhjelm (1980) & Kim (1989) | Lestrade et al. (1993), p. 809 |
| $R_B$ 3.53 $R_\odot$ | Soderhjelm (1980) and Kim (1989) | Lestrade et al. (1993) p. 809, Table 1-A |
| $R_B (R_\odot)$ 3.4 Roche<br>3.6 ± 0.2 Russell-Merrille<br>3.23 Roche<br>3.53 Roche<br>3.5 ±0.1 Roche | Hill et al. (1971)<br>Eaton (1975)<br>Wilson et al (1972)<br>Soderhjelm (1980)<br>Richards (1986) | Richards, M. T. et al. (1988) p. 335, Table VIII |
| 3.5 ± 0.04 | Soderhjelm (1980) p. 110, Table 9 | |
| 3.6 | Eaton (1975) | Cugier & Chen (1977) p.169 |
| 3.4* | Hill et al. (1971), p. 459 | *derived from the polars values R & TagTe |
| Semiaxes of Algol B<br>$a_B$.291<br>$b_B$.258<br>$c_B$.244 | Gillet, D. (1989) p. 222, Table 3 | |
| $R_B = r_g$ 0.248-0.253 at 3428 Å | Chen, K. –Y. et al. (1977) p. 73, Table III | |
| $R_B = r_g$ 0.262 at 3700 Å<br>$R_B = r_g$ 0.244 at 2980 Å | Herczeg (1959)<br>Eaton (1975) | Chen, K.-Y. et al. (1977) p. 73, Table III (in terms of fraction of separation.) |

TABLE 29

Previously published estimates of Radius ($R_\odot$) C

| $R_C$ | Source | Notes/Comments |
|---|---|---|
| 1.5 | Hill et al. (1971) | Kim et al. (1989) |
| 1.56 | Eaton (1975) | P. 1067, Table 4 |
| 1.5 | Soderhjelm (1980) | |
| 1.7 | Kim et al. (1989), Present Study | |
| $R_\odot = 1.60 \pm 0.2$ | Kim (1989) and Richards et al. (1988) | Sarna, M. J. (1993) p. 535, Table 1 |
| $R(R_\odot)$ 1.4 ± 0.1 | Tomkin & Tan (1985) | Richards, M. T. (1993) p. 257, 1-B |
| $R_C$ 1.5 $R_\odot$ | Soderhjelm (1980) & Kim (1989) | Lestrade et al. (1993), p. 809 |
| 1.5 ± 0.1 $R_\odot$ | Soderhjelm, S. (1980) p. 110, Table 9 | |
| 1.5 $R_\odot$ | Hill et al. (1971), p. 459 | |

TABLE 30

Previously published estimates of Temp $T_A$

| $T_A$ | Source | Notes/Comments |
|---|---|---|
| $T_{eff}^{pol}$ (Algol A) = 1300 K | Polubek, G. (1998), p. 22 | See Figure 2, p. 21<br>See model description, p. 19 |
| Ti(K) = 12000 K | Mukherjee et al. (1996)<br>p. 620, Table 2 | See Table 2, p. 620<br>For other comparison |
| $T_{eff}$ (K) 12500 ± 500 | Eaton (1975)<br>Kim (1989)<br>Richards (1988) | Sarna, M. J.<br>p. 535, Table 1 |
| T (K) 13,000 ± 500 | Cugier & Molaro (1983) | Richards, M. T. (1993)<br>p. 257, 1-B |
| $T_A$ (eq) 12,500<br>$T_A$ (Pol) 12550 | Guinan et al. (1976) using<br>Wood's (1972) WINK computer<br>model | Gillet D. et al. (1989)<br>p. 222, Table 3 |
| $T_A$ 12500 K | Flower (1977)<br>Bohm-Vetense (1981) | Richards, M. T. et al. (1988)<br>p. 328, Table 1 |
| T = $10^4$ K | Cugier, H. (1992), p. 393 | |
| $T_{eff}$ (K) 12,500 ± 500 | Soderhjelm, S. (1980)<br>p. 110, Table 9 | |
| $T_{eff}$ 12000 K | Cugier, H. (1979), p. 549 | |
| Te 12,000 K | Chen, K.-Y. et al. (1977), p. 71 | |
| Te 12,000 K | Kurucz (1969)<br>Carbon & Gingerich (1969) | Chen, K.-Y. et al. (1977), p. 72 |
| log Te 4.03 | Hill et al. (1971), p. 459 | |

TABLE 31

Previously published estimates of Temp $T_B$

| $T_B$ | Source | Notes/Comments |
|---|---|---|
| (K) 4500 ± 300<br>B (Algol B) | Richards et al. (1988)<br>Eaton (1975)<br>Kim (1989) | |
| $T_{eff}^{pole}$ (Algol B) = 5000 K | Polubek, G. (1998) | See Figure 2, p. 21<br>See model description, p. 19 |
| $T_2$(K) = 4888 K | Mukherjee et al. (1996)<br>p. 620, Table 2 | |
| $T_{eff}$ (K) 4500 ± 300 | Eaton (1975)<br>Kim (1989)<br>Richards et al. (1988) | Sarna, M. J.<br>p. 535, Table 1 |
| T (K) 4500 ± 300 | Hill et al. (1971) | Richards, M. T. (1993)<br>p. 257, 1-B |
| $T_B$ (eq) 5330<br>$T_B$ (Pol) 5590 | Guinan et al. (1976) using<br>Wood's (1972) WINK computer<br>model | Gillet, D. et al. (1989)<br>p. 222, Table 3 |
| $T_B$ 5000 K | Soderhjelm (1980) | Richards, M. T. et al. (1988)<br>p. 328, Table 1 |
| $T_B$(K) 5080<br>4700<br>5330 ± 180<br>4600 ± 800<br>5300<br>5000<br>5250 ± 250<br>4500 ± 300<br>5000 ± 500 | Chen & Reuning (1966)<br>Hill & Hutchings (1970)<br>Wilson et al. (1972)<br>Eaton (1975)<br>Al-Naimiy & Budding (1977)<br>Chen et al. (1977)<br>Murad & Budding (1984)<br>Richards (1986)<br>Soderhjelm (1980) | Richards, M. T. et al. (1988)<br>p. 336, Table IX |
| 5000 ± 500 K | Soderhjelm (1980)<br>p. 110, Table 9<br>primary (original) | |
| log Te 3.66 | Hill et al. (1971)<br>p. 459 | |

TABLE 32

Previously published estimates of Temp $T_C$

| $T_C$ | Source | Notes/Comments |
|---|---|---|
| $T_{eff}^{pole}$ (Algol C) = 7000 K | Polubek, G. (1998) p. 22 | See Figure 2, p. 21 See model description, p. 19 |
| $T_{eff}$ (K) 8000 ± 700 | Eaton (1975) Kim (1989) Richards et al. (1988) | Sarna, M. J. p. 535, Table 1 |
| T (K) 7000 ± 200 | Richards, M. T. (1993) p. 257, 1-B | |
| $T_{eff}$ (K) 8400 ± 500 | Soderhjelm, S. (1980) p. 110, Table 9 | |
| 8500 K | Chen et al. (1977) p. 71 | |
| 8500 K | Kurucz (1969) & Carbon & Gingerich (1969) | Chen, K.-Y. et al. (1977), p. 72 |
| log Te 3.92 | Hill et al. (1971) p. 459 | |

TABLE 33

Previously published estimates of log of surface gravity, log g of Algol A

| Log $g_A$ | Source | Notes/Comments |
|---|---|---|
| (log) 4.03 4.11 4.10 4.08 | Hill et al. (1971) Eaton (1975) Soderhjelm (1980) Kim et al. (1989) | Kim et al. (1989) p. 1067, Table 4 |
| log g  4.08 | Richards et al. (1988) Tomkin & Lambert (1978) | Richards, M. T. (1993) p. 257, Table 1-B |
| log $g_A$ = 4.0 | Flower (1977) Bohm-Vilense (1981) | Richards, M. T. et al. (1988) p. 328, Table 1 |
| log $g_A$ = 4.0 | Chen, K.-Y. et al. (1977) p. 71 | |
| log $g_A$ = 4.0 | Kurucz (1969) Carbon & Gingerich (1969) | Chen, K.-Y. et al. (1977) |

TABLE 34

Previously published estimates of log of surface gravity, log g of Algol B

| Log g$_B$ | | Source | Notes/Comments |
|---|---|---|---|
| (log) | 3.66 | Hill et al. (1971) | Kim et al. (1989) |
| | 3.66 | Eaton (1975) | p. 1067, Table 4 |
| | 3.70 | Soderhjelm | |
| | 3.69 | Kim et al. (1989) | |
| Log g | 3.2 | Tomkin & Tan (1985) | Richards, M. T. (1993) |
| | | Tomkin & Lambert (1978) | p. 257, Table 1-B |
| Log g $_B$ = 3.0 | | Soderhjelm (1980) | Richards, M. T. et al. (1988) |
| | | | p. 328, Table 1 |
| Log g $_B$ = 3.0 | | Chen, K.-Y. et al. (1977) | |
| | | p. 71 | |

TABLE 35

Previously published estimates of log of surface gravity, log g of Algol C

| Log g$_C$ | | Source | Notes/Comments |
|---|---|---|---|
| (log) | 3.92 | Hill et al. (1971) | Kim et al. (1989) |
| | 3.94 | Eaton (1975) | p. 1067, Table 4 |
| | 3.92 | Soderhjelm (1980) | |
| | 3.93 | Kim et al. (1989) | |
| Log g | 4.4 | Tomkin & Tan (1985) | Richards, M. T. (1993) |
| | | | p. 257, Table 1-B |
| Log g | 4.0 | Chen, K.-Y. et al. (1977) | |
| | | p. 71 | |
| Log g | 4.0 | Kurucz (1969) | Chen, K.-Y. et al. (1977) |
| | | Carbon & Gingerich (1969) | p. 72 |

TABLE 36

Previously published estimates of distance to Algol AB-C system (parasecs)

| Distance, d | Source | Notes/Comments |
|---|---|---|
| pc 28.5 ± 0.8 | ESA (1997) | Richards et al. (2003) p. 338, Chart |
| pc 29.0 | Perryman et al. (1997) | Drake (2003) p. 497 (chart), Table 1 |
| d/pc 28.0 | Perryman et al. (1977) | Ness et al (2002) p. 913 |
| (31 ± 3) pc | Richards cites Labeyrie et al. (1974) | Richards et al. (1996), p. 249 |
| d (pc) 31 ± 3 | Bachmann & Hershey (1975) Labeyrie et al. (1974) | Richards, M. T. (1993) p. 257, Table 1-B |
| d (pc) 28.2 ± 0.8 pc | Pan, X. et al. (1993) p. L129 | |
| ~ 25 pc | Gillet, D. et al. (1989), p. 219 | |
| 30 pc ~ cf | Cugier, H. (1979), p. 550 | |

TABLE 37

Previously published estimates of spectral Type Algol A

| Spectral Type | Source | Notes/Comments |
|---|---|---|
| B8 | Kim et al. (1989), p. 1062 | |
| B8 V + K2IV = 2.8673 | Richards, M. T. (2001) p. 287, Table 2 | |
| B8 V | Mutel et al. (1998), p. 371 | |
| B8 V | Richards cites Batten (1989) Richards (1993) Antunes, Nagase & White (1994) Ottman (1994) Morgan (1935) | Richards et al. (1996), p. 249 Richards, M. T. (1993) p. 257, Table 1 |
| B8 V | Gimenez, A. (1996), p. 23; #2 Case of Algol | |
| Algol A B8 V | Sarna, M. J. (1993) p. 1 (intro) | |
| B8 V | Pan, X. et al. (1993), p. L129 | |
| B8 V | Lestrade et al. (1993), p. 808 | |
| B8 V | Guinan, E. F. (1989), p. 37 | |
| B8 V | Gillet, D. et al. (1989), p. 219 | |
| B8 V | Sahade, J. & Hernandez, C.A. (1985), p. 257 | |
| B8 V | Cugier, H. & Molaro, P. (1985) p. 105 | |
| B8 V | Soderhjelm, S. (1980), p. 100 | |
| B8 V | Cugier, H. (1979), p. 549 | |
| B8 V | Tomkin & Lambert (1978) p. 119 | |
| B8 V | Cugier & Chen (1977), p. 169 | |
| B8 V | Frieboes-Conde, H. et al. (1970) p. 78 | |
| B8 | Struve, O. & Sahade, J. (1957) p. 41 | |

TABLE 38

Previously published estimates of spectral Type Algol B

| Spectral Type | Source | Notes/Comments |
|---|---|---|
| K0 | Kim et al. (1989), p. 1062 | |
| SecondaySpec Type K4 | Zavala et al. (2002), p. 456 | |
| K2 IV | Ness et al. (2002) | |
| K2IV | Mutel et al. (1998), p. 371 | |
| K2 subgiant | Batten (1989) Richards (1993) Antunes, Nagase & White (1994) Ottman (1994) | Richards et al. (1996), p. 249 |
| K0 – 2IV | Gimenez, A. (1996), p. 23 #2 The Case of Algol | |
| K2IV | Richards, M. T. (1993) p. 257, Table 1 | |
| KIV | Pan, X. et al. (1993), p. L129 | |
| K0 – 3IV | Lestrade et al. (1993), p. 808 | |
| K2-3 IV | Guinan, E. F. (1989), p. 37 | |
| Type G5 – K0IV | Gillet, D. et al. (1989), p. 219 | |
| Late B or early K Subgiant | Tomkin & Lambert (1978) | Sahade, J. & Hernandez, C. A. (1985), p. 257 |
| Late G or K type star | Cugier, H. & Molaro, P. (1985) p. 105 | |
| G-type of sub-giant | Soderhjelm, S. (1980), p. 100 | |
| Type K star | Cugier, H. (1979), p. 549 | |
| (B) Cool mass subgiant 1.2 mag deep on V | Tomkin & Lambert (1978) p. L119 | |
| G-K range | Cugier & Chens (1977), p. 169 | |
| Late type G or Early K | Frieboes-Conde, H. et al. (1970) | |

TABLE 39

Previously published estimates of spectral Type Algol C

| Spectral Type | Source | Notes/Comments |
|---|---|---|
| A7M[a] | Kim et al. (1989), p. 1062 | |
| F2V Tertiary | Batten (1989)<br>Richards (1993)<br>Antunes, Nagase & White (1994)<br>Ottman (1994) | Richards et al. (1996), p. 249 |
| A9 / FO V | Gimenez, A. (1996), p. 23<br>#2 Case of Algol | |
| Late A or Early F<br>Luminosity class V | Sarna, M. J. (1993)<br>p. 1 (intro.) | |
| F1 V | Richards et al. (1988) | Richards, M. T. (1993)<br>p. 257, Table 1 |
| A7m | Lestrade et al. (1993), 808 | |
| A5-9 | Guinan, E. F. (1989), p. 37 | |
| AM type star | Gillet, D. et al. (1989), p. 219 | |
| AM star | Huang (1957) | Sahade, J. & Hernandez, C.A.<br>(1985), p. 257 |
| Late spectral A | Cugier, H. & Molaro, P. (1984)<br>p. 105 | |
| Late spectral A | Cugier, H. (1979), p. 550 | |
| Late (A or F) luminosity<br>Class IV or V | Struve & Sahade (1957) | Tomkin & Lambert (1978)<br>p. L119 |
| A – metallic star | Fletcher (1964) | Frieboes-Conde, H. et al. (1970) |
| Metallic line – A star | Struve, O. & Sahade, J. (1957)<br>p. 41 | |

TABLE 40

Previously published estimates of mass ratio of Algol A-B system q= mb/ma

| $q = M_B/M_A$ | Source | Notes/Comments |
|---|---|---|
| Superhumps $q = M_{donor}/M_{receiver} = 0.29 \pm 0.01$ Assuming that the observed peak out ~ 3 days is a true superhump, the mass ratio of the binary system of β per can be estimated from the superhump period excess over the orbital pd. about (6%, §§ 1-2) | Osaki (1985) | Retter et al. (2005), p. 423 (3.2) |
| Permitted range for superhumps Binary mass ratio of β per $q = 0.217 \pm 0.03$ $q = 0.22 \pm 0.03$ | Hill et al. (1988) Tomkin & Lambert (1978) Richards et al. (1988) | Retter et al. (2005), p. 423 (3.2) |
| $q = M_s/M_p$ 0.22 | Richards et al. (2003) p. 338, Chart | |
| $q = M_2/M_1$ 0.227 | Mukherjee et al. (1996) p. 620, Table 2 | |
| $q = 0.22$ | Gimenez, A. (1996), p. 23 Last paragraph | |
| Mass ratio $M_A/M_B$ = ~ 4.5 days | Lestrade et al. (1993), 809 Paragraph 1 | |
| $q = .217$ | Gillet, D. et al. (1989) p. 222, Table 3 | |
| $q = .22$ | Tomkin & Lambert (1978) | Richards, M. T. et al. (1988) p. 328, Table 1 |
| Mass ratio $M_A/M_B = 4.6 \pm 0.1$ | Tomkin & Lambert (1978) | Soderhjelm, S. (1980) |
| Mass ratio $M_A/M_B = 4.6 \pm 0.1$ | Tomkin & Lambert (1978) p. L119 abstract their original | |
| Mass ratio $M_A/M_B = 4.6$ | Hill et al. (1971) | Cugier & Chen (1977) p. 169 |
| Mass ratio 4.6 | Hill et al. (1971), p. 443 original | |

TABLE 40 (A)

Previously published estimates of mass ratio of AB-C system $q = M_{AB}/M_C$

| $q = M_{AB}/M_C$ | Source | Notes/Comments |
|---|---|---|
| $M_{AB}/M_C = 2.63 \pm 0.20$ | Hill et al. (1971), p. 443 | |

TABLE 41

Previously published estimates of separation ($R_\odot$) of center of A to center of B
Sep. ab ($R_\odot$)

| sep. AB | Source | Notes/Comments |
|---|---|---|
| Sep. radius $R_\odot$ 14.0 | Richards et al. (2003) p. 338, Chart | |
| Separation ($10^6$ km) | | |
| ~ 0.03 binary separation | Bonneau (1979) | Richards et al. (1996), p. 249 |
| 14 $R_\odot$ | Gillet, D. et al. (1989), p. 219 | |
| 1 x $10^{12}$ CM = approx. 14 $R_\odot$ | Cugier & Chen (1977), p. 169 | |
| A/$R_\odot$ 13.99 | Mukherjee et al. (1996) p. 620, Table 2 | |
| 14.03 $R_\odot$ | Chung et al. (2004) p. 1193, Figure 10 | |

TABLE 42

Previously published estimates of separation of center of AB system to center of C
(astronomical units AU)

| sep. AB-C | Source | Notes/Comments |
|---|---|---|
| a = 2.67 ± 0.08 AU | Pan, X. et al. (1993) p. L131, Table 2 | |
| a = 2.71 au ± 0.003 au | Bonneau, D. (1979), p. L12 | |

TABLE 43

Previously published estimates of $\Omega$ omega A-B system

| $\Omega_{A-B}$ | Source | Notes/Comments |
|---|---|---|
| $\Omega$ 52 | Soderhjelm (1980) Lestrade et al. (1993) | Borkovits et al. (2003) p. 1096, Table 1 |
| $\Omega 1$ = 5.151 | Mukherjee et al. (1996) p. 620, Table 2 | |

TABLE 43 (A)

Previously published estimates of Ω omega of AB-C system

| $\Omega_{AB-C}$ | Source | Notes/Comments |
|---|---|---|
| $\Omega_2 = 2.299$ | Mukherjee et al. (1996) p. 620, Table 2 | |
| $\Omega = 312°.26 \pm 0.13$ | Pan, X. et al (1993) p. L129 | |
| $\Omega = 313° \pm 5°$ | Bonneau, D. (1979), p. L12 | |
| $Node_{AB-C} = 132°$ | Lestrade et al. (1993) p. 809, Table 1-B | |

TABLE 44

Previously published estimates of mass function of AB system f(m) AB

| $f(m)_{A-B}$ | Source | Notes/Comments |
|---|---|---|
| f(m) (M$_\odot$)<br><br>Algol A  $0.0254 \pm 0.0008$<br>     B  $2.42 \pm 0.22$<br>     C  $2.05 \pm 0.20$ | Hill et al. (1971) | Richards, M. T. (1993) p. 257, Table 1-A |

TABLE 45

Previously published estimates of rotational velocity ($V_{rot}$) (km/s)

| $V_{rot}$ | Source | Notes/Comments |
|---|---|---|
| Vrot 55 kms$^{-1}$ | Cugier, H. & Molaro, P. (1984) p. 111 in discussion | |
| Rotational Velocity<br>Algol A = $55 \pm 4$ kms$^{-1}$ | Hill et al. (1971), p. 443 | |

TABLE 46

Previously published estimates of Roche Lobe (fill out factor) $F_A$ or ratio of Algol A

| $F_A$ | Source | Notes/Comments |
|---|---|---|
| $F_A$ 0.5 | Richards, M. T. et al. (1988) p. 328, Table 1 | |

TABLE 47

Previously published estimates of Roche Lobe (fill out factor) $F_B$ or ratio of Algol B

| $F_B$ | Source | Notes/Comments |
|---|---|---|
| $F_B$ 1.0 | Richards, M. T. et al. (1988) p. 328, Table 1 | |

TABLE 48

Previously published estimates of $\beta_A$, Beta gravity darkening exponent or coefficient for Algol A

| $\beta_A$ | Source | Notes/Comments |
|---|---|---|
| $\beta_A = .25$ | Guinan et al. (1976) using Wood's 1972 WINK computer model | Gillet, D. et al. (1989) p. 222, Table 3 |
| $\beta_A = 0.25$ | Von Zeipel (1924); Lucy (1967) | Richards, M. T. et al. (1988) p. 328, Table 1 |

TABLE 49

Previously published estimates of $\beta_B$, Beta gravity darkening exponent or coefficient for Algol B

| $\beta_B$ | Source | Notes/Comments |
|---|---|---|
| $\beta_B = .25$ | Guinan et al. (1976) using Wood's 1972 WINK computer model | Gillet, D. et al. (1989) p. 222, Table 3 |
| $\beta_B = 0.08$ | Lucy (1967) | Richards, M. T. et al. (1988) p. 328, Table 1 |

TABLE 50

Previously published estimates of bolometric Albedo (or reflection coefficient) for Algol A

| $A_A$ | Source | Notes/Comments |
|---|---|---|
| $A_A = 1.0$ | Rucinski (1969) Milne (1926) | Richards, M. T. et al. (1988) p. 328, Table 1 |
| $W_A = 1.0$ | Guinan et al. (1976) using Wood's 1972 WINK computer model | Gillet, D. (1989) p. 222, Table 3 |

TABLE 51

Previously published estimates of bolometric Albedo (or reflection coefficient) for Algol B

| $A_B$ | Source | Notes/Comments |
|---|---|---|
| $A_B = 0.5$ | Rucinski (1969) Milne (1926) | Richards, M. T. et al. (1988) p. 328, Table 1 |
| $W_B = 0.4$ | Guinan et al. (1976) using Wood's 1972 WINK computer model | Gillet, D. (1989) p. 222, Table 3 |

TABLE 52

Previously published estimates of Luminosity of A

| $L_A$ | Source | Notes/Comments |
|---|---|---|
| (log L/L$_\odot$) 2.41 | Eaton (1975) | Kim et al.(1989) p. 1067, Table 4 |
| (log L/L$_\odot$) 2.26 | Soderhjelm (1980) | |
| (log L/ L$_\odot$) 2.19 | Kim et al. (1989) | |
| $L\chi$ = 1.4 x 10$^{31}$ erg/s | Ness et al. (2002), p. 1045 | Total luminosity sited in a phases b/w 0.74 to 1.06 of LETGS observations. See conclusions, starting p. 1045 |
| log (L/L$_\odot$) 2.265 ± 0.075 | Sarna, M. J. (1993) p. 535, Table 1 | Calculated |
| 2.2 ≤ V ≤ 3.5 | Gillet, D. et al. (1989), p. 219 | |
| x-ray luminosity prob. Varies in the range $L\chi \approx 10^{31} - 10^{33}$ ergs/s | Sahade, J. & Hernandez, C.A. (1985), p. 258 | |
| Lg (L/L$_\odot$) 2.26 ± 0.07 | Soderhjelm, S. (1980) p. 110, Table 9 | |
| X-ray is ~ 2 x 10$^{30}$ ergs/s$^{-1}$ | Harnden, F. R., Jr., et al. (1989) p. 418 | |
| $L_A$ = $L_s$ 0.839 - 0.839 at 3428 Å | Chen, K.-Y. et al. (1977) | In terms of fraction of total luminosity of AB-C system. (g is the greater star, Algol B) Chen, K.-Y. et al. (1977) p. 73, Table III; (s is the smaller star.) |
| $L_A$ = $L_s$ 0.888 at 3700 Å | Herczeg (1959) | |
| $L_A$ = $L_s$ 0.93 at 2980 Å | Eaton (1975) | |
| $L_A$ = .734 fractional luminosity | Guinan et al. (1976), using Wood's 1972 WINK computer model | |

TABLE 53

Previously published estimates of luminosity of B

| $L_B$ | Source | Notes/Comments |
|---|---|---|
| (log L/L$_\odot$) 0.68 | Eaton (1975) from | Kim et al. (1989) p. 1067, Table 4 |
| (log L/L$_\odot$) 0.84 | Soderhjelm (1980) | |
| (log L/ L$_\odot$) 0.81 | Kim et al. p. (1989) | |
| log (L/L$_\odot$) 0.65 ± 0.14 | Sarna, M. J. (1993) | Calculated |
| log (L/L$_\odot$) 0.84 ± 0.16 | Soderhjelm, S. (1980) p. 110, Table 9 | |
| $L_B = L_g$ 0.061– 0.063 at 3428 Å | Chen, K.-Y. et al. (1977) | Chen, K.-Y. et al. (1977) p. 73, Table III In terms of fraction of total luminosity of AB-C system. (g is the greater star, Algol B) |
| $L_B = L_g$ 0.077 at 3700 Å | Herczeg (1959) | |
| $L_B = L_g$ 0.03 at 2980 Å | Eaton (1975) | |
| $L_B$ = .136 fractional luminosity | Guinan et al. (1976) using Wood's 1972 WINK computer model | Gillet, D. et al (1989) p. 222, Table 3 |

TABLE 54

Previously published estimates of luminosity C

| $L_C$ | Source | Notes/Comments |
|---|---|---|
| (log L/L$_\odot$) 1.11 | Eaton (1975) from | Kim et al. (1989) p. 1067, Table 4 |
| (log L/L$_\odot$) 1.00 | Soderhjelm (1980) | |
| (log L/ L$_\odot$) 1.15 | Kim et al. (1989) | |
| Log (L/L$_\odot$) 0.97 ± 0.17 | Sarna, M. J. (1993) p. 535, Table I | Calculated |
| Log (L/L$_\odot$) 1.00 ± 0.12 | Soderhjelm, S. (1980) p. 110, Table 9 | |

TABLE 55

Previously published estimates of limb darkening coefficient (1st order) for Algol A ($\mu_{1A}$)

| $\mu_{1A}$ | Source | Notes/Comments |
| --- | --- | --- |
| $\mu_{1A}$  .35 | Guinan et al. (1976) using Wood's 1972 WINK computer model | Gillet, D. et al (1989) p. 222, Table 3 |
| $\chi_s = 0.4$ at 3428 Å  $\chi_s = 0.5$ at 3700 Å  $\chi_s = 0.7$ at 2980 Å | Chen, K.-Y. et al. (1977)  Herczeg (1959) Eaton (1975) | Chen, K.-Y. et al. (1977) p. (1977), Table III $\chi$ in this paper is the 1st order; s is the smaller star Algol A. |

TABLE 56

Previously published estimates of limb darkening coefficient (2nd order) for Algol A

| $\mu_{2A}$ | Source | Notes/Comments |
| --- | --- | --- |
| $[\mu_1, \mu_2]_A = 1.000; - 0.096$ at 1920 Å; $T_A = 12{,}500$ K | Richards, M. T. et al. (1988) p. 328, Table 1-B | |

TABLE 57

Previously published estimates of limb darkening coefficient (1st order) for Algol B

| $\mu_{1B}$ | Source | Notes/Comments |
| --- | --- | --- |
| $\mu_B = .60$* fixed par. | Guinan et al. (1976), using Woods (1972) Computer model | Gillet et. al. (1989) p. 222, Table 3 |
| $\chi_g = 1.0$ at 3428 Å | Chen, K.-Y. et al. (1977) p. 73, Table 111 | $\chi$ is this paper is 1st order; g is the greater, meaning Algol B. |

TABLE 58

Previously published estimates of limb darkening coefficient (2$^{nd}$ order) of Algol B

| $\mu_{2B}$ | Source | Notes/Comments |
|---|---|---|
| $[\mu_1, \mu_2]_B$ = 1.199; - 0.343 at 1920 Å; $T_B$ = 5,000 K | Richards, M.T. et al. (1988) p. 328, Table 1 (B) | |

TABLE 59

Previously published estimates of position angle of the nodal point between 0° and 180°
Also the position angle of the line of nodes

| $\Omega$ | Source | Notes/Comments |
|---|---|---|
| $\Omega°$ ABC = 132 ± 2<br>$\Omega°$ AB = 132 ± 4 | Soderhjelm, S. (1980) p. 110, Table 8 | |

# APPENDIX B

# Algol Timeline

| Date | Milestone / Event | Reference |
|---|---|---|
| Ancient times | Inhabitants of Arabian peninsula gave the second brightest star in the constellation of Perseus the name, Al Ghūl, meaning "changing spirit." | (2), p. 3, paragraph 3 |
| Ancient times | The Hebrews knew Algol as Rōsh-ha-Satan (Satan's head). | (2), p. 12, bibliographic note No. 12 |
| Ancient times | The Chinese gave it the title Tseih She (the Piled-up Corpses)! | (2), p. 12, bibliographic note No. 12 |
| 205 B.C. - 25 A.D. | "The Chinese had related the observed brightness changes of Tsi-chi. (There are some Chinese character in parentheses) or Algol with death of the population or good or bad fortune recorded in <u>Hsing Ching.</u> | (1), p. 131 |
| 2$^{nd}$ Century A.D. (127-141 A.D.) | The term "double star" was used by Ptolemy, and this is not necessarily a binary system. | (2), p. 1, paragraph 2 (7), p. xi, paragraph 3 |
| 1850s | Father Angelo Secchi and astronomers discovered an unknown element which was named Helium after Helios, the Greek sun god. This discovery was made while studying the body of the sun and cataloguing the chemicals in the sun's spectral bands. These bands also revealed hydrogen, calcium, and iron. | (3) |
| 1609 | Galileo first pointed a telescope in the sky and instantly recorded the first scientific milestone – sunspots. | (3) |
| 1650 circa | The first double star which we know to form a binary system, ζ Ursae Majoris discovered by Father Jean Baptista Riccioli. | (2), p. 1, 2$^{nd}$ paragraph (7), p. 1 |
| 1669 | Geminiano Montanari recorded that Algol's light varied at times. | (1), pp. 130-131 |

# APPENDIX B

# Algol Timeline
continued

| 1670 Nov. 8 | The first known recorded minimum of Algol's light, though its hour or minute is unknown. | (2), p. 3, last paragraph, and p. 13, bibliographic note No. 13 |
|---|---|---|
| 1695 | Maraldi confirmed the variability of Algol. | (2), p. 3, last paragraph |
| 1761 | Lambert published a brief in which it is maintained that the stars are suns and are accompanied by retinues of planets. | (7), p. 2, paragraph 2 |
| 1767 | Reverend John Michell gave the first scientific argument that probably many double stars, then known, were the result of a physical rather than optical association. | (2), p. 2, paragraph 1 |
| 1767 | Michell used a probability argument to conclude that it is highly probable in particular and next to certainty in general that such double stars, etc., as appear to consist of two or more stars placed very close together do really consist of stars, placed near together. | (4), p. 7, paragraph 1 |
| 1779 | Hershel conceived the idea of utilizing double stars to test a method of measuring stellar parallax suggested long before by Galileo in early 1600s. | (7), p. 5, paragraph 1 |
| 1783 | John Goodricke<br>1) discovered that Algol's variability was periodic, thus he discovered the first short-period variable ever known.<br>2) estimated its period to be two days, 20 hours, and 45 minutes.<br>3) reasoned that Algol was an eclipsing binary. | (2), p. 4 |
| 1783 | Goodricke suggested, "The periodic passage between us and the star of a dark object." | (1), p. 2, paragraph 3<br>(5), p. 4, paragraph 2 |
| 1784 | Goodricke discovered that β Lyrae was an eclipsing variable. | (1), p. 2, paragraph 3 |

# APPENDIX B

# Algol Timeline
continued

| | | |
|---|---|---|
| 1802 | The term "binary star" was first used by Sir William Herschel. | (2), p. 1, 1$^{st}$ sentence, paragraph 2 |
| 1802 | William Herschel, in collaboration with his sister, Caroline, published observations over 40 years attempting to establish distances between double stars via the inverse-square law of light propagation. | (4), p. 7, paragraph 1 |
| 1827 | The mathematical method required to determine the geometry of the relative orbit of one component with respect to its companion from a set of observations of position angle and separation was first developed by Savary. | (4), p. 8, paragraph 2 |
| 1850s | Father Angelo Secchi, the Vatican's chief astronomer, pioneered the branch of science called spectroscopy. Spectroscopy revealed a complexity of the sun, activity around the edges of the sun. | (3) |
| 1857 | Mizar (Ursae Majoris) first double star to be observed photographically measurable images being secured by G. P. Bond at the Harvard Observatory. | (7), p. 1, paragraph 1 |
| 1862 | Father Secchi found that the sun must also be a star since the stars and the sun were composed of the same chemicals as revealed through spectroscopy. | (3) |
| 1889 | ζ Ursae Majoris' principal component was recognized as the first spectroscopic binary by E. C. Pickering | (2), p. 1, 2$^{nd}$ paragraph |
| 1889 | Vogel's spectroscopic observations confirmed that Algol was an eclipsing binary. | (2), p. 5, paragraph 2 |
| 1890 | Vogel was first to prove the star Algol was indeed a binary star, by demonstrating that the primary star was receding from the observer before primary eclipse, and approaching the observer after primary eclipse. | (4), p. 10, paragraph 2 |
| 1893 | Belopolsky found that the radial velocities of the H$_\beta$ emission of β Lyrae gave a velocity curve opposite in phase to, and with smaller amplitude than that from the stellar absorption lines. | (1), p. 6, paragraph 6 |

# APPENDIX B

# Algol Timeline
continued

| | | |
|---|---|---|
| 1893 | Dr. Kristian Birkeland, a Norwegian Scientist, drew his own conclusions about the sun. Birkeland's hunch was that the magnetic storms accompanying the Northern Lights (Aurora) were caused by a stream of electrically charged particles buffeting the earth's magnetic field. (His ideas were never taken.) | (3) |
| 1896 | Pickering reported Bailey's discovery of the variation in line intensities of $\mu^1$ scorpii – the lines of one of the components are weaker when its velocities are of recession. | (1), p. 2, paragraph 1 |
| 1898 | Pickering (1904) and Wendell (1909) independently observed a strange increase of brightness immediately after primary on the light curve of R Canis Majoris. | (1), p. 5, last paragraph |
| 1898 | Myers proposed the existence of a gaseous envelope around the system of β Lyrae. | (1), pp. 6-7 |
| 1907-1912 | Schlessinger detected narrow lines near the primary minimum. Later attributed to Algol C by Pearce and Meltzer. | (1), p. 131<br>(1), p. 131<br>(11), (12) |
| 1908 | Curtiss reported variations of the velocity of the center of mass of the eclipsing pair of Algol. | (1), p. 131 |
| 1908 | Barr's analysis of the available orbits and radial velocity curves of spectroscopic binaries found that:<br>A) very often, the velocity curves were unsymmetrical,<br>B) these asymmetries with only a few exceptions, were similar, in the sense that the "ascending branch of the curve" was of greater length than the "descending branch," and<br>C) a logical consequence of A and B, that the values of the longitude of Periastron (ω) did not show a random distribution, but rather were concentrated in the first quadrant. This peculiar distribution of the curves, known as the Barr effect in spectroscopic binaries, already suggested as Barr himself pointed out, that the "observed radial velocities" were contaminated by some extraneous effect that, at that time, was difficult to access. | (1), p. 2, paragraph 6 |
| 1908 & 1911 | Dugan & Stebbins independently discovered the reflection effect. | (1), p. 2, paragraph 5 |

# APPENDIX B

# Algol Timeline
continued

| | | |
|---|---|---|
| 1908 (June) | Hale, while studying absorption lines of the sun's surface, discovered that sunspots were caused by magnetic distortion. These distortions were 4,000 times greater than the earth's magnetic field. | (3) |
| 1910 | "Primitive use of photoelectric effects." | (1), p. 2, paragraph 4 |
| 1910 | Stebbins detected Algol's shallow secondary minimum and variations of light between eclipses, using a selenium conductive cell. | (1), p. 131 |
| 1912 | Hans Geitel & Julius Elster in Germany, and Jakob Kunz in Illinois marked the first use of photoelectric photometry, using a potassium hydride cell. | (8) |
| 1912 | Norris Russel published a general solution of light curves. (The scatter in the data due to the lower precision at that time covered up the transient changes in observed brightness of components now known to interact.) "Instrumental difficulties were such that unusual variations were rarely attributed to the star itself." Astronomers were biased toward periodic unchanging light curves. | (1), pp. 2-3, Quote reference is: (1), p. 5, paragraph 3 |
| 1918 | Maggini obtained the first two-color light curves for Algol at 6450 Å and 4120 Å using a wedge photometer and filters. | (1), p. 131 |
| 1921 | Stebbins obtained the first complete photoelectric light curve for Algol, and he found fluctuations of the period of Algol using light minima from 1852 through 1887. | (1), p. 131 |
| 1922 | Shapley noted a difference in the appearance of the ionized calcium lines on the leading and following sides of the brighter component of β Aurigae. | (1), p. 6, paragraph 2 |
| 1922 | The first suggestion of a third body in the Algol system by Stebbins studying period variations. | (1), p. 131 |
| 1922 | Hellerich presented spectroscopic elements of Algol. | (1), p. 131 |
| 1924 | McLaughlin computed absolute dimensions of Algol. | (1), p. 131 |
| 1924 | Rossiter discovered the rotation effect. | Rossiter (1924) |
| 1924 | Dugan, after comparing his observations with earlier work regarding the strange increase in brightness after primary of R Canis Majoris concluded that it was a real but transient feature of the light curve. | (1), p. 5, last paragraph |
| | | |

# APPENDIX B

# Algol Timeline
continued

| | | |
|---|---|---|
| 1931 | Struve & Elvey were first to observe the doubling of Mg II 4481 (confirmed in 1935 by Morgan). | Article |
| 1935 | Morgan confirmed double lines of Mg II, 4481 Å found near center of primary of Algol. | (1), p. 134 Morgan (1935) |
| 1937 | Dugan and Wright presented the first general study of periodic changes in eclipsing systems and found relatively sudden period changes with no periodic pattern unaccountable for by rotation of the aps or third body effects. (no explanation offered) | (1), p. 6, paragraph 5 |
| 1937 | McLaughlin suggested that earlier periods of motion for Algol C found spectroscopically must be ruled out due to systematic errors in reductions. | (13), p. 598 paragraph 2 |
| 1939 | Hall's precise times of minimum in two colors was probably the first strong evidence against the Tikhoff-Nordmann effect, which is "the alleged phenomenon whereby the time of minimum light is a function of the wavelength." | (1), p. 130, (14) p. 460 |
| 1940s | Algol Paradox noted by astronomers. Term used to describe a discrepancy in our theories of stellar evolution. | (6), p. 25, paragraph 2 |
| 1940s (late) | Rockets discovered fringes of space scorched by radiation from x-rays, gamma rays, and ultraviolet light which had to be coming from the sun. The rays didn't penetrate the earth's atmosphere. | (3) |
| 1941 | Struve postulated a gas stream in β Lyrae. This may be the first time that a gas stream was suggested in a close binary system.) | (1), p. 7 paragraph 5 |
| 1941 | Struve postulated the existence of gaseous streams between the components of β Lyrae and of an ultra expanding general gaseous envelope. | (1), p. 7, paragraph 5 |
| 1942 | Discovery by A. H. Joy that the double H emissions of RW Tauri undergo eclipses as well, and therefore must arise from a gaseous ring around the primary component. | (1), p. 7, paragraph 6 (5), p. 1, paragraph 5 |
| 1947 | Ludwig Biermann calculated that something more than pressure from sunlight pushed the tails of comets *(assumed earlier by Chinese astronomers)*. Biermann called it "solar corpuscular radiation" (solar wind). His idea was immediately rejected. | (3) |

# APPENDIX B

# Algol Timeline
continued

| | | |
|---|---|---|
| 1948 | Eggen found photometrically that the value of the period of Motion of Algol C was satisfied by a period 1.873 years, the value determined by McLaughlin in 1937, spectroscopically. | (13) |
| 1948 | Eggen suggested the Algol system has four components, the fourth component with an orbital period 188.4 years. | Eggen (1948), p. 1, abstract |
| 1948 | Sidney Chapman, a physicist, scorned Biermann's solar wind theory and felt that the corona of the sun extended far beyond where it appeared during a solar eclipse, yet still thought the corona was firmly bound to the sun. | (3) |
| 1950 | Wood was the first to try to explain period changes in a number of eclipsing systems probably indicative of mass loss. | (1) p. 8, paragraph 7 (5), p. 1, paragraph 6 |
| 1950s (late) | Eugene Parker worked on resolving the Biermann/Chapman contradiction in their two theories and found them both to be right. He took it further and found that the only answer would be a supersonic solar wind. He calculated that the sun's corona did have enough thermal energy to escape the sun's gravity stream out into space at 500 km/sec. He published his findings in 1958. (They were thought to be absurd.) | (3) |
| 1954 | Walker and Herbig suggested the existence of a "hot spot" in the nebular ring around the hot component in UX Ursae Majoris. | (5), p. 1 (Intro.), paragraph 7 |
| 1955 | J. A. Crawford and Z. Kopal, independently suggested an explanation of the Algol paradox in terms of post-main sequence evolution. | (1), p. 8, paragraph 2 (5), p. 1, paragraph 6 |
| 1957 | Struve and Sahade interpreted emission features near phase .75 as gas streaming. | (1), p. 134 |
| 1959 | Kopal defines a "close binary system" as those whose separation is comparable to dimensions of the components lending well to their discovery by spectroscopy. | (2), p. 3 |
| 1960 | Herczeg with a photometric solution suggested systematic changes in the depths of primary eclipse of Algol. | (1), p. 133 |

# APPENDIX B

# Algol Timeline
continued

| 1962 | R. Giaconni and his collaborators discovered the first x-ray sources outside our solar system. This discovery began the field now known as X-ray astronomy – The first point x-ray source discovered was an x-ray binary, known as Sco X-1. | (1), p. 13, paragraph 2, (9) |
|---|---|---|
| 1962 | Mariner II's probe to Venus found that space was awash with solar wind, exceeding Eugene Parkers expectations (1958). The probe detected winds of 300-800 kms/sec. These electrically charged hurricanes are ferocious and relentless. The violent gusts break free at times from the sun's gravitational and magnetic forces. These were the flares and coronal mass ejections first witnessed by Skylab in 1970s. | (3) |
| 1964 | Fletcher deduced presence of circumstellar helium around Algol A, which was not distributed symmetrically around the line of centers. | (1), p. 134 |
| 1966 | Algol is the first eclipsing system to have a light curve observed in the infrared, and the orange spectral regions. Chen and Reuning, 1966 | (1), p. 130 |
| 1966 | Paczynski & Plavec independently coined the term "interacting binaries." | (5), p. 1 (Intro.), paragraph 3 |
| 1967 | Glushneva & Esipov detected Algol B spectrographically. | (1), p. 131, (16) |
| 1967 | Plavec and Paczyński independently were the first to suggest a change in the way close binary is defined by taking the evolutionary history into account. | (1), p. 1, paragraph 2 |
| 1967 | International colloquium considering evolution of double stars. | (1), p. 12, paragraph 2 |
| 1968 | Lucy assumes a common convective envelope for W Ursae Majoris systems. | (1), p. 36 |
| 1970 | Frieboes-Conde, Herczeg & Hog determined that only three components are needed to explain the phenomena of Algol | (1), p. 130 (17), p. 1 |
| 1970 | Frieboes-Conde et al. confirmed Algol to be three components only. | (1), p. 132 (17), p. 1 |

# APPENDIX B

# Algol Timeline
continued

| 1971 | Concept of hot spot<br>- Warner and Nather<br>- Photoelectric (Krzemiński and Smak)<br><br>Warner & Nather, and Krzemiński & Smak almost simultaneously developed the concept of "hot spots." Warner & Nather believed that hot spots were created by ejected matter falling on a disk surrounding the secondary star. Krzemiński & Smak suggested that the hot spots on the disk surrounding the primary, white dwarf star are the source of the S-wave emission lines. | (1), p. 13<br>pp. 104 – 105 |
|---|---|---|
| 1971 | Hill et al. confirmed the 32-year periodicity of Algol's $\omega$ to be apsidal. | (1), p. 132 |
| 1971 | Paczyński proposed specific application of the term "close" to such interacting binary systems. | (19), p. 242 |
| 1972 | Wade and Hjellming detected Algol as a quiescent radio source. | (1), p. 134 |
| 1972 | Algol's radio flare activity observed by Hjellming, Wade, and Webster might be correlated with optical spectrum variations observed by Bolton. | (1), p. 134 |
| 1972 | Wilson and Devinney introduced their method for determining "accurate close binary light curves" which is an improvement over the Russel Method for "binaries exhibiting even modest proximity effects." | (1), p. 133, (18) |
| 1973 | Skylab, the first solar space laboratory, was launched beyond earth's atmosphere. This was when the most extensive period of solar observation is history began. Sunspots, networks, filaments, and solar flares were clearly detected. | (3) |
| 1973 | Skylab astronauts discovered coronal mass ejections, huge outbursts of material on a scale much greater than solar flares. | (3) |
| 1973 | Rule & Elsmore used speckle interferometry and accurately located to barycentre of Algol | (19), p. 244 |
| 1974 | Labegrie et al. observed Algol C as an astrometric binary, believing they resolved it by speckle interferometry. | (1), p. 131 |

# APPENDIX B

# Algol Timeline
continued

|  |  |  |
|---|---|---|
| 1975 | Eaton published far ultraviolet (UV) observations of Algol from the orbiting astronomical observatory (OAO-2) taken August 1969 and July 1972. The wavelengths are 2930 Å, 1920 Å, and 1550 Å. | (1), p. 135, paragraph 2 |
| 1975 | The SAS-3 satellite detected x-ray emission from Algol. Shnopper et al., 1976. The intensity equaled $1.6 \times 10^{31}$ ergs/s | (1), p. 82, p. 130 (5) p. 80, paragraph 4.3 |
| 1975 | Soft x-rays were detected from Algol in the range .15-2 KeV to 2 KeV by Harnden et al., indicating a mass loss rate approximately equal to $10^{-10}$ $M_\odot$/year. | Harnden (1977) p. 422 |
| 1975 | Longmore & Jameson found that IR excess was attributed to gas stream. | (19), p. 244 |
| 1975 | Bachmann & Hershey, with inclusion of accurate astrometry, had more confident element specification. | (19), p. 244 |
| 1975 | Gibson et al., through radio observations (1400, 2695, 6000, and 8085 $MH_z$), detected a large flare from Algol. | (19), p. 244 |
| 1975 & 1976 | Chen and Wood obtained UV observations of Algol taken on September 7th, 8th, and 9th, 1973, and January 4, 1974, from the Copernicus satellite. They observed that the Mg II resonance lines near 2800Å appeared to be doubled. | (1), p. 135 |
| 1976 | Algol itself was first detected as an x-ray source by Schnopper et al. (1976). | (1), p. 130 (5), p. 80, paragraph 4.3 |
| 1976 | Guinan et al. estimates the extent of circumstellar material in Algol for $H_A$ and $H_\beta$ light curves to be approximately 2.6 $R_A$ (*I have to check that number because there was an incorrect unit of measure in there.*) | (1), p. 134 |
| 1977 | Kondo et al. used UV observations of Algol taken on October 9th through 10th, 1974, from a balloon-borne spectrometer. The Mg II lines gave evidence of mass flow. | (1), pp. 135-136 |
| 1978 | Tomkin & Lambert used Reticon spectroscopy and detected Na II lines of Algol's secondary. | (19), p. 244 |
| 1979 | Bonneau, using speckle interferometry, revised absolute elements of the triple system of Algol. | (19), p. 244 |

# APPENDIX B

# Algol Timeline
continued

| 1979 | White et al. through x-ray spectrometry found no eclipses. An active corona of Algol's subgiant was inferred. | (19), p. 244 |
|------|---|---|
| 1979 | Cugier, using UV spectrometry, found high resolution Mg II line profiles which allowed detailed modeling. | (19), p. 244 |
| 1981 | Chen et al., using UV spectrometry, found high resolution Mg II line profiles, like Cugier in 1979, allowing detailed modeling. | (19), p. 244 |
| 1983 | Kemp et al. (1983) pointed out the Chandrasekhar Effect from their intensive polarization studies of β Per | (19), p. 242 |
| 1988 | Lestrade et al. observed β per at 2.3 and 8.4 $GH_z$ with the Very Large Array. The brightness temperature of the radio source varies from $3 - 50 \times 10^8$ K and is indicative of gyrosynchrotron emission of mildly relativistic electrons in an active coronal region with magnetic fields of about 30 Gauss at unit optical depth. The size of the source is estimated to be about three cool radii. Two outbursts occurred. One was a high brightness ($\approx 1.5 \times 10^{10}$K), broadband burst, and the other was a short ($\approx 15$ minutes) outburst at 1.66 $GH_z$ with a brightness temperature of about $3 \times 10^{10}$K and a magnetic field of approximately 300 Gauss. Both of these are very likely associated with coronal loops. | (1), p. 130 (5), p. 45, paragraph 1 |
| 1999 | Kempner & Richards (1999) were the first to calculate UV difference profiles for an Algol-type system (U Sge). | (10) |

**References:** (1) Sahade and Wood (1978); (2) Kopal (1959); (3) "The Planets" (DVD); (4) Hilditch (2001); (5) Sahade/McCluskey/Kondo (1993); (6) Pringle & Wade (1985); (7) "The Binary Stars" Robert Grant Aitken, McGraw-Hill, N. Y. (1935); (8) http://www.phys-astro.sonoma.edu/people/faculty/tenn/asphistory/1994.html. The Astronomical Society of the Pacific "106th Annual Meeting" History Sessions: presented by ASP History Committee History II: Historic & Prehistoric Astronomy Tuesday, 28 June 1994, North ArizUniv. Flagstaff, AZ, Invited lecture 1:30p
Photoelectric photometry: The First Fifty Years – John B. Hearnshaw, University of Canterbury; (9) [online document http://lheawww.gsfc.nasa.gov/users/white/xrb.html] Cited: 3/13/06, High energy Astrophysics Learning Center: x-ray binaries; (10) Kempner & Richards (1999); (11) Struve and Sahade (1957b); (12) I. Barney (1923);

# APPENDIX C

## IONIZATION FRACTIONS

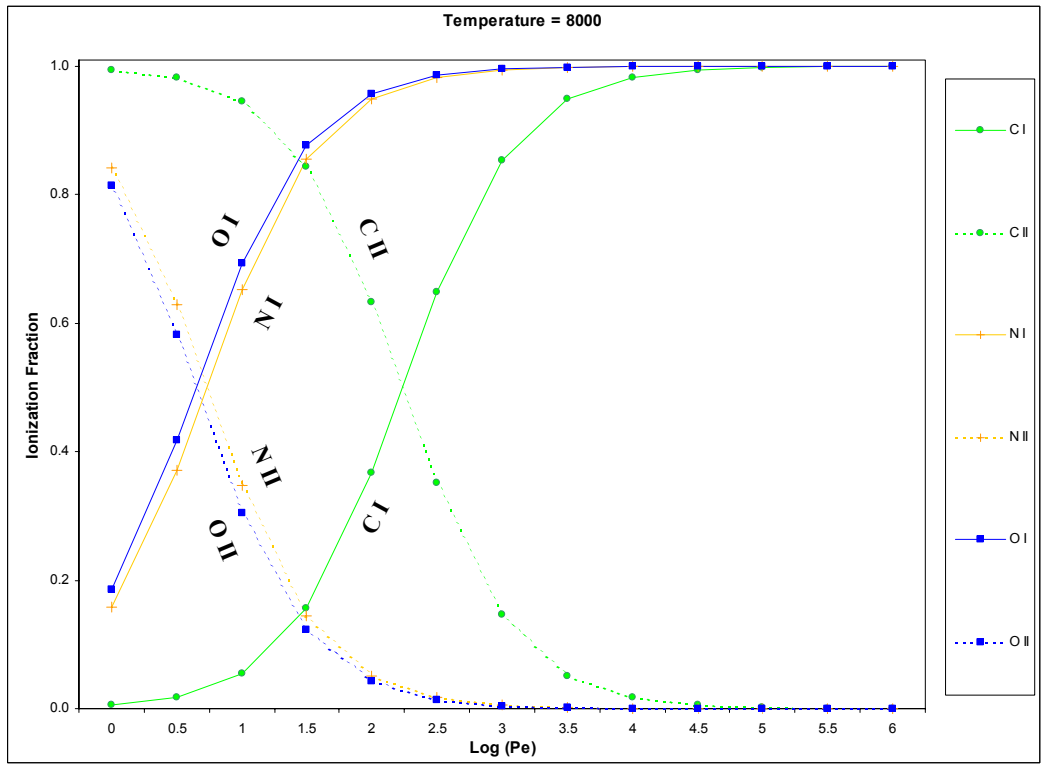

FIG. C.1 - *Ionization Fraction of Carbon, Nitrogen, and Oxygen at 8,000 K*

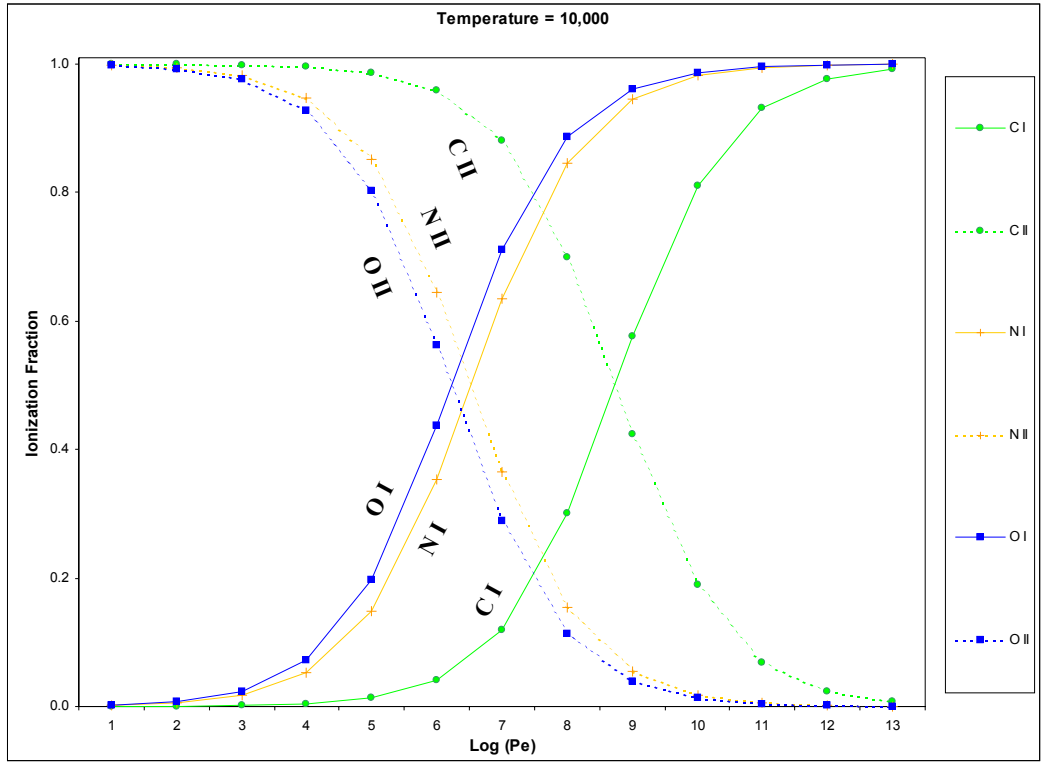

FIG. C.2 - *Ionization Fraction of Carbon, Nitrogen, and Oxygen at 10,000 K*

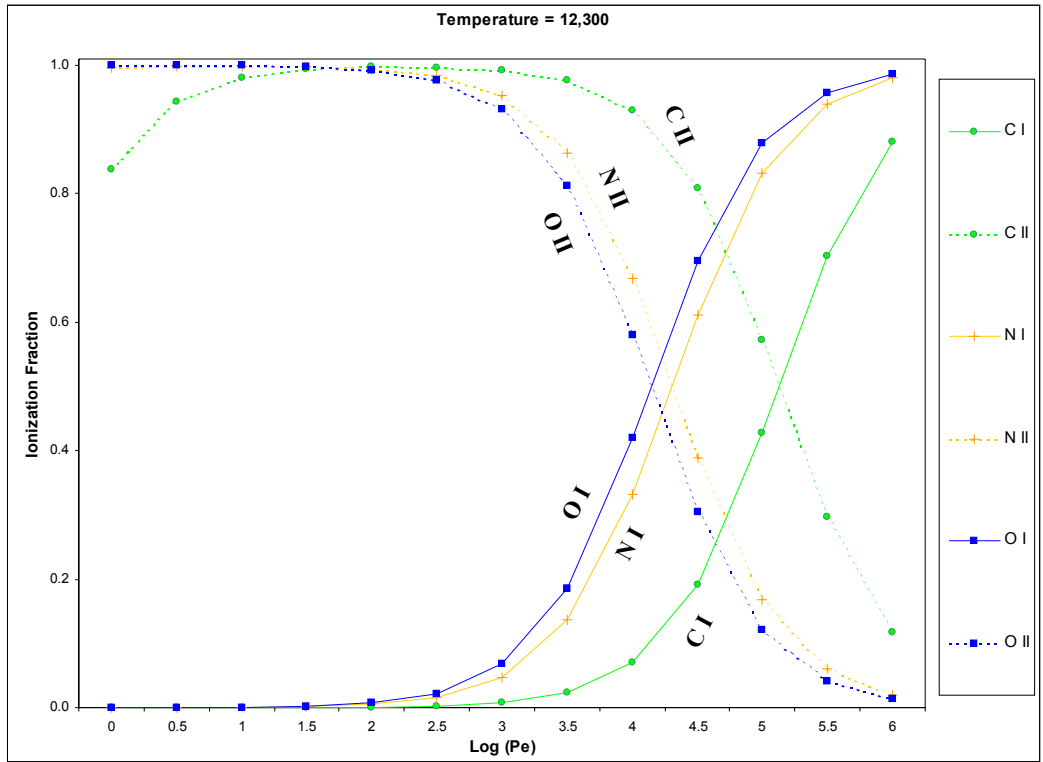

FIG. C.3 - *Ionization Fraction of Carbon, Nitrogen, and Oxygen at 12,300 K*

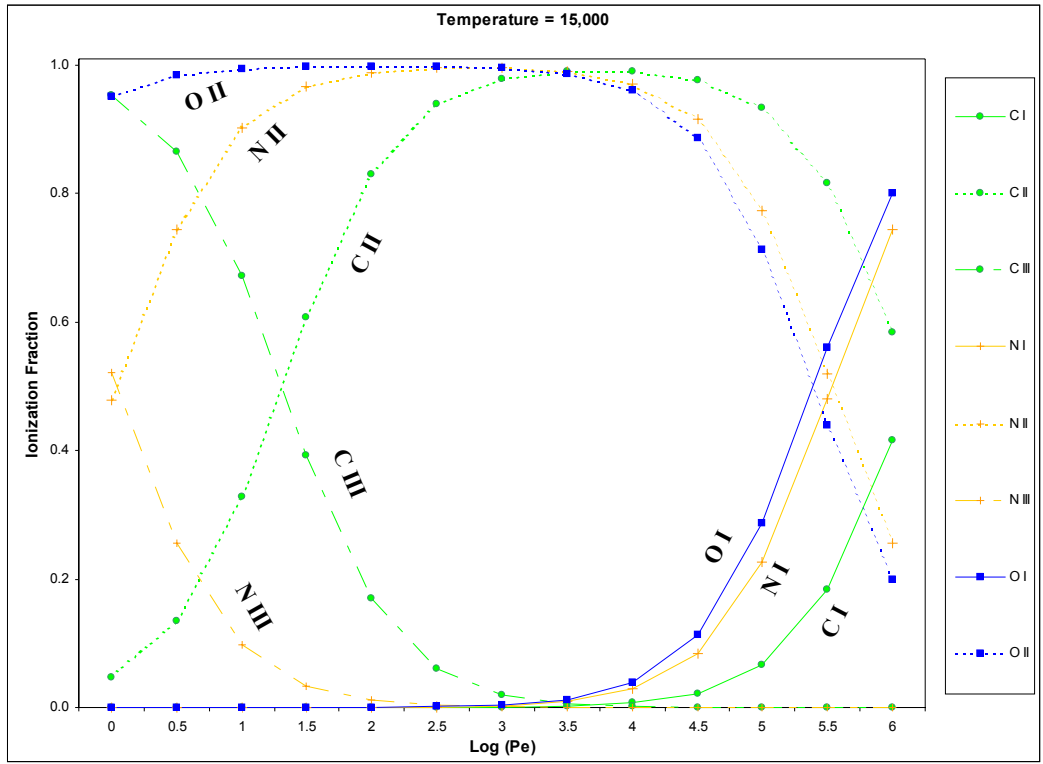

FIG. C.4 - *Ionization Fraction of Carbon, Nitrogen, and Oxygen at 15,000 K*

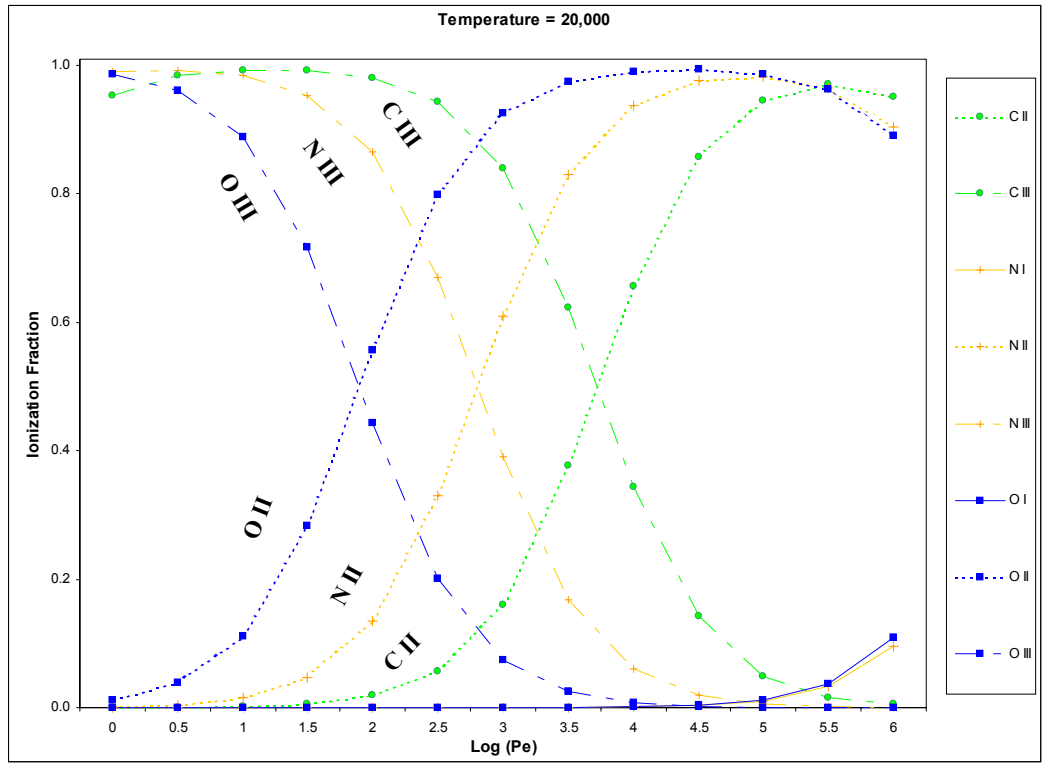

FIG. C.5 - *Ionization Fraction of Carbon, Nitrogen, and Oxygen at 20,000 K*

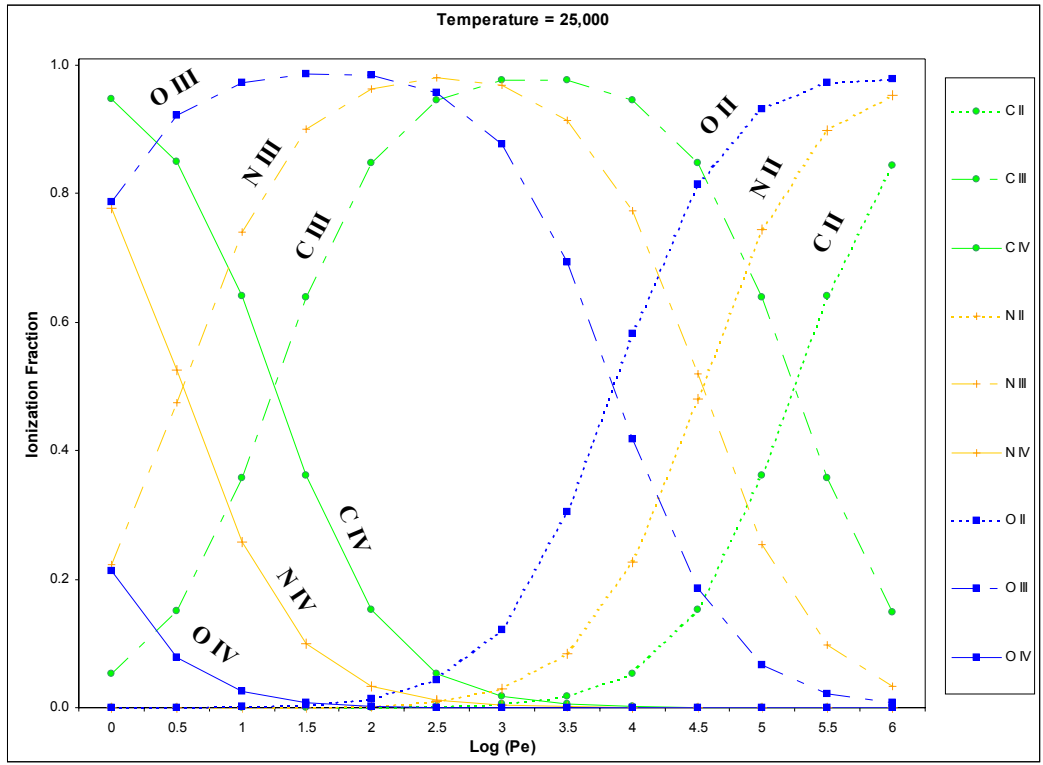

FIG. C.5- *Ionization Fraction of Carbon, Nitrogen, and Oxygen at 25,000 K*

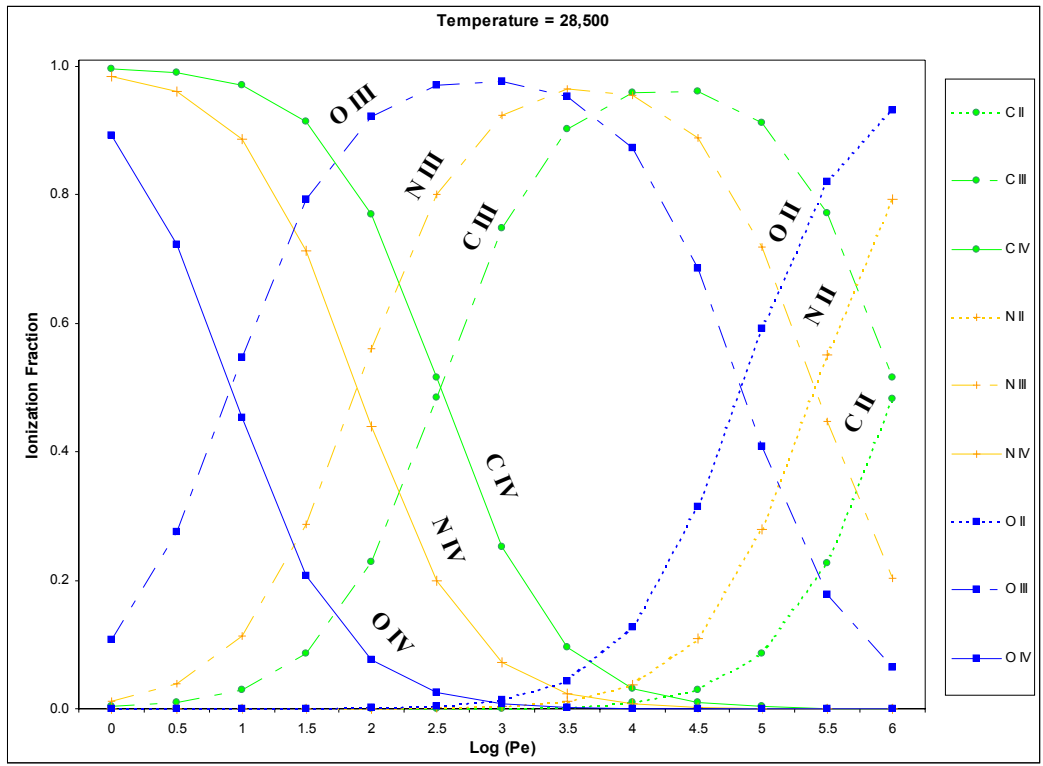

FIG. C.6 - *Ionization Fraction of Carbon, Nitrogen, and Oxygen at 28,500 K*

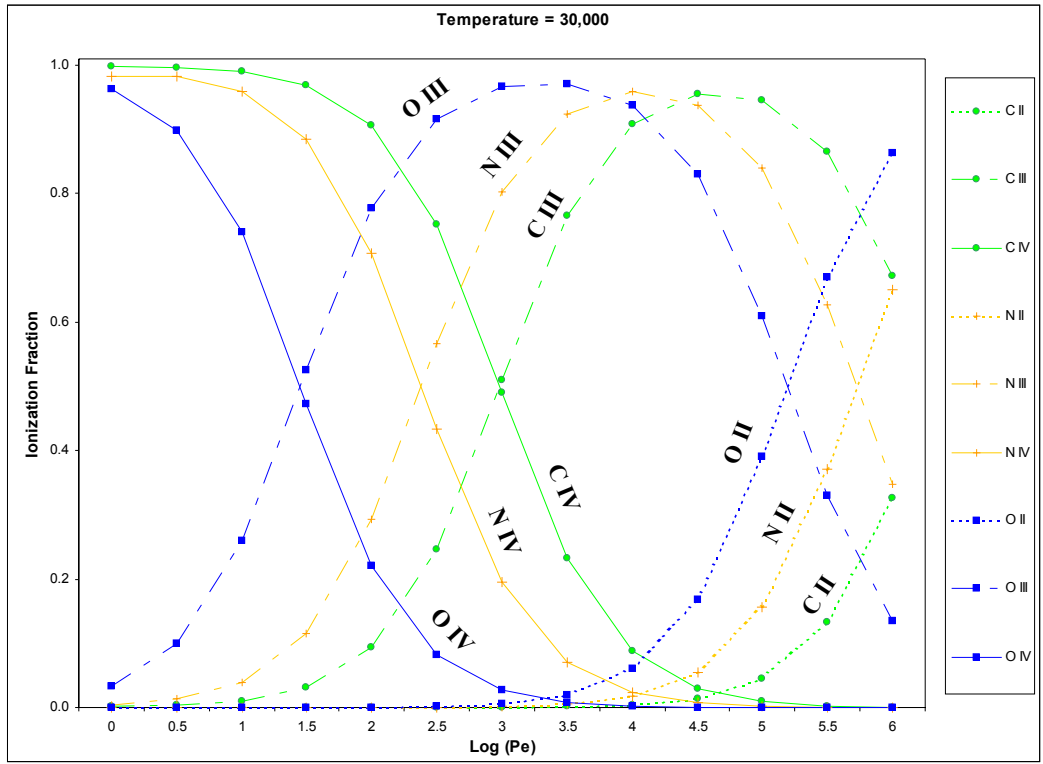

FIG. C.7 - *Ionization Fraction of Carbon, Nitrogen, and Oxygen at 30,000 K*

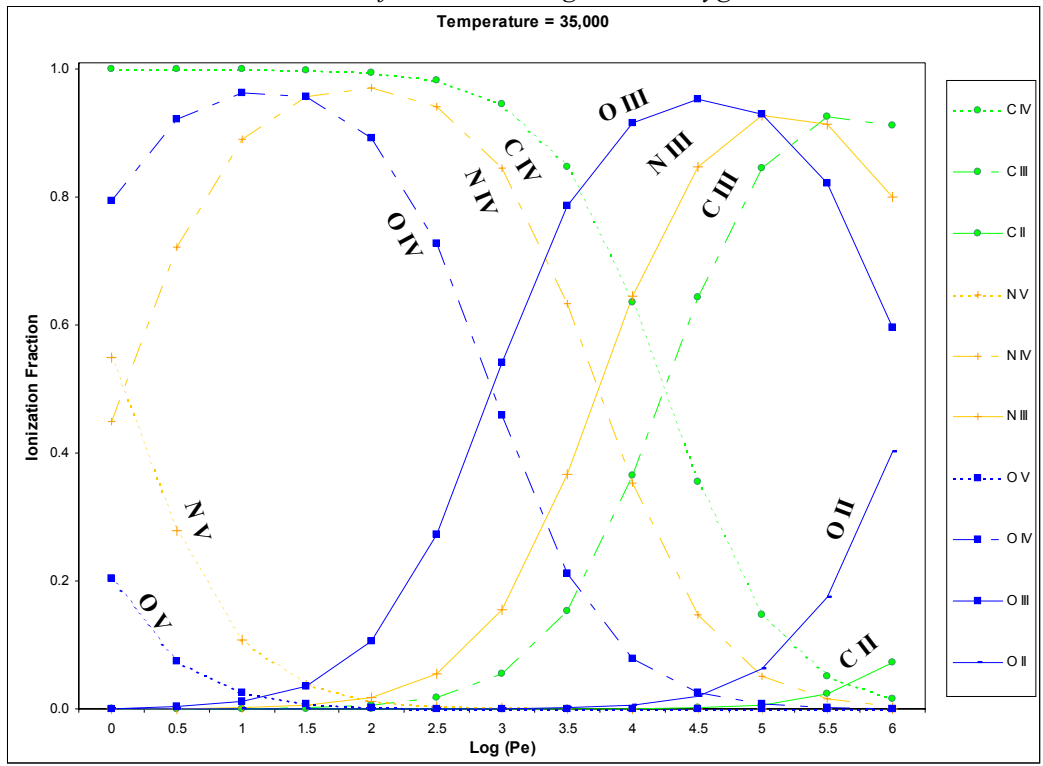

FIG. C.8 - *Ionization Fraction of Carbon, Nitrogen, and Oxygen at 35,000 K*

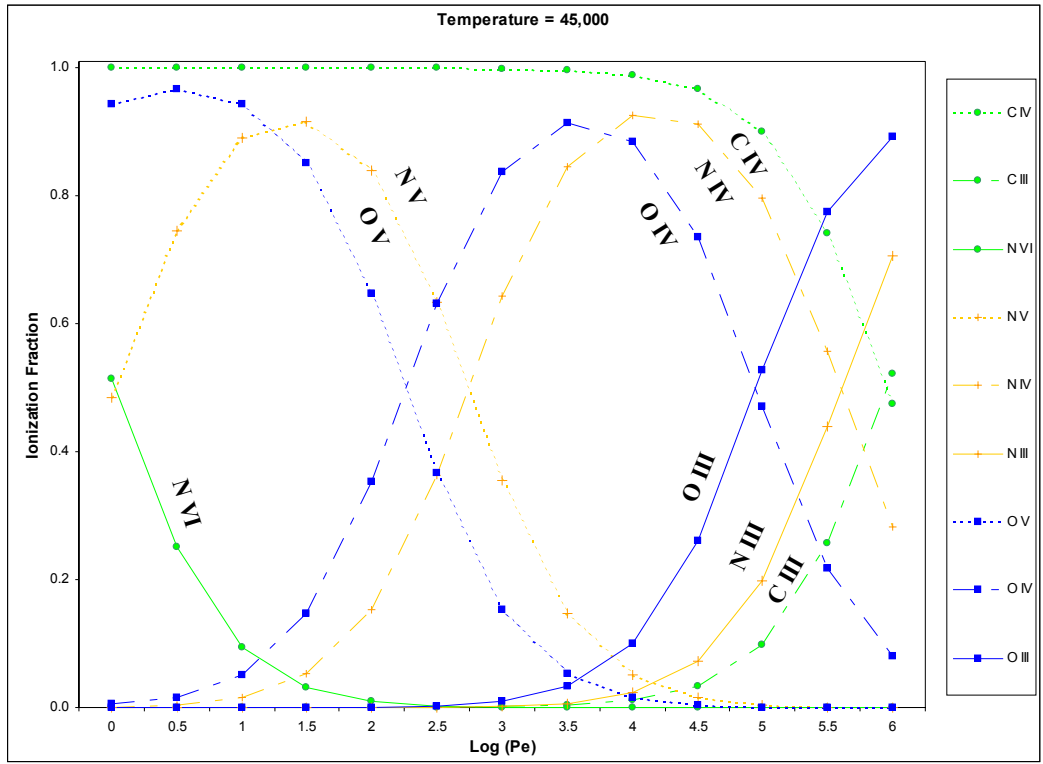

FIG. C.9 - *Ionization Fraction of Carbon, Nitrogen, and Oxygen at 45,000 K*

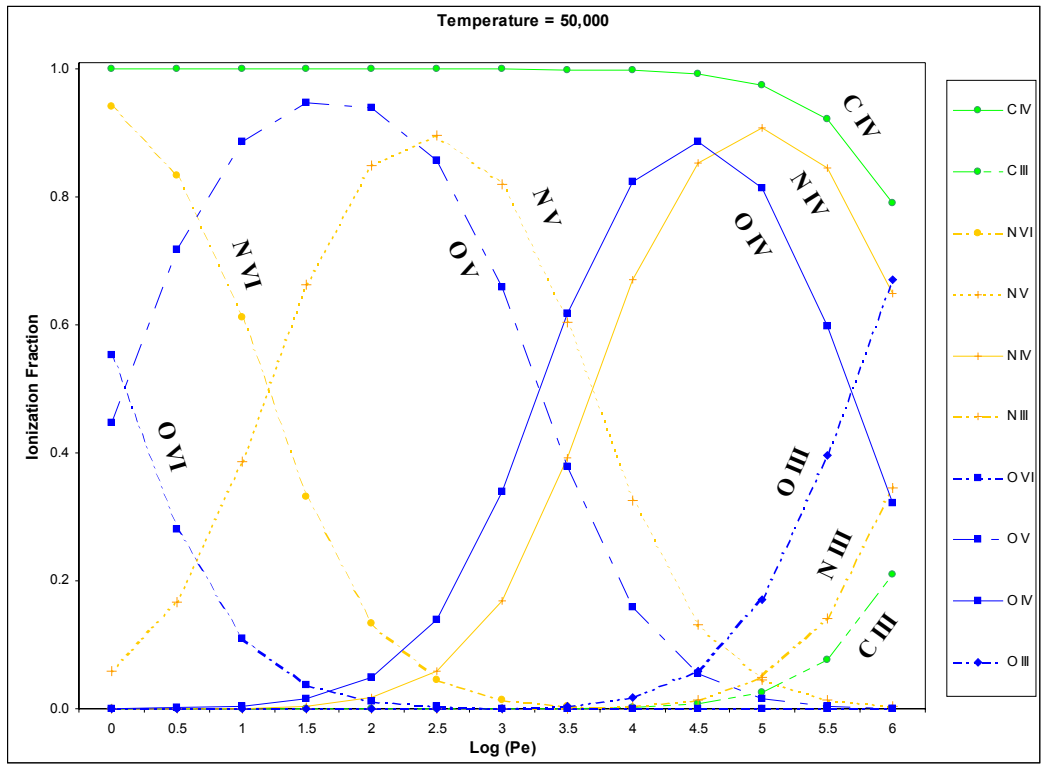

FIG. C.10 - *Ionization Fraction of Carbon, Nitrogen, and Oxygen at 50,000 K*

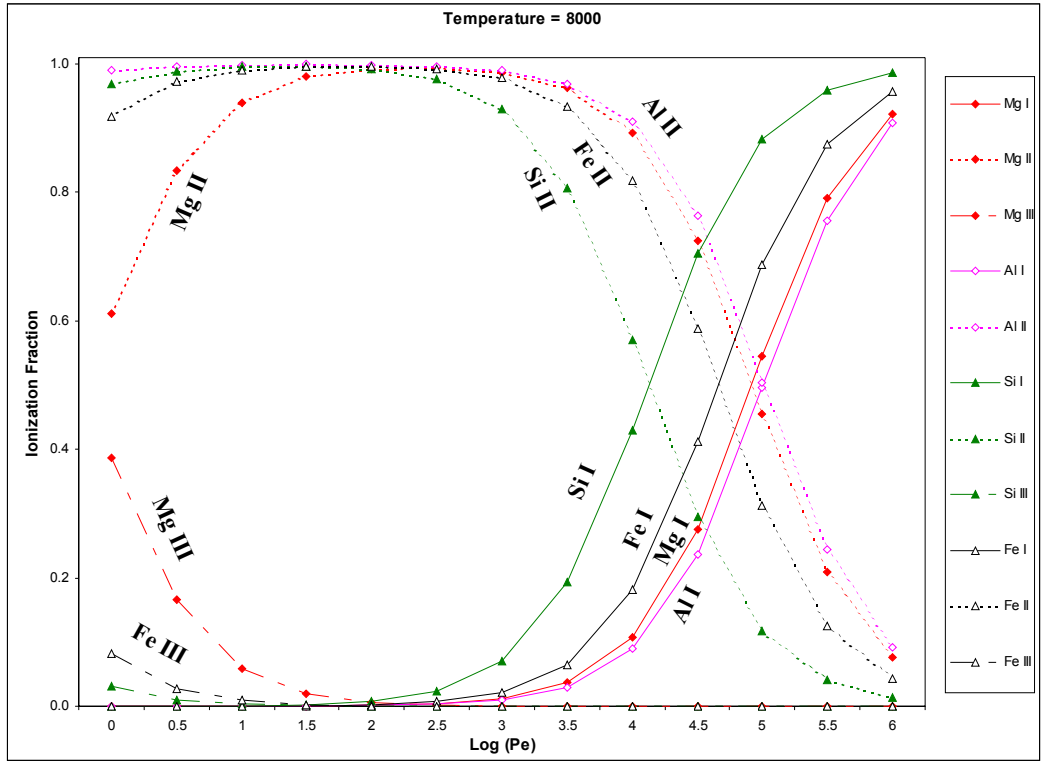

FIG. C.11 - *Ionization Fraction of Mg, Al, Si, and Fe at 8,000 K*

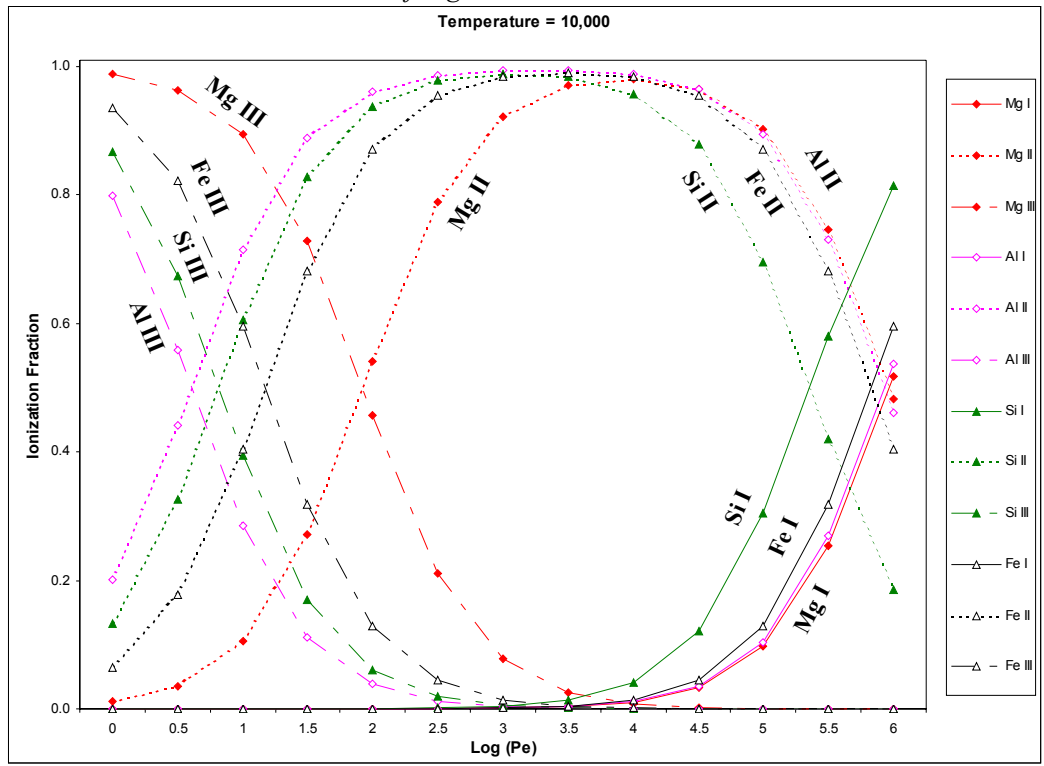

FIG. C.12 - *Ionization Fraction of Mg, Al, Si, and Fe at 10,000 K*

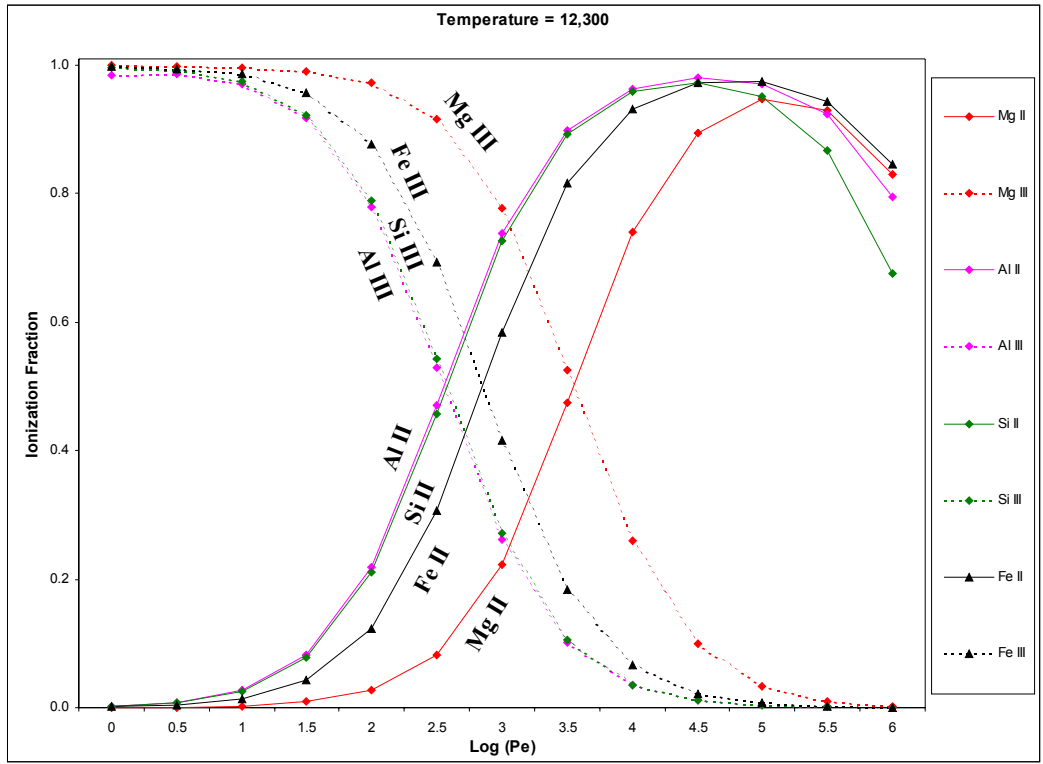

FIG. C.13 - *Ionization Fraction of Mg, Al, Si, and Fe at 12,300 K*

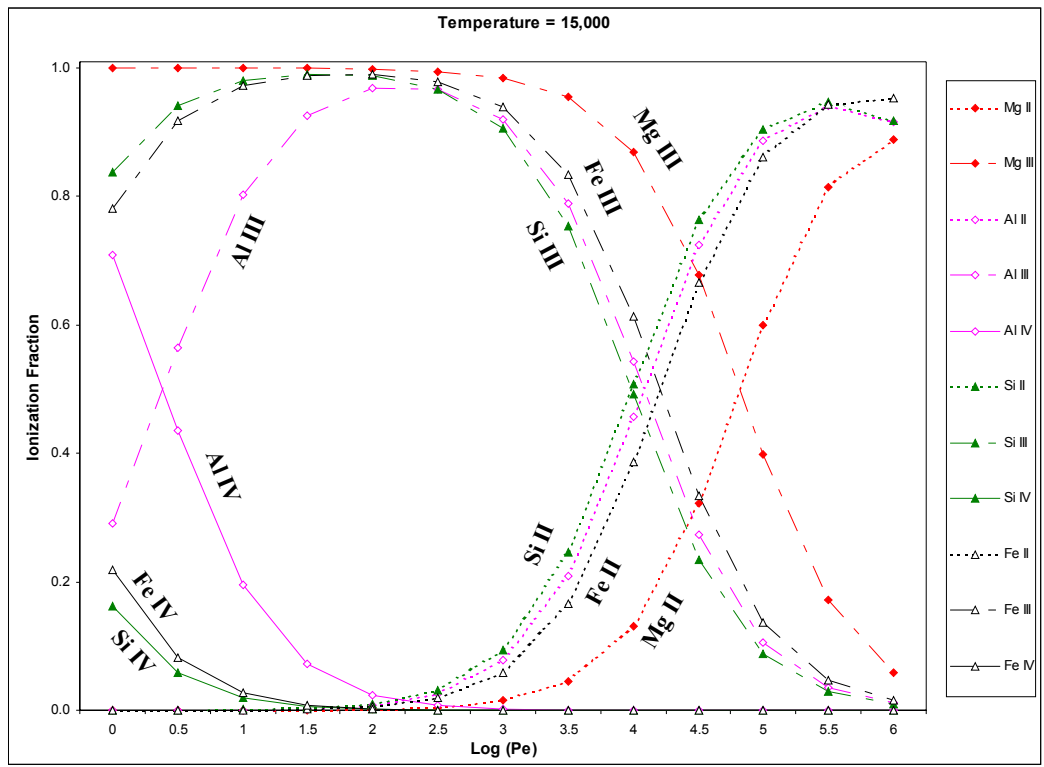

FIG. C.14 - *Ionization Fraction of Mg, Al, Si, and Fe at 15,000 K*

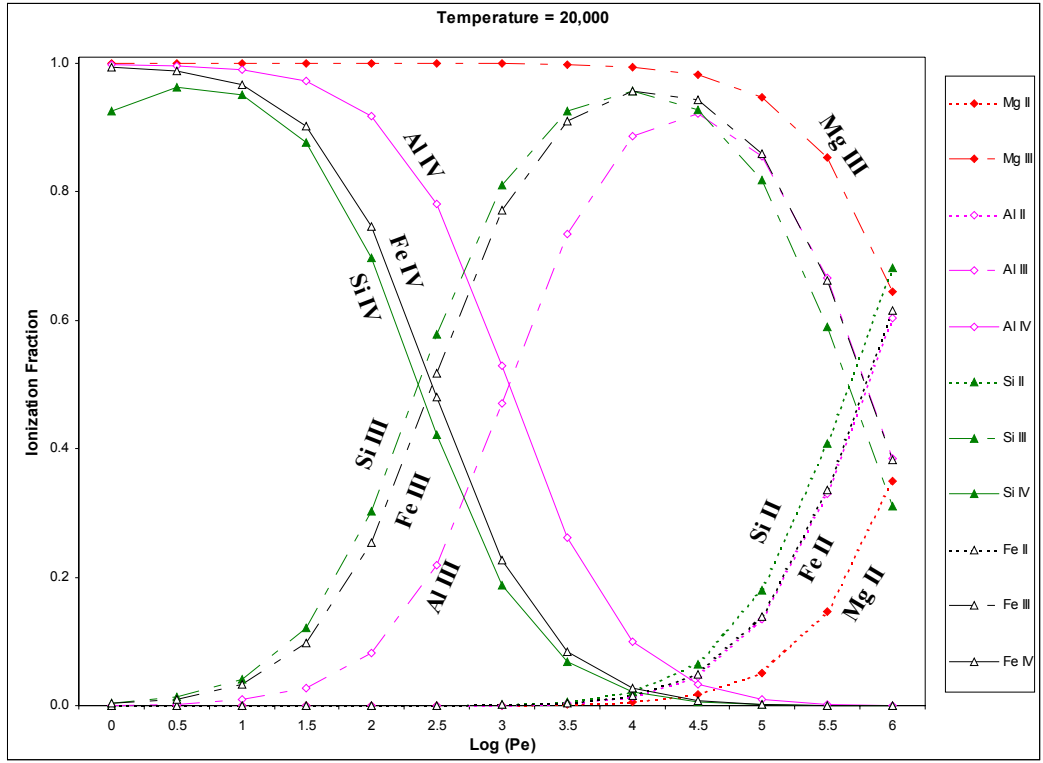

FIG. C.15 - *Ionization Fraction of Mg, Al, Si, and Fe at 20,000 K*

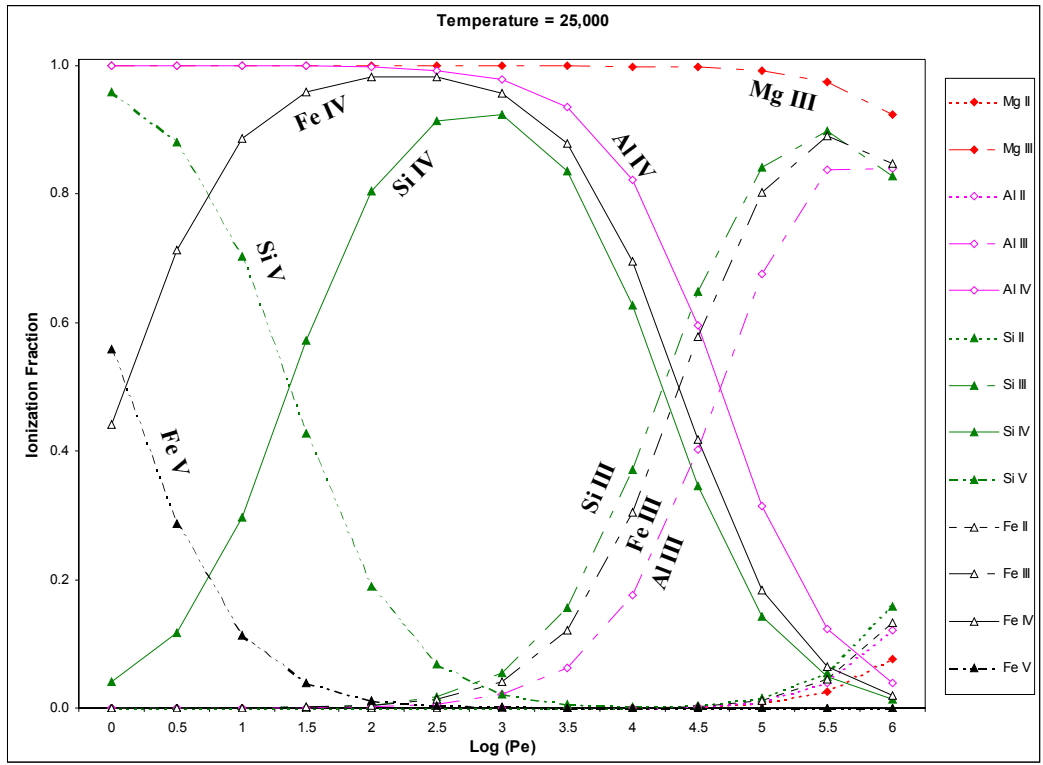

FIG. C.16 - *Ionization Fraction of Mg, Al, Si, and Fe at 25,000 K*

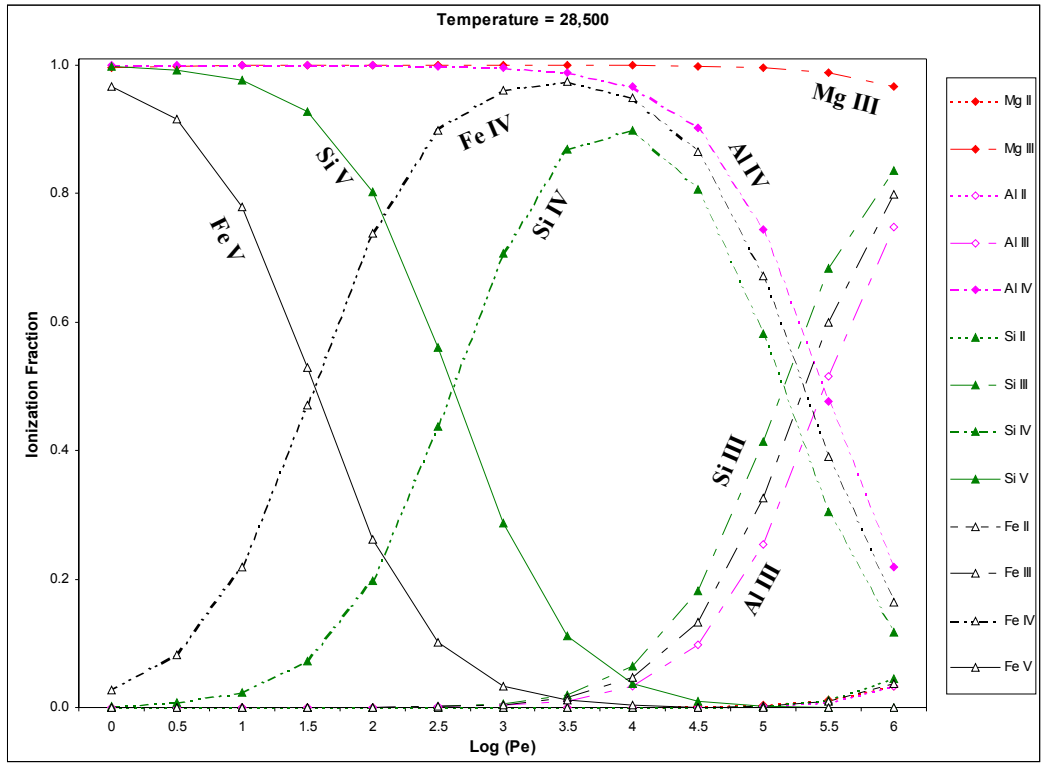

FIG. C.17 - *Ionization Fraction of Mg, Al, Si, and Fe at 28,500 K*

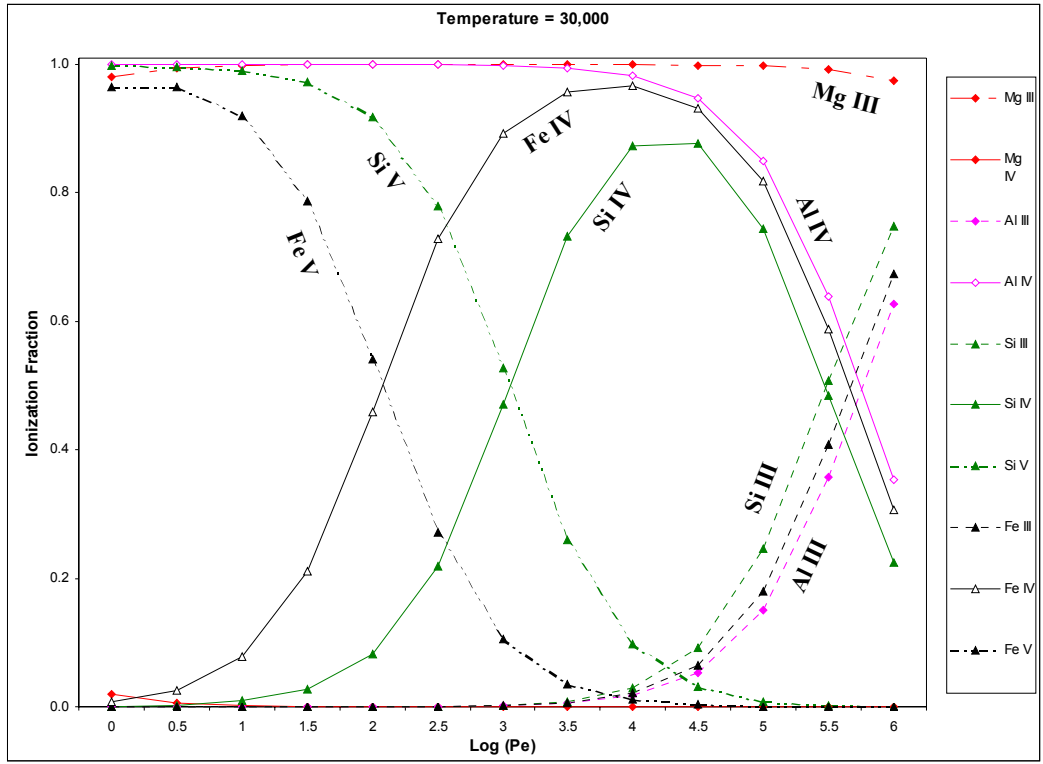

FIG. C.18 - *Ionization Fraction of Mg, Al, Si, and Fe at 30,000 K*

# APPENDIX D.1

## Continuum Flux

TABLE D.1.1.
Large Aperture SWP Continuum Flux Measurements (x 10E-10) (ergs/cm²/s/Å)

| SWP Graph ID | Aper | Phase | 1194 Å | 1243 Å | 1263 Å | 1308 Å | 1350-1400 Å | 1400-1450 Å | 1450-1500 Å | 1500-1550 Å | 1550-1600 Å | 1600-1650 Å | 1650-1700 Å | 1700-1750 Å | 1750-1800 Å | 1800-1850 Å | 1850-1900 Å | 1900-1950 Å | 1950-2000 Å |
|---|---|---|---|---|---|---|---|---|---|---|---|---|---|---|---|---|---|---|---|
| 3  | L | 0.0052 | 1.95  | 4.90  | 7.50  | 8.65  | 8.49  | 8.60  | 7.90  | 7.50  | 7.01  | 6.60  | 6.40  | 6.29  | 5.99  | 6.15  | 6.13  | 6.00  | 5.60  |
| 5  | L | 0.0141 | 3.50  | 6.85  | 8.92  | 10.27 | 10.20 | 10.10 | 10.00 | 9.80  | 9.00  | 9.00  | 8.40  | 8.00  | 7.40  | 7.50  | 7.10  | 7.30  | 6.80  |
| 7  | L | 0.0204 | 3.53  | 8.18  | 11.50 | 13.10 | 12.90 | 13.00 | 12.10 | 11.00 | 10.30 | 9.60  | 9.48  | 9.00  | 8.50  | 8.70  | 8.60  | 8.49  | 8.00  |
| 8  | L | 0.0378 | 4.80  | 12.90 | 19.10 | 22.17 | 20.79 | 22.00 | 19.60 | 19.50 | 17.40 | 16.40 | 15.90 | 15.50 | 14.00 | 14.70 | 14.40 | 14.20 | 13.00 |
| 10 | L | 0.0855 | 12.00 | 23.40 | 31.20 | 33.60 | 31.90 | 33.30 | 31.40 | 30.80 | 28.00 | 27.50 | 26.40 | 24.60 | 23.00 | 23.40 | 22.70 | 22.30 | 20.30 |
| 11 | L | 0.1317 | 13.00 | 23.90 | 30.70 | 34.97 | 31.90 | 34.30 | 32.20 | 31.20 | 28.30 | 27.80 | 26.50 | 24.50 | 23.40 | 23.80 | 22.60 | 22.20 | 20.50 |
| 12 | L | 0.1538 | 13.00 | 21.60 | 29.90 | 34.54 | 31.90 | 32.90 | 31.90 | 30.80 | 27.80 | 26.90 | 26.30 | 24.00 | 22.40 | 22.10 | 22.30 | 22.30 | 20.00 |
| 13 | L | 0.1625 | 10.60 | 20.60 | 27.80 | 31.50 | 31.80 | 32.00 | 30.80 | 30.70 | 27.90 | 26.50 | 26.50 | 23.60 | 21.90 | 21.80 | 21.80 | 21.80 | 18.80 |
| 14 | L | 0.1889 | 9.29  | 22.80 | 31.20 | 34.80 | 34.00 | 35.00 | 31.90 | 30.90 | 27.80 | 26.00 | 25.60 | 23.30 | 23.00 | 23.30 | 22.60 | 22.10 | 19.90 |
| 15 | L | 0.3209 | 13.00 | 24.00 | 30.80 | 34.22 | 32.20 | 33.80 | 32.00 | 30.60 | 28.90 | 28.00 | 27.20 | 24.50 | 23.40 | 23.90 | 22.80 | 22.70 | 20.60 |
| 17 | L | 0.4790 | 9.00  | 22.70 | 31.90 | 34.90 | 32.80 | 33.90 | 32.10 | 30.40 | 27.49 | 26.00 | 24.90 | 23.00 | 22.50 | 22.20 | 22.60 | 22.30 | 19.70 |
| 19 | L | 0.5454 | 9.70  | 23.00 | 30.70 | 34.90 | 33.80 | 33.90 | 32.00 | 29.80 | 26.90 | 25.70 | 24.90 | 24.00 | 22.50 | 22.80 | 22.70 | 22.40 | 20.30 |
| 20 | L | 0.6326 | 12.00 | 23.00 | 33.30 | 33.00 | 34.10 | 34.90 | 32.30 | 30.40 | 27.00 | 25.50 | 25.20 | 23.60 | 22.40 | 22.50 | 22.60 | 22.00 | 20.00 |
| 22 | L | 0.6452 | 14.00 | 23.80 | 32.70 | 34.67 | 34.10 | 35.80 | 32.80 | 31.50 | 27.20 | 26.80 | 25.00 | 23.40 | 22.30 | 22.80 | 22.30 | 22.30 | 20.30 |
| 23 | L | 0.6522 | 15.00 | 23.70 | 31.20 | 34.81 | 32.60 | 32.90 | 31.80 | 31.40 | 28.30 | 27.90 | 26.30 | 24.50 | 23.60 | 23.30 | 22.80 | 22.70 | 21.20 |
| 24 | L | 0.6595 | 12.90 | 22.70 | 31.60 | 34.36 | 34.50 | 34.90 | 32.90 | 31.70 | 27.80 | 26.50 | 25.20 | 24.20 | 22.60 | 22.60 | 22.10 | 22.40 | 20.90 |
| 25 | L | 0.6704 | 12.00 | 22.00 | 31.90 | 34.97 | 34.80 | 34.90 | 33.00 | 32.00 | 27.70 | 26.80 | 25.50 | 23.80 | 23.40 | 23.00 | 22.60 | 22.40 | 20.90 |
| 26 | L | 0.6772 | 10.70 | 21.00 | 32.40 | 34.77 | 34.10 | 34.90 | 32.90 | 32.20 | 27.70 | 26.50 | 25.00 | 23.90 | 23.00 | 22.60 | 22.90 | 22.30 | 20.50 |
| 28 | L | 0.7784 | 10.10 | 23.00 | 33.00 | 34.90 | 33.40 | 34.20 | 31.90 | 32.20 | 26.80 | 25.90 | 24.50 | 23.00 | 22.80 | 22.80 | 22.10 | 22.50 | 20.00 |
| 29 | L | 0.7869 | 13.80 | 24.00 | 32.40 | 33.90 | 33.00 | 33.40 | 32.30 | 31.40 | 28.90 | 27.30 | 26.90 | 24.00 | 23.60 | 23.60 | 22.80 | 22.40 | 20.80 |

TABLE D.1.1 -*continued*

| SWP Graph ID | Aper | Phase | 1194 Å | 1243 Å | 1263 Å | 1308 Å | 1350-1400 Å | 1400-1450 Å | 1450-1500 Å | 1500-1550 Å | 1550-1600 Å | 1600-1650 Å | 1650-1700 Å | 1700-1750 Å | 1750-1800 Å | 1800-1850 Å | 1850-1900 Å | 1900-1950 Å | 1950-2000 Å |
|---|---|---|---|---|---|---|---|---|---|---|---|---|---|---|---|---|---|---|---|
| 30 | L | 0.8005 | 13.00 | 22.80 | 31.10 | 32.90 | 34.40 | 34.00 | 32.90 | 32.00 | 27.90 |       | 26.50 | 25.10 | 23.80 | 22.90 | 23.50 | 22.40 | 22.40 | 20.10 |
| 31 | L | 0.8075 | 14.00 | 24.00 | 31.90 | 33.47 | 33.50 | 33.70 | 32.10 | 31.50 | 28.70 |       | 28.00 | 26.90 | 24.70 | 24.00 | 24.00 | 23.00 | 22.70 | 21.00 |
| 32 | L | 0.8813 | 8.20  | 21.70 | 31.20 | 34.90 | 33.20 | 34.00 | 31.90 | 31.00 | 26.00 |       | 25.40 | 24.10 | 22.50 | 22.90 | 22.80 | 22.30 | 22.60 | 20.50 |
| 33 | L | 0.8866 | 13.00 | 22.70 | 30.40 | 33.80 | 31.70 | 31.80 | 30.90 | 30.90 | 27.50 |       | 27.20 | 25.00 | 23.30 | 22.60 | 22.00 | 21.30 | 21.60 | 20.20 |
| 34 | L | 0.8904 | 11.90 | 19.50 | 29.00 | 26.50 | 28.50 | 29.50 | 29.70 | 29.50 | 25.60 |       | 25.10 | 23.90 | 20.60 | 21.90 | 21.50 | 21.30 | 20.50 | 19.60 |
| 35 | L | 0.9009 | 14.00 | 21.90 | 30.00 | 34.20 | 31.00 | 32.00 | 31.00 | 30.90 | 27.90 |       | 27.00 | 26.50 | 23.40 | 22.10 | 22.40 | 21.70 | 21.30 | 19.80 |
| 36 | L | 0.9070 | 13.30 | 21.70 | 29.90 | 32.80 | 30.20 | 32.10 | 31.10 | 29.80 | 26.50 |       | 26.40 | 24.90 | 22.50 | 22.10 | 21.70 | 21.80 | 21.00 | 20.30 |
| 37 | L | 0.9149 | 12.90 | 23.00 | 30.10 | 34.10 | 31.30 | 33.00 | 31.90 | 31.60 | 28.90 |       | 28.00 | 26.50 | 23.60 | 22.80 | 22.60 | 21.90 | 21.90 | 20.50 |
| 38 | L | 0.9217 | 11.30 | 20.90 | 28.30 | 31.20 | 30.30 | 32.00 | 31.90 | 30.90 | 27.90 |       | 27.00 | 26.00 | 23.30 | 22.70 | 22.30 | 21.90 | 21.60 | 20.40 |
| 39 | L | 0.9218 | 12.00 | 21.90 | 29.90 | 33.20 | 31.60 | 32.30 | 31.80 | 30.30 | 26.90 | 26.80 | 25.90 | 23.40 | 22.60 | 22.30 |       | 22.20 | 21.50 | 20.90 |
| 40 | L | 0.9247 | 9.50  | 21.20 | 28.90 | 32.20 | 30.30 | 32.00 | 31.90 | 30.60 | 27.30 | 26.90 | 25.80 | 23.50 | 23.00 | 22.50 |       | 22.60 | 22.00 | 20.50 |
| 41 | L | 0.9299 | 12.90 | 22.60 | 30.80 | 33.90 | 31.30 | 32.90 | 32.00 | 30.70 | 27.80 | 27.20 | 26.10 | 22.90 | 22.40 | 22.00 |       | 21.90 | 22.00 | 20.40 |
| 42 | L | 0.9358 | 11.80 | 21.00 | 27.90 | 32.70 | 31.00 | 31.50 | 30.80 | 29.70 | 27.10 | 25.80 | 25.50 | 22.80 | 22.00 | 22.00 |       | 21.90 | 21.60 | 19.90 |
| 43 | L | 0.9388 | 9.63  | 19.10 | 28.00 | 29.10 | 28.60 | 30.30 | 29.80 | 28.70 | 25.30 | 24.50 | 23.00 | 21.60 | 21.20 | 20.70 |       | 20.40 | 20.50 | 18.90 |
| 44 | L | 0.9397 | 11.00 | 20.90 | 28.40 | 31.80 | 29.60 | 31.50 | 30.00 | 30.00 | 27.20 | 26.20 | 25.60 | 23.00 | 21.80 | 22.10 |       | 22.10 | 21.00 | 19.60 |
| 45 | L | 0.9471 | 11.00 | 20.60 | 28.20 | 31.40 | 28.90 | 30.30 | 29.30 | 29.30 | 25.70 | 25.20 | 24.50 | 22.10 | 21.50 | 20.80 |       | 20.40 | 19.90 | 19.40 |
| 46 | L | 0.9571 | 8.48  | 16.00 | 21.70 | 24.60 | 23.00 | 24.10 | 23.30 | 23.60 | 20.60 | 20.90 | 20.00 | 18.30 | 17.30 | 16.70 |       | 16.80 | 16.50 | 15.60 |
| 47 | L | 0.9573 | 9.00  | 16.70 | 23.00 | 26.20 | 24.30 | 26.40 | 24.70 | 24.70 | 21.90 | 21.60 | 20.30 | 18.90 | 18.00 | 17.60 |       | 17.40 | 17.30 | 16.20 |
| 48 | L | 0.9636 | 10.00 | 15.90 | 21.90 | 24.50 | 23.00 | 23.90 | 23.10 | 22.60 | 20.30 | 19.90 | 19.40 | 17.40 | 16.90 | 16.00 |       | 16.20 | 15.90 | 14.90 |
| 50 | L | 0.9712 | 5.40  | 10.60 | 14.90 | 17.00 | 15.70 | 16.50 | 16.30 | 16.30 | 14.40 | 14.30 | 13.30 | 12.50 | 12.00 | 12.10 |       | 12.00 | 11.70 | 11.00 |
| 51 | L | 0.9777 | 4.73  | 9.22  | 13.00 | 15.10 | 14.00 | 15.20 | 14.90 | 14.50 | 12.90 | 12.90 | 12.40 | 11.60 | 10.80 | 11.00 |       | 10.80 | 10.80 | 9.80 |
| 53 | L | 0.9791 | 5.90  | 10.20 | 14.20 | 16.30 | 15.10 | 15.70 | 15.00 | 15.30 | 14.00 | 13.40 | 12.70 | 11.60 | 11.10 | 10.90 |       | 10.60 | 10.60 | 10.30 |
| 54 | L | 0.9797 | 3.90  | 7.60  | 12.00 | 12.40 | 11.90 | 12.60 | 12.10 | 12.00 | 11.10 | 10.80 | 10.70 | 9.40  | 9.30  | 9.20  |       | 8.80  | 8.70  | 8.50 |
| 58 | L | 0.9947 | 1.60  | 5.40  | 7.73  | 9.10  | 8.60  | 9.00  | 8.60  | 8.70  | 7.80  | 7.80  | 7.30  | 6.90  | 6.60  | 6.70  |       | 6.57  | 6.50  | 6.10 |

TABLE D.1.2
Small Aperture SWP Continuum Flux Measurements (ergs/cm$^2$/s/Å)

| SWP Graph ID | Aper | Phase | 1194 Å | 1243 Å | 1263 Å | 1308 Å | 1350-1400 Å | 1400-1450 Å | 1450-1500 Å | 1500-1550 Å | 1550-1600 Å | 1600-1650 Å | 1650-1700 Å | 1700-1750 Å | 1750-1800 Å | 1800-1850 Å | 1850-1900 Å | 1900-1950 Å | 1950-2000 Å |
|---|---|---|---|---|---|---|---|---|---|---|---|---|---|---|---|---|---|---|---|
| 1  | S | 0.0015 | 1.46 | 2.75  | 4.34  | 4.55  | 4.71  | 4.94  | 4.73  | 4.72  | 4.34  | 4.19  | 4.04  | 3.88  | 3.84  | 3.59  | 3.67  | 3.66  | 3.56  |
| 2  | S | 0.0031 | 0.92 | 1.86  | 2.81  | 3.14  | 3.09  | 3.23  | 3.30  | 3.29  | 2.89  | 2.65  | 2.60  | 2.41  | 2.45  | 2.37  | 2.34  | 2.38  | 2.35  |
| 4  | S | 0.0053 | 0.59 | 1.17  | 1.89  | 2.16  | 2.12  | 2.14  | 2.09  | 2.12  | 1.89  | 1.83  | 1.77  | 1.66  | 1.67  | 1.59  | 1.59  | 1.65  | 1.59  |
| 6  | S | 0.0183 | 0.60 | 1.14  | 1.73  | 1.94  | 1.92  | 1.99  | 1.92  | 1.95  | 1.74  | 1.68  | 1.67  | 1.54  | 1.52  | 1.49  | 1.49  | 1.52  | 1.47  |
| 9  | S | 0.0816 | 6.38 | 12.66 | 16.92 | 17.46 | 16.98 | 17.22 | 16.81 | 16.35 | 14.90 | 14.39 | 13.95 | 12.92 | 12.64 | 12.41 | 12.52 | 12.82 | 11.86 |
| 16 | S | 0.3297 | 5.10 | 9.72  | 14.70 | 16.15 | 15.54 | 15.80 | 14.93 | 14.82 | 13.63 | 12.66 | 12.76 | 11.20 | 11.23 | 10.42 | 10.86 | 11.27 | 10.23 |
| 18 | S | 0.4927 | 4.19 | 7.61  | 10.95 | 11.79 | 11.44 | 12.00 | 11.33 | 11.28 | 10.27 | 9.79  | 9.42  | 8.61  | 8.35  | 8.17  | 8.09  | 8.21  | 7.59  |
| 21 | S | 0.6361 | 7.02 | 12.54 | 18.16 | 17.72 | 20.26 | 21.20 | 19.76 | 20.34 | 17.56 | 17.52 | 16.69 | 15.23 | 15.09 | 14.22 | 14.92 | 14.40 | 13.49 |
| 27 | S | 0.7674 | 6.51 | 11.78 | 17.82 | 17.65 | 18.01 | 18.87 | 17.38 | 17.42 | 15.88 | 15.42 | 14.66 | 13.55 | 12.95 | 12.53 | 12.96 | 12.67 | 12.54 |
|    | S | 0.9669 | 1.72 | 3.18  | 4.43  | 4.53  | 4.64  | 4.91  | 4.76  | 4.81  | 4.23  | 4.07  | 3.96  | 3.59  | 3.63  | 3.44  | 3.56  | 3.60  | 3.47  |
| 52 | S | 0.9791 | 1.43 | 3.29  | 4.76  | 5.56  | 5.35  | 5.78  | 5.51  | 5.59  | 4.99  | 4.78  | 4.68  | 4.14  | 4.32  | 4.03  | 4.13  | 4.16  | 4.09  |
| 55 | S | 0.9874 | 1.21 | 2.55  | 3.81  | 4.40  | 4.41  | 4.65  | 4.53  | 4.58  | 3.97  | 3.88  | 3.73  | 3.42  | 3.54  | 3.36  | 3.47  | 3.50  | 3.41  |
| 56 | S | 0.9880 | 1.28 | 2.67  | 4.32  | 4.65  | 4.79  | 5.10  | 4.74  | 4.87  | 4.32  | 4.31  | 4.10  | 3.72  | 3.80  | 3.64  | 3.78  | 3.65  | 3.59  |
| 57 | S | 0.9915 | 0.43 | 0.80  | 1.18  | 1.34  | 1.34  | 1.41  | 1.36  | 1.32  | 1.23  | 1.15  | 1.06  | 0.97  | 1.00  | 0.96  | 0.95  | 0.97  | 0.92  |
| 59 | S | 0.9951 | 0.92 | 2.28  | 3.25  | 3.75  | 3.81  | 3.95  | 3.78  | 3.95  | 3.48  | 3.32  | 3.28  | 2.96  | 3.10  | 3.00  | 3.02  | 2.99  | 2.89  |

TABLE D.1.3
Large APERTURE SWP Normalized Continuum Flux Measurements (ergs/cm$^2$/s/Å)

| SWP Graph ID | Aper | Phase | 1194 Å | 1243 Å | 1263 Å | 1308 Å | 1350-1400 Å | 1400-1450 Å | 1450-1500 Å | 1500-1550 Å | 1550-1600 Å | 1600-1650 Å | 1650-1700 Å | 1700-1750 Å | 1750-1800 Å | 1800-1850 Å | 1850-1900 Å | 1900-1950 Å | 1950-2000 Å |
|---|---|---|---|---|---|---|---|---|---|---|---|---|---|---|---|---|---|---|---|
| 3  | L | 0.0052 | 0.15 | 0.21 | 0.24 | 0.25 | 0.25 | 0.25 | 0.25 | 0.24 | 0.25 | 0.24 | 0.25 | 0.26 | 0.26 | 0.26 | 0.27 | 0.27 | 0.28 |
| 5  | L | 0.0141 | 0.27 | 0.29 | 0.29 | 0.30 | 0.30 | 0.30 | 0.31 | 0.32 | 0.32 | 0.32 | 0.32 | 0.33 | 0.32 | 0.32 | 0.32 | 0.33 | 0.33 |
| 7  | L | 0.0204 | 0.27 | 0.35 | 0.37 | 0.38 | 0.39 | 0.38 | 0.38 | 0.36 | 0.37 | 0.35 | 0.36 | 0.37 | 0.37 | 0.37 | 0.38 | 0.38 | 0.39 |
| 8  | L | 0.0378 | 0.38 | 0.55 | 0.61 | 0.64 | 0.62 | 0.64 | 0.61 | 0.63 | 0.62 | 0.59 | 0.61 | 0.64 | 0.61 | 0.62 | 0.64 | 0.64 | 0.64 |
| 10 | L | 0.0855 | 1.02 | 1.00 | 1.00 | 0.94 | 0.96 | 0.98 | 0.98 | 1.00 | 1.01 | 1.01 | 1.01 | 1.02 | 1.00 | 0.99 | 0.99 | 1.00 | 1.00 |
| 11 | L | 0.1317 | 1.03 | 1.02 | 0.99 | 1.01 | 0.95 | 1.01 | 1.00 | 0.99 | 1.00 | 0.98 | 1.01 | 1.01 | 1.01 | 1.01 | 1.00 | 1.00 | 1.01 |
| 12 | L | 0.1538 | 1.00 | 0.92 | 0.96 | 1.00 | 0.95 | 0.96 | 0.99 | 0.99 | 1.00 | 0.98 | 1.01 | 0.99 | 0.97 | 0.94 | 0.98 | 1.00 | 0.98 |
| 13 | L | 0.1625 | 0.83 | 0.86 | 0.89 | 0.92 | 0.95 | 0.94 | 0.96 | 0.98 | 0.99 | 0.97 | 1.01 | 0.98 | 0.95 | 0.92 | 0.95 | 0.98 | 0.93 |
| 14 | L | 0.1889 | 0.72 | 0.99 | 1.00 | 1.00 | 1.02 | 1.01 | 0.99 | 0.99 | 0.99 | 0.95 | 0.98 | 0.97 | 1.00 | 0.99 | 1.00 | 0.99 | 0.98 |
| 15 | L | 0.3209 | 1.02 | 1.02 | 0.98 | 0.99 | 0.99 | 0.98 | 1.00 | 1.00 | 1.03 | 1.03 | 1.04 | 1.01 | 1.01 | 1.01 | 1.01 | 1.02 | 1.01 |
| 17 | L | 0.4790 | 0.71 | 0.96 | 1.02 | 1.00 | 1.01 | 0.99 | 0.99 | 0.99 | 0.98 | 0.95 | 0.96 | 0.95 | 0.97 | 0.95 | 1.00 | 1.00 | 0.97 |
| 19 | L | 0.5454 | 0.78 | 0.98 | 0.99 | 1.02 | 1.00 | 0.99 | 0.99 | 0.96 | 0.96 | 0.93 | 0.96 | 0.99 | 0.97 | 0.97 | 0.98 | 1.01 | 0.99 |
| 20 | L | 0.6326 | 0.95 | 0.99 | 1.04 | 0.96 | 1.02 | 1.01 | 1.00 | 0.98 | 0.97 | 0.93 | 0.96 | 0.98 | 0.97 | 0.96 | 1.00 | 0.98 | 0.98 |
| 22 | L | 0.6452 | 1.10 | 1.00 | 1.06 | 1.01 | 1.03 | 1.04 | 1.01 | 1.02 | 0.97 | 0.97 | 0.96 | 0.98 | 0.96 | 0.96 | 0.98 | 1.00 | 0.99 |
| 23 | L | 0.6522 | 1.18 | 1.00 | 1.00 | 1.01 | 0.98 | 0.96 | 0.98 | 1.01 | 1.02 | 1.01 | 1.02 | 1.01 | 1.02 | 0.99 | 1.00 | 1.02 | 1.04 |
| 24 | L | 0.6595 | 1.00 | 0.96 | 1.01 | 0.99 | 1.03 | 1.01 | 1.02 | 1.03 | 1.00 | 0.97 | 0.97 | 1.00 | 0.98 | 0.96 | 0.98 | 1.01 | 1.01 |
| 25 | L | 0.6704 | 0.90 | 0.93 | 1.02 | 1.01 | 1.04 | 1.01 | 1.02 | 1.03 | 1.00 | 0.97 | 0.97 | 0.99 | 1.01 | 0.97 | 1.01 | 1.01 | 1.02 |
| 26 | L | 0.6772 | 0.84 | 0.90 | 1.05 | 1.00 | 1.02 | 1.02 | 1.02 | 1.04 | 0.99 | 0.97 | 0.97 | 0.99 | 0.99 | 0.96 | 1.01 | 1.00 | 1.00 |
| 28 | L | 0.7784 | 0.78 | 0.98 | 1.03 | 1.01 | 1.00 | 1.00 | 0.99 | 0.97 | 0.96 | 0.95 | 0.94 | 0.96 | 0.97 | 0.97 | 0.98 | 1.01 | 0.99 |
| 29 | L | 0.7869 | 1.08 | 1.02 | 1.04 | 0.98 | 0.99 | 0.98 | 1.00 | 1.01 | 1.03 | 1.00 | 1.02 | 1.00 | 1.02 | 1.00 | 1.00 | 1.00 | 1.01 |
| 30 | L | 0.8005 | 0.98 | 0.98 | 0.99 | 0.95 | 1.03 | 0.99 | 1.02 | 1.03 | 1.00 | 0.97 | 0.97 | 0.99 | 0.99 | 0.99 | 0.99 | 1.01 | 0.98 |
| 31 | L | 0.8075 | 1.09 | 1.03 | 1.02 | 0.96 | 1.00 | 0.98 | 1.00 | 1.02 | 1.02 | 1.03 | 1.03 | 1.03 | 1.04 | 1.02 | 1.01 | 1.02 | 1.03 |
| 32 | L | 0.8813 | 0.66 | 0.93 | 1.01 | 1.01 | 1.00 | 0.99 | 0.99 | 1.00 | 0.93 | 0.93 | 0.93 | 0.94 | 0.98 | 0.97 | 0.98 | 1.01 | 1.00 |
| 33 | L | 0.8866 | 1.04 | 0.96 | 0.98 | 0.97 | 0.95 | 0.94 | 0.96 | 0.99 | 0.99 | 0.99 | 0.97 | 0.96 | 0.97 | 0.94 | 0.94 | 0.97 | 0.99 |
| 34 | L | 0.8904 | 0.94 | 0.82 | 0.95 | 0.77 | 0.85 | 0.86 | 0.92 | 0.94 | 0.92 | 0.92 | 0.91 | 0.85 | 0.94 | 0.90 | 0.94 | 0.92 | 0.96 |
| 35 | L | 0.9009 | 1.06 | 0.94 | 0.98 | 0.99 | 0.91 | 0.93 | 0.97 | 0.98 | 1.00 | 0.99 | 1.01 | 0.97 | 0.96 | 0.96 | 0.96 | 0.96 | 0.97 |
| 36 | L | 0.9070 | 1.01 | 0.92 | 0.96 | 0.90 | 0.91 | 0.93 | 0.97 | 0.96 | 0.95 | 0.96 | 0.96 | 0.93 | 0.96 | 0.92 | 0.96 | 0.95 | 0.99 |
| 37 | L | 0.9149 | 1.01 | 0.99 | 0.97 | 0.99 | 0.94 | 0.96 | 0.98 | 1.02 | 1.03 | 1.02 | 1.02 | 0.98 | 0.99 | 0.95 | 0.97 | 0.99 | 1.00 |
| 38 | L | 0.9217 | 0.90 | 0.89 | 0.91 | 0.90 | 0.91 | 0.94 | 0.99 | 0.99 | 0.99 | 0.99 | 0.99 | 0.96 | 0.98 | 0.95 | 0.96 | 0.97 | 1.00 |

TABLE D.1.3 - continued

| SWP Graph ID | Aper | Phase | 1194 Å | 1243 Å | 1263 Å | 1308 Å | 1350-1400 Å | 1400-1450 Å | 1450-1500 Å | 1500-1550 Å | 1550-1600 Å | 1600-1650 Å | 1650-1700 Å | 1700-1750 Å | 1750-1800 Å | 1800-1850 Å | 1850-1900 Å | 1900-1950 Å | 1950-2000 Å |
|---|---|---|---|---|---|---|---|---|---|---|---|---|---|---|---|---|---|---|---|
| 39 | L | 0.9218 | 0.87 | 0.92 | 0.96 | 0.97 | 0.95 | 0.94 | 0.99 | 0.98 | 0.96 | 0.98 | 0.99 | 0.97 | 0.97 | 0.95 | 0.98 | 0.97 | 1.02 |
| 40 | L | 0.9247 | 0.76 | 0.91 | 0.93 | 0.94 | 0.91 | 0.94 | 0.99 | 0.99 | 0.97 | 0.98 | 0.99 | 0.97 | 0.99 | 0.96 | 1.00 | 0.99 | 1.00 |
| 41 | L | 0.9299 | 0.99 | 0.96 | 0.99 | 0.98 | 0.94 | 0.95 | 0.99 | 0.99 | 1.00 | 0.99 | 1.00 | 0.95 | 0.97 | 0.93 | 0.97 | 0.99 | 1.00 |
| 42 | L | 0.9358 | 0.90 | 0.88 | 0.90 | 0.94 | 0.93 | 0.91 | 0.95 | 0.95 | 0.97 | 0.94 | 0.97 | 0.95 | 0.96 | 0.93 | 0.97 | 0.97 | 0.97 |
| 43 | L | 0.9388 | 0.76 | 0.81 | 0.90 | 0.84 | 0.85 | 0.88 | 0.92 | 0.93 | 0.91 | 0.89 | 0.89 | 0.90 | 0.91 | 0.88 | 0.89 | 0.92 | 0.93 |
| 44 | L | 0.9397 | 0.87 | 0.88 | 0.92 | 0.92 | 0.89 | 0.92 | 0.94 | 0.96 | 0.97 | 0.96 | 0.97 | 0.96 | 0.94 | 0.94 | 0.97 | 0.95 | 0.96 |
| 45 | L | 0.9471 | 0.87 | 0.88 | 0.91 | 0.90 | 0.86 | 0.89 | 0.91 | 0.94 | 0.92 | 0.92 | 0.94 | 0.92 | 0.92 | 0.88 | 0.90 | 0.90 | 0.95 |
| 46 | L | 0.9571 | 0.65 | 0.68 | 0.70 | 0.71 | 0.69 | 0.70 | 0.72 | 0.76 | 0.74 | 0.75 | 0.76 | 0.76 | 0.74 | 0.71 | 0.74 | 0.74 | 0.76 |
| 47 | L | 0.9573 | 0.72 | 0.71 | 0.74 | 0.76 | 0.73 | 0.77 | 0.77 | 0.80 | 0.78 | 0.79 | 0.78 | 0.78 | 0.78 | 0.75 | 0.77 | 0.78 | 0.79 |
| 48 | L | 0.9636 | 0.77 | 0.68 | 0.70 | 0.71 | 0.69 | 0.70 | 0.72 | 0.73 | 0.73 | 0.72 | 0.74 | 0.72 | 0.72 | 0.68 | 0.72 | 0.72 | 0.73 |
| 50 | L | 0.9712 | 0.42 | 0.45 | 0.48 | 0.49 | 0.47 | 0.48 | 0.51 | 0.52 | 0.52 | 0.52 | 0.51 | 0.52 | 0.52 | 0.51 | 0.53 | 0.52 | 0.53 |
| 51 | L | 0.9777 | 0.37 | 0.39 | 0.42 | 0.43 | 0.42 | 0.45 | 0.46 | 0.46 | 0.47 | 0.47 | 0.48 | 0.48 | 0.47 | 0.47 | 0.48 | 0.49 | 0.48 |
| 53 | L | 0.9791 | 0.47 | 0.43 | 0.46 | 0.47 | 0.45 | 0.46 | 0.46 | 0.50 | 0.50 | 0.49 | 0.49 | 0.48 | 0.48 | 0.47 | 0.47 | 0.48 | 0.50 |
| 54 | L | 0.9797 | 0.30 | 0.32 | 0.36 | 0.36 | 0.36 | 0.37 | 0.38 | 0.39 | 0.40 | 0.40 | 0.41 | 0.39 | 0.40 | 0.39 | 0.39 | 0.39 | 0.42 |
| 58 | L | 0.9947 | 0.19 | 0.23 | 0.25 | 0.27 | 0.26 | 0.26 | 0.27 | 0.28 | 0.28 | 0.29 | 0.28 | 0.29 | 0.29 | 0.28 | 0.29 | 0.29 | 0.30 |

TABLE D.1.4
Large APERTURE LWP/LWR Continuum Flux Measurements (ergs/cm$^2$/s/Å)

| LWPR Graph ID | Aper | Phase | 1850-1900 Å | 1900-1950 Å | 1950-2000 Å | 2000-2050 Å | 2025-2075 Å | 2075-2125 Å | 2125-2175 Å | 2175-2225 Å | 2225-2275 Å | 2275-2325 Å | 2325-2375 Å | 2375-2425 Å | 2425-2475 Å | ... |
|---|---|---|---|---|---|---|---|---|---|---|---|---|---|---|---|---|
| 2  | L | 0.0034 | 5.25  | 4.95  | 5.35  | 5.58  | 5.58  | 5.60  | 5.03  | 4.59  | 4.51  | 4.38  | 3.77  | 3.61  | 3.61  |  |
| 6  | L | 0.0220 | 6.13  | 7.50  | 8.10  | 7.96  | 8.58  | 8.38  | 7.55  | 6.98  | 6.59  | 6.37  | 5.52  | 5.21  | 5.37  |  |
| 7  | L | 0.0390 | 15.00 | 16.30 | 15.30 | 15.08 | 14.41 | 14.02 | 12.40 | 10.98 | 10.20 | 9.90  | 8.40  | 8.10  | 8.00  |  |
| 9  | L | 0.0867 | 20.25 | 18.80 | 19.70 | 20.60 | 20.45 | 19.90 | 18.20 | 16.59 | 15.95 | 15.80 | 13.30 | 12.75 | 12.70 |  |
| 10 | L | 0.1303 | 23.00 | 22.13 | 21.20 | 20.94 | 20.81 | 19.37 | 17.70 | 15.80 | 15.10 | 14.50 | 12.90 | 12.20 | 12.30 |  |
| 11 | L | 0.1547 | 23.50 | 20.98 | 20.15 | 21.12 | 20.50 | 20.78 | 19.30 | 17.80 | 16.30 | 16.50 | 13.85 | 12.70 | 12.75 |  |
| 12 | L | 0.1903 | 16.50 | 19.00 | 19.40 | 20.22 | 20.91 | 19.82 | 18.20 | 16.98 | 16.30 | 16.10 | 13.83 | 13.22 | 13.10 |  |
| 13 | L | 0.3196 | 18.00 | 18.10 | 20.10 | 21.21 | 20.99 | 20.18 | 18.70 | 17.12 | 15.95 | 16.75 | 13.50 | 13.12 | 13.15 |  |
| 16 | L | 0.4778 | 19.25 | 16.90 | 19.30 | 19.14 | 19.80 | 19.90 | 17.60 | 16.59 | 15.60 | 15.30 | 13.40 | 12.65 | 13.00 |  |
| 18 | L | 0.5463 | 19.35 | 18.50 | 20.35 | 20.83 | 20.62 | 20.21 | 18.80 | 17.38 | 16.00 | 16.15 | 13.95 | 13.20 | 13.28 |  |
| 19 | L | 0.6343 | 17.50 | 20.10 | 20.80 | 21.20 | 20.48 | 20.25 | 19.00 | 17.11 | 16.15 | 16.05 | 13.75 | 13.00 | 12.95 |  |
| 22 | L | 0.7798 | 17.50 | 18.40 | 20.00 | 20.19 | 20.71 | 20.22 | 18.60 | 17.20 | 16.05 | 15.95 | 13.25 | 12.75 | 12.95 |  |
| 23 | L | 0.7884 | 17.25 | 18.50 | 20.25 | 19.99 | 19.63 | 19.12 | 18.05 | 16.80 | 15.95 | 15.70 | 13.20 | 12.50 | 12.45 |  |
| 24 | L | 0.8825 | 19.30 | 19.85 | 20.20 | 20.37 | 20.60 | 20.69 | 18.45 | 17.35 | 16.20 | 15.90 | 13.40 | 12.55 | 12.80 |  |
| 25 | L | 0.8878 | 14.70 | 18.50 | 19.30 | 19.33 | 19.89 | 19.51 | 18.20 | 16.59 | 16.50 | 16.05 | 13.63 | 12.45 | 12.40 |  |
| 26 | L | 0.8915 | 16.60 | 18.00 | 19.95 | 20.87 | 20.59 | 20.82 | 19.75 | 16.99 | 17.95 | 17.10 | 13.83 | 13.20 | 12.35 |  |
| 27 | L | 0.9018 | 16.95 | 17.85 | 19.95 | 19.14 | 19.33 | 19.30 | 18.20 | 16.90 | 16.10 | 15.20 | 12.80 | 12.40 | 12.50 |  |
| 28 | L | 0.9082 | 16.13 | 18.10 | 19.90 | 21.13 | 20.72 | 21.51 | 19.82 | 16.50 | 17.20 | 17.10 | 13.40 | 13.60 | 13.60 |  |
| 29 | L | 0.9166 | 19.13 | 19.80 | 20.15 | 19.95 | 19.90 | 19.75 | 18.70 | 17.50 | 16.10 | 16.00 | 13.15 | 11.90 | 12.40 |  |
| 30 | L | 0.9227 | 16.60 | 16.85 | 19.00 | 20.04 | 19.80 | 20.81 | 18.80 | 16.59 | 16.00 | 16.70 | 12.85 | 13.05 | 13.15 |  |
| 31 | L | 0.9259 | 18.50 | 17.00 | 19.30 | 20.22 | 20.67 | 20.51 | 18.70 | 16.65 | 15.85 | 15.75 | 12.70 | 12.10 | 12.45 |  |
| 32 | L | 0.9323 | 18.70 | 17.40 | 19.70 | 19.31 | 18.86 | 19.00 | 18.00 | 17.20 | 15.85 | 15.00 | 12.65 | 11.60 | 12.10 |  |
| 33 | L | 0.9367 | 15.90 | 16.30 | 18.25 | 19.13 | 18.79 | 19.90 | 18.07 | 16.89 | 15.85 | 15.60 | 12.30 | 11.85 | 12.70 |  |
| 34 | L | 0.9409 | 24.25 | 24.60 | 22.70 | 21.67 | 21.23 | 19.81 | 17.60 | 16.45 | 14.65 | 13.90 | 12.13 | 11.10 | 12.05 |  |
| 35 | L | 0.9488 | 15.05 | 16.50 | 18.05 | 17.90 | 18.19 | 17.79 | 16.20 | 15.80 | 14.45 | 13.95 | 11.45 | 10.50 | 10.90 |  |
| 36 | L | 0.9581 | 14.05 | 14.00 | 13.95 | 14.39 | 14.31 | 14.89 | 13.50 | 15.58 | 11.50 | 11.45 | 9.25  | 8.80  | 9.65  |  |
| 37 | L | 0.9590 | 20.00 | 19.65 | 18.00 | 17.57 | 17.62 | 16.81 | 14.55 | 12.01 | 11.30 | 11.30 | 9.70  | 9.00  | 9.40  |  |
| 38 | L | 0.9650 | 12.50 | 14.50 | 14.25 | 14.48 | 14.45 | 14.28 | 13.08 | 13.10 | 11.30 | 11.30 | 9.50  | 8.40  | 8.65  |  |
| 40 | L | 0.9721 | 7.00  | 8.50  | 9.80  | 10.34 | 10.46 | 10.80 | 9.80  | 11.98 | 8.45  | 8.50  | 6.64  | 6.50  | 6.79  |  |
| 41 | L | 0.9794 | 8.13  | 8.70  | 9.60  | 9.54  | 9.49  | 9.60  | 8.40  | 8.59  | 7.30  | 6.90  | 5.95  | 5.45  | 5.86  |  |
| 42 | L | 0.9801 | 7.60  | 8.50  | 9.70  | 9.35  | 9.44  | 9.35  | 8.90  | 8.08  | 7.80  | 7.35  | 6.10  | 5.78  | 5.90  |  |

TABLE D.1.4 - *continued*

| LWPR Graph ID | Aper | Phase | 2475-2525 Å | 2525-2575 Å | 2575-2625 Å | 2625-2675 Å | 2675-2725 Å | 2725-2775 Å | 2775-2825 Å | 2825-2875 Å | 2875-2925 Å | 2925-2975 Å | 2975-3025 Å | 3025-3075 Å | 3075-3125 Å | 3125-3175 Å | 3175-3225 Å | 3225-3275 Å |
|---|---|---|---|---|---|---|---|---|---|---|---|---|---|---|---|---|---|---|
| 2  | L | 0.0034 | 3.74  | 3.65  | 3.84  | 3.82  | 3.65  | 3.38  | 3.43  | 3.25  | 3.12  | 3.02  | 3.01  | 2.85 | 2.77 | 2.42 | 2.07 | 1.45 |
| 6  | L | 0.0220 | 5.50  | 5.46  | 5.70  | 5.63  | 5.27  | 5.01  | 4.84  | 4.60  | 4.40  | 4.33  | 4.19  | 4.05 | 3.82 | 3.50 | 3.00 | 2.18 |
| 7  | L | 0.0390 | 8.55  | 8.60  | 8.95  | 8.80  | 8.29  | 7.90  | 7.90  | 7.30  | 6.95  | 6.90  | 6.62  | 6.42 | 6.98 | 5.19 | 4.25 | 2.38 |
| 9  | L | 0.0867 | 13.40 | 12.90 | 13.30 | 13.10 | 12.40 | 11.70 | 11.40 | 10.80 | 10.15 | 9.90  | 9.80  | 9.50 | 9.00 | 7.85 | 6.60 | 4.90 |
| 10 | L | 0.1303 | 12.35 | 12.80 | 12.90 | 13.05 | 12.50 | 12.00 | 11.90 | 11.05 | 10.60 | 10.40 | 10.10 | 9.70 | 9.10 | 7.90 | 6.60 | 4.10 |
| 11 | L | 0.1547 | 12.95 | 12.55 | 12.80 | 13.00 | 12.30 | 11.70 | 11.60 | 10.80 | 10.50 | 10.30 | 9.80  | 9.60 | 9.10 | 7.75 | 5.40 | 2.40 |
| 12 | L | 0.1903 | 13.83 | 13.30 | 13.55 | 13.20 | 12.50 | 11.85 | 11.70 | 10.95 | 10.50 | 10.20 | 10.00 | 9.50 | 9.05 | 7.95 | 6.85 | 5.10 |
| 13 | L | 0.3196 | 13.92 | 13.40 | 13.60 | 13.29 | 12.70 | 12.05 | 12.00 | 11.05 | 10.30 | 10.00 | 10.00 | 9.65 | 9.25 | 8.00 | 6.90 | 5.00 |
| 16 | L | 0.4778 | 13.00 | 13.00 | 13.10 | 12.70 | 12.30 | 11.50 | 11.50 | 10.65 | 10.40 | 10.20 | 9.65  | 9.30 | 8.80 | 7.90 | 7.20 | 4.80 |
| 18 | L | 0.5463 | 14.15 | 13.25 | 13.50 | 13.20 | 12.70 | 11.80 | 11.65 | 11.10 | 10.60 | 10.20 | 9.95  | 9.60 | 9.25 | 7.95 | 6.70 | 4.80 |
| 19 | L | 0.6343 | 13.82 | 13.30 | 13.40 | 13.30 | 12.70 | 12.30 | 11.80 | 11.30 | 10.60 | 10.55 | 9.93  | 9.55 | 9.30 | 8.10 | 6.55 | 4.60 |
| 22 | L | 0.7798 | 13.40 | 13.20 | 13.20 | 13.05 | 12.45 | 11.80 | 11.70 | 10.70 | 10.60 | 10.10 | 9.90  | 9.55 | 9.25 | 8.05 | 6.80 | 5.55 |
| 23 | L | 0.7884 | 13.25 | 12.90 | 12.80 | 13.00 | 12.20 | 11.45 | 11.40 | 10.80 | 10.40 | 10.20 | 9.75  | 9.55 | 8.85 | 8.00 | 6.55 | 4.75 |
| 24 | L | 0.8825 | 13.50 | 13.10 | 13.20 | 13.20 | 12.70 | 11.70 | 11.80 | 10.80 | 10.55 | 10.30 | 10.00 | 9.70 | 9.20 | 8.30 | 6.80 | 5.00 |
| 25 | L | 0.8878 | 12.80 | 12.30 | 12.45 | 12.70 | 12.30 | 11.35 | 11.00 | 10.50 | 10.30 | 9.95  | 9.55  | 9.20 | 8.60 | 7.80 | 5.50 | 3.40 |
| 26 | L | 0.8915 | 12.70 | 12.15 | 12.90 | 12.75 | 11.90 | 11.00 | 10.70 | 10.45 | 10.35 | 10.00 | 9.55  | 9.20 | 8.70 | 7.80 | 6.70 | 5.40 |
| 27 | L | 0.9018 | 13.10 | 12.58 | 12.55 | 12.60 | 12.10 | 11.60 | 11.25 | 10.65 | 10.30 | 9.90  | 9.55  | 9.20 | 8.80 | 7.45 | 5.90 | 3.10 |
| 28 | L | 0.9082 | 13.35 | 12.60 | 13.10 | 13.05 | 11.90 | 11.70 | 11.80 | 10.70 | 10.40 | 10.10 | 9.85  | 9.20 | 9.00 | 7.90 | 6.70 | 4.80 |
| 29 | L | 0.9166 | 12.65 | 12.65 | 12.45 | 12.50 | 11.90 | 11.30 | 11.50 | 10.65 | 10.20 | 9.90  | 9.65  | 9.35 | 8.55 | 7.45 | 5.75 | 2.75 |
| 30 | L | 0.9227 | 13.15 | 13.00 | 13.20 | 13.20 | 12.20 | 11.70 | 11.70 | 10.90 | 10.50 | 10.20 | 10.00 | 9.30 | 8.85 | 8.00 | 5.80 | 4.35 |
| 31 | L | 0.9259 | 13.00 | 12.60 | 13.00 | 12.70 | 12.40 | 11.30 | 11.60 | 10.75 | 10.10 | 10.10 | 9.75  | 9.60 | 8.80 | 7.95 | 6.45 | 4.90 |
| 32 | L | 0.9323 | 12.45 | 11.90 | 12.30 | 12.20 | 11.80 | 11.10 | 10.80 | 10.45 | 10.10 | 9.70  | 9.55  | 9.20 | 8.35 | 7.20 | 5.35 | 3.10 |
| 33 | L | 0.9367 | 12.75 | 12.30 | 12.70 | 12.60 | 12.00 | 11.00 | 11.10 | 10.50 | 10.00 | 9.60  | 9.60  | 8.85 | 8.75 | 7.35 | 5.50 | 3.85 |
| 34 | L | 0.9409 | 12.15 | 12.05 | 12.60 | 12.30 | 11.89 | 11.00 | 11.25 | 10.25 | 9.85  | 9.90  | 9.55  | 9.30 | 8.65 | 7.75 | 6.00 | 3.90 |
| 35 | L | 0.9488 | 11.55 | 11.30 | 11.30 | 11.40 | 11.00 | 10.20 | 10.00 | 9.50  | 9.10  | 8.90  | 8.70  | 8.50 | 7.90 | 6.55 | 5.70 | 3.30 |
| 36 | L | 0.9581 | 9.50  | 9.10  | 9.20  | 9.35  | 9.15  | 8.30  | 8.35  | 7.80  | 7.55  | 7.30  | 7.18  | 6.75 | 6.65 | 5.40 | 3.83 | 2.45 |
| 37 | L | 0.9590 | 9.65  | 9.60  | 9.90  | 9.90  | 9.70  | 8.70  | 8.95  | 8.30  | 8.10  | 7.89  | 7.65  | 7.40 | 6.85 | 6.20 | 4.56 | 2.90 |
| 38 | L | 0.9650 | 8.90  | 8.55  | 8.70  | 8.90  | 8.50  | 7.90  | 7.95  | 7.42  | 7.15  | 7.08  | 6.95  | 6.60 | 6.00 | 5.14 | 3.88 | 2.40 |
| 40 | L | 0.9721 | 6.84  | 6.70  | 6.70  | 6.79  | 6.55  | 6.03  | 6.00  | 5.68  | 5.60  | 5.31  | 5.31  | 4.89 | 4.69 | 4.19 | 3.38 | 2.39 |
| 41 | L | 0.9794 | 6.18  | 5.94  | 6.10  | 5.95  | 5.88  | 5.26  | 5.50  | 5.08  | 4.95  | 4.86  | 4.82  | 4.51 | 4.27 | 3.65 | 2.90 | 2.23 |
| 42 | L | 0.9801 | 6.14  | 5.95  | 6.14  | 6.11  | 6.01  | 5.46  | 5.53  | 5.23  | 5.18  | 5.02  | 4.83  | 4.62 | 4.30 | 3.70 | 2.90 | 2.10 |

TABLE D.1.5
Small APERTURE LWP/LWR Continuum Flux Measurements (ergs/cm2/s/Å)

| LWPR Graph ID | Aper | Phase | 1850-1900 Å | 1900-1950 Å | 1950-2000 Å | 2000-2050 Å | 2025-2075 Å | 2075-2125 Å | 2125-2175 Å | 2175-2225 Å | 2225-2275 Å | 2275-2325 Å | 2325-2375 Å | 2375-2425 Å | 2425-2475 Å | 2475-2525 Å | 2525-2575 Å | ... |
|---|---|---|---|---|---|---|---|---|---|---|---|---|---|---|---|---|---|---|
| 1  | S | 0.0030 | 3.25  | 2.78  | 3.00  | 3.04  | 2.99  | 3.00  | 2.81  | 2.57  | 2.36  | 2.39  | 2.07 | 2.04 | 2.05 | 2.09 | 2.02 | |
| 3  | S | 0.0034 | 2.08  | 1.81  | 1.84  | 1.87  | 1.81  | 1.78  | 1.80  | 1.61  | 1.55  | 1.56  | 1.28 | 1.30 | 1.33 | 1.35 | 1.26 | |
| 4  | S | 0.0049 | 3.01  | 2.50  | 2.82  | 2.84  | 2.84  | 2.81  | 2.69  | 2.35  | 2.09  | 2.28  | 1.86 | 1.70 | 1.83 | 1.88 | 1.76 | |
| 5  | S | 0.0165 | 0.83  | 0.65  | 0.69  | 0.65  | 0.59  | 0.58  | 0.54  | 0.46  | 0.45  | 0.47  | 0.40 | 0.42 | 0.43 | 0.42 | 0.41 | |
| 8  | S | 0.0801 | 10.95 | 8.75  | 9.30  | 9.72  | 9.26  | 9.51  | 9.05  | 7.91  | 7.65  | 6.85  | 5.90 | 6.09 | 6.07 | 6.01 | 5.92 | |
| 14 | S | 0.3200 | 9.50  | 8.25  | 8.60  | 8.41  | 8.24  | 8.52  | 7.90  | 7.10  | 6.81  | 6.67  | 5.78 | 5.77 | 5.83 | 5.69 | 5.34 | |
| 15 | S | 0.3280 | 10.50 | 8.95  | 9.60  | 9.63  | 9.49  | 9.71  | 8.95  | 8.11  | 7.85  | 7.56  | 6.59 | 6.55 | 6.57 | 6.40 | 6.40 | |
| 17 | S | 0.4939 | 12.90 | 9.95  | 10.90 | 11.34 | 10.80 | 11.41 | 10.95 | 9.45  | 9.10  | 8.75  | 7.65 | 7.60 | 7.20 | 7.25 | 7.03 | |
| 20 | S | 0.6520 | 13.95 | 12.00 | 11.20 | 12.56 | 11.55 | 12.60 | 12.10 | 10.84 | 10.20 | 9.90  | 8.70 | 8.50 | 8.40 | 8.65 | 8.30 | |
| 21 | S | 0.7685 | 12.70 | 10.15 | 10.95 | 11.27 | 10.90 | 11.21 | 10.55 | 9.30  | 8.75  | 8.20  | 7.13 | 7.30 | 6.70 | 7.05 | 6.90 | |
| 39 | S | 0.9683 | 5.13  | 4.30  | 4.84  | 4.77  | 4.62  | 4.87  | 4.58  | 3.98  | 3.78  | 3.74  | 3.05 | 3.17 | 3.03 | 3.22 | 3.12 | |
| 43 | S | 0.9895 | 3.08  | 2.60  | 2.84  | 2.98  | 2.92  | 3.10  | 2.82  | 2.53  | 2.45  | 2.44  | 2.14 | 2.07 | 2.08 | 2.12 | 2.20 | |
| 44 | S | 0.9901 | 0.74  | 0.37  | 0.35  | 0.44  | 0.45  | 0.51  | 0.51  | 0.51  | 0.54  | 0.50  | 0.42 | 0.37 | 0.41 | 0.41 | 0.40 | |

TABLE D.1.5 - *continued*

| LWPR Graph ID | Aper | Phase | 2575-2625 Å | 2625-2675 Å | 2675-2725 Å | 2725-2775 Å | 2775-2825 Å | 2825-2875 Å | 2875-2925 Å | 2925-2975 Å | 2975-3025 Å | 3025-3075 Å | 3075-3125 Å | 3125-3175 Å | 3175-3225 Å | 3225-3275 Å |
|---|---|---|---|---|---|---|---|---|---|---|---|---|---|---|---|---|
| 1  | S | 0.0030 | 1.99 | 2.08 | 1.91 | 1.90 | 1.84 | 1.74 | 1.73 | 1.62 | 1.61 | 1.59 | 1.62 | 1.56 | 1.33 | 1.40 |
| 3  | S | 0.0034 | 1.27 | 1.33 | 1.26 | 1.16 | 1.20 | 1.12 | 1.11 | 1.06 | 1.06 | 1.06 | 1.06 | 1.03 | 0.97 | 1.19 |
| 4  | S | 0.0049 | 1.81 | 1.86 | 1.79 | 1.66 | 1.68 | 1.63 | 1.62 | 1.53 | 1.54 | 1.54 | 1.56 | 1.47 | 1.43 | 1.48 |
| 5  | S | 0.0165 | 0.40 | 0.42 | 0.38 | 0.36 | 0.34 | 0.32 | 0.32 | 0.32 | 0.30 | 0.30 | 0.30 | 0.29 | 0.33 | 0.38 |
| 8  | S | 0.0801 | 5.90 | 6.11 | 5.71 | 5.55 | 5.45 | 5.05 | 5.01 | 4.85 | 4.69 | 4.62 | 4.73 | 4.52 | 4.15 | 4.40 |
| 14 | S | 0.3200 | 5.05 | 5.47 | 4.99 | 4.91 | 4.98 | 4.60 | 4.58 | 4.28 | 4.28 | 4.11 | 4.12 | 3.94 | 3.55 | 4.02 |
| 15 | S | 0.3280 | 6.30 | 6.36 | 5.81 | 5.71 | 5.80 | 5.21 | 5.21 | 5.00 | 4.89 | 4.68 | 4.80 | 4.52 | 3.99 | 4.12 |
| 17 | S | 0.4939 | 6.88 | 7.10 | 6.68 | 6.58 | 6.40 | 5.89 | 5.78 | 5.55 | 5.52 | 5.37 | 5.30 | 5.39 | 5.05 | 5.25 |
| 20 | S | 0.6520 | 8.05 | 8.25 | 7.70 | 7.70 | 7.60 | 6.85 | 6.80 | 6.51 | 6.37 | 6.28 | 6.15 | 5.85 | 5.90 | 5.80 |
| 21 | S | 0.7685 | 6.74 | 6.83 | 6.48 | 6.22 | 6.20 | 5.75 | 5.61 | 5.35 | 5.35 | 5.17 | 5.18 | 5.15 | 4.49 | 5.35 |
| 39 | S | 0.9683 | 3.14 | 3.14 | 3.02 | 2.88 | 2.94 | 2.67 | 2.72 | 2.59 | 2.60 | 2.56 | 2.55 | 2.51 | 2.37 | 2.42 |
| 43 | S | 0.9895 | 1.96 | 2.05 | 1.90 | 1.86 | 1.86 | 1.69 | 1.66 | 1.62 | 1.59 | 1.56 | 1.54 | 1.48 | 1.34 | 1.41 |
| 44 | S | 0.9901 | 0.40 | 0.39 | 0.38 | 0.37 | 0.36 | 0.36 | 0.34 | 0.33 | 0.32 | 0.34 | 0.36 | 0.40 | 0.41 | 0.61 |

TABLE D.1.6
Large APERTURE LWP/LWR Normalized Continuum Flux Measurements (ergs/cm2/s/Å)

| LWPR Graph ID | Aper | Phase | 1850-1900 Å | 1900-1950 Å | 1950-2000 Å | 2000-2050 Å | 2025-2075 Å | 2075-2125 Å | 2125-2175 Å | 2175-2225 Å | 2225-2275 Å | 2275-2325 Å | 2325-2375 Å | 2375-2425 Å | 2425-2475 Å | 2475-2525 Å | 2525-2575 Å | ... |
|---|---|---|---|---|---|---|---|---|---|---|---|---|---|---|---|---|---|---|
| 2 | L | 0.0034 | 0.29 | 0.27 | 0.26 | 0.27 | 0.27 | 0.28 | 0.27 | 0.27 | 0.28 | 0.27 | 0.28 | 0.28 | 0.28 | 0.27 | 0.28 | |
| 6 | L | 0.0220 | 0.33 | 0.41 | 0.40 | 0.38 | 0.41 | 0.41 | 0.41 | 0.41 | 0.41 | 0.39 | 0.41 | 0.40 | 0.41 | 0.40 | 0.41 | |
| 7 | L | 0.0390 | 0.82 | 0.88 | 0.76 | 0.73 | 0.69 | 0.69 | 0.67 | 0.64 | 0.64 | 0.61 | 0.62 | 0.62 | 0.62 | 0.62 | 0.65 | |
| 9 | L | 0.0867 | 1.10 | 1.02 | 0.97 | 0.99 | 0.99 | 0.99 | 0.98 | 0.97 | 1.00 | 0.98 | 0.98 | 0.98 | 0.98 | 0.97 | 0.97 | |
| 10 | L | 0.1303 | 1.25 | 1.20 | 1.05 | 1.01 | 1.01 | 0.96 | 0.95 | 0.92 | 0.94 | 0.90 | 0.95 | 0.94 | 0.95 | 0.89 | 0.97 | |
| 11 | L | 0.1547 | 1.28 | 1.13 | 1.00 | 1.02 | 1.02 | 1.03 | 1.04 | 1.04 | 1.02 | 1.02 | 1.02 | 0.98 | 0.98 | 0.94 | 0.95 | |
| 12 | L | 0.1903 | 0.90 | 1.03 | 0.96 | 0.97 | 0.97 | 0.98 | 0.98 | 0.99 | 1.00 | 1.00 | 0.99 | 1.02 | 1.01 | 1.00 | 1.00 | |
| 13 | L | 0.3196 | 0.98 | 0.98 | 0.99 | 1.02 | 1.02 | 1.00 | 1.00 | 1.00 | 1.02 | 1.04 | 1.01 | 1.01 | 1.01 | 1.01 | 1.01 | |
| 16 | L | 0.4778 | 1.05 | 0.91 | 0.95 | 0.92 | 0.95 | 0.99 | 0.95 | 0.97 | 0.98 | 0.95 | 0.99 | 0.97 | 1.00 | 0.94 | 0.98 | |
| 18 | L | 0.5463 | 1.05 | 1.00 | 1.00 | 1.00 | 0.99 | 1.00 | 1.01 | 1.02 | 1.00 | 1.00 | 1.03 | 1.02 | 1.02 | 1.03 | 1.00 | |
| 19 | L | 0.6343 | 0.95 | 1.09 | 1.03 | 1.02 | 1.02 | 1.00 | 1.02 | 1.00 | 1.00 | 0.99 | 1.01 | 1.00 | 1.00 | 1.00 | 1.00 | |
| 22 | L | 0.7798 | 0.95 | 0.99 | 0.99 | 0.97 | 0.97 | 1.00 | 1.00 | 1.01 | 1.00 | 0.99 | 0.97 | 0.98 | 1.00 | 0.97 | 1.00 | |
| 23 | L | 0.7884 | 0.94 | 1.00 | 1.00 | 1.00 | 0.96 | 0.95 | 0.97 | 0.98 | 1.00 | 0.97 | 0.97 | 0.96 | 0.96 | 0.96 | 0.97 | |
| 24 | L | 0.8825 | 1.05 | 1.07 | 1.00 | 0.98 | 0.98 | 1.02 | 0.99 | 1.01 | 1.01 | 0.99 | 0.99 | 0.97 | 0.98 | 0.98 | 0.99 | |
| 25 | L | 0.8878 | 0.80 | 1.00 | 0.95 | 0.93 | 0.93 | 0.97 | 0.96 | 0.99 | 1.03 | 0.99 | 1.00 | 0.96 | 0.95 | 0.93 | 0.93 | |
| 26 | L | 0.8915 | 0.90 | 0.97 | 0.99 | 1.01 | 1.01 | 1.03 | 0.99 | 0.99 | 1.12 | 1.06 | 1.02 | 1.02 | 0.95 | 0.92 | 0.92 | |
| 27 | L | 0.9018 | 0.92 | 0.96 | 0.99 | 0.92 | 0.92 | 0.96 | 0.98 | 0.96 | 1.01 | 0.94 | 0.94 | 0.96 | 0.95 | 0.95 | 0.95 | |
| 28 | L | 0.9082 | 0.88 | 0.98 | 0.98 | 1.02 | 1.02 | 1.06 | 1.06 | 1.02 | 1.08 | 1.06 | 0.99 | 1.05 | 1.05 | 0.97 | 0.95 | |
| 29 | L | 0.9166 | 1.04 | 1.07 | 1.00 | 0.96 | 0.96 | 0.98 | 0.98 | 1.07 | 1.01 | 0.99 | 0.97 | 0.92 | 0.92 | 0.92 | 0.96 | |
| 30 | L | 0.9227 | 0.90 | 0.91 | 0.94 | 0.97 | 0.97 | 1.03 | 1.01 | 1.01 | 1.00 | 1.04 | 0.94 | 1.00 | 1.01 | 0.95 | 0.98 | |
| 31 | L | 0.9259 | 1.01 | 0.92 | 0.95 | 0.97 | 1.00 | 1.02 | 1.01 | 1.01 | 0.99 | 0.98 | 0.93 | 0.93 | 0.96 | 0.94 | 0.95 | |
| 32 | L | 0.9323 | 1.02 | 0.94 | 0.97 | 0.93 | 0.91 | 0.94 | 0.97 | 0.99 | 0.99 | 0.93 | 0.93 | 0.89 | 0.93 | 0.90 | 0.90 | |
| 33 | L | 0.9367 | 0.86 | 0.88 | 0.90 | 0.92 | 0.91 | 0.99 | 0.97 | 0.96 | 0.99 | 0.97 | 0.90 | 0.91 | 0.98 | 0.92 | 0.93 | |
| 34 | L | 0.9409 | 1.32 | 1.33 | 1.12 | 1.04 | 1.02 | 0.98 | 0.95 | 0.92 | 0.92 | 0.86 | 0.89 | 0.85 | 0.93 | 0.88 | 0.91 | |
| 35 | L | 0.9488 | 0.82 | 0.89 | 0.89 | 0.86 | 0.88 | 0.88 | 0.87 | 0.91 | 0.90 | 0.86 | 0.84 | 0.81 | 0.84 | 0.84 | 0.85 | |
| 36 | L | 0.9581 | 0.76 | 0.76 | 0.69 | 0.69 | 0.69 | 0.74 | 0.73 | 0.70 | 0.72 | 0.71 | 0.68 | 0.68 | 0.74 | 0.69 | 0.69 | |
| 37 | L | 0.9590 | 1.09 | 1.06 | 0.89 | 0.85 | 0.85 | 0.83 | 0.78 | 0.77 | 0.71 | 0.70 | 0.71 | 0.69 | 0.72 | 0.70 | 0.73 | |
| 38 | L | 0.9650 | 0.68 | 0.78 | 0.70 | 0.70 | 0.70 | 0.71 | 0.70 | 0.70 | 0.71 | 0.70 | 0.70 | 0.65 | 0.67 | 0.64 | 0.65 | |
| 40 | L | 0.9721 | 0.38 | 0.46 | 0.48 | 0.50 | 0.50 | 0.53 | 0.53 | 0.50 | 0.53 | 0.53 | 0.49 | 0.50 | 0.52 | 0.50 | 0.51 | |
| 41 | L | 0.9794 | 0.44 | 0.47 | 0.47 | 0.46 | 0.46 | 0.48 | 0.45 | 0.46 | 0.46 | 0.43 | 0.44 | 0.42 | 0.45 | 0.45 | 0.45 | |
| 42 | L | 0.9801 | 0.41 | 0.46 | 0.48 | 0.45 | 0.45 | 0.46 | 0.48 | 0.47 | 0.49 | 0.46 | 0.45 | 0.44 | 0.45 | 0.44 | 0.45 | |

TABLE D.1.6 -continued

| LWPR Graph ID | Aper | Phase | 2575-2625 Å | 2625-2675 Å | 2675-2725 Å | 2725-2775 Å | 2775-2825 Å | 2825-2875 Å | 2875-2925 Å | 2925-2975 Å | 2975-3025 Å | 3025-3075 Å | 3075-3125 Å | 3125-3175 Å | 3175-3225 Å | 3225-3275 Å |
|---|---|---|---|---|---|---|---|---|---|---|---|---|---|---|---|---|
| 2  | L | 0.0034 | 0.29 | 0.29 | 0.29 | 0.28 | 0.29 | 0.29 | 0.30 | 0.29 | 0.30 | 0.30 | 0.30 | 0.30 | 0.30 | 0.30 |
| 6  | L | 0.0220 | 0.42 | 0.43 | 0.42 | 0.42 | 0.41 | 0.42 | 0.42 | 0.42 | 0.42 | 0.42 | 0.42 | 0.44 | 0.44 | 0.44 |
| 7  | L | 0.0390 | 0.67 | 0.67 | 0.66 | 0.66 | 0.66 | 0.66 | 0.66 | 0.67 | 0.67 | 0.67 | 0.76 | 0.65 | 0.63 | 0.49 |
| 9  | L | 0.0867 | 0.99 | 0.99 | 0.99 | 0.98 | 0.96 | 0.98 | 0.97 | 0.97 | 0.99 | 0.99 | 0.98 | 0.98 | 0.97 | 1.00 |
| 10 | L | 0.1303 | 0.96 | 0.99 | 1.00 | 1.01 | 1.00 | 1.00 | 1.01 | 1.02 | 1.02 | 1.01 | 1.00 | 0.99 | 0.97 | 0.84 |
| 11 | L | 0.1547 | 0.95 | 0.98 | 0.98 | 0.98 | 0.97 | 0.98 | 1.00 | 1.01 | 0.99 | 1.00 | 1.00 | 1.00 | 0.79 | 0.49 |
| 12 | L | 0.1903 | 1.01 | 1.00 | 1.00 | 0.99 | 0.98 | 0.99 | 1.00 | 0.98 | 1.01 | 0.99 | 0.99 | 0.99 | 1.01 | 1.04 |
| 13 | L | 0.3196 | 1.01 | 1.01 | 1.01 | 1.01 | 1.01 | 1.00 | 1.01 | 1.00 | 1.01 | 1.01 | 1.01 | 1.00 | 1.01 | 1.02 |
| 16 | L | 0.4778 | 0.97 | 0.96 | 0.98 | 0.96 | 0.97 | 0.96 | 0.98 | 0.98 | 0.97 | 0.97 | 0.96 | 0.99 | 1.06 | 0.98 |
| 18 | L | 0.5463 | 1.00 | 1.00 | 1.01 | 0.99 | 0.98 | 1.00 | 0.99 | 1.00 | 1.00 | 1.00 | 1.01 | 0.99 | 0.99 | 0.98 |
| 19 | L | 0.6343 | 1.00 | 1.01 | 1.01 | 1.03 | 0.99 | 1.02 | 1.01 | 1.03 | 1.00 | 0.99 | 1.02 | 1.01 | 0.96 | 0.94 |
| 22 | L | 0.7798 | 0.98 | 0.99 | 0.99 | 0.99 | 0.98 | 0.97 | 1.01 | 0.99 | 1.00 | 0.99 | 1.01 | 1.01 | 1.00 | 1.13 |
| 23 | L | 0.7884 | 0.95 | 0.98 | 0.97 | 0.96 | 0.96 | 0.98 | 0.99 | 1.00 | 0.98 | 0.99 | 0.97 | 1.00 | 0.96 | 0.97 |
| 24 | L | 0.8825 | 0.98 | 1.00 | 1.01 | 0.98 | 0.99 | 0.98 | 1.01 | 1.01 | 1.01 | 1.01 | 1.01 | 1.04 | 1.00 | 1.02 |
| 25 | L | 0.8878 | 0.93 | 0.96 | 0.98 | 0.95 | 0.92 | 0.95 | 0.98 | 0.97 | 0.96 | 0.96 | 0.94 | 0.98 | 0.81 | 0.69 |
| 26 | L | 0.8915 | 0.96 | 0.97 | 0.95 | 0.92 | 0.90 | 0.95 | 0.99 | 0.98 | 0.96 | 0.96 | 0.95 | 0.98 | 0.99 | 1.10 |
| 27 | L | 0.9018 | 0.93 | 0.95 | 0.96 | 0.97 | 0.95 | 0.96 | 0.98 | 0.97 | 0.96 | 0.96 | 0.96 | 0.93 | 0.87 | 0.63 |
| 28 | L | 0.9082 | 0.97 | 0.99 | 0.95 | 0.98 | 0.99 | 0.97 | 0.99 | 0.99 | 0.99 | 0.96 | 0.98 | 0.99 | 0.99 | 0.98 |
| 29 | L | 0.9166 | 0.93 | 0.95 | 0.97 | 0.95 | 0.97 | 0.96 | 0.98 | 0.97 | 0.97 | 0.97 | 0.94 | 0.93 | 0.85 | 0.56 |
| 30 | L | 0.9227 | 0.98 | 1.00 | 1.01 | 0.98 | 0.98 | 0.99 | 1.00 | 1.00 | 1.01 | 0.97 | 0.97 | 1.00 | 0.85 | 0.89 |
| 31 | L | 0.9259 | 0.97 | 0.96 | 0.99 | 0.95 | 0.97 | 0.97 | 0.97 | 0.99 | 0.98 | 1.00 | 0.96 | 0.99 | 0.95 | 1.00 |
| 32 | L | 0.9323 | 0.92 | 0.92 | 0.94 | 0.93 | 0.91 | 0.95 | 0.97 | 0.95 | 0.96 | 0.96 | 0.91 | 0.90 | 0.79 | 0.63 |
| 33 | L | 0.9367 | 0.94 | 0.95 | 0.96 | 0.92 | 0.93 | 0.95 | 0.96 | 0.94 | 0.97 | 0.92 | 0.96 | 0.92 | 0.81 | 0.79 |
| 34 | L | 0.9409 | 0.94 | 0.93 | 0.95 | 0.92 | 0.95 | 0.93 | 0.94 | 0.97 | 0.96 | 0.97 | 0.95 | 0.97 | 0.88 | 0.80 |
| 35 | L | 0.9488 | 0.84 | 0.86 | 0.88 | 0.86 | 0.84 | 0.86 | 0.87 | 0.87 | 0.88 | 0.89 | 0.86 | 0.82 | 0.84 | 0.67 |
| 36 | L | 0.9581 | 0.68 | 0.71 | 0.73 | 0.70 | 0.70 | 0.71 | 0.72 | 0.71 | 0.72 | 0.70 | 0.73 | 0.68 | 0.56 | 0.50 |
| 37 | L | 0.9590 | 0.74 | 0.75 | 0.77 | 0.73 | 0.75 | 0.75 | 0.77 | 0.77 | 0.77 | 0.77 | 0.75 | 0.78 | 0.67 | 0.59 |
| 38 | L | 0.9650 | 0.65 | 0.67 | 0.68 | 0.66 | 0.67 | 0.67 | 0.68 | 0.69 | 0.70 | 0.69 | 0.66 | 0.64 | 0.57 | 0.49 |
| 40 | L | 0.9721 | 0.50 | 0.51 | 0.52 | 0.51 | 0.50 | 0.51 | 0.54 | 0.52 | 0.54 | 0.51 | 0.51 | 0.52 | 0.50 | 0.49 |
| 41 | L | 0.9794 | 0.45 | 0.45 | 0.47 | 0.44 | 0.46 | 0.46 | 0.47 | 0.47 | 0.49 | 0.47 | 0.47 | 0.46 | 0.43 | 0.46 |
| 42 | L | 0.9801 | 0.46 | 0.46 | 0.48 | 0.46 | 0.46 | 0.47 | 0.50 | 0.49 | 0.49 | 0.48 | 0.47 | 0.46 | 0.43 | 0.43 |

# APPENDIX D.2

# Radial Velocities

TABLE D.2.1
OBSERVED SWP RADIAL VELOCITIES (km/s)

| SWP Graph ID | Al II 1670.7867 | Al III 1854.716 | *Al III 1862.7895 | C II 1334.5323 | C II 1335.7077 | C IV 1548.185 |
|---|---|---|---|---|---|---|
| 1 | 1.49 | 0.65 | -12.79 | 3.98 | -19.68 | -10.65 |
| 2 | -3.00 | 11.96 | -3.14 | 3.98 | -15.19 | 6.78 |
| 3 | -9.28 | -15.52 | -28.89 | -9.50 | -26.42 | -33.89 |
| 4 | -11.97 | -25.22 | -32.11 | -7.26 | -19.68 | 2.90 |
| 5 | -8.38 | -15.52 | -25.67 | -7.26 | -35.39 | -22.27 |
| 6 | -19.15 | -28.45 | -36.94 | -11.75 | -35.39 | -35.82 |
| 7 | -24.53 | -31.68 | -36.94 | -20.73 | -44.37 | -24.21 |
| 8 | -24.53 | -28.45 | -33.72 | -27.47 | -39.88 | -39.70 |
| 9 | -19.15 | -12.28 | -14.40 | -16.24 | -35.39 | -8.71 |
| 10 | -25.43 | -25.22 | -29.69 | -18.49 | -42.13 | -35.82 |
| 11 | -38.88 | -34.91 | -43.37 | -36.46 | -53.35 | -4.84 |
| 12 | -35.29 | -34.91 | -41.76 | -38.71 | -51.11 | -45.51 |
| 13 | -65.80 | -67.24 | -65.90 | -52.18 | -82.53 | -74.55 |
| 14 | -50.55 | -46.23 | -54.64 | -40.95 | -60.08 | -66.81 |
| 15 | -33.50 | -38.15 | -51.42 | -22.98 | -48.86 | -31.95 |
| 16 | -22.73 | -26.83 | -33.72 | -16.24 | -44.37 | -45.51 |
| 17 | -10.17 | -12.28 | -27.28 | -11.75 | -24.17 | -39.70 |
| 18 | 5.98 | 8.73 | -6.36 | 8.47 | -12.95 | -12.59 |
| 19 | 5.98 | 2.26 | -9.58 | -0.52 | -8.46 | -16.46 |
| 20 | 43.66 | 36.21 | 29.05 | 44.41 | 16.23 | 8.71 |
| 21 | 31.99 | 29.74 | 27.44 | 24.19 | 16.23 | -2.90 |
| 22 | 40.07 | 36.21 | 33.88 | 38.80 | 22.96 | 6.78 |
| 23 | 45.45 | 32.97 | 33.88 | 44.41 | 11.74 | 12.59 |
| 24 | 48.14 | 32.97 | 33.88 | 46.66 | 25.21 | 6.78 |
| 25 | 45.45 | 36.21 | 33.07 | 39.92 | 22.96 | 2.90 |
| 26 | 49.04 | 42.67 | 41.92 | 42.17 | 27.45 | 6.78 |
| 27 | 56.22 | 55.60 | 49.97 | 46.66 | 38.67 | 20.33 |
| 28 | 38.27 | 39.44 | 22.61 | 41.04 | 31.94 | 10.65 |
| 29 | 49.04 | 45.91 | 43.53 | 51.15 | 25.21 | 26.14 |
| 30 | 49.04 | 45.91 | 41.92 | 48.90 | 34.18 | 16.46 |
| 31 | 43.66 | 45.91 | 38.71 | 51.15 | 34.18 | 26.14 |
| 32 | 27.51 | 23.28 | 16.98 | 26.44 | 11.74 | -0.97 |
| 33 | 29.30 | 23.28 | 14.56 | 21.95 | 11.74 | -4.84 |
| 34 | 22.12 | 23.28 | 27.44 | 19.70 | 9.49 | -10.65 |
| 35 | 20.33 | 11.96 | 8.13 | 3.98 | 5.01 | -0.97 |

TABLE D.2.1 - *continued*

| SWP Graph ID | Al II 1670.7867 | Al III 1854.716 | *Al III 1862.7895 | C II 1334.5323 | C II 1335.7077 | C IV 1548.185 |
|---|---|---|---|---|---|---|
| 36 | 7.77 | 11.96 | 8.13 | 8.47 | -1.73 | -4.84 |
| 37 | 20.33 | 21.66 | 6.52 | 8.47 | 2.76 | 22.27 |
| 38 | 31.10 | 42.67 | 32.27 | 36.55 | 11.74 | -4.84 |
| 39 | 13.15 | 15.19 | 1.69 | 8.47 | -1.73 | 22.27 |
| 40 | 20.33 | 28.13 | 17.78 | 21.95 | 2.76 | 24.21 |
| 41 | 16.74 | 16.81 | 8.13 | 11.84 | 5.01 | 12.59 |
| 42 | 9.56 | 2.26 | -6.36 | 3.98 | -1.73 | -10.65 |
| 43 | -1.20 | 29.74 | 17.78 | -0.52 | -10.71 | 26.14 |
| 44 | 16.74 | 16.81 | 11.35 | 12.96 | 0.52 | 22.27 |
| 45 | 14.05 | 10.34 | -3.14 | -0.52 | -6.22 | -8.71 |
| 46 | 16.74 | 16.81 | 8.13 | 12.96 | -3.97 | -0.97 |
| 47 | 23.92 | 18.43 | 12.96 | 17.45 | -3.97 | 10.65 |
| 48 | 21.23 | 11.96 | 3.30 | -0.52 | -3.97 | 2.90 |
| 49 | 33.79 | 53.99 | 49.97 | 33.18 | 11.74 | 18.40 |
| 50 | 14.95 | 8.73 | 1.69 | 15.21 | 0.52 | 2.90 |
| 51 | 18.54 | 21.66 | 8.13 | 3.98 | -8.46 | 10.65 |
| 52 | 16.74 | 42.67 | 33.88 | 28.69 | 9.49 | 26.14 |
| 53 | 14.95 | 20.04 | 1.69 | -0.52 | 0.52 | 6.78 |
| 54 | 10.46 | 8.73 | -3.14 | 12.96 | 0.52 | -10.65 |
| 55 | 5.98 | 26.51 | 27.44 | 17.45 | -6.22 | 14.52 |
| 56 | 15.84 | 18.43 | 11.35 | 19.70 | 9.49 | 0.97 |
| 57 | 4.18 | 13.58 | 1.69 | 8.47 | 15.11 | 0.97 |
| 58 | 5.98 | 3.88 | -14.40 | -2.76 | -12.95 | 18.40 |
| 59 | 2.39 | 23.28 | 17.78 | 10.72 | -19.68 | 45.51 |

TABLE D.2.2
OBSERVED SWP RADIAL VELOCITIES (km/s)

| SWP Graph ID | Fe III 1895.456 | Fe III 1914.056 | Fe III 1926.304 | Fe II 1608.45106 | Fe II 1635.401 | Fe II 1639.40124 | Fe II 1640.15205 |
|---|---|---|---|---|---|---|---|
| 1 | 18.03 | 5.33 | -3.74 | 7.26 | -0.18 | 5.26 | 1.45 |
| 2 | 14.87 | -11.90 | -19.30 | 9.12 | 16.31 | 8.92 | -7.69 |
| 3 | 3.80 | 0.63 | -22.41 | -13.24 | -13.02 | -17.60 | -7.69 |
| 4 | 10.12 | -24.43 | -6.85 | 12.85 | -1.10 | -2.06 | -2.20 |
| 5 | -2.53 | -30.70 | -30.19 | -13.24 | -9.35 | -18.51 | -9.51 |
| 6 | -0.95 | -21.30 | -25.52 | -9.52 | -20.35 | -16.68 | -10.43 |
| 7 | -9.65 | -36.96 | -38.75 | -30.02 | -33.18 | -36.80 | -31.45 |
| 8 | -21.51 | -35.40 | -37.97 | -26.29 | -35.01 | -33.14 | -27.79 |
| 9 | -4.90 | -18.17 | -23.97 | -11.38 | -11.18 | -18.51 | -10.43 |
| 10 | -13.60 | -26.78 | -34.86 | -24.43 | -16.68 | -16.68 | -22.31 |
| 11 | -27.84 | -38.53 | -49.65 | -43.07 | -38.68 | -36.80 | -33.28 |
| 12 | -29.42 | -33.83 | -45.76 | -39.34 | -35.01 | -25.83 | -45.16 |
| 13 | -61.84 | -58.89 | -73.77 | -71.02 | -69.84 | -61.49 | -68.00 |
| 14 | -35.75 | -44.80 | -71.43 | -48.66 | -38.68 | -56.92 | -51.55 |
| 15 | -19.93 | -35.40 | -64.43 | -29.09 | -31.35 | -34.97 | -42.41 |
| 16 | -16.77 | -32.27 | -37.97 | -16.97 | -24.01 | -18.51 | -24.14 |
| 17 | 3.01 | -13.47 | -27.08 | -0.20 | -11.18 | -17.60 | -27.79 |
| 18 | 18.03 | 8.46 | 0.16 | 8.19 | 16.31 | 16.23 | 6.94 |
| 19 | 11.70 | 3.76 | -8.40 | -3.93 | -3.85 | -5.71 | -4.03 |
| 20 | 55.99 | 46.05 | 28.95 | 31.49 | 45.65 | 34.52 | 27.04 |
| 21 | 51.25 | 40.17 | 24.28 | 31.49 | 38.31 | 36.35 | 30.70 |
| 22 | 48.08 | 31.95 | 35.95 | 31.49 | 40.15 | 36.35 | 50.80 |
| 23 | 48.08 | 37.43 | 18.05 | 26.83 | 36.48 | 45.49 | 45.32 |
| 24 | 57.57 | 40.57 | 24.28 | 31.49 | 43.81 | 41.83 | 39.84 |
| 25 | 55.20 | 42.92 | 27.39 | 34.28 | 41.98 | 45.49 | 38.01 |
| 26 | 55.99 | 42.13 | 32.06 | 37.08 | 47.48 | 39.09 | 48.98 |
| 27 | 70.22 | 46.05 | 38.29 | 65.04 | 56.64 | 50.98 | 48.98 |
| 28 | 57.57 | 35.08 | 24.28 | 30.56 | 43.81 | 30.86 | 45.32 |
| 29 | 56.78 | 44.48 | 49.18 | 33.35 | 45.65 | 50.98 | 39.84 |
| 30 | 54.41 | 36.65 | 32.84 | 31.49 | 38.31 | 39.09 | 48.98 |
| 31 | 55.99 | 43.70 | 35.17 | 33.35 | 51.14 | 47.32 | 45.32 |
| 32 | 38.59 | 36.65 | 24.28 | 10.99 | 35.56 | 16.23 | 14.25 |
| 33 | 35.43 | 17.86 | 18.05 | 12.85 | 30.98 | 32.69 | 27.04 |
| 34 | 39.54 | 30.39 | 16.50 | 18.44 | 18.15 | 16.23 | 17.90 |
| 35 | 26.73 | 10.02 | 7.16 | 14.71 | 19.98 | 3.43 | 8.76 |
| 36 | 22.78 | 9.24 | 4.82 | -2.06 | 1.65 | 3.43 | -4.03 |
| 37 | 24.36 | 17.86 | 17.28 | 9.12 | 23.65 | 25.37 | 10.59 |
| 38 | 43.34 | 35.08 | 33.62 | 20.30 | 38.31 | 32.69 | 27.04 |
| 39 | 19.61 | 13.94 | 14.47 | 9.12 | -0.18 | 6.17 | 6.94 |
| 40 | 32.27 | 22.55 | 25.83 | 10.99 | 10.82 | 14.40 | 30.70 |
| 41 | 30.68 | 16.29 | 13.38 | 5.39 | 12.65 | 18.06 | 6.94 |
| 42 | 22.78 | 7.67 | 8.72 | -0.20 | -0.18 | -1.14 | -4.03 |
| 43 | 11.70 | 7.67 | 1.71 | -7.65 | -0.18 | -0.23 | -11.34 |
| 44 | 29.10 | 19.42 | 12.61 | 9.12 | 16.31 | 18.06 | 7.85 |
| 45 | 21.98 | 6.11 | -9.18 | -0.20 | 8.98 | 12.57 | -4.03 |
| 46 | 22.78 | 16.29 | 7.16 | 12.85 | 14.48 | 9.83 | 5.11 |

TABLE D.2.2 - *continued*

| SWP Graph ID | Fe III 1895.456 | Fe III 1914.056 | Fe III 1926.304 | Fe II 1608.45106 | Fe II 1635.401 | Fe II 1639.40124 | Fe II 1640.15205 |
|---|---|---|---|---|---|---|---|
| 47 | 33.85 | 30.39 | 10.27 | 16.58 | 21.81 | 17.15 | 28.87 |
| 48 | 25.94 | 19.42 | 10.27 | 14.71 | 18.15 | 16.23 | 10.59 |
| 49 | 46.50 | 50.75 | 39.84 | 37.08 | 49.31 | 41.83 | 39.84 |
| 50 | 27.52 | 11.59 | 3.27 | 14.71 | 16.31 | 14.40 | 10.59 |
| 51 | 33.85 | 25.69 | 18.05 | 12.85 | 16.31 | 18.06 | 27.04 |
| 52 | 46.50 | 46.83 | 32.06 | 33.35 | 34.65 | 29.03 | 27.04 |
| 53 | 27.52 | 16.29 | -9.96 | 24.03 | 23.65 | 19.89 | 17.90 |
| 54 | 23.57 | 8.46 | 8.72 | 1.67 | 5.32 | 7.09 | 7.85 |
| 55 | 29.10 | 36.65 | 25.83 | 26.83 | 23.65 | 14.40 | 25.21 |
| 56 | 32.27 | 24.12 | 13.38 | 14.71 | 43.81 | 21.72 | 27.04 |
| 57 | 25.94 | 24.12 | 13.38 | 25.90 | 21.81 | 54.63 | 19.73 |
| 58 | 18.82 | 11.59 | -14.63 | 5.39 | 19.06 | 14.40 | 6.94 |
| 59 | 22.78 | -0.94 | -6.85 | 18.44 | 18.15 | 2.52 | 14.25 |

TABLE D.2.3
OBSERVED SWP RADIAL VELOCITIES (km/s)

| SWP Graph ID | Si II 1260.4212 | Si II 1264.7374 | *Si II 1304.37 | Si II 1305.59 | Si II 1309.2769 | Si II 1526.7076 | Si II 1533.432 | *Si II 1808.0117 | *Si II 1816.9796 |
|---|---|---|---|---|---|---|---|---|---|
| 1 | -10.99 | 5.36 | 66.65 | 5.74 | 30.48 | 4.40 | 1.56 | -5.26 | 38.84 |
| 2 | -22.88 | 10.10 | 36.77 | -9.18 | 19.03 | -1.49 | -10.17 | -26.81 | 23.17 |
| 3 | -10.99 | 16.02 | 64.35 | 4.59 | 21.32 | -5.42 | -10.17 | -26.81 | 28.12 |
| 4 | -16.93 | 0.62 | 50.56 | -4.59 | 23.61 | -3.46 | -4.30 | -11.89 | 36.36 |
| 5 | -28.83 | 0.62 | 62.06 | 0.00 | 16.74 | -11.31 | -17.99 | -25.15 | 18.22 |
| 6 | -19.31 | 5.36 | 36.77 | -9.18 | 9.87 | -19.17 | -17.99 | -19.35 | 23.17 |
| 7 | -34.77 | 0.62 | 60.91 | -12.63 | -3.87 | -21.13 | -27.76 | -28.47 | 5.02 |
| 8 | -28.83 | -6.49 | 32.18 | -25.26 | -1.58 | -19.17 | -29.72 | -31.79 | 11.62 |
| 9 | -40.72 | -7.68 | 27.58 | -14.93 | -1.58 | -15.24 | -25.81 | -21.84 | 19.87 |
| 10 | -25.85 | -12.42 | 36.77 | -16.07 | -6.16 | -21.13 | -29.72 | -16.03 | 9.97 |
| 11 | -52.61 | -23.09 | 25.28 | -39.27 | -24.48 | -28.98 | -51.22 | -28.47 | 1.72 |
| 12 | -31.21 | -23.09 | 26.43 | -9.18 | -8.45 | -30.95 | -39.49 | -40.91 | 19.87 |
| 13 | -57.37 | -57.46 | -9.19 | -58.55 | -47.38 | -62.37 | -74.68 | -61.63 | -13.13 |
| 14 | -40.72 | -33.75 | 16.09 | -40.18 | -31.35 | -44.69 | -57.09 | -41.74 | -13.13 |
| 15 | -25.85 | -18.35 | 27.58 | -41.33 | -6.16 | -32.91 | -14.08 | -40.08 | -3.23 |
| 16 | -28.83 | -10.05 | 25.28 | -47.07 | 0.71 | -21.13 | -25.81 | -30.13 | 26.47 |
| 17 | -25.85 | 5.36 | 63.21 | 0.00 | 12.16 | -11.31 | -14.08 | -18.52 | 19.87 |
| 18 | 3.88 | 19.58 | 64.35 | 5.74 | 21.32 | 8.33 | 9.38 | 17.96 | 53.69 |
| 19 | -16.93 | 18.39 | 71.25 | 10.33 | 19.03 | -1.49 | -4.30 | 6.35 | 51.21 |
| 20 | 42.53 | 52.76 | 94.23 | 39.04 | 67.11 | 37.78 | 32.84 | 44.49 | 79.26 |
| 21 | 18.74 | 45.65 | 91.93 | 32.15 | 48.79 | 35.82 | 25.02 | 37.86 | 99.06 |
| 22 | 40.15 | 61.06 | 97.68 | 29.85 | 69.40 | 37.78 | 36.75 | 47.80 | 85.86 |
| 23 | 37.77 | 44.47 | 109.17 | 48.22 | 64.82 | 43.67 | 36.75 | 42.83 | 77.61 |
| 24 | 30.64 | 55.14 | 104.58 | 43.63 | 64.82 | 39.74 | 36.75 | 54.44 | 77.61 |
| 25 | 61.56 | 57.51 | 105.73 | 27.55 | 67.11 | 41.71 | 40.66 | 39.51 | 87.51 |
| 26 | 54.42 | 52.76 | 112.62 | 48.22 | 71.69 | 41.71 | 40.66 | 51.12 | 82.56 |
| 27 | 42.53 | 52.76 | 110.32 | 47.07 | 83.14 | 53.49 | 50.44 | 61.07 | 95.76 |
| 28 | 33.61 | 52.76 | 103.43 | 49.37 | 62.53 | 37.78 | 36.75 | 34.54 | 82.56 |
| 29 | 52.04 | 55.14 | 113.77 | 66.59 | 64.82 | 39.74 | 42.62 | 46.15 | 85.86 |
| 30 | 42.53 | 57.51 | 103.43 | 50.52 | 71.69 | 39.74 | 42.62 | 41.17 | 71.01 |
| 31 | 48.47 | 52.76 | 99.98 | 55.11 | 71.69 | 39.74 | 34.80 | 49.46 | 82.56 |
| 32 | 23.50 | 52.76 | 87.34 | 41.33 | 69.40 | 24.04 | 25.02 | 22.93 | 64.41 |
| 33 | 18.74 | 21.95 | 91.93 | 34.44 | 41.93 | 24.04 | 17.20 | 4.69 | 62.76 |
| 34 | 13.99 | 38.54 | 82.74 | 34.44 | 37.35 | 24.04 | 17.20 | 14.64 | 61.11 |
| 35 | 3.88 | 16.02 | 77.00 | 36.74 | 39.64 | 20.11 | 11.34 | 9.67 | 64.41 |
| 36 | -5.04 | 21.95 | 66.65 | 18.37 | 23.61 | 8.33 | 5.47 | 3.03 | 52.86 |
| 37 | 2.09 | 18.39 | 74.70 | 36.74 | 37.35 | 18.14 | 13.29 | 14.64 | 54.51 |
| 38 | 33.61 | 39.73 | 85.04 | 29.85 | 53.37 | 24.04 | 17.20 | 34.54 | 59.46 |
| 39 | 6.85 | 24.32 | 68.95 | 32.15 | 30.48 | 12.25 | 9.38 | -5.26 | 36.36 |
| 40 | 24.69 | 38.54 | 87.34 | 40.18 | 41.93 | 18.14 | 19.16 | 19.62 | 57.81 |
| 41 | -5.04 | 19.58 | 78.14 | 24.11 | 32.77 | 22.07 | 15.25 | -0.28 | 42.96 |
| 42 | 6.85 | 19.58 | 80.44 | 33.30 | 28.19 | 10.29 | 9.38 | -1.94 | 38.01 |
| 43 | -7.42 | 14.84 | 48.27 | 9.18 | 12.16 | -3.46 | -6.26 | 11.33 | 38.01 |
| 44 | 4.47 | 24.32 | 80.44 | 34.44 | 32.77 | 16.18 | 5.47 | 11.33 | 56.16 |
| 45 | 4.47 | 23.14 | 64.35 | 27.55 | 30.48 | 12.25 | 9.38 | -1.94 | 33.06 |

TABLE D.2.3 - *continued*

| SWP Graph ID | Si II 1260.4212 | Si II 1264.7374 | *Si II 1304.37 | Si II 1305.59 | Si II 1309.2769 | Si II 1526.7076 | Si II 1533.432 | *Si II 1808.0117 | *Si II 1816.9796 |
|---|---|---|---|---|---|---|---|---|---|
| 46 | 3.88 | 29.06 | 85.04 | 35.59 | 28.19 | 12.25 | 15.25 | 14.64 | 51.21 |
| 47 | 6.85 | 33.80 | 80.44 | 41.33 | 39.64 | 22.07 | 15.25 | 30.39 | 62.76 |
| 48 | -5.04 | 13.65 | 73.55 | 28.70 | 28.19 | 14.22 | 15.25 | 4.69 | 42.96 |
| 49 | 18.74 | 52.76 | 96.53 | 24.80 | 48.79 | 33.85 | 28.93 | 41.17 | 69.36 |
| 50 | 4.47 | 24.32 | 72.40 | 29.85 | 25.90 | 10.29 | 11.34 | -0.28 | 42.96 |
| 51 | 13.99 | 26.69 | 80.44 | 43.63 | 32.77 | 14.22 | 11.34 | 27.91 | 67.71 |
| 52 | 27.07 | 36.17 | 78.14 | 18.37 | 39.64 | 27.96 | 21.11 | 22.93 | 66.06 |
| 53 | 11.61 | 13.65 | 78.14 | 39.04 | 28.19 | 20.11 | 17.20 | 11.33 | 44.61 |
| 54 | 4.47 | 13.65 | 73.55 | 20.67 | 23.61 | 8.33 | -4.30 | -1.94 | 44.61 |
| 55 | 30.64 | 26.69 | 71.25 | 13.78 | 32.77 | 14.22 | 9.38 | 14.64 | 59.46 |
| 56 | 27.07 | 21.95 | 72.40 | 6.89 | 48.79 | 22.07 | 19.16 | 22.93 | 61.11 |
| 57 | 6.85 | 40.91 | 56.31 | 18.37 | 46.50 | 14.22 | 3.52 | 27.91 | 49.56 |
| 58 | 3.88 | 16.02 | 58.61 | 16.07 | 30.48 | 4.40 | 1.56 | 6.35 | 57.81 |
| 59 | -5.04 | 14.84 | 51.71 | 4.59 | 23.61 | 6.36 | 1.56 | 1.38 | 41.31 |

TABLE D.2.4 - OBSERVED SWP RADIAL VELOCITIES (km/s)

| SWP Graph ID | Si III 1294.545 | Si III 1296.726 | Si III 1298.96 | Si III 1301.149 | Si III 1302.2 | Si III 1303.323 | Si IV 1393.755 | extra comp. | Si IV 1402.77 | extra comp. |
|---|---|---|---|---|---|---|---|---|---|---|
| 1 | 2.32 | -6.01 | -32.31 | -13.59 | -4.60 | -0.69 | -34.42 | | -10.69 | |
| 2 | 1.16 | -3.70 | -20.77 | -18.20 | -16.12 | -12.19 | -23.66 | | 27.78 | |
| 3 | 1.16 | -24.51 | -33.47 | -22.81 | -29.93 | -19.09 | -15.06 | | 2.14 | |
| 4 | 4.63 | -15.26 | -25.39 | -27.42 | -4.60 | -21.39 | -39.79 | | -49.15 | |
| 5 | -15.05 | -31.44 | -41.54 | -33.18 | -31.08 | -25.99 | -23.66 | | 2.14 | |
| 6 | -15.05 | -24.51 | -36.93 | -28.57 | -3.45 | -24.84 | -46.25 | | -47.02 | |
| 7 | -19.68 | -41.85 | -47.31 | -34.33 | -44.89 | -42.09 | -30.11 | | -32.06 | |
| 8 | -22.00 | -33.75 | -48.47 | -38.94 | -39.14 | -30.59 | -33.34 | | -35.26 | |
| 9 | -17.37 | -33.75 | -36.93 | -35.48 | -27.63 | -29.44 | -21.51 | | -3.21 | |
| 10 | -22.00 | -38.38 | -47.31 | -34.33 | -35.68 | -29.44 | -29.04 | | -27.78 | |
| 11 | -33.58 | -48.78 | -56.54 | -48.15 | -54.10 | -51.29 | -44.09 | | -23.51 | |
| 12 | -28.95 | -47.63 | -62.31 | -48.15 | -55.25 | -51.29 | -37.64 | | -37.40 | |
| 13 | -59.05 | -74.21 | -92.32 | -75.80 | -79.43 | -76.60 | -66.68 | | -68.39 | |
| 14 | -45.16 | -58.03 | -76.16 | -50.46 | -57.55 | -62.80 | -44.09 | | -47.02 | |
| 15 | -38.21 | -54.56 | -65.78 | -52.76 | -43.74 | -51.29 | -41.94 | | -36.33 | |
| 16 | -26.63 | -36.07 | -60.01 | -38.94 | -40.29 | -38.64 | -41.94 | | -23.51 | |
| 17 | -5.79 | -17.57 | -41.54 | -22.81 | -31.08 | -30.59 | -29.04 | | 0.00 | |
| 18 | 12.74 | -1.39 | -24.23 | -28.57 | -9.21 | -6.44 | 3.23 | | 51.29 | |
| 19 | -3.47 | -10.63 | -21.93 | -6.68 | -11.51 | -8.74 | -11.83 | | 36.33 | |
| 20 | 54.42 | 33.29 | 16.16 | 23.27 | 17.27 | 29.21 | 35.49 | | 68.39 | |
| 21 | 26.63 | 12.48 | 5.77 | 16.36 | 13.81 | 10.81 | 24.74 | | 81.21 | |
| 22 | 52.11 | 35.60 | 18.46 | 26.73 | 16.12 | 22.31 | 37.64 | | 57.70 | |
| 23 | 52.11 | 37.92 | 16.16 | 11.75 | 11.51 | 20.01 | 41.94 | | 55.57 | |
| 24 | 45.16 | 30.98 | 23.08 | 32.49 | 19.57 | 22.31 | 44.09 | | 47.02 | |
| 25 | 42.84 | 33.29 | 19.62 | 47.46 | 18.42 | 26.91 | 39.79 | | 66.25 | |
| 26 | 55.58 | 35.60 | 20.77 | 30.18 | 18.42 | 31.51 | 50.55 | | 91.90 | |
| 27 | 59.05 | 42.54 | 25.39 | 46.31 | 16.12 | 36.11 | 59.15 | | 102.58 | |
| 28 | 54.42 | 28.67 | 13.85 | 25.58 | 12.66 | 24.61 | 33.34 | | 27.78 | |
| 29 | 56.74 | 37.92 | 27.70 | 47.46 | 13.81 | 38.41 | 48.40 | | 51.29 | |
| 30 | 44.00 | 30.98 | 20.77 | 41.70 | 6.91 | 43.01 | 41.94 | | 37.40 | |
| 31 | 47.47 | 26.36 | 20.77 | 41.70 | 12.66 | 17.71 | 48.40 | | 44.88 | |
| 32 | 28.95 | 5.55 | 6.92 | 12.90 | 0.00 | 24.61 | 39.79 | | 38.47 | |
| 33 | 33.58 | 3.24 | -6.92 | 7.14 | 2.30 | -5.29 | 27.96 | | 28.85 | |
| 34 | 24.32 | 14.80 | 0.00 | 0.23 | 4.60 | 22.31 | 22.59 | | 0.00 | |
| 35 | 28.95 | | -3.46 | 2.53 | -3.45 | 1.61 | 13.98 | | 10.69 | |
| 36 | 13.89 | -6.01 | -20.77 | -6.68 | -10.36 | -4.14 | 11.83 | | 12.82 | |
| 37 | 15.05 | 7.86 | 0.00 | 0.23 | 0.00 | -9.89 | 25.81 | | 23.51 | |
| 38 | 38.21 | 28.67 | -4.62 | 10.60 | 18.42 | 10.81 | 37.64 | | 38.47 | |
| 39 | 12.74 | -3.70 | -16.16 | -3.23 | -9.21 | -1.84 | 10.75 | | 14.96 | |
| 40 | 45.16 | -1.39 | -13.85 | 12.90 | 1.15 | 3.91 | 10.75 | | 38.47 | |
| 41 | 19.68 | 5.55 | -9.23 | 0.23 | -9.21 | -4.14 | 22.59 | | 19.23 | |
| 42 | 12.74 | -4.86 | -15.00 | -4.38 | -13.81 | -8.74 | 5.38 | | 11.75 | |
| 43 | 4.63 | -15.26 | -32.31 | -10.14 | -23.02 | -8.74 | 5.38 | | 19.23 | |
| 44 | 13.89 | 0.92 | -12.69 | -0.92 | -2.30 | 6.21 | 16.13 | | 12.82 | |
| 45 | 19.68 | 5.55 | 0.00 | 2.53 | -13.81 | -7.59 | 9.68 | | 17.10 | |

TABLE D.2.4 - *continued*

| SWP Graph | Si III | Si III | Si III | Si III | Si III | Si III | Si IV | extra comp. | Si IV | extra comp. |
|---|---|---|---|---|---|---|---|---|---|---|
| ID | 1294.545 | 1296.726 | 1298.96 | 1301.149 | 1302.2 | 1303.323 | 1393.755 | | 1402.77 | |
| 46 | 23.16 | 3.24 | -11.54 | -3.23 | 2.30 | 7.36 | 16.13 | | 19.23 | |
| 47 | 32.42 | 11.33 | -13.85 | 5.99 | 0.00 | 10.81 | -2.15 | | 14.96 | |
| 48 | 17.37 | 3.24 | 8.08 | 4.84 | -6.91 | -4.14 | 8.60 | | 14.96 | |
| 49 | 49.79 | 21.73 | -9.23 | 11.75 | 23.02 | 38.41 | 30.11 | 100.02 | 57.70 | |
| 50 | 22.00 | 7.86 | -13.85 | -4.38 | -9.21 | 3.91 | 21.51 | | 19.23 | |
| 51 | 3.47 | 14.80 | -9.23 | 4.84 | -6.91 | 11.96 | 1.08 | | 25.65 | |
| 52 | 35.90 | 14.80 | 6.92 | 11.75 | 9.21 | 16.56 | 23.02 | 94.64 | 47.02 | |
| 53 | 24.32 | 12.48 | -3.46 | 7.14 | -8.06 | -5.29 | -2.15 | | 11.75 | |
| 54 | 17.37 | 4.39 | -21.93 | 0.23 | -11.51 | 1.61 | -1.08 | | 16.03 | |
| 55 | 31.26 | 10.17 | -5.77 | 16.36 | 0.00 | 10.81 | 13.98 | 90.34 | 0.00 | 82.28 |
| 56 | 26.63 | 17.11 | -5.77 | 0.23 | 4.60 | 8.51 | 23.66 | | 38.47 | |
| 57 | 23.16 | 5.55 | -8.08 | -18.20 | 9.21 | 37.26 | 22.59 | | 47.02 | |
| 58 | 5.79 | 0.92 | -23.08 | -6.68 | -18.42 | -6.44 | -2.15 | | 19.23 | |
| 59 | 24.32 | 0.92 | -11.54 | -5.53 | -12.66 | 2.76 | 7.53 | 100.02 | -17.10 | 82.28 |

TABLE D.2.5 - OBSERVED LWP/LWR RADIAL VELOCITIES (km/s)

| LWP/LWR Graph ID | Fe II 2586.65 | Fe II 2599.147 | Fe II 2600.173 | Fe II 2750.13 | Mg II 2791.6 | Mg II 2796.352 | *Mg II 2798.823 | Mg II 2803.531 |
|---|---|---|---|---|---|---|---|---|
| 1 | -4.64 | -11.19 | 0.81 | -20.71 | -4.30 | -2.89 | -12.43 | -2.25 |
| 2 | -5.79 | -3.11 | 6.57 | -3.27 | -2.15 | 5.15 | -1.72 | 15.93 |
| 3 | -15.07 | -4.27 | -5.53 | -20.71 | -15.03 | -5.57 | -31.17 | -3.31 |
| 4 | -4.64 | -7.73 | 4.27 | -20.71 | -25.77 | 2.47 | -17.79 | -3.31 |
| 5 | -4.06 | -18.11 | -28.59 | -41.42 | -35.44 | -61.86 | -44.56 | -1.18 |
| 6 | -24.34 | -29.30 | -22.25 | -22.89 | -22.55 | -16.30 | -23.14 | -7.59 |
| 7 | -19.70 | -29.64 | -21.10 | -19.62 | -22.02 | -15.01 | -23.14 | -18.29 |
| 8 | -5.22 | -15.80 | -7.26 | -31.61 | -23.63 | -15.01 | -28.50 | -17.22 |
| 9 | -18.54 | -23.30 | -17.64 | -21.80 | -19.33 | -15.01 | -21.86 | -11.87 |
| 10 | -24.34 | -18.11 | -18.79 | -18.53 | -27.92 | -16.30 | -23.14 | -16.15 |
| 11 | -38.25 | -39.45 | -45.31 | -35.97 | -42.96 | -51.14 | -52.60 | -40.74 |
| 12 | -31.29 | -49.25 | -34.94 | -32.70 | -37.59 | -27.02 | -25.82 | -34.33 |
| 13 | -41.72 | -50.40 | -45.31 | -42.51 | -49.94 | -45.78 | -44.56 | -41.81 |
| 14 | -27.82 | -51.56 | -36.09 | -37.06 | -57.99 | -43.10 | -60.63 | -28.98 |
| 15 | -37.09 | -52.13 | -38.39 | -39.24 | -53.70 | -37.74 | -44.56 | -28.98 |
| 16 | -3.48 | -4.84 | 4.27 | -2.18 | -4.30 | 2.47 | -1.72 | 9.52 |
| 17 | 4.64 | 7.27 | 22.71 | 5.45 | 3.22 | 7.83 | 10.28 | 12.73 |
| 18 | 19.70 | 17.07 | 26.17 | 25.07 | 22.02 | 23.91 | 22.38 | 38.39 |
| 19 | 31.29 | 20.53 | 34.24 | 39.24 | 31.14 | 26.59 | 30.42 | 30.90 |
| 20 | 26.66 | 19.95 | 19.25 | 9.81 | 21.48 | 18.55 | 14.35 | 22.35 |
| 21 | 42.30 | 48.79 | 50.96 | 46.87 | 48.33 | 37.31 | 42.41 | 29.83 |
| 22 | 39.41 | 46.48 | 43.47 | 54.51 | 46.18 | 39.99 | 43.81 | 45.87 |
| 23 | 52.15 | 49.94 | 46.93 | 54.51 | 52.62 | 48.03 | 62.55 | 49.08 |
| 24 | 46.36 | 40.14 | 40.01 | 46.87 | 45.10 | 45.35 | 41.13 | 46.94 |
| 25 | 30.71 | 23.41 | 26.75 | 27.25 | 22.55 | 26.59 | 25.06 | 33.04 |
| 26 | 17.62 | | | | | 26.59 | | 28.77 |
| 27 | 17.38 | 19.95 | 21.56 | 19.62 | 20.40 | 15.87 | 14.35 | 26.63 |
| 28 | 15.07 | | | | | 26.59 | 22.38 | 27.70 |
| 29 | 20.28 | 17.65 | 23.87 | 25.07 | 17.18 | 26.59 | 18.63 | 35.18 |
| 30 | 17.38 | 11.30 | 21.56 | 20.71 | 18.26 | 26.59 | 17.03 | 28.77 |
| 31 | 23.18 | 23.41 | 26.17 | 28.34 | 23.63 | 27.87 | 30.42 | 31.97 |
| 32 | 18.54 | 17.07 | 22.71 | 17.44 | 10.74 | 23.91 | 17.03 | 31.97 |
| 33 | 15.65 | 9.92 | 15.80 | 19.62 | 18.26 | 21.23 | 14.35 | 31.97 |
| 34 | 26.66 | 18.80 | 26.17 | 23.98 | 23.63 | 27.87 | 22.38 | 39.46 |
| 35 | 16.23 | 7.84 | 14.64 | 16.35 | 10.74 | 13.19 | 8.99 | 22.35 |
| 36 | 24.34 | 19.38 | 23.87 | 21.80 | 22.55 | 26.05 | 19.70 | 28.77 |
| 37 | 26.08 | 30.91 | 28.48 | 27.25 | 19.33 | 26.59 | 19.70 | 34.11 |
| 38 | 21.44 | 18.22 | 20.41 | 20.71 | 15.03 | 15.87 | 19.70 | 26.63 |
| 39 | 19.70 | 28.60 | 28.48 | 8.72 | 23.63 | 26.59 | 27.74 | 18.07 |
| 40 | 24.92 | 19.95 | 23.87 | 21.80 | 24.70 | 17.15 | 19.70 | 26.63 |
| 41 | 28.40 | 19.38 | 25.60 | 29.43 | 19.33 | 18.55 | 19.70 | 28.77 |
| 42 | 23.18 | 16.49 | 16.95 | 15.81 | 15.03 | 7.83 | 25.06 | 22.35 |
| 43 | 13.91 | 16.49 | 15.80 | 10.90 | 21.48 | 5.15 | 17.03 | 11.66 |
| 44 | -6.95 | 15.92 | 19.25 | 18.53 | 23.63 | 2.47 | -1.72 | 7.38 |

# APPENDIX D.3
## Line Widths

TABLE D.3.1— SWP FWHM (Å) Al II, Al III

| SWP Graph ID | Phase | Exposure ID | Time Group Labels | Al II 1670 | Al III 1862 Short ward Width | Al III 1862 FWHM net | Al III 1862 Long ward Width | Al III 1854 Short ward Width | Al III 1854 FWHM net | Al III 1854 Long ward Width |
|---|---|---|---|---|---|---|---|---|---|---|
| 2  | 0.0031 | SWP03262 | Sep-Nov'78 | 1.00 | … | 1.30 | 0.00 | … | 1.64 | … |
| 9  | 0.0816 | SWP03303 |  | 0.86 | … | 1.24 | 0.00 | … | 1.22 | 0.24 |
| 21 | 0.6361 | SWP02643 |  | 0.80 | … | 0.90 | … | … | 0.68 | 0.00 |
| 49 | 0.9669 | SWP03298 |  | 1.10 | … | 1.36 | 0.42 | … | 1.27 | 0.13 |
| 52 | 0.9791 | SWP03259 |  | 1.00 | … | 1.40 | 0.00 | … | 1.44 | 0.30 |
| 55 | 0.9874 | SWP03260 |  | 0.98 | … | 1.42 | 0.46 | … | 1.54 | 0.50 |
| 59 | 0.9951 | SWP03261 |  | 0.96 | … | 1.44 | 0.37 | … | 1.38 | 0.22 |
| 1  | 0.0015 | SWP03772 | Dec'78-Jan'79 | 0.92 | 0.18 | 0.85 | … | 0.21 | 0.91 | … |
| 16 | 0.3297 | SWP03752 |  | 0.80 | … | 0.94 | … | … | 0.86 | … |
| 18 | 0.4927 | SWP03818 |  | 0.73 | … | 0.84 | … | … | 0.66 | … |
| 27 | 0.7674 | SWP03794 |  | 0.90 | … | 0.74 | … | … | 0.67 | … |
| 56 | 0.9880 | SWP03771 |  | 0.82 | 0.19 | 0.93 | … | … | 1.72 | … |
| 4  | 0.0053 | SWP07111 | Nov'79 | 0.87 | 0.26 | 1.00 | … | 0.51 | 1.41 | … |
| 6  | 0.0183 | SWP07112 |  | 0.97 | … | 1.04 | … | … | 0.96 | … |
| 57 | 0.9915 | SWP07110 |  | 0.86 | … | 0.97 | … | … | 0.84 | … |
| 34 | 0.8904 | SWP21452 | Nov'83 | 0.90 | … | 0.44 | … | … | 0.81 | … |
| 36 | 0.9070 | SWP21453 |  | 0.85 | … | 0.54 | … | … | 0.74 | … |
| 39 | 0.9218 | SWP21454 |  | 0.85 | 0.11 | 0.73 | … | … | 0.85 | … |
| 42 | 0.9358 | SWP21455 |  | 0.92 | 0.20 | 0.98 | … | … | 0.90 | … |
| 46 | 0.9571 | SWP21456 |  | 0.98 | 0.19 | 0.99 | … | … | 0.82 | … |
| 50 | 0.9712 | SWP21457 |  | 1.04 | 0.14 | 0.90 | … | … | 0.90 | … |
| 54 | 0.9797 | SWP21458 |  | 1.14 | 0.13 | 0.93 | … | … | 0.92 | … |
| 12 | 0.1538 | SWP22417 | Mar'84 | 0.85 | … | 0.68 | 0.00 | … | 0.84 | … |
| 13 | 0.1625 | SWP22418 |  | 0.81 | … | 0.72 | 0.00 | … | 0.80 | … |
| 38 | 0.9217 | SWP22411 |  | 1.02 | … | 1.36 | 0.18 | … | 1.55 | 0.25 |
| 43 | 0.9388 | SWP22439 |  | 1.02 | … | 1.56 | 0.32 | … | 1.65 | 0.55 |

TABLE D.3.1— *(continued)*

| SWP Graph ID | Phase | Exposure ID | Time Group Labels | Al II 1670 | Al III 1862 | | | Al III 1854 | | |
|---|---|---|---|---|---|---|---|---|---|---|
| | | | | | Short ward Width | FWHM net | Long ward Width | Short ward Width | FWHM net | Long ward Width |
| 33 | 0.8866 | SWP26472 | Jul'85 | 0.80 | … | 0.85 | … | … | 0.73 | … |
| 35 | 0.9009 | SWP26473 | | 0.85 | … | 0.98 | … | … | 0.82 | … |
| 37 | 0.9149 | SWP26474 | | 0.85 | … | 0.85 | … | … | 0.80 | … |
| 41 | 0.9299 | SWP26475 | | 0.84 | 0.07 | 0.93 | … | … | 0.92 | … |
| 45 | 0.9471 | SWP26476 | | 0.98 | 0.15 | 1.05 | … | 0.21 | 0.95 | … |
| 48 | 0.9636 | SWP26477 | | 0.93 | 0.21 | 0.95 | … | 0.23 | 0.97 | … |
| 53 | 0.9791 | SWP26478 | | 1.02 | 0.17 | 0.95 | … | 0.22 | 0.90 | … |
| 20 | 0.6326 | SWP27781 | Feb-Mar'86 | 0.88 | … | 1.22 | … | 0.25 | 1.09 | … |
| 22 | 0.6452 | SWP27782 | | 0.83 | … | 0.94 | … | … | 1.02 | … |
| 23 | 0.6522 | SWP27783 | | 0.85 | … | 0.98 | … | … | 1.17 | … |
| 24 | 0.6595 | SWP27784 | | 0.94 | … | 0.96 | … | … | 1.00 | … |
| 25 | 0.6704 | SWP27785 | | 0.88 | … | 1.04 | … | … | 1.06 | … |
| 26 | 0.6772 | SWP27786 | | 0.82 | 0.08 | 0.98 | … | … | 0.97 | … |
| 29 | 0.7869 | SWP27830 | | 0.82 | 0.18 | 1.12 | … | 0.19 | 1.19 | … |
| 30 | 0.8005 | SWP27831 | | 0.91 | … | 1.12 | … | … | 1.32 | … |
| 31 | 0.8075 | SWP27832 | | 0.91 | 0.09 | 0.99 | … | … | 1.30 | … |
| 3 | 0.0052 | SWP36983 | Sep'89 | 0.88 | … | 0.94 | … | 0.11 | 0.85 | … |
| 5 | 0.0141 | SWP37019 | | 0.82 | … | 0.82 | … | … | 0.74 | … |
| 7 | 0.0204 | SWP36984 | | 0.81 | … | 0.84 | … | … | 0.90 | … |
| 8 | 0.0378 | SWP36985 | | 0.87 | … | 0.84 | … | … | 0.80 | … |
| 10 | 0.0855 | SWP37021 | | 0.80 | … | 0.90 | … | … | 0.84 | … |
| 11 | 0.1317 | SWP36989 | | 0.76 | … | 0.76 | … | … | 0.76 | … |
| 14 | 0.1889 | SWP36992 | | 0.82 | 0.09 | 0.79 | … | … | 0.80 | … |
| 15 | 0.3209 | SWP36996 | | 0.81 | … | 0.84 | … | … | 0.80 | … |
| 17 | 0.4790 | SWP37001 | | 0.73 | … | 0.86 | … | … | 0.74 | … |
| 19 | 0.5454 | SWP37003 | | 0.74 | … | 0.84 | … | … | 0.74 | … |
| 28 | 0.7784 | SWP37011 | | 0.76 | 0.12 | 0.92 | … | 0.08 | 0.74 | … |
| 32 | 0.8813 | SWP37014 | | 0.86 | 0.25 | 1.10 | … | 0.30 | 1.14 | … |
| 40 | 0.9247 | SWP37016 | | 0.96 | … | 1.28 | 0.16 | … | 1.20 | … |
| 44 | 0.9397 | SWP36979 | | 1.03 | … | 1.20 | … | … | 1.14 | … |
| 47 | 0.9573 | SWP36980 | | 1.07 | … | 1.20 | … | … | 1.15 | … |
| 51 | 0.9777 | SWP36981 | | 1.05 | … | 1.21 | 0.11 | … | 1.18 | … |
| 58 | 0.9947 | SWP36982 | | 0.93 | 0.13 | 0.90 | … | … | 0.94 | … |

TABLE D.3.2— SWP FWHM (Å), C II, Fe II, Fe III, Si II, Si III

| SWP Graph ID | Phase | Exposure ID | Time Group Labels | C II 1334 | C II 1335 | Fe II 1608 | Fe III 1926 | Si II 1533 | Si III 1294 |
|---|---|---|---|---|---|---|---|---|---|
| 2 | 0.0031 | SWP03262 | Sep-Nov'78 | 0.80 | 1.44 | 0.45 | 0.85 | 1.12 | 0.56 |
| 9 | 0.0816 | SWP03303 | | 0.98 | 1.32 | 0.35 | 0.92 | 0.82 | 0.55 |
| 21 | 0.6361 | SWP02643 | | 0.76 | 1.04 | 0.45 | 0.96 | 0.64 | 0.56 |
| 49 | 0.9669 | SWP03298 | | 0.84 | 1.40 | 0.47 | 1.16 | 0.96 | 0.57 |
| 52 | 0.9791 | SWP03259 | | 0.88 | 1.30 | 0.37 | 1.00 | 0.93 | 0.61 |
| 55 | 0.9874 | SWP03260 | | 0.90 | 1.30 | 0.46 | 1.24 | 0.88 | 0.56 |
| 59 | 0.9951 | SWP03261 | | 0.88 | 1.35 | 0.44 | 0.93 | 1.01 | 0.58 |
| 1 | 0.0015 | SWP03772 | Dec'78-Jan'79 | 0.80 | 1.06 | 0.43 | 0.80 | 0.80 | 0.51 |
| 16 | 0.3297 | SWP03752 | | 0.88 | 1.02 | 0.42 | 0.96 | 0.68 | 0.44 |
| 18 | 0.4927 | SWP03818 | | 0.74 | 1.10 | 0.50 | 0.75 | 0.69 | 0.44 |
| 27 | 0.7674 | SWP03794 | | 0.80 | 1.08 | 0.59 | 0.80 | 0.72 | 0.50 |
| 56 | 0.9880 | SWP03771 | | 0.76 | 1.05 | 0.36 | 0.87 | 0.74 | 0.54 |
| 4 | 0.0053 | SWP07111 | Nov'79 | 0.88 | 0.94 | 0.40 | 1.00 | 0.88 | 0.46 |
| 6 | 0.0183 | SWP07112 | | 0.90 | 1.08 | 0.49 | 0.90 | 0.97 | 0.48 |
| 57 | 0.9915 | SWP07110 | | 0.94 | 0.92 | 0.55 | 0.92 | 0.98 | 0.55 |
| 34 | 0.8904 | SWP21452 | Nov'83 | 0.70 | 0.96 | 0.53 | 0.63 | 0.75 | 0.48 |
| 36 | 0.9070 | SWP21453 | | 0.82 | 0.94 | 0.49 | 0.80 | 0.76 | 0.51 |
| 39 | 0.9218 | SWP21454 | | 0.80 | 1.02 | 0.48 | 0.79 | 0.77 | 0.50 |
| 42 | 0.9358 | SWP21455 | | 0.94 | 1.04 | 0.57 | 0.84 | 0.82 | 0.50 |
| 46 | 0.9571 | SWP21456 | | 0.81 | 1.06 | 0.45 | 0.81 | 0.78 | 0.46 |
| 50 | 0.9712 | SWP21457 | | 0.88 | 1.14 | 0.41 | 0.84 | 0.90 | 0.48 |
| 54 | 0.9797 | SWP21458 | | 0.90 | 1.09 | 0.46 | 0.87 | 0.94 | 0.51 |
| 12 | 0.1538 | SWP22417 | Mar'84 | 0.92 | 1.02 | 0.47 | 0.78 | 0.77 | 0.44 |
| 13 | 0.1625 | SWP22418 | | 0.97 | 1.10 | 0.61 | 0.82 | 0.76 | 0.48 |
| 38 | 0.9217 | SWP22411 | | 0.92 | 1.37 | 0.60 | 1.00 | 0.94 | 0.54 |
| 43 | 0.9388 | SWP22439 | | 1.01 | 1.53 | 0.61 | 1.06 | 1.06 | 0.55 |

TABLE D.3.2— *(continued)*

| SWP Graph ID | Phase | Exposure ID | Time Group Labels | C II 1334 | C II 1335 | Fe II 1608 | Fe III 1926 | Si II 1533 | Si III 1294 |
|---|---|---|---|---|---|---|---|---|---|
| 33 | 0.8866 | SWP26472 | Jul'85 | 0.88 | 1.00 | 0.52 | 0.69 | 0.70 | 0.44 |
| 35 | 0.9009 | SWP26473 |  | 0.74 | 0.98 | 0.49 | 0.82 | 0.84 | 0.50 |
| 37 | 0.9149 | SWP26474 |  | 0.82 | 1.00 | 0.49 | 0.69 | 0.75 | 0.47 |
| 41 | 0.9299 | SWP26475 |  | 0.84 | 1.05 | 0.60 | 1.00 | 0.84 | 0.47 |
| 45 | 0.9471 | SWP26476 |  | 0.79 | 1.06 | 0.58 | 0.91 | 0.82 | 0.48 |
| 48 | 0.9636 | SWP26477 |  | 0.72 | 1.11 | 0.41 | 0.83 | 0.74 | 0.52 |
| 53 | 0.9791 | SWP26478 |  | 0.80 | 1.08 | 0.40 | 0.82 | 0.91 | 0.46 |
| 20 | 0.6326 | SWP27781 | Feb-Mar'86 | 0.92 | 1.00 | 0.61 | 0.73 | 0.69 | 0.53 |
| 22 | 0.6452 | SWP27782 |  | 0.97 | 0.96 | 0.58 | 0.93 | 0.72 | 0.45 |
| 23 | 0.6522 | SWP27783 |  | 0.80 | 1.20 | 0.55 | 0.80 | 0.77 | 0.51 |
| 24 | 0.6595 | SWP27784 |  | 0.84 | 1.03 | 0.58 | 0.69 | 0.70 | 0.47 |
| 25 | 0.6704 | SWP27785 |  | 0.68 | 1.18 | 0.68 | 0.91 | 0.78 | 0.48 |
| 26 | 0.6772 | SWP27786 |  | 0.84 | 1.07 | 0.44 | 0.70 | 0.78 | 0.50 |
| 29 | 0.7869 | SWP27830 |  | 1.05 | 1.12 | 0.40 | 0.93 | 0.76 | 0.49 |
| 30 | 0.8005 | SWP27831 |  | 0.88 | 1.10 | 0.50 | 0.77 | 0.81 | 0.52 |
| 31 | 0.8075 | SWP27832 |  | 1.04 | 1.14 | 0.45 | 0.86 | 0.77 | 0.51 |
| 3 | 0.0052 | SWP36983 | Sep'89 | 0.79 | 0.94 | 0.59 | 0.70 | 0.80 | 0.43 |
| 5 | 0.0141 | SWP37019 |  | 0.84 | 0.95 | 0.49 | 0.71 | 0.90 | 0.40 |
| 7 | 0.0204 | SWP36984 |  | 0.82 | 1.00 | 0.48 | 0.64 | 0.79 | 0.45 |
| 8 | 0.0378 | SWP36985 |  | 0.82 | 1.00 | 0.47 | 0.69 | 0.76 | 0.47 |
| 10 | 0.0855 | SWP37021 |  | 0.88 | 1.02 | 0.57 | 0.82 | 0.72 | 0.45 |
| 11 | 0.1317 | SWP36989 |  | 0.94 | 0.94 | 0.45 | 0.86 | 0.76 | 0.47 |
| 14 | 0.1889 | SWP36992 |  | 0.94 | 1.05 | 0.61 | 0.76 | 0.72 | 0.39 |
| 15 | 0.3209 | SWP36996 |  | 0.90 | 0.98 | 0.70 | 0.71 | 0.66 | 0.53 |
| 17 | 0.4790 | SWP37001 |  | 0.82 | 0.95 | 0.50 | 0.94 | 0.72 | 0.45 |
| 19 | 0.5454 | SWP37003 |  | 0.72 | 1.04 | 0.57 | 1.06 | 0.64 | 0.46 |
| 28 | 0.7784 | SWP37011 |  | 0.87 | 1.01 | 0.69 | 0.64 | 0.77 | 0.46 |
| 32 | 0.8813 | SWP37014 |  | 0.84 | 1.20 | 0.50 | 1.01 | 0.81 | 0.54 |
| 40 | 0.9247 | SWP37016 |  | 0.90 | 1.23 | 0.64 | 1.00 | 0.88 | 0.62 |
| 44 | 0.9397 | SWP36979 |  | 0.86 | 1.24 | 0.50 | 0.91 | 0.95 | 0.48 |
| 47 | 0.9573 | SWP36980 |  | 0.86 | 1.14 | 0.55 | 0.90 | 1.00 | 0.61 |
| 51 | 0.9777 | SWP36981 |  | 0.80 | 1.20 | 0.52 | 0.84 | 0.82 | 0.54 |
| 58 | 0.9947 | SWP36982 |  | 0.84 | 1.00 | 0.58 | 0.92 | 0.82 | 0.55 |

TABLE D.3.3— SWP FWHM (Å) Si IV

| SWP Graph ID | Phase | Exposure ID | Time Group Labels | Si IV 1393 | | | Si IV 1402 | | |
|---|---|---|---|---|---|---|---|---|---|
| | | | | Short ward Width | FWHM net | Long ward Width | Short ward Width | FWHM net | Long ward Width |
| 2 | 0.0031 | SWP03262 | Sep-Nov'78 | … | 1.43 | … | … | 1.35 | … |
| 9 | 0.0816 | SWP03303 | | … | 1.00 | 0.19 | … | 1.03 | 0.16 |
| 21 | 0.6361 | SWP02643 | | … | 0.62 | … | … | 0.84 | … |
| 49 | 0.9669 | SWP03298 | | … | 1.17 | 0.25 | … | 1.23 | 0.21 |
| 52 | 0.9791 | SWP03259 | | … | 1.57 | 0.36 | … | 1.32 | 0.22 |
| 55 | 0.9874 | SWP03260 | | … | 1.73 | 0.27 | … | 0.62 | 0.25 |
| 59 | 0.9951 | SWP03261 | | … | 1.42 | 0.21 | … | 0.64 | 0.18 |
| 1 | 0.0015 | SWP03772 | Dec'78-Jan'79 | … | 1.00 | … | … | 0.93 | … |
| 16 | 0.3297 | SWP03752 | | … | 0.90 | … | … | 0.66 | 0.12 |
| 18 | 0.4927 | SWP03818 | | … | 0.71 | … | … | 0.90 | … |
| 27 | 0.7674 | SWP03794 | | … | 0.48 | … | … | 0.61 | … |
| 56 | 0.9880 | SWP03771 | | 0.31 | 0.79 | … | … | 0.82 | … |
| 4 | 0.0053 | SWP07111 | Nov'79 | … | 0.96 | … | … | 0.91 | … |
| 6 | 0.0183 | SWP07112 | | … | 0.93 | … | … | 0.79 | 0.15 |
| 57 | 0.9915 | SWP07110 | | … | 1.25 | … | … | 0.85 | … |
| 34 | 0.8904 | SWP21452 | Nov'83 | … | 0.46 | … | … | #N/A | … |
| 36 | 0.9070 | SWP21453 | | 0.14 | 0.72 | … | … | 0.48 | … |
| 39 | 0.9218 | SWP21454 | | … | 0.65 | … | … | 0.54 | … |
| 42 | 0.9358 | SWP21455 | | 0.19 | 0.91 | … | … | 0.65 | … |
| 46 | 0.9571 | SWP21456 | | … | 0.78 | … | … | 0.70 | … |
| 50 | 0.9712 | SWP21457 | | 0.24 | 0.76 | … | … | 0.62 | … |
| 54 | 0.9797 | SWP21458 | | 0.29 | 0.99 | … | … | 0.75 | … |
| 12 | 0.1538 | SWP22417 | Mar'84 | … | 0.73 | … | … | 0.54 | … |
| 13 | 0.1625 | SWP22418 | | 0.10 | 0.71 | … | … | 0.60 | 0.10 |
| 38 | 0.9217 | SWP22411 | | … | 1.15 | … | … | 0.96 | … |
| 43 | 0.9388 | SWP22439 | | … | 1.85 | 0.54 | … | 1.86 | 0.65 |

TABLE D.3.3— *(continued)*

| SWP Graph ID | Phase | Exposure ID | Time Group Labels | Si IV 1393 | | | Si IV 1402 | | |
|---|---|---|---|---|---|---|---|---|---|
| | | | | Short ward Width | FWHM net | Long ward Width | Short ward Width | FWHM net | Long ward Width |
| 33 | 0.8866 | SWP26472 | Jul'85 | … | 0.66 | … | … | 0.67 | … |
| 35 | 0.9009 | SWP26473 | | 0.11 | 0.67 | … | … | 0.63 | … |
| 37 | 0.9149 | SWP26474 | | 0.11 | 0.69 | … | … | 0.62 | … |
| 41 | 0.9299 | SWP26475 | | 0.06 | 0.86 | … | … | 0.74 | … |
| 45 | 0.9471 | SWP26476 | | 0.23 | 0.81 | … | 0.15 | 0.77 | … |
| 48 | 0.9636 | SWP26477 | | 0.12 | 0.75 | … | … | 0.75 | … |
| 53 | 0.9791 | SWP26478 | | … | 0.80 | … | … | 0.66 | … |
| 20 | 0.6326 | SWP27781 | Feb-Mar'86 | … | 1.00 | … | … | 1.19 | … |
| 22 | 0.6452 | SWP27782 | | 0.11 | 0.85 | … | … | 1.00 | … |
| 23 | 0.6522 | SWP27783 | | … | 1.08 | … | … | 1.00 | … |
| 24 | 0.6595 | SWP27784 | | … | 0.90 | … | … | 0.95 | … |
| 25 | 0.6704 | SWP27785 | | … | 0.92 | … | … | 0.96 | … |
| 26 | 0.6772 | SWP27786 | | … | 0.93 | … | … | 0.89 | … |
| 29 | 0.7869 | SWP27830 | | … | 0.92 | … | … | 0.85 | … |
| 30 | 0.8005 | SWP27831 | | … | 1.00 | … | … | 0.79 | 0.17 |
| 31 | 0.8075 | SWP27832 | | … | 0.96 | … | … | 0.90 | … |
| 3 | 0.0052 | SWP36983 | Sep'89 | … | 0.76 | … | … | 0.95 | … |
| 5 | 0.0141 | SWP37019 | | 0.12 | 0.82 | … | … | 0.94 | … |
| 7 | 0.0204 | SWP36984 | | 0.15 | 0.69 | … | … | 0.81 | 0.15 |
| 8 | 0.0378 | SWP36985 | | … | 0.53 | … | … | 0.59 | 0.14 |
| 10 | 0.0855 | SWP37021 | | … | 0.68 | … | … | 0.65 | 0.07 |
| 11 | 0.1317 | SWP36989 | | … | 0.58 | … | … | 0.84 | … |
| 14 | 0.1889 | SWP36992 | | … | 0.82 | … | … | 0.66 | 0.12 |
| 15 | 0.3209 | SWP36996 | | … | 0.45 | … | … | 0.73 | 0.13 |
| 17 | 0.4790 | SWP37001 | | 0.09 | 0.60 | … | … | 0.60 | … |
| 19 | 0.5454 | SWP37003 | | … | 0.62 | … | … | 0.84 | … |
| 28 | 0.7784 | SWP37011 | | … | 0.68 | … | … | 0.60 | … |
| 32 | 0.8813 | SWP37014 | | 0.34 | 0.91 | … | … | 1.14 | … |
| 40 | 0.9247 | SWP37016 | | … | 1.26 | … | … | 1.21 | … |
| 44 | 0.9397 | SWP36979 | | … | 1.37 | … | … | 0.94 | 0.24 |
| 47 | 0.9573 | SWP36980 | | … | 1.34 | … | … | 1.07 | 0.38 |
| 51 | 0.9777 | SWP36981 | | … | 1.23 | … | … | 1.19 | 0.19 |
| 58 | 0.9947 | SWP36982 | | 0.15 | 0.83 | … | … | 1.20 | … |

TABLE D.3.4— LWP/LWR FWHM (Å) Mg II λ2791, Mg II λ2796

| LWP/LWR ID | Phase | Exposure ID | Time Group Labels | Fe II 2600 FWHM net | Mg II 2791 Short ward Width | Mg II 2791 FWHM net | Mg II 2791 Long ward Width | Mg II 2796 Short ward Width | Mg II 2796 FWHM net | Mg II 2796 Long ward Width |
|---|---|---|---|---|---|---|---|---|---|---|
| 4  | 0.0049 | LWR02856 | Sep-Nov'78   | 0.95 | …    | 1.16 | …    | 0.42 | 1.39 | …    |
| 8  | 0.0801 | LWR02911 |              | 0.89 | …    | 1.06 | …    | …    | 1.40 | …    |
| 20 | 0.6520 | LWR02345 |              | 0.91 | …    | 1.10 | …    | …    | 1.08 | …    |
| 39 | 0.9683 | LWR02906 |              | 0.98 | …    | 1.14 | …    | …    | 1.48 | 0.47 |
| 1  | 0.0030 | LWR03350 | Dec'78-Jan'79| 1.17 | …    | 1.26 | 0.20 | 0.41 | 1.21 | …    |
| 14 | 0.3200 | LWR03328 |              | 0.84 | …    | 0.87 | 0.49 | …    | 1.30 | 0.10 |
| 15 | 0.3280 | LWR03329 |              | 1.13 | …    | 1.30 | …    | …    | 1.35 | …    |
| 17 | 0.4939 | LWR03398 |              | 0.83 | …    | 1.13 | …    | …    | 1.03 | …    |
| 21 | 0.7685 | LWR03374 |              | 0.95 | …    | 1.00 | …    | …    | 1.23 | 0.18 |
| 43 | 0.9895 | LWR03349 |              | 0.88 | 0.28 | 0.88 | …    | …    | 1.20 | …    |
| 3  | 0.0034 | LWR06047 | Nov'79       | 0.83 | …    | 1.00 | …    | 0.53 | 1.38 | …    |
| 5  | 0.0165 | LWR06048 |              | 1.11 | …    | 0.84 | …    | …    | 1.79 | 0.34 |
| 44 | 0.9901 | LWR06046 |              | 0.45 | …    | 0.80 | …    | 0.18 | 0.98 | …    |
| 26 | 0.8915 | LWP02226 | Nov'83       | …    | …    | …    | …    | …    | 1.00 | …    |
| 28 | 0.9082 | LWP02227 |              | …    | …    | …    | …    | …    | 1.10 | …    |
| 30 | 0.9227 | LWP02228 |              | 0.89 | …    | 0.92 | 0.12 | …    | 1.30 | …    |
| 33 | 0.9367 | LWP02229 |              | 0.95 | …    | 0.98 | …    | …    | 1.53 | …    |
| 36 | 0.9581 | LWP02230 |              | 0.86 | …    | 0.86 | …    | …    | 1.50 | …    |
| 40 | 0.9721 | LWP02231 |              | 0.83 | 0.07 | 0.89 | …    | …    | 1.60 | …    |
| 11 | 0.1547 | LWP02895 | Mar'84       | 0.91 | …    | 0.93 | 0.16 | …    | 1.23 | 0.28 |
| 25 | 0.8878 | LWP06484 | Jul'85       | 0.82 | …    | 0.94 | …    | …    | 1.28 | …    |
| 27 | 0.9018 | LWP06485 |              | 0.85 | …    | 1.00 | …    | …    | 1.20 | …    |
| 29 | 0.9166 | LWP06486 |              | 0.73 | 0.14 | 0.80 | …    | …    | 1.25 | …    |
| 32 | 0.9323 | LWP06487 |              | 0.78 | 0.25 | 1.05 | …    | …    | 1.38 | …    |
| 35 | 0.9488 | LWP06488 |              | 0.89 | 0.16 | 1.02 | …    | …    | 1.40 | …    |
| 38 | 0.9650 | LWP06489 |              | 0.74 | 0.15 | 0.79 | …    | …    | 1.45 | …    |
| 42 | 0.9801 | LWP06490 |              | 0.82 | 0.18 | 0.82 | …    | 0.25 | 1.70 | …    |

TABLE D.3.4—*(continued)*

|  |  |  |  |  | | Mg II 2791 | | | Mg II 2796 | |
| --- | --- | --- | --- | --- | --- | --- | --- | --- | --- | --- |
| LWP/ LWR ID | Phase | Exposure ID | Time Group Labels | Fe II 2600 FWHM net | Short ward Width | FWHM net | Long ward Width | Short ward Width | FWHM net | Long ward Width |
| 19 | 0.6343 | LWP07717 | Feb-Mar'86 | 0.78 | … | 0.84 | … | … | 1.20 | … |
| 23 | 0.7884 | LWP07738 |  | 0.80 | … | 0.80 | … | … | 1.32 | 0.32 |
| 2 | 0.0034 | LWP16323 | Sep'89 | 1.00 | … | 0.90 | … | 0.33 | 1.28 | … |
| 6 | 0.0220 | LWP16324 |  | 0.91 | … | 0.80 | … | … | 1.50 | … |
| 7 | 0.0390 | LWP16325 |  | 0.90 | … | 0.67 | … | … | 1.48 | … |
| 9 | 0.0867 | LWP16356 |  | 0.99 | … | 0.90 | … | … | 1.35 | … |
| 10 | 0.1303 | LWP16327 |  | 0.75 | … | 0.70 | … | … | 1.33 | … |
| 12 | 0.1903 | LWP16329 |  | 0.98 | 0.19 | 0.93 | … | … | 1.30 | 0.30 |
| 13 | 0.3196 | LWP16333 |  | 0.85 | 0.28 | 0.90 | … | … | 1.26 | 0.16 |
| 16 | 0.4778 | LWP16338 |  | 0.80 | … | 1.00 | … | … | 1.18 | … |
| 18 | 0.5463 | LWP16340 |  | 0.75 | … | 0.98 | … | … | 1.05 | … |
| 22 | 0.7798 | LWP16347 |  | 0.73 | … | 0.86 | 0.22 | … | 1.20 | 0.20 |
| 24 | 0.8825 | LWP16350 |  | 0.80 | … | 1.02 | 0.32 | … | 1.35 | … |
| 31 | 0.9259 | LWP16352 |  | 0.88 | … | 1.14 | … | … | 1.50 | … |
| 34 | 0.9409 | LWP16320 |  | 0.86 | … | 1.00 | … | … | 1.53 | … |
| 37 | 0.9590 | LWP16321 |  | 0.76 | … | 0.88 | … | … | 1.50 | … |
| 41 | 0.9794 | LWP16322 |  | 0.95 | 0.12 | 0.96 | … | … | 1.68 | 0.15 |

TABLE D.3.5— LWP/LWR FWHM (Å) Mg II λ2798, Mg II λ2803

| LWP/LWR ID | Phase | Exposure ID | Time Group Labels | Mg II 2798 | | | Mg II 2803 | | |
|---|---|---|---|---|---|---|---|---|---|
| | | | | Shortward Width | FWHM net | Longward Width | Shortward Width | FWHM net | Longward Width |
| 4 | 0.0049 | LWR02856 | Sep-Nov'78 | … | 0.70 | … | 0.24 | 1.35 | … |
| 8 | 0.0801 | LWR02911 | | … | 0.75 | … | … | 1.43 | … |
| 20 | 0.6520 | LWR02345 | | … | 0.60 | … | … | 0.91 | … |
| 39 | 0.9683 | LWR02906 | | … | 0.80 | … | … | 1.63 | 0.27 |
| 1 | 0.0030 | LWR03350 | Dec'78-Jan'79 | 0.13 | 0.83 | … | 0.10 | 1.27 | … |
| 14 | 0.3200 | LWR03328 | | … | 0.85 | 0.35 | … | 1.28 | … |
| 15 | 0.3280 | LWR03329 | | … | 0.71 | … | … | 1.41 | … |
| 17 | 0.4939 | LWR03398 | | … | 0.65 | … | … | 0.97 | … |
| 21 | 0.7685 | LWR03374 | | … | 0.73 | … | … | 1.15 | 0.50 |
| 43 | 0.9895 | LWR03349 | | … | 0.68 | … | … | 1.02 | … |
| 3 | 0.0034 | LWR06047 | Nov'79 | … | 0.91 | … | 0.37 | 1.33 | … |
| 5 | 0.0165 | LWR06048 | | … | 0.89 | 0.19 | 0.21 | 0.93 | … |
| 44 | 0.9901 | LWR06046 | | … | 0.45 | … | 0.08 | 0.38 | … |
| 26 | 0.8915 | LWP02226 | Nov'83 | … | … | … | … | 1.05 | … |
| 28 | 0.9082 | LWP02227 | | … | 0.65 | … | … | 1.05 | … |
| 30 | 0.9227 | LWP02228 | | … | 0.90 | … | … | 1.06 | … |
| 33 | 0.9367 | LWP02229 | | … | 0.93 | … | 0.23 | 1.35 | … |
| 36 | 0.9581 | LWP02230 | | … | 0.80 | … | … | 1.35 | … |
| 40 | 0.9721 | LWP02231 | | … | 0.79 | … | … | 1.38 | … |
| 11 | 0.1547 | LWP02895 | Mar'84 | … | 0.77 | 0.06 | … | 1.24 | 0.16 |
| 25 | 0.8878 | LWP06484 | Jul'85 | … | 0.85 | … | … | 0.94 | … |
| 27 | 0.9018 | LWP06485 | | … | 0.90 | … | … | 1.13 | … |
| 29 | 0.9166 | LWP06486 | | … | 0.73 | … | … | 1.02 | … |
| 32 | 0.9323 | LWP06487 | | … | 0.87 | … | … | 1.13 | … |
| 35 | 0.9488 | LWP06488 | | … | 0.95 | … | 0.06 | 1.21 | … |
| 38 | 0.9650 | LWP06489 | | … | 0.80 | … | … | 1.31 | … |
| 42 | 0.9801 | LWP06490 | | 0.20 | 0.80 | … | 0.25 | 1.56 | … |

TABLE D.3.5—*(continued)*

|   |   |   |   | Mg II 2798 | | | Mg II 2803 | | |
|---|---|---|---|---|---|---|---|---|---|
| LWP/ LWR ID | Phase | Exposure ID | Time Group Labels | Short ward Width | FWHM net | Long ward Width | Short ward Width | FWHM net | Long ward Width |
| 19 | 0.6343 | LWP07717 | Feb-Mar'86 | … | 0.90 | … | … | 0.99 | 0.29 |
| 23 | 0.7884 | LWP07738 |  | … | 0.75 | … | … | 1.17 | 0.33 |
| 2  | 0.0034 | LWP16323 | Sep'89 | … | 0.85 | … | … | 1.20 | … |
| 6  | 0.0220 | LWP16324 |  | … | 0.71 | … | … | 1.40 | … |
| 7  | 0.0390 | LWP16325 |  | … | 0.70 | … | … | 1.21 | 0.25 |
| 9  | 0.0867 | LWP16356 |  | … | 0.80 | … | … | 1.18 | 0.18 |
| 10 | 0.1303 | LWP16327 |  | … | 0.70 | … | … | 1.13 | 0.18 |
| 12 | 0.1903 | LWP16329 |  | … | 1.00 | … | … | 1.17 | 0.29 |
| 13 | 0.3196 | LWP16333 |  | … | 0.80 | … | … | 1.15 | 0.21 |
| 16 | 0.4778 | LWP16338 |  | … | 0.98 | … | … | 1.10 | … |
| 18 | 0.5463 | LWP16340 |  | … | 0.91 | … | … | 0.90 | … |
| 22 | 0.7798 | LWP16347 |  | … | 0.75 | … | … | 0.92 | 0.16 |
| 24 | 0.8825 | LWP16350 |  | … | 1.03 | … | … | 1.15 | … |
| 31 | 0.9259 | LWP16352 |  | … | 1.10 | … | … | 1.17 | … |
| 34 | 0.9409 | LWP16320 |  | … | 1.18 | … | 0.15 | 1.35 | … |
| 37 | 0.9590 | LWP16321 |  | … | 0.95 | … | … | 1.42 | … |
| 41 | 0.9794 | LWP16322 |  | … | 0.93 | … | … | 1.32 | … |

# APPENDIX D.4

## Residual Intensities

TABLE D.4.1 - SWP Residual Intensities

| SWP Graph ID | Phase | Time Group Labels | Al II 1670 | Al III 1862 | Al III 1854 | C II 1334 | C II 1335 | Fe III 1926 | Si II 1533 | Si III 1294 | Si IV 1393 | Si IV 1402 |
|---|---|---|---|---|---|---|---|---|---|---|---|---|
| 2 | 0.0031 | Sep-Nov'78 | -0.01 | 0.38 | 0.33 | 0.18 | 0.19 | 0.50 | 0.23 | 0.47 | 0.35 | 0.50 |
| 9 | 0.0816 | | 0.07 | 0.28 | 0.21 | 0.20 | 0.18 | 0.46 | 0.18 | 0.35 | 0.28 | 0.35 |
| 21 | 0.6361 | | 0.09 | 0.55 | 0.46 | 0.21 | 0.16 | 0.67 | 0.21 | 0.39 | 0.69 | 0.80 |
| 49 | 0.9669 | | -0.03 | 0.25 | 0.21 | 0.15 | 0.16 | 0.34 | 0.09 | 0.37 | 0.25 | 0.33 |
| 52 | 0.9791 | | 0.03 | 0.34 | 0.27 | 0.18 | 0.17 | 0.45 | 0.15 | 0.37 | 0.30 | 0.40 |
| 55 | 0.9874 | | 0.08 | 0.38 | 0.34 | 0.16 | 0.16 | 0.49 | 0.12 | 0.42 | 0.37 | 0.49 |
| 59 | 0.9951 | | 0.03 | 0.41 | 0.32 | 0.26 | 0.18 | 0.50 | 0.16 | 0.45 | 0.43 | 0.53 |
| 1 | 0.0015 | Dec'78-Jan'79 | 0.08 | 0.48 | 0.42 | 0.19 | 0.15 | 0.61 | 0.13 | 0.46 | 0.49 | 0.72 |
| 16 | 0.3297 | | 0.13 | 0.54 | 0.44 | 0.25 | 0.22 | 0.63 | 0.18 | 0.42 | 0.56 | 0.72 |
| 18 | 0.4927 | | 0.13 | 0.52 | 0.47 | 0.18 | 0.18 | 0.64 | 0.18 | 0.45 | 0.59 | 0.79 |
| 27 | 0.7674 | | 0.18 | 0.55 | 0.47 | 0.19 | 0.16 | 0.62 | 0.18 | 0.41 | 0.58 | 0.80 |
| 56 | 0.9880 | | 0.09 | 0.46 | 0.44 | 0.19 | 0.17 | 0.61 | 0.12 | 0.45 | 0.55 | 0.78 |
| 4 | 0.0053 | Nov'79 | 0.11 | 0.43 | 0.57 | 0.22 | 0.19 | 0.58 | 0.10 | 0.35 | 0.43 | 0.58 |
| 6 | 0.0183 | | 0.03 | 0.35 | 0.27 | 0.20 | 0.20 | 0.53 | 0.15 | 0.37 | 0.32 | 0.43 |
| 57 | 0.9915 | | 0.00 | 0.38 | 0.41 | 0.23 | 0.19 | 0.41 | 0.16 | 0.39 | 0.48 | 0.55 |
| 34 | 0.8904 | Nov'83 | 0.04 | 0.64 | 0.22 | 0.19 | 0.16 | 0.54 | 0.13 | 0.37 | 0.39 | 1.00 |
| 36 | 0.9070 | | 0.08 | 0.44 | 0.27 | 0.19 | 0.15 | 0.50 | 0.16 | 0.32 | 0.32 | 0.44 |
| 39 | 0.9218 | | 0.03 | 0.34 | 0.27 | 0.18 | 0.15 | 0.49 | 0.14 | 0.34 | 0.32 | 0.41 |
| 42 | 0.9358 | | 0.03 | 0.30 | 0.21 | 0.20 | 0.14 | 0.48 | 0.13 | 0.33 | 0.29 | 0.39 |
| 46 | 0.9571 | | 0.00 | 0.29 | 0.20 | 0.16 | 0.13 | 0.50 | 0.10 | 0.34 | 0.26 | 0.39 |
| 50 | 0.9712 | | 0.02 | 0.28 | 0.22 | 0.15 | 0.14 | 0.48 | 0.11 | 0.34 | 0.30 | 0.40 |
| 54 | 0.9797 | | 0.06 | 0.31 | 0.23 | 0.15 | 0.13 | 0.53 | 0.14 | 0.39 | 0.30 | 0.43 |
| 12 | 0.1538 | Mar'84 | 0.08 | 0.39 | 0.30 | 0.19 | 0.17 | 0.55 | 0.17 | 0.40 | 0.39 | 0.46 |
| 13 | 0.1625 | | 0.10 | 0.46 | 0.27 | 0.20 | 0.18 | 0.60 | 0.21 | 0.36 | 0.42 | 0.54 |
| 38 | 0.9217 | | 0.03 | 0.22 | 0.19 | 0.18 | 0.15 | 0.41 | 0.17 | 0.36 | 0.19 | 0.33 |
| 43 | 0.9388 | | 0.02 | 0.23 | 0.16 | 0.20 | 0.16 | 0.40 | 0.16 | 0.34 | 0.25 | 0.29 |
| 33 | 0.8866 | Jul'85 | 0.08 | 0.35 | 0.25 | 0.17 | 0.16 | 0.53 | 0.19 | 0.40 | 0.35 | 0.47 |
| 35 | 0.9009 | | 0.01 | 0.31 | 0.26 | 0.16 | 0.13 | 0.55 | 0.19 | 0.43 | 0.33 | 0.42 |
| 37 | 0.9149 | | 0.04 | 0.29 | 0.22 | 0.14 | 0.12 | 0.48 | 0.18 | 0.39 | 0.30 | 0.44 |
| 41 | 0.9299 | | 0.03 | 0.27 | 0.22 | 0.14 | 0.11 | 0.47 | 0.16 | 0.36 | 0.31 | 0.39 |
| 45 | 0.9471 | | 0.05 | 0.24 | 0.18 | 0.15 | 0.13 | 0.48 | 0.16 | 0.40 | 0.28 | 0.37 |
| 48 | 0.9636 | | 0.03 | 0.23 | 0.15 | 0.15 | 0.11 | 0.45 | 0.10 | 0.33 | 0.24 | 0.30 |
| 53 | 0.9791 | | 0.02 | 0.21 | 0.16 | 0.16 | 0.13 | 0.46 | 0.16 | 0.37 | 0.24 | 0.30 |
| 20 | 0.6326 | Feb-Mar'86 | 0.12 | 0.51 | 0.44 | 0.27 | 0.20 | 0.57 | 0.12 | 0.48 | 0.51 | 0.62 |
| 22 | 0.6452 | | 0.11 | 0.46 | 0.39 | 0.24 | 0.19 | 0.61 | 0.17 | 0.47 | 0.45 | 0.65 |
| 23 | 0.6522 | | 0.17 | 0.46 | 0.41 | 0.23 | 0.21 | 0.67 | 0.23 | 0.43 | 0.50 | 0.62 |
| 24 | 0.6595 | | 0.14 | 0.43 | 0.39 | 0.24 | 0.18 | 0.56 | 0.13 | 0.39 | 0.47 | 0.59 |
| 25 | 0.6704 | | 0.09 | 0.45 | 0.38 | 0.26 | 0.20 | 0.59 | 0.17 | 0.45 | 0.45 | 0.60 |

TABLE D.4.1 - *continued*

| SWP Graph ID | Phase | Time Group Labels | Al II 1670 | Al III 1862 | Al III 1854 | C II 1334 | C II 1335 | Fe III 1926 | Si II 1533 | Si III 1294 | Si IV 1393 | Si IV 1402 |
|---|---|---|---|---|---|---|---|---|---|---|---|---|
| 26 | 0.6772 |  | 0.11 | 0.41 | 0.39 | 0.24 | 0.19 | 0.65 | 0.16 | 0.42 | 0.48 | 0.59 |
| 29 | 0.7869 |  | 0.13 | 0.37 | 0.31 | 0.21 | 0.17 | 0.48 | 0.20 | 0.41 | 0.38 | 0.46 |
| 30 | 0.8005 |  | 0.14 | 0.30 | 0.28 | 0.21 | 0.18 | 0.45 | 0.17 | 0.41 | 0.34 | 0.43 |
| 31 | 0.8075 |  | 0.14 | 0.30 | 0.26 | 0.22 | 0.17 | 0.50 | 0.18 | 0.41 | 0.37 | 0.41 |
| 3 | 0.0052 | Sep'89 | -0.02 | 0.41 | 0.28 | 0.19 | 0.16 | 0.56 | 0.08 | 0.35 | 0.43 | 0.54 |
| 5 | 0.0141 |  | 0.05 | 0.43 | 0.35 | 0.14 | 0.13 | 0.56 | 0.17 | 0.30 | 0.51 | 0.62 |
| 7 | 0.0204 |  | -0.05 | 0.30 | 0.24 | 0.18 | 0.14 | 0.55 | 0.07 | 0.39 | 0.39 | 0.50 |
| 8 | 0.0378 |  | -0.03 | 0.26 | 0.19 | 0.16 | 0.13 | 0.48 | 0.02 | 0.29 | 0.25 | 0.45 |
| 10 | 0.0855 |  | 0.08 | 0.33 | 0.24 | 0.13 | 0.13 | 0.52 | 0.13 | 0.37 | 0.35 | 0.45 |
| 11 | 0.1317 |  | 0.07 | 0.45 | 0.35 | 0.18 | 0.15 | 0.63 | 0.17 | 0.31 | 0.51 | 0.63 |
| 14 | 0.1889 |  | 0.08 | 0.50 | 0.44 | 0.20 | 0.17 | 0.66 | 0.10 | 0.36 | 0.62 | 0.67 |
| 15 | 0.3209 |  | 0.09 | 0.52 | 0.46 | 0.19 | 0.15 | 0.63 | 0.21 | 0.34 | 0.67 | 0.70 |
| 17 | 0.4790 |  | 0.09 | 0.52 | 0.44 | 0.18 | 0.15 | 0.63 | 0.11 | 0.39 | 0.65 | 0.72 |
| 19 | 0.5454 |  | 0.09 | 0.49 | 0.44 | 0.17 | 0.13 | 0.69 | 0.10 | 0.40 | 0.61 | 0.76 |
| 28 | 0.7784 |  | 0.13 | 0.52 | 0.42 | 0.22 | 0.18 | 0.57 | 0.12 | 0.44 | 0.66 | 0.72 |
| 32 | 0.8813 |  | 0.03 | 0.29 | 0.24 | 0.16 | 0.16 | 0.47 | 0.08 | 0.39 | 0.37 | 0.40 |
| 40 | 0.9247 |  | 0.08 | 0.32 | 0.25 | 0.14 | 0.12 | 0.50 | 0.18 | 0.37 | 0.34 | 0.39 |
| 44 | 0.9397 |  | 0.07 | 0.31 | 0.24 | 0.13 | 0.15 | 0.43 | 0.17 | 0.36 | 0.29 | 0.36 |
| 47 | 0.9573 |  | 0.04 | 0.30 | 0.22 | 0.14 | 0.13 | 0.46 | 0.14 | 0.31 | 0.28 | 0.35 |
| 51 | 0.9777 |  | 0.03 | 0.33 | 0.25 | 0.15 | 0.13 | 0.48 | 0.10 | 0.37 | 0.30 | 0.42 |
| 58 | 0.9947 |  | 0.06 | 0.39 | 0.31 | 0.13 | 0.12 | 0.61 | 0.15 | 0.37 | 0.39 | 0.49 |

TABLE D.4.2
LWP/LWR Residual Intensities

| LWPR Graph ID | Phase | Time Group Labels | Mg II 2790 | Mg II 2795 | Mg II 2797 | Mg II 2802 |
|---|---|---|---|---|---|---|
| 4  | 0.0049 | Sep-Nov'78 | 0.52 | 0.92 | 0.63 | 0.26 |
| 8  | 0.0801 |             | 0.53 | 0.23 | 0.62 | 0.26 |
| 20 | 0.6520 |             | 0.62 | 0.30 | 0.68 | 0.36 |
| 39 | 0.9683 |             | 0.52 | 0.16 | 0.63 | 0.18 |
| 1  | 0.0030 | Dec'78-Jan'79 | 0.60 | 0.23 | 0.73 | 0.29 |
| 14 | 0.3200 |             | 0.70 | 0.37 | 0.70 | 0.43 |
| 15 | 0.3280 |             | 0.59 | 0.40 | 0.71 | 0.43 |
| 17 | 0.4939 |             | 0.61 | 0.31 | 0.66 | 0.40 |
| 21 | 0.7685 |             | 0.62 | 0.34 | 0.73 | 0.45 |
| 43 | 0.9895 |             | 0.63 | 0.25 | 0.61 | 0.30 |
| 3  | 0.0034 | Nov'79 | 0.66 | 0.27 | 0.75 | 0.28 |
| 5  | 0.0165 |             | 0.57 | 0.15 | 0.56 | 0.24 |
| 44 | 0.9901 |             | 0.67 | 0.11 | 0.61 | 0.38 |
| 26 | 0.8915 | Nov'83 | 1.00 | 0.19 | 1.00 | 0.18 |
| 28 | 0.9082 |             | 1.00 | 0.18 | 0.64 | 0.18 |
| 30 | 0.9227 |             | 0.59 | 0.15 | 0.57 | 0.17 |
| 33 | 0.9367 |             | 0.61 | 0.13 | 0.59 | 0.14 |
| 36 | 0.9581 |             | 0.59 | 0.14 | 0.56 | 0.10 |
| 40 | 0.9721 |             | 0.56 | 0.15 | 0.56 | 0.13 |
| 11 | 0.1547 | Mar'84 | 0.73 | 0.27 | 0.67 | 0.25 |
| 25 | 0.8878 | Jul'85 | 0.77 | 0.19 | 0.64 | 0.17 |
| 27 | 0.9018 |             | 0.72 | 0.18 | 0.62 | 0.23 |
| 29 | 0.9166 |             | 0.65 | 0.15 | 0.59 | 0.13 |
| 32 | 0.9323 |             | 0.67 | 0.14 | 0.59 | 0.14 |
| 35 | 0.9488 |             | 0.68 | 0.12 | 0.57 | 0.12 |
| 38 | 0.9650 |             | 0.60 | 0.09 | 0.51 | 0.11 |
| 42 | 0.9801 |             | 0.59 | 0.12 | 0.55 | 0.11 |
| 19 | 0.6343 | Feb-Mar'86 | 0.74 | 0.58 | 0.65 | 0.39 |
| 23 | 0.7884 |             | 0.67 | 0.28 | 0.63 | 0.41 |
| 2  | 0.0034 | Sep'89 | 0.67 | 0.25 | 0.61 | 0.32 |
| 6  | 0.0220 |             | 0.65 | 0.22 | 0.53 | 0.30 |
| 7  | 0.0390 |             | 0.62 | 0.22 | 0.54 | 0.21 |
| 9  | 0.0867 |             | 0.69 | 0.18 | 0.55 | 0.21 |
| 10 | 0.1303 |             | 0.66 | 0.33 | 0.66 | 0.34 |
| 12 | 0.1903 |             | 0.70 | 0.39 | 0.72 | 0.45 |
| 13 | 0.3196 |             | 0.74 | 0.37 | 0.68 | 0.43 |
| 16 | 0.4778 |             | 0.69 | 0.36 | 0.66 | 0.44 |
| 18 | 0.5463 |             | 0.73 | 0.28 | 0.65 | 0.40 |
| 22 | 0.7798 |             | 0.71 | 0.33 | 0.66 | 0.42 |
| 24 | 0.8825 |             | 0.64 | 0.20 | 0.60 | 0.28 |
| 31 | 0.9259 |             | 0.66 | 0.16 | 0.56 | 0.26 |
| 34 | 0.9409 |             | 0.63 | 0.16 | 0.55 | 0.22 |
| 37 | 0.9590 |             | 0.62 | 0.15 | 0.57 | 0.20 |
| 41 | 0.9794 |             | 0.62 | 0.19 | 0.57 | 0.26 |

# APPENDIX D.5
Equivalent Widths

TABLE D.5.1— SWP EQUIVALENT WIDTHS

| SWP ID | Phase | Exposure ID | Time Group Labels | Al II 1670 | Al III 1862 | Al III 1854 | C II 1334 | C II 1335 | Fe II 1608 | Fe III 1926 | Si II 1533 | Si III 1294 | Si IV 1393 | Si IV 1402 |
|---|---|---|---|---|---|---|---|---|---|---|---|---|---|---|
| 2  | 0.0031 | SWP03262 | Sep — | 1.07 | 0.86 | 1.17 | 0.70 | 1.24 | 0.21 | 0.46 | 0.92 | 0.32 | 0.98 | 0.72 |
| 9  | 0.0816 | SWP03303 | Nov'78 | 0.85 | 0.95 | 1.03 | 0.83 | 1.16 | 0.15 | 0.53 | 0.72 | 0.38 | 0.77 | 0.71 |
| 21 | 0.6361 | SWP02643 |  | 0.78 | 0.43 | 0.39 | 0.64 | 0.93 | 0.17 | 0.33 | 0.54 | 0.37 | 0.21 | 0.18 |
| 49 | 0.9669 | SWP03298 |  | 1.21 | 1.09 | 1.07 | 0.76 | 1.25 | 0.25 | 0.81 | 0.93 | 0.38 | 0.94 | 0.88 |
| 52 | 0.9791 | SWP03259 |  | 1.04 | 0.99 | 1.12 | 0.76 | 1.15 | 0.20 | 0.59 | 0.84 | 0.41 | 1.16 | 0.84 |
| 55 | 0.9874 | SWP03260 |  | 0.96 | 0.94 | 1.08 | 0.80 | 1.16 | 0.22 | 0.68 | 0.83 | 0.35 | 1.16 | 0.55 |
| 59 | 0.9951 | SWP03261 |  | 0.99 | 0.91 | 1.00 | 0.69 | 1.18 | 0.22 | 0.49 | 0.90 | 0.34 | 0.86 | 0.47 |
| 1  | 0.0015 | SWP03772 | Dec'78 — | 0.91 | 0.47 | 0.56 | 0.69 | 0.96 | 0.18 | 0.34 | 0.74 | 0.29 | 0.54 | 0.28 |
| 16 | 0.3297 | SWP03752 | Jan'79 | 0.74 | 0.46 | 0.51 | 0.70 | 0.85 | 0.16 | 0.38 | 0.60 | 0.27 | 0.42 | 0.19 |
| 18 | 0.4927 | SWP03818 |  | 0.68 | 0.43 | 0.38 | 0.64 | 0.96 | 0.21 | 0.29 | 0.60 | 0.26 | 0.31 | 0.20 |
| 27 | 0.7674 | SWP03794 |  | 0.78 | 0.36 | 0.38 | 0.69 | 0.97 | 0.26 | 0.32 | 0.63 | 0.31 | 0.21 | 0.13 |
| 56 | 0.9880 | SWP03771 |  | 0.79 | 0.53 | 1.02 | 0.65 | 0.93 | 0.17 | 0.37 | 0.70 | 0.31 | 0.38 | 0.19 |
| 4  | 0.0053 | SWP07111 | Nov'79 | 0.82 | 0.61 | 0.65 | 0.73 | 0.81 | 0.16 | 0.45 | 0.84 | 0.32 | 0.58 | 0.41 |
| 6  | 0.0183 | SWP07112 |  | 1.00 | 0.72 | 0.74 | 0.76 | 0.92 | 0.19 | 0.45 | 0.88 | 0.32 | 0.68 | 0.48 |
| 57 | 0.9915 | SWP07110 |  | 0.91 | 0.64 | 0.53 | 0.77 | 0.79 | 0.27 | 0.58 | 0.88 | 0.36 | 0.69 | 0.41 |
| 34 | 0.8904 | SWP21452 | Nov'83 | 0.92 | 0.17 | 0.67 | 0.60 | 0.86 | 0.23 | 0.31 | 0.69 | 0.32 | 0.30 | … |
| 36 | 0.9070 | SWP21453 |  | 0.83 | 0.32 | 0.58 | 0.70 | 0.85 | 0.24 | 0.42 | 0.68 | 0.37 | 0.52 | 0.28 |
| 39 | 0.9218 | SWP21454 |  | 0.87 | 0.51 | 0.66 | 0.70 | 0.92 | 0.24 | 0.43 | 0.70 | 0.35 | 0.47 | 0.34 |
| 42 | 0.9358 | SWP21455 |  | 0.95 | 0.73 | 0.76 | 0.80 | 0.95 | 0.29 | 0.47 | 0.76 | 0.36 | 0.69 | 0.42 |
| 46 | 0.9571 | SWP21456 |  | 1.04 | 0.75 | 0.70 | 0.72 | 0.99 | 0.24 | 0.43 | 0.75 | 0.33 | 0.61 | 0.46 |
| 50 | 0.9712 | SWP21457 |  | 1.09 | 0.69 | 0.75 | 0.80 | 1.05 | 0.22 | 0.47 | 0.85 | 0.34 | 0.57 | 0.40 |
| 54 | 0.9797 | SWP21458 |  | 1.14 | 0.68 | 0.75 | 0.81 | 1.00 | 0.24 | 0.44 | 0.86 | 0.33 | 0.74 | 0.46 |
| 12 | 0.1538 | SWP22417 | Mar'84 | 0.83 | 0.44 | 0.62 | 0.79 | 0.90 | 0.20 | 0.37 | 0.68 | 0.28 | 0.47 | 0.31 |
| 13 | 0.1625 | SWP22418 |  | 0.78 | 0.41 | 0.63 | 0.82 | 0.96 | 0.26 | 0.35 | 0.64 | 0.33 | 0.44 | 0.30 |
| 38 | 0.9217 | SWP22411 |  | 1.05 | 1.13 | 1.33 | 0.80 | 1.24 | 0.27 | 0.63 | 0.83 | 0.37 | 0.99 | 0.68 |
| 43 | 0.9388 | SWP22439 |  | 1.07 | 1.28 | 1.48 | 0.86 | 1.37 | 0.31 | 0.67 | 0.95 | 0.38 | 1.47 | 1.40 |
| 33 | 0.8866 | SWP26472 | Jul'85 | 0.79 | 0.59 | 0.59 | 0.78 | 0.89 | 0.20 | 0.35 | 0.60 | 0.28 | 0.45 | 0.38 |
| 35 | 0.9009 | SWP26473 |  | 0.90 | 0.72 | 0.65 | 0.66 | 0.90 | 0.21 | 0.39 | 0.72 | 0.30 | 0.48 | 0.39 |
| 37 | 0.9149 | SWP26474 |  | 0.87 | 0.65 | 0.66 | 0.75 | 0.94 | 0.21 | 0.38 | 0.66 | 0.30 | 0.51 | 0.37 |
| 41 | 0.9299 | SWP26475 |  | 0.86 | 0.72 | 0.76 | 0.77 | 0.99 | 0.27 | 0.56 | 0.75 | 0.32 | 0.63 | 0.48 |
| 45 | 0.9471 | SWP26476 |  | 0.99 | 0.85 | 0.83 | 0.71 | 0.99 | 0.26 | 0.50 | 0.74 | 0.30 | 0.61 | 0.52 |
| 48 | 0.9636 | SWP26477 |  | 0.96 | 0.78 | 0.87 | 0.65 | 1.05 | 0.23 | 0.49 | 0.71 | 0.37 | 0.60 | 0.55 |
| 53 | 0.9791 | SWP26478 |  | 1.06 | 0.80 | 0.80 | 0.71 | 1.00 | 0.21 | 0.47 | 0.81 | 0.31 | 0.64 | 0.49 |

TABLE D.5.1— *(continued)*

| SWP ID | Phase | Exposure ID | Time Group Labels | Al II 1670 | Al III 1862 | Al III 1854 | C II 1334 | C II 1335 | Fe II 1608 | Fe III 1926 | Si II 1533 | Si III 1294 | Si IV 1393 | Si IV 1402 |
|---|---|---|---|---|---|---|---|---|---|---|---|---|---|---|
| 20 | 0.6326 | SWP27781 | Feb–Mar'86 | 0.83 | 0.64 | 0.65 | 0.72 | 0.85 | 0.28 | 0.33 | 0.65 | 0.29 | 0.52 | 0.49 |
| 22 | 0.6452 | SWP27782 | | 0.79 | 0.54 | 0.66 | 0.79 | 0.82 | 0.22 | 0.39 | 0.64 | 0.26 | 0.50 | 0.37 |
| 23 | 0.6522 | SWP27783 | | 0.75 | 0.57 | 0.73 | 0.66 | 1.01 | 0.20 | 0.28 | 0.63 | 0.31 | 0.57 | 0.41 |
| 24 | 0.6595 | SWP27784 | | 0.86 | 0.58 | 0.65 | 0.68 | 0.89 | 0.25 | 0.32 | 0.65 | 0.30 | 0.51 | 0.42 |
| 25 | 0.6704 | SWP27785 | | 0.85 | 0.61 | 0.70 | 0.54 | 1.00 | 0.35 | 0.40 | 0.69 | 0.28 | 0.54 | 0.40 |
| 26 | 0.6772 | SWP27786 | | 0.78 | 0.62 | 0.63 | 0.68 | 0.92 | 0.19 | 0.26 | 0.69 | 0.31 | 0.51 | 0.39 |
| 29 | 0.7869 | SWP27830 | | 0.76 | 0.75 | 0.88 | 0.88 | 0.99 | 0.15 | 0.52 | 0.65 | 0.31 | 0.61 | 0.49 |
| 30 | 0.8005 | SWP27831 | | 0.83 | 0.83 | 1.01 | 0.74 | 0.96 | 0.24 | 0.45 | 0.72 | 0.33 | 0.71 | 0.48 |
| 31 | 0.8075 | SWP27832 | | 0.83 | 0.73 | 1.03 | 0.87 | 1.00 | 0.18 | 0.46 | 0.67 | 0.32 | 0.64 | 0.57 |
| 3 | 0.0052 | SWP36983 | Sep'89 | 0.95 | 0.59 | 0.65 | 0.68 | 0.84 | 0.29 | 0.33 | 0.79 | 0.30 | 0.46 | 0.47 |
| 5 | 0.0141 | SWP37019 | | 0.83 | 0.50 | 0.51 | 0.77 | 0.88 | 0.23 | 0.33 | 0.80 | 0.30 | 0.43 | 0.38 |
| 7 | 0.0204 | SWP36984 | | 0.91 | 0.62 | 0.73 | 0.72 | 0.91 | 0.27 | 0.30 | 0.78 | 0.29 | 0.44 | 0.43 |
| 8 | 0.0378 | SWP36985 | | 0.96 | 0.66 | 0.69 | 0.73 | 0.93 | 0.31 | 0.38 | 0.80 | 0.36 | 0.43 | 0.35 |
| 10 | 0.0855 | SWP37021 | | 0.79 | 0.64 | 0.68 | 0.81 | 0.95 | 0.29 | 0.42 | 0.67 | 0.30 | 0.47 | 0.38 |
| 11 | 0.1317 | SWP36989 | | 0.75 | 0.45 | 0.53 | 0.82 | 0.85 | 0.20 | 0.34 | 0.67 | 0.35 | 0.31 | 0.33 |
| 14 | 0.1889 | SWP36992 | | 0.81 | 0.42 | 0.47 | 0.80 | 0.93 | 0.26 | 0.27 | 0.69 | 0.27 | 0.33 | 0.23 |
| 15 | 0.3209 | SWP36996 | | 0.78 | 0.43 | 0.46 | 0.77 | 0.89 | 0.27 | 0.28 | 0.55 | 0.37 | 0.16 | 0.23 |
| 17 | 0.4790 | SWP37001 | | 0.71 | 0.44 | 0.44 | 0.71 | 0.86 | 0.21 | 0.37 | 0.68 | 0.29 | 0.22 | 0.18 |
| 19 | 0.5454 | SWP37003 | | 0.71 | 0.46 | 0.44 | 0.64 | 0.96 | 0.34 | 0.35 | 0.61 | 0.29 | 0.26 | 0.22 |
| 28 | 0.7784 | SWP37011 | | 0.70 | 0.47 | 0.45 | 0.72 | 0.89 | 0.33 | 0.29 | 0.72 | 0.28 | 0.25 | 0.18 |
| 32 | 0.8813 | SWP37014 | | 0.88 | 0.83 | 0.93 | 0.75 | 1.07 | 0.30 | 0.57 | 0.79 | 0.35 | 0.61 | 0.73 |
| 40 | 0.9247 | SWP37016 | | 0.94 | 0.93 | 0.96 | 0.82 | 1.15 | 0.29 | 0.54 | 0.77 | 0.42 | 0.88 | 0.79 |
| 44 | 0.9397 | SWP36979 | | 1.02 | 0.88 | 0.92 | 0.79 | 1.13 | 0.25 | 0.55 | 0.84 | 0.33 | 1.04 | 0.63 |
| 47 | 0.9573 | SWP36980 | | 1.09 | 0.89 | 0.95 | 0.78 | 1.06 | 0.29 | 0.52 | 0.91 | 0.45 | 1.03 | 0.73 |
| 51 | 0.9777 | SWP36981 | | 1.09 | 0.86 | 0.94 | 0.72 | 1.12 | 0.27 | 0.47 | 0.78 | 0.36 | 0.91 | 0.74 |
| 58 | 0.9947 | SWP36982 | | 0.93 | 0.58 | 0.69 | 0.77 | 0.94 | 0.28 | 0.38 | 0.74 | 0.37 | 0.54 | 0.65 |

TABLE D.5.2– LWP/LWR EQUIVALENT WIDTHS

| LWP/LWR ID | Phase | Exposure ID | Time Group Labels | Fe II 2600 | Mg II 2791 | Mg II 2796 | Mg II 2798 | Mg II 2803 |
|---|---|---|---|---|---|---|---|---|
| 4 | 0.0049 | LWR02856 | Sep-Nov'78 | 0.63 | 0.59 | 1.18 | 0.28 | 1.06 |
| 8 | 0.0801 | LWR02911 | | 0.54 | 0.53 | 1.15 | 0.30 | 1.14 |
| 20 | 0.6520 | LWR02345 | | 0.49 | 0.44 | 0.80 | 0.21 | 0.62 |
| 39 | 0.9683 | LWR02906 | | 0.65 | 0.59 | 1.32 | 0.32 | 1.43 |
| 1 | 0.0030 | LWR03350 | Dec'78-Jan'79 | 0.68 | 0.53 | 0.99 | 0.24 | 0.97 |
| 14 | 0.3200 | LWR03328 | | 0.33 | 0.28 | 0.87 | 0.27 | 0.78 |
| 15 | 0.3280 | LWR03329 | | 0.54 | 0.57 | 0.87 | 0.22 | 0.86 |
| 17 | 0.4939 | LWR03398 | | 0.45 | 0.46 | 0.75 | 0.23 | 0.62 |
| 21 | 0.7685 | LWR03374 | | 0.53 | 0.40 | 0.86 | 0.21 | 0.67 |
| 43 | 0.9895 | LWR03349 | | 0.49 | 0.35 | 0.96 | 0.28 | 0.76 |
| 3 | 0.0034 | LWR06047 | Nov'79 | 0.46 | 0.37 | 1.07 | 0.25 | 1.02 |
| 5 | 0.0165 | LWR06048 | | 0.66 | 0.39 | 1.62 | 0.41 | 1.40 |
| 44 | 0.9901 | LWR06046 | | 0.29 | 0.59 | 0.93 | 0.19 | 0.79 |
| 26 | 0.8915 | LWP02226 | Nov'83 | … | … | 0.86 | … | 0.92 |
| 28 | 0.9082 | LWP02227 | | … | … | 0.96 | 0.25 | 0.92 |
| 30 | 0.9227 | LWP02228 | | 0.56 | 0.41 | 1.17 | 0.41 | 0.93 |
| 33 | 0.9367 | LWP02229 | | 0.65 | 0.41 | 1.40 | 0.40 | 1.24 |
| 36 | 0.9581 | LWP02230 | | 0.60 | 0.37 | 1.37 | 0.38 | 1.29 |
| 40 | 0.9721 | LWP02231 | | 0.58 | 0.45 | 1.45 | 0.37 | 1.27 |
| 11 | 0.1547 | LWP02895 | Mar'84 | 0.44 | 0.26 | 0.95 | 0.27 | 0.98 |
| 25 | 0.8878 | LWP06484 | Jul'85 | 0.45 | 0.23 | 1.09 | 0.33 | 0.83 |
| 27 | 0.9018 | LWP06485 | | 0.50 | 0.29 | 1.04 | 0.36 | 0.92 |
| 29 | 0.9166 | LWP06486 | | 0.43 | 0.30 | 1.13 | 0.31 | 0.94 |
| 32 | 0.9323 | LWP06487 | | 0.49 | 0.37 | 1.26 | 0.38 | 1.04 |
| 35 | 0.9488 | LWP06488 | | 0.54 | 0.34 | 1.31 | 0.43 | 1.14 |
| 38 | 0.9650 | LWP06489 | | 0.51 | 0.36 | 1.41 | 0.41 | 1.24 |
| 42 | 0.9801 | LWP06490 | | 0.57 | 0.36 | 1.60 | 0.39 | 1.48 |
| 19 | 0.6343 | LWP07717 | Feb-Mar'86 | 0.36 | 0.23 | 0.84 | 0.33 | 0.65 |
| 23 | 0.7884 | LWP07738 | | 0.40 | 0.28 | 1.01 | 0.30 | 0.74 |
| 2 | 0.0034 | LWP16323 | Sep'89 | 0.54 | 0.32 | 1.02 | 0.36 | 0.88 |
| 6 | 0.0220 | LWP16324 | | 0.56 | 0.30 | 1.24 | 0.36 | 1.05 |
| 7 | 0.0390 | LWP16325 | | 0.55 | 0.27 | 1.22 | 0.34 | 1.02 |
| 9 | 0.0867 | LWP16356 | | 0.59 | 0.30 | 1.18 | 0.38 | 1.00 |
| 10 | 0.1303 | LWP16327 | | 0.38 | 0.25 | 0.95 | 0.26 | 0.80 |
| 12 | 0.1903 | LWP16329 | | 0.44 | 0.29 | 0.85 | 0.30 | 0.69 |
| 13 | 0.3196 | LWP16333 | | 0.40 | 0.24 | 0.84 | 0.28 | 0.70 |
| 16 | 0.4778 | LWP16338 | | 0.38 | 0.33 | 0.81 | 0.36 | 0.65 |
| 18 | 0.5463 | LWP16340 | | 0.41 | 0.28 | 0.80 | 0.34 | 0.57 |
| 22 | 0.7798 | LWP16347 | | 0.37 | 0.27 | 0.86 | 0.27 | 0.56 |
| 24 | 0.8825 | LWP16350 | | 0.49 | 0.39 | 1.15 | 0.44 | 0.88 |
| 31 | 0.9259 | LWP16352 | | 0.53 | 0.41 | 1.33 | 0.51 | 0.93 |
| 34 | 0.9409 | LWP16320 | | 0.57 | 0.39 | 1.37 | 0.57 | 1.12 |
| 37 | 0.9590 | LWP16321 | | 0.48 | 0.36 | 1.35 | 0.43 | 1.21 |
| 41 | 0.9794 | LWP16322 | | 0.61 | 0.41 | 1.45 | 0.42 | 1.04 |

# APPENDIX D.6
## Difference Spectra Results and Ion Column Densities

TABLE D.6.1

| Line | $f_{ik}$ | Lab Wavelength |
|---|---|---|
| Mg II 2796 | 6.08E-01 | 2796.352 |
| Mg II 2803 | 3.03E-01 | 2803.531 |
| | | |
| | | |
| | | |

TABLE D.6.2 - Mg II 2796

| Pair ID | far Left A | FWHM | Depth | Eq. Width | Ni, atoms along l.o.s | Vrad | Left Width (km/s) | Right Width (km/s) | Inner Left B | FWHM | Depth | Eq. Width | Ni, atoms along l.o.s | Vrad | Left Width (km/s) | Right Width (km/s) |
|---|---|---|---|---|---|---|---|---|---|---|---|---|---|---|---|---|
| 1 | 2794.72 | 0.40 | 0.250 | 0.21 | 4.90E+12 | -175.0 | 61.6 | 21.4 | 2795.45 | 0.70 | 0.365 | 0.27 | 6.46E+12 | -96.7 | 37.5 | 37.5 |
| 2 | | | | | | | | | 2795.43 | 0.65 | 0.199 | 0.14 | 3.27E+12 | -98.8 | 34.8 | 34.8 |
| 3 | 2794.82 | 0.59 | 0.355 | 0.39 | 9.29E+12 | -164.2 | 79.3 | 31.6 | 2795.67 | 0.78 | 0.400 | 0.33 | 7.89E+12 | -73.1 | 41.8 | 41.8 |
| 4 | 2794.53 | 0.60 | 0.360 | 0.44 | 1.06E+13 | -195.3 | 92.2 | 32.2 | 2795.44 | 1.00 | 0.460 | 0.49 | 1.16E+13 | -97.8 | 53.6 | 53.6 |
| 5 | 2794.92 | 0.25 | 0.400 | 0.24 | 5.72E+12 | -153.5 | 47.2 | 13.4 | 2795.48 | 0.85 | 0.575 | 0.52 | 1.24E+13 | -93.5 | 45.6 | 45.6 |
| 6 | | | | | | | | | 2795.64 | 0.70 | 0.285 | 0.21 | 5.05E+12 | -76.3 | 37.5 | 37.5 |
| 7 | | | | | | | | | 2795.78 | 0.70 | 0.230 | 0.17 | 4.07E+12 | -61.3 | 37.5 | 37.5 |
| 8 | | | | | | | | | 2795.77 | 0.75 | 0.260 | 0.27 | 6.41E+12 | -62.4 | 64.3 | 40.2 |
| 9 | | | | | | | | | 2796.00 | 0.60 | 0.115 | 0.07 | 1.75E+12 | -37.7 | 32.2 | 32.2 |
| 10 | | | | | | | | | 2796.35 | 0.00 | 0.00 | 0.00 | 0.00E+00 | 0.0 | 0.0 | 0.0 |
| 11 | | | | | | | | | 2795.62 | 0.80 | 0.162 | 0.14 | 3.28E+12 | -78.5 | 42.9 | 42.9 |
| 12 | | | | | | | | | 2796.35 | 0.00 | 0.00 | 0.00 | 0.00E+00 | 0.0 | 0.0 | 0.0 |
| 13 | | | | | | | | | 2796.35 | 0.00 | 0.00 | 0.00 | 0.00E+00 | 0.0 | 0.0 | 0.0 |
| 14 | | | | | | | | | 2795.58 | 0.60 | 0.180 | 0.11 | 2.73E+12 | -82.8 | 32.2 | 32.2 |
| 15 | | | | | | | | | 2795.44 | 0.70 | 0.230 | 0.17 | 4.07E+12 | -97.8 | 37.5 | 37.5 |
| 16 | | | | | | | | | 2796.35 | 0.00 | 0.00 | 0.00 | 0.00E+00 | 0.0 | 0.0 | 0.0 |
| 17 | | | | | | | | | 2796.35 | 0.00 | 0.00 | 0.00 | 0.00E+00 | 0.0 | 0.0 | 0.0 |
| 18 | | | | | | | | | 2796.35 | 0.00 | 0.00 | 0.00 | 0.00E+00 | 0.0 | 0.0 | 0.0 |
| 19 | | | | | | | | | 2796.35 | 0.00 | 0.00 | 0.00 | 0.00E+00 | 0.0 | 0.0 | 0.0 |
| 20 | | | | | | | | | 2795.60 | 0.79 | 0.172 | 0.14 | 3.44E+12 | -80.6 | 42.3 | 42.3 |
| 21 | | | | | | | | | 2795.73 | 0.59 | 0.108 | 0.07 | 1.61E+12 | -66.7 | 31.6 | 31.6 |
| 22 | | | | | | | | | 2796.35 | 0.00 | 0.00 | 0.00 | 0.00E+00 | 0.0 | 0.0 | 0.0 |
| 23 | | | | | | | | | 2796.35 | 0.00 | 0.00 | 0.00 | 0.00E+00 | 0.0 | 0.0 | 0.0 |
| 24 | | | | | | | | | 2795.49 | 0.5 | 0.110 | 0.06 | 1.39E+12 | -92.4 | 26.8 | 26.8 |
| 25 | | | | | | | | | 2796.10 | 0.55 | 0.060 | 0.04 | 8.35E+11 | -27.0 | 29.5 | 29.5 |
| 26 | | | | | | | | | | | | | | | | |
| 27 | | | | | | | | | 2796.06 | 0.64 | 0.098 | 0.07 | 1.59E+12 | -31.3 | 34.3 | 34.3 |
| 28 | | | | | | | | | | | | | | | | |
| 29 | | | | | | | | | 2795.58 | 0.49 | 0.105 | 0.05 | 1.30E+12 | -82.8 | 26.3 | 26.3 |
| 30 | | | | | | | | | 2795.42 | 0.7 | 0.162 | 0.12 | 2.87E+12 | -99.9 | 37.5 | 37.5 |
| 31 | | | | | | | | | 2795.6 | 0.63 | 0.210 | 0.14 | 3.35E+12 | -80.6 | 33.8 | 33.8 |
| 32 | | | | | | | | | 2795.46 | 0.9 | 0.215 | 0.21 | 4.89E+12 | -95.6 | 48.2 | 48.2 |
| 33 | | | | | | | | | 2795.58 | 0.9 | 0.200 | 0.19 | 4.55E+12 | -82.8 | 48.2 | 48.2 |
| 34 | | | | | | | | | 2795.38 | 0.93 | 0.230 | 0.23 | 5.41E+12 | -104.2 | 49.9 | 49.9 |
| 35 | | | | | | | | | 2795.5 | 0.83 | 0.230 | 0.20 | 4.83E+12 | -91.3 | 44.5 | 44.5 |
| 36 | | | | | | | | | 2795.32 | 0.73 | 0.145 | 0.11 | 2.68E+12 | -110.6 | 39.1 | 39.1 |
| 37 | | | | | | | | | 2795.38 | 1 | 0.240 | 0.26 | 6.07E+12 | -104.2 | 53.6 | 53.6 |
| 38 | | | | | | | | | 2795.42 | 0.95 | 0.251 | 0.25 | 6.03E+12 | -99.9 | 50.9 | 50.9 |
| 39 | | | | | | | | | 2795.6 | 0.65 | 0.260 | 0.29 | 7.00E+12 | -80.6 | 79.3 | 34.8 |
| 40 | | | | | | | | | 2795.3 | 1.1 | 0.265 | 0.31 | 7.37E+12 | -112.8 | 59.0 | 59.0 |
| 41 | | | | | | | | | 2795.3 | 1.08 | 0.242 | 0.28 | 6.61E+12 | -112.8 | 57.9 | 57.9 |
| 42 | | | | | | | | | 2795.5 | 1 | 0.310 | 0.33 | 7.84E+12 | -91.3 | 53.6 | 53.6 |
| 43 | | | | | | | | | 2795.11 | 1.3 | 0.202 | 0.28 | 6.64E+12 | -133.2 | 69.7 | 69.7 |
| 44 | | | | | | | | | 2795.48 | 1.15 | 0.330 | 0.40 | 9.60E+12 | -93.5 | 61.6 | 61.6 |

TABLE D.6.3 - Mg II 2796

| Pair ID | Extra Width L or R | far Left A | FWHM | Depth | Eq. Width | Ni, atoms along l.o.s | Vrad w | Extra Width L or R | Inner Left B | FWHM | Depth | Eq. Width | Ni, atoms along l.o.s | Vrad w |
|---|---|---|---|---|---|---|---|---|---|---|---|---|---|---|
| 1 | L | 2794.15 | 1.15 | 0.250 | 0.31 | 7.27E+12 | -236.6 | n | | | | | | |
| 2 | n | | | | | | | n | | | | | | |
| 3 | L | 2794.08 | 1.48 | 0.355 | 0.56 | 1.33E+13 | -243.6 | n | | | | | | |
| 4 | L | 2793.67 | 1.72 | 0.360 | 0.66 | 1.57E+13 | -287.5 | n | | | | | | |
| 5 | L | 2794.48 | 0.88 | 0.400 | 0.37 | 8.90E+12 | -200.7 | n | | | | | | |
| 6 | n | | | | | | | n | | | | | | |
| 7 | n | | | | | | | n | | | | | | |
| 8 | n | | | | | | | L | 2795.17 | 1.2 | 0.260 | 0.33 | 7.89E+12 | -126.7 |
| 9 | n | | | | | | | n | | | | | | |
| 10 | n | | | | | | | n | | | | | | |
| 11 | n | | | | | | | n | | | | | | |
| 12 | n | | | | | | | n | | | | | | |
| 13 | n | | | | | | | n | | | | | | |
| 14 | n | | | | | | | n | | | | | | |
| 15 | n | | | | | | | n | | | | | | |
| 16 | n | | | | | | | n | | | | | | |
| 17 | n | | | | | | | n | | | | | | |
| 18 | n | | | | | | | n | | | | | | |
| 19 | n | | | | | | | n | | | | | | |
| 20 | n | | | | | | | n | | | | | | |
| 21 | n | | | | | | | n | | | | | | |
| 22 | n | | | | | | | n | | | | | | |
| 23 | n | | | | | | | n | | | | | | |
| 24 | n | | | | | | | n | | | | | | |
| 25 | n | | | | | | | n | | | | | | |
| 26 | n | | | | | | | n | | | | | | |
| 27 | n | | | | | | | n | | | | | | |
| 28 | n | | | | | | | n | | | | | | |
| 29 | n | | | | | | | n | | | | | | |
| 30 | n | | | | | | | n | | | | | | |
| 31 | n | | | | | | | n | | | | | | |
| 32 | n | | | | | | | n | | | | | | |
| 33 | n | | | | | | | n | | | | | | |
| 34 | n | | | | | | | n | | | | | | |
| 35 | n | | | | | | | n | | | | | | |
| 36 | n | | | | | | | n | | | | | | |
| 37 | n | | | | | | | n | | | | | | |
| 38 | n | | | | | | | n | | | | | | |
| 39 | n | | | | | | | L | 2794.86 | 1.48 | 0.260 | 0.41 | 9.73E+12 | -159.954 |
| 40 | n | | | | | | | n | | | | | | |
| 41 | n | | | | | | | n | | | | | | |
| 42 | n | | | | | | | n | | | | | | |
| 43 | n | | | | | | | n | | | | | | |
| 44 | n | | | | | | | n | | | | | | |

TABLE D.6.4 - Mg II 2796

| Pair ID | Inner Right C | FWHM | Depth | Eq. Width | Ni, atoms along l.o.s | Vrad | Left Width (km/s) | Right Width (km/s) | Far Right D | FWHM | Depth | Eq. Width | Ni, atoms along l.o.s | Vrad | Left Width (km/s) | Right Width (km/s) |
|---|---|---|---|---|---|---|---|---|---|---|---|---|---|---|---|---|
| 1 | 2797.02 | 0.38 | 0.152 | 0.06 | 1.46E+12 | 71.6 | 20.4 | 20.4 | | | | | | | | |
| 2 | 2796.85 | 0.58 | 0.17 | 0.10 | 2.42E+12 | 53.4 | 31.1 | 31.1 | | | | | | | | |
| 3 | 2797.50 | 1.00 | 0.19 | 0.20 | 4.81E+12 | 123.1 | 53.6 | 53.6 | | | | | | | | |
| 4 | 2797.23 | 1.20 | 0.270 | 0.45 | 1.06E+13 | 94.1 | 64.3 | 101.8 | | | | | | | | |
| 5 | 2797.02 | 0.84 | 0.13 | 0.15 | 3.60E+12 | 71.6 | 45.0 | 72.4 | | | | | | | | |
| 6 | 2796.80 | 0.64 | 0.19 | 0.13 | 3.04E+12 | 48.0 | 34.3 | 34.3 | | | | | | | | |
| 7 | 2796.82 | 0.56 | 0.21 | 0.13 | 2.97E+12 | 50.2 | 30.0 | 30.0 | | | | | | | | |
| 8 | 2796.80 | 0.60 | 0.198 | 0.19 | 4.48E+12 | 48.0 | 32.2 | 63.8 | | | | | | | | |
| 9 | 2796.86 | 0.62 | 0.270 | 0.18 | 4.23E+12 | 54.5 | 33.2 | 33.2 | | | | | | | | |
| 10 | 2796.90 | 0.41 | 0.230 | 0.10 | 2.38E+12 | 58.7 | 22.0 | 22.0 | | | | | | | | |
| 11 | 2796.99 | 0.44 | 0.155 | 0.07 | 1.72E+12 | 68.4 | 23.6 | 23.6 | | | | | | | | |
| 12 | 2796.94 | 0.37 | 0.17 | 0.07 | 1.61E+12 | 63.0 | 19.8 | 19.8 | | | | | | | | |
| 13 | 2796.91 | 0.30 | 0.180 | 0.06 | 1.37E+12 | 59.8 | 16.1 | 16.1 | | | | | | | | |
| 14 | 2796.74 | 0.27 | 0.110 | 0.03 | 7.51E+11 | 41.6 | 14.5 | 14.5 | | | | | | | | |
| 15 | 2796.71 | 0.28 | 0.080 | 0.02 | 5.67E+11 | 38.4 | 15.0 | 15.0 | | | | | | | | |
| 16 | 2796.35 | 0.00 | 0.00 | 0.00 | 0.00E+00 | 0.0 | 0.0 | 0.0 | | | | | | | | |
| 17 | 2796.35 | 0.00 | 0.00 | 0.00 | 0.00E+00 | 0.0 | 0.0 | 0.0 | | | | | | | | |
| 18 | 2796.35 | 0.00 | 0.00 | 0.00 | 0.00E+00 | 0.0 | 0.0 | 0.0 | | | | | | | | |
| 19 | 2796.35 | 0.00 | 0.00 | 0.00 | 0.00E+00 | 0.0 | 0.0 | 0.0 | | | | | | | | |
| 20 | 2797.28 | 0.62 | 0.108 | 0.07 | 1.69E+12 | 99.5 | 33.2 | 33.2 | | | | | | | | |
| 21 | 2796.35 | 0.00 | 0.00 | 0.00 | 0.00E+00 | 0.0 | 0.0 | 0.0 | | | | | | | | |
| 22 | 2796.35 | 0.00 | 0.00 | 0.00 | 0.00E+00 | 0.0 | 0.0 | 0.0 | | | | | | | | |
| 23 | 2797.08 | 0.7 | 0.185 | 0.14 | 3.28E+12 | 78.0 | 37.5 | 37.5 | | | | | | | | |
| 24 | 2796.9 | 0.65 | 0.298 | 0.21 | 4.90E+12 | 58.7 | 34.8 | 34.8 | | | | | | | | |
| 25 | 2796.87 | 0.60 | 0.215 | 0.14 | 3.26E+12 | 55.5 | 32.2 | 32.2 | | | | | | | | |
| 26 | | | | | | | | | | | | | | | | |
| 27 | 2796.84 | 0.83 | 0.148 | 0.13 | 3.11E+12 | 52.3 | 44.5 | 44.5 | | | | | | | | |
| 28 | | | | | | | | | | | | | | | | |
| 29 | 2796.8 | 0.73 | 0.298 | 0.23 | 5.50E+12 | 48.0 | 39.1 | 39.1 | | | | | | | | |
| 30 | 2796.83 | 0.9 | 0.318 | 0.30 | 7.24E+12 | 51.2 | 48.2 | 48.2 | | | | | | | | |
| 31 | 2796.99 | 0.93 | 0.295 | 0.29 | 6.94E+12 | 68.4 | 49.9 | 49.9 | | | | | | | | |
| 32 | 2796.89 | 0.82 | 0.338 | 0.30 | 7.01E+12 | 57.7 | 44.0 | 44.0 | | | | | | | | |
| 33 | 2796.92 | 0.98 | 0.35 | 0.37 | 8.67E+12 | 60.9 | 52.5 | 52.5 | | | | | | | | |
| 34 | 2797.02 | 1.02 | 0.37 | 0.40 | 9.54E+12 | 71.6 | 54.7 | 54.7 | | | | | | | | |
| 35 | 2796.83 | 0.95 | 0.28 | 0.28 | 6.73E+12 | 51.2 | 50.9 | 50.9 | | | | | | | | |
| 36 | 2797.02 | 0.93 | 0.412 | 0.41 | 9.69E+12 | 71.6 | 49.9 | 49.9 | | | | | | | | |
| 37 | 2797.1 | 1.03 | 0.385 | 0.42 | 1.00E+13 | 80.2 | 55.2 | 55.2 | | | | | | | | |
| 38 | 2796.98 | 1.05 | 0.375 | 0.42 | 9.96E+12 | 67.3 | 56.3 | 56.3 | | | | | | | | |
| 39 | 2797 | 0.84 | 0.37 | 0.50 | 1.19E+13 | 69.5 | 45.0 | 91.1 | | | | | | | | |
| 40 | 2797.1 | 1.1 | 0.408 | 0.48 | 1.14E+13 | 80.2 | 59.0 | 59.0 | | | | | | | | |
| 41 | 2797.2 | 1.2 | 0.385 | 0.49 | 1.17E+13 | 90.9 | 64.3 | 64.3 | | | | | | | | |
| 42 | 2796.93 | 0.93 | 0.388 | 0.38 | 9.13E+12 | 62.0 | 49.9 | 49.9 | | | | | | | | |
| 43 | 2796.98 | 0.52 | 0.15 | 0.08 | 1.97E+12 | 67.3 | 27.9 | 27.9 | | | | | | | | |
| 44 | 2797.5 | 1.23 | 0.385 | 0.50 | 1.20E+13 | 123.1 | 65.9 | 65.9 | | | | | | | | |

# TABLE D.6.5 - Mg II 2796

| Pair ID | Extra Width L or R | Inner Right C | FWHM | Depth | Eq. Width | Ni, atoms along l.o.s | Vrad w | Extra Width L or R | Far Right D | FWHM | Depth | Eq. Width | Ni, atoms along l.o.s | Vrad w |
|---|---|---|---|---|---|---|---|---|---|---|---|---|---|---|
| 1 | n | | | | | | | n | | | | | | |
| 2 | n | | | | | | | n | | | | | | |
| 3 | n | | | | | | | n | | | | | | |
| 4 | R | 2798.18 | 1.9 | 0.270 | 0.55 | 1.30E+13 | 196.0 | n | | | | | | |
| 5 | R | 2797.70 | 1.35 | 0.130 | 0.19 | 4.44E+12 | 144.0 | n | | | | | | |
| 6 | n | | | | | | | n | | | | | | |
| 7 | n | | | | | | | n | | | | | | |
| 8 | R | 2797.40 | 1.19 | 0.198 | 0.25 | 5.96E+12 | 111.8 | n | | | | | | |
| 9 | n | | | | | | | n | | | | | | |
| 10 | n | | | | | | | n | | | | | | |
| 11 | n | | | | | | | n | | | | | | |
| 12 | n | | | | | | | n | | | | | | |
| 13 | n | | | | | | | n | | | | | | |
| 14 | n | | | | | | | n | | | | | | |
| 15 | n | | | | | | | n | | | | | | |
| 16 | n | | | | | | | n | | | | | | |
| 17 | n | | | | | | | n | | | | | | |
| 18 | n | | | | | | | n | | | | | | |
| 19 | n | | | | | | | n | | | | | | |
| 20 | n | | | | | | | n | | | | | | |
| 21 | n | | | | | | | n | | | | | | |
| 22 | n | | | | | | | n | | | | | | |
| 23 | n | | | | | | | n | | | | | | |
| 24 | n | | | | | | | n | | | | | | |
| 25 | n | | | | | | | n | | | | | | |
| 26 | n | | | | | | | n | | | | | | |
| 27 | n | | | | | | | n | | | | | | |
| 28 | n | | | | | | | n | | | | | | |
| 29 | n | | | | | | | n | | | | | | |
| 30 | n | | | | | | | n | | | | | | |
| 31 | n | | | | | | | n | | | | | | |
| 32 | n | | | | | | | n | | | | | | |
| 33 | n | | | | | | | n | | | | | | |
| 34 | n | | | | | | | n | | | | | | |
| 35 | n | | | | | | | n | | | | | | |
| 36 | n | | | | | | | n | | | | | | |
| 37 | n | | | | | | | n | | | | | | |
| 38 | n | | | | | | | n | | | | | | |
| 39 | R | 2797.85 | 1.7 | 0.370 | 0.67 | 1.59E+13 | 160.6 | n | | | | | | |
| 40 | n | | | | | | | n | | | | | | |
| 41 | n | | | | | | | n | | | | | | |
| 42 | n | | | | | | | n | | | | | | |
| 43 | n | | | | | | | n | | | | | | |
| 44 | n | | | | | | | n | | | | | | |

# TABLE D.6.6 - Mg II 2796

| Pair ID | Ni Total Blue | Ni Total Red |
|---|---|---|
| 1 | 1.14E+13 | 1.46E+12 |
| 2 | 3.27E+12 | 2.42E+12 |
| 3 | 1.72E+13 | 4.81E+12 |
| 4 | 2.22E+13 | 1.06E+13 |
| 5 | 1.81E+13 | 3.60E+12 |
| 6 | 5.05E+12 | 3.04E+12 |
| 7 | 4.07E+12 | 2.97E+12 |
| 8 | 6.41E+12 | 4.48E+12 |
| 9 | 1.75E+12 | 4.23E+12 |
| 10 | 0.00E+00 | 2.38E+12 |
| 11 | 3.28E+12 | 1.72E+12 |
| 12 | 0.00E+00 | 1.61E+12 |
| 13 | 0.00E+00 | 1.37E+12 |
| 14 | 2.73E+12 | 7.51E+11 |
| 15 | 4.07E+12 | 5.67E+11 |
| 16 | 0.00E+00 | 0.00E+00 |
| 17 | 0.00E+00 | 0.00E+00 |
| 18 | 0.00E+00 | 0.00E+00 |
| 19 | 0.00E+00 | 0.00E+00 |
| 20 | 3.44E+12 | 1.69E+12 |
| 21 | 1.61E+12 | 0.00E+00 |
| 22 | 0.00E+00 | 0.00E+00 |
| 23 | 0.00E+00 | 3.28E+12 |
| 24 | 1.39E+12 | 4.90E+12 |
| 25 | 8.35E+11 | 3.26E+12 |
| 26 | | |
| 27 | 1.59E+12 | 3.11E+12 |
| 28 | | |
| 29 | 1.30E+12 | 5.50E+12 |
| 30 | 2.87E+12 | 7.24E+12 |
| 31 | 3.35E+12 | 6.94E+12 |
| 32 | 4.89E+12 | 7.01E+12 |
| 33 | 4.55E+12 | 8.67E+12 |
| 34 | 5.41E+12 | 9.54E+12 |
| 35 | 4.83E+12 | 6.73E+12 |
| 36 | 2.68E+12 | 9.69E+12 |
| 37 | 6.07E+12 | 1.00E+13 |
| 38 | 6.03E+12 | 9.96E+12 |
| 39 | 7.00E+12 | 1.19E+13 |
| 40 | 7.37E+12 | 1.14E+13 |
| 41 | 6.61E+12 | 1.17E+13 |
| 42 | 7.84E+12 | 9.13E+12 |
| 43 | 6.64E+12 | 1.97E+12 |
| 44 | 9.60E+12 | 1.20E+13 |